Блецкан Д.І.

# Кристалічні та склоподібні телуриди кремнію

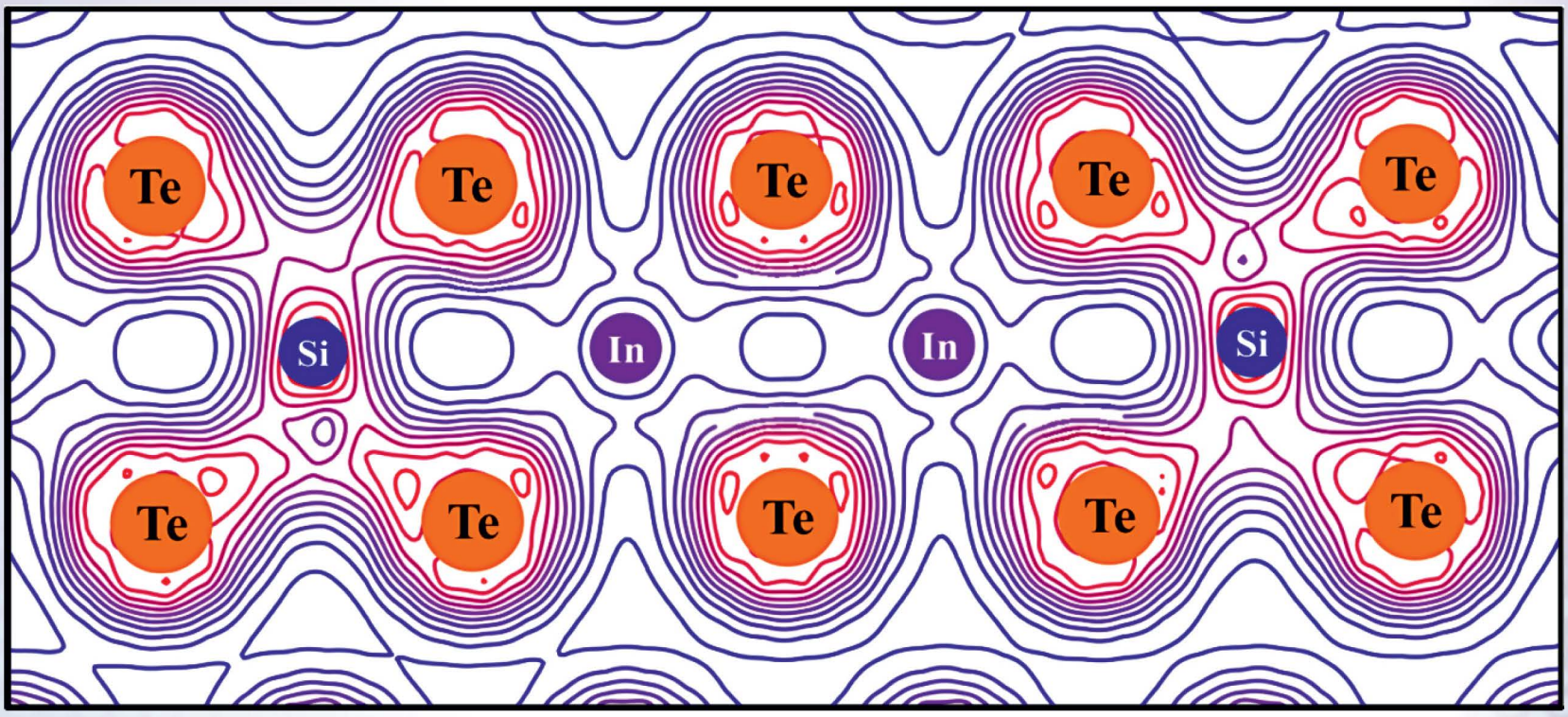

Міністерство освіти і науки України
Ужгородський національний університет

**Д. І. Блецкан**

# КРИСТАЛІЧНІ ТА СКЛОПОДІБНІ ТЕЛУРИДИ КРЕМНІЮ

## Синтез, властивості, використання

У даній монографії приведені результати експериментальних і теоретичних досліджень бінарних й потрійних кристалічних та склоподібних телуридів кремнію. Докладно описані методи синтезу й вирощування обємних і наноструктурованих бінарних кристалів $Si_2Te_3$, $SiTe_2$ та потрійних кристалів, відомих у системах M–Si–Te (M = Na, K, Cu, Ag, Al, In), а також натрій-кремнієвих і телур-кремнієвих клатратів. Значна увага приділена результатам дослідження їх електронної структури, оптичних, електричних, фотоелектричних і фотолюмінесцентних властивостей.

Видання розраховано для наукових співробітників та фахівців у галузі напівпровідникового матеріалознавства, фізики і техніки напівпровідників, викладачів, аспірантів і студентів відповідних спеціальностей.

# ЗМІСТ

## ПЕРЕЛІК УМОВНИХ ПОЗНАЧЕНЬ

| | | |
|---|---|---|
| **ХТР** | – | метод хімічно транспортних реакцій |
| **РФЕС** | – | рентгенівська фотоелектронна спектроскопія |
| **РЕС** | – | рентгенівська емісійна спектроскопія |
| **ПРК** | – | пара-рідина-кристал |
| **ННЗ** | – | нерівноважні носії заряду |
| **НК** | – | ниткоподібні кристали |
| **НВ** | – | нановіскери |
| **ЗБ** | – | зона Брилюена |
| **ДТА** | – | диференціально термічний аналіз |
| **РФА** | – | рентгенофазовий аналіз |
| **ФЛ** | – | фотолюмінесценція |
| **ФП** | – | фотопровідність |
| **ТАФ** | – | термічна активація фотопровідності |
| **ТСС** | – | термостимульований струм |
| **ВАХ** | – | вольт-амперні характеристики |
| **ДСК** | – | диференціальна скануюча калометрія |
| **ХСН** | – | халькогенідні склоподібні напівпровідники |
| **ADSC** | – | модульована диференціальна скануюча калометрія |
| **ПГ** | – | просторова група |
| **DFT** | – | density-functional theory |
| **LDA** | – | local density approximations |
| **GGA** | – | generalizes gradient approximations |
| **HSE06** | – | Heyd-Scuseria-Ernzerhaf hibrid functional |
| **UPS** | – | ultraviolet photoelectron spectra |
| **XPS** | – | X-ray photoelectron spectra |
| **XES** | – | X-ray emission spectra |
| **XAS** | – | X-ray absorption spectroscopy |
| **CVD** | – | chemical vapor deposition |
| **SEM** | – | скануючий електронний мікроскоп |
| **XANES** | – | X-ray absorption near edge structure |

# ПЕРЕДМОВА

В останні роки інтерес до бінарних і потрійних кристалічних та склоподібних телуридів кремнію неухильно зростає у зв'язку із все більш чітко виявленими широкими можливостями їх практичного використання у різних приладах і пристроях сучасної електроніки, оптоелектроніки й акустооптики.

У монографії узагальнено експериментальний матеріал по діаграмам стану і областям склоутворення у бінарній Si–Te та потрійних системах M–Si–Te (M = Na, K, Cu, Ag, Al, In). Важливою особливістю системи Si–Te є наявність у ній крім двох бінарних сполук $Si_2Te_3$ і $SiTe_2$, ще й двох типів клатратних сполук $Te_{16}Si_{38}$ і $Te_{7+x}Si_{20-x}$ ($x \sim 2.5$) синтезованих при високих тисках 3–7 ГПа і високих температурах 973–1123 K. Інтерес до клатратів зумовлений унікальними властивостями, які були виявлені в цих матеріалах включаючи низьку теплопровідність і надпровідність. Ці властивості є наслідком структури і хімічного зв'язку в клатратах.

Описані сучасні методи синтезу даних сполук і вирощування монокристалів різного габітусу (пластинчастих, ниткоподібних, об'ємних), а також клатратів. Приведені дані про кристалічну структуру поліморфних фаз бінарних сполук $Si_2Te_3$ і $SiTe_2$ та потрійних сполук, відомих у системах M–Si–Te. Шаруваті кристали $Si_2Te_3$ володіють унікальними властивостями, які не притаманні іншим матеріалам, а саме обертання димерів кремнію (Si–Si), що дає ще одну свободу для налаштування електронних властивостей даного матеріалу. Зважаючи на гігроскопічність кристалів $Si_2Te_3$, особлива увага приділяється аналізу стану їх поверхні, оскільки стан останньої сильно впливає на фізичні властивості.

Для всіх розглядуваних бінарних і потрійних кристалів телуридів кремнію, а також натрій-кремнієвих і телур-кремнієвих клатратів, приведені оригінальні результати першопринципних розрахунків електронних зонних структур, повних та парціальних густин електронних станів, а також карт розподілу густин заряду валентних електронів, що дало можливість визначити ширини заборонених зон та природу хімічного зв'язку.

Вперше узагальнені результати експериментальних досліджень спектрів крайового поглинання, фотопровідності й фотолюмінесценції шаруватих кристалів $Si_2Te_3$ і $SiTe_2$ в широкому інтервалі температур.

# РОЗДІЛ 1

# ДІАГРАМА СТАНУ БІНАРНОЇ СИСТЕМИ Si–Te. СИНТЕЗ ОБ'ЄМНИХ ТА НАНОСТРУКТУРОВАНИХ КРИСТАЛІВ $Si_2Te_3$ І $SiTe_2$ ТА ЇХ СТРУКТУРА

## 1.1. ФАЗОВІ РІВНОВАГИ В СИСТЕМІ Si–Te

Вибір складу сполук, методів та оптимальних умов їх синтезу, вирощування кристалів, отримання склоподібних речовин і композитів неможливий без знання діаграм стану відповідних систем. Особливо велика їхня роль при розробці методів вирощування кристалів, оскільки в цьому випадку потрібно не тільки потрапити в область кристалізації необхідної фази, але й підібрати умови, які забезпечують відсутність паразитного зародкоутворення, сталість пересичення в часі і т. д. Крім того, діаграма стану дає повну інформацію щодо отримання наноматеріалів, розвитку нанотехнологій, оскільки розглядається один із важливих фізико-хімічних процесів – кристалізація [1, 2].

У даному параграфі узагальнено дані стосовно стабільності фазових рівноваг у системі Si–Te. Незважаючи на численні вивчення фазової діаграми Si–Te [3–7], існують певні розбіжності щодо кількості стабільних фаз (числа сполук) та їх складу, а також по границям області гомогенності фази $Si_2Te_3$ та складу і температури плавлення евтектики. Вперше діаграму стану системи Si–Te в області концентрацій 35–100 ат. % Te за результатами диференціального термічного аналізу (ДТА) та металографічного аналізу побудував автор [3] (рис. 1.1, *а*). Встановлено, що в твердому стані утворюється тільки одна сполука – сесквітелурид кремнію ($Si_2Te_3$), який плавиться інконгруентно при 1165 K. Однофазний полікристалічний $Si_2Te_3$ автор [3] отримував шляхом загартування розплаву від температури вище лінії ліквідусу (~ 1523 K) з наступним відпалом при температурі нижче перетектичної реакції (~ 1063 K) протягом 90 год.

Характер розташування ліній на *T*−*x*-діаграмах, побудованих авторами [5, 6] (рис. 1.1, *б* і 1.2), досить близький до представлених на рис. 1.1 *а*. Підтверджено утворення в системі Si–Te тільки однієї сполуки $Si_2Te_3$, яка плавиться інконгруентно при 1168 K [4], 1158 K [5] та 1159 K [6]. Тільки в єдиній роботі [9] вказується, що сполука $Si_2Te_3$ плавиться конгруентно при 1162 K.

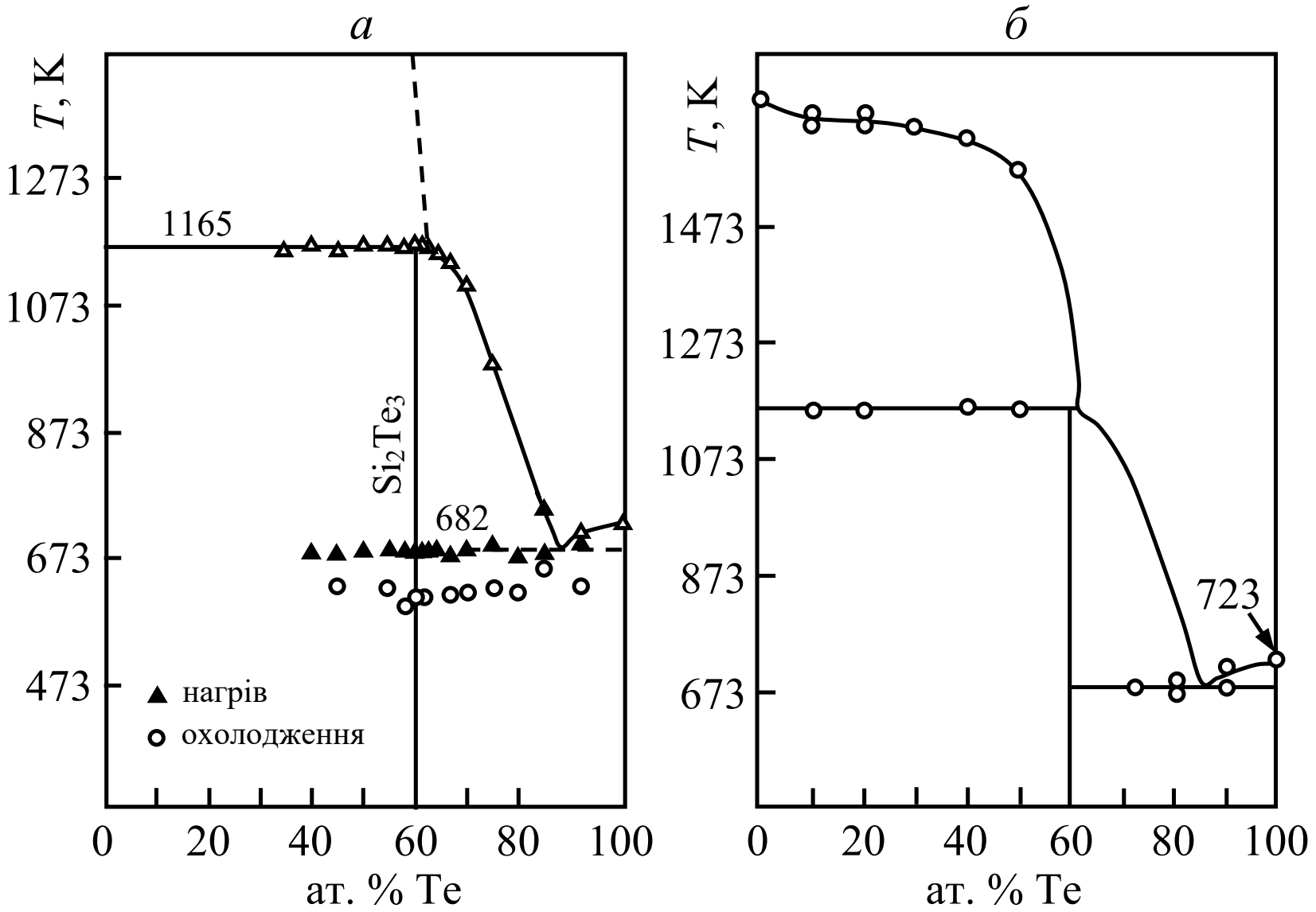

Рис. 1.1. Діаграма стану системи Si–Te: *а* – [3]; *б* – [5].

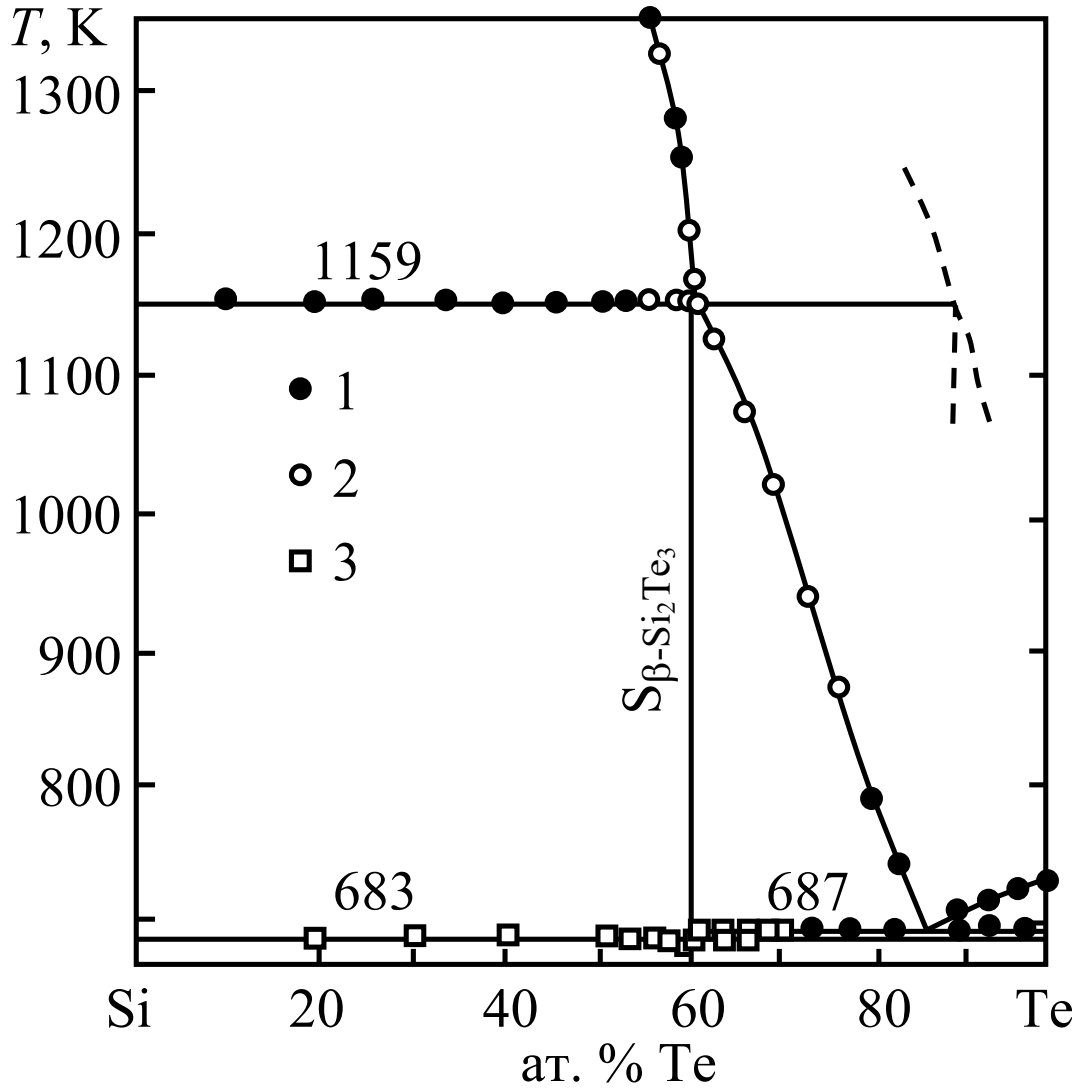

Рис. 1.2. Діаграма стану системи Si–Te [6]: 1 – дані кривих нагріву ДТА; 2 – дані ДТА при одночасному визначенні тиску парів; 3 – дані ДСК.

Сполука $Si_2Te_3$ зазнає $\alpha \rightarrow \beta$ поліморфне перетворення при 673 К [3, 10]. Спроба загартувати високотемпературну β-фазу $Si_2Te_3$ шляхом різкого охолодження у воду з льодом від температури 873 або 963 К (попередня витримка при цих температурах 24 год) не увінчалася успіхом [6]. Рентгенофазовий аналіз загартованих зразків показав наявність лише ліній низькотемпературної α-фази $Si_2Te_3$, отже, процес поліморфного переходу $\beta \rightarrow \alpha$ загальмувати у зазначених умовах важко. Експериментальна густина α-$Si_2Te_3$, визначена пікнометричним методом у бромоформі $\rho_{екс}$ = 4.56 ± 0.1 г/см$^3$, а розрахована – $\rho_{роз}$= 4.566 г/см$^3$ [6].

Більшість бінарних напівпровідникових сполук має вузьку область гомогенності і на звичайній діаграмі стану її, як правило, не зображують. Однак наявність надлишкової понад стехіометричну кількість атомів одного з компонентів може приводити до істотної зміни їх електричних, фотоелектричних та оптичних властивостей. Експериментально встановити область гомогенності $Si_2Te_3$ досить складно через мале відхилення граничних складів фаз на основі даної сполуки від стехіометричного. Очевидно, це є головною причиною протиріч, наявних відомостей щодо границь області гомогенності фази $Si_2Te_3$. При вивченні $p_{Te_2} - T$-проекції системи Si–Te авторами [11] встановлено, що область гомогенності не виходить за межі складів 59.45 та 60.50 ат. % Te. Іншої думки дотримуються автори [10], вважаючи, що область гомогенності фази на основі $Si_2Te_3$ знаходиться в межах 60÷66.6 ат. % Te, тобто від $Si_2Te_3$ до $SiTe_2$. Істотна різниця у визначенні меж області гомогенності $Si_2Te_3$ спонукала авторів [6] знову провести ретельніші дослідження, використавши для цих цілей три незалежні методи: РФА, мікроструктурний аналіз та тензиметричний. Найбільш широку область гомогенності від 59.6 до 60.25 ат. % Te отримано з використанням даних РФА. Проте, як стверджують самі автори [6], отримані границі орієнтовні, оскільки метод РФА має чутливість лише на рівні 0.5 мас. % стосовно реєстрації другої фази. За допомогою мікроструктурного аналізу, границі області гомогенності знаходяться в межах від 59.80 до 60.18 з точністю визначення 0.15 ат. % Te. Більш точні дані про границі області гомогенності $Si_2Te_3$ отримані тензиметричним методом: від 59.85 ± 0.06 ат. % до 60.14 ± 0.04 ат. % Te (при 1023 К). Тензиметричне визначення меж області гомогенності має низку переваг. По-перше, це більш висока точність (для $Si_2Te_3$ близько 0.05 ат. %), по-друге, експерименти проводяться безпосередньо при температурі, при якій

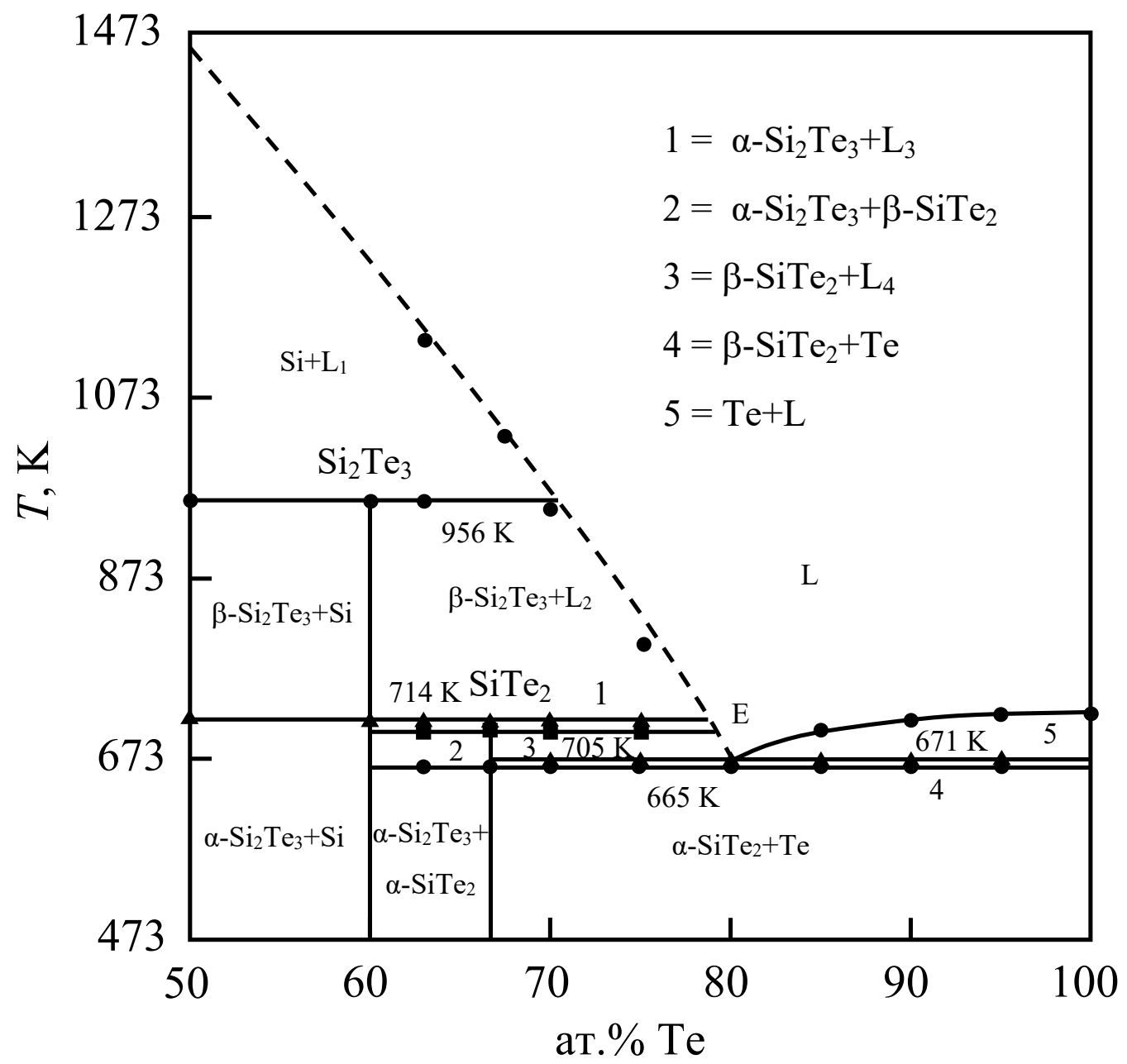

Рис. 1.3. Діаграма стану системи Si–Te [7].

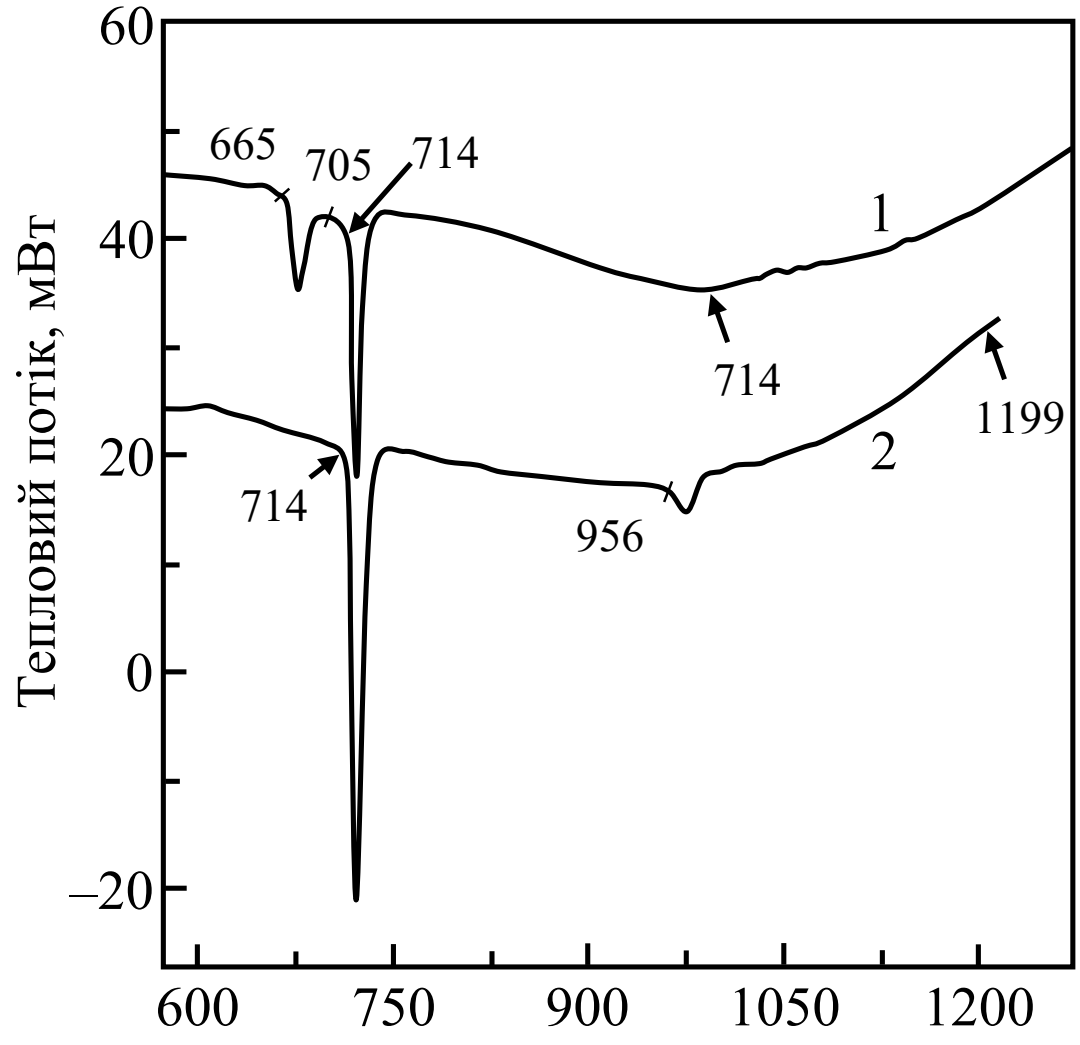

Рис. 1.4. ДТА фаз $SiTe_2$ (1) і $Si_2Te_3$ (2) [7].

визначається межа області гомогенності, і тиску пари, що відповідає дослідженим рівновагам [6].

У літературі наявні розбіжності щодо складу евтектики. Так, згідно [3] евтектика між Te та $Si_2Te_3$ знаходиться в інтервалі концентрацій від 10 до 20 ат. % Si, а за даними [12] можливий інтервал існування евтектики складає 15–19 ат. % Si. Автор [13] вважає, що евтектика між Te та $Si_2Te_3$ відповідає орієнтовному складу 17 ат. % Si і має відносно низьку температуру плавлення (682 К). На наявність евтектики у системі Si–Te з параметрами: складом 82.5 ат. % Te та температурою плавлення 679 К вказується в [4]. У цій області складів в'язкість розплавів досить значна, тому в цих умовах можливе затвердіння розплаву з утворенням стекол. Термічні ефекти поблизу 683 К пов'язуються з фазовим перетворенням у твердому стані в сполуці $Si_2Te_3$.

Незважаючи на те, що на діаграмах стану системи Si–Te, побудованих трьома групами авторів [3, 5, 6] (рис. 1.1 та 1.2), представлена лише одна стабільна сполука $Si_2Te_3$, у літературі повідомляється про існування також інших телуридів кремнію: SiTe [14–17] та $SiTe_2$ [18–22]. Автори [3, 14] виявили SiTe лише у пароподібному стані. Отримання SiTe у твердому стані шляхом вакуумного термічного розкладання $SiTe_2$ повідомляється в роботі [15].

Для підтвердження наявності у системі Si–Te, крім сполуки $Si_2Te_3$ ще й інших можливих сполук $SiTe_2$ та SiTe, автори [7] знову провели детальне вивчення діаграми стану системи Si–Te (рис. 1.3) із залученням ширшого кола методів дослідження. У результаті надійно встановлено, що крім $Si_2Te_3$ у цій системі існує ще одна сполука $SiTe_2$. Для обох сполук встановлено $\alpha \rightarrow \beta$ фазовий перехід при 714 і 665 К відповідно.

Результати диференціального термічного аналізу $SiTe_2$ та $Si_2Te_3$ наведені на рис. 1.4. На кривій ДТА сполуки $Si_2Te_3$ (крива 2) спостерігаються два ендотермічні піки: інтенсивний при 714 К і слабкий за інтенсивністю при 956 К. Пік при 714 К зумовлений $\alpha \rightarrow \beta$ фазовим переходом у $Si_2Te_3$, тоді як пік при 956 К викликаний перитектичним розкладанням згідно реакції [8]:

$$Si_2Te_3\,(тв) = (2\text{–}x)Si(тв) + (3\text{–}x)/2\ Te_2(газ) + xSiTe(газ).$$

На кривій ДТА сполуки $SiTe_2$ (рис. 1.4, крива 1) спостерігаються три ендотермічні піки при 665 К, 705 К і 714 К. Інтенсивний пік при 665 К відповідає $\alpha \rightarrow \beta$ фазовому переходу $SiTe_2$. Ендотермічні піки при 705 К і 714 К, які перекриваються, утворюючи плече поблизу

714 К, зумовлені перитектичним розкладанням $SiTe_2$ згідно реакції:

$$2\beta\text{-}SiTe_2(тв) = \alpha\text{-}Si_2Te_3(тв) + Te(р).$$

Термогравіметрична крива (залежність зміни маси від температури, рис. 1.5, *а*) для чистої сполуки $SiTe_2$ показує, що дана сполука розкладається з утворенням Si в кілька стадій. На першій стадії β-$SiTe_2$(тв) розкладається на α-$Si_2Te_3$(тв) у діапазоні температур від 689 К до 833 К згідно реакції:

$$\beta\text{-}SiTe_2(тв) \rightarrow \alpha\text{-}Si_2Te_3(тв) + Te\uparrow(газ).$$

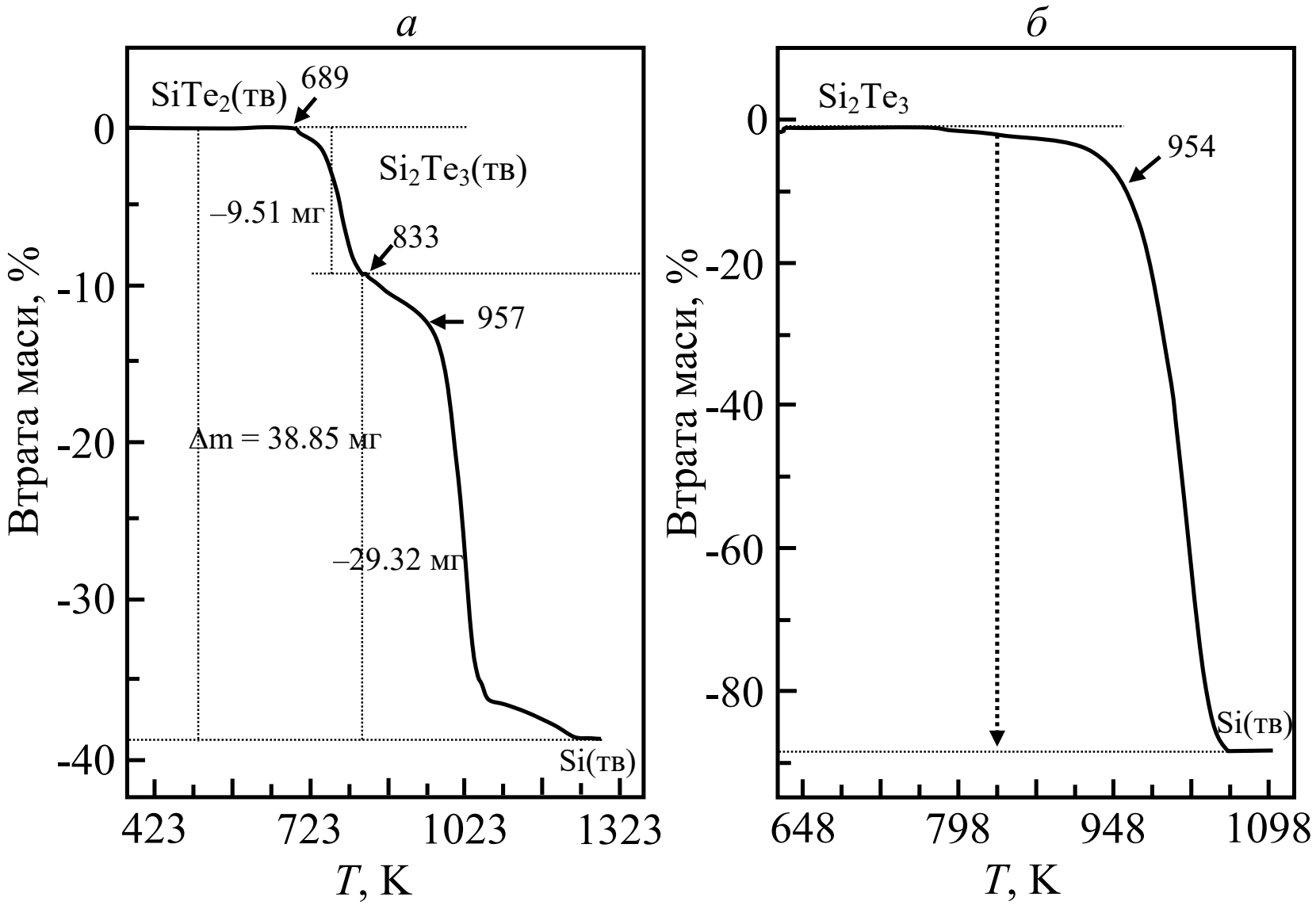

Рис. 1.5. Термогравіметричні криві $SiTe_2$ (*а*) і $Si_2Te_3$ (*б*) [7].

При цьому для вихідного зразка масою 43.3 мг автори [7] фіксували втрату маси, яка склала 9.51 мг і зумовлена втратою Te↑(газ) з $SiTe_2$ та утворенням $Si_2Te_3$(тв). Проте очікувана втрата маси у цьому процесі має становити 9.7 мг. Різницю в 0.19 мг автори [7] пов'язують з неповним розкладанням сполуки $SiTe_2$. Втрата маси в наступному температурному інтервалі 833−957 К відповідає випаровуванню телуру з β-$Si_2Te_3$(тв). Це супроводжується різким збільшенням втрати маси після 957 К за рахунок випаровування Te(р), утвореного внаслідок перитектичного розкладання $Si_2Te_3$ згідно реакції:

$$\beta\text{-}Si_2Te_3(тв) = 2Si(тв) + 3Te(р).$$

На останньому етапі нагрівання (1073–1273 К) зменшення швидкості випаровування автори [7] пов'язують з кінетичними ефектами. Загальна втрата маси, що спостерігається для реакції

$$Si_2Te_3(тв) = 2Si(тв) + 3Te\uparrow(газ),$$

склала 29.32 мг проти очікуваної втрати маси 29.30 мг. Надмірна втрата 0.2 мг зумовлена через втрату Si в SiTe(газ). На рис. 1.5 *б* наведена термогравіметрична крива чистого $Si_2Te_3$, записана в аналогічних умовах. Із порівняння кривих на рис. 1.5 *а* і *б* видно, що починаючи з другого етапу характер залежності для $Si_2Te_3$ аналогічний до другого етапу $SiTe_2$, підтверджуючи цим процес розкладання.

## 1.2. СИНТЕЗ РЕЧОВИНИ ТА ВИРОЩУВАННЯ КРИСТАЛІВ $SiTe_2$ І $Si_2Te_3$

**1.2.1. Синтез полікристалічного $Si_2Te_3$.** Найбільш простим і широко вживаним способом синтезу полікристалічного $Si_2Te_3$ є пряме сплавлення вихідних елементарних компонентів, взятих у стехіометричному співвідношенні, в евакуйованих кварцових ампулах [6, 25]. В якості вихідних компонентів використовують монокристалічний кремній і спеціально очищений від оксидів телур шляхом плавлення у вакуумі та прокапування через кварцовий капіляр [6]. Розраховані наважки компонентів завантажують у попередньо очищену графітизовану кварцову ампулу довжиною 160–180 мм і діаметром 18–20 мм. Ампулу з речовиною відкачують до тиску залишкових газів 133 Па і запаюють. Через велику пружність парів телуру при високих температурах, синтез речовини $Si_2Te_3$ здійснюють у два етапи. На першій стадії кварцову ампулу з вихідною шихтою поміщають у горизонтальну трубчасту резистивну піч і нагрівають до температур 850–900 K зі швидкістю 0.1–0.2 K/год з наступною витримкою при цій температурі протягом 15–20 год., після чого температуру в печі підвищують до 1200 K зі швидкістю 0.05–0.1 K/c. При цій температурі витримують розплав протягом 24–48 год. з метою забезпечення синтезу та гомогенізації розплаву. По завершенню процесу синтезу включають програмне зниження температури із швидкістю 0.2 K/год і отримують полікристалічний злиток.

У процесі синтезу під час витримок при фіксованих температурах бажано вмикати пристрій для віброперемішування синтезованої суміші. Використання примусового вібраційного перемішування при синтезі $Si_2Te_3$ у порівнянні з традиційним конвективним сплавлен-ням компонентів дає наступні переваги: відпадає необхідність дроблення компонентів перед завантаженням вихідної шихти для вирощування монокристалів, що в свою чергу зменшує ймовірність забруднення одержуваних кристалів чужорідними домішками; покращується якість синтезованого полікристалічного секвітелуриду кремнію (менша зернистість, відсутність непрореагованих компонентів, підвищується однорідність суміші).

Залежно від методу та умов вирощування кристали $Si_2Te_3$ можуть набувати різної форми: одновимірні (1D), двовимірні (2D) та тривимірні (3D).

**1.2.2. Вирощування об'ємних кристалів $Si_2Te_3$ і $SiTe_2$ методом Бріджмена.** Однофазні об'ємні монокристалічні зливки $Si_2Te_3$ отримують спрямованою кристалізацією, шляхом опускання ампули з розплавом у холодну зону трубчастої вертикальної печі. Цей метод приваблює відносною простотою та можливістю отримувати відносно великі кристали. Для вирощування монокристалів цим методом використовують вертикальні ростові печі, які складаються з двох незалежних зон, що дозволяє змінювати поздовжній температурний градієнт у необхідних межах і проводити відпал у найбільш прийнятних умовах [1]. Для переміщення фронту кристалізації застосовують спеціальний підйомний механізм, що дозволяє переміщати печі або ампулу з речовиною. В якості ростових контейнерів використовують кварцові ампули з конічним дном або подовженням у нижній частині її у вигляді капіляра діаметром 2÷3 мм і довжиною 10÷15 мм.

Кварцову ампулу, що містить вихідну речовину $Si_2Te_3$, припаюють верхнім кінцем до кварцового супорта та поміщають у верхню високотемпературну зону печі. Після розплавлення речовини в кварцовій ампулі та недовготривалій витримці вмикається механізм переміщення супорта, й ампула поступово входить у низькотемпературну зону. Для усунення конвекції та радіації у просторі печі, а також для досягнення більш різкої зміни градієнта температури, зони печі повинні бути розділені діафрагмою. Електроживлення печі має бути стабілізовано. У багатьох випадках цього достатньо для підтримки температурного режиму. Коливання підведеної потужності підвищують або знижують температуру в обох зонах печі, внаслідок чого змінюється швидкість росту кристалів. Це може викликати утворення кристалографічних недосконалостей і неоднорідностей за складом, особливо при вирощуванні змішаних або легованих кристалів. Тому для вирощування більш досконалих кристалів потрібно ретельно регулювати температуру. Оскільки рушійною силою процесу росту кристала за цим методом є температурний градієнт на межі розділу фаз, то величина та форма цього градієнта визначає швидкість і положення фронту кристалізації. Якщо фронт кристалізації увігнутий (наприклад, при занадто швидкому опусканні ампули), то охолодження через бічну поверхню призводить до утворення випадкових зародків, внаслідок чого зливок виходить полікристалічним.

При вирощуванні кристалів $Si_2Te_3$ із розплаву швидкість росту може бути відносно високою, але отримання якісних кристалів вимагає спеціальних заходів, таких як запобігання перегріву розплаву, зменшення флуктуації температури, тощо. При використанні розплавних методів стійкість фронту кристалізації і структурна досконалість сильно залежать від мікроколивань температури. Склад кристала змінюється по його довжині із-за постійної зміни складу розплаву і зміни температури.

Дослідження [9, 10] показали, що для вирощування об’ємних (3D) монокристалів $Si_2Te_3$ за методом Бріджмена найбільш оптимальними умовами є: температура гарячої зони 1273 K, температура холодної зони 1073 K, швидкість переміщення фронту кристалізації 0.125 мм/год [9] (1.375 мм/год [10]), градієнт температури в області фронту кристалізації 3–5 K/мм. Використання таких малих швидкостей переміщення ампули забезпечує повноту протікання перитектичної реакції навіть у разі отримання монокристалів сполук, які інконгруентно плавляться, включно й $Si_2Te_3$.

Методом Бріджмена також були вирощені монокристали $SiTe_2$ циліндричної форми довжиною 1 см та діаметром 1 см [22]. Оптимальні умови вирощування монокристали $SiTe_2$ цим методом: температура гарячої зони $1348 \pm 50$ K, швидкість переміщення фронту кристалізації 1 мм/год.

До недоліків методу вирощування кристалів $Si_2Te_3$ і $SiTe_2$ з розплаву слід віднести: малу швидкість цього процесу, необхідність застосування високих температур, що призводять до підвищених концентрацій дефектів (вакансій і дислокацій) у кристалах, факт локалізації в кристалі всіх домішок, що містяться у вихідних речовинах. Оскільки при вирощуванні монокристалів $Si_2Te_3$ з розплаву методом Бріджмена не завжди забезпечується висока однорідність їх властивостей, то для отримання більш якісних кристалів використовуються методи вирощування з газоподібної фази: сублімації, хімічно-транспортних реакцій (ХТР) і Піццарелло.

**1.2.3. Вирощування монокристалів $Si_2Te_3$ із газової фази.** Кристалізація з газової фази широко використовується для вирощування об'ємних, пластинчастих та ниткоподібних кристалів $Si_2Te_3$. Вирощування кристалів $Si_2Te_3$ з газової фази можна проводити при значно менших температурах, істотно нижче температури плавлення речовини [1]. Це означає, що концентрацію вакансій та дислокацій у кристалі можна звести до мінімуму (концентрація вакансій експоненційно залежить від температури). Крім цього, методи низькотемпературної кристалізації стають найбільш прийнятними при отриманні кристалів речовин, що плавляться інконгруентно. Основна вимога при вирощуванні кристалів $Si_2Te_3$ з газової фази – неперервна подача парів елементів кремнію та телуру в зону реакції і росту. Джерелом парів елементарних компонентів може слугувати або дисоціація попередньо синтезованої полікристалічної сполуки $Si_2Te_3$, або самі елементи. Пари елементів, дифундуючи або переносячись в область, де вони стають пересиченими, утворюють кристал. У залежності від способу транспортування кристалоутворюючих елементів кристалізацію з газової фази можна розділити на статичний та динамічний методи. У динамічному методі використовується газ-носій, який безперервно протікає в системі. Таким методом вирощують новіскери і нанопластини $Si_2Te_3$, де в якості носія використовують газ аргон (Ar). У статичному методі вирощування транспорт забезпечується дифузією через газову фазу. У цьому методі джерелом є певна кількість полікристалічної сполуки, що знаходиться в гарячій зоні (при високій температурі). Пари, що утворюються при дисоціації бінарної сполуки дифундують в область нижчих температур (у холодну зону), де відбувається зародження кристалів та їх зростання. Досконалість кристалів при вирощуванні з парової фази визначається швидкістю перенесення маси і термодинамічною рівновагою між зростаючим кристалом і паровою фазою, зокрема перенасиченням в області кристалізації. Значення цієї швидкості залежить від процесів, які відбуваються в паровій фазі, і визначаються концентрацією (тиском) компонентів у системі та температурним градієнтом між зонами джерела і кристалізації. Габітус кристалів визначається температурними умовами росту, парціальними тисками кожного з компонентів сполуки, які впливають на стехіометрію зростаючого кристала, та наявність домішок у системі. При вирощуванні з газової фази швидкості зростання кристалів невеликі, тому газофазні методи доцільно використовувати для отримання невеликих огранених кристалів хорошої якості.

Таблиця 1.1. Умови вирощування пластинчастих кристалів $Si_2Te_3$ та $SiTe_2$ з газової фази

| Сполука | Метод вирощування | Температура в зоні випаровування, K | Температура в зоні конденсування, K | Тривалість процесу росту, *год.* | Форма кристалів | Максимальні розміри кристалів, *мм* | Література |
|---|---|---|---|---|---|---|---|
| $Si_2Te_3$ | Статична сублімація | 1000 | 900 | 40÷50 | пластини | 10×8×0.2 | [25] |
| $Si_2Te_3$ | Статична сублімація | 1103 | 1023÷1073 | 70÷80 | пластини | діаметр – 20, товщина – 0.2÷1 | [3] |
| $Si_2Te_3$ | Статична сублімація | 1100 | 980 | 120 | пластини | 6×6 товщина – 0.2 | [26] |
| $Si_2Te_3$ | Піццарелло | 1173 | 1130 | | об'ємні | діаметр – 20, довжина – 30 | [9] |
| $SiTe_2$ | Статична сублімація | 1123 | 1023 | 48÷72 | пластини | діаметр – 10, товщина – 0.5 | [20] |
| $SiTe_2$ | ХТР, $J_2$ | 923 | 863 | 120 | пластини | діаметр – 10, товщина – 0.5 | [20] |

**Вирощування пластинчастих (2D) кристалів $Si_2Te_3$ методом сублімації.** Двовимірні пластинчасті кристали $Si_2Te_3$ одержують методом статичної сублімації [3, 25, 26]. Для вирощування кристалів $Si_2Te_3$ методом статичної сублімації запаяну кварцову ампулу з полікристалічною речовиною поміщають у горизонтальну двозонну трубчасту електропіч. У статичному методі вирощування кристалів за допомогою сублімації пересичення створюється різницею температур ($\Delta T$) між зонами випаровування і конденсації. Оптимальні умови вирощування кристалів $Si_2Te_3$ методом сублімації: температура гарячої зони $T_{гар}$ = 1000–1100 K; температура холодної зони $T_{хол}$ = 900–1000 K; тривалість процесу зростання 50–120 год. (табл. 1.1). При вирощуванні монокристалів $Si_2Te_3$ шляхом статичної сублімації, тобто в умовах спонтанної кристалізації, на внутрішній поверхні кварцової трубки виростають пластинчасті кристали розміром до 2 см в діаметрі і товщиною 0.2 мм, ребра яких утворюють кути 120°. Кристали мають природні дзеркальні поверхні (001) з віссю *c* перпендикулярною площині сколу. Таким чином, осадження із парової фази дає змогу отримати високоякісні кристали $Si_2Te_3$ достатніх розмірів, які не потребують додаткової механічної та хімічної обробки поверхні, для дослідження фізичних властивостей та практичного використання.

**Вирощування об'ємних (3D) кристалів $Si_2Te_3$ з газової фази методом Піццарелло.** Процес вирощування об'ємних кристалів $Si_2Te_3$ з газової фази автори [9] здійснювали у вертикальній печі Бріджмена з оберненим температурним профілем. Вертикальна конфігурація ростової печі дає змогу розмістити кварцову ампулу з вихідною речовиною по осі печі для досягнення симетричного радіального температурного розподілу. Таке розміщення ростової ампули дозволяє послабити конвекційні потоки у ній. При цьому також різко сповільнюється швидкість росту зливка. Вперше цей метод був використаний Піццарелло [23] і детально описаний у монографії [1]. Речовину, що підлягає перекристалізації через газову фазу (сублімація-конденсація або за допомогою оборотних хімічних транспортних реакцій), поміщають у кварцову ампулу, верхня частина якої відтягнута у вигляді конуса з перетяжкою. У конусі вживаються заходи забезпечення переважного зростання одного зародка кристалізації. З цією метою ампулу з речовиною переміщують через піч із температурним градієнтом. Кристали хорошої якості $Si_2Te_3$ (рис. 1.6) виростали коли температура печі в зоні випаровування становила

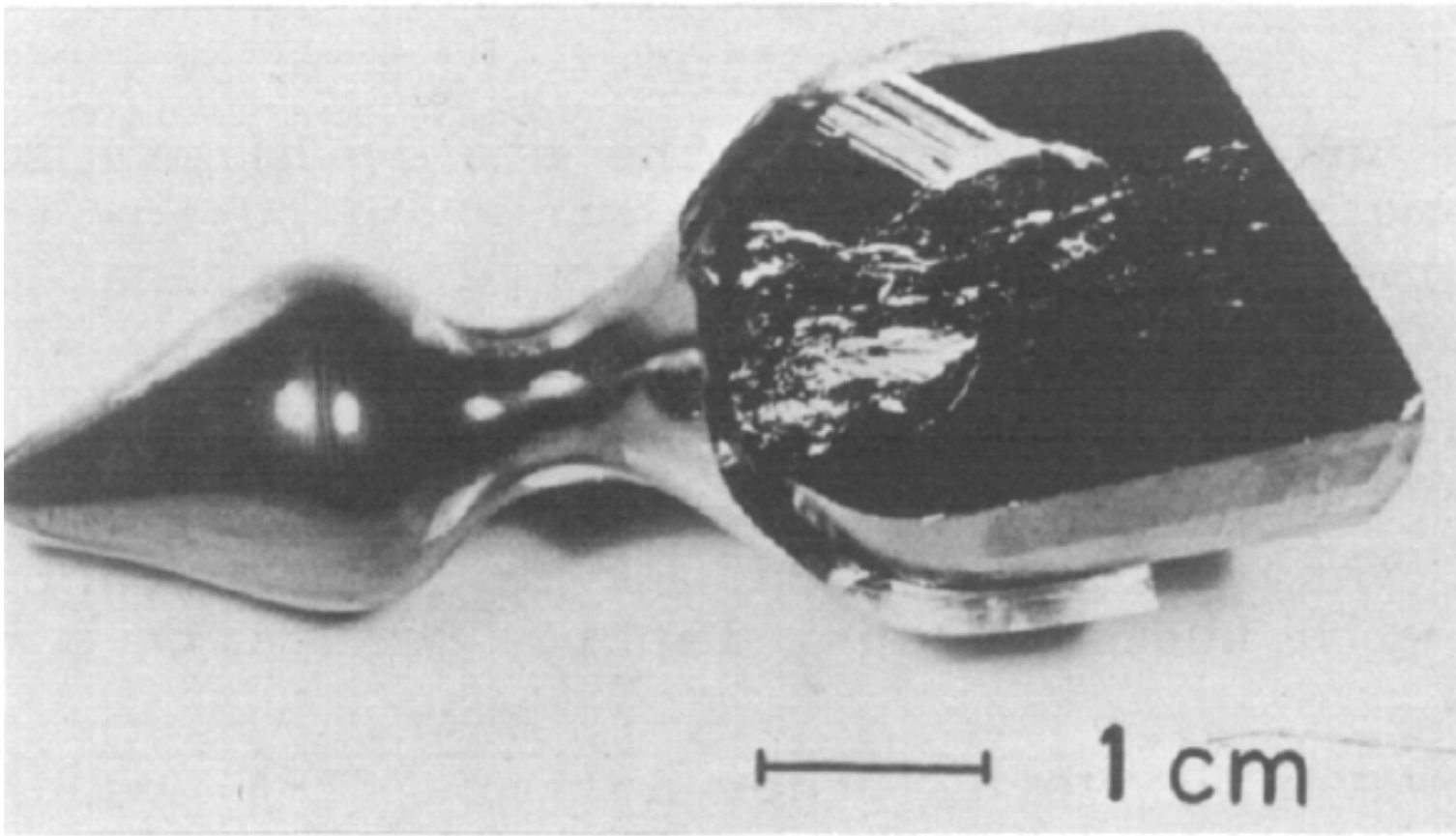

Рис. 1.6. Загальний вид природного сколу монокристала $Si_2Te_3$, вирощеного методом Піццарелло [9].

1173 K, а температура росту була на 40 K нижче температури випаровування. Швидкість переміщення ампули у верхню зону печі становила 0.04 мм/*год*. Дослідження кристалів, отриманих сублімацією за методом Піццарелло, показали, що їх склад відповідає складу шихти тільки у випадку «квазітермічних» умов їх вирощування, тобто у випадку малих градієнтів температур на фронті кристалізації і малих швидкостей переміщення ампули. Вирощені кристали мали площини спайності, паралельні напрямку росту. Густина кристалів становила 4.42 г/см$^{-3}$ [9].

**1.2.4. Вирощування шаруватих кристалів $SiTe_2$ із газової фази.** Умови синтезу речовини та вирощування кристалів $SiTe_2$ описані в роботах [18–22, 24] та узагальнені у табл.1.1. Полікристалічний $SiTe_2$ автори [18–21] синтезували прямим сплавленням стехіометричних кількостей елементів в евакуйованих кварцових ампулах при 1323–1348 K протягом 48–72 год. Вирощування кристалів автори [20] здійснювали методами сублімації та хімічних транспортних реакцій (ХТР) у горизонтальній трубчастій електропечі. Контейнером слугували кварцові ампули довжиною 15 см і внутрішнім діаметром 15 мм. Основною вимогою при вирощуванні кристалів з газової фази є неперервна подача парів кремнію і телуру в зону реакції й росту. У статичному методі вирощування транспорт забезпечується дифузією компонентів через газову фазу. У цьому методі джерелом парів є певна кількість полікристалічної сполуки $SiTe_2$, що знаходиться в гарячій зоні. Пари компонентів, що утворюються при дисо-ціації бінарної сполуки, дифундують в область більш низьких температур (у холодну зону), де відбувається процес зародкоутворення і ріст кристалів.

Оптимальні умови вирощування кристалів $SiTe_2$ методом сублімації: температура гарячої зони – 1123 K, а холодної – 1023 K; у випадку вирощування методом ХТР – $T_{гар}$ = 923 K та $T_{хол}$ = 863 K; тривалість процесу росту обома методами становила 5 діб. Кристали виростали у вигляді гексагональних темновато-червоних пластин, розмірами до 1 см і товщиною 0.5 мм [20]. Кристали $SiTe_2$, вирощені авторами [21], за технологією, описаною вище, містили велику кількість аніонних вакансій, на що вказував хімічний аналіз. За даними [21] склад кристалів змінювався від $SiTe_{1.996}$ до $SiTe_{1.886}$, а згідно з [24] від $SiTe_{1.999}$ до $SiTe_{1.908}$. Пластинчасті кристали $SiTe_2$ окислюються на повітрі з виділенням $H_2Te$ і згодом їх поверхня покривається сірим нальотом телуру, який уповільнює їх подальше розкладання. Кристали $SiTe_2$ при 1493 K випаровуються у вакуумі, минаючи процес плавлення. Сублімат містить суміш $SiTe_2$, Si, Te та сірого монотелуриду SiTe [19].

Характерною особливістю шаруватих сполук $SiTe_2$ і $Si_2Te_3$ є анізотропія властивостей, обумовлена будовою їх кристалічних ґраток. Особливо яскраво ця властивість проявляється при вирощуванні монокристалів із газової фази. Форма і розміри зростаючих кристалів залежать не тільки від зовнішніх факторів, але і від анізотропії швидкості росту кристалів у різних кристалографічних напрямках. Швидкість зростання кристалів даних сполук, які кристалізуються в шаруватій структурі, значно менша в напрямку осі *c* у порівнянні з напрямком, перпендикулярним до осі *c*. Внаслідок цього кристали $SiTe_2$ і $Si_2Te_3$ при вирощуванні методом сублімації та ХТР мають форму тонких пластин.

Слабкий зв'язок між сусідніми тришаровими пакетами приводить до виникнення в шаруватих кристалах $SiTe_2$ і $Si_2Te_3$ крайових і часткових дислокацій, які пов’язані зі зміщенням частини кристала за базисним площинами. З частковими дислокаціями зв’язані дефекти упаковки, які є одним із основних типів дефектів у шаруватих телуридах кремнію [10, 24]. Дефект упаковки полягає у зміщенні шару з правильного положення у структурі. Оскільки напрям зміщення є паралельним до шарів, то роль дефектів зводиться до порушення регулярності структури в напрямку, перпендикулярному до зміщення, який в шаруватих кристалах співпадає з *c*-віссю. Поява дефектів упаковки не змінює ні числа ближніх сусідів, ні відстані до них. Але із-за зміни в розташуванні наступних шарів (не найближчих) у випадку шаруватих кристалів появляється одномірне розупорядкування у напрямку перпендикулярному до шарів, що сильно відображається на анізотропії явищ переносу.

## 1.3. ІДЕНТИФІКАЦІЯ СПОЛУК $SiTe_2$ ТА $Si_2Te_3$ ЗА ДОПОМОГОЮ СПЕКТРОСКОПІЧНИХ МЕТОДІВ ДОСЛІДЖЕННЯ

Більшість фізико-хімічних властивостей у конденсованому стані значною мірою визначається локальною структурою матеріалу. В даний час для аналізу структури кристалів поряд із прямими рентгеноструктурними методами: дифракція рентгенівських променів, електронів та нейтронів; широко використовується непрямі спектроскопічні методи до яких належать: XANES, рентгенівська фотоелектронна спектроскопія (РФЕС) остовних рівнів атомів.

**1.3.1.Спектри рентгенівського поглинання.** Потужним інформативним методом вивчення електронної підсистеми та локальної структури твердих тіл є дослідження ближньої тонкої структури рентгенівського поглинання (X-ray Absorption Near Edge Structure, XANES) [28, 29]. Тонка біляпорогова структура поглинання XANES є дуже чутливою як до електронного стану поглинаючого атома, так і до його локального оточення. Тому метод XANES застосовують для отримання даних про електронну та кристалічну будову речовини. Поглинання рентгенівського випромінювання речовиною пов'язане з взаємодією фотонів із електронами внутрішніх оболонок атомів. Внаслідок такої взаємодії відбувається вибивання електронів з атома, що призводить до різкого зростання поглинання рентгенівського випромінювання (стрибка) при перевищенні енергії фотонів енергії зв'язку електрона з ядром (порога збудження). Поріг збудження є характеристичною величиною для кожного хімічного елемента, що дозволяє однозначно визначати хімічний елемент за положенням порога збудження.

У методах рентгенівської спектроскопії поглинання вимірюється залежністю коефіцієнта рентгенівського поглинання ( $\mu_E$) від енергії рентгенівських фотонів. Використання режиму пропускання рентгенівського вимірювання є найбільш поширеним, оскільки він передбачає лише вимірювання потоку рентгенівського випромінювання до і після того, як промінь проходить через зразок. Тому лінійний коефіцієнт рентгенівського поглинання визначається за формулою:

$$\mu_E = \ln \frac{I_0}{I_t}, \tag{1.1}$$

де $I_0$ і $I_t$ – відповідно інтенсивність падаючого випромінювання та випромінювання, яке пройшло через зразок. Основна частина сучасних досліджень у галузі спектроскопії рентгенівського поглинання проводиться з використанням джерел синхротронного випромінювання тому, що в таких дослідженнях потрібно варіювання енергії

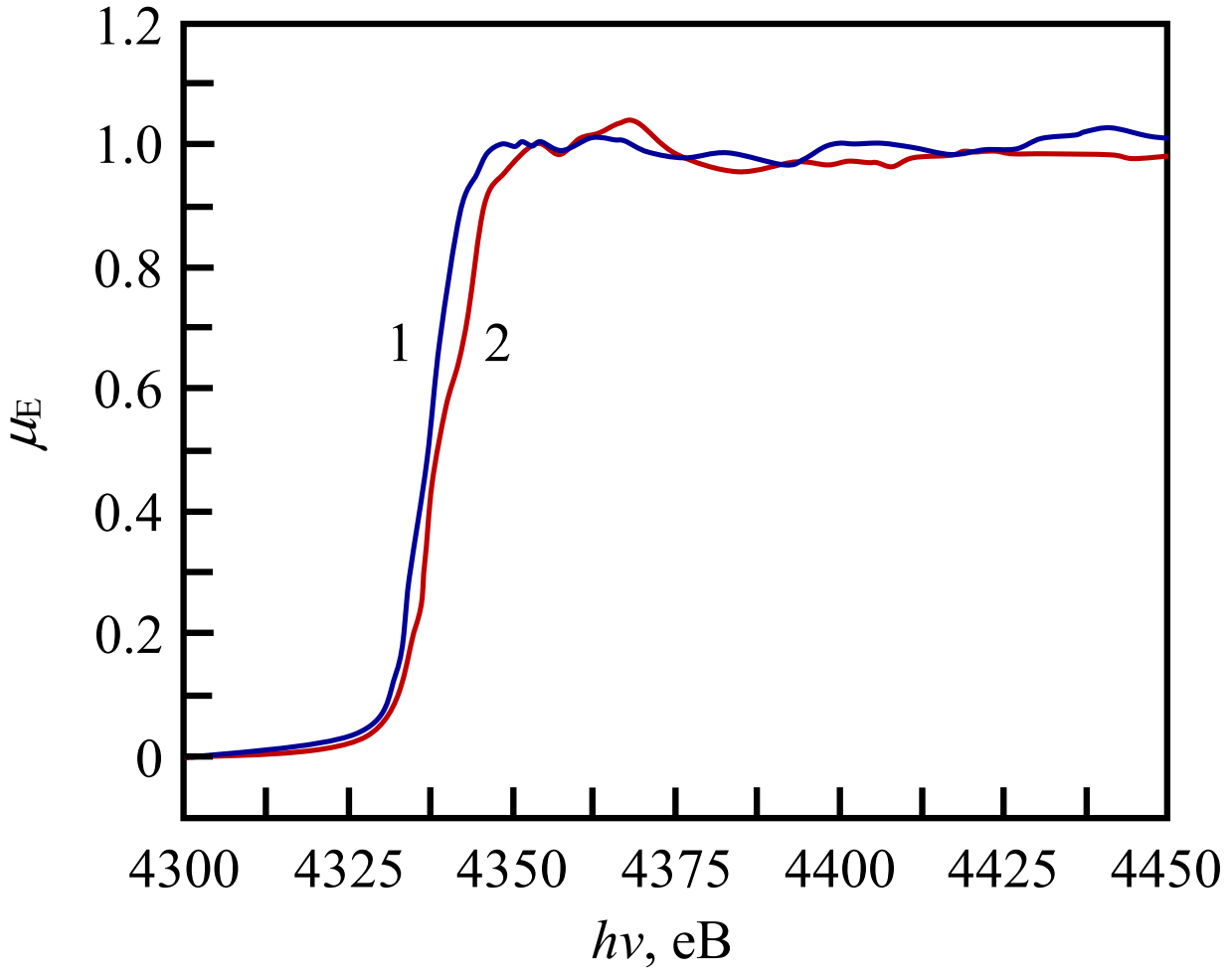

Рис. 1.8. Спектральна залежність коефіцієнта поглинання ($\mu$) у XANES вимірюваннях для $SiTe_2$ (1) і $Si_2Te_3$ (2) [7].

падаючого випромінювання в широкому діапазоні. Результати вимірювання XANES на зразках $SiTe_2$ та $Si_2Te_3$ наведено на рис. 1.8 [7]. На цьому рисунку чітко видно явне зміщення краю рентгенівського поглинання $Si_2Te_3$ (крива 2) в область більших енергій, у порівнянні з $SiTe_2$ (крива 1).

**1.3.2. Рентгенівська фотоелектронна спектроскопія остовних рівнів атомів поверхні кристалів і нанопластин $SiTe_2$ і $Si_2Te_3$.** Метою РФЕС спектроскопії остовних рівнів є забезпечення аналізу складу поверхні кристалів $SiTe_2$ і $Si_2Te_3$. Оскільки енергія остовних рівнів в основному визначає сорт атома, спостереження РФЕС піків, положення яких на шкалі енергій відповідає певній енергії зв'язку, можна розглядати як індикатор наявності в приповерхневій області даного елемента. Таким чином, спектри РФЕС містять інформацію, за якою можна визначити склад поверхневої області. Для кожного

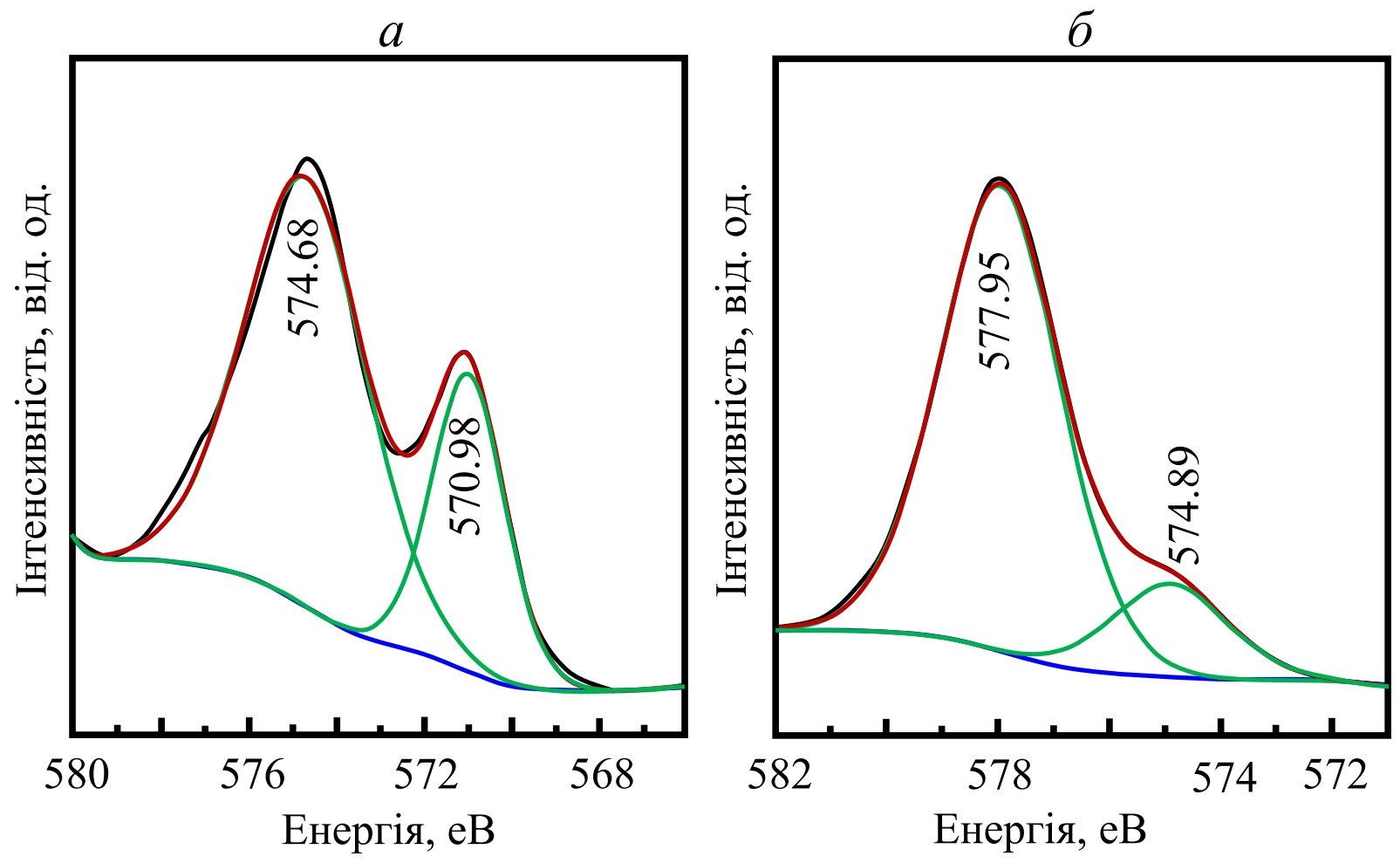

Рис. 1.9. Рентгенівські фотоелектронні спектри $Te3d_{5/2}$ для $SiTe_2$ (*а*) і $Si_2Te_3$ (*б*) [7].

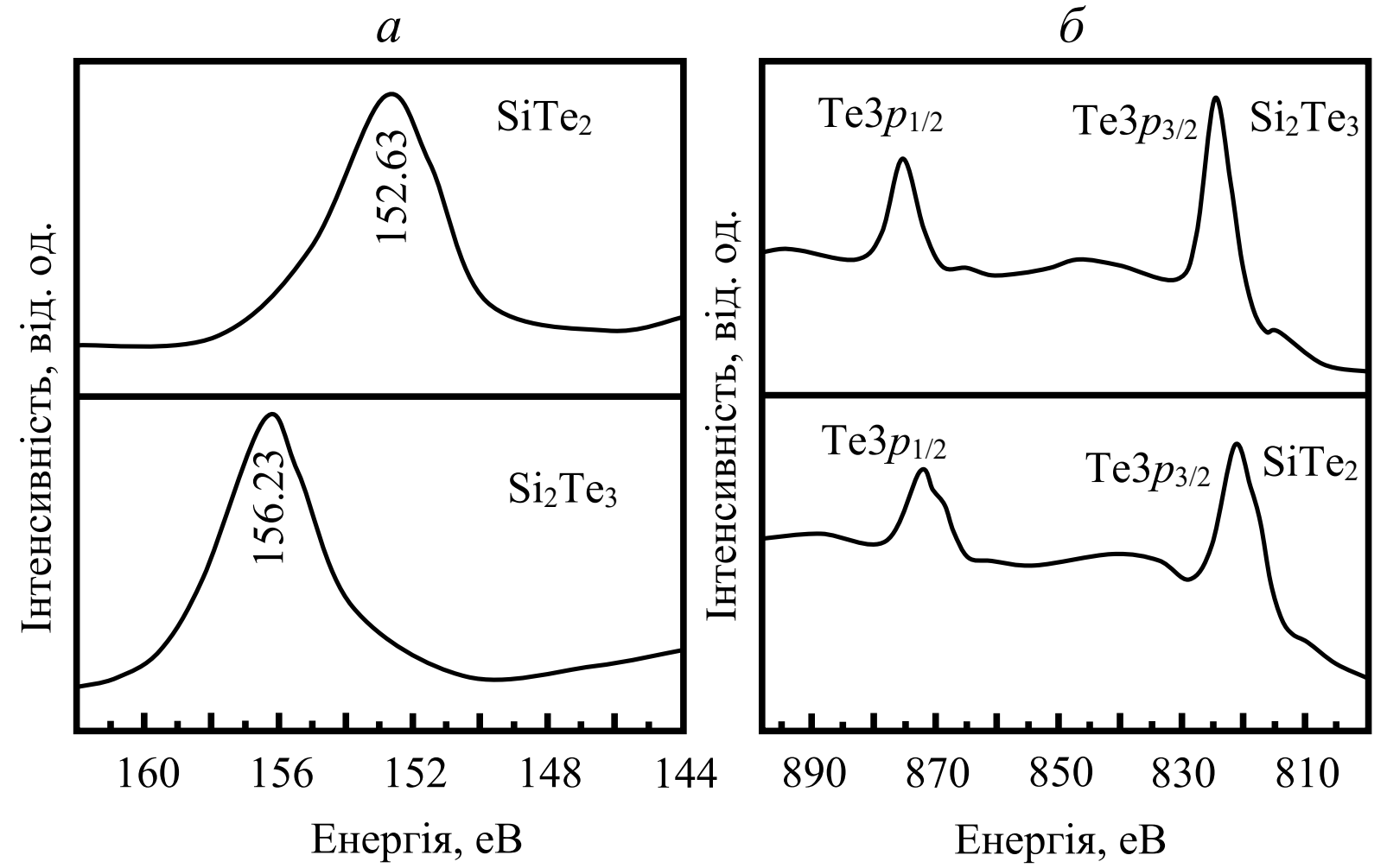

Рис. 1.10. Рентгенівські фотоелектронні спектри Si2*p*-рівнів (*а*) і Te3*p*-рівнів (*б*) для $SiTe_2$ і $Si_2Te_3$ [7].

елемента існує свій набір енергій остовних рівнів, при цьому енергії остовних рівнів, які відповідають різним елементам, досить добре енергетично розділені. Дана методика дозволяє ідентифікувати різні елементи по фотоелектронним спектрам, тобто отримати інформа-цію про елементний склад даного кристала. Визначаючи енергії піків у рентгенівських фотоелектронних спектрах, можна отримати інформацію не тільки про те, атоми якого елемента знаходяться у даному кристалі, але і в якому вони хімічному стані. Формування хімічного зв'язку між атомами твердого тіла (кристала), яке супроводжується перерозподілом електронної густини, приводить до зміни енергії зв'язку електронів, що, природно, буде проявлятися і в зміні кінетичної енергії фотоелектронів.

Енергія зв'язку $E_j$ остовних електронів (електронів внутрішніх оболонок атома) специфічні для кожного хімічного елемента, та їх аналіз знаходить широке застосування у дослідженнях елементного складу твердого тіла. Величина енергії зв'язку залежить також від фізико-хімічного стану атома і тому вимірювання енергетичного зсуву остовних рівнів використовується для ідентифікації хімічних сполук [28] і, зокрема, кристалів і нанопластин $SiTe_2$ і $Si_2Te_3$ [7, 30]. Величина ефекту може досягати кількох (іноді навіть десять і більше) електрон-вольт. Реєстрація таких зсувів не вимагає надто високої енергетичної роздільної здатності приладу і проводиться за допомогою стандартних рентгеноелектронних спектрометрів.

Метод фотоелектронної спектроскопії полягає в опроміненні досліджуваної речовини потоком монохроматичних фотонів та аналізі енергетичного спектра фотоелектронів. Спектроскопія остовних рівнів, є одним з різновидів цього методу, заснована на процесі збудження остовних електронів атомів. Спектр остовних фотоелектронів є дискретним. Кожна його лінія пов'язана із збудженням електронів із певного рівня атома.

***Хімічний зсув.*** Незважаючи на сталість енергії остовних рівнів атома, в різних речовинах є певна різниця в енергіях зв'язку для даного атома при переході від однієї речовини до іншої [31]. Як випливає з експериментальних даних, енергія зв'язку $E_{зв}$ електронів остова дещо змінюється при зміні характеру хімічного оточення атома, спектр якого вивчається. Зміну енергії зв'язку ($E_{зв}$) для електронного рівня одного і того ж елемента в різних сполуках прийнято називати *хімічним зсувом*. Хімічний зсув – це зсув першопочаткової енергії зв'язку, викликаний зміною електронного оточення атома. Зсув енергії внутрішніх електронів залежно від хімічного оточення ілюструє рис. 1.9 для $SiTe_2$ та $Si_2Te_3$. Для $SiTe_2$ характерні два добре виражені Te $3d_{5/2}$ піки при енергіях зв'язку 574.68 еВ і 570.98 еВ (рис. 1.9, *а*). Енергія зв'язку чистого телуру $3d_{5/2}$ 573 еВ. У сполуці $Si_2Te_3$ ці два піки зміщуються в область високих енергій зв'язку з одночасним різким зменшенням інтенсивності низькоенергетичного піка (рис. 1.9, *б*).

На відміну від рентгенівських фотоелектронних спектрів валентних смуг $SiTe_2$ і $Si_2Te_3$, спектри поглинання та фотоелектронні спектри внутрішніх *p*- і *d*-рівнів Si і Te становлять значний інтерес для аналізу формування хімічного зв'язку в цих сполуках. Із рис. 1.9 і 1.10 видно, що відбувається зменшення енергії зв'язку внутрішніх 2*p*-ліній кремнію, а також Te3*d*- і 3*p*-ліній $SiTe_2$ в порівнянні з $Si_2Te_3$. Спостережуваний зсув свідчить, частково, на користь зменшення іонного вкладу в хімічний зв'язок кремнію з його оточенням. Цей факт важливий, оскільки самі собою енергетичні зсуви мають складну природу, і саме з цієї причини важливо розглядати одночасно поведінку внутрішніх рівнів і спектрів поглинання. Як випливає з аналізу рис. 1.8–1.10, обидві серії експериментальних результатів дають узгоджену картину.

### 1.4. КРИСТАЛІЧНА СТРУКТУРА Te, $SiTe_2$ ТА $Si_2Te_3$

Тип кристалічної структури характеризується заданням просторової групи та розподілом атомів за позиціями Вайкоффа. У реальних кристалічних структурах кратність займаних атомами позицій пов'язана зі стехіометричним складом кристала, який задається хімічною формулою. При описі типу кристалічної структури наводиться також число формульних одиниць у кристалографічній комірці.

**1.4.1. Кристалічна структура телуру.** Телур кристалізується в сильно анізотропній тригональній структурі, симетрія якої описується просторовою групою $P3_121$ ( $D_3^4$ ), з параметрами ґратки: *a* = 4.456 Å, *c* = 5.921 Å (позиція Вайкоффа 3*a* з координатами (0.2636, 0.0000, 0.3333)) [32] та *a* = 4.457 Å, *c* = 5.929 Å (позиція Вайкоффа 3*a* з координатами (0.2633, 0.0000, 0.3333)) [33]. Кристалічна структура тригонального телуру складається з нескінченних спіральних ланцюжків ковалентно зв'язаних атомів, витягнутих уздовж осі *c* (вздовж напрямку [001]). Його симетрія приводить до існування двох дзеркально-ізомерних (енантіоморфних) форм, що містять як право- ($D_3^4$ ( $P3_121$)), так і лівосторонні ($D_3^6$ ($P3_221$)) спіралі (на рис. 1.11, *a* наведено правосторонній варіант) [32, 33]. Нижній індекс 3 у символах $D_3^4$ або $D_3^6$ вказує на те, що вісь найвищої симетрії – це вісь третього порядку (тригональна); верхні індекси 4 і 6 – довільні символи, що вказують, що просторова група містить праві гвинтові осі

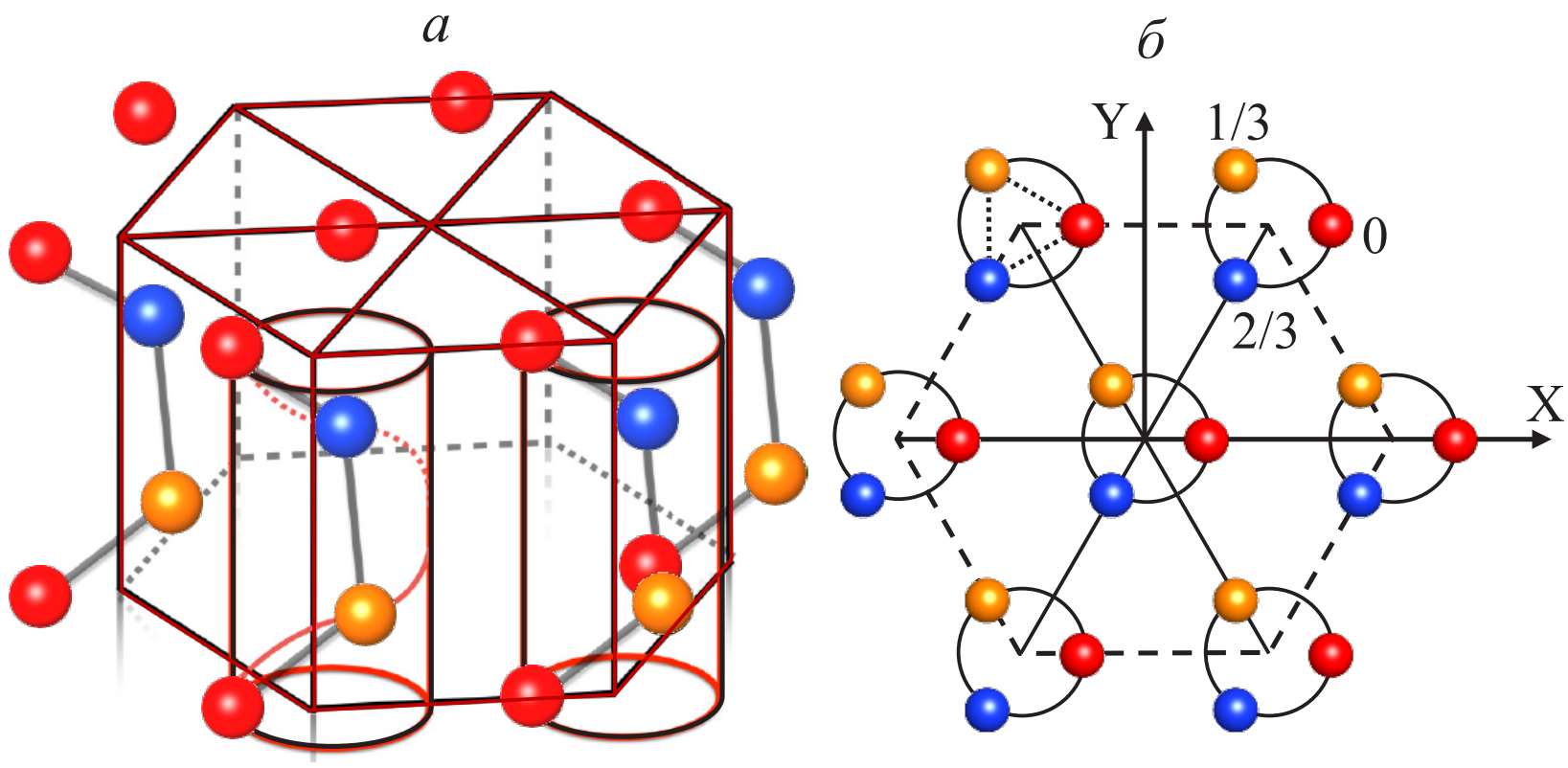

Рис. 1.11. *а* – Кристалічна ґратка тригонального телуру; *б* – проекція атомів тривимірної кристалічної структури телуру на площину, перпендикулярну до осі *z*. Числа біля кожного атома позначають *z*-координату в одиницях *c*.

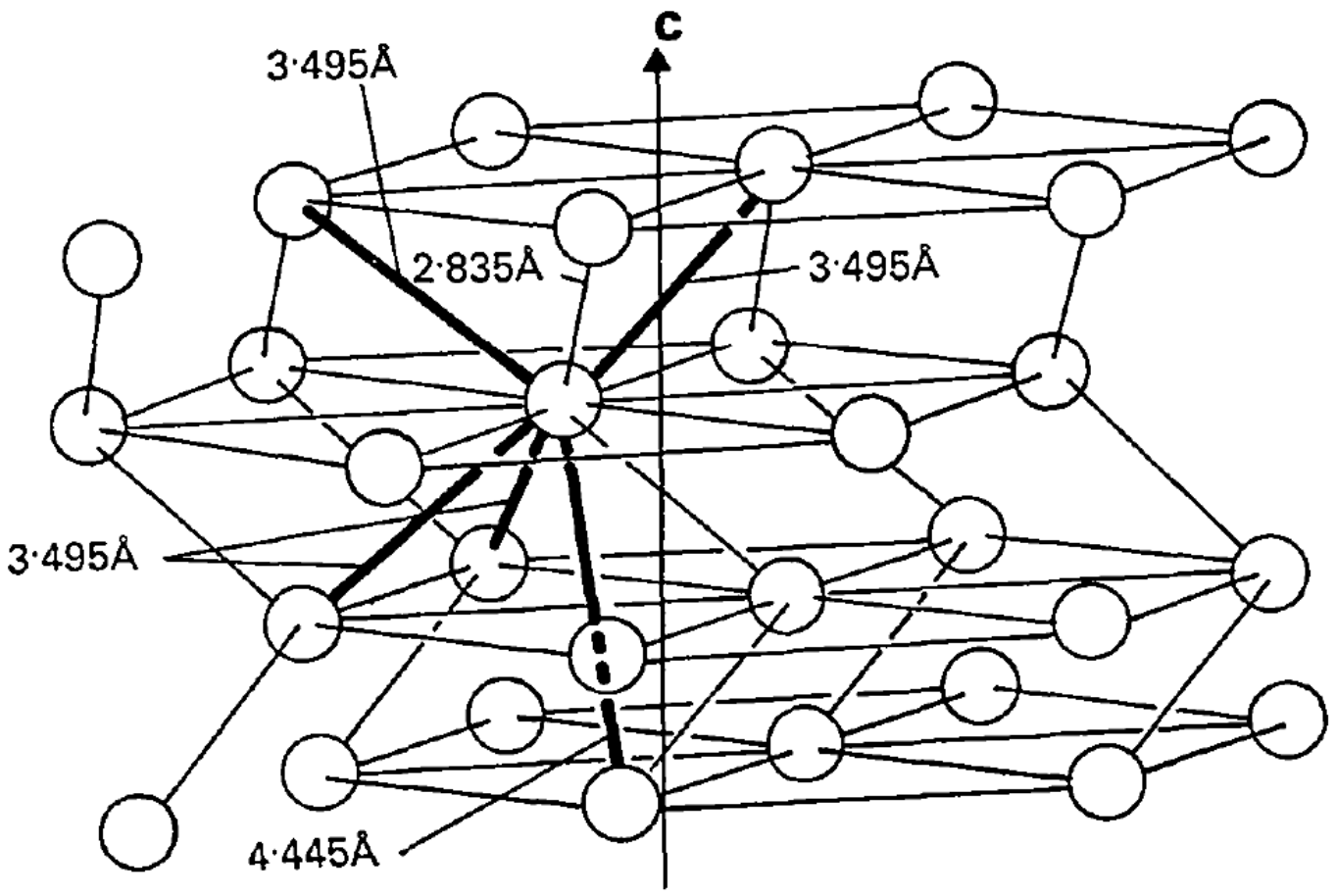

Рис. 1.12. Міжатомні відстані у тригональному телурі.

( $D_3^4$ ) і ліві ( $D_3^6$ ). Спіральні ланцюжки розташовані паралельно один до одного по вершинах та у центрі правильного шестикутника (рис. 1.11, *а*). Тому телур відносять до гексагональної ґратки. Три атоми, що входять в елементарну комірку становлять один «виток» нескінченної спіралі. При цьому кожен четвертий атом розташований точно над першим усередині ланцюжка, в результаті чого проекція на площину, яка перпендикулярна до осі *c*, являє собою рівносторонній трикутник (рис. 1.11, *б*). Кожен атом в ланцюжковій модифікації телуру має два найближчих сусіди у власному ланцюжку на відстані 2.835 Å і чотири атоми, що належать сусіднім ланцюжках, на більш віддалених відстанях 3.495, 4.445 Å (рис. 1.12). Радіус спірального ланцюжка рівний 1.1756 Å, кут між атомами Te–Te–Te 103.14° [32]. Кожен атом телуру зв'язаний з двома найближчими атомами по ланцюжку ковалентними зв'язками. Між однотипними атомами сусідніх ланцюжків діють слабкі сили ван-дер-Ваальса, що є причиною низької температури плавлення телуру. Структуру Te можна розглядати також як складову з трьох простих зв'язаних гексагональних ґраток, кожна з яких поставляє в елементарну комірку по одному атому (рис. 1.11).

Ця структура надає Te притаманну хіральність, а також сильну тенденцію до одновимірного (1D) зростання, так що його кристали завжди ростуть вздовж осі *c*, і мають тенденцію утворювати одновимірні структури.

**1.4.2. Кристалічна структура $SiTe_2$.** Дителурид кремнію належить до родини шаруватих сполук типу $A^{IV}B_2^{VI}$ ( A = Si, Ge, Sn; B = S, Se, Te), характерною ознакою яких є наявність поліморфізму і політипізму [1]. Кристалічна структура $SiTe_2$ складається з ідентичних тришарових пакетів «сендвічів» Te–Si–Te із сильним іонно-ковалентним зв'язком, зв'язаних між собою слабкими силами ван-дер-Ваальса. Існують два типи координації атомів кремнію в сендвічі: тригональна призматична, якщо шари халькогену займають однакові положення, та октаедрична, при якій атоми халькогену верхнього і нижнього шару займають положення, яке відповідає щільній упаковці. Перші результати рентгенографічного дослідження (дифрактометричним методом порошку, рентгенограми обертання і вайсенбергограми) червоних шаруватих кристалів $SiTe_2$, вирощених методом сублімації [18, 19], показали, що дителурид кремнію кристалізується в гексагональній шаруватій структурі з параметрами ґратки $a$ = 4.28, $c$ = 6.71 Å, просторова група $D_{3d}^3$–$C\bar{3}m$. Пікномет-Таблиця 1.2. Кристалоструктурні дані телуридів кремнію.

| Сполука | Тип кристалічної структури, | Просторова група | Число фор. од. | Параметры гратки, Å | | | Густина, г/см$^3$ | | Література |
|---|---|---|---|---|---|---|---|---|---|
| | | | | *a* | *b* | *c* | експе-римент | розра-хунок | |
| $SiTe_2$ | Гексагональна | $C\bar{3}\,\mathrm{m}-D^3_{3\mathrm{d}}$ | Z = 1 | 4.28 | | 6.71 | 4.39 | | [18, 19] |
| | Тригональна | $P\bar{3}\,\mathrm{m}1-D^2_{3d}$ | Z = 1 | 4.2887 | | 6.733 | 4.38 | 4.41 | [22] |
| | Тригональна | $P\bar{3}\,\mathrm{m}1-D^2_{3d}$ | Z = 1 | 4.2858 | | 6.7286 | | | [7] |
| | Триклінна | $P1-D^2_{3d}$ | Z = 1 | 3.945 | | 7.076 | | | [34] |
| $Si_2Te_3$ | Тригональна | $P\bar{3}\,1c-D^2_{3d}$ | Z = 4 | 7.430 | | 13.482 | 4.42 | 4.52 | [8] |
| | Тригональна | $P\bar{3}\,1c-D^2_{3d}$ | Z = 4 | 7.425 | | 13.467 | | | [25] |
| | Тригональна | $P\bar{3}\,1c-D^2_{3d}$ | Z = 4 | 7.429 | | 13.471 | 4.5 | 4.53 | [35] |
| | Тригональна | $P\bar{3}\,1c-D^2_{3d}$ | Z = 4 | 7.422 | | 13.465 | | | [4] |
| | Тригональна | $P\bar{3}\,1c-D^2_{3d}$ | Z = 4 | 7.427 | | 13.475 | 4.56 | 4.566 | [ 6] |
| | Тригональна | $P\bar{3}\,1c-D^2_{3d}$ | Z = 4 | 7.4216 | | 13.4521 | | | [7] |

рична густина кристалів $SiTe_2$ $\rho_{вим} = 4.39$ г/см$^3$ [19].

За даними рентгеноструктурних досліджень кристалів $SiTe_2$, вирощених методом Бріджмена [22], дителурид кремнію кристалізується в тригональній структурі типу $CdI_2$ з параметрами ґратки $a$ = 4.2887 Å, $c$ = 6.7330 Å. Довжини зв'язків Si–Te та Te–Te рівні 3.013 і 4.125 Å відповідно.

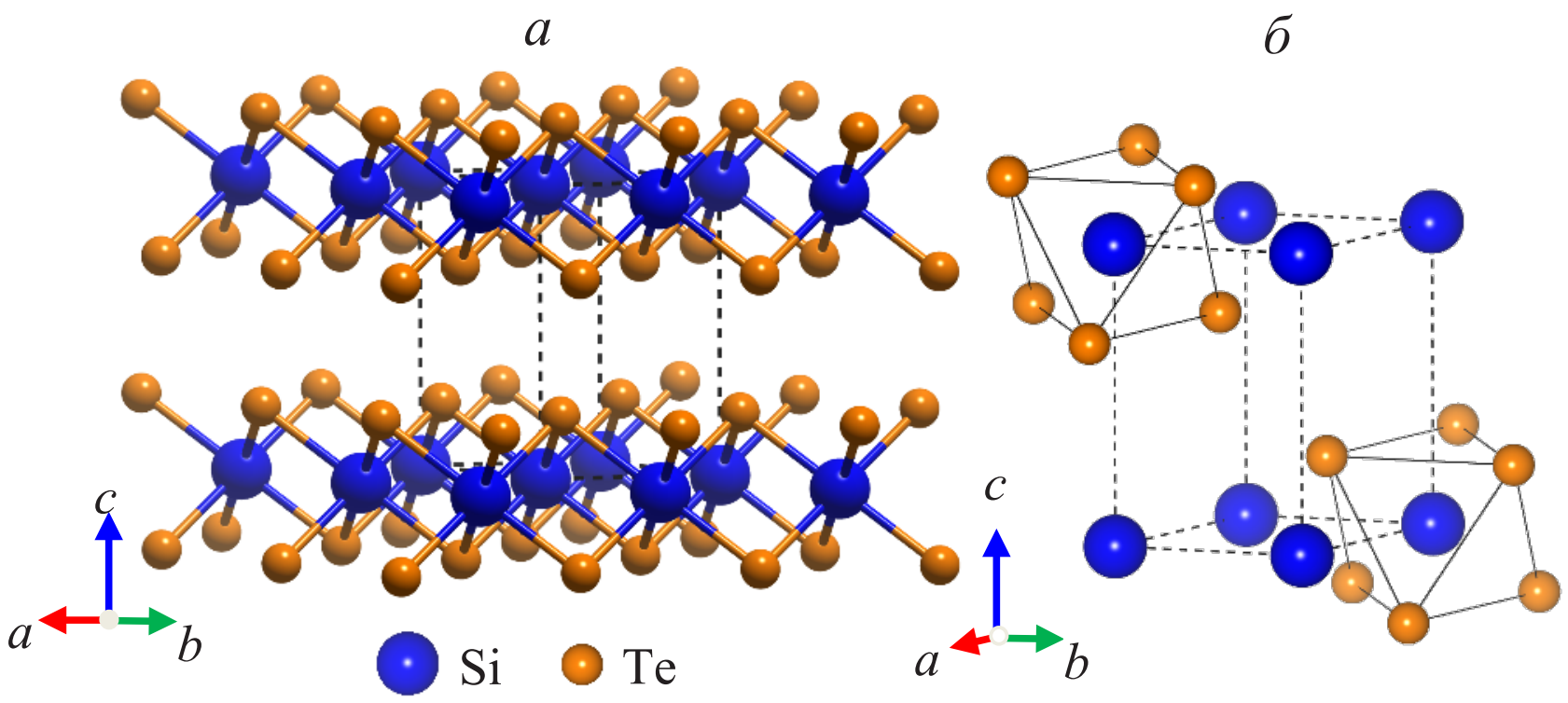

Рис. 1.13. Кристалічна структура (*а*) та елементарна комірка з виділеними октаедрами [$SiTe_6$] (*б*) тригонального $SiTe_2$.

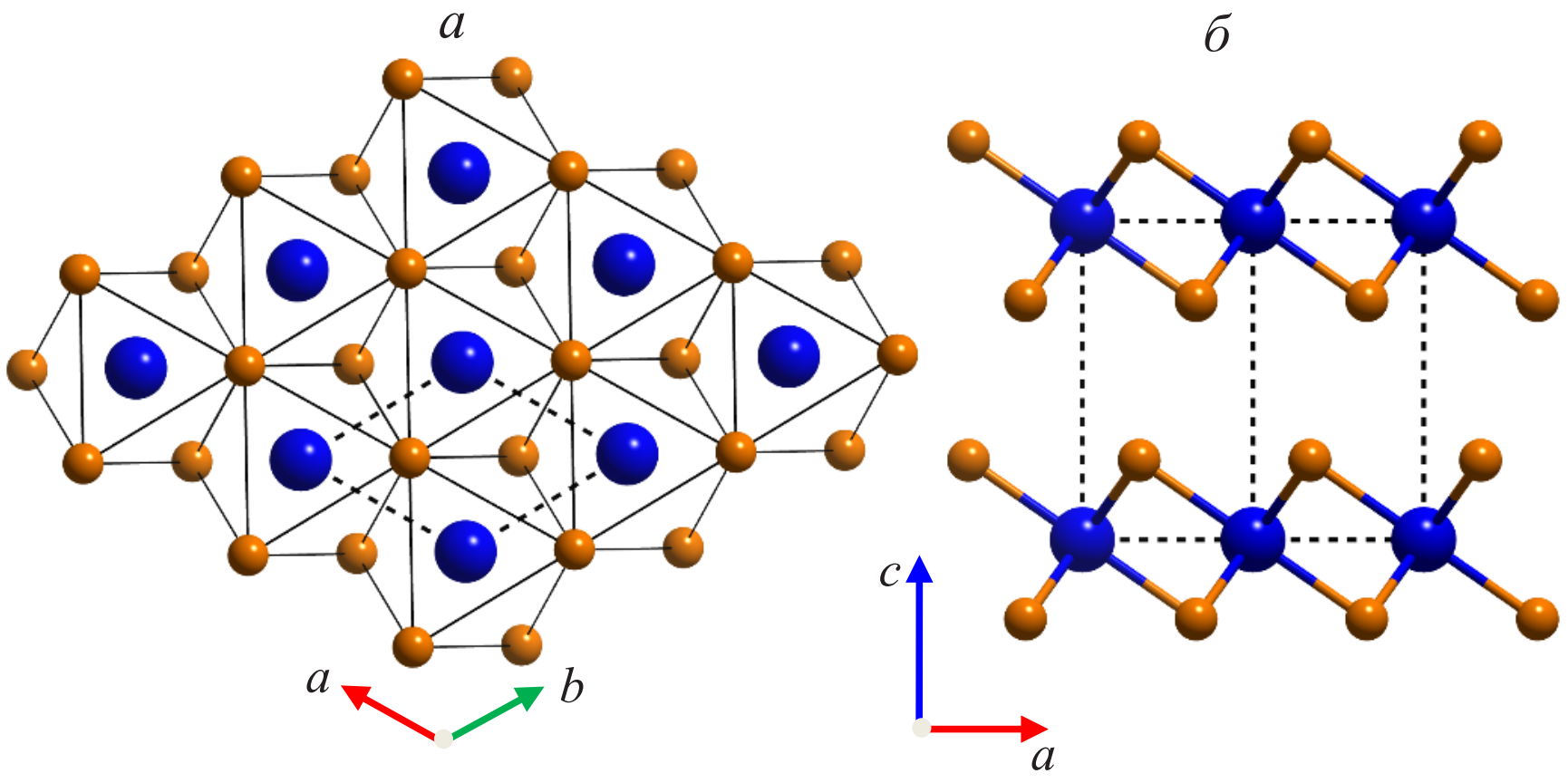

Рис. 1.14. Проекції кристалічної структури тригонального $SiTe_2$ на площини XY (*а*) і XZ (*б*).

Авторами [7] знову проведено дослідження кристалічної структури дителуриду кремнію і підтверджена його шарувата будова. У

$SiTe_2$ чотиривалентний кремній ($Si^{4+}$) шестикратно координований атомами телуру ($Te^{2-}$). Найближчими сусідами атома кремнію є шість атомів телуру, які розташовані у вершинах октаедра [$SiTe_6$]. Октаедри [$SiTe_6$], зв'язані між собою спільними ребрами, формують тришарові пакети – «сендвічі» Te–Si–Te, паралельні площині (001) (рис. 1.13). Симетрія кристалічної ґратки $SiTe_2$ описується просторовою групою $D_{3d}^3$ ($P\bar{3}m1$), а кристалічний клас – точковою групою $D_{3d}$ ($\bar{3}$ 2/m і $\bar{3}$ m). Параметри кристалічної ґратки: $a = b = 4.2858$ Å; $c = 6.7286$ Å; $\gamma = 120°$ [7] є близькими до наведених у роботах [18, 19]. Просторова група $P\bar{3}m1$ характеризується інверсійною віссю 3-го порядку, спрямованої вздовж нормалі до площини «сендвічів». Зниження порядку симетрії до третього, незважаючи на гексагональну симетрію всіх моношарів, що входять до складу «сендвіча» пов'язана з різним зсувом халькогенних площин щодо поверхні атомів кремнію. Такий зсув можна розглядати як розворот халькогенних площин на кут 60° однієї відносно одної. Оскільки кожен атом кремнію координований трьома атомами телуру з кожної площини, то такий розворот приводить до нееквівалентності верхньої та нижньої площин халькогену і, як наслідок, до зниження симетрії.

Структуру $SiTe_2$ можна представити також як таку, що складається з трьох плоских гексагональних сіток – двох, у яких атоми телуру гексагонально щільноупаковані та однієї сітки атомів кремнію між ними, які займають половину октаедричних позицій, які періодично повторюються вздовж осі *c* кристала, вони чергуються в послідовності –Te–Si–Te– (рис. 1.14, *б*).

Відстань між тришаровими пакетами (3.176 Å) перевищує відстань (1.776 Å) між атомними моношарами всередині одного «сендвіча». Усередині тришарових пакетів зв'язок має іонно-ковалентний характер, а зв'язок між тришаровими пакетами здійснюється переважно силами ван-дер-Ваальса, чим і обумовлена велика анізотропія фізичних властивостей кристалів $SiTe_2$.

В елементарній комірці $SiTe_2$ містяться три атоми, що належать лише одному тришаровому пакету. Елементарна комірка $SiTe_2$, якщо користуватися при її описі звичайною координатною системою з двома векторами $\boldsymbol{a}_1$ і $\boldsymbol{a}_2$ в площині XY, розташованими один по відношенню до іншого під кутом в 120° і вектором $\boldsymbol{c}$ в Z напрямку, складається з атома Si, що знаходиться в позиції Вайкоффа 1*a* (0, 0, 0), та двох атомів телуру, що знаходяться в позиції Вайкоффа 2*d* (1/3, 2/3, *z*) з координатами (1/3, 2/3, *z*), (2/ 3, 1/3, –*z*) с $z = 0.265$ [19];

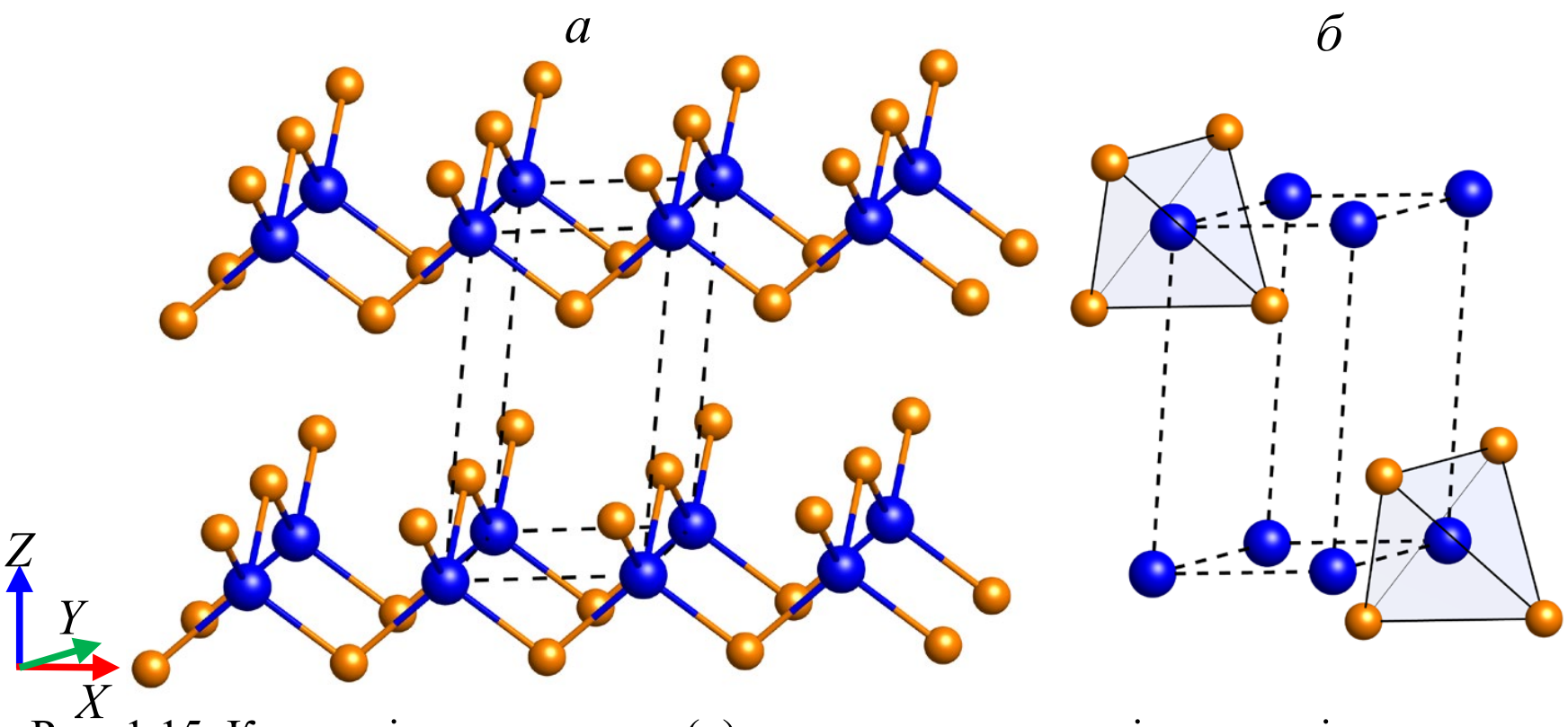

Рис. 1.15. Кристалічна структура (*а*) та елементарна комірка з виділеними тетраедрами [$SiTe_4$] (*б*) триклінного $SiTe_2$

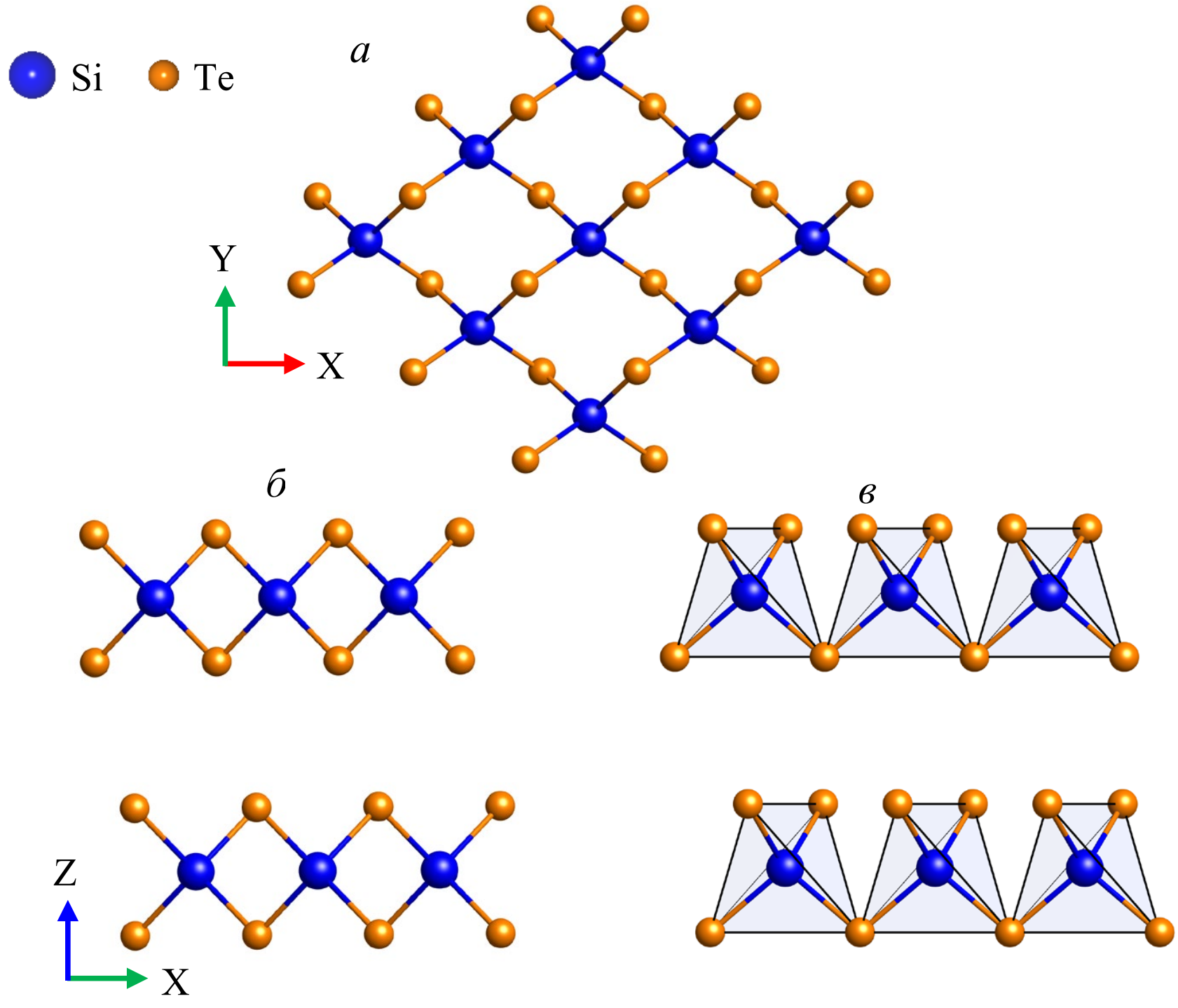

Рис. 1.16. Проекції кристалічної структури $SiTe_2$ на площини XY (*а*) і XZ (*б*).

0.255 [22]; 0.264 [7]. Симетрія зазначених позицій описується локальними групами $3m$ і $3m$ відповідно.

Використовуючи результати еволюційного алгоритму та першопринципні розрахунки автори [34] передбачили нову шарувату кристалічну структуру $SiTe_2$, яка є енергетично більш стабільною, ніж структура типу $CdI_2$. Прогнозована структура має триклінну кристалічну ґратку з просторовою групою $P1$. Кристалічна структура триклінної фази $SiTe_2$ і проекції структури на площини XY і XZ наведені на рис. 1.15 та 1.16 відповідно. Примітивна елементарна комірка складається з трьох атомів (один атом Si (0, 0, 0) і два атома Te1 (–0.0383, 0.4632, 0.2330) і Te2 (0.5360, 0.0391, –0.2330)). Параметри ґратки наведені в табл. 1.2. Основним структурним елементом даної фази є тетраедр [$SiTe_4$]. Автори [34] аналізують чому прогнозована ними триклінна структура $SiTe_2$ є більш стабільною ніж структура типу $CdI_2$. Як видно з рис. 1.15, *б* атом кремнію має чіткий тетраедричний зв'язок, що вказує на ковалентний зв'язок, який викликаний гібридизацію 3*s*- і 3*p*-орбіталей Si. Натомість у структурі типу $CdI_2$ катіон (Si) шестикратно координований атомами Te – рис. 1.13, *б*. Враховуючи сильну тенденцію Si до утворення таких ковалентних зв'язків, логічно стверджувати, що утворення шестикоординованого Si пов'язане з втратою енергії порівняно з чотирикоординованим Si. З іншого боку, тетраедри [$SiTe_4$] сильно спотворені. Причиною цього спотворення є той факт, що через великий розмір атом Te віддає перевагу гексганальній щільній упаковці (ГЩУ) структури, яка не повністю сумісна з тетраедричним зв'язком Si. Таким чином, прогнозована структура $SiTe_2$ має спотворений тетраедричний зв'язок для атомів Si та спотворену гексганальну щільну упаковку атомів Te, що вказує на те, що структура формується як компроміс цих двох конкуруючих вимог.

**1.4.3. Кристалічна структура $Si_2Te_3$.** Сесквітелурид кремнію належить до сполук, яким властива наявність природних дефектів, викликаних особливостями їх кристалохімії. Кристали $Si_2Te_3$ містять велику кількість природних стехіометричних вакансій (~$10^{27}$ м$^3$), пов'язаних з природою самої речовини, в якій третина вузлів катіонної підґратки вакантна. Сам факт існування таких вакансій та їх наявність у кристалі не залежить від методу та умов його вирощування, а визначається виключно виконанням умови рівності вузлів катіонної та аніонної підґраток у тригональній структурі.

Таким чином, $Si_2Te_3$ належить до групи напівпровідників із позиційною невпорядкованістю. Це відповідає ситуації, коли число пози-

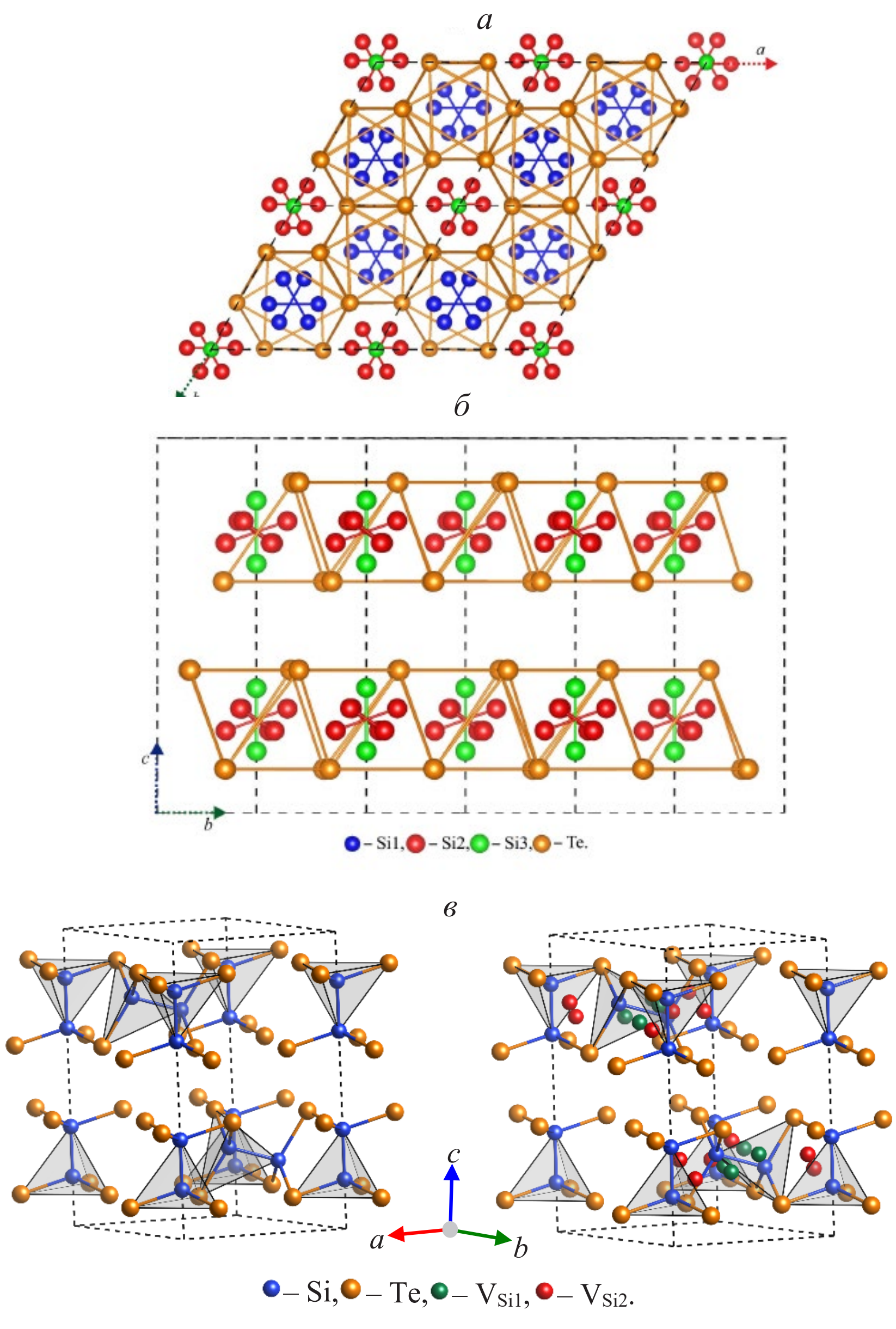

Рис. 1.17. Кристалічна структура (*а*) та проекції структури на площини XY (*б*) і XZ (*в*) $Si_2Te_3$.

цій атомів певного сорту (у разі кремнію) перевищує число самих атомів, а розподіл атомів по цих позиціях носить, принаймні (частково), випадковий характер.

Структура сполуки $Si_2Te_3$ тригональна, належить до просторової групи $P\overline{3}1c$ [9, 25, 27]. Кристалічна структура та проекції структури на площині XY та XZ наведені на рис. 1.17. За звичайних умов для $Si_2Te_3$ характерна структура, похідна від структурного типу $CdI_2$ з атомами Te, що утворюють щільну гексагональну упаковку та пари атомів Si–Si, які заселяють половину октаедричних порожнин [9]. В основі структури гексагональна щільна упаковка атомів телуру у двопакетному вигляді, кожен з яких представляє два шари атомів телуру, між якими знаходяться атоми кремнію у вигляді гантельних утворень $Si_2$ у порожніх міжпакетних просторах, відстані між найближчими атомами телуру двох сусідніх шарів 4.02 Å. Гантелі атомів кремнію можуть розташовуватися двояко: або вертикально по всіх чотирьох ребрах «*c*» комірки з міжатомною відстанню Si–Si = 2,27 Å, або близько до горизонтального (~18° до горизонтальної площини) з відстанню Si–Si = 2.35 Å. Пари атомів кремнію розміщуються всередині майже правильних октаедрів з атомів телуру [$Te_6$]. Атоми кремнію вертикальних гантелей знаходяться на відстані 2.53 Å від трьох найближчих атомів телуру з валентними кутами Te–Si–Te = 113.8°. Атоми кремнію «горизонтальних» гантелей також координовані трьома найближчими атомами телуру на відстані 2.45; 2.13 і 2.66 Å і з валентними кутами Te–Si–Te 112.4; 114.6 та 118.5° або 2.46; 2.56 та 2.61° Å і з кутами 113.5; 114.9 та 116.9°. Таким чином, всі атоми кремнію координовані тетраедрами [$SiTe_3Si$], де четвертою вершиною тетраедра є найближчий по гантелі партнер – атом кремнію.

Найважливішою особливістю структури є статистичне розміщення 8 атомів кремнію у двох позиціях 12*i* та однієї 4*e*. Вказані позиції заповнені з дефіцитом в 71%, оскільки в них замість 28 атомів кремнію розміщуються лише 8. Причому, в першій позиції 12*i* розміщуються 4 атоми, у другій – 2 атоми і в позиції 4*e* – також 2 атоми. Іншими словами, обидві позиції 12*i* зайняті на 1/3 або 1/6 відповідно, а позиція 4*e* – наполовину. Таким чином, виявляється, що гантелі $Si_2$ розподіляються у структурі на «вертикальні» та «горизонтальні» у співвідношенні 1:3.

Така особливість – гантельне спарювання атомів компонентів сполук – досить широко відома, особливо для структур халькогеногіподифосфатів двовалентних металів (гантелі P–P), сульфіду галію (Ga–Ga) та ін. Але з частково заселеними спареними атомами координатними позиціями структура $Si_2Te_3$ становить виняткову рідкість, що залишає широкі можливості обговорення властивостей цієї сполуки.

Слабкий зв'язок між сусідніми тришаровими пакетами приводить до виникнення в шаруватих кристалах $SiTe_2$ та $Si_2Te_3$ крайових і часткових дислокацій, які пов'язані зі зміщенням частини кристала за базисними площинами. З частковими дислокаціями зв'язані дефекти пакування, які є одними із основних типів дефектів у шаруватих телуридах кремнію [10, 24]. Дефект пакування полягає у зміщенні шару з правильного положення в структурі. Оскільки напрям зміщення є паралельним до шарів, то роль дефектів зводиться до порушення регулярності структури в напрямі, перпендикулярному до зміщення, який у шаруватих кристалах співпадає з *c*-віссю. Звідси випливає, що існування дефектів упаковки приводить до одновимірного розупорядкування структури вздовж кристалографічної осі *c*. Поява дефектів упаковки не змінює ні числа ближніх сусідів, ні відстані до них. Але із-за зміни в розташуванні наступних шарів (не найближчих) у випадку шаруватих кристалів появляється одновимірне розупорядкування у напрямку перпендикулярному до шарів, що сильно відображається на анізотропії явищ переносу.

**1.4.4. Фазовий перехід в $Si_2Te_3$ індукований температурою і тиском.** Електронографічні дослідження характеру зміни структури кристалів $Si_2Te_3$ в процесі нагрівання показали, що при температурі 673 К відбивання від тригональної ґратки $Si_2Te_3$ набувають дифузного характеру [10], що свідчить про часткове розупорядкування структури. Аналіз інтенсивності дифузного розсіювання показав, що в діапазоні температур 673–723 К має місце дисоціація димерів Si–Si з міграцією частини атомів Si в тетраедричні положення. При температурі 723 К дифузність рефлексів зникає і виникає надструктура з подвоєним параметром *a*. Структура цієї високотемпературної впорядкованої β-фази характеризується розміщенням Si в одній чверті тетраедричних порожнин гексагональної щільної упаковки з атомів Te. Таким чином, зміни структури $Si_2Te_3$ при нагріванні стосуються тільки підґрати з атомів Si з повним збереженням будови підґратки з атомів Te. Автори [10] відзначають, що всі явища, які спостерігаються, повною мірою виявляються і при порушенні стехіометрії $Si_2Te_3$ – для складів $Si_{2-x}Te_3$ с $0.5 < x < 1$, що пояснюється статистичним заміщенням атомів Te додатковими атомами Si.

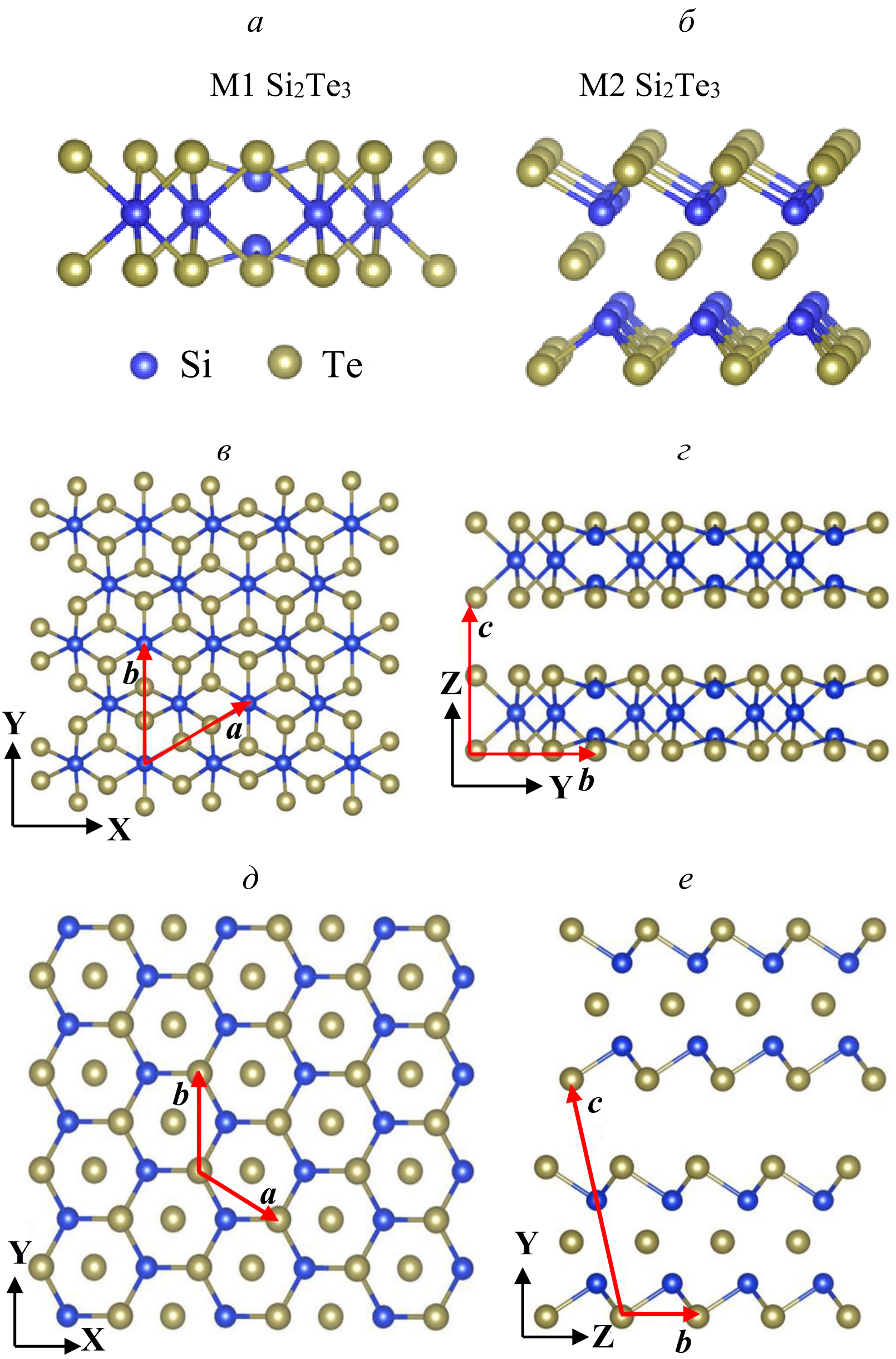

Рис. 1.18. Кристалічна структура (*а*, *б*) та проекції металічних фаз $M_1$ (*в*, *г*) і $M_2$ (*д*, *е*) високого тиску $Si_2Te_3$ [36].

Спеціально нелеговані та інтеркальовані марганцем нанопластини $Si_2Te_3$ авторами [35] були піддані тиску до 12 ГПа з використанням технології комірки з алмазною наковальнею. Під час стиснення спостерігалась зміна кольору нанопластин від прозорого червоного до непрозорого чорного, що вказує на фазовий перехід напівпровідник −метал. Дослідження спектрів комбінаційного розсіювання нанопластин $Si_2Te_3$ під дією зовнішнього тиску показали, що фазовий перехід відбувається при 9.5 ± 0.5 ГПа, про що свідчить зникнення $A_{1g}$ моди. Інтеркаляція нанопластин $Si_2Te_3$ марганцем до 1ат.% приводить до зниження тиску фазового переходу до 7.5 ± 1 ГПа [35]. Однак кристалічна структура металічного $Si_2Te_3$ не була досліджена.

Використовуючи першопринципні методи розрахунку автори [36] запропонували дві високотемпературні металічні фази $Si_2Te_3$: гексагональну $M_1$ з параметрами ґратки: $a$ = 6.82 Å, $b$ =6.82 Å, $c$ = 7.48 Å і триклінну $M_2$ з параметрами ґратки: $a$ = 3.90 Å, $b$ = 3.90 Å, $c$ = 10.93 Å. Обидві фази $M_1$ і $M_2$ $Si_2Te_3$ стехіометричні. Проекції кристалічних структур фаз $M_1$ і $M_2$ високого тиску наведені на рис. 1.18. На відміну від напівпровідникової тригональної фази $Si_2Te_3$ низького тиску, в якій присутні димери Si–Si, в обох металічних фазах $M_1$ і $M_2$ наявні окремі атоми кремнію, які відіграють важливу роль у металізації.

За результатами першопринципних розрахунків з використанням методу теорії функціонала електронної густини (DFT) моношар $Si_2Te_3$ може витримувати одновісну деформацію розтягу до 38%, що є самим високим показником серед усіх відомих двовимірних матеріалів [37]. Навіть при такій високій критичній деформації міцність на розрив складає 8.59 Н/м, що вказує на надзвичайну гнучкість даного матеріалу.

За даними [38] при тиску 12.9 ГПа та кімнатній температурі $Si_2Te_3$ стає аморфним. Ці автори не виявили фазового переходу при тиску 9.5 ГПа і кімнатній температурі про який повідомлялось у роботі [35]. Разом з тим, встановлено, що при нагріванні при різних тисках і температурах $Si_2Te_3$ починає розкладатися при 6 ГПа і 700К, 7.5 ГПа і 650 К і 11.5 ГПа і 500 К з утворенням фази типу $Mn_5Si_3$ (hexl), якій передує поява більш щільної гексагональної фази (hex2). Сесквітелурид кремнію $a$-$Si_2Te_3$ розкладається на суміш фаз, яка включає клатратну ($Te_8$@($Si_{38}Te_8$)) та гексагональну фази при високому тиску та високих температурах. Чим вищий тиск, тим нижча температура для двох фаз. При подальшому підвищенні температури починає синтезуватися клатрат типу-I $Te_8$@($Si_{38}Te_8$) [38].

## 1.5. ДОСЛІДЖЕННЯ ПОВЕРХНІ ШАРУВАТИХ КРИСТАЛІВ $Si_2Te_3$ МЕТОДОМ ОЖЕ-ЕЛЕКТРОННОЇ СПЕКТРОСКОПІЇ

Зважаючи на гігроскопічність кристалів $Si_2Te_3$, особлива увага приділяється аналізу стану їх поверхні, оскільки стан останньої сильно впливає на фізичні властивості. Інформативним методом дослідження якісного та кількісного елементного складу поверхні твердих тіл є метод оже-електронної спектроскопії (ОЕС) [39, 40]. Головною перевагою ОЕС порівняно з багатьма іншими методами є дуже мала глибина аналізу, що робить цю методику придатною для дослідження поверхні.

**1.5.1. Фізичні засади оже-електронної спектроскопії.** Емісія оже-електронів зумовлена оже-ефектом, який було відкрито в 1925 р. французьким ученим П. Оже. Оже-ефект є наслідком іонізації однієї з внутрішніх оболонок атома під дією первинного електронного пучка. Суть оже-процесу полягає в тому, що на незаповнений рівень *K* атомного остова переходить електрон із зовнішньої оболонки, а вся енергія, яка звільнилася, передається електрону, що знаходиться на іншій орбіталі. Цей електрон вилітає із зразка з характерною енергією і називається оже-електроном. При цьому енергія випущеного оже-електрона ніяк не залежить від енергії падаючого електрона та повністю визначається спектром енергетичних рівнів у кристалі.

На рис. 1.19 наведена схема, яка ілюструє процес утворення оже-електронів для атома кремнію. У цій схемі розглядається іонізація *K*-рівня падаючим електроном з енергією $E_p$, що перевищує енергію зв'язку $E_K$ електрона на *K*-рівні. Після іонізації остовного *K*-рівня на ньому утворюється вакансія (дірка), яка за час $\tau = 10^{-14} \div 10^{-16}$ с заповнюється електроном з якогось верхнього рівня (у схемі рис. 1.19, *а* – з рівня $L_1$). Енергія $E_K - E_{L_1}$, яка виділяється при цьому або переходить в енергію фотона характерного рентгенівського випромінювання, або передається ще одному (іншому) електрону, що знаходиться на тому ж або більш високому рівні, внаслідок чого цей електрон залишає атом. Перший процес називається рентгенівською флуоресценцією, на ньому заснований метод рентгенівської емісійної спектроскопії. Другий – оже-емісія. Енергія оже-електронів, що випускаються на прикладі, представленому на рис. 1.19, *а* дорівнює [39]:

$$E_A = E_K - E_{L_1} - E_{L_2} - U\left(L_1, L_2\right), \qquad (1.2)$$

де $E_A$, $E_K$, $E_{L_1}$, $E_{L_2}$ – енергія оже-електрона й енергії зв'язку електронів на рівнях $K$, $L_1$, $L_2$. Доданок $U(L_1, L_2)$ враховує зміну енергії зв'язку електрона лише на рівні $L_1$ (або $L_2$) за наявності вакансії лише на рівні $L_2$ (або $L_1$).

Електрон з енергією $E_A$, розрахований із співвідношення (1.2), позначається у відповідність з рівнями, задіяними при оже-емісії (рис. 1.19, *а*), як $KL_1L_{2,3}$ – оже-електрон. Якщо дірки, що утворилися при оже-емісії, знаходяться у валентній зоні, то вони позначаються буквою $V$ з індексами відповідних рівнів, наприклад $L_{2,3}VV$ – оже-перехід з початковою діркою в $L_{2,3}$-оболонці та дві кінцеві дірки у валентній зоні.

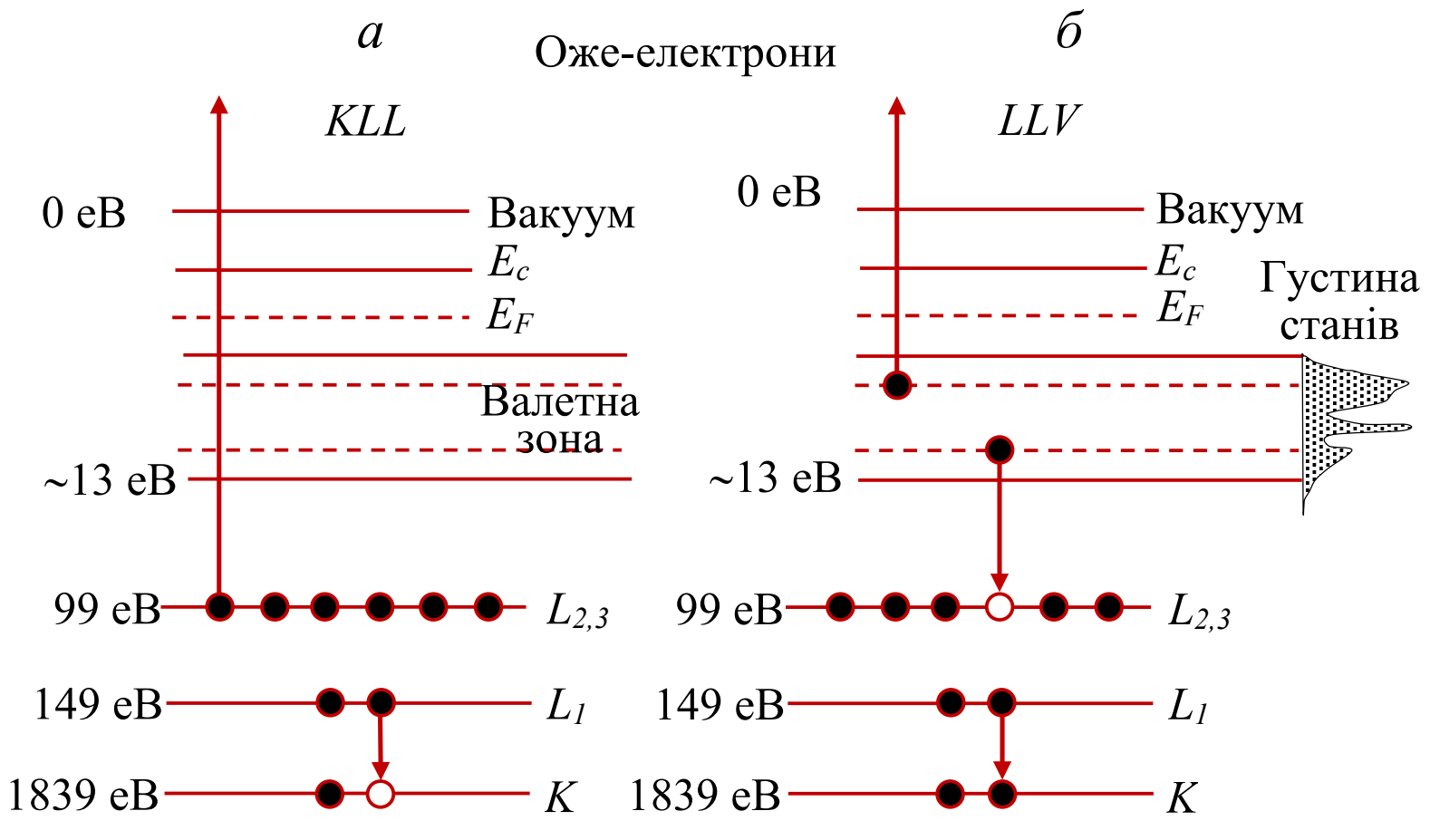

Рис. 1.19. Схема оже-процесів $KL_1L_{2,3}$ (*а*) і $LV_1V_2$ (*б*) при знятті збудження в атомі кремнію [40].

Оже-спектроскопія побудована на аналізі розподілу вилетівших оже-електронів по енергіях. Оскільки більша частина електронних рівнів носить дискретний характер, метод дає інформацію про енергетичне розташування рівнів, а отже, про хімічний склад речовини. Розподіл оже-електронів по кінетичних енергіях є оже-спектр. Зазвичай експериментальні оже-спектри представляють у вигляді першої похідної від кривої розподілу вторинних електронів по енергії $dN(E)/dE$. Це зв'язано з тим, що частка оже-электронів у загаль-ному потоці вторинних електронів незначна і оже-піки проявляються у вигляді слабких особливостей в інтегральному спектрі вторинних електронів. Диференціювання кривої *N(E)* дозволяє позбутися безструктурного фону вторинних електронів і точніше визначити положення оже-піка. При цьому диференціювання здійснюється електричними методами безпосередньо в процесі запису спектра.

Із співвідношення (1.2) випливає, що кінетична енергія оже-електрона, що випускається, визначається енергіями зв'язку відповідних атомних рівнів даного елемента і не залежить від енергії збуджуючих електронів. Таким чином, для кожного елемента періодичної таблиці існує певний, характерний тільки для даного елемента набір енергій оже-електронів. По характерному набору піків в енергетичному спектрі оже-електронів можна однозначно ідентифікувати елементний склад досліджуваної речовини. Для цього використовують атласи оже-спектрів, у яких наведено оже-спектри чистих елементів. Оже-спектри чистого кремнію, телуру і кисню, взяті з робіт [41, 42] наведено на рис. 1.20 та 1.21. Форма оже-ліній для Si і O в $SiO_2$ показана на рис. 1.22.

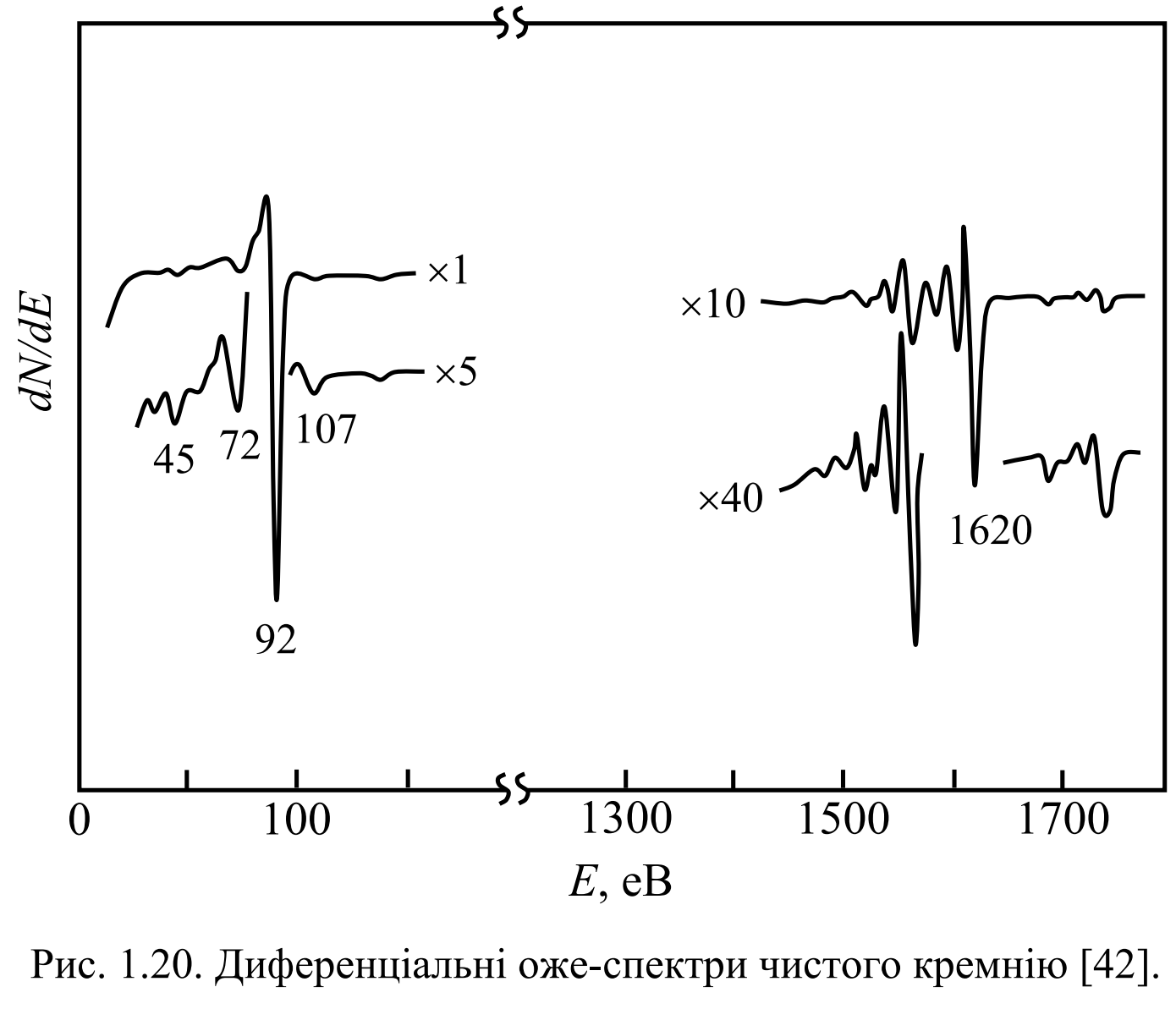

Рис. 1.20. Диференціальні оже-спектри чистого кремнію [42].

Найбільш інтенсивний оже-пік у чистому Si з'являється при 92 еВ і походить від переходу $L_{2,3}VV$ (рис. 1.20). Цей пік зазвичай контролюється для ідентифікації елементарного Si. На рис. 1.20 також

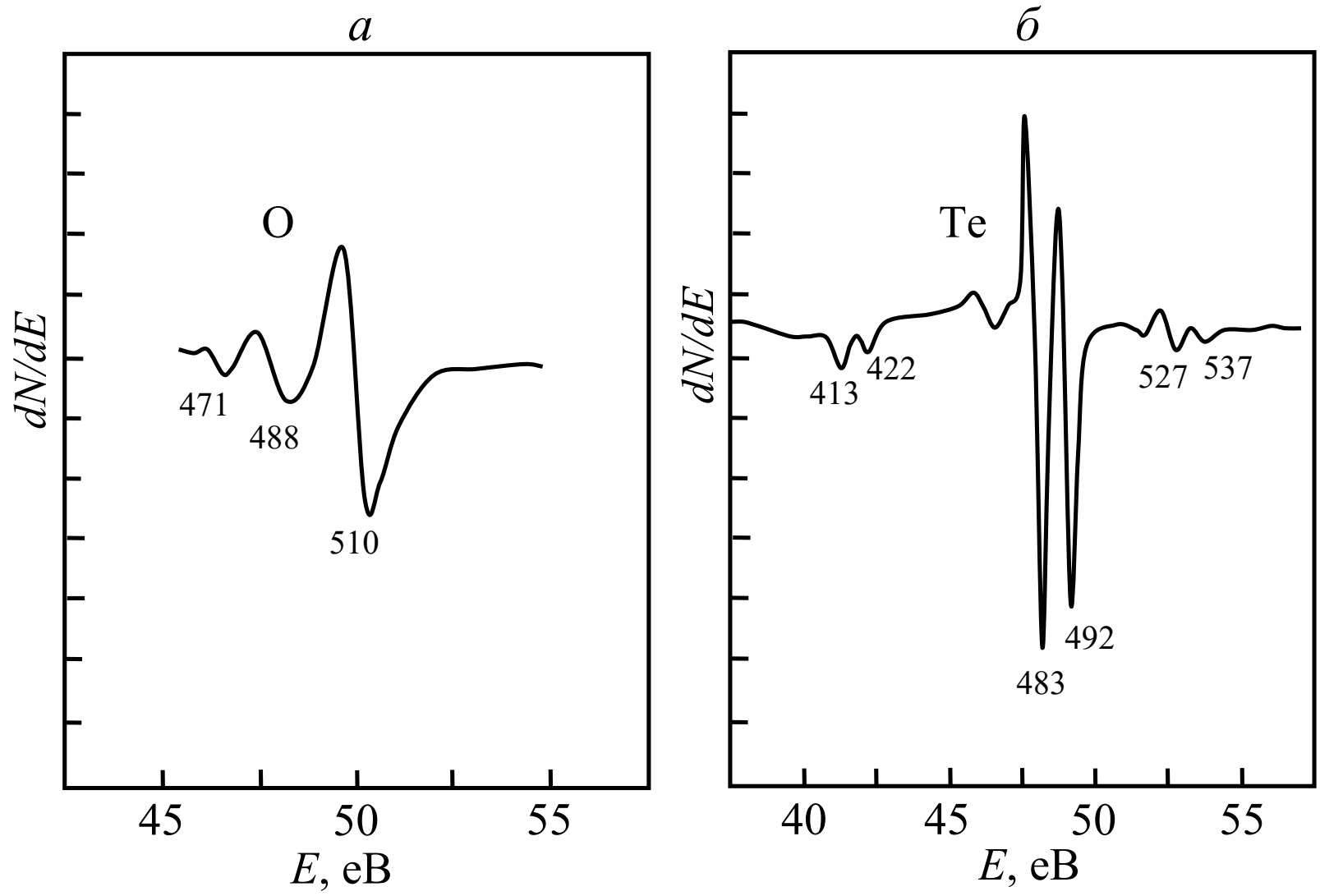

Рис. 1.21. Диференціальні оже-спектри чистого кисню (*а*) і телуру (*б*) [41].

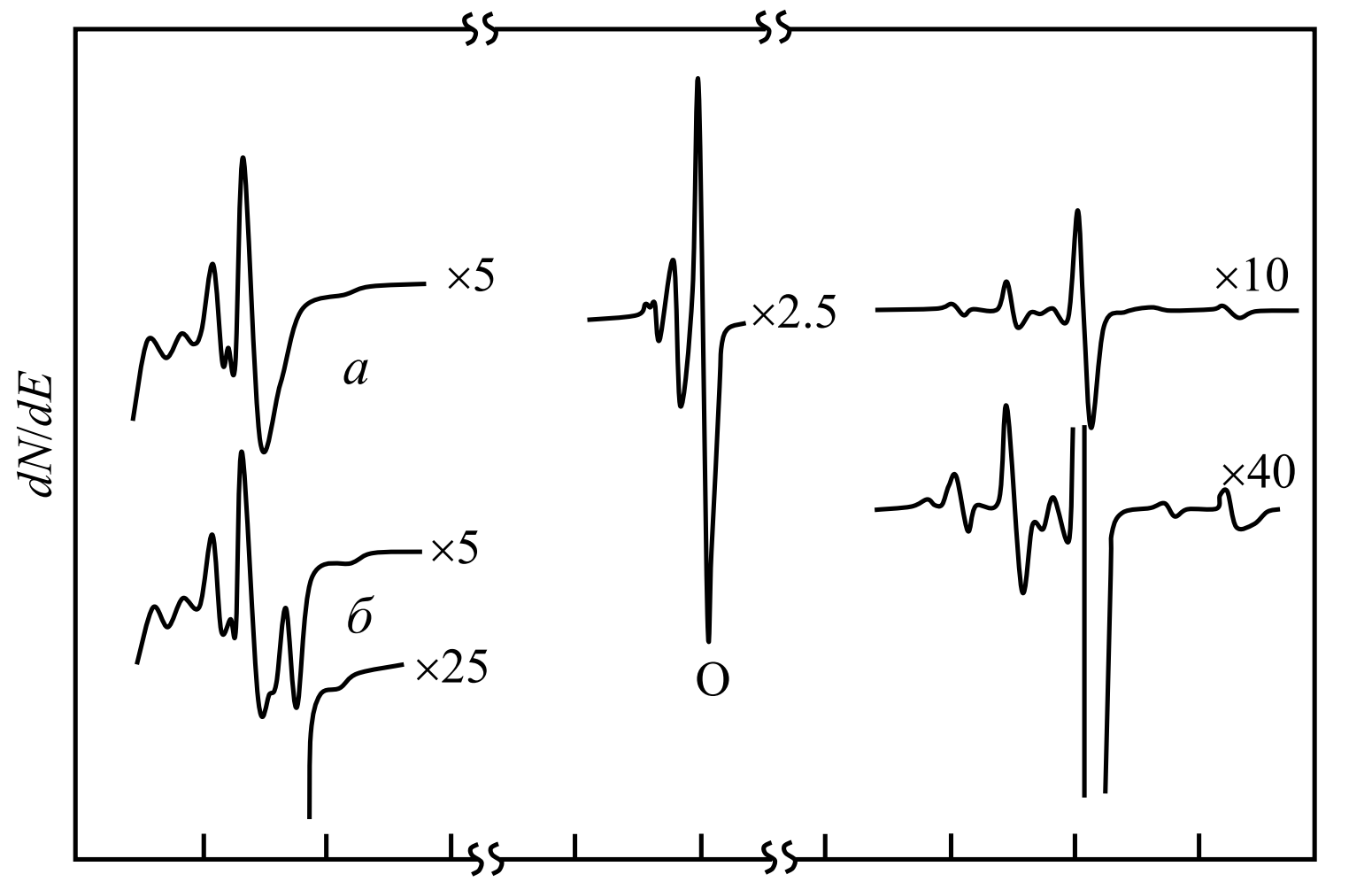

Рис. 1.22. *а* – Диференціальний оже-спектр термічно вирощеного $SiO_2$; *б* – після 5-хвилинного електронного бомбардування $E_p$ = 3 кеВ і $I_p$ = 20 мкА [42].

наведено високоенергетичну частину оже-спектра. Найбільш інтенсивний пік у цій зв'язці спостерігається при 1620 еВ і походить від переходу $KL_{2,3}L_{2,3}$. Точність вимірювання енергії становить близько ±1 еВ. Коли існує проблема перекриття поблизу піка 92 еВ, переважно відстежувати пік 1620 еВ – хоча і менш інтенсивний – для ідентифікації Si. На низькоенергетичній стороні кожного із зазначених оже-піків наявні ще два довільних піка, рівномірно розподілених по енергії. Ці піки пов'язані з втратою плазмонів першого і другого порядку та служить додатковим засобом ідентифікації Si.

Калібрування спектра чистого Si по мінімуму при 92 еВ дозволяє ідентифікувати на ньому дві відомі особливості: добре виражений мінімум при $E$ = 72 (74) еВ, віддалений від основної лінії на величину 20 (~17.8 еВ), рівну енергії об'ємного плазмона в чистому кремнії (~17 еВ), і досить слабко виражений злам при 81 еВ, який в [43] був приписаний $L_{2,3}VV$-переходу і, можливо, пов'язаний з існуванням двох піків густини станів у валентній зоні чистого кремнію [44]. Слід зазначити, що іншою причиною появи особливості профілю смуги $L_{2,3}VV$ оже-спектра чистого кремнію в області $E \sim 80–84$ еВ може бути збудження поверхневого плазмону з енергією ~ 11 еВ [44].

На рис. 1.22 наведені оже-спектри поверхні $SiO_2$. З порівняння рис. 1.20 та 1.22 видно, що низькоенергетичні оже-спектри Si та $SiO_2$ помітно відрізняються один від одного. Найбільш інтенсивний пік при 92 еВ, характерний для чистого Si, змінюється по формі та зсувається в область менших енергій до 75 еВ в оже-спектрі $SiO_2$. Низькоенергетичне зміщення основного піка Si на 17 еВ пов'язане з тим, що коли атом кремнію вступає в хімічну взаємодію з киснем він віддає свої валентні електрони атомам кисню, так як він має велику електронну спорідненість. При цьому енергетичні рівні електронів, особливо зовнішні валентні рівні, зміщуються в бік менших енергій і відповідно енергія оже-піка кремнію, що з'являється в результаті оже-переходу $L_{2,3}VV$ стає меншою, ніж енергія оже-піка чистого кремнію. Крім того, в оже-піка кремнію в сполуці $SiO_2$ виявляються сателітні оже-піки з енергіями 66, 60 і 50 еВ.

Таким чином, хімічний зсув, що спостерігається в $SiO_2$ (від 92 до 75 еВ), не пов'язаний головним чином зі зсувом енергій зв'язку в Si, а викликаний зміною густини стану у валентній зоні (зміни в густині переходу). На противагу цьому, форма високоенергетичних *KLL* оже-ліній практично не змінюється. Порівнюючи спектр Si-*KLL* з елементарним Si і $SiO_2$, найяскравіша й одразу очевидна особливість

– це послаблення в $SiO_2$ піків втрат плазмону, пов'язаних з оже-піками. Значний хімічний зсув у низькочастотному $L_{2,3}VV$ оже-піка та відносно слабкі піки плазмонних втрат у спектрі *KLL* зазвичай використовуються для встановлення відмінності $SiO_2$ від елементарного Si.

Важливою особливістю оже-електронної спектроскопії є дуже мала глибина виходу оже-електронів, яка, наприклад, для оже-переходу в кремнії (92 еВ) становить 5 Å. Дана обставина робить метод оже-аналізу вкрай чутливим до атестації хімічного складу поверхні шаруватих кристалів $Si_2Te_3$.

**1.5.2. Дослідження поверхні шаруватих кристалів $Si_2Te_3$ методом оже-електронної спектроскопії.** Незважаючи на велику стійкість порівняно з відомими сполуками в системах Si–S і Si–Se, кристали $Si_2Te_3$ все ж таки повільно гідролізуються. У результаті хімічної реакції свіжосколотої поверхні кристала $Si_2Te_3$ з парами води, які міститься в атмосфері, на поверхні утворюється шар $SiO_2$, збагачений Te, згідно реакції [46]:

$$Si_2Te_3 + 4H_2O \rightarrow 2SiO_2 + Te + 2H_2Te\uparrow + 2H_2. \quad (1.3)$$

Зразок, що піддається впливу парів води, має металічну сіру поверхню, тоді як чистий кристал є темно-червоним і напівпрозорим.

Для вивчення складу поверхневого шару кристалів $Si_2Te_3$, автори [47, 48] використовували метод Оже-електронної спектроскопії. Важливою особливістю Оже-електронної спектроскопії є її чутливість до хімічного стану аналізованих елементів на поверхні. Хімічний стан елементів досліджуваного зразка відображається на формі та положенні особливостей спектра Оже-електронів. Зупинимося більш докладно на результатах аналізу поверхні та поверхневих шарів, вивчення якісного та кількісного складу та мікроморфології поверхонь розділу фаз шаруватих кристалів $Si_2Te_3$, підданих впливу парів води, кисню та атмосфери повітря, за допомогою оже-електронної спектроскопії, викладених у роботах [43, 47, 48]. На кожній стадії впливу при енергії первинних електронів 450 еВ записувалися оже-спектри. Запис спектрів здійснювалася у вигляді першої похідної від функції розподілу енергії електронів $dN/dE$.

На рис. 1.23 наведені оже-спектри взяті з роботи [48] для чистого кристала $Si_2Te_3$ і підданого впливу $O_2$, парів $H_2O$ або лабораторної атмосфери. В оже-спектрі чистого кристала $Si_2Te_3$ (крива 1, рис. 1.23) пік Si спостерігається при тій самій енергії, що і пік у спектрі ОЕС чистого кремнію (92 еВ, рис. 1.20). Як зазначають автори [47,

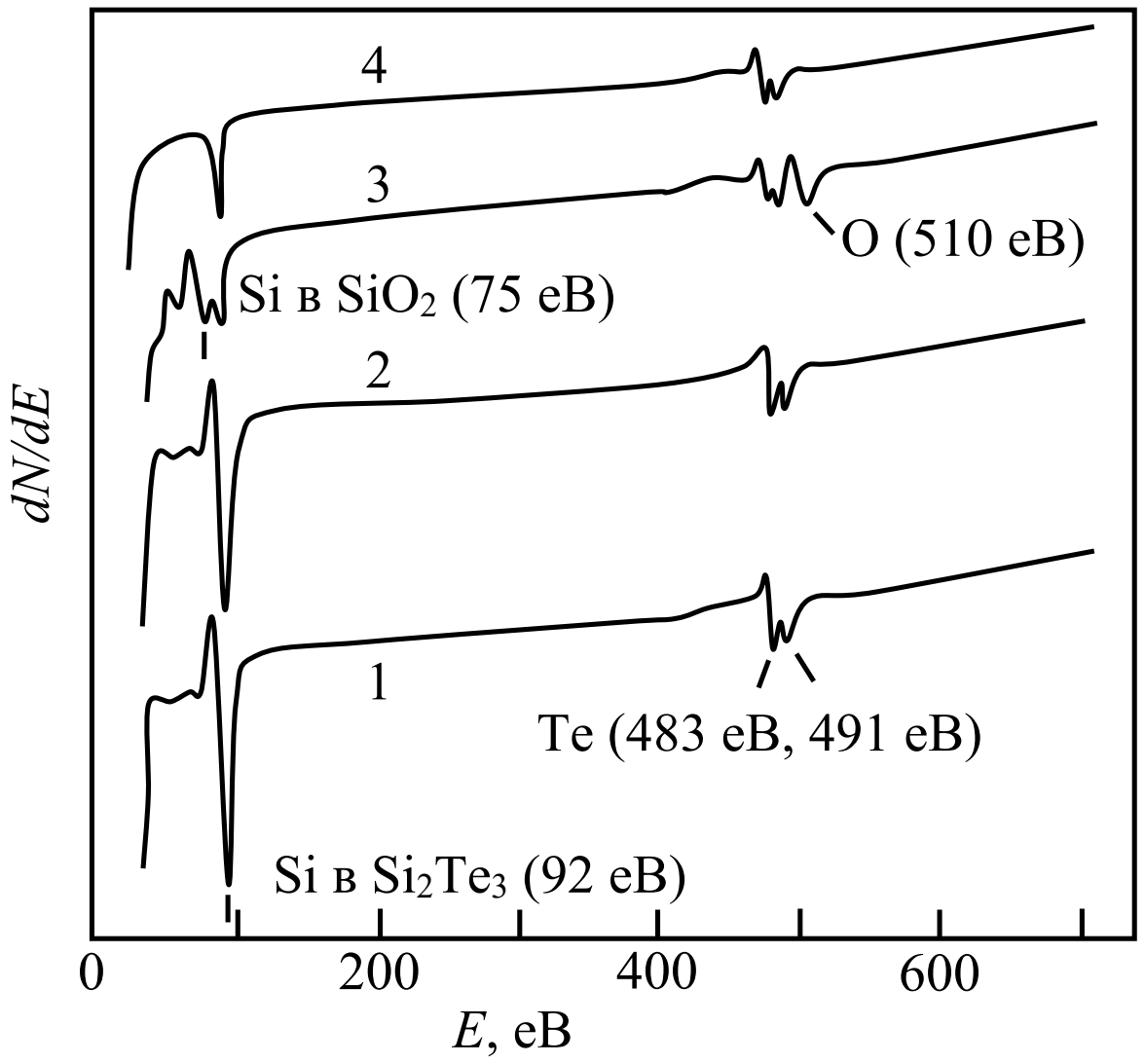

Рис. 1.23. Диференціальні оже-спектри кристала $Si_2Te_3$:
1 – вихідного чистого зразка; 2 – зразка, підданого дії 2 Торр $O_2$ протягом 15 *хв*; 3 – підданого парам води протягом 15 *хв*; 4 – що знаходився в атмосфері лабораторії протягом 15 *хв* [48].

48], вплив кисню не чинить помітного впливу на склад поверхні сколотого у вакуумі зразка, тоді як вплив парів води або вологого повітря значно змінює оже-спектр.

Як видно із рис. 1.23 (крива 3), оже-пік, характерний для Si в чистому $Si_2Te_3$, сильно загасає, у той час як оже-піки, які відповідають кисню (510 еВ) і кремнію в $SiO_2$ (75 еВ), з'являються в спектрах після впливу парів $H_2O$ та атмосфери. Таким чином, у шарі $SiO_2$, який утворився на поверхні кристала $Si_2Te_3$, оже-пік Si проявляється при енергії 75 еВ, тобто його енергетичне положення в спектрі оже-електронів зміщено на 17 еВ в область менших енергій, що характерно для діоксиду кремнію (рис. 1.22). Як зазначалося вище, цей зсув зумовлений тим, що при хімічній взаємодії атома Si з киснем, він віддає свої валентні електрони атомам кисню, оскільки має велику електронну спорідненість. При цьому енергетичні рівні електронів, особливо зовнішні валентні рівні, зміщуються в область менших енергій і відповідно енергія оже-піка Si, що з'являється в результаті оже-переходу, наприклад, *LMN* стає меншою, ніж енергія оже-піка Si за відсутності хімічного зв'язку з киснем.

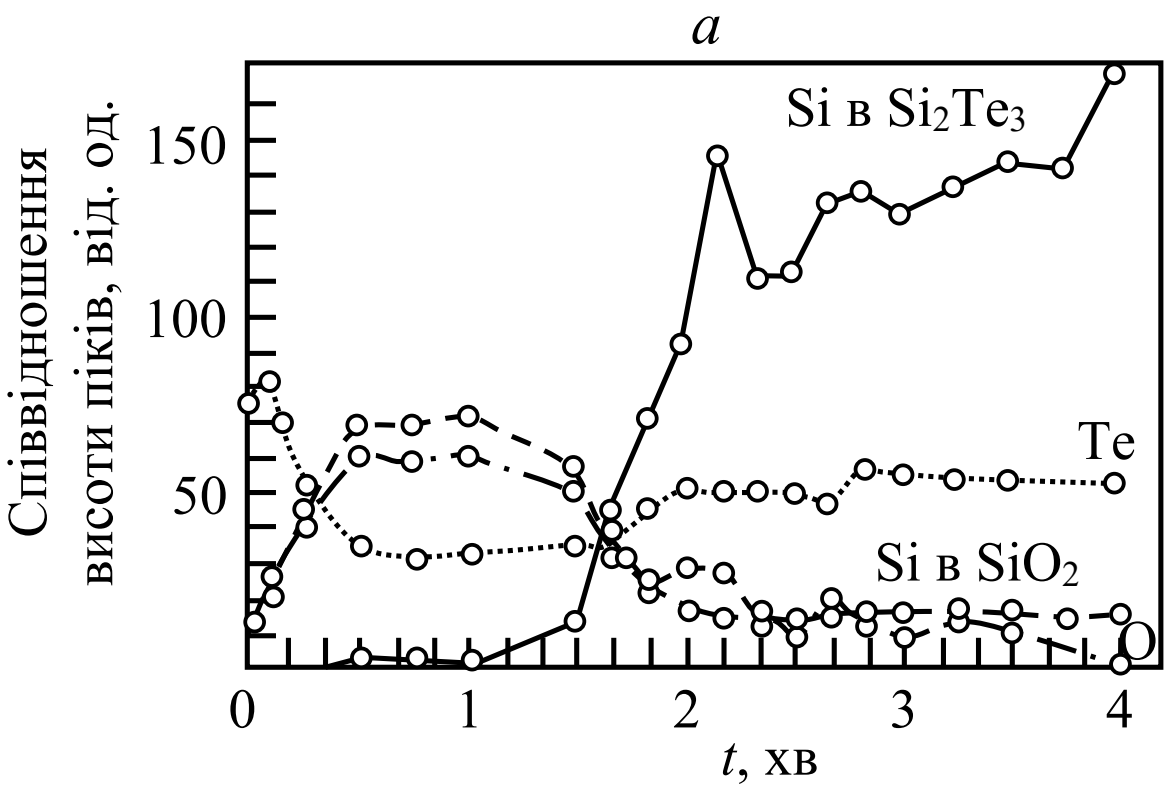

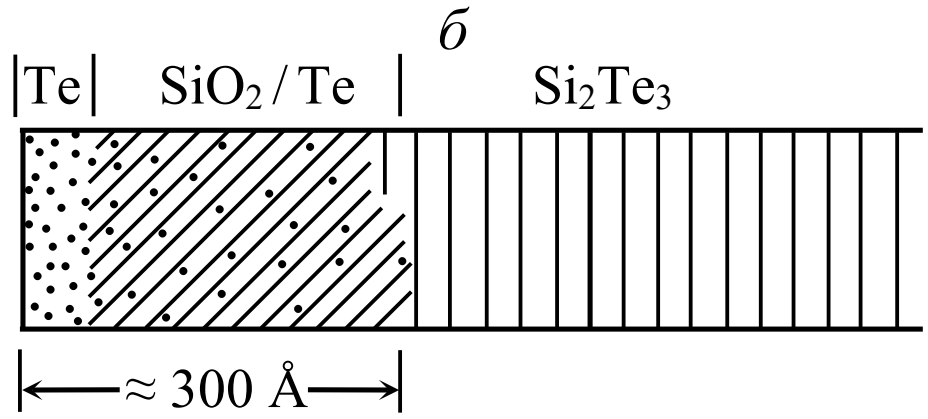

Рис. 1.24. Структура поверхневого шару кристала $Si_2Te_3$, підданого впливу повітря протягом кількох тижнів: *а*) глибина профілю згідно з даними ОЕС [47, 48]; *б*) схематичний розподіл складу в поверхневому шарі згідно [43].

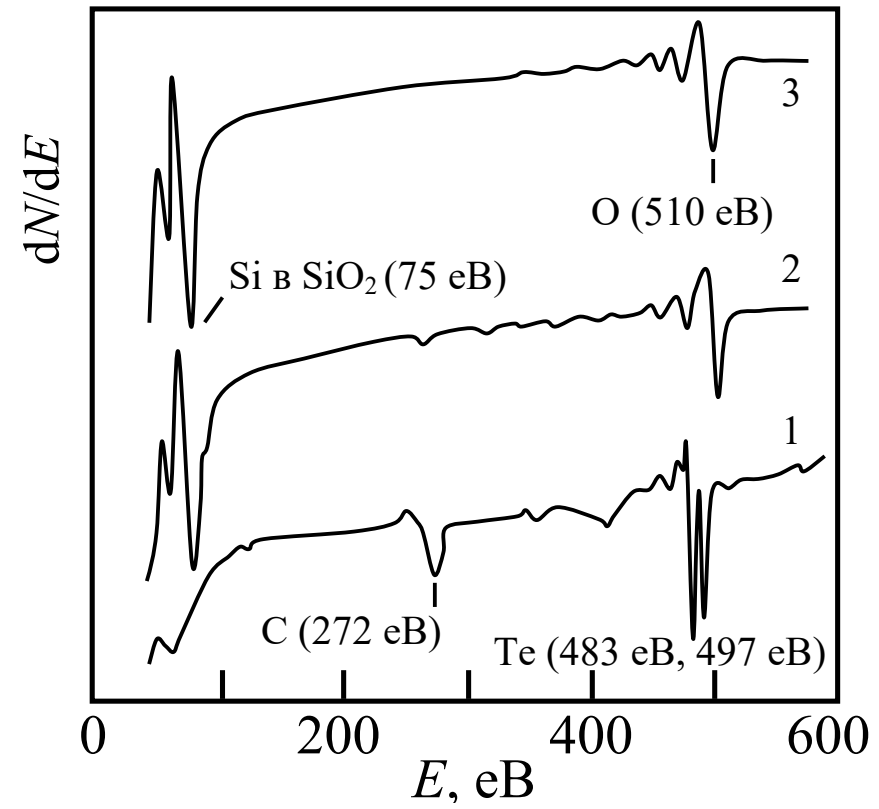

Рис. 1.25. Диференціальні оже-спектри кристала $Si_2Te_3$, що знаходився тривалий час в атмосфері повітря : крива 1 – до початку термообробки; 2 – після термообробки при 673 K протягом 2-х годин та подальшого охолодження до кімнатної температури; 3 – після наступного перебування в атмосфері лабораторії протягом 1 години [48].

Аналіз складу поверхневого шару кристала $Si_2Te_3$ по глибині за допомогою оже-спектроскопії показав, що поверхневий шар неоднорідний і збагачений телуром поблизу поверхні (рис. 1.24). Телур, що міститься в поверхневому шарі, можна видалити термічною обробкою, після якої залишається щільний шар $SiO_2$. На рис. 1.25 наведено оже-спектри кристала $Si_2Te_3$, що знаходився тривалий час в атмосфері за нормальних умов, і поміщеного потім усередину спектрометра, до початку термообробки (крива 1), підданого термообробці всередині спектрометра при $T$ = 673 K з подальшим охолодженням до кімнатної температури (крива 2) і після перебування термообробленого зразка в атмосфері повітря протягом однієї години (крива 3). Із зіставлення кривих 1 і 2 на рис. 1.25 чітко видно, що внаслідок термообробки кристала $Si_2Te_3$ при 673 K зовнішній шар телуру зникає, залишаючи поверхню, оже-спектр якої – типовий спектру $SiO_2$. Привнесення термічно очищеного кристала в атмосферу лабораторії з 53% відносною вологістю і наступною витримкою протягом одної години не мало помітного впливу на зміну складу поверхні. Таким чином, наявність шару $SiO_2$ запобігає подальшій реакції з парами води, що містяться в атмосфері. Зразки, пасивовані таким способом, можуть бути використані для дослідження напівпровідникових властивостей об’ємного матеріалу.

## 1.6. ОТРИМАННЯ НАНОСТРУКТУРОВАНИХ Si, Te ТА $Si_2Te_3$ МЕТОДОМ ХІМІЧНОГО ОСАДЖЕННЯ ІЗ ПАРОВОЇ ФАЗИ

У зв'язку з розвитком наноелектроніки зріс інтерес до дослідження структур зі зниженою розмірністю, зокрема таких квантоворозмірних структур як ниткоподібні кристали (НК) або нановіскери (НВ). Унікальні властивості НВ роблять їх перспективними для застосування у сучасних електронних та оптичних приладах – польових транзисторах тощо. Останні досягнення в галузі технологій вирощування напівпровідників дозволили синтезувати новий клас наноструктур сесквітелуриду кремнію ($Si_2Te_3$), таких як нанопластинки (nanoplates) [49–51], нанострічки (nanobelts (nanoribbons)) [49, 52] нанодроти (nanowires) та наношпильки [52] методом хімічного осадження із парової фази (CVD процес).

**1.6.1. Вирощування ниткоподібних кристалів Si методом CVD у відкритій проточній системі.** Найбільш поширеним механізмом вирощування ниткоподібних кристалів Si є механізм пара-рідина-кристал (ПРК), запропонований Вагнером та Елісом у 1964 році [53].

Використання золота зумовлене його здатністю утворювати евтектичні розплави з низькою температурою плавлення. Так, бінарна система Si–Au має просту фазову діаграму евтектичного типу (рис. 1.26) [54]. Температура евтектики складає 636 К. Нижче температури евтектики розчинність золота та кремнію у твердому стані практично відсутня, а вище температури евтектики у рівновазі знаходяться змішана рідка фаза Si+Au (яка не утворює хімічних сполук) і тверда фаза Si. Таким чином, при виборі температури росту НВ діапазон температур росту вибирається за межами евтектичних температур (648–748 К). Це повинно забезпечити більш високу дифузію між золотом і кремнієм.

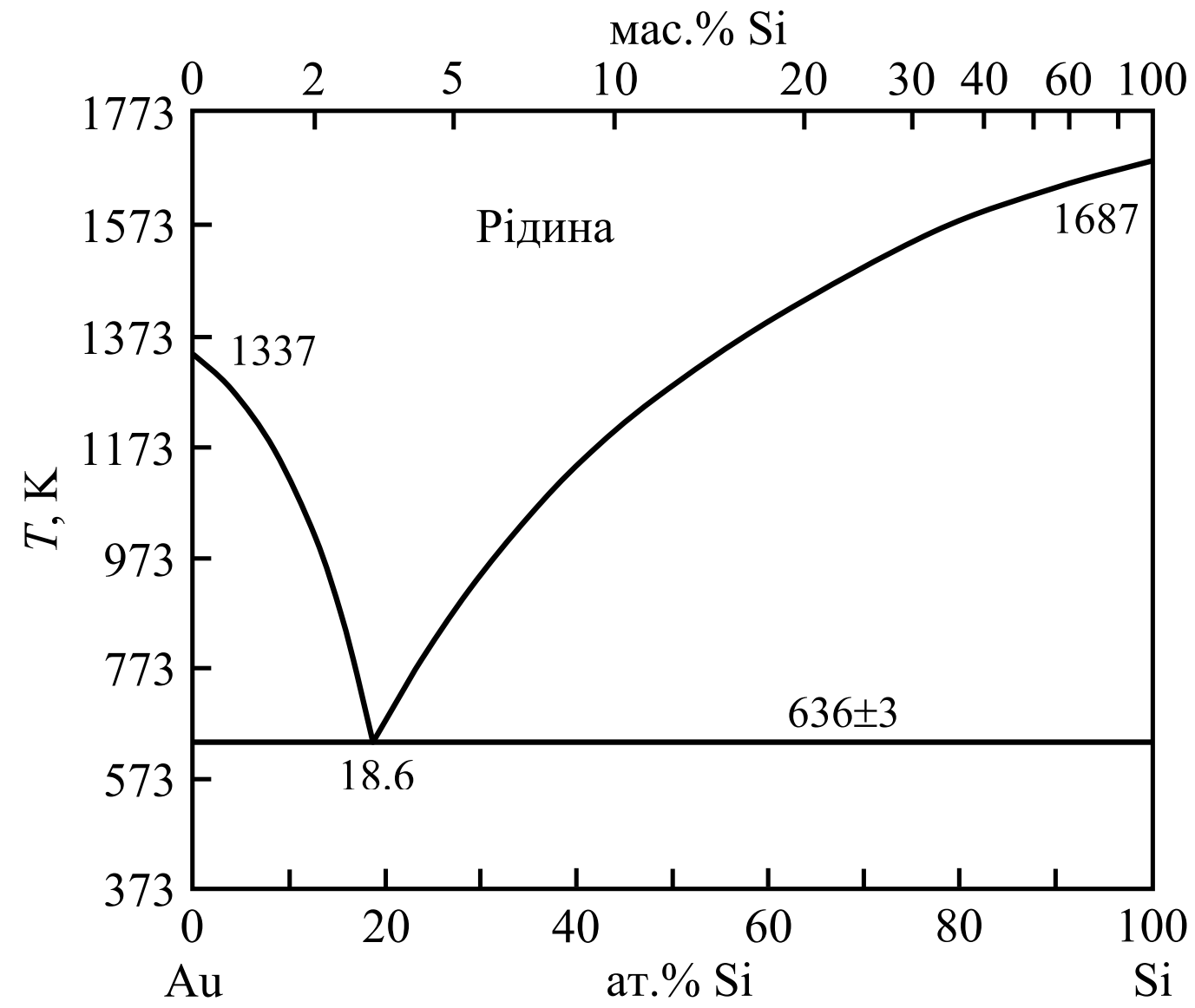

Рис. 1.26. Діаграма стану системи Si–Au [54].

Схема росту напівпровідникового нановіскера за механізмом ПРК наведена на рис. 1.27 на прикладі осадження Si на поверхню Si (111), активовану золотом. Процес вирощування нановіскерів включає три етапи. На першому етапі на кремнієві підкладки орієнтації (111) наносять тонкі плівки золота товщиною 4 нм. Тонкі шари золота можуть наносити різними способами: або в ростовій камері епітаксійної установки, або в окремій установці вакуумного розпилення. На другому етапі кремнієва підкладка з тонким нанесеним шаром золота нагрівається вище евтектичної температури плавлення розчину матеріалу підкладки з каталізатором (Au–Si). У результаті на поверхні відбувається коагуляція золота в нанокраплі (рис. 1.27), що знаходяться в термодинамічній рівновазі з підкладкою, які відіграють роль ініціаторів (каталізаторів) росту за механізмом пара-рідина-кристал. Середній діаметр коагульованих після відпалу плівки золота нанокрапель Si–Au залежить від товщини осадженої плівки золота. На третьому етапі при фіксованій температурі поверхні *T* проводиться осадження напівпровідникового матеріалу (Si) з відомою швидкістю осадження *V* протягом часу *t*. Середній діаметр нанодротин залежить від діаметра коагульованих нанокрапель Si–Au.

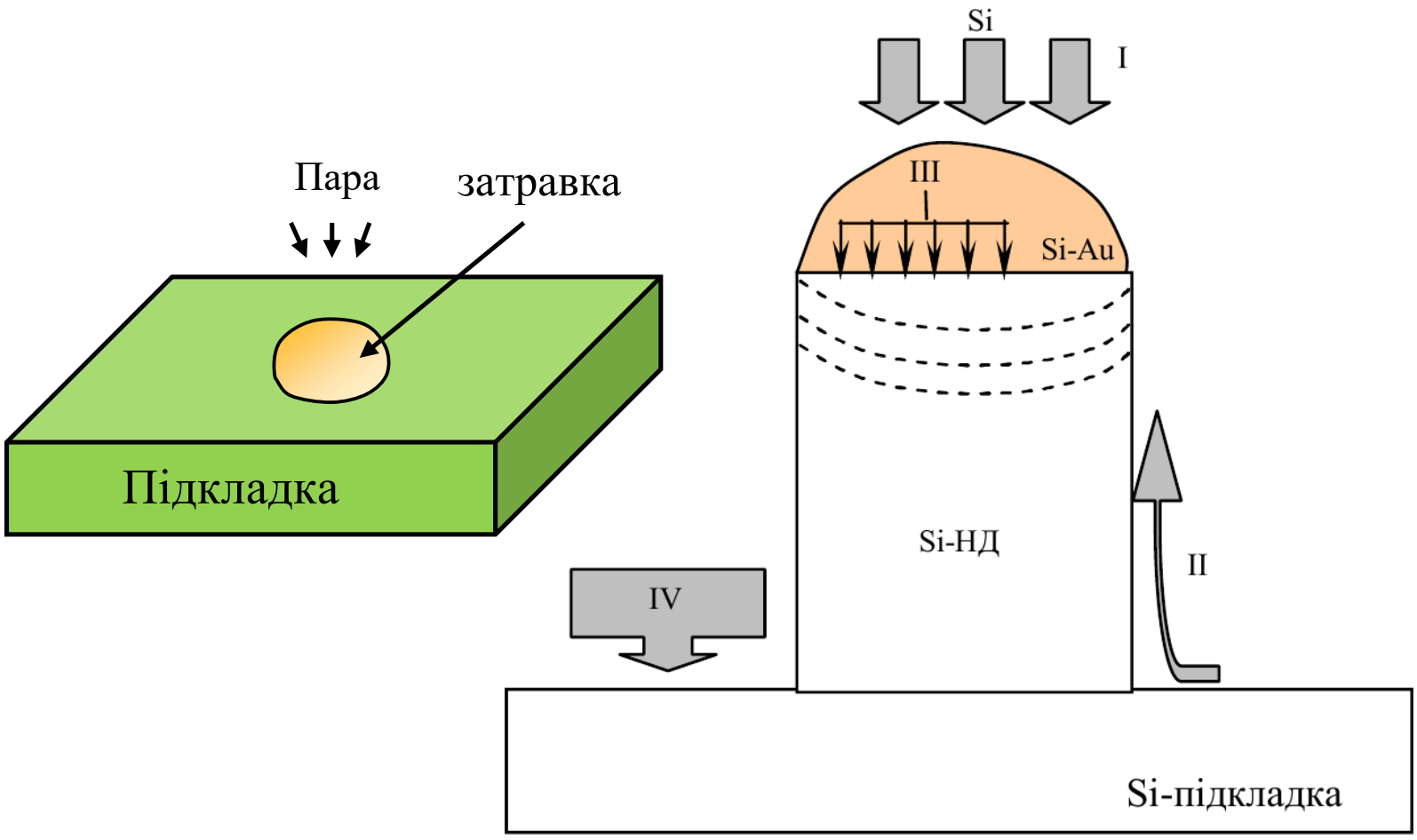

Рис. 1.27. Схематичне зображення формування кремнієвої нанодротини [56].

Згідно нестаціонарної теорії комбінованого росту [55, 56] нанокристалів за механізмом "пара–рідина–кристал" процес осадження кремнію можна описати з урахуванням наступних кінетичних процесів (рис. 1.27):

I. Адсорбція й десорбція на поверхні краплі.
II. Дифузійний потік у краплю, що є сумою потоків часток, адсорбованих безпосередньо на бічних стінках й тих, що мігрували з поверхні підкладки.
III. Пошаровий ріст із рідкого розчину на границі рідина-кристал під краплею з урахуванням кінцевого розміру грані.
IV. Ріст на неактивованій поверхні підкладки.

Середній діаметр НВ залежить від діаметра коагульованих нано-крапель Au–Si, а діаметр останніх – від товщини осадженої плівки золота, при цьому збільшення часу росту приводить до зростання діаметра нанодротин.

**1.6.2. Вирощування ниткоподібних кристалів кремнію методом хімічних транспортних реакцій.** Віскери кремнію автори [57] вирощували в закритих кварцових ампулах довжиною 220–240 мм і внутрішнім діаметром 30 мм, вакуумованих до тиску $10^{-3}$ Па. В якості джерела використовували монокристалічний кремній (2 г), бром (200 мг) і золото (5–10 мг).

У запропонованій технології компоненти газової фази, необхідні для вирощування віскерів, подаються із зони джерела в зону росту за допомогою хімічних транспортних реакцій внаслідок дифузії бромідів кремнію та золота. В основі методу хімічних транспортних реакцій для росту кристалів, як відомо, лежать реакції диспропорціонування утворених сполук основної речовини. Такі реакції відбуваються гетерогенно і для бромідів кремнію та золота мають вигляд:

$$2SiBr_2 \leftrightarrow SiBr_4 + Si, \qquad (1.4)$$

$$3AuBr_2 \leftrightarrow 2Au + AuBr_4 + Br_2. \qquad (1.5)$$

Вирощування віскерів проводили у трубчатій однозонній печі з певним температурним профілем. Проведені дослідження показали, що оптимальними температурами зони джерела є 1453–1473 К, а зони росту – 1323–1223 К.

На першому етапі завантажену кварцову ампулу розміщають у ростовій печі так, щоб зона джерела знаходилася при температурі 1453 К, а зона росту – при температурі близько 1473 К. У такому режимі ампулу витримують протягом 15 *хв*. Головне призначення першого етапу полягає у формуванні в ростовій ампулі парової фази бромідів кремнію та золота. Оскільки на цьому етапі ампула знаходиться в умовах, близьких до ізотермічних, у паровій фазі переважають нижчі броміди – $SiBr_2$ та $AuBr_2$. Для досягнення стаціонарного співвідношення між нижчими і вищими бромідами, що відповідає заданому розподілу температури в ампулі, потрібен певний час. Як було визначено, для цього цілком вистачає витримки протягом 15 *хв*. Крім того, перший етап виконує також допоміжну корисну функцію. Його реалізація дозволяє провести в ампулі додаткове очищення зони росту віскерів.

Основними умовами, за якими розпочинається осадження кремнію та ріст віскерів, є пересичення парової фази та зсув рівноваги реакцій диспропорціонування бромідів кремнію та золота, сформо-ваних на першій стадії, у бік виділення кремнію та золота. Для цього слугує перехідний процес тривалістю 10 хв, при якому проводиться охолодження зони росту ампули до температур $T \approx 1323–1223$ К. Великі рівні пересичення, що виникають при цьому, ведуть до осадження на стінках кварцової ампули кластерів кремнію зі структурою, розвинутою на нанорівні. Віскери виростають із сформованої на стінках кварцової ампули основи, яка є сукупністю зародків. Цей процес відбувається в кінетичному режимі.

**1.6.3. Синтез нанодротів та нанотрубок телуру шляхом фізичного осадження із парової фази.** Нанодроти та нанотрубки Te були вирощені термічним випаровуванням порошку та шматків Te в трубчастій печі в потоці газу аргону [58]. Вирощування нанодротів і нанотрубок Te автори [58] проводили в трубчастій печі в потоці газу аргону 50–200 мл/*хв* при атмосферному тиску без використання будь-якого каталізатора. Для цього порошок Te високої чистоти (99.99%) і шматки (~2 г) завантажували в глиноземний човен і поміщали в кварцову трубку завдовжки 1 м, розташовану в печі (рис. 1.28). Джерело поміщали в гарячу зону печі. Полікристалічні $Al_2O_3$ і Si (111) підкладки були розміщені вздовж кварцової трубки в на-

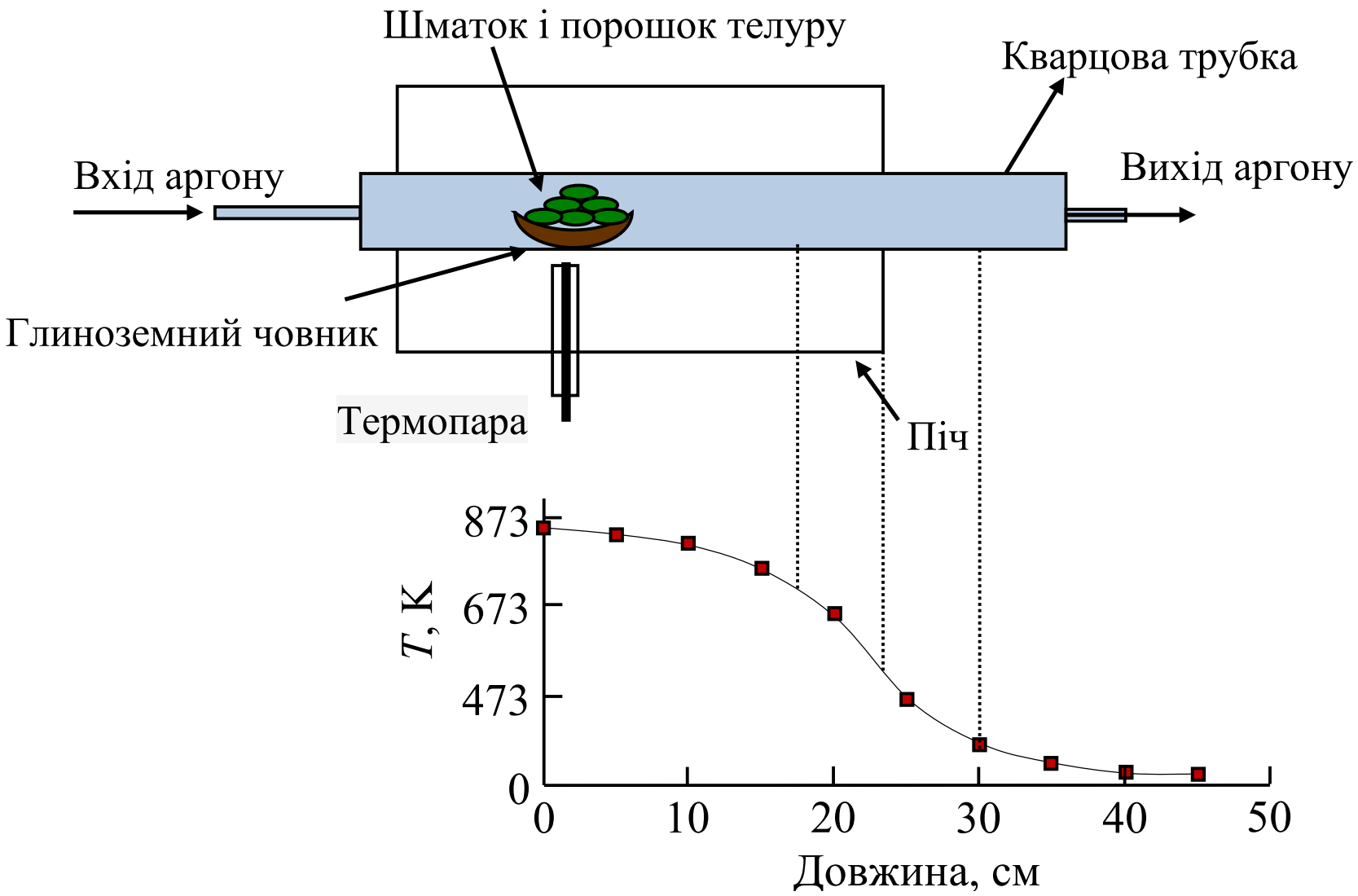

Рис. 1.28. Принципова схема експериментальної установки, що демонструє трубчасту піч з газопроточним розташуванням. Також показано різні температурні зони в печі вздовж потоку газу[58].

прямку потоку аргону. Кварцова трубка була достатньо довгою для створення різних температурних зон уздовж її довжини, як показано на рис. 1.28. Температуру печі підвищували (зі швидкістю 200 К/*год*) у присутності потоку газу аргону та підтримували потрібну температуру (473–1173 К) протягом 2 *год*.

Встановлено, що мікроструктура Te залежить від температур випаровування та конденсації та швидкості потоку газу. При низьких температурах, нижче точки плавлення Te, спостерігається зростання нанодротів Te на поверхні шматків Te. При температурі печі 823 К нанотрубки Te осаджувалися на стінках кварцової трубки в низькотемпературній зоні печі. Зростання цих структур стає зрозумілим на основі кристалічної структури Te та процесу росту твердого тіла в парі. Механізм росту включає зародження сферичних частинок з наступним зростанням одновимірних структур завдяки анізотропним властивостям. Початковий ріст 1D нанострижнів супроводжується еволюцією в нанотрубки при меншому потоці Te. Таким чином, температуру джерела і зростання, а також швидкість потоку газу потрібно критично відрегулювати, щоб отримати якісне осадження нанотрубок.

**1.6.4. Вирощування нанодротин і наношпильок $Si_2Te_3$.** Методом вирощування з парової фази з використанням нанокластерів Au в якості каталізатора, який ініціював і супроводжував ріст, успішно отримано орієнтовані нанодроти $Si_2Te_3$ на твердих підкладках SiO/Si [52, 59, 60] та гетероструктуровані нановіскери $Si_2Te_3$/Si [61].

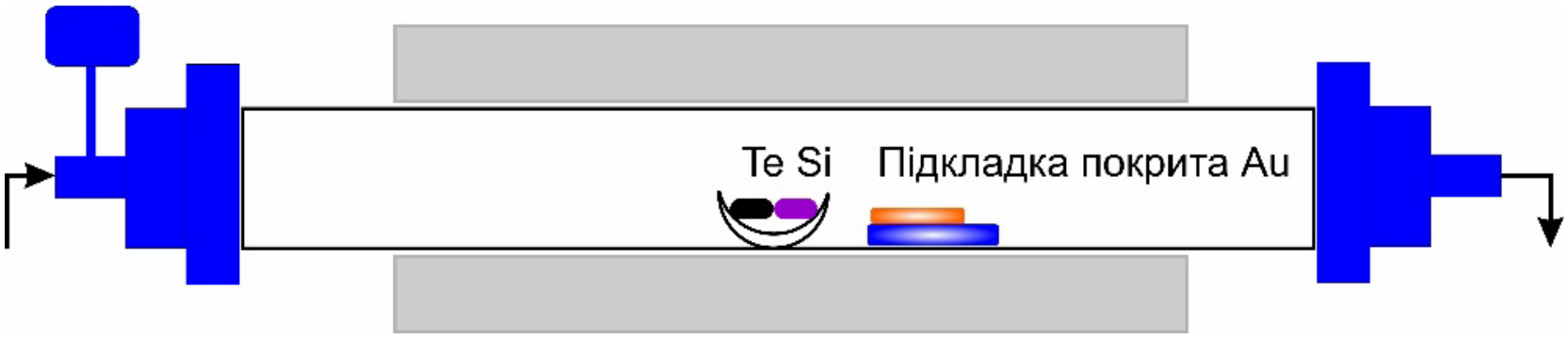

Рис. 1.29. Схема печі для синтезу нанодротин $Si_2Te_3$ [61] .

В якості вихідних матеріалів для приготування зразків $Si_2Te_3$ автори [61] використовували порошки телуру і кремнію, які були поміщені в керамічний тигель і завантажені в високотемпературну трубчасту піч. Підкладки $SiO_2$/Si з нанесеним тонким шаром Au товщиною від 40 до 80 нм, розміщували в печі після тигля у напрямку потоку газу, як показано на рис. 1.29. Кварцову трубку спочатку відкачували, а потім у камеру вводили газ-носій (азот) високої чистоти для підтримки тиску на рівні 9.12 Торр. Швидкість потоку азоту становила 15 см$^3$/хв. Потім піч нагрівали до 1123 K зі швидкістю 20 K/хв. Ріст проводили від 823 K до 923 K на протязі 3–5 *хв*, після чого керамічний тигель і субстрати охолоджували до кімнатної температури.

Існує залежність форми (габітусу) експериментальних зразків від температури підкладки та човника з вихідним матеріалом. Зміна температури та часу росту під час синтезу наноструктур $Si_2Te_3$ дозволяє контролювати морфологію отриманих продуктів росту [52]. Зображення гетеростуктурованих нанодротин $Si_2Te_3$ (H-NW), отримане за допомогою скануючого електронного мікроскопа (СЕМ), наведено на рис. 1.30.

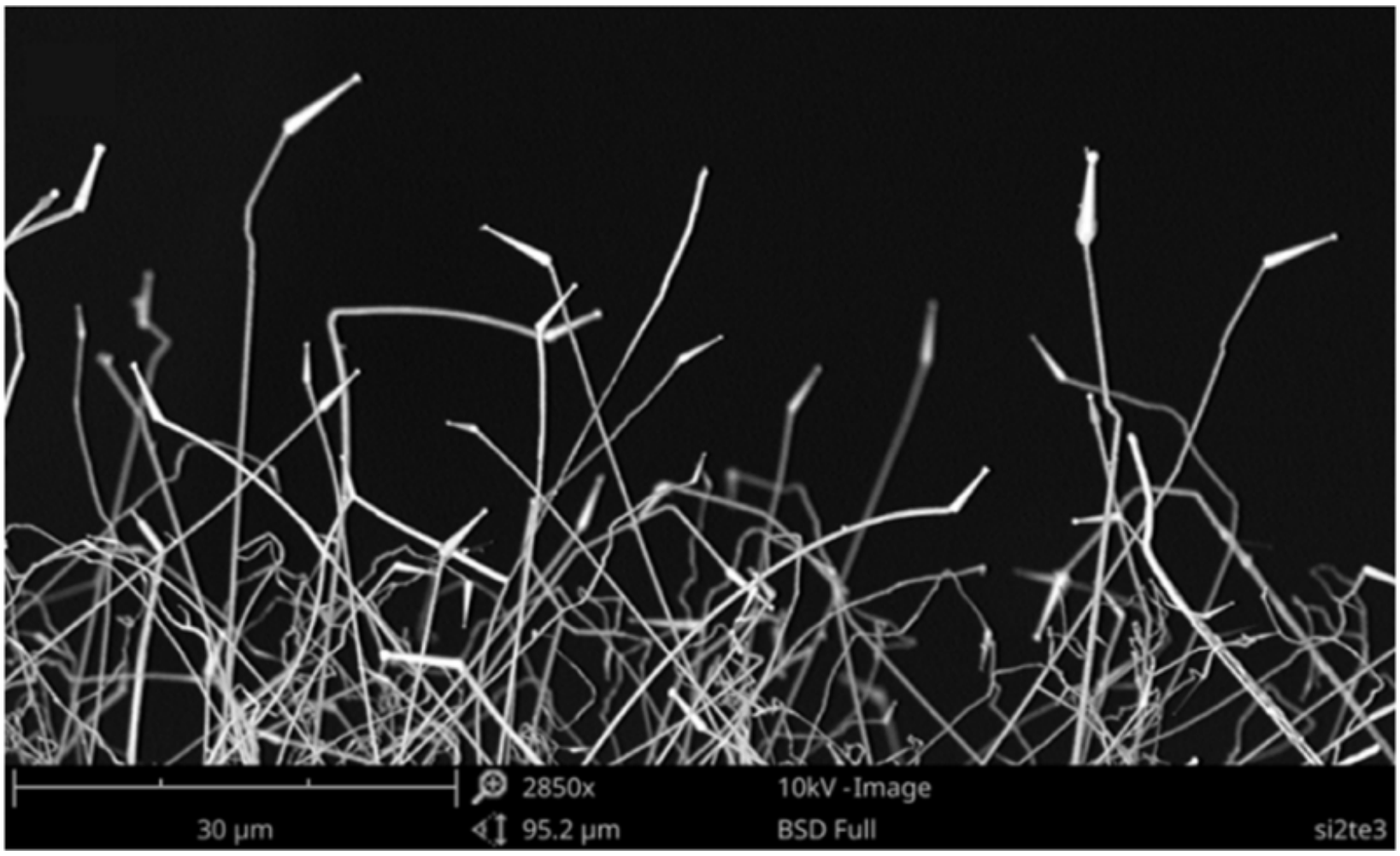

Рис. 1.30. СЕМ зображення гетероструктурних нанодротин $Si_2Te_3$ [61].

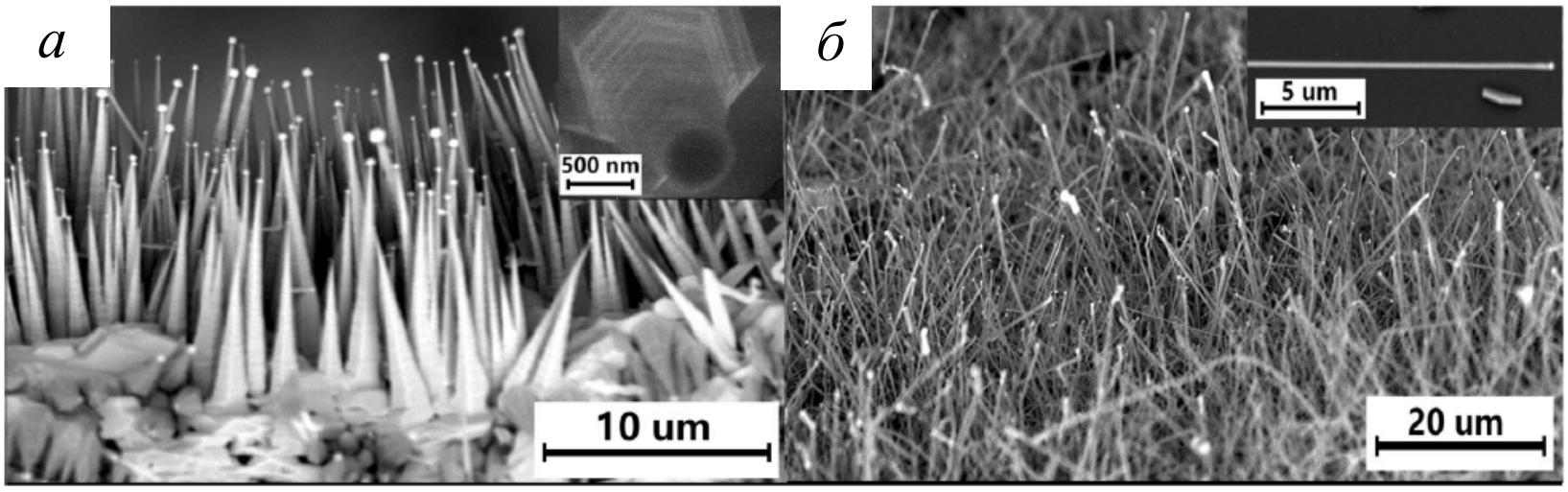

Рис. 1.31. СЕМ зображення наношпильок (*а*) і нанодротин (*б*) вирощених на кремнієвій підкладці покритої каталізатором Au [52].

На рис. 1.31 показано СЕМ-зображення наношпильок $Si_2Te_3$ і нанодротів, вирощених на підкладках $SiO_2$/Si покритих Au. На рис. 1.31, *а* показано ріст наношпильок у вигляді конуса, які були отримані при 853 К протягом ~4 *хв*. Конструкції мають товсте дно діаметром близько 3 мкм і гострі кінці у верху. Видно, що на кожному з конусів наявні круглі краплі, які є каталізаторами Au. Це свідчить про те, що ріст на кінчику структури відбувається швидше, ніж у горизонтальному напрямку через наявність каталізаторів. Конічні структури формуються шляхом укладання нанопластин уздовж вертикального напрямку, про що свідчить СЕМ-зображення на вставці рис. 1.31, *а*. Методом енергодисперсійної рентгенівської спектроскопії (EDX) підтверджено, що хімічний склад отриманих наноструктур відповідає відношенню Si:Te як 2:3.

На рис. 1.31, *б* показано СЕМ-зображення нанодротів $Si_2Te_3$, вирощених при температурі підкладки 873 К протягом 4 *хв*. Отримано нанодротини довжиною до 50 мкм і діаметром сотні нанометрів. Товщина нанодротин дуже однорідна на відміну від наношпильок. Також, каталізатори Au видно на кінчиках нанодротів. На вставці рис. 1.31, *б* показана окрема нанодротина з Au на кінчику. Наявність наночастинки золота на кінці нанодротини свідчить про те, що механізм ПРК домінує в процесі росту, в якому каталізатор Au індукує зародження та ріст нанодротини.

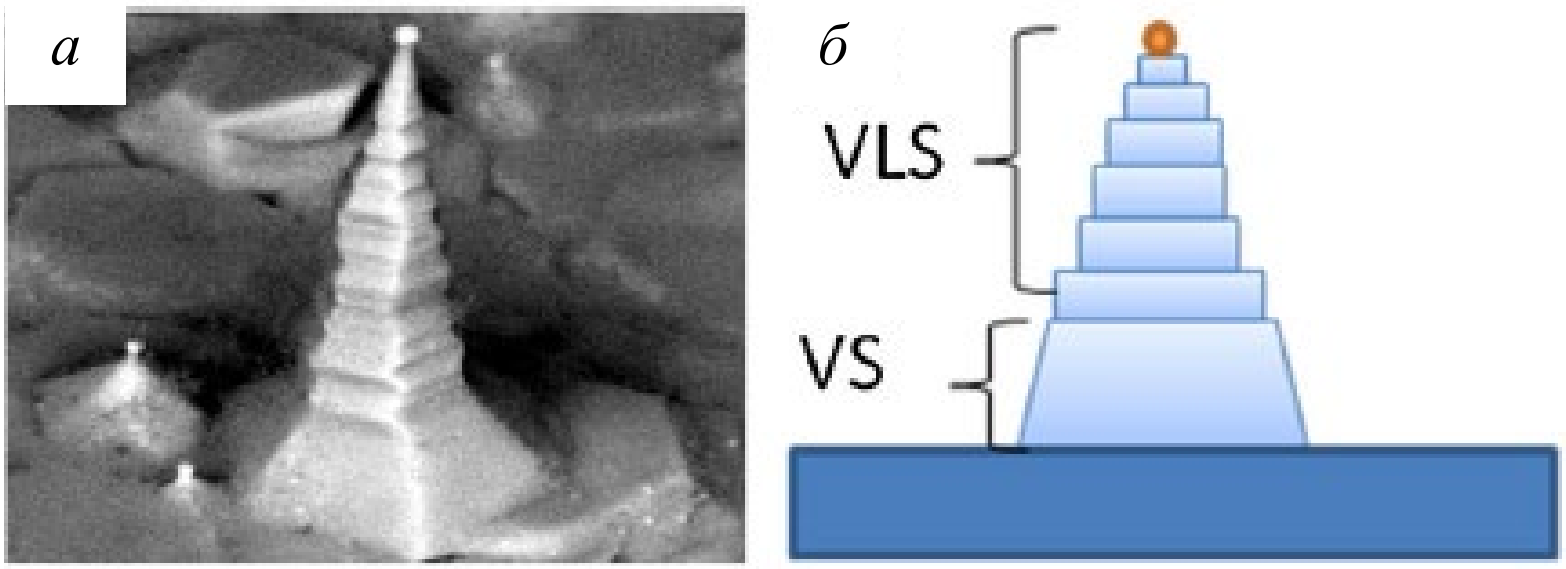

Рис. 1.32. СЕМ зображення (*а*) та схематична діаграма механізму росту (*б*) наношпильки $Si_2Te_3$ [52].

НВ $Si_2Te_3$, вирощені при різних температурах підкладки, як показано на рис. 1.32, демонструють, що морфологія $Si_2Te_3$ змінюється від наношпильки до нанодротини зі збільшенням температури підкладки. Цей факт автори [52] пояснюють наступним чином: у методі вирощування ПРК рідкі нанорозмірні краплі утворюються на під-кладці через евтектичну реакцію між наночастинками Au та розчинами $Si_2Te_3$, отриманими з парофазних прекурсорів у системі CVD. Постійна подача реагентів $Si_2Te_3$ у краплі рідини перенасичує евтектику, що приводить до зародження твердого напівпровідника. Поверхня розділу кристал/рідина утворює границю росту, яка діє як поглинач. Твердий напівпровідник вбудовується в інтерфейс для формування нанодротів, а краплі Au знаходяться на вершині. Насправді існують дві конкуруючі границі розділу під час росту нанодротини, тобто межа розділу рідина/кристал між евтектикою та нанодротиною і границя розділу пар/кристал між реагентами та відкритою поверхнею нанодротини, що росте. Перший інтерфейс приводить до зростання ПРК і осьового подовження нанодротин, тоді як другий інтерфейс приводить до росту ПК і потовщення нанодротин у радіальному напрямку. Обидва механізми залежать від умов росту, таких як тиск, швидкість потоку, температура та співвідношення видів реагентів. При низькій температурі підкладки ріст наношпильок $Si_2Te_3$ сприяє високе співвідношення Te/Si, оскільки Te має низьку температуру випаровування, що приводить до росту ПК над ПРК.

Однак ріст ПКР поступово домінує, оскільки співвідношення Te/Si зменшується при високій температурі, що приводить до росту нанодротин. Слід зазначити, що температура впливає на тиск парів Si і Te в ростовій камері, а отже, і на режим росту.

На рис. 1.32 показано СЕМ-зображення наноструктури $Si_2Te_3$ конічної форми та її схематичну діаграму механізму росту. Це комбінація росту ПК і ПРК. На початку морфологія росту наноконусів $Si_2Te_3$ демонструє пірамідальну форму, що зумовлено моделлю росту ПК. Далі процес росту ПК замінюється режимом ПРК. Конічна форма $Si_2Te_3$ складається з нанопластин із дуже малим розміром і зазвичай обмежена каталізаторами.

**1.6.5. Вирощування нанострічок та нанопластин $Si_2Te_3$.** Синтез кристалітів $Si_2Te_3$ нанометрової товщини (так званих «малошарових») також викликає підвищений інтерес у зв’язку з можливістю отримання нових фізичних характеристик при переході від об’ємних кристалів до нанокристалів, які складаються тільки з невеликої кількості молекулярних шарів, а також до моношарів. Моно- та кількашаровий $Si_2Te_3$ може бути виготовлений з використанням двох підходів – «зверху вниз» і «знизу вгору». Підходи «зверху вниз» передбачають видалення шарів механічним або хімічним шляхом із об’ємного матеріалу. Такі методи включають: механічне відлущування, рідкофазне відлущування за допомогою ультразвуку та мето-ди інтеркаляції. До недавна більшість вивчених ультратонких 2D $Si_2Te_3$ були виготовлені шляхом механічного відлущування з високоякісних монокристалів, які були отримані за допомогою методу хімічного переносу парів (CVD). Для отримання моношарових або одношарових матеріалів об'ємні монокристали притискають до клейкої стрічки, і в результаті тавтологічного розщеплення утворюються лусочки, наклеєні на стрічку, які можна перенести на різні підкладки. Підходи «знизу вгору», за допомогою яких шари $Si_2Te_3$ вирощуються з його складових елементів, включають хімічне осадження з парової фази, атомно-шарове осадження та молекулярно-променева епітаксія.

Автори [49] повідомляють про синтез високоякісних монокристалічних двовимірних шаруватих наноструктур $Si_2Te_3$ у кількох морфологіях, що контролюються температурою підкладки та затравкою Te. Морфології включають нанострічки, утворені ростом ПРК з крапель Te, вертикальні гексагональні нанопластини через кристалографічно орієнтований ріст у парах і твердому тілі на підкладках з аморфного оксиду, та плоскі гексагональні нанопластини, утворені шляхом росту ПРК великої площі в рідких краплях Te. В залежності від вибору субстрату й умов росту можливим є процес легування. Наприклад, вертикальні нанопластини, вирощені на сапфірових підкладках, можуть включати рівномірну густину атомів Al з підкладки.

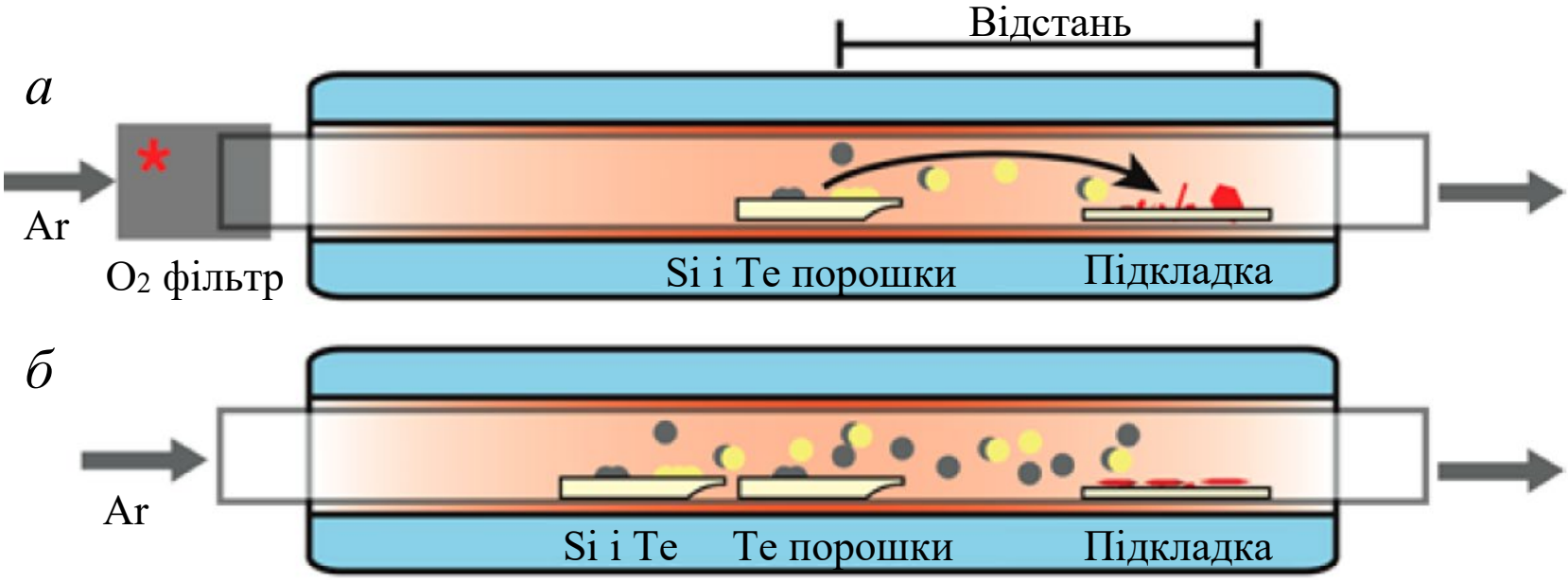

Рис. 1.33. Вирощування нанострічок і вертикальних нанопластин (*а*) та плоских нанопластин з використанням посіву підкладки за допомогою Te (*б*) [49].

Нанокристали $Si_2Te_3$ одержують методом хімічного парового осадження у відкритій проточній системі за механізмом «пара-рідина-кристал» (ПРК). Синтез нанокристалів $Si_2Te_3$ автори [49–51] проводили у трубчастій печі опору в потоці Ar. Ростова установка, схема якої зображена на рис. 1.33, складається з горизонтальної трубчастої печі, в яку поміщений кварцовий реактор з керамічним (алундовим) човником. Для вирощування нанокристалів $Si_2Te_3$ вихідні порошкоподібні телур і кремній завантажують в керамічний тигель, який потім поміщають у високотемпературну трубчасту піч [49]. Реактор приєднували до системи очищення та подачі газів. Морфологічні властивості ансамблів нанокристалів (форма, довжина, діаметр, поверхнева густина) визначаються способом підготовки поверхні підкладки та умовами процесу росту (табл. 1.3).

Нанострічки ростуть на підкладках з кремнію, плавленого кварцу або сапфіру, нагрітих до температури 953 K і розташованих на відстані ~12 см нижче центральної зони печі, де розташований керамічний тигель з вихідними речовинами, температура якої складає 1073 K (рис. 1.33, *б*). При температурі 1073 K кремній ще не випаровується (оскільки температура плавлення кремнію складає 1673 K) про що свідчить відсутність транспорту Si у відсутності Te, але може реагувати з парами $Te_2$. Телур випаровується при відносно низькій температурі ($T_{пл}$ = 723 K) і переноситься у вигляді $Te_2$, а в результаті реакції з порошком Si, SiTe, транспортує кремній. При більш низьких температурах (< 1023 K) телур випаровується, але не реагує істотно із кремнієм, так що на підкладках утворюється тільки відкладання телуру. Транспортований надлишок телуру утворює сферичні частинки на підкладках що може привести до самокаталізованого ПРК росту нанострічок та плоских нанопластин.

**Ріст нанопластин $Si_2Te_3$ паралельно підкладці.** Заповнивши підкладки великим масивом телуру можна вирощувати пластинки великої площі, паралельних підкладці. Для цього керамічний тигель заповнений порошками Si і Te розміщують при температурі 1073 K, як показано на рис. 1.33, *б*. Порошок телуру розміщують в центральній частині трубчастої печі з максимальною температурою 1123 K. У результаті, спочатку телур випаровується в центральній зоні печі й осаджується на підкладках розташованих при меншій температурі за течією. Нанопластини ростуть щляхом пошарового росту паратверде тіло, або виділяючись із підкладки, де орієнтація контролюється початковою взаємодією парів прекурсора з підкладкою, або плоско на підкладці за механізмом ПРТ великої площі.

**Ріст вертикальних нанопластин.** Вертикальні нанопластини $Si_2Te_3$ ростуть завдяки зв'язуванні $Si_2Te_3$ з оксидними підкладками, починаючи приблизно на 12.5 см нижче за потоком від центру печі, що відповідає піковій температурі 923 K. Цей режим росту чітко

Таблиця 1.3 Морфологія нанокристалів $Si_2Te_3$, отриманих за різних умов.

| Морфологія росту | Температура джерела Si/Te, K | Положення підкладки, см | Температура підкладки, K | Каталізатор | Час росту, хв | Література |
|---|---|---|---|---|---|---|
| Вертикальні нанопластини | 1073 | ~12.5 | ~923 | Te | | [49] |
| | 1123 | | 923 | | 3–5 | [52] |
| Нанопластини | 1123 | | 873 | | 3–5 | [52] |
| Плоскі пластини | 1073 | >12 | < 953 | | | [49] |
| | 1123 | | 823 | | 3–5 | [52] |
| Наношпильки | 1123 | | 853 | Au | 3–4 | [52] |
| Нанодротини | 1123 | | 873 | Au | 4 | [52] |
| Нанострічки | 1073 | ~12.5 | ~ 953 | | | [49] |
| Макрокристали | 1073 | >15 | < 698 | | | [49] |

кристалографічно орієнтований таким чином, що грань, зв'язана з підкладкою, завжди є гранню $\{1\bar{1}00\}$, тоді як інші краї нанопластинки мають (як і в інших режимах росту) $\{2\bar{1}\bar{1}0\}$ грань, що приводить до п'ятикутної форми з двома кутами 90° і трьома кутами 120°. При вирощуванні вертикальних нанопластин на сапфірових підкладках відбувається їх легування алюмінієм за рахунок його потрапляння із підкладки $Al_2O_3$.

Активний розвиток нових перспективних напрямків нанотехнологій ініціює не тільки удосконалення сучасних аналітичних методів дослідження, але й пошуки принципіально нових підходів до їх аналізу. Однією із важливіших характеристик наноструктури є її склад. Оскільки мова йде про параметри системи пониженої розмірності, то і просторове розділення інструмента, який вимірює склад повинно бути, як мінімум субмікронним, а ще краще нанометровим.

## 1.7. ОТРИМАННЯ ПОЛІКРИСТАЛІЧНИХ ТОНКИХ ПЛІВОК $Si_2Te_3$

Однорідні полікристалічні тонкі плівки $Si_2Te_3$ великої площі ~ 2×8 см$^2$ автори [62] вирощували на підкладках $SiO_2$/Si методом хімічного осадження з парової фази (chemical vapor deposition (CVD)). Процес вирощування однорідних тонких плівок $Si_2Te_3$ на поверхні підкладок $SiO_2$/Si методом CVD проводився в 1-дюймовій кварцовій трубці (рис.1.34), яка була поміщена в 3-дюймову кварцову трубку горизонтальної трубчастої печі.

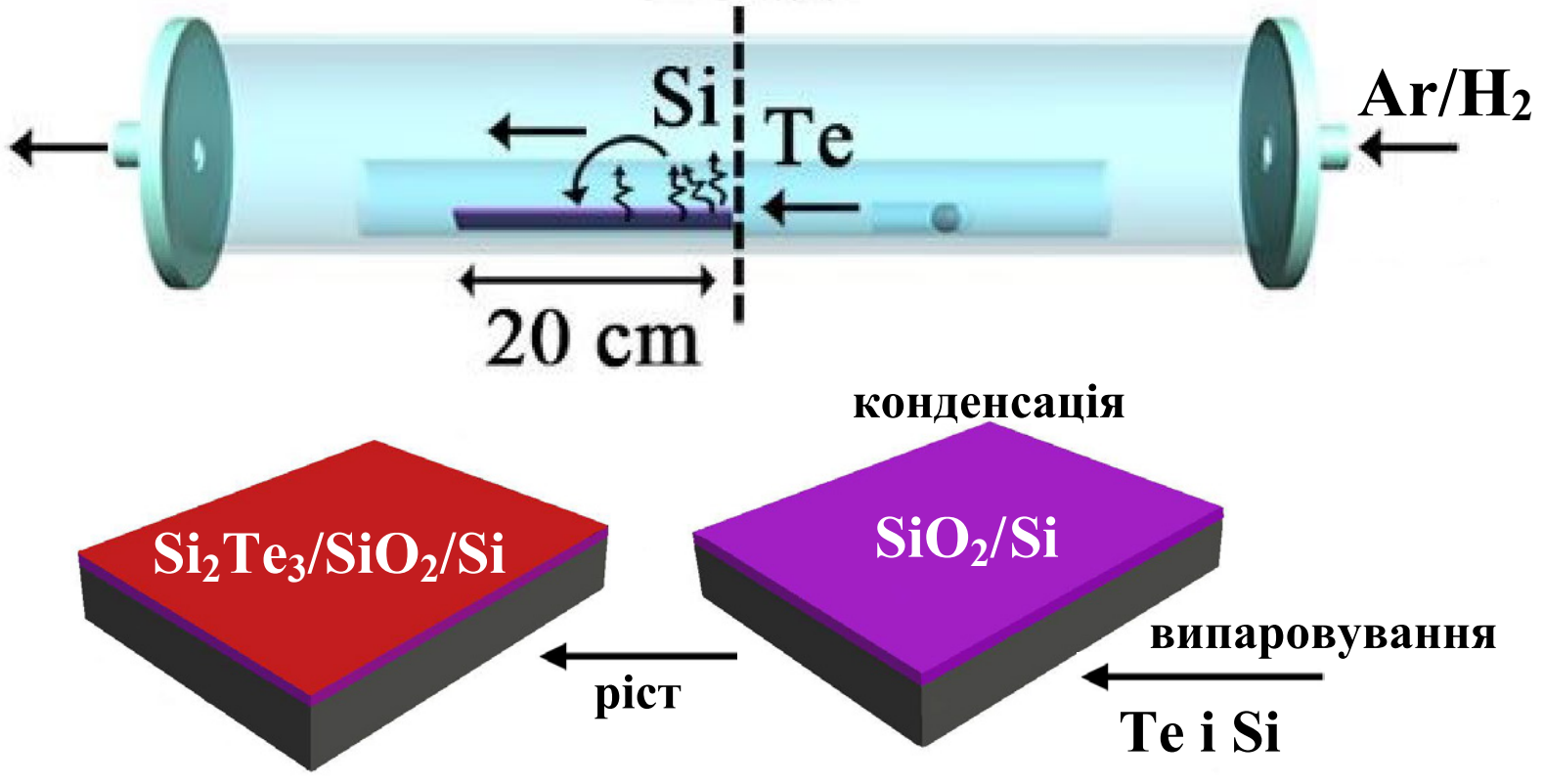

Рис. 1.34. Схематична ілюстрація процесу росту тонких плівок $Si_2Te_3$ CVD [62].

Ріст плівок відбувається за механізмом пара – рідина −тверде тіло, який дозволяє контролювати товщину плівки та кристалічну структуру за температурою підкладки. Температура підкладки задавалась температурною зоною, у якій розміщувалась підкладка. Спочатку підкладки поміщали в температурну зону близько 773–973 К. Потім систему термічного вирощування нагрівали до 1023 К в атмосфері суміші Ar і $H_2$ (50:5 см$^3$) при тиску 20 мТорр. Джерело Te, розміщене вище за течією, випарувалося при температурі близько 723 К і транспортувалося за течією. Тривалість процесу росту плівок складала від 30 до 60 *хв*. Після вирощування температуру в печі повільно охолоджували до 773 К зі швидкістю 25 К/*хв*, а потім швидко охолоджували до кімнатної температури, використовуючи Ar як захисний газ. У цій системі температура підкладки та час вирощування є двома ключовими параметрами, якими можна контролювати товщину тонких плівок $Si_2Te_3$.

Джерело Te і підкладки $SiO_2$/Si були розміщені відповідно до, та в центрі зони постійної температури відповідно. Порошок Te випарувався і транспортувався вниз за течією за допомогою газу-носія (аргон (Ar)/водень ($H_2$) = 50 см$^3$ / 5 см$^3$), коли температура підвищувалась до 1023 К. Для отримання стабільної подачі попередників у експерименті, порошок Te поміщали в невелику кварцову трубку з герметичним одним кінцем. Ідеальна низька температура плавлення джерела Te при 20 мТорр може знизити температуру плавлення Si до 1023 К, або навіть нижче. У цій системі підкладка діє як опора, так і джерело Si для росту плівки $Si_2Te_3$. У температурній зоні 773 – 973 К була отримана суцільна багатошарова тонка плівка $Si_2Te_3$ при тривалості росту 60 *хв*. Під час росту атоми Si і Te реагували один з одним у зоні постійної температури (1023 К), а потім транспортувалися вниз за течією газом-носієм. При відповідній температурі підкладки продукти в суміші були зібрані разом і під час повільного охолодження утворилася тонка плівка. Для отримання якісних однорідних плівок найбільш оптимальною є температура підкладки в діапазоні 873 – 923 К. Рентгеноструктурні дослідження показали, що плівки $Si_2Te_3$ є полікристалічними з параметрами гратки $a$ = 7.421 Å і $c$ = 13.504 Å.

# РОЗДІЛ 2

## СКЛОУТВОРЕННЯ В СИСТЕМАХ Si–Te ТА Si–M–Te (M = Se, Ge, Pb, Sn, Cu, Ag, Al, In)

Телуридні стекла, тобто стекла на основі телуру, становлять особливий клас матеріалів, що використовуються в багатьох технологічних пристроях. Хоча багато телуридних систем не є хорошими склоутворюювачами, було розроблено широкий спектр композицій на основі кремній-телуридних стекол, які мають унікальний набір акустооптичних і електричних властивостей. Кремній-телуридні стекла мають дуже широку прозорість в інфрачервоному діапазоні, яка може досягати понад 20 мкм, що робить їх особливо привабливими для оптичних застосувань в далекому інфрачервоному діапазоні. Кілька сімейств кремній-телуридних стекол (наприклад, потрійних на основі Te, Si та Se, Ge, Sn, Cu, Ag, Al, In) були досліджені та оптимізовані для подальших різноманітних застосувань. Бінарні і потрійні телуридні стекла (та аморфні плівки) знаходять застосування у виготовлені перезаписуваних оптичних дисків і пристроїв пам'яті зі зміною фази, оскільки деякі композиції демонструють швидке та зворотне перетворення між кристалічною та склоподібною (аморфною) фазами.

### 2.1. СКЛОУТВОРЕННЯ В БІНАРНІЙ СИСТЕМІ Si–Te

**2.1.1. Одержання об'ємних стекол методом гартування розплаву.** Головна особливість, яка відрізняє склоподібний стан від інших аморфних станів – це те, що у скла існує зворотний перехід із склоподібного стану в розплав та із розплаву в склоподібний стан. Ця властивість характерна тільки для скла. В інших типів аморфних станів при нагріванні відбувається перехід речовини спочатку в кристалічний стан і тільки при підвищенні температури до температури плавлення – у рідкий стан. У склоутворюючих розплавах поступове зростання в'язкості розплаву перешкоджає кристалізації речовини, тобто переходу в термодинамічно більш стійкий стан з меншою вільною енергією. Процес склування характеризується температурним інтервалом $\Delta T$ – інтервалом склування.

Склоутворення в бінарній системі Si–Te досліджувалось багатьма авторами [64–82], але результати цих робіт істотно різняться стосовно концентраційних меж областей склоутворення, що викликано різними умовами одержання скла – загальною масою наважки, швидкістю охолодження розплаву, температурою, від якої проходило загартування і видом охолоджуючої речовини. Традиційно стекла $Si_xTe_{100-x}$ синтезують методом прямого сплавлення елементарних компонентів у вакуумованих кварцових ампулах. При приготуванні цих стекол особливу увагу необхідно приділяти виключенню контакту з парами води або охолоджувача.

Про можливість склоутворення в системі Si–Te вперше згадується в роботі [64]. З порівняння області склоутворення в бінарній системі Si–Te та діаграми стану цієї системи (рис.1.2, розділ 1) випливає, що стекла утворюються в області евтектики з боку телуру. Евтектика між Te та $Si_2Te_3$ має відносно низьку температуру плавлення (682 K) і відповідає складу з 17÷18 ат. % Si. Положення області склоутворення у системі Si–Te та максимальна склоутворююча здатність згідно з [65] пов'язані з існуванням максимуму в'язкості розплавів для складу поблизу евтектики, що зумовлено посиленням ковалентного характеру зв'язку Te–Si поблизу евтектики. При розгляді процесів склоутворення важливу роль відіграє швидкість охолодження розплаву, що подавляє процес кристалізації при фазовому переході рідина – тверде тіло.

Синтез сплавів системи Si–Te автори [66, 67] проводили із вихідних матеріалів високої ступені чистоти (~99.999%) у відкачаних і запаяних кварцових ампулах зі сплощеним (товщиною 0.5–2 мм) або видовженим конусним кінцем. З метою гомогенізації здійснювалось багаторазове перемішування розплаву у процесі синтезу. Охолодження розплавів проводилось у конусній частині ампули у режимі остигання на повітрі (швидкість охолодження 50–70 К/хв, маса наважки 10 г). Це дозволило авторам [66, 67] оцінити склоутворюючу здатність розплаву за величиною діаметра перерізу конуса, в якому утворилося скло без кристалізації (рис. 2.1). Склоподібний стан ідентифікувався за характерним раковистим зломом та відсутністю ліній на дебаєграмах. Як видно із рис. 2.1, найбільшу склоутворюючу здатність має не евтектичний склад, а $Si_{20}Te_{80}$.

Згідно [68], експериментально виявлене концентраційне зміщення мінімумів критичної швидкості охолодження склоутворюючих розплавів щодо евтектичних складів викликано тим, що склади переохолоджених розплавів, у яких нерівноважні процеси зародження і росту кристалів різних фаз відбуваються з однаковими швидкостями, не збігаються з рівноважними евтектичними складами, а також утворенням при вимірюваннях критичної швидкості охолодження метастабільних кристалічних фаз та евтектичних структур.

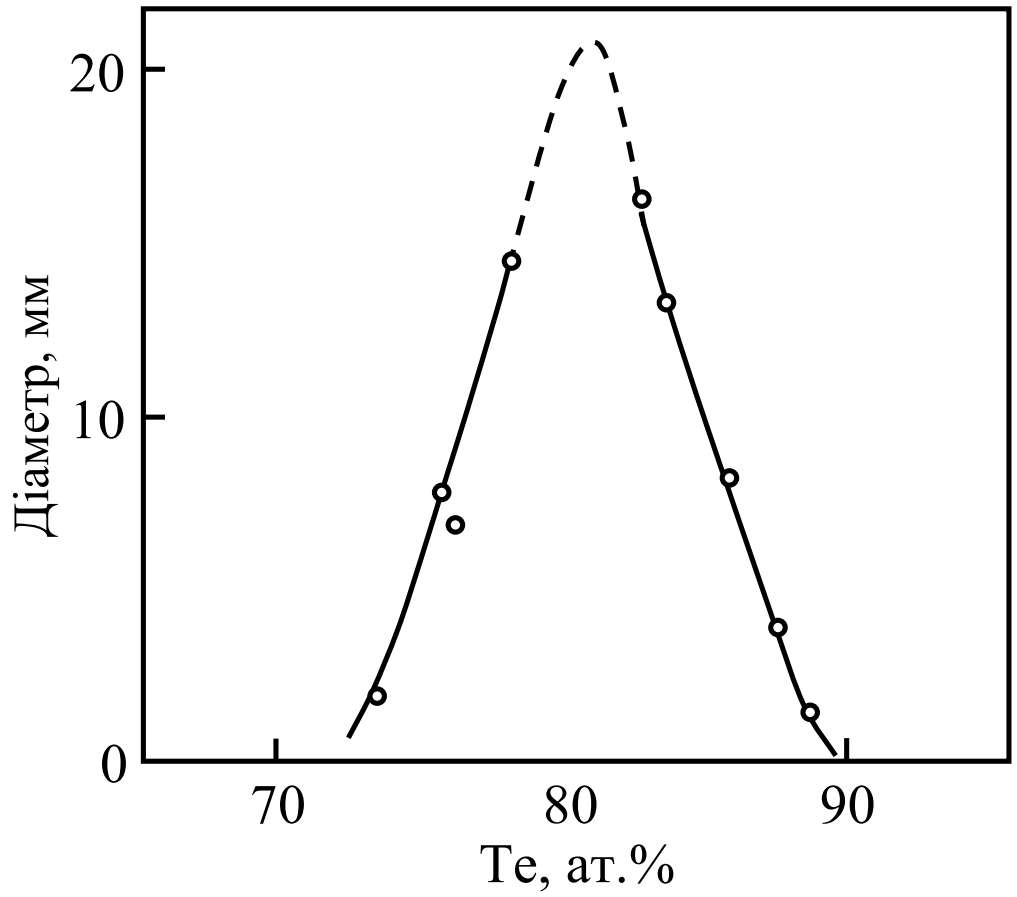

Рис. 2.1. Залежність склоутворюючої здатності скла (діаметр конуса в мм) від складу системи кремній–телур [66].

Використовуючи загартування розплаву від температури 1273 K в рідкому азоті, автори [70] встановили область склоутворення у цій системі в межах складів від 15 до 25 ат. % Si. Близьку за розмірами область склоутворення (15÷23(25) ат. % Si) отримала інша група авторів [71–73], використовуючи загартування ампул із розплавом на повітрі та у воді. За більшої швидкості охолодження розплаву (180 K/c) область склоутворення у цій системі лежить в інтервалі концентрацій від 10 до 22 ат.% Si [65]. Використовуючи тонкостінні сплощені ампули і провівши загартування розплаву від температур, що перевищують температури ліквідусу на 150÷200 K для відповідного складу, в крижану воду (швидкість охолодження ~ 250 K/c) вдалося ще більше розширити межі області склоутворення в системі Si–Te від 10 до 27.5 ат.% Si [66, 67, 74–77]. Ці дані підтверджують результати робіт [78–82], де стекла $Si_xTe_{100-x}$ були отримані звичайним методом швидкого загартування, а стекла на границі області склоутворення були отримані з використанням сплощених кварцових ампул і загартуванням розплаву в суміші NaOH і крижаної води.

Стекла $Si_{20}Te_{80}$ з різним ступенем структурної досконалості одержані авторами [69]. З цією метою синтез стекол $Si_{20}Te_{80}$ проводився в кварцових ампулах різної конфігурації: 1) зі сплощеним кінцем, товщиною ~ 1 мм; 2) циліндричних, із внутрішнім діаметром 10–50 мм; 3) конусних, з діаметром 0.5–20 мм. Процес загартування роз-плавів проводився у крижаній воді або на повітрі. У результаті були отримані зразки у вигляді пластинок товщиною 0.8–1 мм, злитків діаметром 10–50 мм та вагою від 2 до 300 г, які мали раковистий злом і не містили видимих під мікроскопом кристалічних включень. Швидкість охолодження не тільки визначає межі області склоутворення, але й істотно позначається на формуванні структури скла. Рентгеноструктурні дослідження отриманих стекол $Si_{20}Te_{80}$ показали, що в результаті різних режимів охолодження сплаву його структура терпить наступні зміни: охолодження у крижаній воді дає однорідне скло, а при охолодженні на повітрі у сітці скла виникає система нанокристалітів Te розміром, як правило, ~ 100 Å.

**2.1.2. Одержання аморфних стрічок $Si_xTe_{100-x}$ методом спінінгувавння розплаву (melt-spinning technigue).** Одним з найбільш перспективних методів отримання аморфних халькогенідних напівпровідників, що забезпечує швидкість охолодження $10^6$ К/с, вважається метод спінінгування розплаву на зовнішній поверхні обертального диска чи барабана. У даному методі матеріал розплавляється у тиглі, після чого струмина розплавленого матеріалу під тиском інертного газу видавлюється через сопло і потрапляє на зовнішню поверхню диска, що обертається, де твердне у вигляді тонкої стрічки, яка потім відокремлюється від диска під дією відцентрової сили. Найбільш детально умови синтезу аморфних $Si_xTe_{100-x}$ методом спінінгування розплаву описані в роботах [83–85]. Попередньо сплавлений злиток нагрівають до температури, яка на 50 К перевищує температуру ліквідусу. Струмина розплаву діаметром 1–1.5 мм виливається на поверхню водоохолоджуваного диска, який обертається з частотою 900–1500 об/с, що забезпечує швидкість охолодження $10^6$ К/с. Процес проводиться в атмосфері аргону при надмірному тиску 0.2 МПа. Сили поверхневого натягу не дають розплаву вільно виливатися. При додаванні надлишкового тиску аргону струмина рідкого матеріалу падає на барабан (диск), охолоджується і твердне у формі стрічки. Схематичне розташування сопла над поверхнею диска, що гартує, показано на рис. 2.2. Для прискорення процесу плавлення та досягнення максимальних температур сопло повинно задовольняти наступним вимогам [85]: діаметр кварцової трубки 12–14 мм, товщина стінок трубки до 1 мм, довжина звуженої частини сопла 3–5 мм. Що стосується діаметра сопла, то для металевих систем існує емпірична залежність між діаметром круглого отвору сопла та шириною одержуваної стрічки: $b = (1.1–2.5)\cdot d$, де $b$ – ширина стрічки в мм; $d$ – діаметр сопла в мм [85].

Для сплавів на основі телуру коефіцієнт пропорційності становить ~5. Тобто при діаметрі сопла ~1 мм ширина одержуваної стрічки досягає значення ~5 мм. Для визначення оптимальної кутової швидкості обертання диска, автор [85] скористався моделлю безперервної течії струменя, вихідними даними про розмір сопла, часу проходження спінінгування та ін.

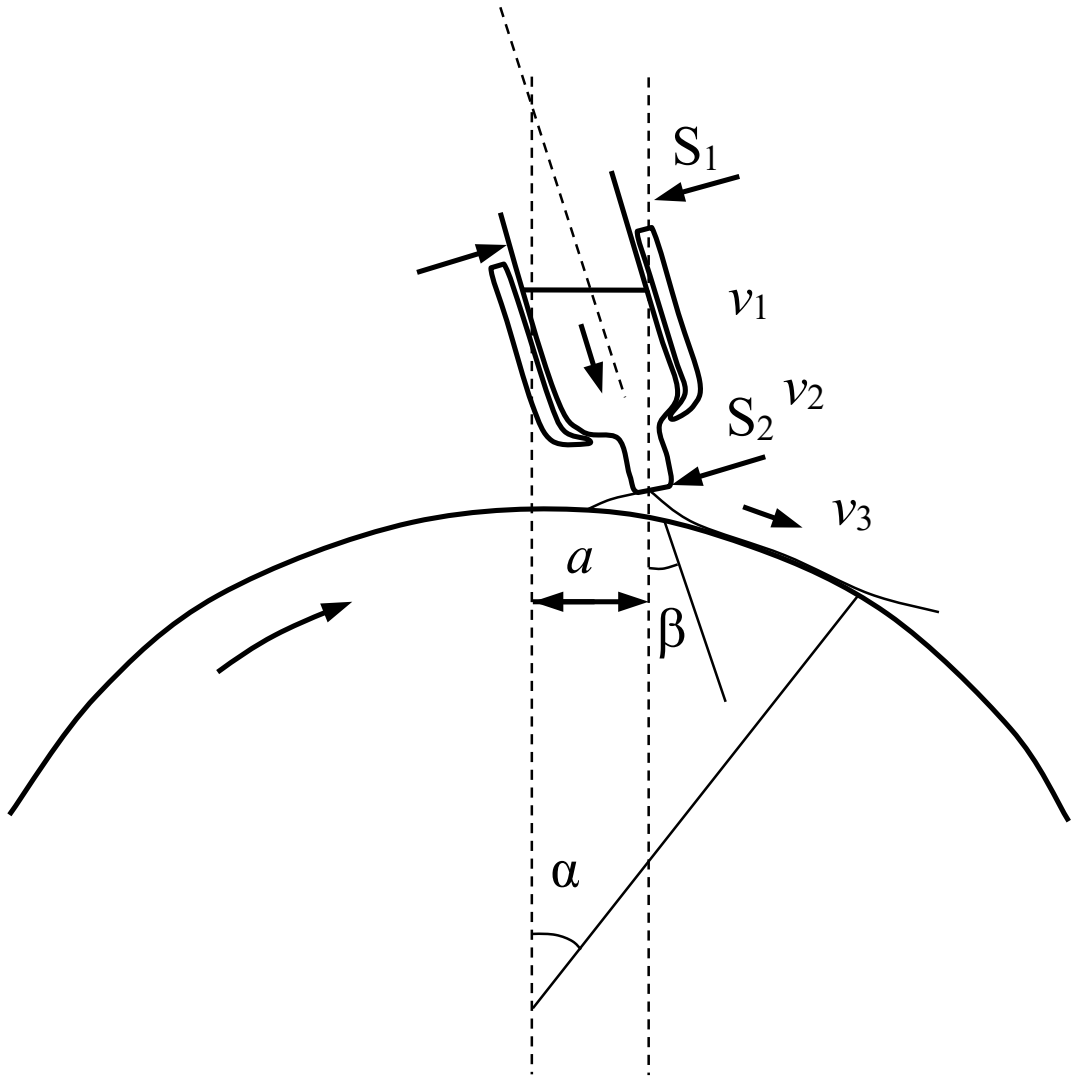

Рис. 2.2. Схематичне розміщення сопла поблизу диска [85].

Як приклад, нижче наведено розрахункові дані для процесу спінінгування сплаву $Si_{0.07}Te_{0.93}$ за результатами роботи [85]. При висоті розплаву в кварцовій ампулі $h$ = 20 мм і тривалості процесу спінінгування $t$ = 3 с, середня швидкість руху рідини в ампулі дорівнюватиме $\upsilon_1 = h/t = 7{\cdot}10^3$ м/с. Враховуючи, що $S_1$ = 113 мм$^2$, а $S_2$ = 0.5 мм$^2$, швидкість витікання рідини із сопла ампули: $\upsilon_2 = \upsilon_1{\cdot}S_1/S_2$ = 1.58 м/с. Без урахування зміни об'єму при переході з рідкого в аморфний стан розрахована швидкість руху стрічки на поверхні закалюючого диска в процесі спінінгування $\upsilon_3 = \upsilon_2{\cdot}S_2/S_3$ = 13.2 м/с, де $S_3$ – площа поперечного перерізу стрічки, що дорівнює добутку очікуваних значень ширини стрічки на товщину стрічки. Швидкість $\upsilon_3$ – не що інше, як лінійна швидкість руху точок поверхні диска, радіус якого $R$ = 80 мм. Звідси, кутова швидкість обертання диска $\omega = \upsilon_3/R$ = 165 рад/с. Відповідно, частота або кількість обертів диска за секунду буде 27

Гц. Отже, для підтримки оптимальних умов спінінгування потрібно, щоб частота обертання диска була 27 Гц, а надлишковий тиск аргону забезпечував швидкість витікання рідини із сопла 1.58 м/с. Окрім того, на якість одержуваної стрічки в процесі спінінгування впливає положення та кут нахилу сопла. Як видно із рис. 2.2, параметрами, що задають положення і кут нахилу сопла, є відстань між вертикальною віссю симетрії диска і кінцем сопла, та кут $\beta$ нахилу осі симетрії ампули щодо вертикальної осі. Ці параметри підбираються емпірично, і для спінінгування сплавів системи Si–Te оптимальні значення становлять: $a = 2$ см; $\beta = 15°$.

Застосовуючи метод спінінгування розплаву авторам [83–85] вдалося розширити інтервал отримання об'ємних аморфних сплавів у системі $Si_xTe_{100-x}$ до $5 < x < 40$. Зважаючи на збереження ковалентного типу зв'язків і в некристалічному стані, стрічки виходять крихкими, і відбувається їх часткове руйнування при зіткненні з внутрішніми стінками установки. Крім того, довжина стрічки обмежена кількістю матеріалу, який завантажується у кварцову ампулу, обтиснуту танталовим нагрівником. Тому метод є квазінеперервний. Цим і зумовлені труднощі отримання безперервної стрічки. Тому оптимальні розміри шматків стрічок, отриманих авторами [83–85] при швидкості охолодження ~$10^6$ К/с становили: товщина 20–40 мкм, ширина до 6 мм і довжина 10–20 см.

Автори [83–85] вказують на ряд технологічних труднощів отримання аморфних стрічок $Si_xTe_{100-x}$: неможливість використання високочастотного індуктора; висока схильність матеріалу до окислення, що вимагає додаткових запобіжних заходів; спрямованість ковалентних зв'язків і, як наслідок, крихкість одержуваних стрічок; вплив матеріалу диска або барабана на якість одержуваної стрічки. Матеріал барабана (диска) та обробка його поверхні мають важливе значення, оскільки визначають швидкість тепловідведення, адгезію розплаву та теплопередачу від нього до охолоджувача. При необхідності можна використовувати барабани, виготовлені з різних матеріалів (латуні, міді, дюралі та сталі). Але найбільш ідеальним матеріалом диска для спінінгування розплаву $Si_xTe_{100-x}$ є мідь.

Геометрія, структура та інші характеристики одержуваних стрічок визначаються сукупністю низки технологічних параметрів процесу спінінгування, до найважливіших з яких можна віднести такі: температура розплаву; форма, переріз, нахил та швидкість подачі струменя розплаву; матеріал, температура, швидкість руху та стан поверхні охолоджуваного диска; відстань між соплом та цією повер-хнею; склад та тиск навколишнього газового середовища.

**2.1.3. Одержання склоподібного сплаву $Si_{20}Te_{80}$ у відсутності гравітації.** Синтез склоподібних сплавів в умовах мікрогравітації або повної відсутності гравітації призводить до покращення ряду харатеристик стекол. Відсутність гравітації – це фактично, безконтейнерний спосіб синтезу. Розплав, внаслідок поверхневого натягу, збирається у сферичні утворення і зависає в середині ампули, не контактуючи з її стінками. Невагомість і відсутність впливу стінок ампули на процес синтезу приводить до того, що у розплаві усуваються [86]: 1) термічні конвекційні потоки, які призводять до утворення флуктуацій густини; 2) забруднення розплаву киснем та іншими контейнерними домішками, які погіршують оптичні властивості речовини; 3) гетерогенне зародкоутворення при стиканні розплаву зі стінками контейнера.

Вперше в умовах мікрогравітації (у космосі, на станції «МИР») автори [87–90] отримали склоподібний $Si_{20}Te_{80}$, який виявився більш однорідним по структурі й менш дефектним, ніж його земний аналог. Цей факт засвідчує зменшення ймовірності зародження кластерів при затвердіванні в умовах мікрогравітації внаслідок ефекту відриву розплаву від внутрішніх стінок ампули. Ренгеноструктурний мікроаналіз складу показав, що скло $Si_{20}Te_{80}$, отримане в космосі, є однофазним з вмістом Si 20 ± 1 ат.%. Натомість у склі, отриманому на землі, концентрація Si коливалась в межах 3 ат.%. Середня густина «космічного» скла (5.033 г/м$^3$) була трохи вищою ніж «земного» (5.029 г/м$^3$). Мікротвердість «космічного» скла (136 кг/мм$^2$) менша, ніж «земного» (150 кг/мм$^2$), що вказує на більш високу мікрооднорідність «космічного» зразка.

Порівняльне дослідження виявлених газових бульбашок (пухирців) в стеклах $Si_{20}Te_{80}$ показало, що розміри пор (до 40 мкм) та їх розподіл по шліфам в основному не залежать від гравітаційних умов затвердівання. Аналіз можливостей газовиділення у розплаві $Si_{20}Te_{80}$ дозволив авторам [87–90] встановити, що воно зумовлено випаровуванням телуру – зародження нової фази відбувається в умовах термодинамічної рівноваги і піддається оцінкам у тій мірі, в якій відомі фізичні параметри розплаву.

**2.1.4. Методи одержання тонких аморфних плівок $Si_xTe_{100-x}$.** Тонкі аморфні плівки $Si_xTe_{100-x}$ отримують різними методами: термічним випаром у вакуумі попередньо синтезованих полікристалічних сплавів складу 2 ÷ 25 ат. % Si [71]; вакуумним дискретним термічним випаром полікристалів (склади 5 ÷ 50 ат. % Si) на непідігріті підкладки із шліфованого скловуглецю [91, 92]; спільним випаровуванням кремнію і телуру у вакуумі $1.33\cdot10^{-6}$ Па при температурі підкладок 323 K (складів 0 ÷ 82 ат. % Te) [93] та ВЧ-розпиленням полікристалічних сплавів в атмосфері Ar на підкладки з каптону або монокристалічного $SiO_2$ [94].

## 2.2. КРИСТАЛІЗАЦІЯ СТЕКОЛ $Si_xTe_{100-x}$, ОДЕРЖАНИХ РІЗНИМИ МЕТОДАМИ

Кристалізація заевтектичних, евтектичних та доевтектичних сплавів $Si_xTe_{100-x}$ є багатостадійним процесом і у кожному випадку має свої характерні особливості, що залежать від умов одержання скла.

**2.2.1. Кристалізація стекол, отриманих загартуванням розплаву у воду.** Дослідження залежності температури початку кристалізації $T_к$ стекол із бінарних сплавів, близьких до евтектичного складу $A^{IV}_{15}Te_{85}$, де $A^{IV}$ = Si, Ge, Sn, Pb, від атомного номера Z елемента $A^{IV}$ показали, що $T_к$ зменшується із зростанням Z за лінійним законом [95]. Ця проста залежність здається дуже примітивною, оскільки розглянуті подвійні системи не володіють властивостями ідеальних розчинів ні в рідкому, ні в твердому стані й вказують на існування деякої домінуючої властивості, що визначається елементом $A^{IV}$ і не залежить від природи та розмірів ближнього порядку. Аналіз термодинамічних даних цих речовин показує, що властивості сплаву в склоподібному стані визначаються властивостями продуктів дисоціації, присутніми в розплаві, а $T_к$ зростає за лінійним законом зі зростанням ентропії плавлення, яка припадає на один атом елементу $A^{IV}$ [95].

Визначення наявності в стеклах $Si_xTe_{100-x}$ термічно-індукованого фазового переходу із склоподібного у кристалічний стан проводили автори [78–82, 96, 97] методами ДТА, диференціальної скануючої калометрії (ДСК), рентгеноструктурного аналізу та месбауерівської спектроскопії. Однак, результати цих досліджень суттєво різняться. Так, у роботі [78] експерименти з ДСК проводилися при швидкості нагрівання 20 К/хв двічі: до проходження першого екзотермічного піка кристалізації та після швидкого повторного охолодження, коли проявляється ендотермічний ефект «повторного» розсклування. На кривих ДСК (криві А, рис. 2.3, *а*, *б*) свіжоприготовлених

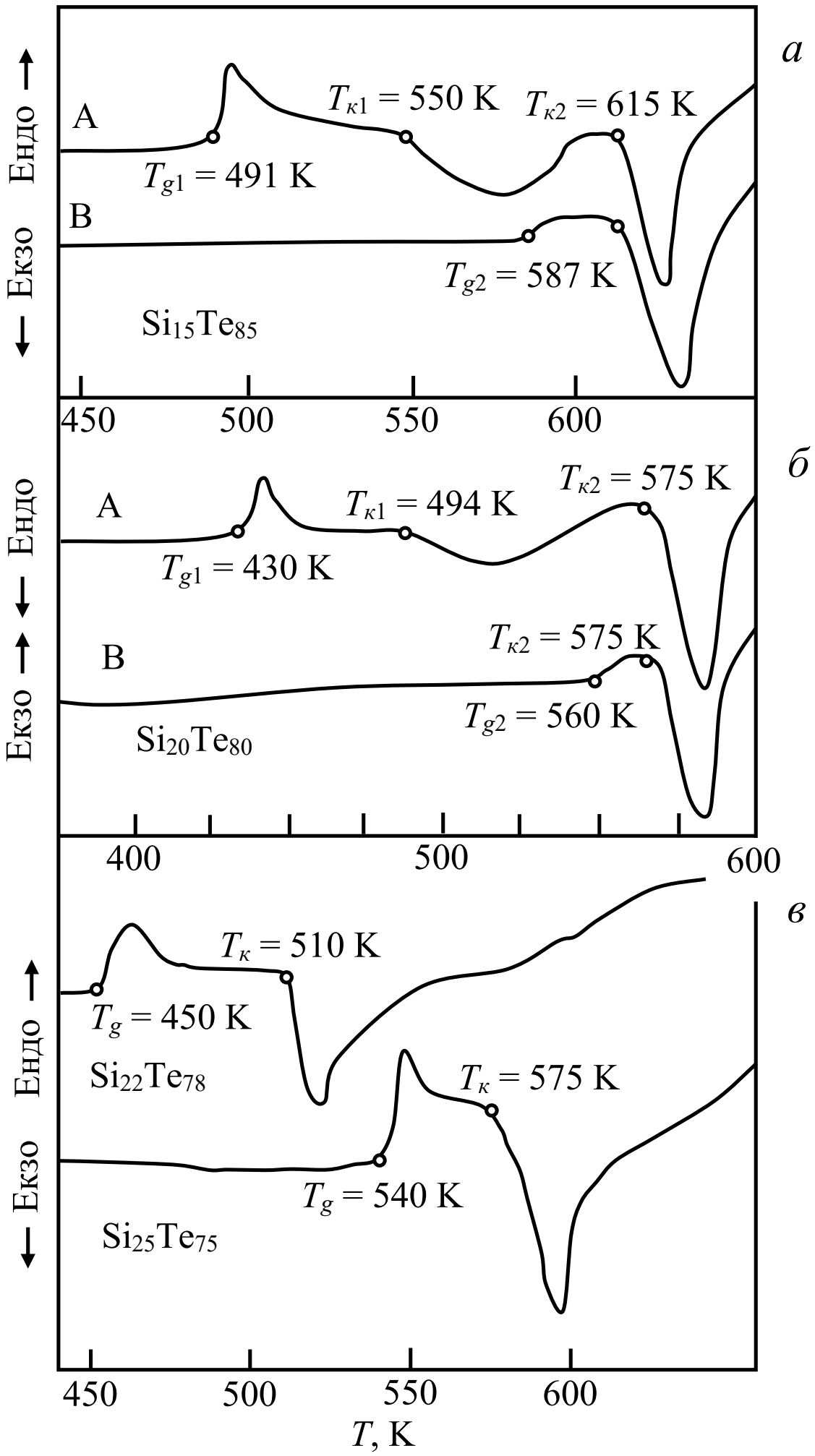

Рис. 2.3. Криві ДСК стекол $Si_xTe_{100-x}$. *x:* 15 (*а*); 20 (*б*); 22 і 25 (*в*). Швидкість нагрівання 20 К/хв; *dH/dt* = 10 мкал/с. Криві: А – безперервне нагрівання до повної кристалізації; В – безперервне нагрівання зразків, попередньо нагрітих до закінчення першої стадії кристалізації та охолоджених до кімнатної температури [79].

стекол $Si_xTe_{100-x}$ складів $10 \le x \le 20$ при нагріванні спостерігається один ендотермічний пік, пов'язаний з ефектом склування, і два екзотермічні піки, пов'язані з двоступінчатою кристалізацією. При охолодженні сплаву після нагрівання до температури, що лежить між двома кристалізаційними піками, при наступному нагріванні спостерігався новий ендотермічний пік (криві В, рис. 2.3, *а, б*), пов'язаний з ефектом склування, що свідчить про розшарування матриці на першій стадії кристалізації. Дифракційні дані показують, що на першій стадії кристалізації виділяється тригональна фаза Te, а на другій – гексагональна фаза $Si_2Te_3$ із аморфної матриці, що залишалася, після її повторної кристалізації. Стекла складів з $20 < x \le 28$ мають одну температуру склування (рис. 2.3, *в*) і кристалізуються одностадійно за евтектичною реакцією з утворенням гексагональних фаз Te і $Si_2Te_3$. Отримані закристалізовані фази мають дефектну структуру. При збільшені концентрації Si температура плавлення збільшується, температура другої стадії кристалізації і друга температура склування зменшуються, а температура першої стадії кристалізації і перша температура склування змінюються немонотонно і досягають мінімуму при $x = 20$ (рис. 2.4, *а*) [79]. Ці дані підтверджуються ще одним незалежним методом визначення температур склування $T_g$ і кристалізації $T_к$ – фотоакустичним (рис. 2.4, *б*) [98].

Незважаючи на те, що подальші дослідження [82, 96, 97] процесу кристалізації стекол $Si_xTe_{100-x}$, отриманих загартуванням розплаву у воду, підтвердили двостадійний процес кристалізації доевтектичних і евтектичних складів сплавів та одностадійний заевтектичних, проте температури плавлення, склування й кристалізації та їх концентраційні залежності суттєво відрізняються від результатів [79] (порівняй рис. 2.4 і 2.5). Автори [82, 96] не виявили мінімуму $T_g$ при $x = 20$, які спостерігали автори [79], та інтерпретували його наявність у термінах структурних змін скла для даного складу. Отримані результати кристалізації стекол $Si_xTe_{100-x}$ автори [82] інтерпретують наступним чином. При $x \le 21$ кристалізація стекол починається відразу після склування і внаслідок кристалізації телуру вміст Si у рідкій фазі збільшується. Оскільки $T_g$ збільшується зі збільшенням $x$, залишкова рідина стає більш в'язкою при сталій температурі $T$ і знову перетворюється у скло, по крайній мірі, у безпосередній близькості від зростаючих кристалів Te. При температурах, які значно перевищують початкову $T_g$, кристалізація відроджується і через деякий час дає пік при 510 K. Під час першої кристалізації в ізохронному

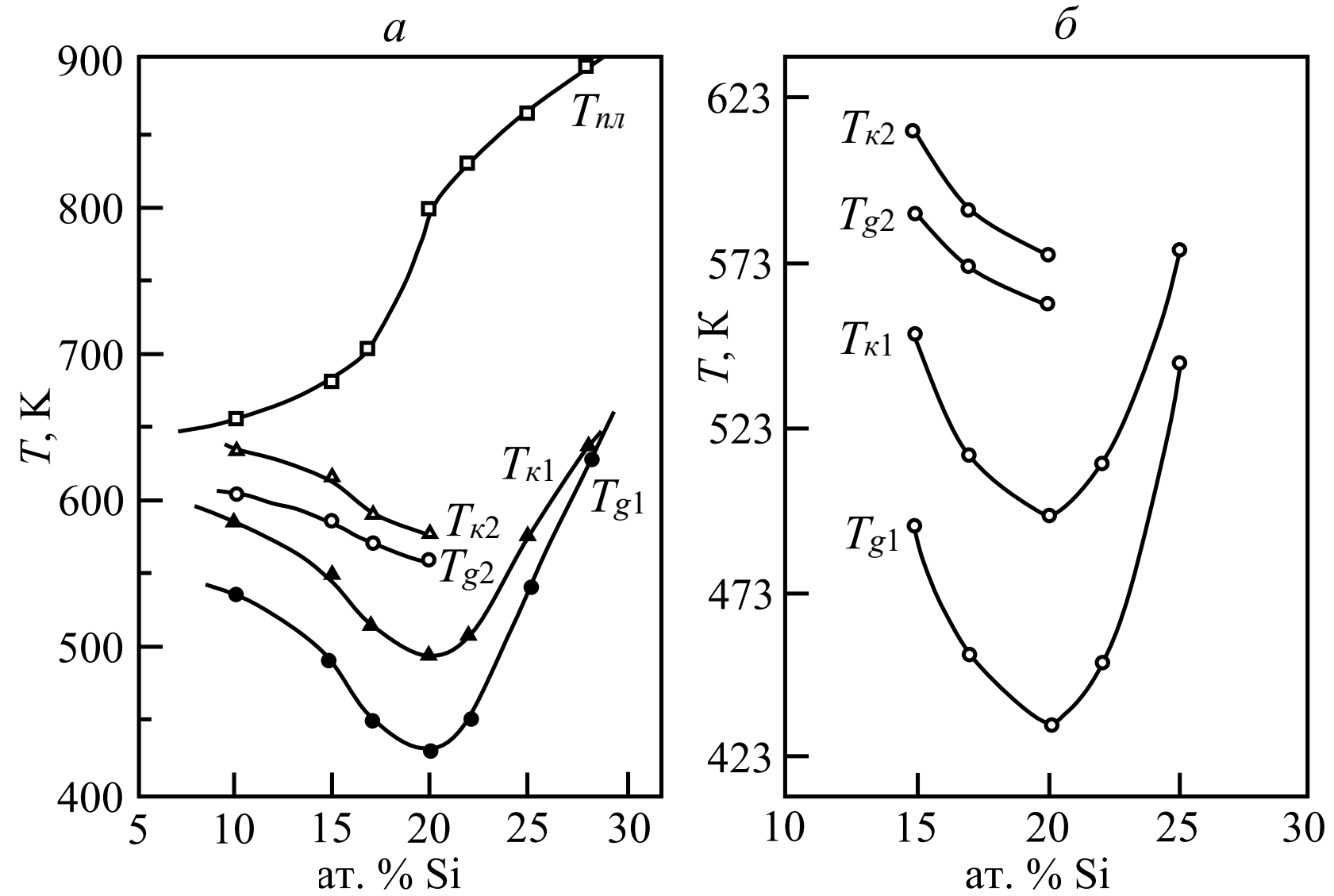

Рис. 2.4. Концентарційні залежності температур склування ($T_g$), кристалізації ($T_к$) та плавлення ($T_{пл}$) стекол $Si_xTe_{100-x}$, виміряні методом ДСК (*а*) [79] та фотоакустичним методом (*б*) [98].

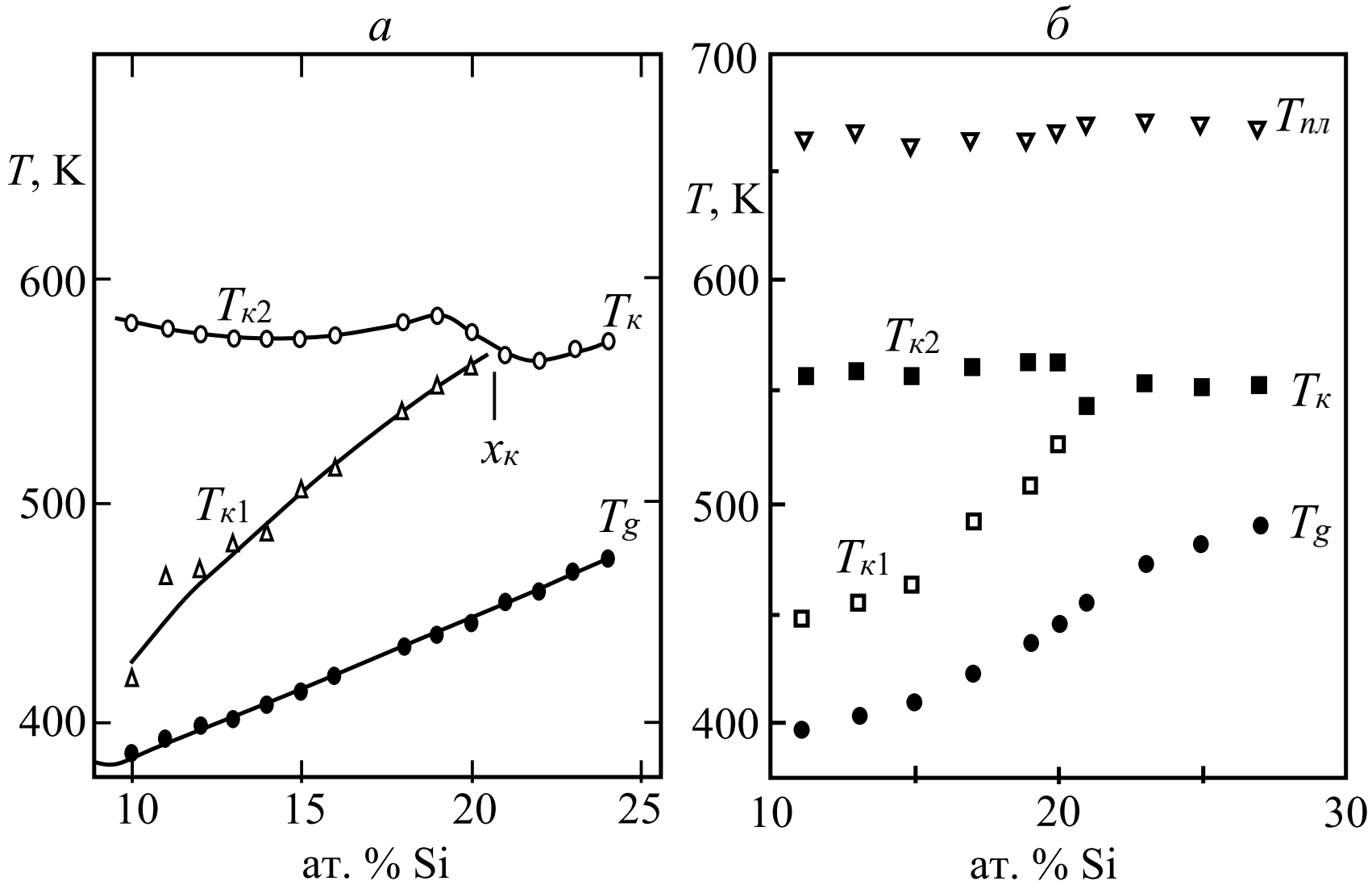

Рис. 2.5. Концентраційні залежності температур склування ($T_g$), кристалізації ($T_{к1}$, $T_{к2}$) та плавлення ($T_{пл}$) закристалізованого сплаву стекол $Si_xTe_{100-x}$. *а* – [96]; *б* – [82].

Таблиця 2.1. Термічні параметри стекол $Si_xTe_{100-x}$, одержаних загартуванням розплаву у воду.

| Склад | $T_{g1}$, K | $T_{g2}$, K | $T_{к1}$, K | $T_{к2}$, K | $T_m$, K | Література |
|---|---|---|---|---|---|---|
| $Si_{10}Te_{90}$ | 535<br>389 | 605<br>- | 586<br>442 | 636<br>582 | 656<br>- | [79]<br>[96] |
| $Si_{13}Te_{87}$ | 401 | - | 482 | 574 | - | [96] |
| $Si_{15}Te_{85}$ | 491 | 587 | 550 | 615 | 680 | [79] |
| $Si_{16}Te_{84}$ | 421 | - | 527 | 576 | - | [96] |
| $Si_{17}Te_{83}$ | 450 | 570 | 516 | 590 | 702 | [79] |
| $Si_{18}Te_{82}$ | 435 | - | 547 | 583 | - | [96] |
| $Si_{20}Te_{80}$ | 430<br>444 | 560<br>- | 494<br>563 | 575<br>577 | 798<br>- | [79]<br>[96] |
| $Si_{21}Te_{79}$ | 452 | - | 563 | - | - | [96] |
| $Si_{22}Te_{78}$ | 450<br>458 | - | 510<br>565 | - | 828<br>- | [79]<br>[96] |
| $Si_{23}Te_{77}$ | 468 | - | 570 | - | - | [96] |
| $Si_{25}Te_{75}$ | 540 | - | 575 | - | 860 | [79] |
| $Si_{28}Te_{72}$ | 626 | - | 633 | - | 890 | [79] |

Таблиця 2.2. Термічні параметри стекол $Si_xTe_{100-x}$, одержаних методом спінінгування розплаву [84].

| Склад | $T_g$, K | $T_{к1}$, K | $T_{к2}$, K |
|---|---|---|---|
| $Si_{10}Te_{90}$ | 381 | 417 | 554 |
| $Si_{15}Te_{85}$ | 400.5 | 472 | 553 |
| $Si_{20}Te_{80}$ | 428 | 547 | - |
| $Si_{25}Te_{75}$ | 437.5 | 546 | 605 |

експерименті склад рідини, що відновлюється спочатку швидко наближається до складу рідини з $T_g$, що відповідає реальній температурі, після чого слідує за кривою $T_g(x)$ (рис. 2.5, *а*). Насамкінець, квазіевтектика також кристалізується.

**2.2.2. Вплив домішки кисню на процес кристалізації стекол $Si_xTe_{100-x}$.** Легування стекол $Si_xTe_{100-x}$ ($10 < x < 25$) киснем автори [97] здійснювали шляхом додавання домішки $TeO_2$ в якості вихідного матеріалу в поєднанні з елементарними Si і Te й домішкою. Процес синтезу здійснювали в попередньо вакуумованих запаяних кварцових ампулах з наступним загартуванням розплаву у воду. Перші ознаки того, що кисень справді відіграє важливу роль у визначенні температури склування $T_g$ стекол $Si_xTe_{100-x}$, особливо для складів $x < 20$, видно при порівнянні результатів концентраційної залежності $T_g$ (рис. 2.6), отриманих двома групами авторів [79, 96, 97]. Крива 1 відображає температури склування стекол $Si_xTe_{100-x}$, отриманих окисленням елементів у процесі нагрівання, у ймовірно кисневовміс-

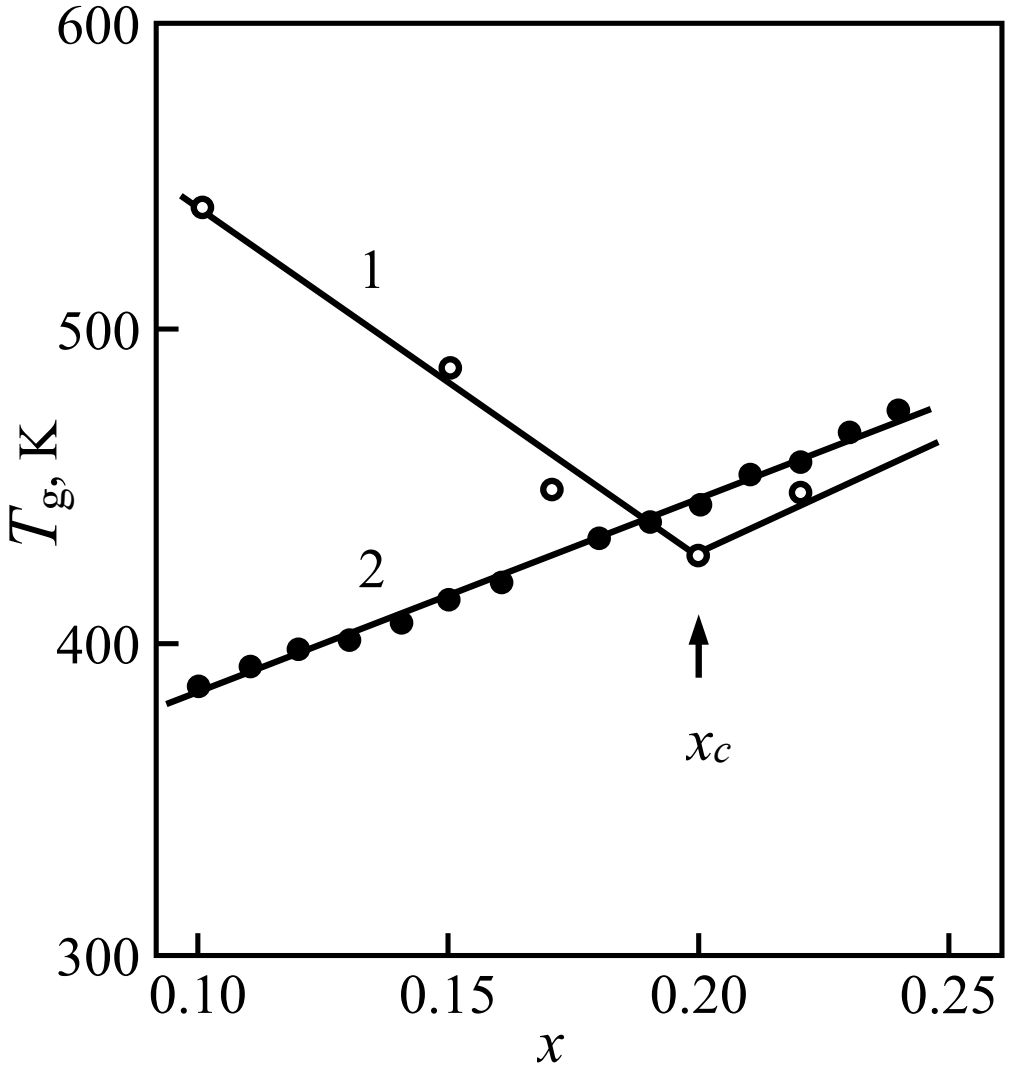

Рис. 2.6. Концентраційна залежність температури склування $T_g$ стекол $Si_xTe_{100-x}$ за даними: 1 – [79], 2 – [96,97].

ному аргоні [79]. Особливо важливим у цих результатах є той факт, що ефекти легування киснем різко зростають при $x \leq 20$, викликаючи збільшення $T_g$ майже на 130 К, наприклад, при $x = 10$, при цьому практично не відбувається зміна $T_g$ після концентрації Si $x \geq 20$. При $x \geq 20$ роль кисню зводиться до прискорення процесу кристалізації цих стекол.

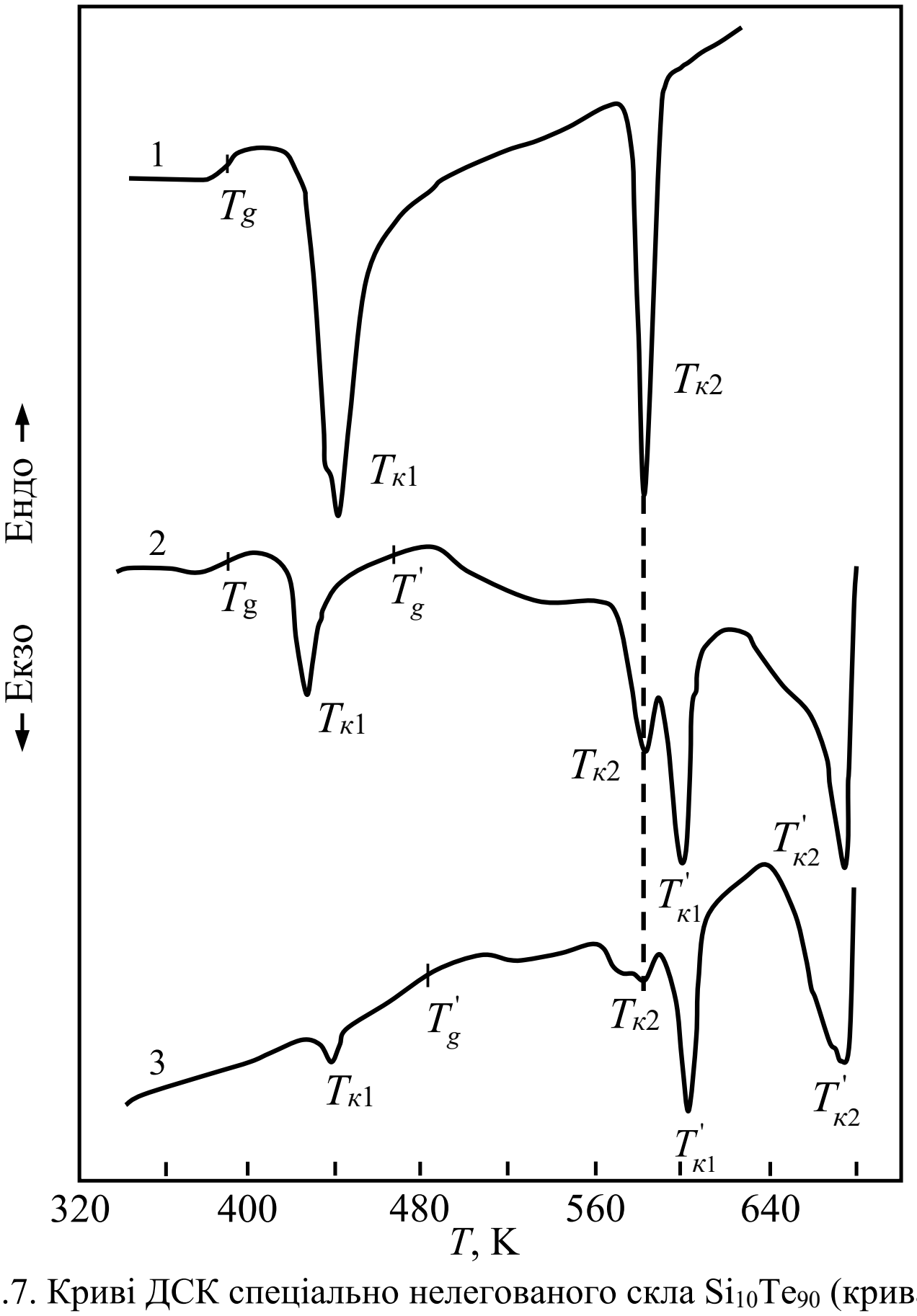

Рис. 2.7. Криві ДСК спеціально нелегованого скла $Si_{10}Te_{90}$ (крива 1) та стекол, легованих киснем у вакуумі (крива 2) та в парах аргону (крива 3) [97].

На рис. 2.7 наведені криві ДСК спеціально нелегованого чистого об'ємного скла $Si_{10}Te_{90}$ (крива 1), скла $Si_{10}Te_{89.5}O_{0.5}$, отриманого шляхом легування матеріалу у вакуумі (крива 2), і скла $Si_{10}Te_{89.5}O_{0.5}$+Ar , легованого парціальним тиском 300 мм аргону високої чистоти (крива 3). На кривій ДСК скла, легованого киснем (рис. 2.7, крива 2), спостерігаються три нові ефекти, позначені як $T_g'$ (≈ 460 K),$T_{к}'_1$ (≈ 600 K) і $T_{к}'_2$ (≈ 670 K). Екзотерми $T_g'$ і ендотерми $T_{к}'_1$ і $T_{к}'_2$ відображають характерні температури $Si_{10}Te_{90}$, описані в роботі [79] і ці значення характеризують кисневовмісне скло $Si_{10}Te_{90}$. Основні зміни між ДСК (криві 2 та 3 на рис. 2.7) полягають у тому, що присутність аргону в процесі легування та загартування у розплаві сприяє зростанню кисневовмісної фази $Si_{10}Te_{90}$, що проявляється у значному збільшенні теплоти кристалізації при екзотермах $T_{к}'_1$ і $T_{к}'_2$ по відношенню до $T_{к1}$ і $T_{к2}$ [97].

Однією з можливих причин збільшення $T_g$ (від $T_g'$ = 390 K до $T_g'$ = 480 K) при легуванні киснем полягає в наступному. Мікроскопічно можна візуалізувати кисень для розриву ланцюжків Te та/або для вибіркової заміни Te на внутрішніх поверхнях характерних кластерів скловидної мережі. Міжфазний натяг у кластерах, оточених киснем, ймовірно, збільшується за рахунок ефектів перенесення заряду і зміщується до вищих значень температури розм'якшення скла.

**2.2.3. Кінетика кристалізації стекол $Si_{15}Te_{85}$ і $Si_{20}Te_{80}$.** Важливою частиною вивчення процесу кристалізації склоподібних телуридів кремнію є дослідження їхньої кінетики кристалізації [99–101]. Інформацію про термічну стабільність та динамічний процес кристалізації автори [99] отримували за допомогою формул кінетики кристалізації. В якості об'єктів дослідження було обрано два склади стекол поблизу евтектичної точки $Si_{15}Te_{85}$ та $Si_{20}Te_{80}$. Об'ємні стекла були отримані шляхом загартування розплавів у рідкому азоті. Кінетика кристалізації цих стекол була досліджена з використанням підходу неізотермічної кристалізації. Температури склування та кристалізації стекол $Si_{15}Te_{85}$ та $Si_{20}Te_{80}$ були визначені з використанням диференціального скануючого калориметра при різних швидкостях нагрівання 5, 10, 20, 30 К/хв. Методом ДСК вивчено вплив швидкості нагрівання β стекол $Si_{15}Te_{85}$ та $Si_{20}Te_{80}$ на величини температур склування ($T_g$), кристалізації ($T_к$) і плавлення закристалізованого сплаву ($T_{пл}$).

На рис. 2.8 наведена крива кристалізації склоподібного $Si_{20}Te_{80}$ при швидкості нагрівання 20 К/хв, яка ілюструє визначення таких параметрів як температура склування $T_g$, температура початку кристалізації $T_к$, яка є точкою перетину двох дотичних, температура початку кристалізації $T_0$, яка є точкою перетину кривої кристалізації й лінії між початковою та кінцевою точками, і піковою температурою кристалізації $T_p$. Таким чином, можна визначити область переохолодженої рідини $\Delta T = T_p - T_g$, що показує термостабільність, та зниже-

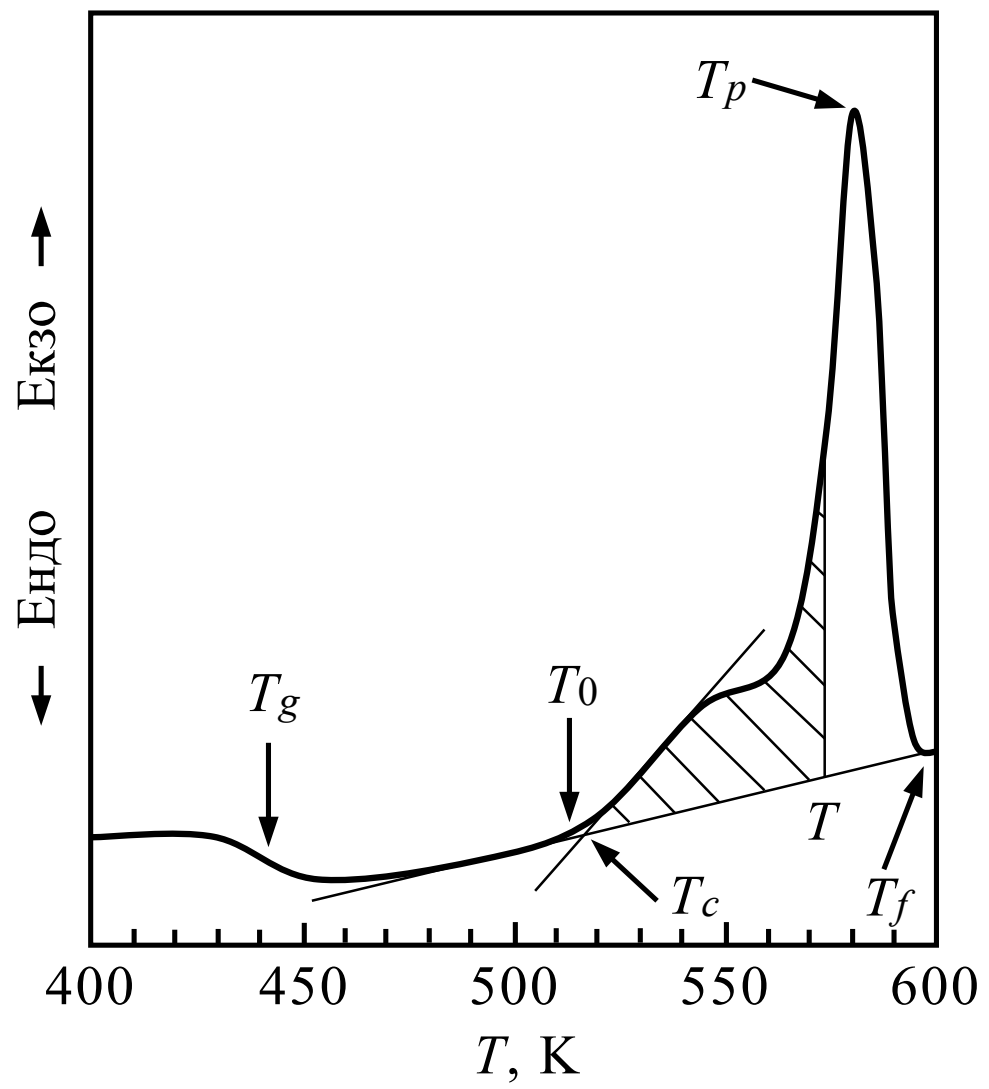

Рис. 2.8. Крива ДСК для скла $Si_{20}Te_{80}$ при швидкості нагрівання $\beta = 20$ K/хв [99].

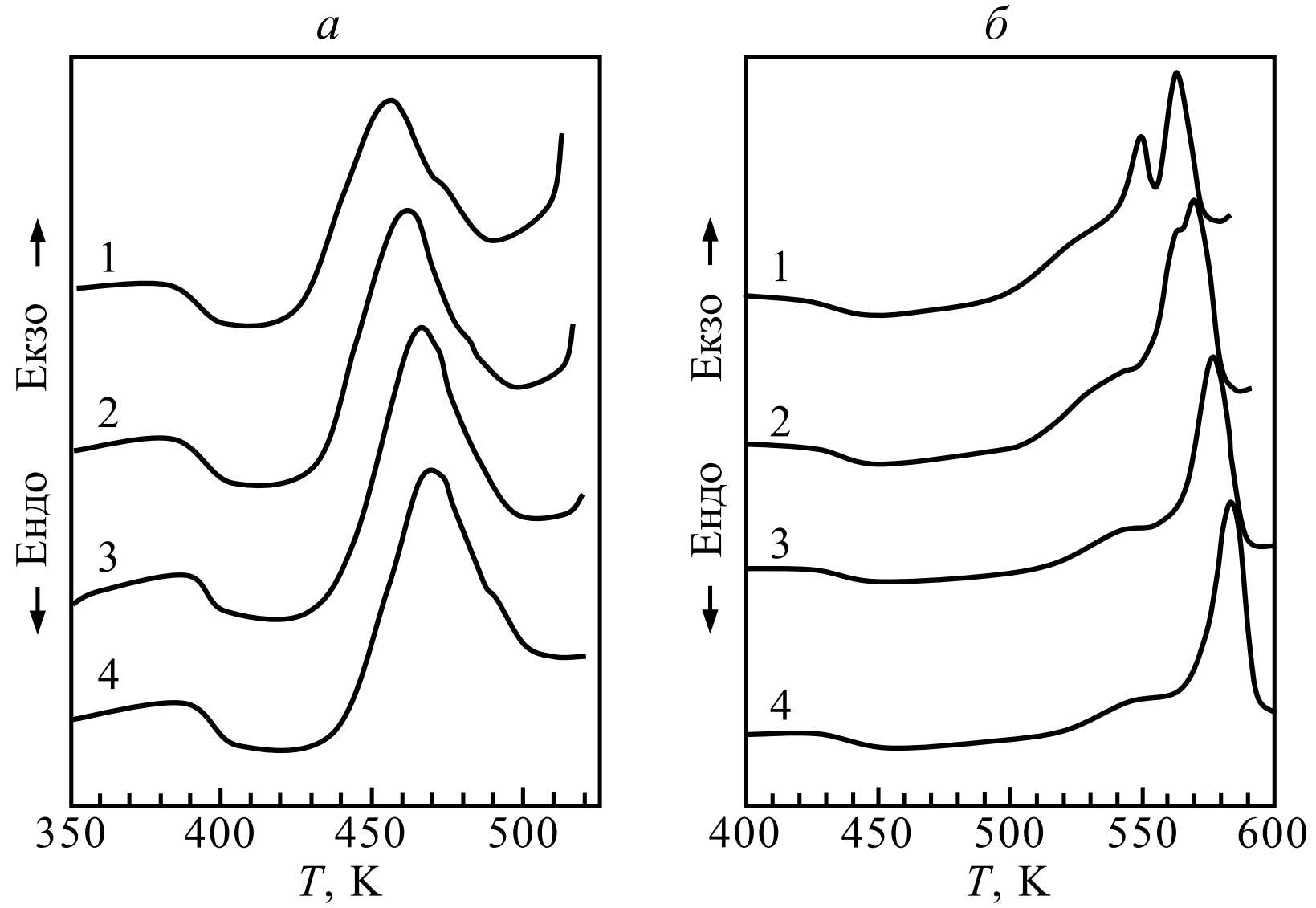

Рис. 2.9. Криві ДСК стекол $Si_{15}Te_{85}$(*а*) та $Si_{20}Te_{80}$ (*б*), виміряні при різних швидкостях нагрівання. β, K/хв: крива 1 – 5, 2 – 10, 3 – 20, 4 – 30 [99].

ну температуру склування $T_{rg} = T_g / T_p$, що має здатність до кристалізації. Всі ці параметри наведені в табл. 2.3. Встановлено, що порівняно із склом $Si_{15}Te_{85}$, значення $T_g$, $T_к$ та $T_p$ для скла $Si_{20}Te_{80}$ збільшилися приблизно на 30, 60 та 90 К відповідно. Склоподібна система $Si_xTe_{100-x}$, збагачена Si, має покращену термічну стабільність завдяки збільшеній області переохолодженої рідини. Знижені температури склування $T_{rg}$ цих двох стекол становлять 0.82 і 0.75, показуючи, що здатність до утворення стекол $Si_{15}Te_{85}$ та $Si_{20}Te_{80}$ досить високі.

Таблиця 2.3. Температурні параметри $T_g$, $T_к$, $T_p$, $\Delta T$ і $T_{rg}$ стекол $Si_xTe_{100-x}$ за кривими ДСК, відсканованими при 20 К/хв [99].

| Склад | $T_g$, К | $T_к$, К | $T_p$, К | $\Delta T$, К | $T_{rg}$ |
|---|---|---|---|---|---|
| $Si_{15}Te_{85}$ | 406.6 | 454.8 | 495.9 | 48.2 | 0.82 |
| $Si_{20}Te_{80}$ | 438.7 | 508.8 | 580.8 | 72.0 | 0.75 |

На рис. 2.9 наведені криві ДСК для стекол $Si_{15}Te_{85}$ (рис. 2.9, *а*) та $Si_{20}Te_{80}$ (рис. 2.9, *б*) при різних швидкостях нагрівання β. На рис. 2.9, *а* чітко видно, що зі збільшенням швидкості нагрівання скла $Si_{15}Te_{85}$ від 5 до 30 К/хв $T_g$ та $T_p$ зміщуються у бік більш високих температур, а переохолоджена область склоподібного $Si_{15}Te_{85}$, яка є областю між $T_g$ і $T_p$ набагато менша, ніж у скла $Si_{20}Te_{80}$. По мірі збільшення швидкості нагрівання скла $Si_{20}Te_{80}$, як $T_g$, так і $T_p$ також зсуваються в область високих температур (рис. 2.9, *б*), а переохолоджена область набагато більша. Крім того, з'являються два піка кристалізації при швидкостях нагрівання нижче 10 К/хв, що вказує на домінування фазової сегрегації $Si_{20}Te_{80}$. Це підтверджується рентгеноструктурним аналізом. Зі збільшенням швидкості нагрівання обидва піки зливаються.

Використовуючи відомі різні рівняння кінетики кристалізації, автори [99] визначили основні параметри кінетики кристалізації. Енергії активації склування $E_g$ становлять 402–416 і 330–340 кДж/моль для $Si_{15}Te_{85}$ та $Si_{20}Te_{80}$ відповідно, а енергії активації для кристалізації $E_к$ становлять 198–204 та 236–244 кДж/моль відповідно. Аналіз даних кінетики кристалізації показав, що склоподібний $Si_{20}Te_{80}$ має меншу здатність до кристалізації та більш високу термічну стабільність порівняно зі $Si_{15}Te_{85}$. Процеси кристалізації цих двох складів стекол вказують на змішаний механізм дво- і тривимірного росту.

**2.2.4. Кристалізація аморфних сплавів $Si_xTe_{100-x}$, отриманих методом спінінгування розплаву.** Застосування методу спінінгування розплаву для отримання аморфних сплавів $Si_xTe_{100-x}$ розширює межі аморфізації (порівняно із загартуванням у воду) до $5 < x < 40$. Це дозволило авторам [83–85, 102, 103] у широкому інтервалі концентрацій дослідити залежність температур кристалізації $T_к$ від складу скла. На рис. 2.10 наведені криві ДТА аморфних сплавів $Si_xTe_{100-x}$ ($x$ = 10, 14, 37, 20, 25 ат.%), отриманих методом спінінгування розплаву, зняті зі швидкістю нагрівання 10 К/хв. Визначена за кривими ДТА температура склування $T_g$ складає відповідно 381, 400.5, 428 і 437.5 К та попадає у діапазон значень, отриманих авторами [65, 96]. Із рис. 2.10 видно, що окрім сплаву $Si_{20}Te_{80}$, процес кристалізації є двостадійним і характеризується двома екзотермічними піками при температурах $T_{к1}$ і $T_{к2}$. Кристалізаційні процеси в $Si_{25}Te_{75}$ відбуваються при $T_{к1}$ = 546 (583) К і $T_{к2}$ = 605 (628) К (крива 4, рис. 2.10). Зауважимо, що для зразків, отриманих шляхом загартування розплаву у воду (рис. 2.3), кристалізація при $x > 20$ протікає одностадійно.

Дослідження кристалізації аморфних сплавів $Si_xTe_{1-x}$ показало, що характер перетворень для зразків з $x > 20$ і $x < 20$ різний. Відповідно до електронографічних досліджень [102] у сплавах з вмістом Si менше 20 ат.% при $T_{к1}$ з аморфної матриці виділяється кристалічний телур. Лінії другої фази, що виділяється при $T_{к2}$, автори [102], проіндентифікували як ромбоедричний $Si_2Te_3$. Тривала витримка сплавів при $T_{к1}$ не змінює характеру їхньої кристалізації.

Для зразків складу $x = 20$ кристалізація протікає в одну стадію. Спочатку аморфна матриця перетворюється в пересичений твердий розчин кремнію у телурі, який зі збільшенням часу термообробки розпадається з виділенням Te та ромбоедричного $Si_2Te_3$. У сплавах з вмістом Si більше 20 ат.% ступінь пересичення кремнію у телурі зменшується, а другою фазою, яка виділяється при $T_{к2}$, є гексагональний $Si_2Te_3$ ($a$ = 7.429 Å, $c$ = 13.471 Å). Концентраційні залежності температур кристалізації ($T_{к1}$, $T_{к2}$) аморфних сплавів $Si_xTe_{100-x}$, отриманих методом спінінгування, наведені на рис. 2.11.

Встановлено, що залежність температури кристалізації першого ендопіка $T_{к1}$ від концентрації $x$ аморфних сплавів $Si_xTe_{100-x}$ у інтервалі від $x = 6$ до $x = 33$ носить лінійний характер (рис. 2.11). Екстраполяція залежності $T_{к1}(x)$ до нульової концентрації Si ($x = 0$) дозволила визначити температуру кристалізації чистого телуру, яка виявилась

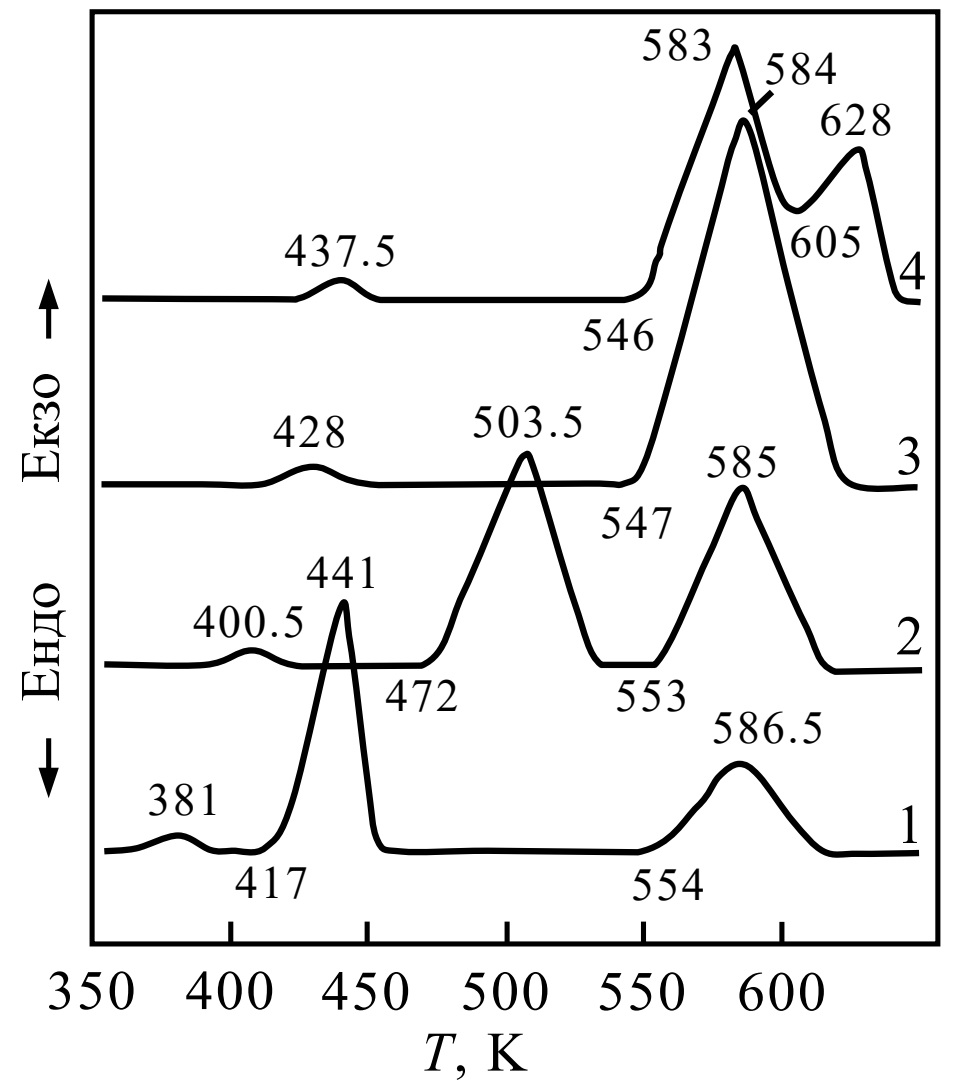

Рис. 2.10. Криві ДТА аморфних сплавів $Si_xTe_{100-x}$, отриманих методом спінінгування. *x,* ат. % : 1 – 10, 2 – 14, 37, 3 – 20, 4 – 25 [103].

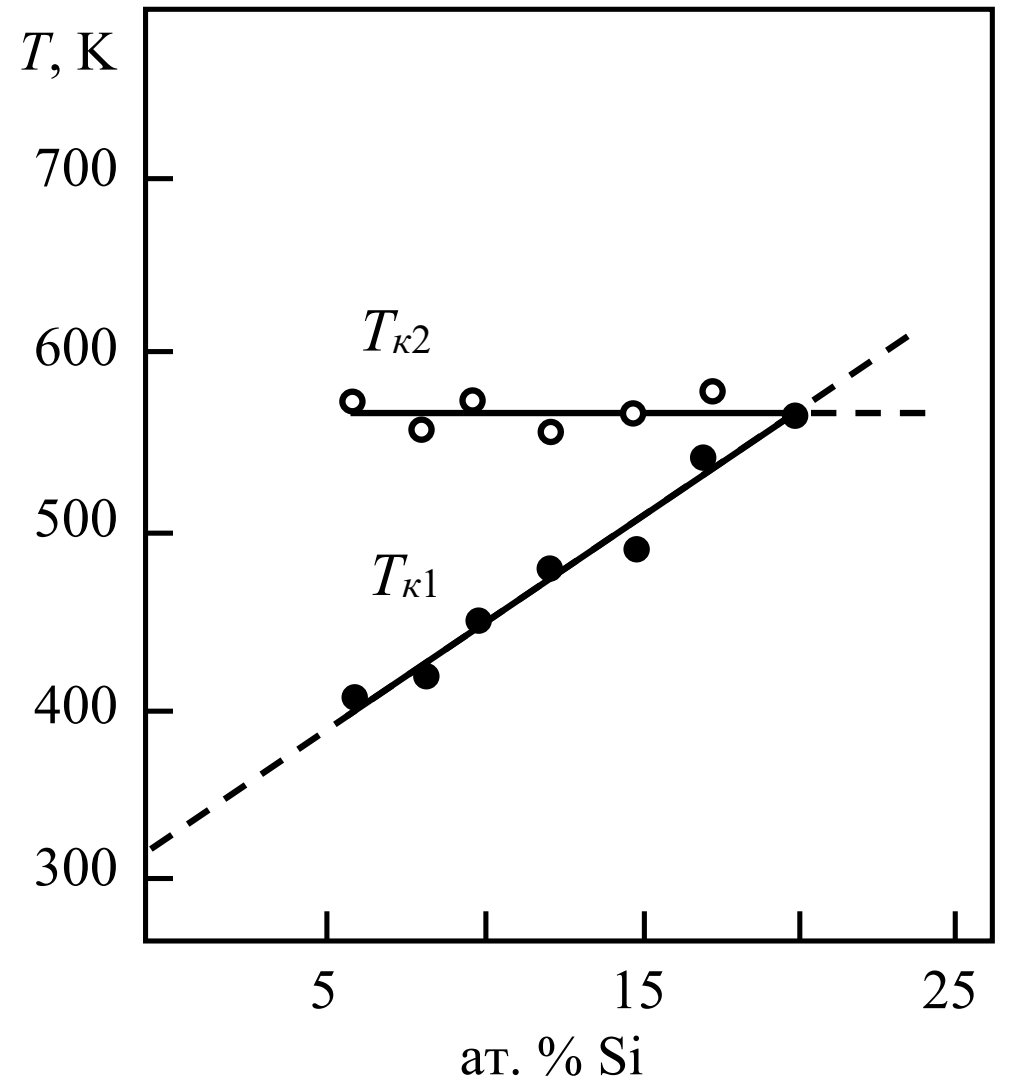

Рис. 2.11. Концентраційні залежності температур кристалізації $T_{к1}$, $T_{к2}$ аморфних сплавів $Si_xTe_{100-x}$, отриманих методом спінінгування [83].

рівною ~ 315.6 К. Температура другого екзопіка $T_{к2}$ при $x < 20$ від концентрації сплавів практично не залежить і складає ~ 552 К. Одержані результати автори [102, 103] пояснюють з точки зору моделі з хімічно впорядкованою сіткою, в рамках якої двостадійний процес кристалізації пов'язується з гетерофазністю структури, наявністю в ній двох аморфних фаз з різним ближнім порядком і різними температурами кристалізації $T_{к1}$ і $T_{к2}$. Враховуючи характер концентраційної залежності $T_{к1}$, хімічне впорядкування у системі Si–Te може бути реалізоване на складі $x = 20$ з утворенням хімічно впорядкованої сполуки $a$-$SiTe_4$.

Результати кінетики кристалізації аморфних сплавів $Si_xTe_{100-x}$, отриманих методом спінінгування показали [84], що процес перетворення сплавів при $x > 20$ істотно відрізняється один від одного. У сплавах із $x < 20$ при температурі $T_{к1}$ з аморфної матриці виділяється кристалічний телур. Лінії другої фази, що виділяється при $T_{к2}$, вкладаються в ромбоедричну ґратку кристалічного $Si_2Te_3$. Наявність у структурі закристалізованих сплавів $к$-$Si_2Te_3$ у двох кристалічних модифікацій (ромбоедричної й гексагональної), очевидно, пов’язано з різними механізмами його виділення із аморфної матриці.

На початку кристалізації аморфного сплаву з $x = 20$ аморфна матриця перетворюється в твердий розчин кремнію у телурі. Зі збільшенням часу витримки, або температури твердий розчин, який формується у гексагональній структурі, розпадається на $к$-Te і ромбоедричний $к$-$Si_2Te_3$.

В аморфних сплавах $20 < x < 33$ при температурі першого екзопіка з аморфної матриці також виділяється твердий розчин Si в Te, проте ступінь пересичення кремнію у ньому помітно менший. При температурі другого екзопіка має місце виділення дещо деформованого $к$-$Si_2Te_3$ відомої гексагональної модифікації. Кристалізація аморфного сплаву з $x = 33$ характеризується виділенням $к$-Te і гексагонального $к$-$Si_2Te_3$. Таким чином, аналіз отриманих результатів показує, що для складу $x = 20$ спостерігаються численні аномалії максимуму приведеної температури $T_p$ і енергії активації кристалізації, утворення пересиченого твердого розчину кремнію у телурі, зміна типу морфології структури. Цей же склад є граничним для існування (чи не існування) ромбоедричного $к$-$Si_2Te_3$. Автори [84] роблять висновок про те, що кристалізація аморфних сплаві при $x = 20$ в одну стадію не пов’язана зі швидкістю гартування та іншими технологічними особливостями експерименту, а обумовлена більш глибокими змінами структури. Тому, враховуючи можливість зміни координаційного числа телуру, приведені численні аномалії можна пояснити утворенням хімічно упорядкованої сполуки *a*-$SiTe_4$.

Одностадійний процес кристалізації спостерігається також і для складу $x = 33$. Слід відзначити, що цей сплав є крайньою межею аморфізації в системі Si–Te і, природно, що характер кристалізації аморфних сплавів з $x > 33$ поки що невідомий. Проте в рамках моделі з хімічно впорядкованою сіткою цей склад відповідає правилу 8*N* для звичайного ковалентного зв'язку. З пониженням температури на ньому спостерігається тенденція до утворення другого мінімуму на кривих електропровідності. Крім того, температура перетворення аморфного сплаву з $x = 33$, яка складає ~ 708 К, перевищує температуру плавлення евтектики на рівноважній діаграмі стану Si–Te. Останній феномен автори [84] пояснюють лише існуванням хімічного упорядкування для складу $x = 33$ з утворенням *a*-$SiTe_2$, хоча, безумовно, цей висновок вимагає додаткових досліджень.

**2.2.5. Вплив тиску на процес кристалізації стекол, отриманих загартуванням розплаву у воду.** Індукована тиском поліморфна кристалізація в об'ємному склі $Si_{20}Te_{80}$ досліджена авторами [104–106]. В інтервалі температур 293 ÷ 640 К та тиску від атмосферного до 8.5 ГПа методами ДСК, електронної мікроскопії та рентгенографії досліджено вплив тиску на кристалізацію склоподібного $Si_{20}Te_{80}$. Рентгенографічні дослідження показали, що при тиску 7 ГПа має місце поліморфна кристалізація, скло $Si_{20}Te_{80}$ переходить у кристалічну фазу гексагональної сингонії із $c/a = 1.5$. Нагрівання цієї кристалічної фази приводить до її розпаду при 586 К на дві стабільні при цій температурі кристалічні фази: Te і $SiTe_2$. Слід зазначити, що індукований тиском при 7 ГПа перехід скло-кристал є також переходом напівпровідник-метал [105]. Скло $Si_{20}Te_{80}$ має дві температури склування та дві стадії кристалізації. При нагріванні спочатку кристалізується надлишок Te, а потім склоподібна фаза, яка залишилась, кристалізується у вигляді $SiTe_2$. Обговорено різницю між кристалізацією, зумовленою зміною температури (первинної кристалізації) і поліморфною або інконгруентною кристалізацією, індукованою тиском.

## 2.3. СКЛОУТВОРЕННЯ В ПОТРІЙНИХ СИСТЕМАХ Si–M–Te (M = Se, Ge, Sn, Pb)

Система Si–Te в багатьох випадках є базовою при синтезі трикомпонентних стекол, які широко використовуються в лазерній техніці, акустооптиці, а також в якості перемикачів та оптичних елементів з високою роздільною здатністю. Використання великої кількості хімічних елементів у якості легуючих домішок при синтезі кремнійтелуридних стекол дало змогу значно розширити спектр їх експлуатаційних характеристик.

**2.3.1. Система Si–Se–Te.** Область склоутворення в системі Si–Se–Te на даний час ще не встановлена. У роботі [107] синтезовані стекла $Si_x(Se_{100-y}Te_y)_{100-x}$ складів $33.3 \leq x \leq 43$ і $0 \leq y \leq 15$, які частково включають два розрізи $SiSe_2$–$SiTe_2$ і $Si_2Se_3$–$Si_2Te_3$ зі сторони збагаченої $SiSe_2$ і $Si_2Se_3$, та вивчена їх структура методом КР спектроскопії. Синтез потрійних стекол проводили з елементарних компонентів (Si та Te чистоти 99.999 % і спеціально очищеного селену) в евакуйованих до $133 \cdot 10^{-3}$ Па кварцових ампулах довжиною 8 см і внутрішнім діаметром 6 мм. Загальна наважка компонентів становила 1÷2 г. При максимальній температурі синтезу 1370 К розплави витримували 60 год, після чого охолоджували загартуванням у воду. Стекла, збагачені селеном, гігроскопічні. У квазібінарній системі $SiSe_2$–$SiTe_2$ область склоутворення за даними [107] існує в інтервалі 0–10 мол.% $SiTe_2$. При режимі гартування від 1370 К розплавів масою 2 г склоутворення у системі $Si_2Se_3$–$Si_2Te_3$ існує в межах 0–60 мол. % $Si_2Te_3$.

Автори [108, 109] виявили, що часткове заміщення телуру селеном знижує склоутворюючу здатність розплавів системи Si–Se–Te, тому для вивчення акустооптичних властивостей були синтезовані стекла тільки двох складів $Si_{19.7}Te_{78.7}Se_{1.6}$ і $Si_{19.2}Te_{76.8}Se_4$, оскільки сплав $Si_{18.7}Te_{74.6}Se_{6.7}$ виявився кристалічним і нестійким на повітрі через сильну гідратацію.

**2.3.2 Система Si–Ge–Te.** Вперше область склоутворення в системі Si–Ge–Te встановлена авторами [110, 111] (рис. 2.12), вона витягнута уздовж лінії, яка з’єднує подвійні евтектики в системах Si–Te і Ge–Te. Фельц [112] вважає, що склоутворенню сприяють відносно низькі температури ліквідусу вздовж евтектичної лінії, яка з'єднує обидві подвійні системи. Сплави синтезували методом прямого сплавлення елементарних компонентів при 1273÷1473 К протягом 24 год в запаяних вакуумованих кварцових ампулах із застосуван-ням вібраційного перемішування. Стекла хорошої якості можуть бути отримані тільки за умови використання вихідних матеріалів високої чистоти, зокрема, необхідно повністю видалити з них окси-ди. З цією метою суміші перед розплавленням нагрівали при 473 К у вакуумі для видалення оксидних плівок Te. Після завершення процесу синтезу ампули з розплавом (загальною наважкою 10 – 20 г) гартували у воду. Методом ДТА визначено температури склування $T_g$ стекол системи Si–Ge–Te.

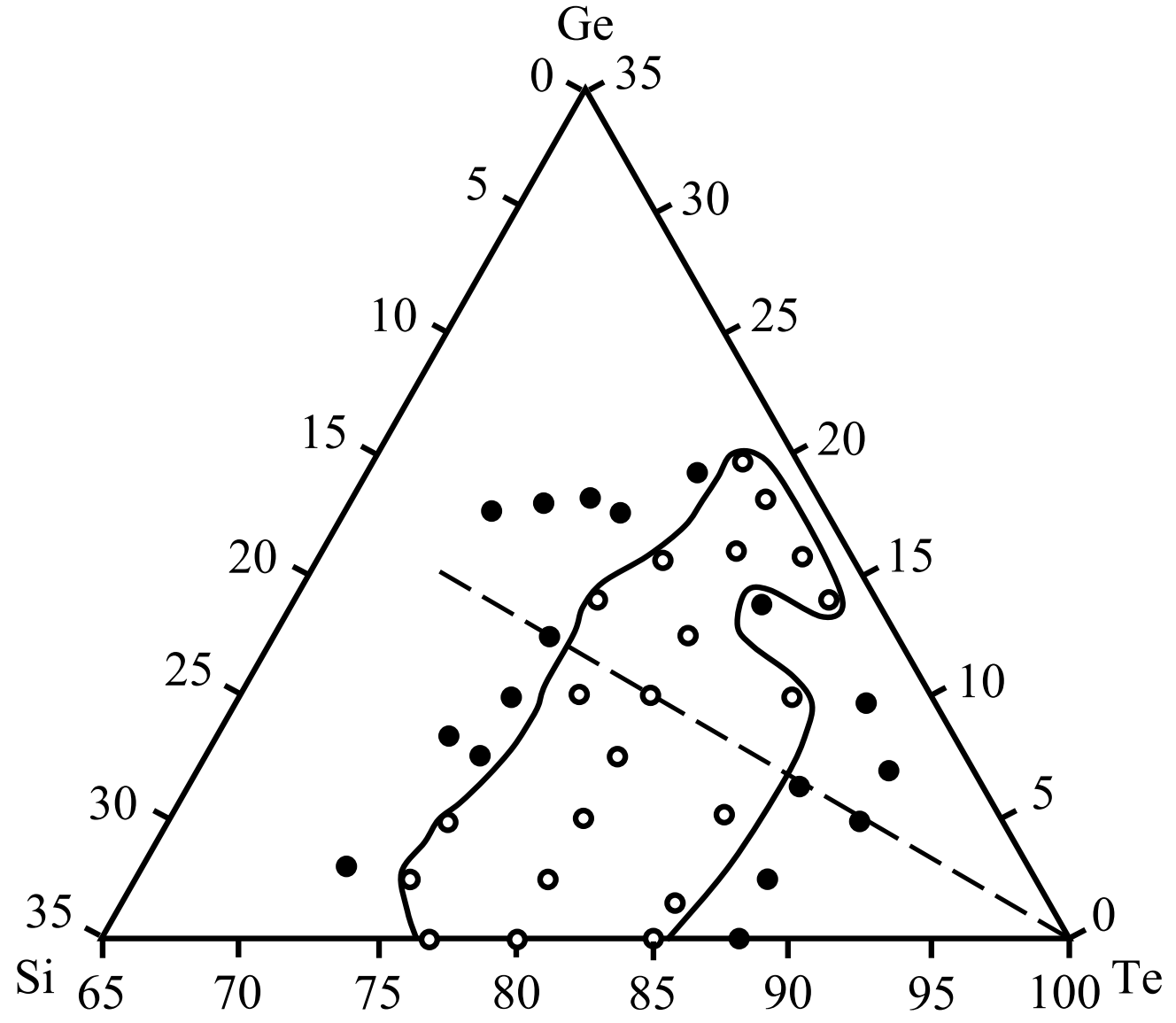

Рис. 2.12. Область склоутворення у системі Si–Ge–Te [110] і $Ge_xSi_xTe_{100-x}$ ($6 < x < 15$) пунктирна лінія [116].

Склоутворення за розрізом $Si_{20}Te_{80}$ – $Ge_{20}Te_{80}$ досліджували також автори [117]. Сплави отримували методом вакуумного синтезу з наступним загартуванням розплаву в крижаній воді. Використовуючи сплощену форму ампул, вдалося досягти швидкостей охолодження розплаву 200 ÷ 250 К/с. В результаті були отримано стекла всіх складів розрізу $Si_{20}Te_{80}$ – $Ge_{20}Te_{80}$.

Синтез стекол за розрізом $Si_xGe_{15}Te_{85-x}$ ($2 \le x \le 12$) автори [113, 114] проводили шляхом повільного нагрівання (~100 К/год.) елементарних компонентів у запаяних вакуумованих кварцових ампулах до 1373 К у горизонтальній печі, яка оберталась. При цій температурі ампули витримували протягом 24 годин і також неперервно оберта-ли зі швидкістю 10 об/хв, з метою забезпечення однорідності розплаву. Для отримання об'ємних склоподібних зразків ампули з розплавом гартували в суміші NaOH і крижаної води.

Авторами [114] методом диференціальної сканучої калориметрії вивчені температури склування $T_g$ та кристалізації $T_к$ стекол $Si_xGe_{15}Te_{85-x}$ ($2 \le x \le 12$). Ці стекла демонструють один процес склування та дві стадії кристалізації при нагріванні. Найменше значення $T_g$ = 373 K зафіксовано для складу скла, збагаченого германієм $Si_5Ge_{10}Te_{85}$. По мірі збільшення вмісту кремнію у стеклах і відповідно зменшення германію, $T_g$ стекол зростає й досягає максимального значення 435 K для скла $Si_{22.5}Ge_{2.5}Te_{75}$. Встановлено, що температура склування зростає майже лінійно зі збільшенням вмісту кремнію (рис. 2.13, крива 1). Температура першої кристалізації ($T_{к1}$) збільшується зі збільшенням вмісту кремнію для складів $x < 5$; $T_{к1}$ залишається майже сталою в діапазоні складів $5 < x \le 10$. Вона збільшується порівняно більш різко при $x > 10$ (рис. 2.13, крива 2). Автори [114] також зазначають, що друга температура кристалізації $T_{к2}$ практично не змінюється від складу стекол $Si_xGe_{15}Te_{85-x}$.

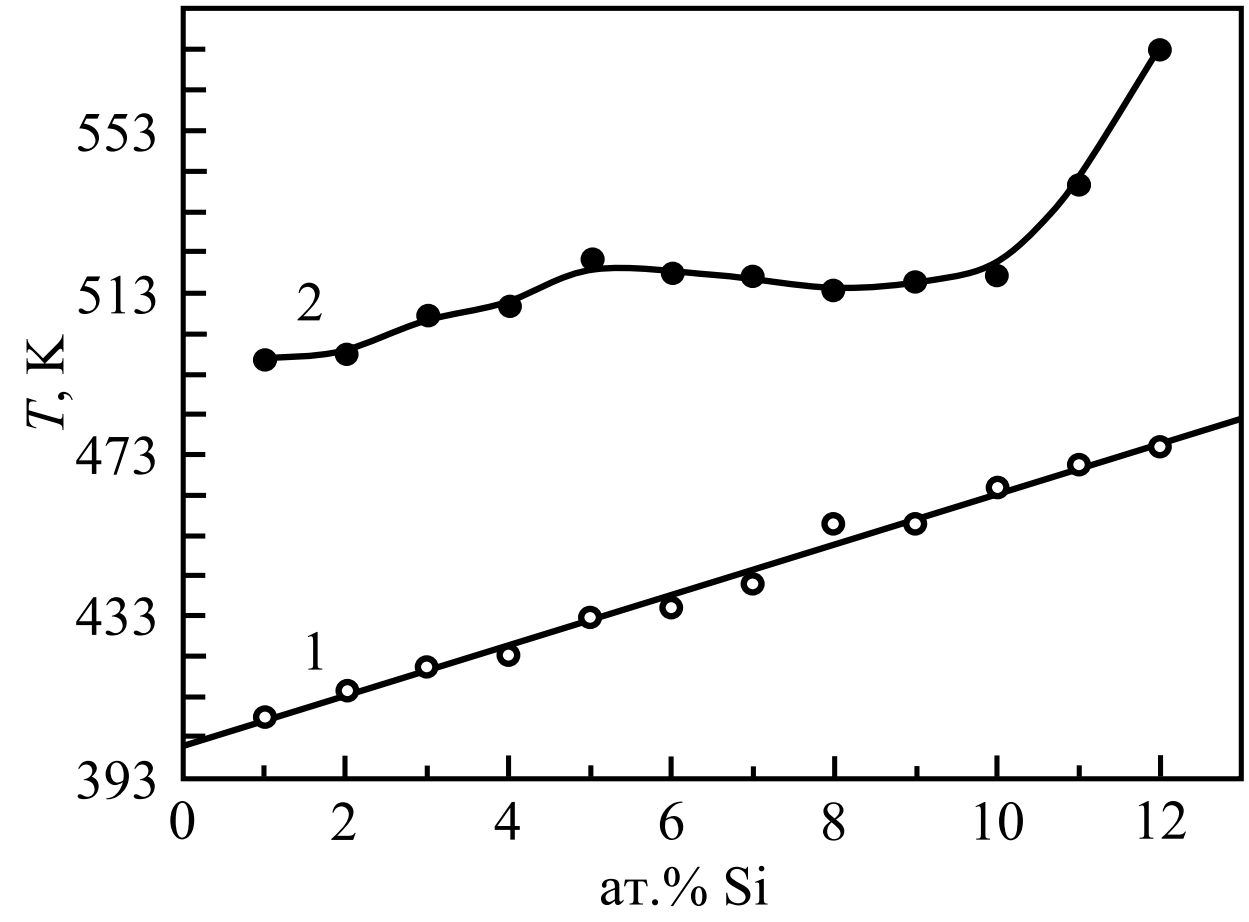

Рис. 2.13. Концентраційні залежності температури склування $T_g$ (1) та температури першої кристалізації $T_{к1}$ (2) стекол $Si_xGe_{15}Te_{85-x}$ [114].

Аналогічні дослідження термічної поведінки для стекол розрізу $Si_{15}Ge_xTe_{85-x}$ ($1 \le x \le 11$) були виконані в роботі [115] з використанням модульованої диференціальної сканучої калориметрії (ADSC), в якій синусоїдальна зміна температури накладається на звичайну лінійну зміну температури DSC. Результати цих досліджень наведено на рис. 2.14. В інтервалі складів $1 \le x < 3$ стекол $Si_{15}Ge_xTe_{85-x}$ спостерігаються два процеси кристалізації при температурах $T_{к1}$ і $T_{к2}$ відповідно, тоді як в інтервалі складів $3 \le x \le 11$ дві реакції кристалізації протікають при одній температурі кристалізації $T_{к1}$. З рис. 2.14 також видно, що при $T_{к3}$ спостерігається нова перколяція кристалічної фази в діапазоні складів $6 \le x \le 11$. Для встановлення фаз, які утворюються в процесі нагрівання стекол Si–Ge–Te, проведено рентгеноструктурні дослідження кристалізації.

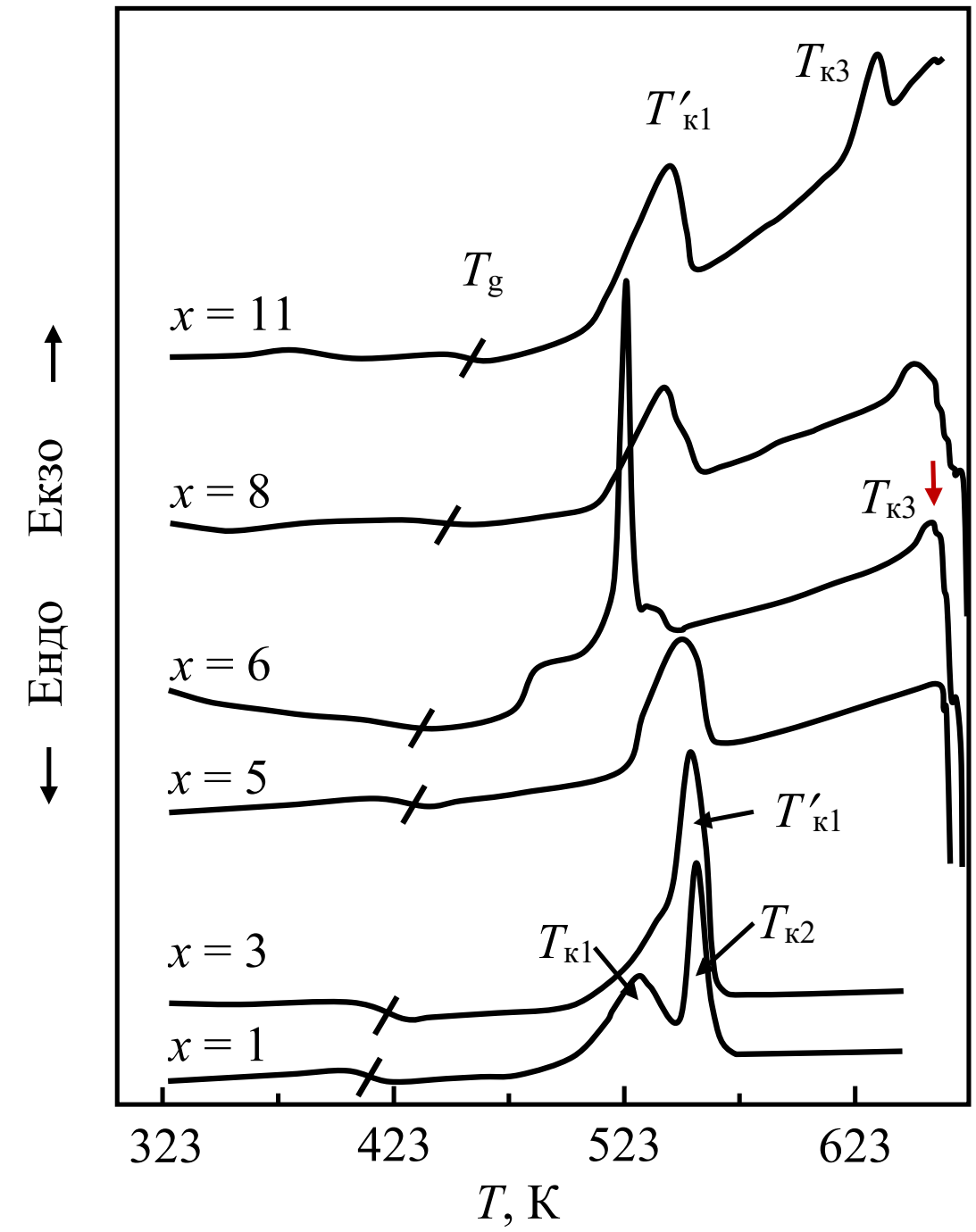

Рис. 2.14. Криві ADSC стекол $Si_{15}Ge_xTe_{85-x}$ [115].

У результаті досліджень встановлено, що в стеклах $Si_{15}Ge_xTe_{85-x}$ ($1 \le x \le 11$) спостерігається аномальний поділ фаз. Поява кристалічної фази $SiTe_2$ спостерігається в діапазоні складів $6 \le x \le 11$. Рентгеноструктурні дослідження виявили структурне перетворення орторомбічного *o*-GeTe (GeTe–I) в орторомбічну *o*-GeTe фазу високого тиску (GeTe–II) у зразках з наявною фазою $SiTe_2$ при $T_{к3}$.

Вплив тривалості процесу синтезу стекол $Si_xGe_xTe_{100-2x}$ на темпе-ратуру їх склування $T_g$ вивчено в роботі [116]. Стекла синтезували прямим сплавленням елементарних компонентів при температурі 1223 К протягом 7 і 14 діб. На рис. 2.15 приведена концентраційна залежність $T_g$ стекол розрізу $Si_xGe_xTe_{100-2x}$ після 7 та 14 діб витримки. Має місце незначне збільшення $T_g$ для зразків з витримкою 14 діб. З рис. 2.15 видно, що незалежно від тривалості процесу синтезу стекол, зі збільшенням вмісту кремнію у діапазоні концентрацій $6 < x < 12$ $T_g$ монотонно збільшується досягаючи максимального значення при $x = 12$, потім різко зменшується і далі практично не змінюється.

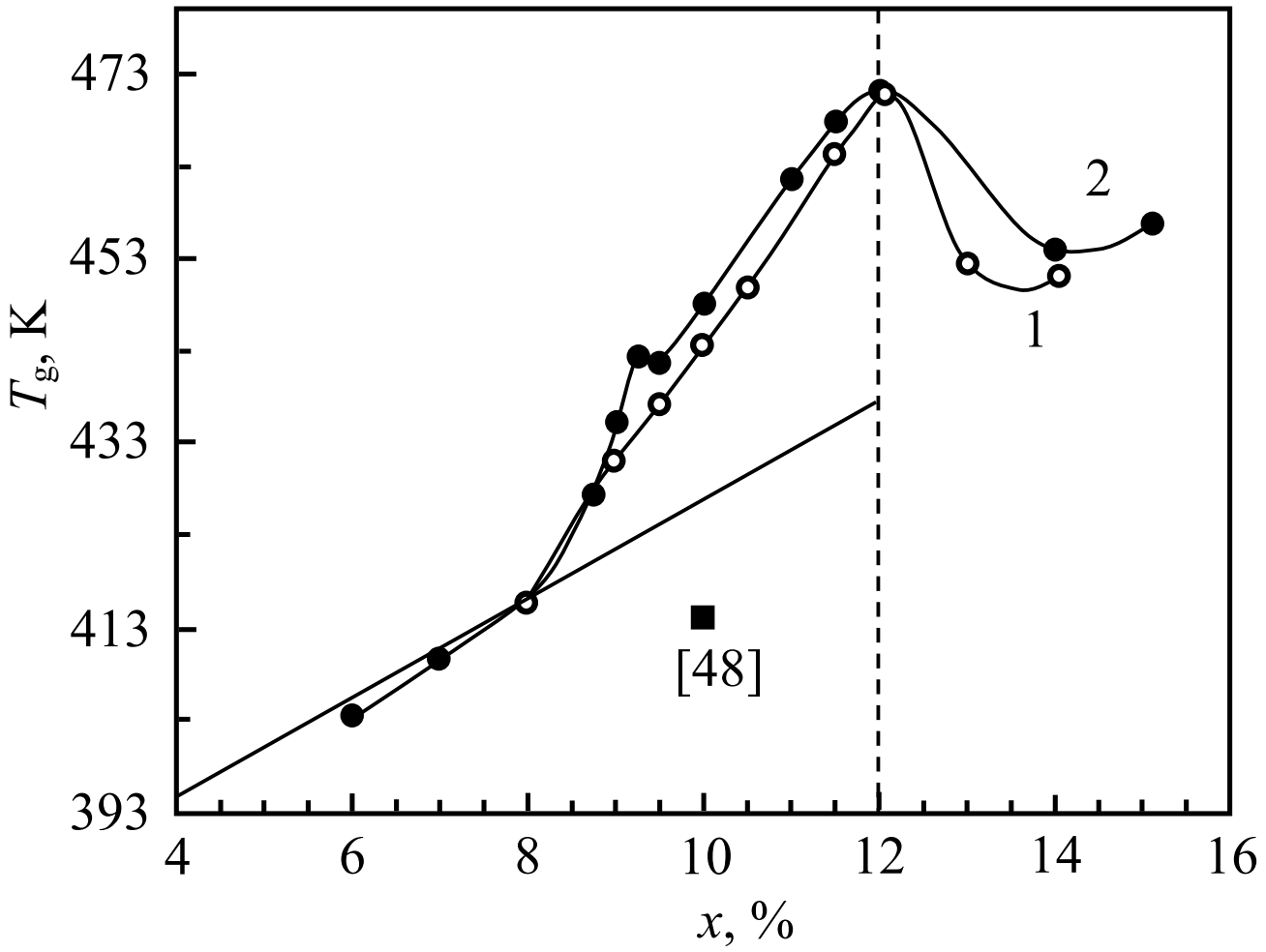

Рис. 2.15. Концентраційні залежності температур склування $T_g$ стекол розрізу $Si_xGe_xTe_{100-2x}$, синтезованих протягом 7 днів (крива 1) та 14 днів (крива 2) [116].

**2.3.3. Система Si–Sn–Te.** Склоутворення у системі Si–Sn–Te за розрізом $Si_{20}Sn_xTe_{80-x}$ $(1 \le x \le 7)$ досліджували автори [118 – 121]. Синтез склоподібних сплавів проводили із елементарних компонентів простих речовин. Синтез здійснювали однотемпературним методом. Вакуумовані кварцові ампули із шихтою повільно нагрівали від кімнатної температури до 1373 К в горизонтальній печі, яка оберталася зі швидкістю 10 об/хв. Для гомогенізації розплаву ампулу витримували в печі, що оберталася при цій температурі протягом 24 год. Після цього проводили загартування розплавів у суміші льодяної води і гідроксиду натрію (NaOH).

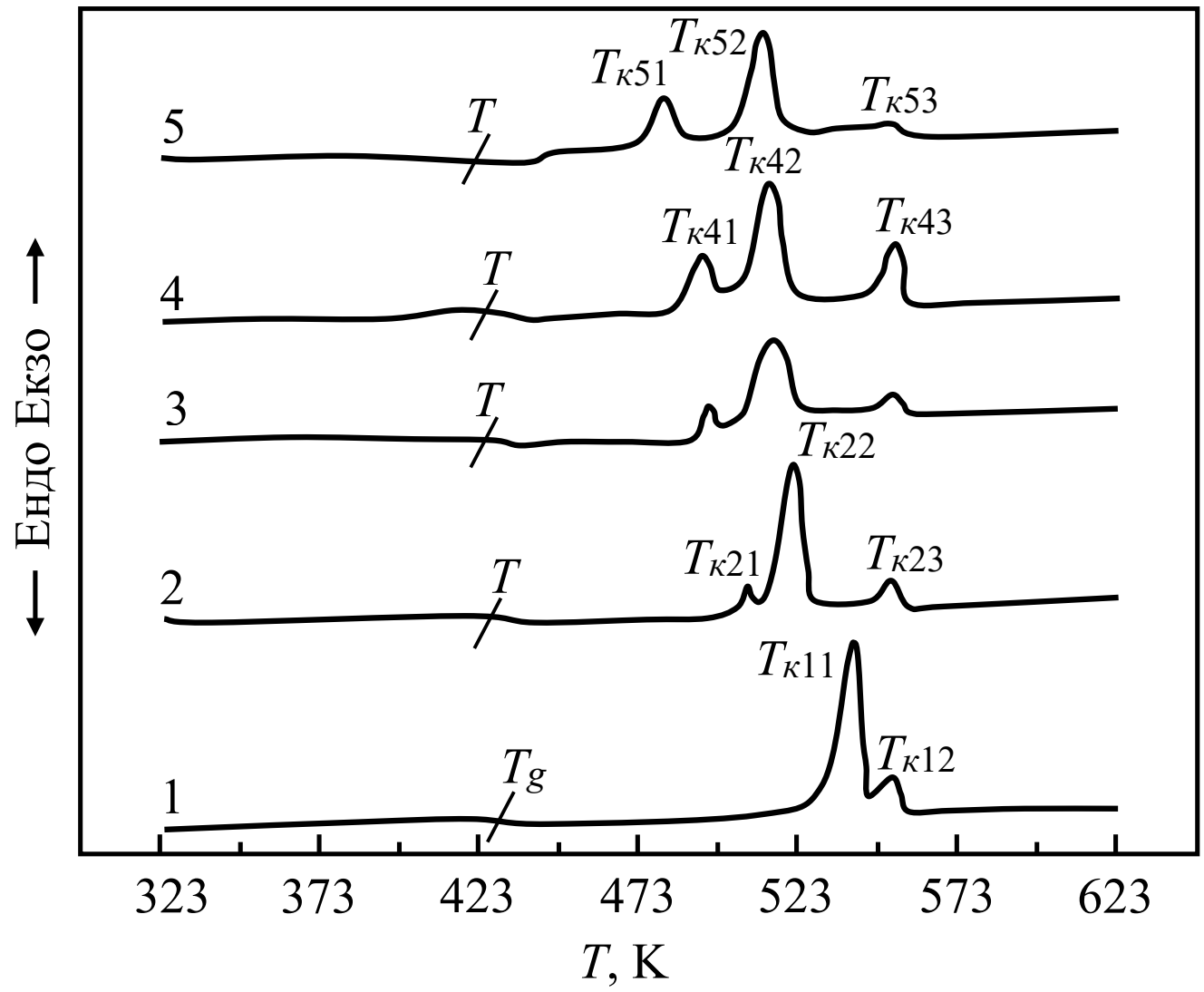

Рис. 2.16. Криві ADSC стекол $Si_{20}Sn_xTe_{80-x}$ [120].
*x*, %: 1 – 1; 2 – 2; 3 – 3; 4 – 4; 5 – 5.

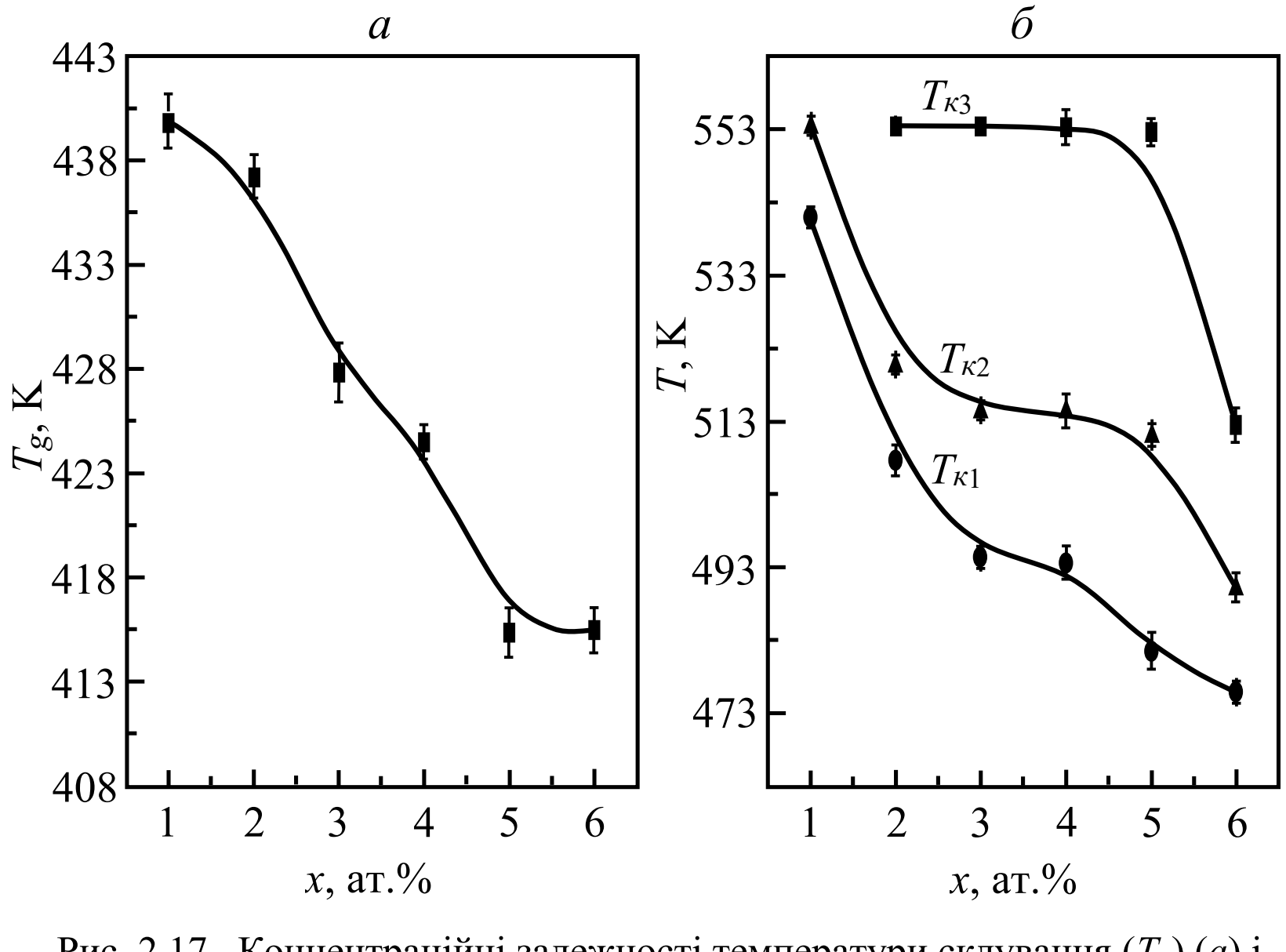

Рис. 2.17. Концентраційні залежності температури склування ($T_g$) (*а*) і температур кристалізації (*б*) стекол $Si_{20}Sn_xTe_{80-x}$ [120].

Склоподібний стан сплавів контролювався за допомогою рентгенофазового і мікроструктурного аналізів. Для одержаних стекол $Si_{20}Sn_xTe_{80-x}$ методом модульованої диференціальної скануючої калориметрії визначено термічні параметри: температуру склування ($T_g$) і температуру кристалізації ($T_к$). На рис. 2.16 наведені криві ADSC для п'яти складів стекол даного розрізу. Як видно з цього рисунка, для всіх складів ($x$ = 1–5) на термограмах ADSC наявні одна ендотермічна температура склування ($T_g$), дві ($T_{к11}$, $T_{к12}$ для $x$ = 1) і три ($T_{к51}$, $T_{к52}$, $T_{к53}$ для $x$ = 5) екзотермічні температури кристалізації $T_к$. Концентраційні залежності температури склування ($T_g$) і температур кристалізації ($T_{к51}$, $T_{к52}$, $T_{к53}$) наведені на рис 2.17.

Для виявлення кристалічних фаз, наявних у стеклах $Si_{20}Sn_xTe_{80-x}$ ($x$ = 3 і 5), автори [119] відпалювали зразки при відповідних температурах кристалізації $T_{к32}$ (533 K) і $T_{к51}$ (500 K) протягом двох годин у відкачаних ампулах. Рентгеноструктурні дослідження показали, що в термічно закристалізованих об'ємних зразках $Si_{20}Sn_3Te_{77}$ і $Si_{20}Sn_5Te_{75}$ наявні тільки бінарні фази гексагонального $SiTe_2$, гексагонального $Si_2Te_3$, гексагонального Te і кубічного $SnTe_2$. Ці дані вказують на те, що атоми Sn не взаємодіють активно з матрицею Si–Te, а отже, не має покращення мережевого з'єднання.

Загартуванням розплавів у видовжених конічних ампулах, які забезпечують різну швидкість охолодження (від 200 К/с і нижче в одному досліді), автори [76] встановили область склоутворення в евтектичній області потрійної системи Si–Sn–Te: Si (7.5 – 20), Sn (0 – 10) і Te (75 – 85) ат. %. Структура ближнього порядку цих потрійних стекол досліджена за допомогою Мессбауерівської спектроскопії на ізотопах $^{119}Sn$.

Синтез ще двох розрізів стекол системи Si–Sn–Te : $Si_{(20-x)}Sn_xTe_{80}$ ($x$ = 0.2–8.0) і $Si_{0.8x}Sn_{0.2x}Te_{(100-x)}$ ($x$ = 14.3–25) провели автори [73]. Синтез об`ємних стекол проводили у вакуумованих кварцових ампулах із елементарних компонент при 1373 К протягом 20 год при неперервному перемішуванні розплаву з наступним загартовуванням на повітрі. Звуження області склоутворення при заміщенні атомів кремнію на атоми олова у склі автори [73] пояснюють значною металізацією хімічного зв`язку атомів олова з атомами телуру. Рентгеноструктурні дослідження закристалізованих стекол виявили наявність у них кристалічних бінарних фаз $Si_2Te_3$ і SnTe, а також кристалічного телуру.

**2.3.4. Система Si–Pb–Te.** У системі Si–Pb–Te склоутворення автори [122–126] досліджували за розрізом $Si_{20-x}Pb_xTe_{80}$ ($2.5 \leq x \leq 20$). Для приготування сплавів використовували високочисті елементарні компоненти. Загальна наважка складала 20 г. Їх суміш сплавляли індукційним методом з використанням високочастотного генератора (потужність 5 кВт) при температурі близько 1073 К протягом 3 хв, а потім повільно охолоджували. Згодом, підготовлені таким чином, сплави швидко охолоджували за допомогою «пістолета», а також способом «рухомого поршня і ковадла», або повільно загартовували у воді. У разі методу «поршня і ковадла» відповідні зразки у вигляді плоскопаралельних пластин завтовшки близько 0.1 мм були отримані сплющенням капель розплаву в момент їх прольоту між двома металевими поверхнями.

Стекла $Si_{20-x}Pb_xTe_{80}$ демонструють двостадійну кристалізацію та подвійну температуру склування у діапазоні складів від 2.5 до 10 ат.%. Pb (рис. 2.18). Друга температура склування $T_{g2}$ проявляється після повторного нагрівання зразка, який попередньо нагрівався до кінця першої стадії кристалізації, а потім охолоджувався. При повторному нагріванні перший екзотермічний ефект кристалізації $T_{к1}$ і $T_{g1}$ зникають (рис. 2.18, крива 2). Збільшення концентрації Pb зменшує температури склування і кристалізації (рис. 2.19). Лінійна залежість $T_{g1}$ і $T_{g2}$ від складу вказує на те, що ці температури відповідають розчинам різного складу в кожному склі даної системи. Двостадійна кристалізація стекол $Si_{20-x}Pb_xTe_{80}$ в інтервалі концентрації $2.5 \leq x \leq 10$ Pb вказує на фазові розділення, які відбуваються у матеріалі. Розділення фаз відбувається у розплаві з миттєвою кристалізацією осадженого Te. Охолоджений сплав являє собою склокераміку, в аморфній матриці якої наявні дендрити телуру. $T_g$ вважається температурою перетворення скла в розплав аморфного матеріалу, який залишається після кристалізації телуру.

Досліджено вплив на температуру другої стадії кристалізації попереднього нагрівання зразків зі швидкістю 20 К/хв до закінчення першої стадії кристалізації і подальшого швидкого (360 К/хв) охолоджування до кімнатної температури. Показано, що цей процес призводить до пониження температури кристалізації та зменшення швидкості зародкоутворення. Методами електронної мікроскопії й рентгенографії показано, що на першій стадії кристалізації утворюються лише кристали Te. На другій стадії відбувається евтектоїдна кристалізація PbTe і твердого розчину Si в телурі [124].

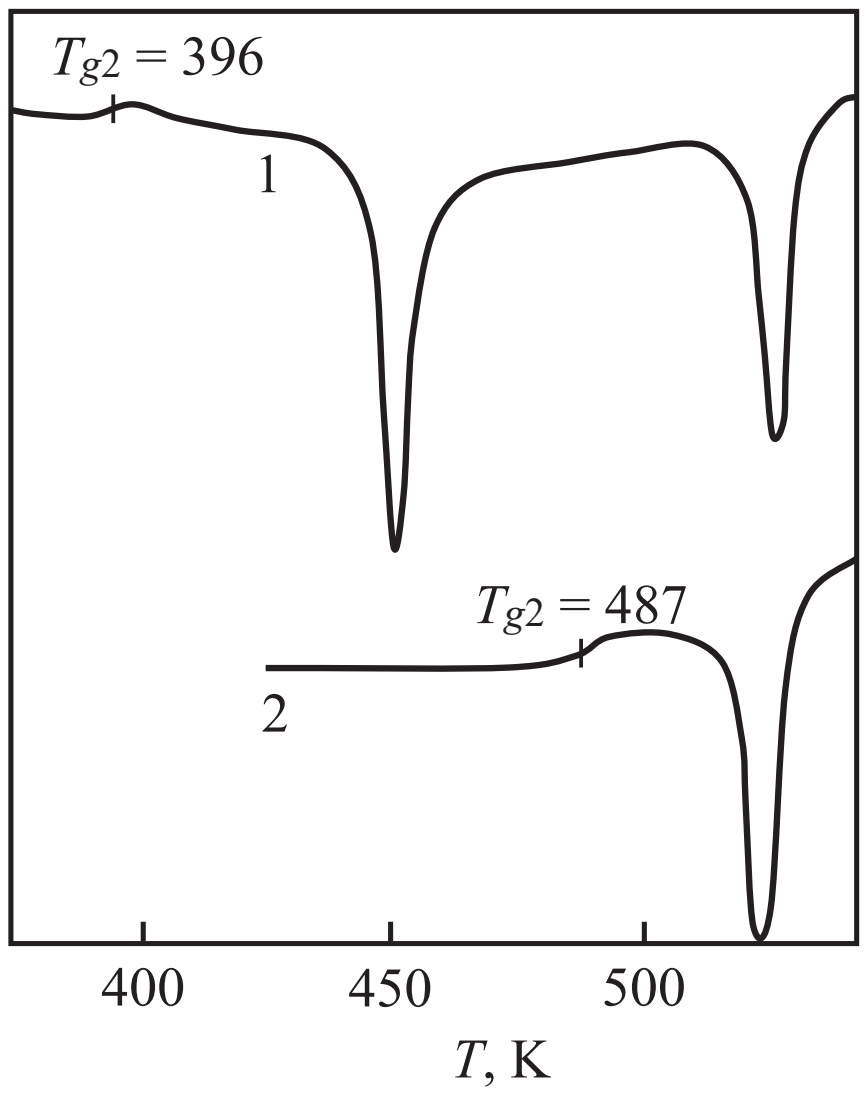

Рис. 2.18. Криві ДСК скла $Si_{12.5}Pb_{7.5}Te_{80}$ [124].

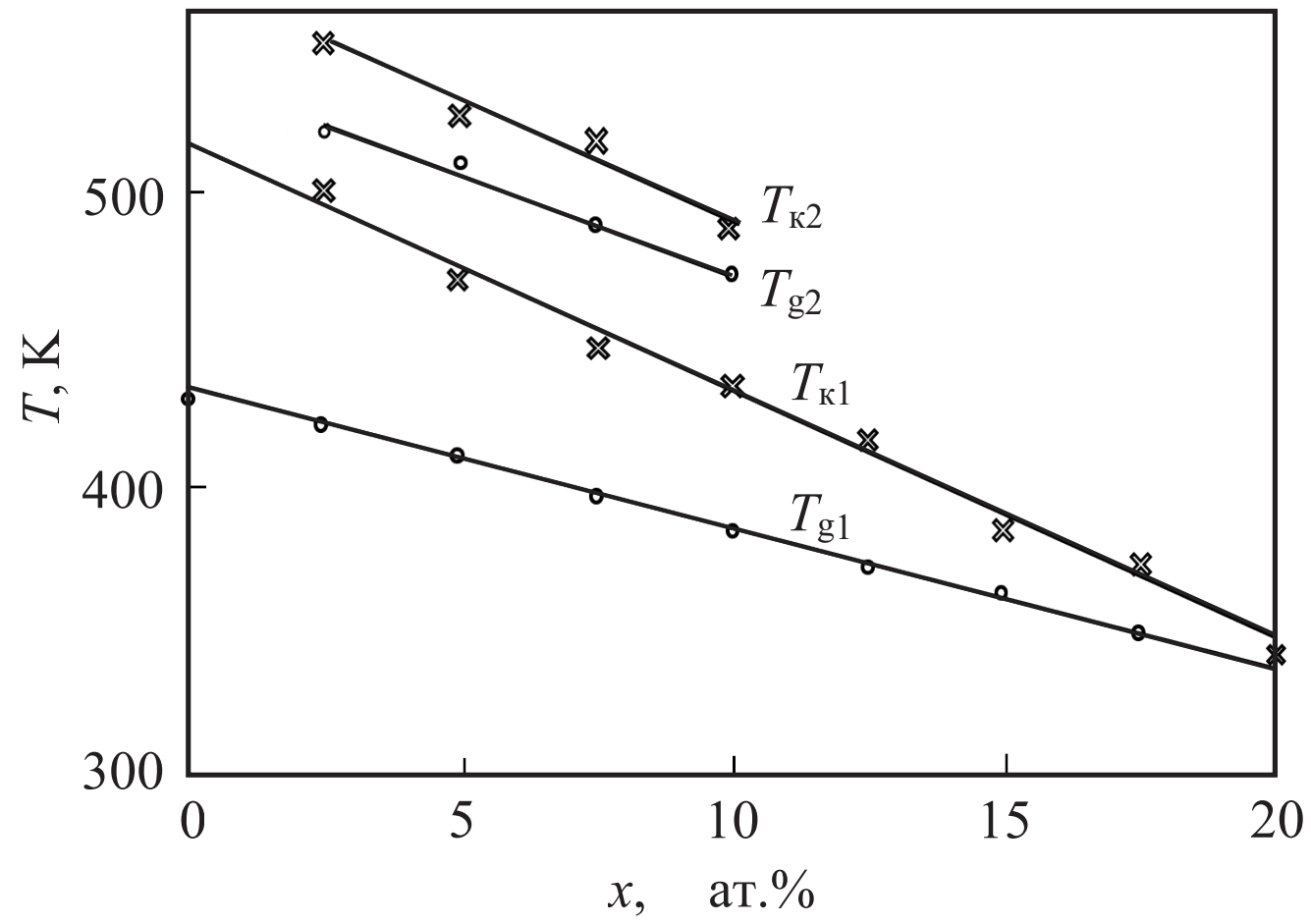

Рис 2.19. Концентраційні залежності температур склування $T_{g1}$ і $T_{g2}$ та кристалізації $T_{к1}$ і $T_{к2}$ [125].

## 2.4. СКЛОУТВОРЕННЯ У ПОТРІЙНИХ СИСТЕМАХ Si–M–Te (M = Cu, Ag, Al, In)

Підвищена склоутворююча здатність міді та срібла з телуридами кремнію обумовлена тим, що ці метали взаємодіють не тільки з одним халькогеном, а з обома компонентами бінарного телуридного скла. При цьому в складі скла утворюються складні структурні одиниці, які містять усі три компоненти. Ковалентна складова хімічного зв'язку в таких потрійних сполуках більша, а ніж у телуридах кремнію, тому складні структурні одиниці, які утворюються, здатні взаємодіяти з ковалентно ув'язаною структурою телуридного скла і впливати на його фізико-хімічні властивості.

Тільки ті метали, які здатні взаємодіяти з обома компонентами скла з утворенням складних структурних одиниць, можуть бути в значній кількості введені до складу бінарних халькогенідних стекол [127]. Утворення складних трикомпонентних структурних одиниць має істотне значення для склоутворення в халькогенідних системах, і, відповідно, до створення склоподібних напівпровідників. При цьому слід мати на увазі, що склоутворенню у трикомпонентних системах сприяє утворення не тільки термодинамічно стійких потрійних сполук, які виділяються у вигляді кристалічних фаз при відпалі стекол. Визначити здатність до склоутворення, відігравати вирішальну роль у формуванні структури скла можуть нестабільні структурні утворення, отримані в результаті загартування розплаву.

**2.4.1. Система Si–Cu–Te.** Склоутворення у системі Si–Cu–Te і фізико-хімічні властивості стекол у цій системі вперше досліджували автори [128, 129]. Синтез сплавів системи Si–Cu–Te проводили з елементів напівпровідникової чистоти у вакуумованих кварцових ампулах при температурі 1273 K протягом 12 год при інтенсивному перемішувані. Гартування розплавів здійснювали в холодну воду. Склоподібний стан ідентифікували за раковистим зломом і відсутністю ліній на дебаєграмах. Область склоутворення у системі Si–Cu–Te наведена на рис. 2.20, з якого видно, що вона витягнута вздовж розрізу, який з'єднує евтектичну точку системи Si–Te зі сторони Te з $Cu_2Te$. До складу стекол даної системи можна ввести до 18 ат.% Cu.

Для стекол розрізів: $Si_{20-x}Cu_xTe_{80}$, $Si_{12}Cu_xTe_{88-x}$ і $Si_xCu_4Te_{96-x}$ автори [128] провели дослідження наступних фізико-хімічних властивостей: густину *d*, виміряну гідростатичним зважуванням, мікротвердість *H*, електропровідність σ і енергію активації електропровідності $E_a$, результати яких наведені в табл. 2.4. На кривих ДТА всіх дослід-

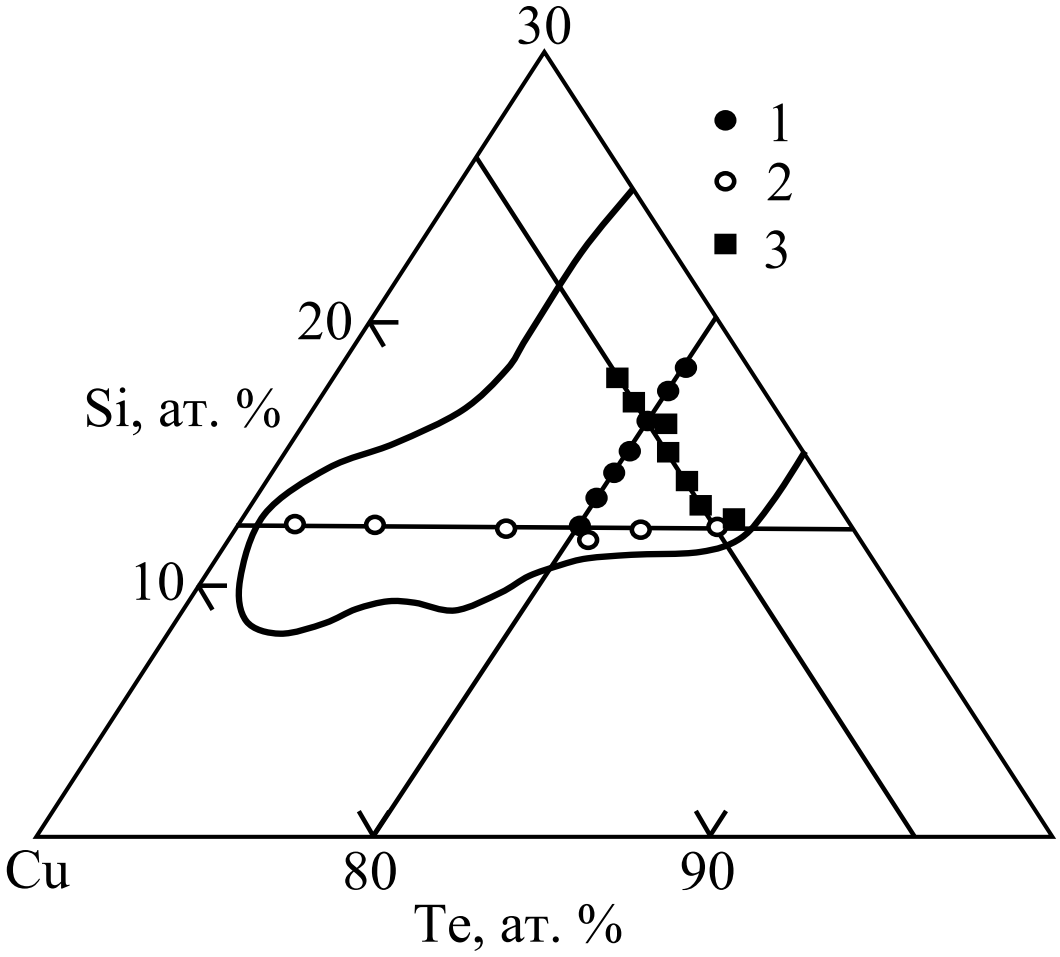

Рис. 2.20. Область склоутворення в системі Si–Cu–Te: сплави розрізів $Si_{20-x}Cu_xTe_{80}$ (1), $Si_{12}Cu_xTe_{88-x}$ (2) і $Si_xCu_4Te_{96-x}$ (3) [128].

Таблиця 2.4. Фізико-хімічні властивості стекол системи Si–Cu–Te [128].

| **Склад стекол** | ***d***, г/см$^3$ | ***H***, кг/мм$^2$ | **–lgσ$_{20°}$**, Ом$^{-1}$·см$^{-1}$ | ***E$_a$***, еВ |
|---|---|---|---|---|
| $Si_{18}Cu_2Te_{80}$ | 5.24±0.05 | 122±6 | 6.3±0.1 | – |
| $Si_{17}Cu_3Te_{80}$ | 5.20 | 128 | 6.2 | 0.31±0.05 |
| $Si_{16}Cu_4Te_{80}$ | 5.35 | 123 | – | – |
| $Si_{15}Cu_5Te_{80}$ | 5.35 | 122 | 5.5 | – |
| $Si_{14}Cu_6Te_{80}$ | 5.34 | 123 | – | – |
| $Si_{13}Cu_7Te_{80}$ | 5.51 | 113 | 4.7 | 0.43 |
| $Si_{12}Cu_8Te_{80}$ | 5.58 | 119 | 4.5 | – |
| $Si_{12}Cu_4Te_{84}$ | 5.49 | 110 | 5.8 | – |
| $Si_{12}Cu_6Te_{82}$ | 5.54 | 106 | – | – |
| $Si_{12}Cu_{10}Te_{78}$ | 5.61 | 120 | 4.8 | – |
| $Si_{12}Cu_{14}Te_{74}$ | 5.7 | 128 | 3.3 | – |
| $Si_{12}Cu_{16}Te_{72}$ | – | 131 | 2.7 | 0.35 |
| $Si_{13}Cu_4Te_{83}$ | 5.43 | 118 | 4.6 | – |
| $Si_{14}Cu_4Te_{82}$ | 5.36 | 125 | 4.8 | 0.48 |
| $Si_{15}Cu_4Te_{81}$ | 5.33 | 120 | 5.4 | – |
| $Si_{17}Cu_4Te_{79}$ | 5.27 | 139 | – | – |
| $Si_{18}Cu_4Te_{78}$ | 5.26 | 138 | 6.6 | – |

жених стекол виявлено два ендотермічні ефекти, які відповідають двом температурам розм'якшення, значення яких лежать у діапазоні 383–433 К. Як видно з табл. 2.4, густина (*d*) стекол системи Si–Cu–Te змінюється в межах від 5.20 до 5.70 $г \cdot см^{-3}$. При збільшенні вмісту міді (зменшенні концентрації Si і Te) у склоподібних сплавах, що відповідають розрізу з постійним вмістом телуру або кремнію, густина лінійно збільшується.

Введення міді до складу склоподібних телуридів кремнію приводить до істотного зміцнення їх структури. Свідченням цього є більш високі значення зміни мікротвердості і величин $T_g$. Мікротвердість стекол *H* змінюється в межах 106–139 $кг/мм^2$. Зі збільшенням вмісту телуру в стеклах спостерігається зменшення мікротвердості внаслідок накопичення у їх складі ланцюжкових структурних одиниць $TeTe_{2/2}$. Збільшенням вмісту міді в склоподібних сплавах Si–Cu–Te призводить до збільшення електропровідності σ від $4.7 \cdot 10^{-6}$ до $2.1 \cdot 10^{-3}$ $Ом^{-1} \cdot см^{-1}$.

Ще один розріз $Si_{15}Cu_xTe_{85-x}$ ($1 \leq x \leq 10$) стекол системи Si–Cu–Te досліджували автори [130, 131]. Синтез стекол $Si_{15}Cu_xTe_{85-x}$ проводили із елементарних компонент з загальною наважкою 1 г у вакуумованих кварцових ампулах. Запаяну кварцову ампулу повільно нагрівали до 1373 К зі швидкістю нагріву 100 К/год у горизонтальній обертовій печі. Тривалість процесу ситнтезу складала 24 год. Після цього амулу із росплавом гартували у крижаному водному розчині NaOH.

Авторами [131] проведено комплексне дослідження температур склування $T_g$ і кристалізації $T_к$, а також теплових параметрів, таких як незворотна зміна ентальпії ($\Delta H$) та зміна питомої теплоємності ($\Delta C_p$), залежно від складу. На рис. 2.21, *а* наведені криві ДСК стекол $Si_{15}Cu_xTe_{85-x}$ ($1 \leq x \leq 10$), на яких для складів $x$ = 2, 4, 6 чітко видно наявність одного піка ендотермічного склування $T_g$ та трьох виражених екзотермічних піків кристалізації $T_{к1}$, $T_{к2}$ та $T_{к3}$. Для складів $x$ = 8 і 10 спостерігаються тільки два піки кристалізації $T_{к1}$ і $T_{к2}$.

Концентраційна залежність температури склування $T_g$ стекол $Si_{15}Te_{85-x}Cu_x$ ($1 \leq x \leq 10$) наведена на рис. 2.22 (крива 1). Як видно з цього рисунка, $T_g$ спочатку зменшується до складу $x \leq 2$, а потім неперервно зростає. Збільшення $T_g$ складів стекол з $x > 2$ вказує на утворення зв'язків Cu–Te. Аналогічна тенденція спостерігається і на концентраційній залежності першої ($T_{к1}$) та третьої ($T_{к3}$) температур кристалізації стекол даного розрізу (рис. 2.21, *б*). У діапазоні складів

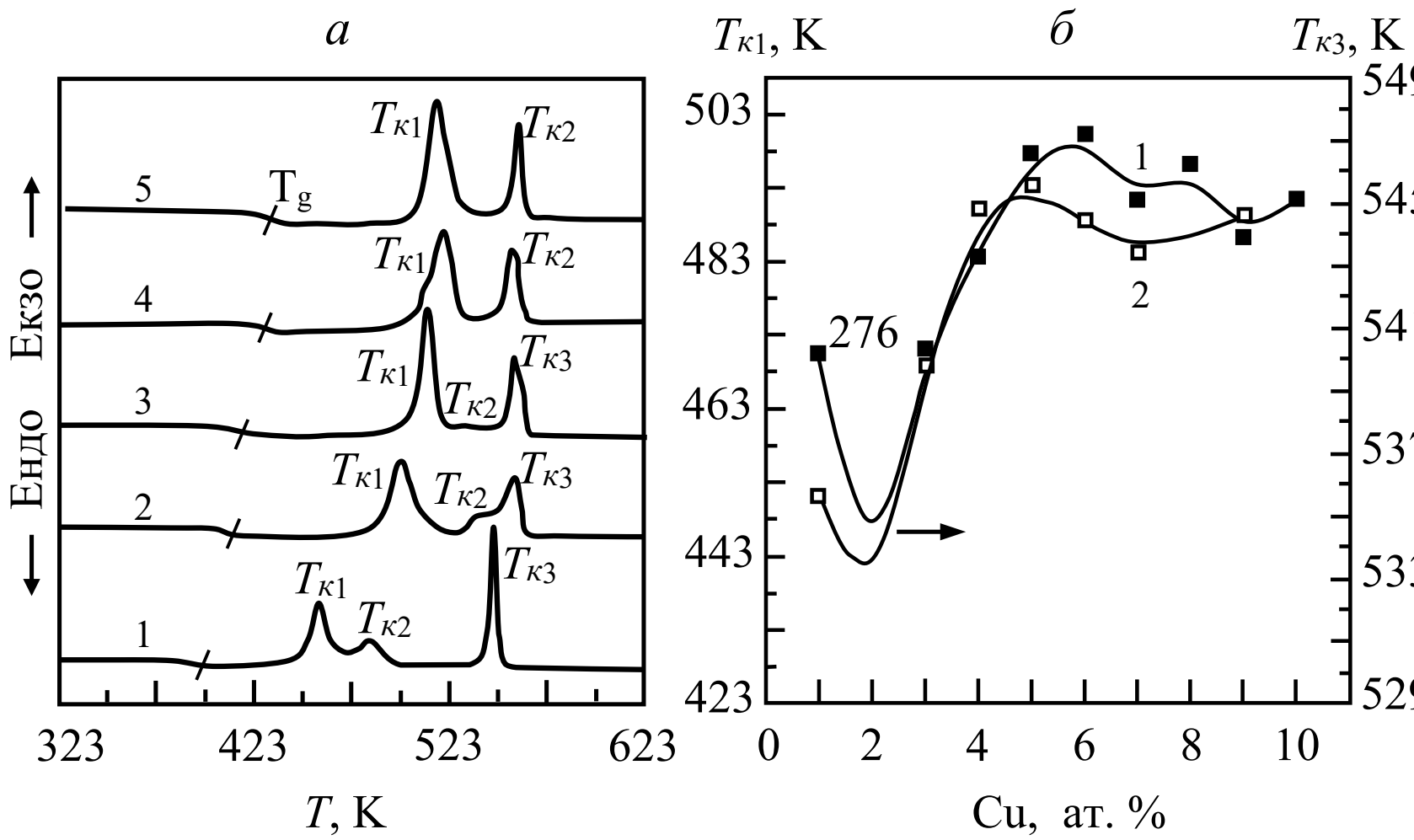

Рис. 2.21. *а* – Криві ДТА; *б* – концентраційні залежності температур кристалізації $T_{к1}$ (1) і $T_{к3}$ (2) стекол $Si_{15}Te_{85-x}Cu_x$ *x* ат.%: 1 – 2, 2 – 4, 3 – 6, 4 – 8, 5 – 10 [131].

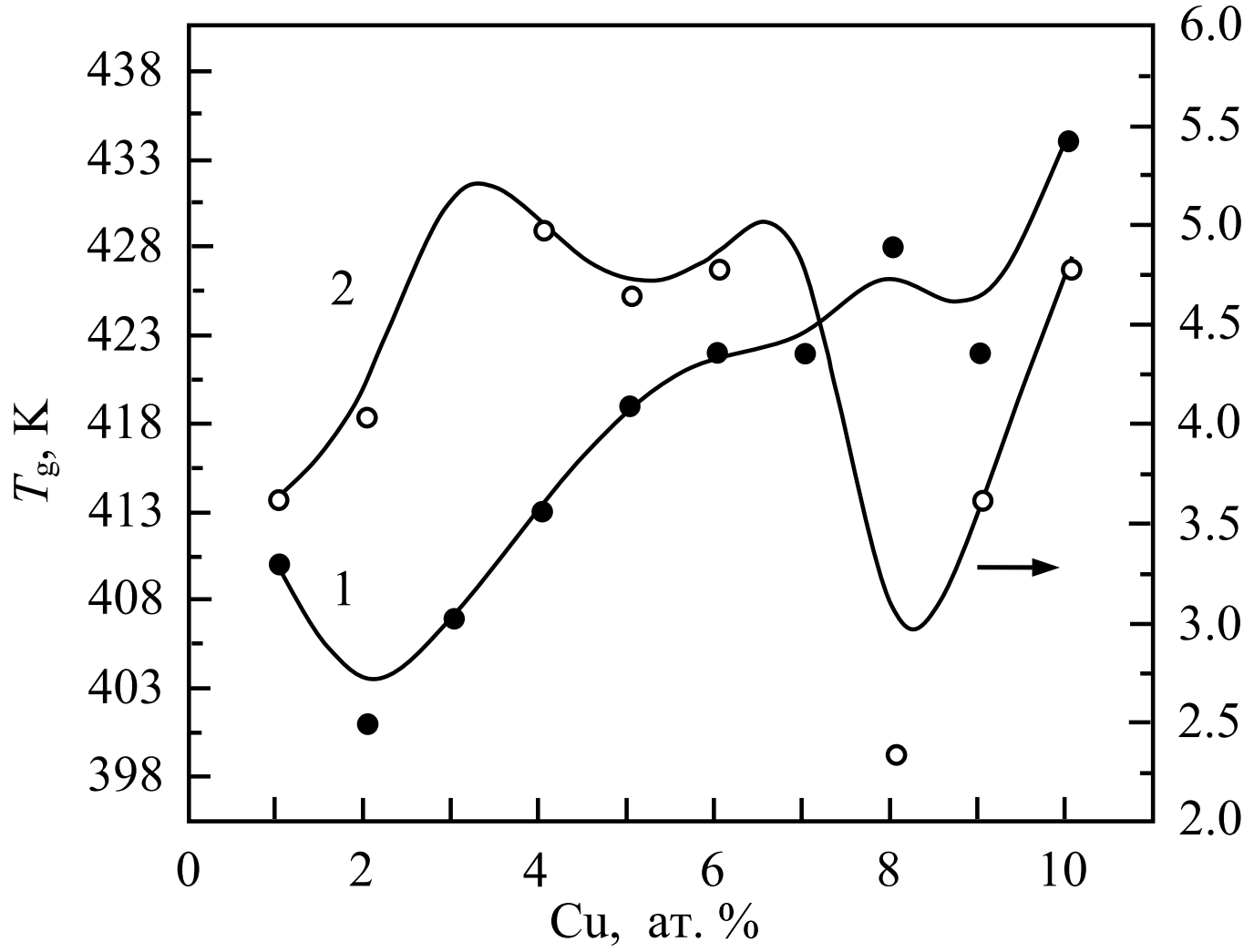

Рис. 2.22. Концентраційні залежності температур склування $T_g$ (1) і густини *d* (2) стекол $Si_{15}Te_{85-x}Cu_x$ [131].

$2 \leq x \leq 6$ відбувається різке зростання $T_{к1}$ і $T_{к3}$, а потім – насичення.

Концентраційна залежність густини стекол $Si_{15}Te_{85\text{-}x}Cu_x$ ($1 \leq x \leq 10$) наведена на рис. 2.22, крива 2. Незважаючи на те, що атомна маса міді (63.546 г/моль) менша, ніж Te (127.6 г/моль), заміщення Te малими концентраціями міді ($x \leq 3$) супроводжується збільшенням густини. При більших концентраціях ($3 \leq x \leq 6$) спостерігається плато, а потім густина різко зменшується при $x = 8$ з наступним різким зростанням при $x > 8$ .

**2.4.2. Система Si–Ag–Te.** Результати перших досліджень області склоутворення та деяких термічних властивостей стекол системи кремній – срібло – телур наведені в роботі [132]. Сплави синтезували з елементів напівпровідникової частоти у вакуумованих кварцових ампулах протягом 6–7 годин з періодичним перемішуванням при температурах, які перевищували на ~100 градусів температуру плавлення найбільш тугоплавкої сполуки в даній системі. Розплав загартовували від температури синтезу зануренням ампули у воду з льодом. Синтез і загартування сплавів автори [132] проводили в подовжених конічних ампулах, що дозволяло моделювати різну швидкість охолодження розплаву даного складу в одному експерименті. Склоподібний стан ідентифікували за характерним блискучим раковистим зломом, відсутностю ліній на дебаєграмах, наявністю ефекту розм'якшення на кривих ДТА.

Область склоутворення у системі Si–Ag–Te приведена на рис. 2.23. За зазначених умов синтезу і гартування вдається ввести в сплави до 22.5 ат. % срібла зі збереженням склоподібного стану, що дещо більше, а ніж у аналогічній системі з міддю (рис. 2.20). Область склоутворення має витягнуту форму вздовж лінії розбавлення сріблом подвійної евтектики в системі Si–Te. Цей факт підтверджує висловлене автором [129] припущення, що найбільшою схильністю до склоутворення у потрійних телуридних системах мають сплави речовин, розташованих у концентраційному трикутнику вздовж лінії розведення подвійних евтектик третім компонентом або стійким з'єднанням третього компонента.

Ще два склади стекол системи Si–Ag–Te ($Si_{15}Ag_{15}Te_{70}$ та $Si_{15}Ag_5Te_{80}$) синтезували автори [133] із високочистих простих речовин (Si, Ag, Te) у вакуумованих кварцових ампулах при 1273 K протягом 6–7 год. при періодичному перемішувані. Гартування розплаву проводили зануренням ампули у воду з льодом. У даних стеклах виявлено перехід у надпровідний стан при високих тисках.

Авторами [132] досліджені термічні властивості стекол системи Si–Ag–Te методом ДТА, результати яких наведені в табл. 2.5. На термограмах більшості стекол спостерігається один ефект розм'якшення і, як правило, два піки кристалізації (табл. 2.5). Слід відзначити, що найменша температура плавлення, яка виявлена в збагачених телуром кристалічних сплавах потрійної системи Si–Ag–Te, становить ~ 653 K, тобто близька до температури евтектики в бінарній системі Si–Te.

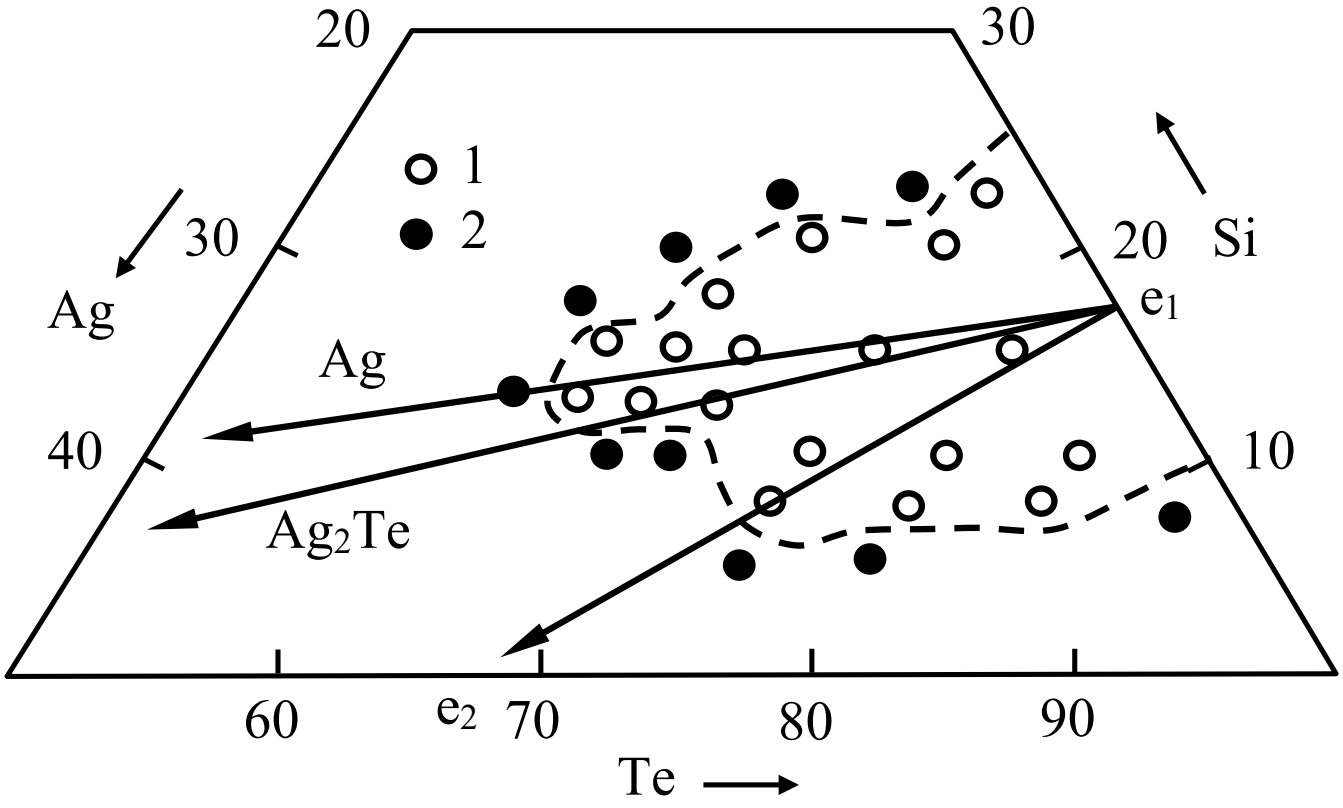

Рис. 2.23. Область склоутворення в системі Si–Ag–Te.
1– склоподібні сплави; 2 – кристалічні сплави [132].

Таблиця 2.5. Термічні властивості стекол системи Si–Ag–Te [132].

| **Склад стекол** | $T_g$, K | $T_{к1}$, K | $T_{к2}$, K | $T_{пл}$, K |
|---|---|---|---|---|
| $Si_{15}Ag_{17.5}Te_{67.5}$ | 402 | 489 | 558 | 654 |
| $Si_{12.5}Ag_{17.5}Te_{70}$ | 418 | 483 | 544 | 654 |
| $Si_{15}Ag_{15}Te_{70}$ | 424 | 495 | 544 | 652 |
| $Si_{10}Ag_{15}Te_{75}$ | 390 | 443 | 548 | 655 |
| $Si_{15}Ag_{10}Te_{75}$ | 399 | 485 | 548 | 652 |
| $Si_{20}Ag_{5}Te_{75}$ | 421 | 548 | – | 658 |
| $Si_{22.5}Ag_{2.5}Te_{75}$ | 453 | 538 | – | 653 |
| $Si_{10}Ag_{10}Te_{80}$ | 382 | 444 | 554 | 656 |
| $Si_{15}Ag_{5}Te_{80}$ | 395 | 486 | 554 | 656 |
| $Si_{10}Ag_{5}Te_{80}$ | 374 | 425 | 540 | 555 |

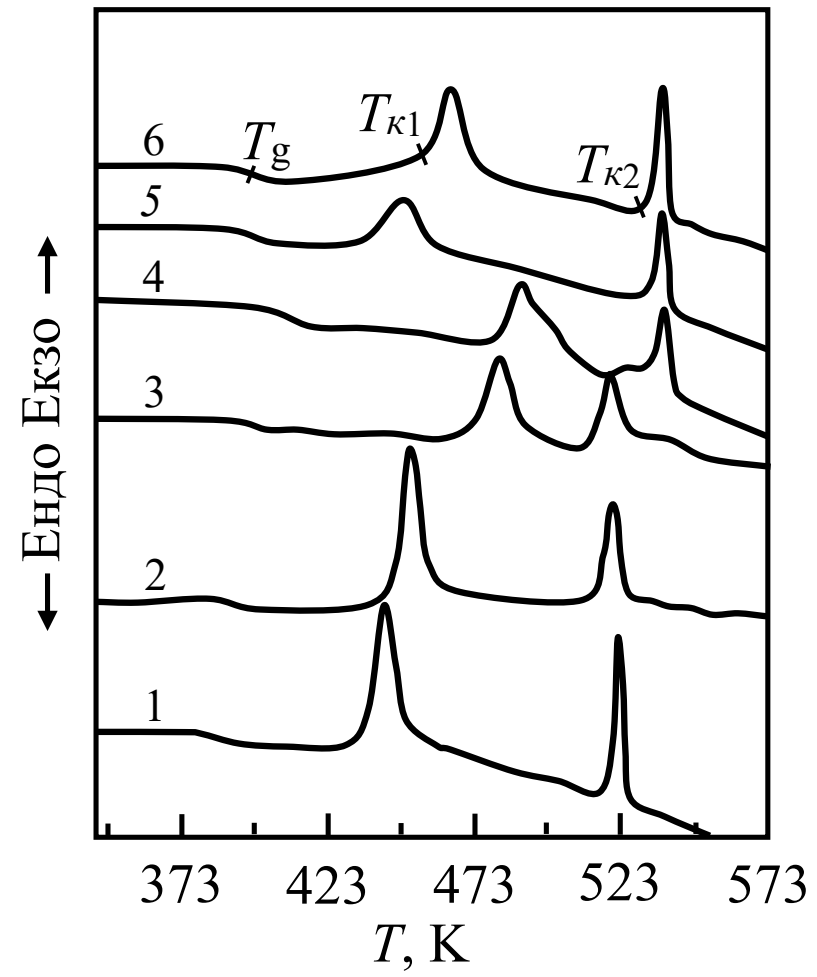

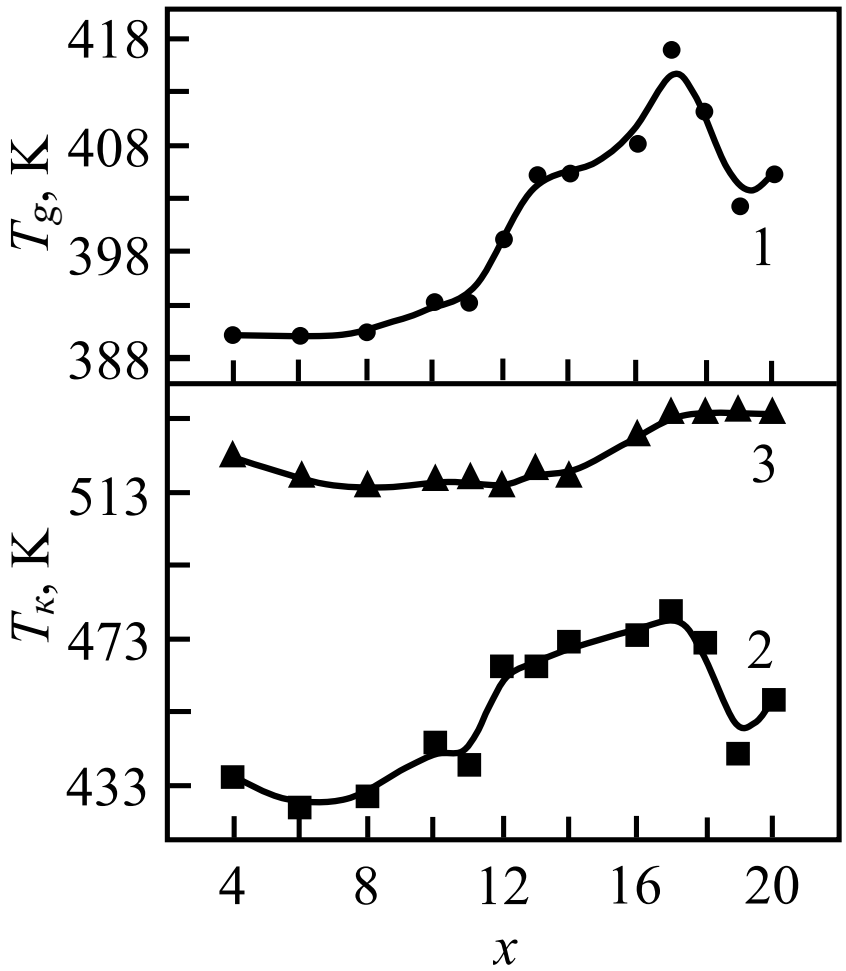

Рис. 2.24. Криві ДСК для стекол системи $Si_{15}Te_{85-x}Ag_x$ [62].

Рис. 2.25. Концентраційні залежності $T_g$ (1), $T_{к1}$ (2) і $T_{к2}$ (3) стекол $Si_{15}Te_{85-x}Ag_x$ [62].

Результати більш детального дослідження термічних властивостей об'ємних стекол $Si_{15}Ag_xTe_{85-x}$ ($4 \leq x \leq 20$) приведені в роботі [134]. Відповідні кількості складових елементів високої чистоти (99.999%) герметизували у відкачаній кварцовій ампулі при $10^{-5}$ Торр і повільно нагрівали (~ 100 К/год) в горизонтальній печі, що оберталась. Ампули витримували при 1373 К протягом 36 год і безперервно обертали задля забезпечення однорідного розплаву. Потім ампули з розплавом гартували в суміші крижаної води і NaOH.

На рис. 2.24 наведено криві ДТА для ряду стекол $Si_{15}Ag_xTe_{85-x}$ ($4 \leq x \leq 20$), які демонструють одну стадію ендотермічного склування і два різні екзотермічні піки кристалізації, як це має місце і для інших стекол потрійної системи Si–Ag–Te (табл. 2.5). Ці результати вказують на те, що дані стекла складаються із двох термодинамічно стабільних фаз, із яких найбільш стабільна починає кристалізуватися при $T_{к1}$. Рентгеноструктурні дослідження утворених фаз при температурах кристалізації $T_{к1}$ і $T_{к2}$ стекол $Si_{15}Ag_xTe_{85-x}$ вказують на наявність гексагонального телуру з параметрами елементарної комірки $a = 4.458$, і $c = 5.925$ Å і потрійної кубічної фази $Ag_8SiTe_6$ ($a = 22.98$ Å). Концентраційні залежності $T_g$, $T_{к1}$ і $T_{к2}$ наведені на рис. 2.25.

Із термічних досліджень випливає, що в склоподібній системі $Si_{15}Ag_xTe_{85-x}$ ($4 \leq x \leq 20$) при концентраціях $x = 12$ і $x = 19$ виникають жорсткість і хімічні пороги, а стекла в області $12 < x < 17$ є більш стійкими у порівнянні з іншими складами стекол цього ж розрізу.

**2.4.3. Система Si–Al–Te.** Враховуючи, що бінарна система Al–Te є хорошою склоутворюючою системою [135], це сприяє отриманню стекол і в потрійній системі Si–Al–Te. Дані про склоутворення і термічні властивості, а також ефект перемикання стекол системи Si–Al–Te приведені в роботах [136, 137]. Склоутворення в системі Si–Al–Te вивчено авторами [136] за розрізом $Si_xAl_{15}Te_{85-x}$ ($2 \leq x \leq 12$). Стекла отримували прямим синтезом елементів високої частоти у відкачаних запаяних кварцових ампулах з наступним загартуванням розплаву у воду.

Термічний аналіз стекол проводився з використанням модульованого диференціального скануючого колориметричного аналізу. На рис. 2.26 наведено криві повного теплового потоку для чотирьох складів стекол $Si_xAl_{15}Te_{85-x}$ ($x = 6, 7, 8, 9$). В інтервалі концентрацій $x < 8$ стекла даного розрізу демонструють одну ендотермічну реакцію склування ($T_g$) і дві екзотермічні реакції кристалізації ($T_{к1}$ і $T_{к2}$) в інтервалі складів $2 \leq x \leq 12$.

На рис. 2.27 наведена залежність температури склування ($T_g$) і температур кристалізації ($T_{к1}$ і $T_{к2}$) від складу стекол $Si_xAl_{15}Te_{85-x}$. На цьому рисунку видно, що температура склування ($T_g$) слабо зростає в діапазоні складів $2 \leq x \leq 7$, після чого відбувається різке збільшення $T_g$ для $x = 8$ і практично залишається сталою для $x > 8$. З рис. 2.27 також видно, що перший пік кристалізації ($T_{к1}$) закономірно зміщується в бік високих температут з підвищенням концентрації кремнію, тоді як $T_{к2}$ залишається практично сталою. Варто також зазначити, що температури кристалізації $T_{к1}$ і $T_{к2}$ співпадають при $x = 8$, вище якого спостерігається лише одна реакція кристалізації.

Для виявлення та ідентифікації кристалічних фаз, які формуються в процесі нагрівання стекол $Si_xAl_{15}Te_{85-x}$ при $T_{к1}$ і $T_{к2,}$ були проведені рентгеноструктурні дослідження термічно відпалених стекол складів $Si_4Al_{15}Te_{81}$ і $Si_8Al_{15}Te_{77}$ [136]. Процес відпалу тривав у вакуумі протягом 2 год. У зразку $Si_4Al_{15}Te_{81}$, відпаленому при $T_{к1}$ = 521 К протягом 2 год, виявлено кристалічний Te з гексагональною структурою ($a = 4.457$ і $c = 5.929$ Å). При відпалі скла цього ж складу при $T_{к2}$ = 550 К виявлено три кристалічні фази: кристалічний Te, гексаго-

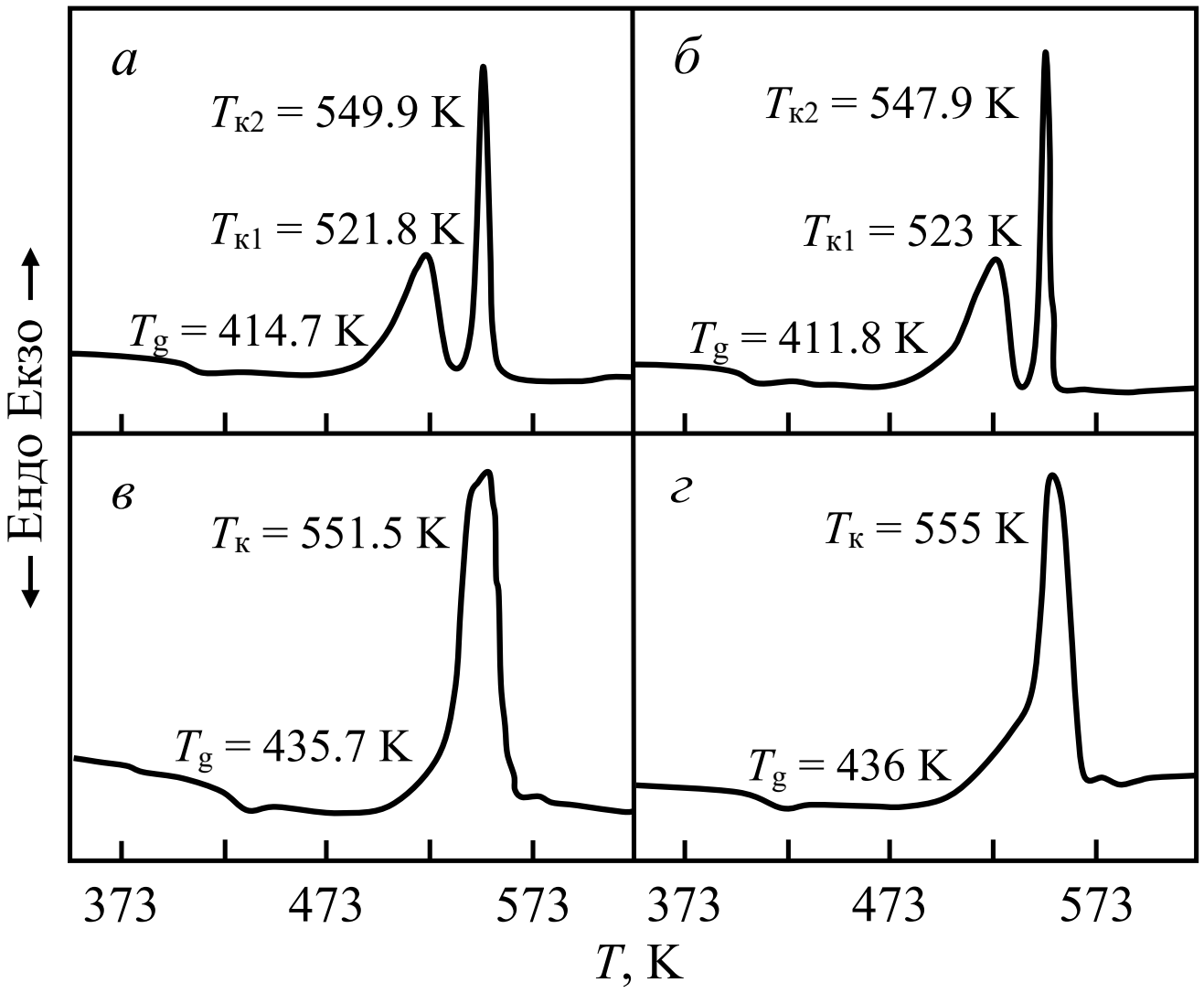

Рис. 2.26. Криві повного теплового потоку ADSC для стекол $Si_xAl_{15}Te_{85-x}$ ($x$ : *а* – 6, *б* – 7, *в* – 8, *г* – 9) [136].

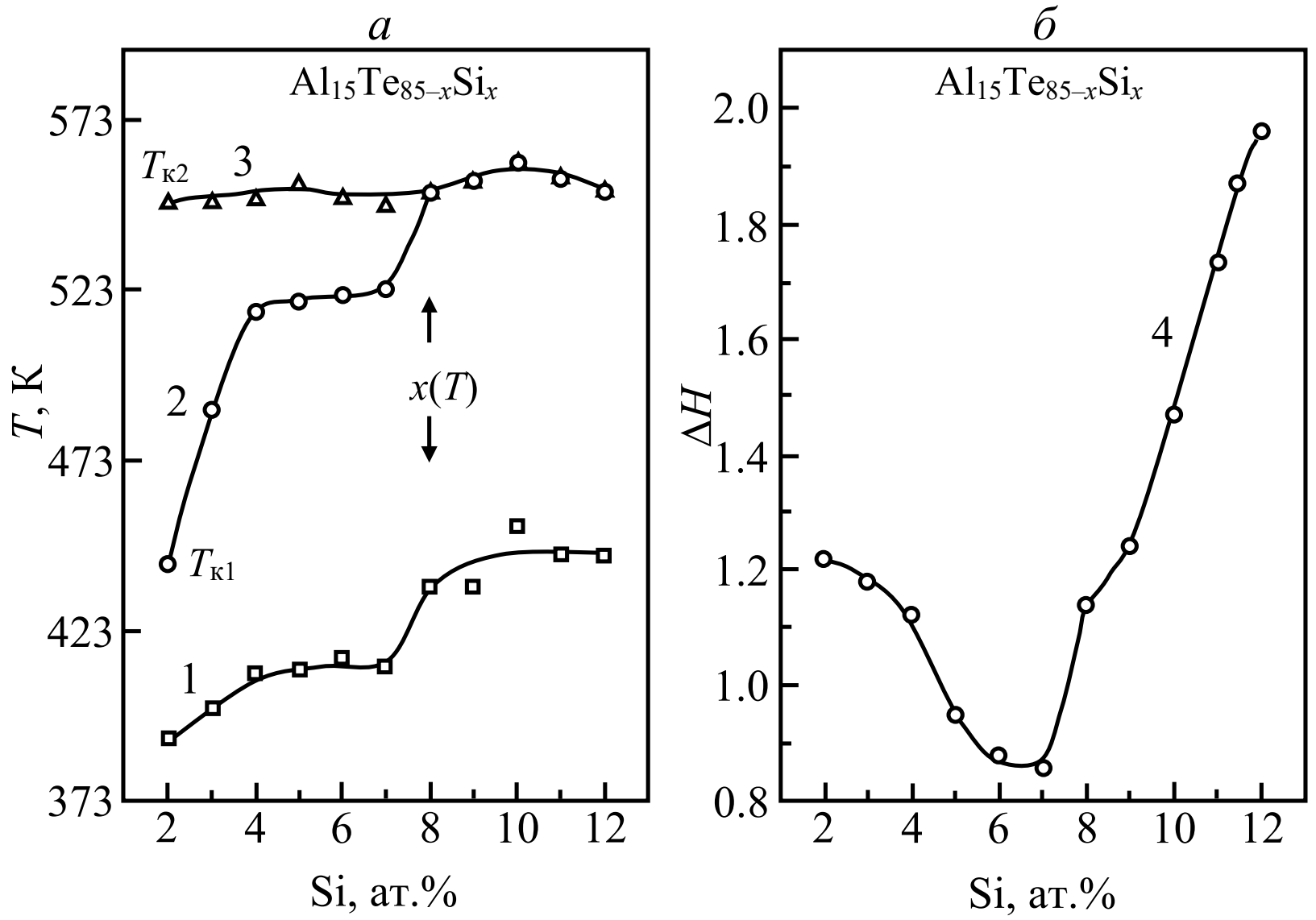

Рис. 2.27. Концентраційні залежності температур склування $T_g$ (1) і кристалізації $T_{к1}$ (2) і $T_{к2}$ (3) і ентальпії $\Delta H$ (4) стекол $Si_xAl_{15}Te_{85-x}$ [136].

нальний $AlSiTe_3$ і моноклінний $Al_2Te_3$. Усі ці три кристалічні фази виявлені і в закристалізованому склі з більшим вмістом кремнію $Si_8Al_{15}Te_{77}$, відпаленому при $T_{к2}$ = 552 К.

Скло з більшим вмістом алюмінію $Si_5Al_{20}Te_{75}$ системи Si–Al–Te синтезували автори [137] і дослідили порогове перемикання. Елементарні компоненти у відповідній кількості завантажували в кварцові ампули, які потім відкачували і запаювали. Запаяні ампули поміщали в горизонтальну піч**.** Температуру печі підвищували до 1223 К зі швидкістю 100 К/год і витримували протягом 48 год. Для забезпечення гомогенності ампули з розплавом неперервно обертали. Потім температуру зменшували до 1073 К і здійснювали гартування розплаву в суміші крижаної води і NaOH.

**2.4.4. Система Si–In–Te.** Синтез склоподібних сплавів у системі Si–In–Te автор [129] проводив із простих речовин Si, In, Te у вакуумованих кварцових ампулах. Вакуумовані ампули із шихтою нагрівалися до 1273 К. При цій температурі витримувалися протягом 12 год. при інтенсивному перемішуванні, після чого проводилося гартування в холодну воду. Склоподібний стан сплавів контролювався за характерним скляним блиском, раковистим зломом та рентгеноструктурним аналізом. За результатами досліджень побудовано область склоутворення в системі Si–In–Te, яка наведена на рис. 2.28.

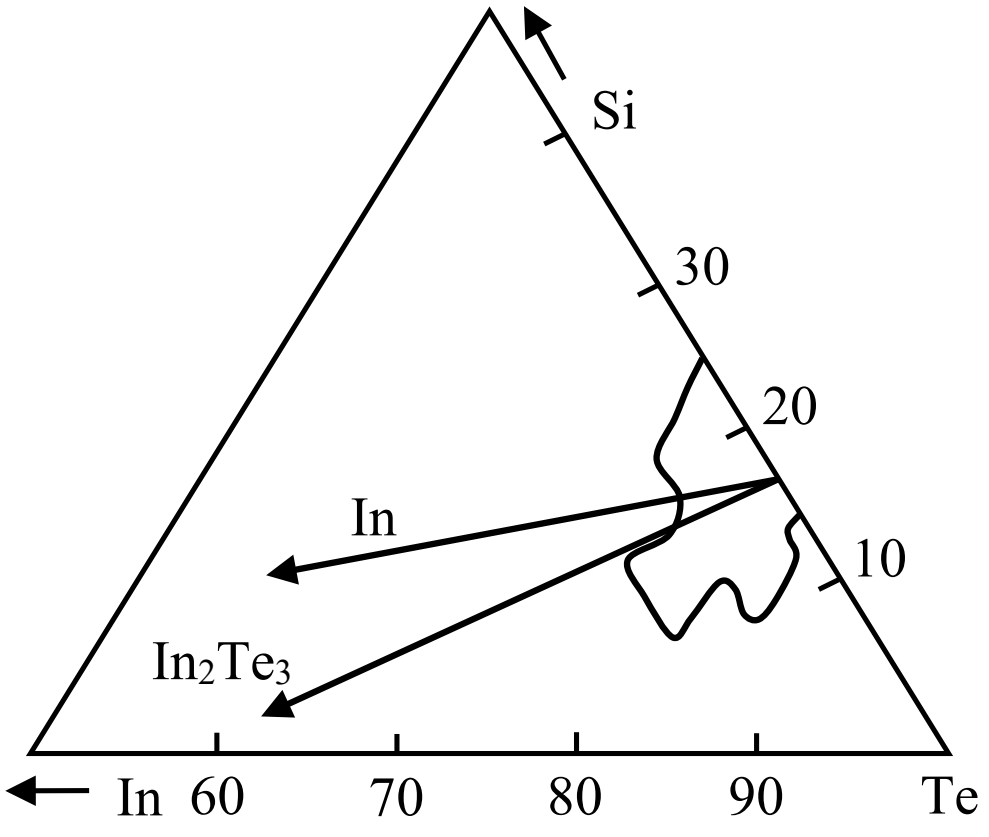

Рис. 2.28. Область склоутворення в системі Si–In–Te [129].

Об'ємні стекла за розрізом $Si_{15}In_xTe_{85-x}$ ($1 \leq x \leq 10$) загальною вагою 1.5 г синтезували автори [138]. Для приготування сплавів використовувалися прості речовини (Si, In, Te) високої чистоти (99.999 %). Синтез проводився у вакуумованих запаяних кварцових ампу-лах, які були поміщені в горизонтально-обертовій печі.

Максимальна температура синтезу становила 1373 K, тривалість процесу синтезу 36 год. З метою гомогенізації здійснювалося безперервне перемішування розплаву шляхом обертання печі зі швидкістю 10 об/хв. Загартування розплавів проводилось в крижаному водному розчині NaOH. Аморфність, природа і гомогенність загартованих зразків підтверджена методами рентгенівської дифракції та модульованої скануючої колометрії. Сканування ADSC усіх досліджуваних зразків проводилось з частотою 3 K/хв. і швидкістю модуляції 1 K/хв. Криві ADSC для ряду складів стекол розрізу $Si_{15}In_xTe_{85-x}$ ($1 \le x \le 10$) наведені на рис. 2.29. Із цього рисунка видно, що на всіх кривих ADSC спостерігається один пік екзотермічного стклування ($T_g$) і три різні екзотермічні піки кристалізації ($T_{к1}$, $T_{к2}$ і $T_{к3}$). Цей факт вказує на те, що при нагріванні в стеклах Si–In–Te, при різних температурах, утворюються стабільні кристалічні фази.

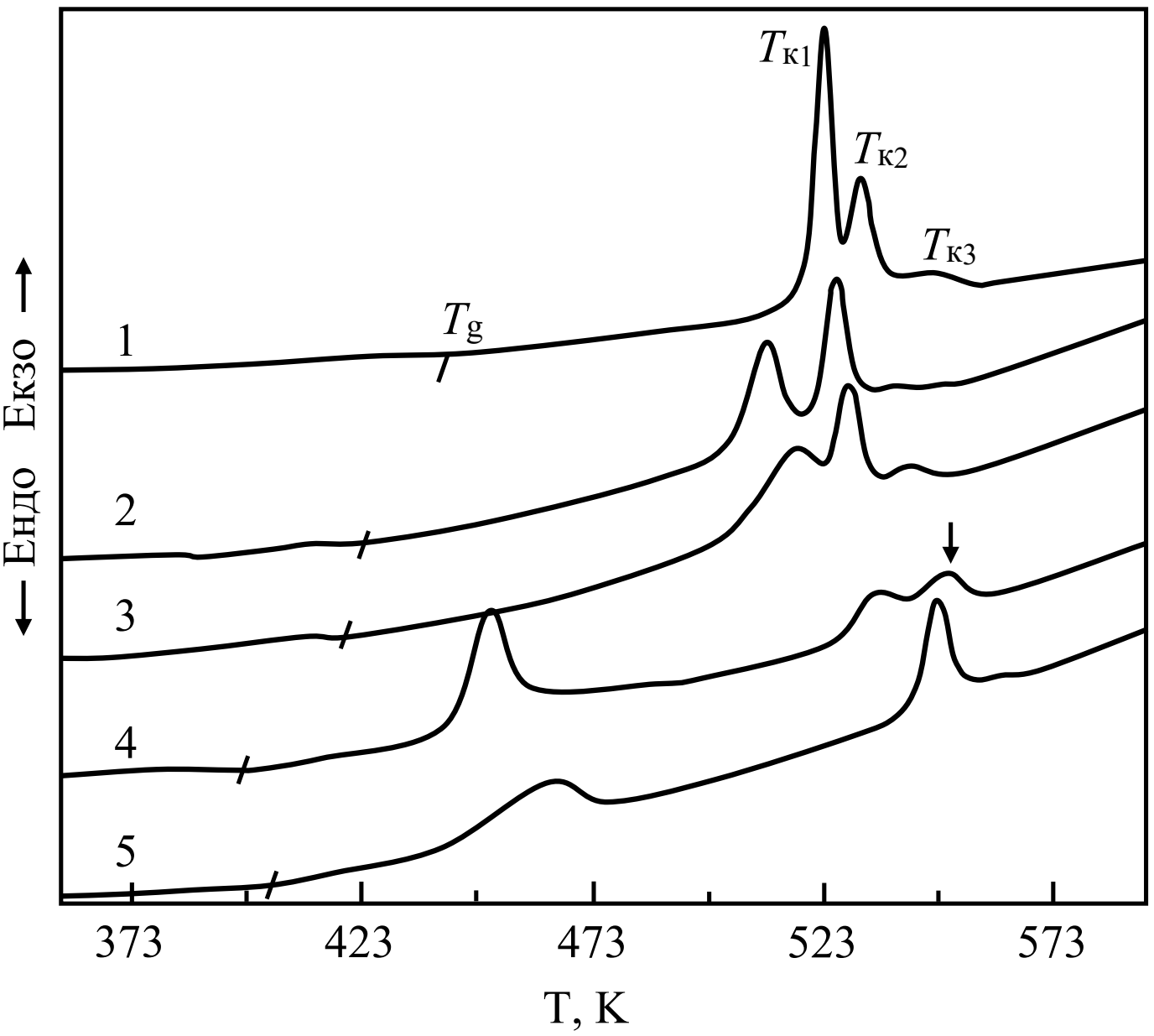

Рис. 2.29. Криві повного теплового потоку ADSC для стекол $Si_{15}Te_{85-x}In_x$ з різною концентрацією індію. $x$, ат.%: 1 – 10; 2 – 7; 3 – 6; 4 – 2; 5 – 1 [2].

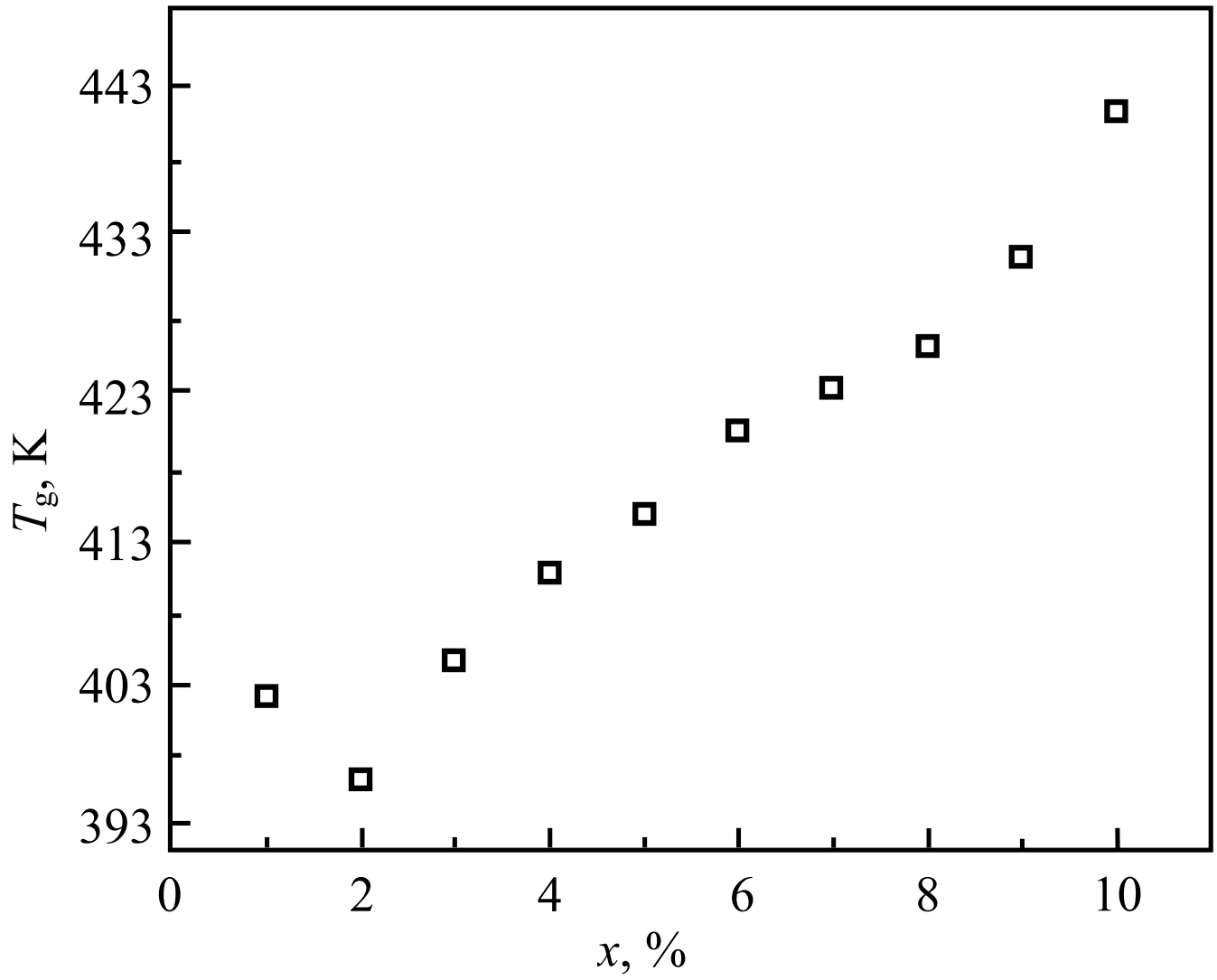

Рис. 2.30. Концентраційна залежність температури склування ($T_g$) стекол $Si_{15}Te_{85-x}In_x$ [138].

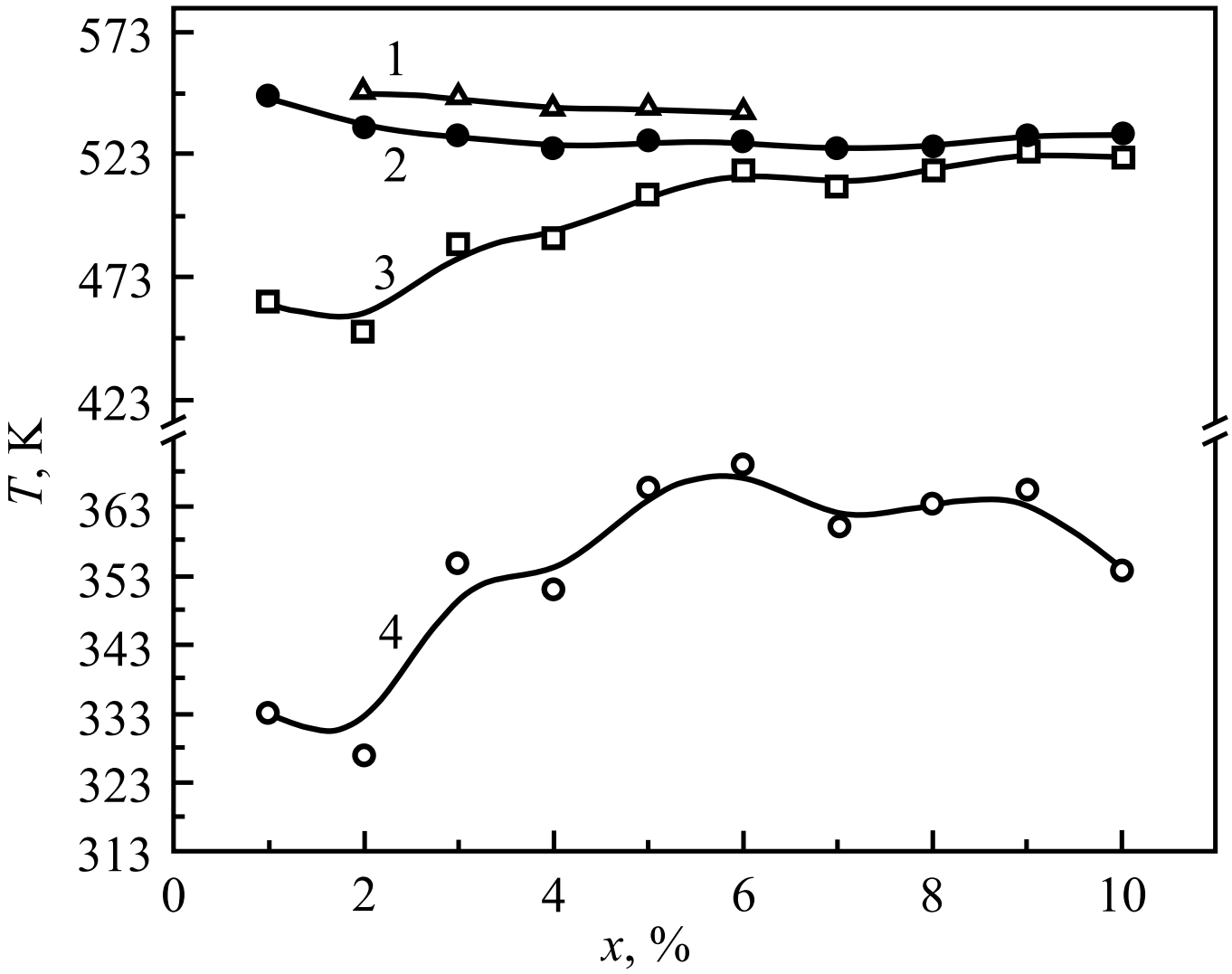

Рис. 2.31. Концентраційні залежності $T_{к1}$ (3), $T_{к2}$ (2), $T_{к3}$ (1) і $\Delta T = T_{к1}-T_g$ (4) стекол $Si_{15}Te_{85-x}In_x$ [138].

Концентраційна залежність температури склування $T_g$ стекол $Si_{15}In_xTe_{85-x}$, наведена на рис. 2.30. Видно, що в інтервалі складів $x \leq 2$ спостерігається спочатку зменшення $T_g$, за яким слідує безперервне зростання. Початкове зниження $T_g$ стекол $Si_{15}In_xTe_{85-x}$ в області складів $1 \leq x \leq 2$ автори [138] пов'язують з сегрегацією гомополярних зв'язків Te–Te в системі Si–Te–In. Тут також цікаво відзначити, що перша температура кристалізації $T_{к1}$ стекол $Si_{15}In_xTe_{85-x}$ збільшується зі збільшенням концентрації індію в області $2 \leq x \leq 6$ і згодом насичується (рис. 2.31, крива 3). Натомість збільшення концентрації індію практично не впливає на температури $T_{к2}$ і $T_{к3}$. З рис. 2.31, крива 4 також видно, що термостабільність і склоутворююча здатність скла, яка прямо пропорційна різниці $T_{к1}$ і $T_g$ ($\Delta T = T_{к1} - T_g$), збільшується зі збільшенням концентрації індію у композиційному інтервалі $2 \leq x \leq 6$. Ці результати переконливо підтверджують ідею про те, що мережне з'єднання і, отже, мережна жорсткість стекол $Si_{15}In_xTe_{85-x}$ збільшуються у діапазоні складів $2 \leq x \leq 6$.

Рентгеноструктурні дослідження закристалізованих стекол $Si_{15}In_xTe_{85-x}$ показали, що фази, які кристалізуються при $T_{к1}$, $T_{к2}$ і $T_{к3}$ представляють собою гексагональний Te ($a$ = 4.488 *і* $c$ = 5.925 Å) і орторомбічний $In_4Te_3$ ($a$ = 15.549, $b$ = 12.7 *і* $c$ = 4.460 Å).

# РОЗДІЛ 3

# ЕЛЕКТРОННА СТРУКТУРА ТА ОПТИЧНІ ВЛАСТИВОСТІ КРИСТАЛІВ $Si_2Te_3$, $SiTe_2$ І СТЕКОЛ $Si_xTe_{100-x}$

Електронна зонна структура $E(\boldsymbol{k})$ є однією з фундаментальних характеристик, яка визначає більшість фізичних властивостей напівпровідникових кристалів, зокрема явища переносу заряду, оптичні та фотоемісійні властивості. Електронні стани в околі рівня Фермі $E_F$ також є одним із головних факторів, які визначають функціонування приладів напівпровідникової твердотільної електроніки. При цьому важливим є не тільки розташування електронних станів по енергії (яке обумовлює, наприклад, ширину забороненої зони), але і характер дисперсії цих станів у $\boldsymbol{k}$ просторі.

## 3.1. СТРУКТУРА ЕНЕРГЕТИЧНИХ ЗОН І ПРИРОДА ЕЛЕКТРОННИХ СТАНІВ $Si_2Te_3$

Розрахунки в рамках теорії функціонала електронної густини (DFT) є потужним методом аналізу структурних і електронних властивостей напівпровідників. Більшість DFT розрахунків використовує наближення локальної густини (*Local Density Approximations*, LDA) або наближення узагальненого градієнта (*Generalized Gradient Approximations*, GGA). Однак відомо, що DFT з використанням стандартних LDA і GGA наближень «недооцінює» ширину забороненої зони напівпровідника, тобто дає сильно занижені значення $E_g$. Причина полягає у тому, що моделі LDA і GGA погано описують збуджені стани, а отже й величину $E_g$ та структуру зони провідності.

Покращити опис зонної структури кристалів у наближенні локальної густини дозволяє врахування поправки Хаббарда $U$ у відповідних гамільтоніанах LDA і GGA. Покажемо це на прикладі кристала $Si_2Te_3$, електронна структура якого розрахована в моделі LDA+$U$ [139, 140]. Величини параметрів прямої кулонівської та обмінної взаємодії складали $U = 7$ еВ і $J = 0,7$ еВ. Зазначимо, що для кристалів $Si_2Te_3$ проведені дослідження краю власного поглинання, що дозволило визначити ширину забороненої зони і порівняти її значення з теоретично розрахованим.

Зонна структура кристала $Si_2Te_3$, розрахована методом LDA+$U$ без врахування спін-орбітальної взаємодії у точках високої симетрії і уздовж симетричних напрямків у незвідній частині зони Бріллюена (рис. 3.1), наведена на рис. 3.2. За нульовий рівень енергії прийнято найвищій заповнений стан. Вершина валентної зони в $Si_2Te_3$ локалізована в точці Г, а дно зони провідності знаходиться в точці K. Отже, тригональний сесквітелурид кремнію є непрямозонним напівпровідником із розрахованою величиною ширини забороненої зони $E_{gi}$ = 2.05 еВ, близькою до експериментально визначеної з аналізу краю власного поглинання $E_{gi}$ = 2.13 еВ.

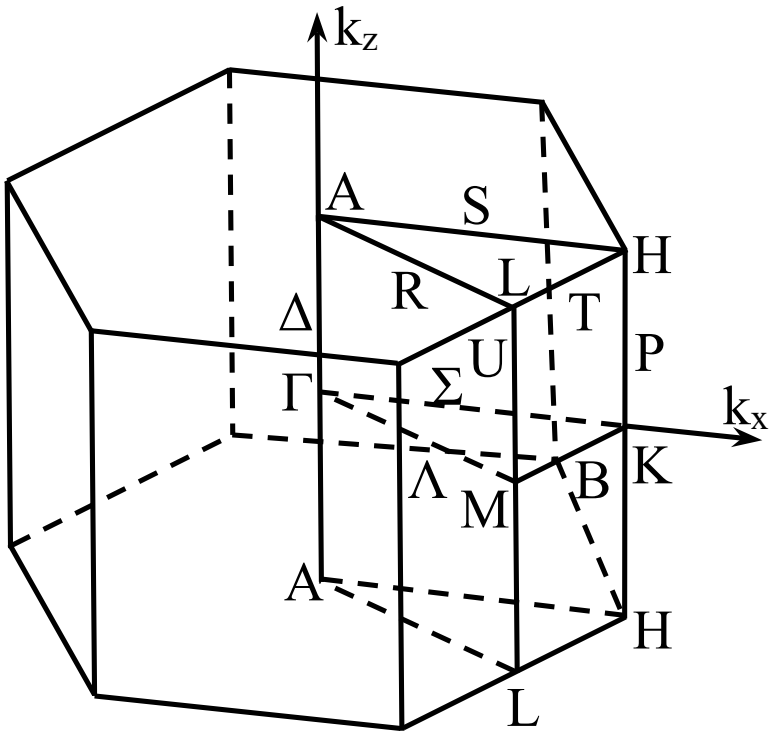

Рис. 3.1. Зона Бріллюена $Si_2Te_3$.

Електронна структура кристала $Si_2Te_3$ також розрахована за допомогою функціоналів TB-mBJ [141], і PBE [142] у рамках узагальненого градієнтного наближення (GGA) та із застосуванням гібридного функціоналу HSE06, без врахування і з врахуванням спін-орбітальної взаємодії (SOC) [142]. Згідно даних [141], максимум валентної зони і мінімум зони провідності локалізовані відповідно в точках Г і H зони Бріллюена, тобто підтверджено, що сесквітелурид кремнію є непрямозонним напівпровідником, а розрахована ширина забороненої зони $E_{gi}$ = 1.78 еВ є очікувано заниженою. Врахування спін-орбітальної взаємодії приводить до ще більшого зменшення розрахованого значення $E_{gi}$ = 1.64 еВ. Однак положення вершини валентної зони і дна зони провідності залишаються без змін при врахуванні спін-орбітальної взаємодії. При використанні гібридного функціоналу HSE06 для розрахунку зонної структури $Si_2Te_3$ без врахування SOC знайдено ширини забороненої зони для об’ємного зразка і моношару рівними 2.29 та 2.65 еВ відповідно [142].

Будова країв енергетичних зон, тобто максимумів валентної зони і мінімумів зони провідності, визначає фундаментальні фізичні властивості напівпровідників, і є досить чутливою до наближень, за

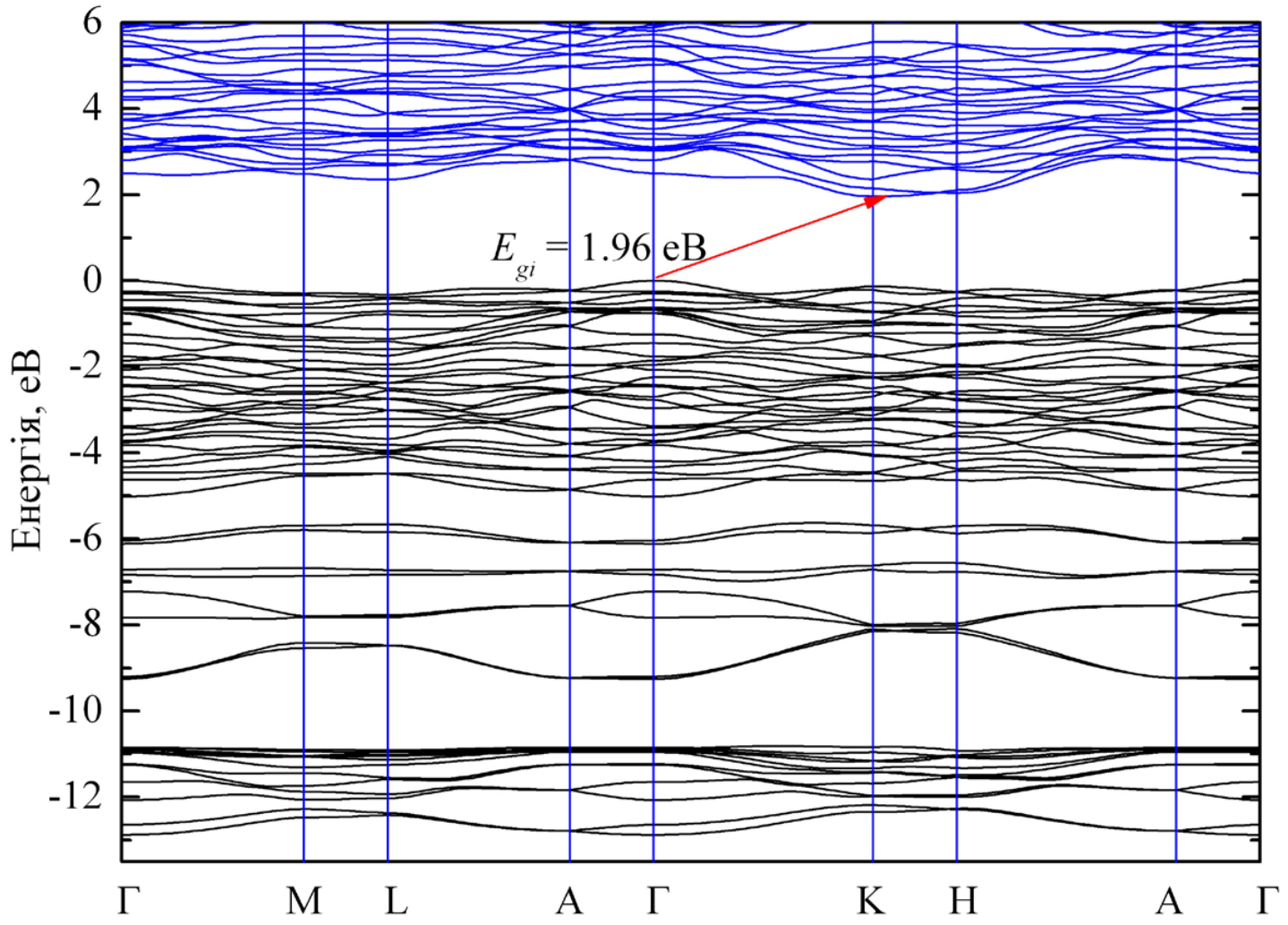

Рис. 3.2. Електронна структура $Si_2Te_3$ [140].

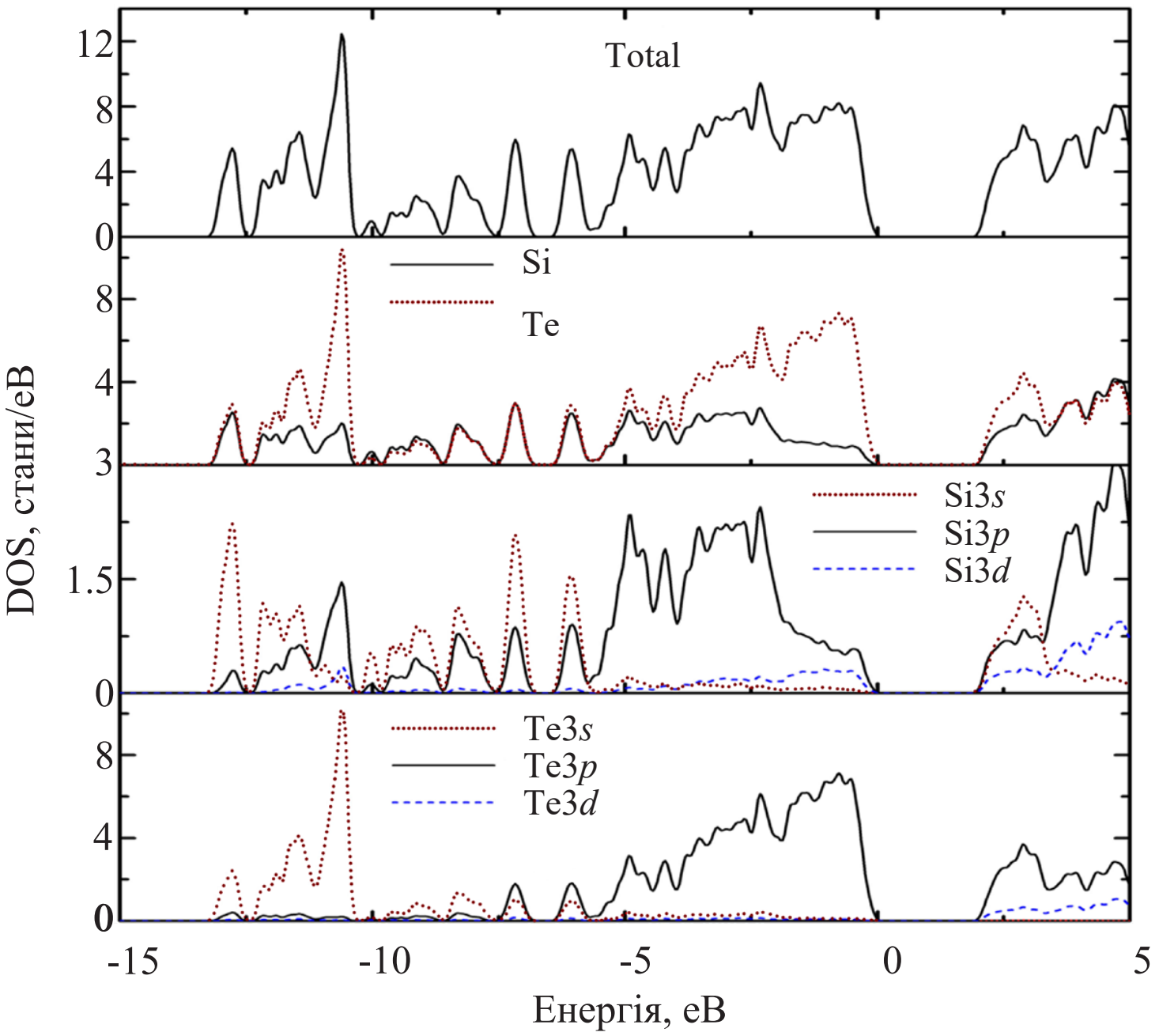

Рис. 3.3. Повна та локальні парціальні густини електронних станів кристала $Si_2Te_3$, розраховані в наближенні LDA+U [140].

допомогою яких проводяться розрахунки, особливо при наявності конкуруючих екстремумів. Як видно з рис. 3.2, краї енергетичних зон $Si_2Te_3$ в околі забороненої зони характеризуються наявністю близьких максимумів і мінімумів. Відносні розташування цих екстремумів у роботах [140–142] виявились різними. Так, за даними [140, 141] вершина валентної зони розташована в центрі зони Бріллюена Г, а згідно [142] – у точці X. Розбіжності також мають місце в будові зони провідності. Згідно [140], дно зони провідності локалізоване в точці K, за даними [141] – у точці H, і в точці Г згідно [142].

Таким чином, правильний вибір параметрів одноцентрового обміну і кореляції в LDA+$U$ розрахунках електронної зонної структури приводить до результатів, які є ближчими до експериментального значення ширини забороненої зони для нелегованого $Si_2Te_3$ у порівнянні з розрахунками з використанням наближень TB-mBJ і HSE06.

В електронній структурі $Si_2Te_3$ знаходить своє відображення шаруватий характер кристала, що проявляється в анізотропії дисперсії віток уздовж і поперек тришарових пакетів. Так, у напрямку Г→M (вздовж тришарового пакету) дисперсія більшості віток валентних зон перевищує дисперсію в напрямку Г→A (поперек шарів). Такий хід дисперсійних віток є відображенням відмінностей у природі хімічного зв'язку вздовж вказаних напрямків, що додатково свідчить про шаруватий характер даного кристала.

Важливими класифікаційними принципами при аналізі електронної структури валентної зони кристалів є число валентних електронів, що дозволяє встановити кількість дисперсійних віток, характер партнерів по хімічному зв'язку (чим визначається взаємне енергетичне розташування валентних підзон) і кристалічна структура речовини (впливає на розподіл станів валентної зони, особливо верхніх підзон) [143]. Так як елементарна комірка $Si_2Te_3$ містить 12 шестивалентних аніонів (Te) і 8 чотиривалентних катіони (Si), то число валентних електронів у ЗБ рівне 104 і відповідний енергетичний спектр валентної зони складається з 52 дисперсійних віток, об'єднаних у три зв'язки заповнених підзон, розділених забороненими щілинами. Сумарна ширина зайнятих зон складає 12.88 еВ.

Повна та парціальні густини електронних станів $N(E)$ визначають значну частину електронних властивостей твердих тіл. Знання повної густини електронних станів необхідне для правильної інтерпретації експериментальних рентгенівських та ультрафіолетових фотоелектронних спектрів кристала, а парціальні густини електронних станів дозволяють ідентифікувати якими саме атомними орбіталями сформовані підзони валентної зони та зони провідності.

Отримані в результаті зонного розрахунку власні функції $\psi_{i,k}(\boldsymbol{r})$ та власні значення енергії $E(\boldsymbol{k})$ автори [139, 140] використовували для розрахунку повної $N(E)$ та парціальних густин електронних станів $Si_2Te_3$. Повна густина електронних станів задається виразом:

$$n(E) = \frac{2}{\Omega_{BZ}} \sum_i \int_{\Omega_{BZ}} \delta(E - E_i(k))dk\,, \tag{3.1}$$

а локальні парціальні густини електронних станів

$$n_{sl}(E) = \frac{2}{\Omega_{BZ}} \sum_i \int_{\Omega_{BZ}} Q_k^{sl}(E - E_i(k))dk\,, \tag{3.2}$$

де $i$ – номер енергетичної зони, $\Omega_{BZ}$ – об’єм першої зони Бріллюена, $Q_k^{sl}$ визначає заряд $l$-типу симетрії всередині атомної сфери, яка охоплює в елементарній комірці атоми $s$-типу.

Профілі розподілу повної густини електронних станів $N(E)$, а також внесків від окремих станів різних атомів для $Si_2Te_3$ наведені на рис 3.3. Аналіз парціальних внесків в електронну густину станів дозволив ідентифікувати походження різних підзон валентної зони і зони провідності $Si_2Te_3$. Співвідношення між інтенсивностями максимумів у парціальних густинах електронних станів різні для різних типів симетрії. У валентній зоні сесквітелуриду кремнію переважають парціальні 5*s*- і 5*p*-стани атомів телуру, причому їх енергетичне положення істотно відрізняється. У глибині валентної зони даної сполуки у повній густині електронних станів $N(E)$ домінує внесок 5*s*-станів телуру, тоді як у верхній частині валентної зони переважає внесок 5*p*-станів атомів Te. Таким чином, найнижча валентна підзона, розташована в енергетичному інтервалі від –12.88 до –10.81 еВ, сформована переважно 5*s*-станами телуру. Незважаючи на переважний характер Te5*s*-станів, для даної підзони істотними є ефекти гібридизації електронних станів атомів кремнію і телуру, що приводить до появи внесків 3*s*-станів атомів кремнію, які виявляються в основному локалізованими у нижній частині цієї підзони та 3*s*-, 3*p*- , 3*d*-станів Si у її верхній частині.

Середню частину валентної зони у діапазоні енергій від –9.26 до –5.63 еВ можна розділити на чотири підгрупи щодо підзон, кожна з яких містить дві дисперсійні вітки. Дві нижні підгрупи з чотирьох валентних зон (–9.26 ÷ –5.63 еВ) формуються гібридизованими Si3*s*-, 3*p* – Te5*s*-станами. Наступні дві верхні підгрупи мають змішаний характер за участю 5*s*- і 5*p*-станів Te і 3*s*- і 3*p*-станів Si.

Найбільш складною є верхня підзона зайнятих станів (–5.02 ÷ 0 еВ), що складається з 32 дисперсійних віток. Верх цієї підзони, розташований безпосередньо поблизу вершини валентної зони (–1.60 ÷ 0 еВ), сформований переважно 5*p*-станами телуру з незначним домішуванням 3*p*-, 3*d*-станів кремнію. Нижня частина цієї підзони (–5.02 ÷ –1.60 еВ) сформована гібридизованими 5*p*-станами телуру та 3*p*-станами кремнію.

Електронна низькоенергетична структура незаповнених електронних станів у сесквітелуриді кремнію формується переважно замішуванням вільних Te *p*-, *d*- і Si *s*-, *p*-, *d*-станів, з основним внеском *p*-станів обох атомів. Таким чином, аналіз повної та парціальних густин електронних станів вказує на значну гібридизацію *s*- та *p*-станів атомів Si та Te, що свідчить про сильно ковалентний характер хімічного зв'язку Si–Te у координаційному октаедрі [$Si_2Te_6$] – структурній одиниці $Si_2Te_3$, а основну роль в оптичних міжзонних переходах відіграє перенесення заряду між Te5*p*-зайнятими станами та Te *p* + Si *s*, *p* незайнятими станами у зоні провідності.

Враховуючи той факт, що в кристалічній структурі $Si_2Te_3$ наявні димери Si–Si, чверть яких орієнтовані вертикально, тоді як решта димерів орієнтовані горизонтально з довільним вибором 0°, 30° або 60°, авторами [144] вперше проведені DFT розрахунки зонної структури об’ємного і моношару сесквітелуриду кремнію у залежності від орієнтації димерів. У результаті виявлено, що конфігурація, де всі димери орієнтовані горизонтально в одному напрямку (рис. 3.4, *а*) має найнижчу енергію, тобто відповідна структура є найстабільнішою. Електронна структура об’ємного $Si_2Te_3$ змінюється при переорієнтації димерів (рис. 3.4). У випадку, коли всі димери орієнтовані горизонтально в одному напрямку, верх валентної зони локалізований у точці S′, а дно зони провідності в точці A, тобто кристал є непрямозонним напівпровідником з $E_{gi}$ = 1.4 еВ. Натомість коли димери приймають всі можливі орієнтації (рис. 3.4, *б*), положення верха валентної зони змінюється до точки Δ, а дно зони провідності залишається в точці A, тобто кристал залишається непрямозонним напівпровідником, але ширина забороненої зони при цьому зменшується до $E_{gi}$ = 1.1 еВ. Таким чином, у результаті проведених розрахунків електронної структури [144] встановлено, що ширина забороненої зони $Si_2Te_3$ варіюється до 30% в залежності від орієнтації димерів. Зменшення ширини забороненої зони при появі вертикальних диме-

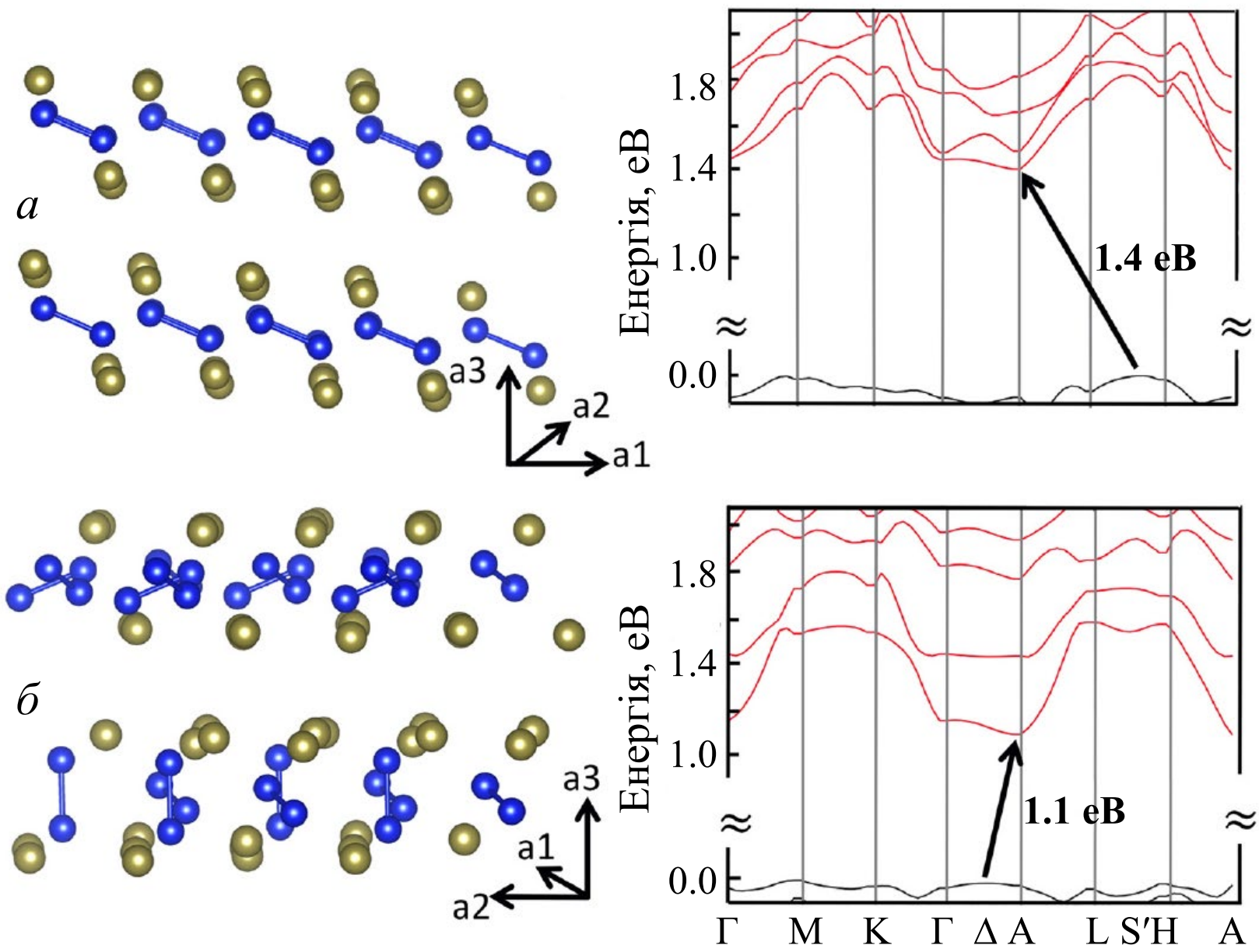

Рис. 3.4. Конфігурації димерів Si–Si в об'ємному $Si_2Te_3$ та відповідні зонні структури, *а* - усі димери орієнтовані горизонтально в одному напрямку; *б* – димери приймають усі можливі орієнтації [144].

рів пояснює зворотню зміну кольору з червоного на чорний при нагріванні кристала $Si_2Te_3$ до 483 K [37], оскільки більшість димерів знаходиться у вертикальному положенні при більш високій температурі.

Авторами [145] згодом були знову проведені першопринципні розрахунки електронної структури об'ємного $Si_2Te_3$ для двох випадків орієнтації усіх димерів: вертикальної та горизонтальної. Результати цих розрахунків дещо відрізняються від приведених у роботі [144]. Це в першу чергу стосується локалізації максимумів вершин валентної зони і дна зони провідності та отриманих величин ширини забороненої зони. Згідно даних [145], у випадку вертикальної орієнтації всіх димерів, максимум валентної зони розташований у точці Г, а дно зони провідності у точці H, і розрахована ширина забороненої зони рівна $E_{gi}$ = 1.3 еВ. У випадку горизонтальної орієнтації димерів вершина валентної зони залишається у точці Г, а дно зони провідності зміщується до точки K, при цьому ширина забороненої зони зменшується до значення $E_{gi}$ = 1.26 еВ.

## 3.2 ВПЛИВ ОДНОВІСНОЇ ДЕФОРМАЦІЇ РОЗТЯГУ НА ЕЛЕКТРОННУ СТРУКТУРУ МОНОШАРУ $Si_2Te_3$.

Використовуючи результати розрахунків із перших принципів, автори [37] показали, що моношар $Si_2Te_3$ може витримати одновісну деформацію розтягу до 38%, яка є однією з найбільших серед усіх відомих 2D матеріалів. Висока механічна гнучкість дозволяє прикладати до $Si_2Te_3$ велику механічну напругу для налаштування електронної структури. На рис. 3.5 приведені зонні структури моношару $Si_2Te_3$ для різних значень одновісних деформацій вздовж напрямку Y, розраховані методом DFT з урахуванням ефекту спін-орбітальної взаємодії. Для недеформованої структури (рис. 3.5, *а*) максимум

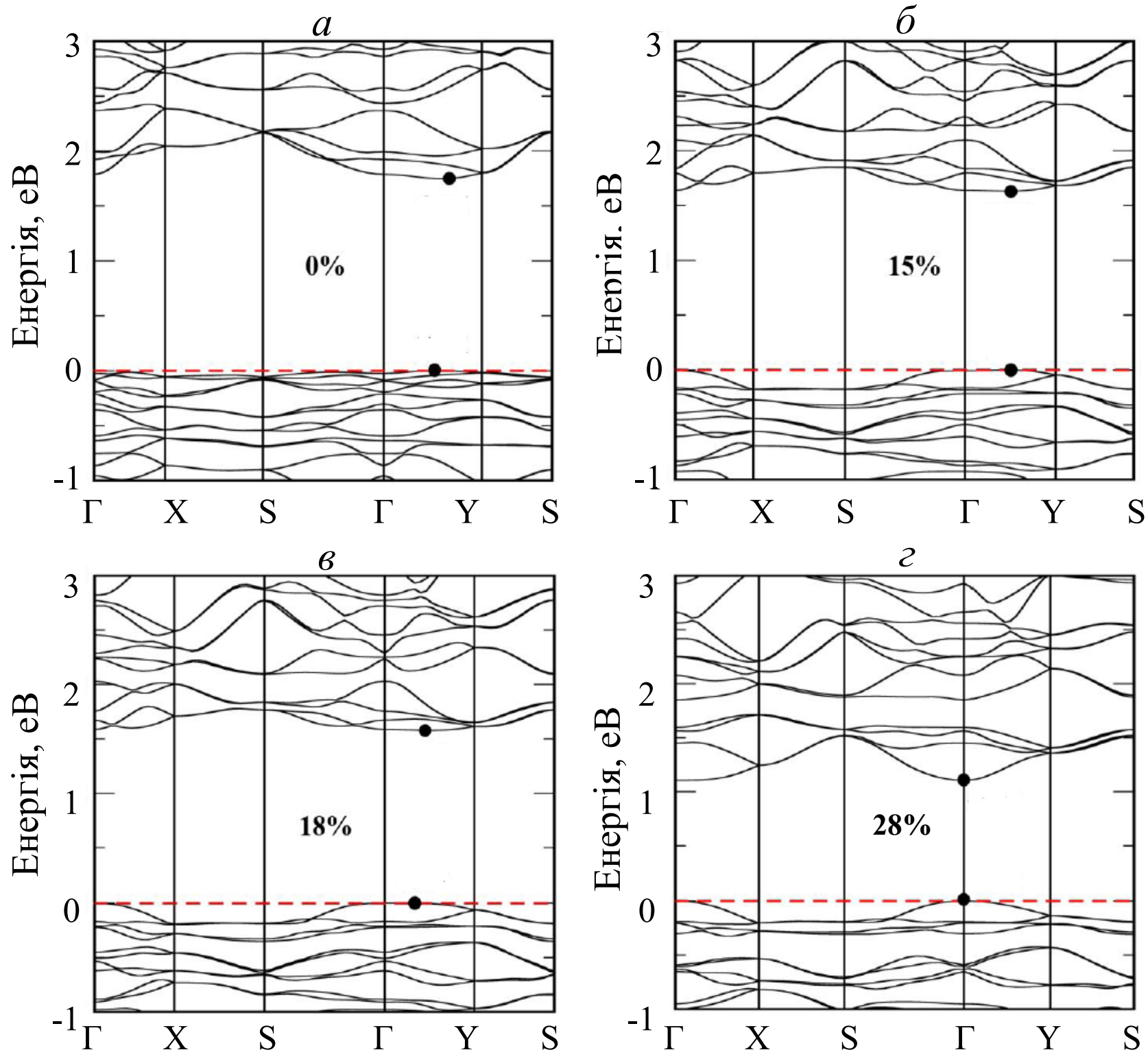

Рис. 3.5. Зонні структури моношару $Si_2Te_3$ при одновісній деформації вздовж осі Y [37].

валентної зони (МВЗ) і мінімум зони провідності (МЗП) знаходяться у різних позиціях між точками Г і Y, тобто заборонена зона моношару є непрямою. При 15% деформації (рис. 3.5, *б*) МВЗ і МЗП зміщуються у бік точки Г і за положенням практично співпадають, що робить заборонену зону прямою. Зі збільшенням деформації заборонена зона знову стає непрямою (18% на рис. 3.5, *в*), а подальше збільшення деформації приводить до того, що МВЗ і МЗП знову збігаються в одній точці, роблячи заборонену зону прямою (27% на рис. 3.5, *г*), форма якої зберігається аж до 38%. Отже, механічна деформація змінює як величину забороненої зони, так і характер зонного переходу. Такий характер трансформації непрямої – прямої – непрямої – прямої забороненої зони під дією механічної деформації автори [37] спостерігали вперше в 2D матеріалах, що пояснюють модифікацією хімічного зв'язку в структурі. Таким чином, механічна деформація потенційно може значно покращити властивості електронного транспорту в $Si_2Te_3$.

Основною причиною такої надзвичайно високої гнучкості моношару $Si_2Te_3$ є наявність димерів Si–Si, оскільки вони несуть основне навантаження, коли до зразка прикладається стиск. Висока механічна гнучкість дозволяє застосовувати механічну деформацію для зменшення ширини забороненої зони моношару. Крім того, збільшення деформації призводить до того, що заборонена зона зазнає нетипової дворазової зміни типу і відповідно характеру переходів від непрямого до прямого. Таким чином, одновісною деформацією можна ефективно контролювати зміну характеру зон за рахунок вирівнювання димерів Si–Si, що є корисним для низки практичних застосувань.

### 3.3. ЕЛЕКТРОННА СТРУКТУРА ФАЗ ВИСОКОГО ТИСКУ $Si_2Te_3$

Нещодавній експеримент авторів [35] показав, що нанопластини сесквителуриду кремнію зазнають зміни кольору з червоного на чорний при гідростатичному тиску 9.5 ГПа, що вказує на фазовий перехід до металевої фази. Такий фазовий перехід можна спостерігати при відносно меншому тиску - близько 7 ГПа, якщо між кристалічними шарами наявні інтеркальовані атоми Mn. Однак кристалічну структуру металічного $Si_2Te_3$ ще належить дослідити. В іншому експерименті автори [59] виявили, що нанодроти $Si_2Te_3$ можуть перемикатися між напівпровідниковим і металевим станами під дією зовнішньої електричної напруги. Крім того, автори [3, 10] повідомили про фазовий перехід при 673 – 723 K, пов'язаний з розпадом ди-

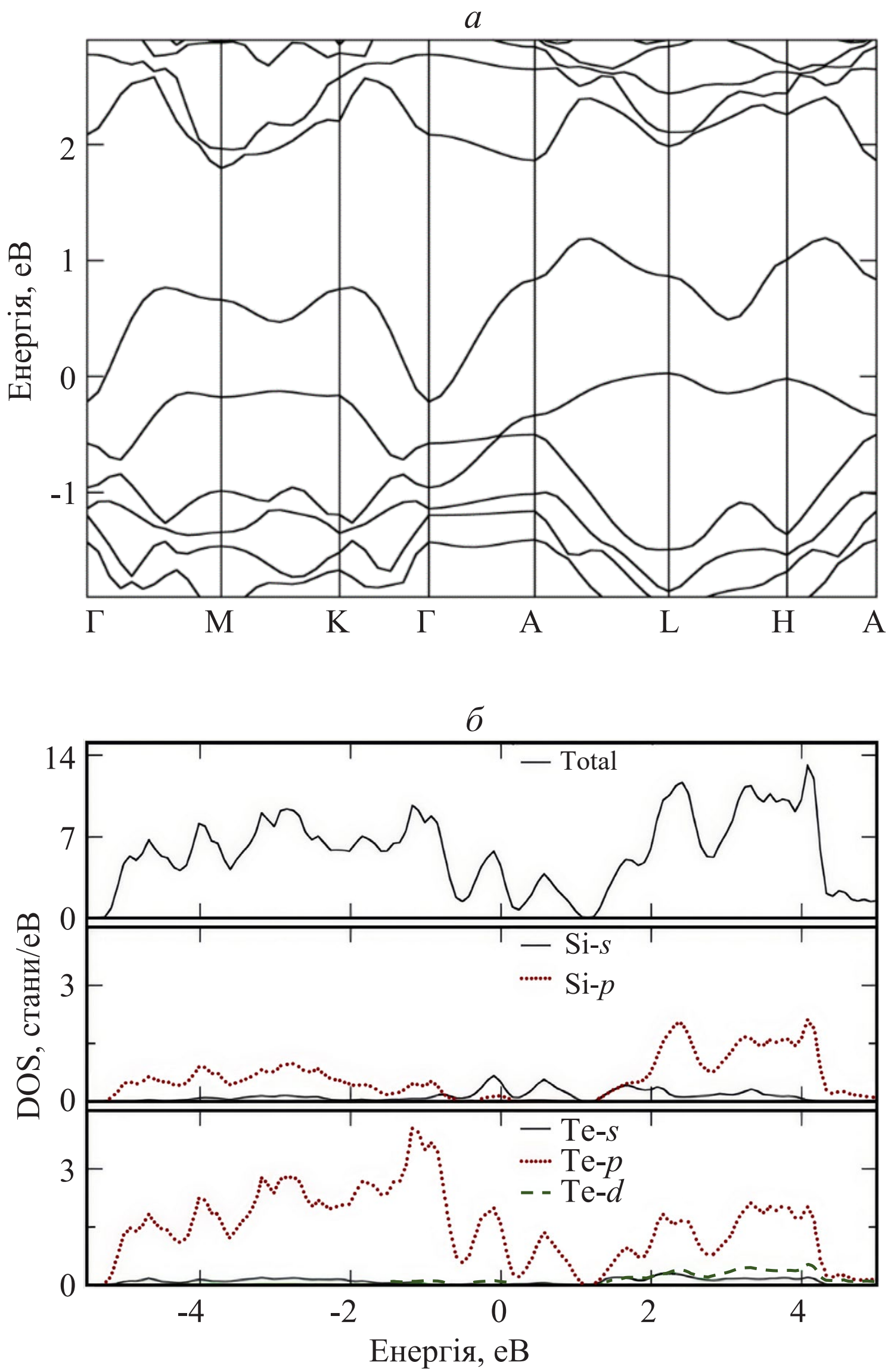

Рис. 3.6. Електронна структура (*а*), повна та парціальні густини електронних станів (*б*) розраховані методом HSE06 з урахуванням спін-орбітальної взаємодії металічної гексагональної фази М1[36]

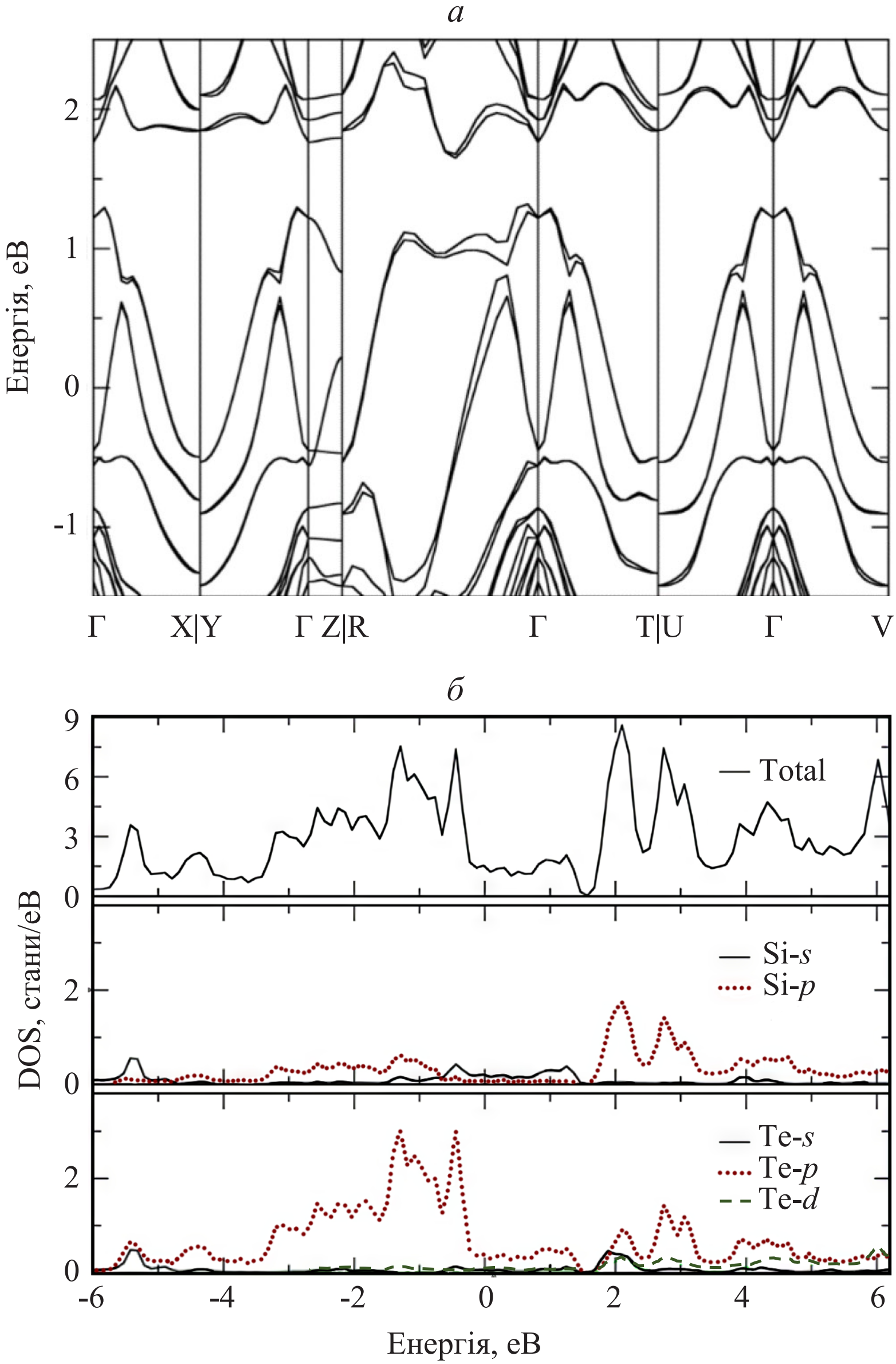

Рис. 3.7. Електронна структура (*а*), повна та парціальні густини електронних станів (*б*), розраховані методом HSE06 з урахуванням спін-орбітальної взаємодії, металічної гексагональної фази M2 $Si_2Te_3$ [36] мерів кремнію. Ці індуковані тиском, електричним полем та температурою фазові зміни роблять особливо цікавим дослідження інших можливих фаз $Si_2Te_3$, зокрема металічних. Провівши комп'ютерне дослідження $Si_2Te_3$ під високим тиском з використанням еволюційного алгоритму в поєднанні з першопринципними розрахунками теорії функціоналу густини (DFT) автори [36] встановили наявність двох металічних фаз M1 і M2, структура яких описана в главі 1. Фаза M1 має гексагональну кристалічну ґратку, тоді як M2 має триклінну кристалічну ґратку.

Розраховані авторами [36] електронні зонні структури, повні та локальні парціальні густини електронних станів першопринципним методом DFT (HSE06) з урахуванням спін-орбітальної взаємодії наведені на рис. 3.6 і 3.7 для M1 і M2 фаз відповідно. Із зонної структури фази M1 (рис. 3.6, *а*) видно, що в ній відсутня заборонена зона; це вказує на те, що дана фаза є металічною. Відповідні повна та парціальні густини електронних станів фази M1 приведені на рис. 3.6, *б*, з якого видно, що *p*-орбіталі атома телуру в основному формують валентну зону, тоді як *p*-орбіталі Si, і атома телуру дають основний внесок у зону провідності. Існують також невеликі внески від *s*-орбіталей атомів Si і Te, а також *d*-орбіталей атома Te. Подібні закономірності спостерігаються на рис. 3.7, *а* і рис 3.7, б для фази M2, де як зонна структура, так і густини електронних станів вказують на металеву природу.

## 3.4. КАРТИ РОЗПОДІЛУ ЕЛЕКТРОННОЇ ГУСТИНИ

Властивості речовини визначаються як просторовою, так і енергетичною електронними структурами. Важливою фізичною характеристикою кристала, аналіз якої дозволяє співвіднести хімічну будову молекули з локальними особливостями електронного розподілу, є електронна густина, тобто густина імовірності просторового розподілу електронів у кристалі. Перерозподіл електронної густини кристала в порівнянні з електронною густиною складових атомів відображає характер хімічного зв'язку і дозволяє судити про взаємний вплив атомів, а також про зміну характеру зв’язків у різних поліморфних формах тієї ж сполуки.

Для точного опису хімічного зв’язку необхідно знати загальну картину просторового розподілу електронної густини. На практиці це можливо реалізувати тільки в результаті дуже трудомістких експериментальних досліджень, наприклад, методом розсіювання рент-генівських променів. Тому широкого використання набули теоретичні розрахунки розподілу електронної густини в твердофазних сполуках, у тому числі й у сесквителуриді кремнію [140].

Метод функціонала густини дозволяє провести розрахунки розподілу електронної густини в кристалі. Зарядова густина *n*-ої зони задається наступним співвідношенням [146]:

$$\rho_{\mathbf{n}}(\mathbf{r}) = \mathbf{e}\sum_{\mathbf{k}}^{\mathbf{ЗБ}} |\Psi_{\mathbf{n,k}}(\mathbf{r})|^2 , \tag{3.3}$$

де $\Psi_{\mathbf{n,k}}(\mathbf{r})$ – хвильова функція стану **k** у зоні *n*, а *e* – заряд електрона. Повна густина заряду валентних електронів визначається як

$$\rho(\mathbf{r}) = \sum_{\mathbf{n}} \rho_{\mathbf{n}}(\mathbf{r}) , \tag{3.4}$$

де сумування ведеться по валентних зонах.

Оскільки густина електронного заряду ρ(**r**) – функція у тривимірному просторі, то зображати результати зручніше у вигляді топографічних карт, на яких для певних перерізів тривимірного простору задаються лінії рівня електронної густини ρ(**r**) = const [146] (ρ(**r**) зображають в одиницях e/V, де V – об’єм елементарної комірки).

Враховуючи те, що основною структурною одиницею кристала $Si_2Te_3$ є октаедр [$Si_2Te_6$], утворений двома тригональними пірамідами [$SiTe_3$], з’єднаними вершинами з атомів кремнію, у цьому випадку найбільш зручно представити контурні карти ρ(**r**) у площинах, які проходять через два атоми телуру і один атом кремнію (рис. 3.8, *а*), вздовж димера Si–Si і двома атомами телуру (рис. 3.8, *б*), у площині *ac*, перпендикулярній двом тришаровим пакетам Te–Si–Te (рис. 3.8, *в*), а також у площині, яка проходить через основу тригональної піраміди [$SiTe_3$], тобто атомарний шар телуру (рис. 3.8, *г*), в октаедрі [$Si_2Te_6$]. Суцільні лінії на контурних картах описують поверхні зі сталою електронною густиною, а густина ліній на рисунку характеризує градієнт електронної густини. При формуванні хімічного зв’язку відбувається перерозподіл електронної густини між взаємодіючими атомами, що наглядно видно із наведених карт електронної густини, при цьому більша густина відображає сильніший міжатомний зв'язок.

З наведених контурних карт видно, що розподіл електронної густини в $Si_2Te_3$ характеризується наявністю спільних контурів електронних оболонок атомів у вказаних структурних одиницях, причому електронні стани атомів телуру займають помітно більшу частину міжатомного простору, ніж внески атомів кремнію. Лінії

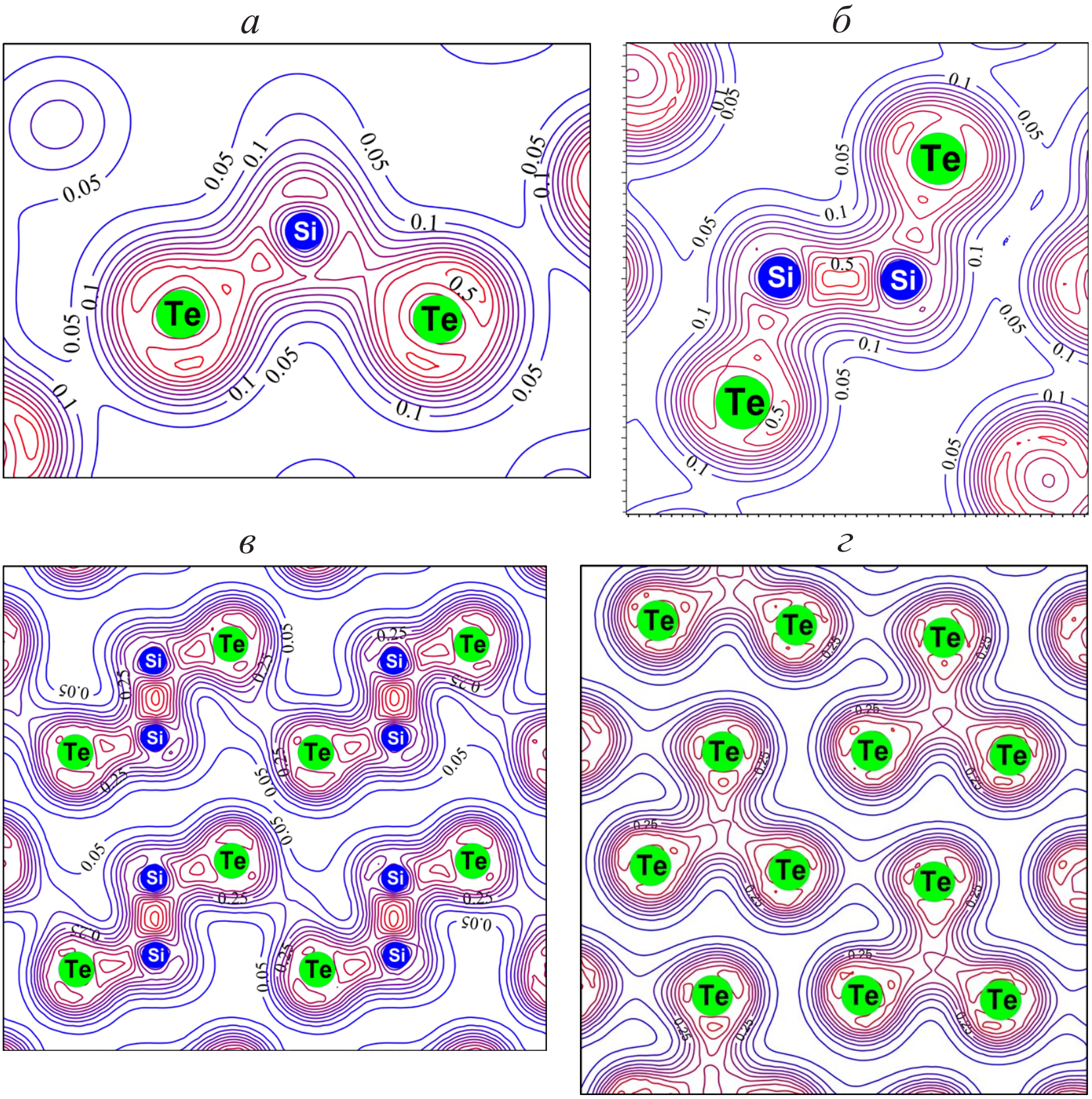

Рис. 3.8. Карти розподілу електронної густини в $Si_2Te_3$ [140].

описують поверхні з сталою електронною густиною, а густина ліній на рисунках характеризує градієнт електронної густини.

Спільні контури ρ(**r**), що охоплюють атоми кремнію і телуру в октаедрах [$Si_2Te_6$], вказують на існування ковалентної складової хімічного зв'язку, за формування якої відповідає гібридизація Si 3*s*-, 3*p*- і Te5*s*-, 5*p*-станів. Поляризація зарядової густини в напрямку Si→Te вказує на наявність окрім ковалентної ще й іонної складової зв’язку. Характерною особливістю хімічного зв’язку в $Si_2Te_3$ є наявність спільних контурів ρ(**r**) між трьома атомами Te в атомному шарі теллуру (рис. 3.8, *г*), що належать окремому октаедру [$Si_2Te_6$], що не властиве іншим шаруватим кристалам, які кристалізуються у структурі типу $CdI_2$, наприклад, $SnSe_2$ [147].

Із рис. 3.8, *б*, *в* видно, що основний максимум електронної густини знаходиться посередині на зв'язках Si–Si у димерах, так само як у кристалічному Si, тобто зв'язок між катіонами носить яскраво виражений ковалентний характер.

Сильна анізотропія оптичних і електричних властивостей шаруватих кристалів $Si_2Te_3$ стає зрозумілою із наведеної на рис. 3.8, *в* карти розподілу густини валентних електронів, проведеної у площині, яка пересікає два тришарові пакети. Електронна густина всередині тришарових пакетів (рис. 3.8, *в*), що відображає хімічний зв'язок атомів кремнію з найближчими сусідами (Te) в октаедрах [$Si_2Te_6$], значно вища, ніж на їх межах. Водночас не спостерігається спільних ліній рівня $\rho(\mathbf{r})$ для сусідніх атомів телуру, що належать двом різним тришаровим пакетам, що свідчить про слабке перекриття їх хвильових функцій. Така просторова анізотропія електронної густини і енергетичного розподілу електронних 5*p*-станів телуру є причиною квазідвовимірності бінарної сполуки $Si_2Te_3$.

Таким чином, характер розподілу електронної густини в $Si_2Te_3$ вказує на змішаний іонно-ковалентний тип зв'язку в тришарових пакетах і наявність слабкої ван-дер-ваальсової складової зв'язку між пакетами.

## 3.5. СПЕКТРИ ФУНДАМЕНТАЛЬНОГО ПОГЛИНАННЯ СЕСКВІТЕЛУРИДУ КРЕМНІЮ

При дослідженні фізичних властивостей напівпровідників у першу чергу звертають увагу на можливість одержання інформації про їх енергетичну структуру шляхом аналізу частотної залежності коефіцієнта поглинання $\alpha$ на краю власного поглинання [148].

Вперше спектри пропускання тонких шаруватих кристалів $Si_2Te_3$ товщиною 2–50 мкм в інтервалі температур 77–300 K були досліджені авторами [149]. З аналізу краю фундаментального поглинання (рис. 3.9) визначено непряму ширину забороненої зони $E_{gi}$ = 1.98 еВ при $T$ = 300 K та її температурний коефіцієнт $dE_g/dT$= $6.5\cdot10^{-4}$ еВ·K$^{-1}$. За інтерференційними смугами у спектрах пропускання десятка тонких зразків товщиною від 38 до 3 мкм визначено дисперсію показника заломлення $n$ (рис. 3.10, крива 1), а за допомогою формули $R = [(1 - n) / (1 + n)]^2$ розраховано дисперсію коефіцієнта відбивання (рис. 3.10, крива 2).

Спектри оптичного поглинання і відбивання кристалів $Si_2Te_3$, вирощених методом сублімації, в області краю власного поглинання

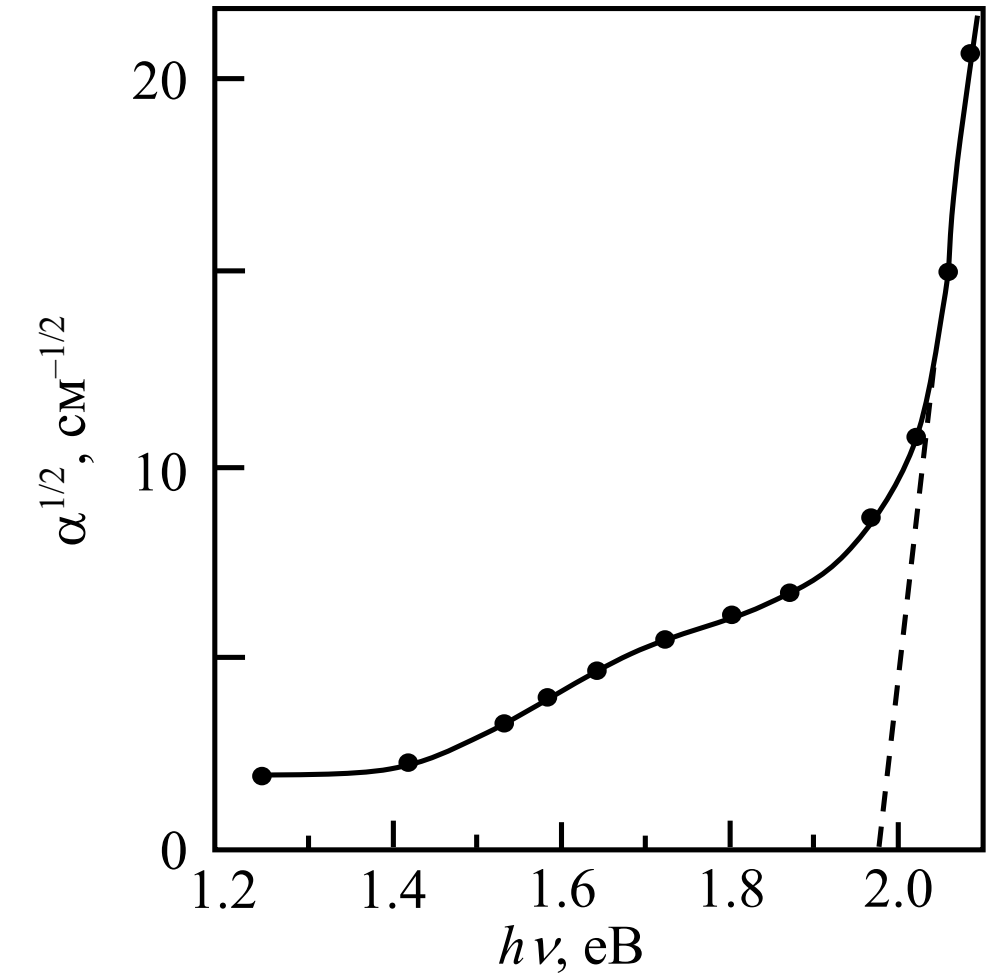

Рис. 3.9. Спектр поглинання кристала $Si_2Te_3$ [149].

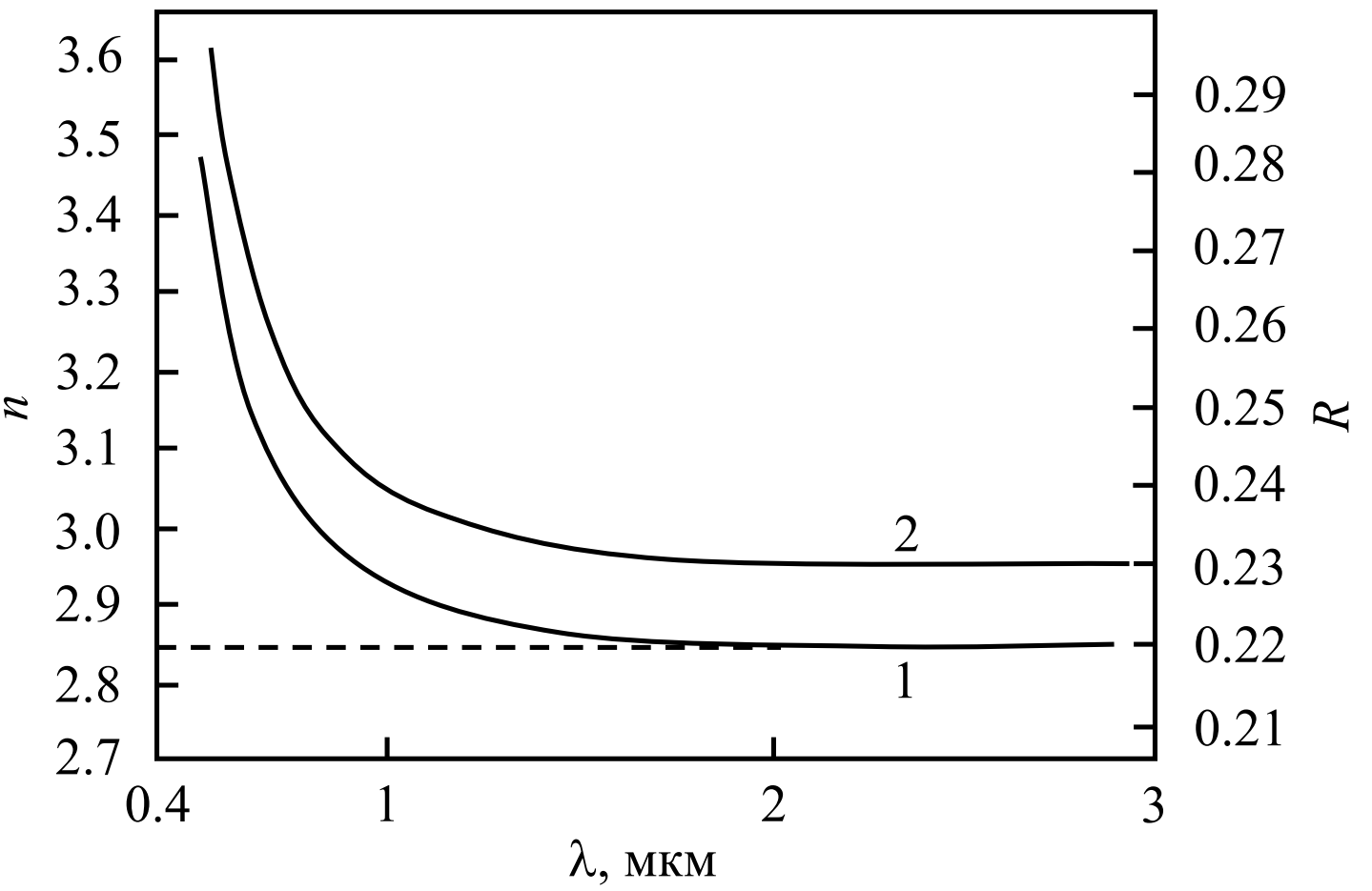

Рис. 3.10. Дисперсія показника заломлення *n* (1) та коефіцієнта відбивання *R* (2) для кристала $Si_2Te_3$[149].

($1.7 \le \hbar\omega \le 2.3$ еВ) і в інтервалі температур $104 \le T \le 476$ К, дослідили також автори [150]. На залежності коефіцієнта поглинання $\alpha(\hbar\omega)$ (рис. 3.11) можна виділити дві області: при більших $\hbar\omega$ спектральна залежність коефіцієнта поглинання $\alpha$ описується правилом Урбаха (3.6), і згідно [150] обумовлена уширенням екситонної лінії у внутрішніх розупорядкованих локальних електричних полях, зв'язаних з наявними дефектами. В області менших енергій особливості спектрів $\alpha(\hbar\omega)$ пояснюються поглинанням на домішкових центрах. Визначено температурний коефіцієнт ширини забороненої зони (при $\alpha = 500$ см$^{-1}$) рівний $dE_g/dT = -9.3\cdot10^{-4}$ еВ/К.

Оскільки шаруваті кристали $Si_2Te_3$ легко сколюються уздовж площини спайності, це дозволяє отримувати тонкі плоскопаралельні пластинки і дослідити на них ефект Франца-Келдиша. На зразках товщиною 20 – 30 мкм з напівпрозорими напиленими золотими електродами авторами [150] досліджено даний ефект в електричних полях до 170 кВ/см на частоті 220 Гц. Наявний «червоний» зсув краю поглинання (рис. 3.12) має квадратичну залежність від поля. Використовуючи відому формулу Франца для приведеної ефективної маси електронів і дірок:

$$\boldsymbol{\mu} = \mathbf{6.4\cdot10^{-17}}\ \frac{\mathbf{еВ^3}}{\mathbf{(В/см)}^{\ 2}}\ \frac{\boldsymbol{S^2F^2}}{\Delta\boldsymbol{E}}\boldsymbol{m}_0, \qquad (3.5)$$

де $S = d\ ln\ \alpha/dE$, $F$ – електричне поле, $\Delta E$ – «червоний» зсув, $m_0$ – маса вільного електрона, автори [150] оцінили значення приведеної ефективної маси вздовж осі $c$ : $\mu = 2.7\ m_0$; ефективні маси електронів

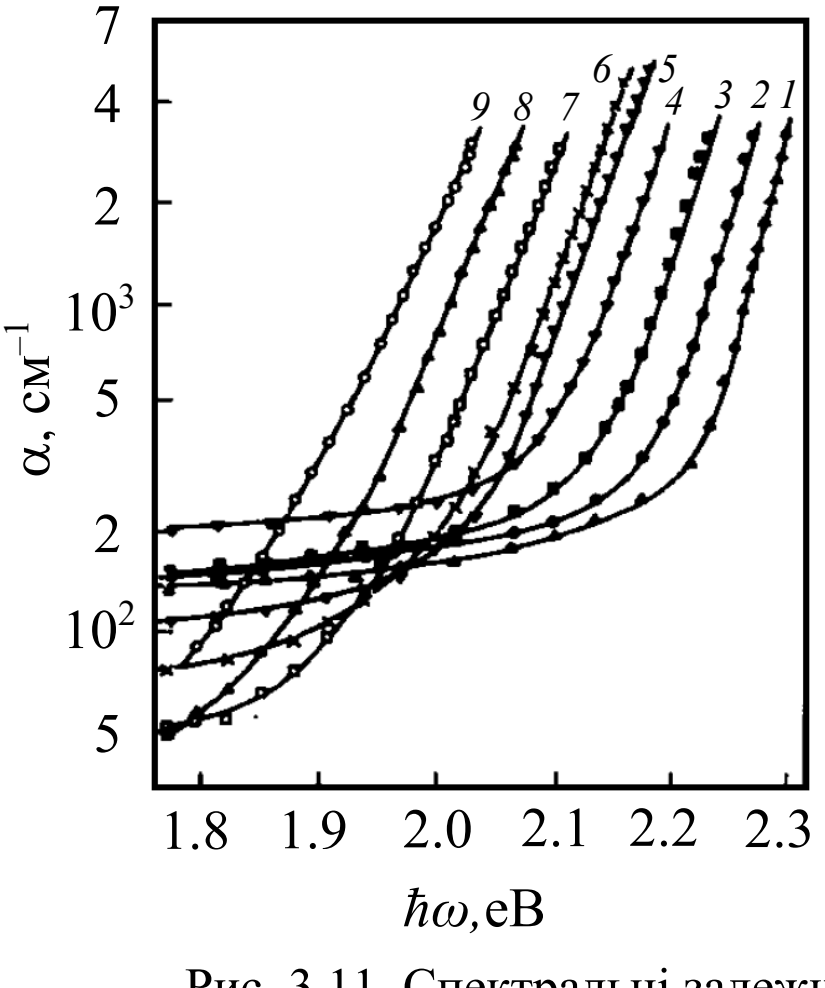

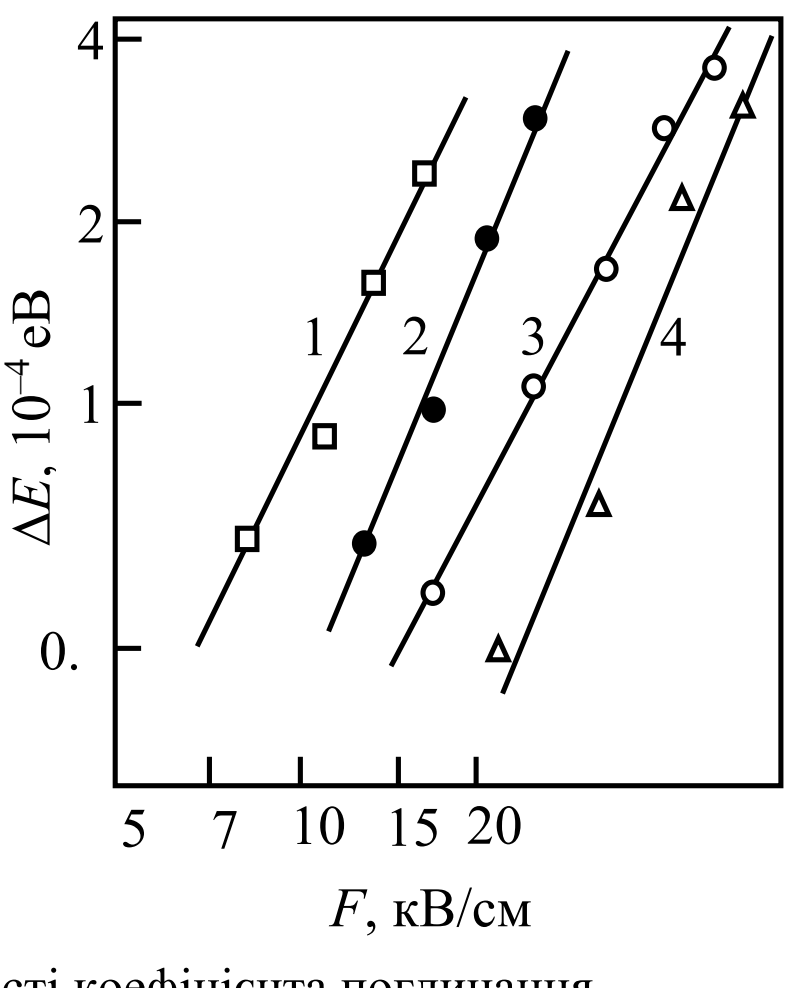

Рис. 3.11. Спектральні залежності коефіцієнта поглинання кристала $Si_2Te_3$ при різних температурах. $T$, К: 1 – 104; 2 – 150; 3 – 200; 4 – 250; 5 – 298; 6 – 323; 7 – 373; 8 – 426 ; 9 – 476 [150].

Рис. 3.12. Польове розширення краю поглинання при різних енергіях кванта $E$, еВ: 1 – 2.084; 2 – 2.108; 3 – 2.121; 4 – 2.126. $F$ – шкала дійсна лише для $E = 2.084$ еВ [150].

і дірок $m_{e,h}$ = 5.4 $m_0$ та ефективну густину електронних станів: $N = 2{\cdot}10^{20}$ см$^{-3}$.

Дослідження краю власного поглинання шаруватих кристалів $Si_2Te_3$, вирощених методом сублімації, проводили також автори [140] на зразках різної товщини ($d$ = 20 – 400 мкм) у широкому інтервалі температур (80 – 293 K). Світловий пучок поширювався вздовж нормалі до площини шарів кристалічної структури. Спектральні залежності коефіцієнта поглинання шаруватих кристалів $Si_2Te_3$ наведені на рис. 3.13. Коефіцієнт поглинання розраховувався за стандартною методикою двох товщин [151]. На експериментальних спектрах $\alpha = f(h\nu)$ можна виділити дві характерні ділянки, формування яких зумовлене різними механізмами взаємодії світла з кристалічною ґраткою $Si_2Te_3$. На першій довгохвильовій ділянці, коефіцієнт поглинання $\alpha$ слабо залежить від енергії фотонів, і для різних зразків знаходиться у межах 30–100 см$^{-1}$. Цю довгохвильову ділянку, як правило, зв'язують із наявністю статичних дефектів ґратки різної природи (неконтрольованих залишкових домішок, пор, дислокацій, тріщин тощо) [152, 153].

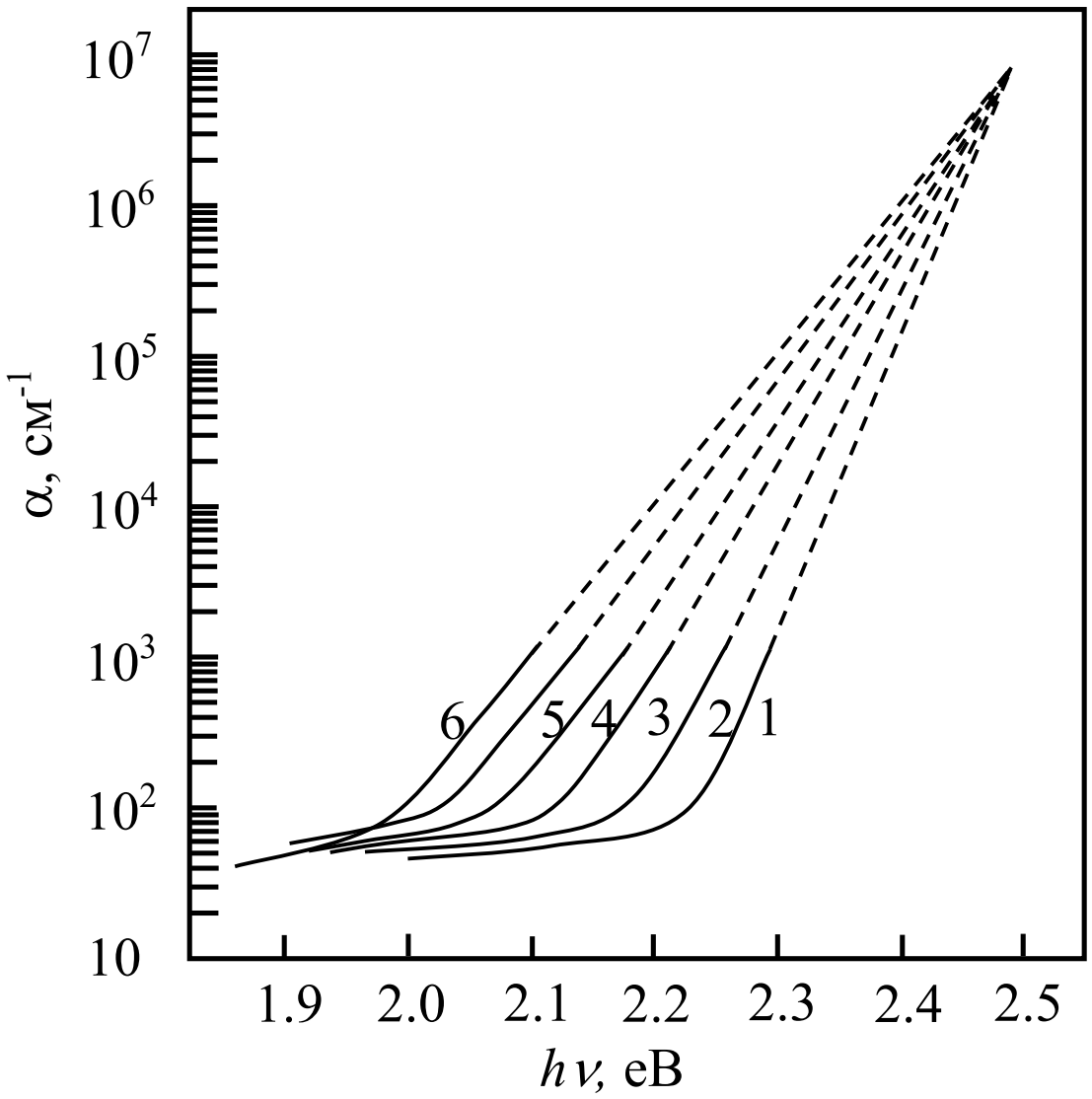

Рис. 3.13. Спектральна залежність крайового поглинання кристала $Si_2Te_3$ при різних температурах. $T$, K: 1 – 80; 2 – 100; 3 – 150; 4 – 200; 5 – 250; 6 – 293 [140].

На другій ділянці, в інтервалі значень коефіцієнта поглинання $80 - 10^3$ см$^{-1}$, спектральна залежність краю власного поглинання $Si_2Te_3$ описується правилом Урбаха. У випадку урбахівської поведінки краю поглинання температурно – спектральна залежність коефіцієнта поглинання описується співвідношенням [154, 156]:

$$\alpha(\hbar\omega,T) = \alpha_0 \cdot \exp\left[\frac{\sigma(\hbar\omega - E_0)}{kT}\right] = \alpha_0 \cdot \exp\left[\frac{(\hbar\omega - E_0)}{E_U(T)}\right], \qquad (3.6)$$

де $E_U$ – урбахівська енергія, яка рівна енергетичній ширині краю поглинання та є величиною, оберненою до нахилу краю поглинання $E_U = \Delta(ln\ \alpha) / \Delta(\hbar\omega)$; $\alpha_0$ та $E_0$ – параметри, які є координатами збіжності урбахівського «віяла», $\sigma$ – параметр крутизни краю.

На рис. 3.13 видно, що високоенергетична ділянка спектра крайового поглинання кристалічного $Si_2Te_3$ у досліджуваному інтервалі температур (80 – 293 K) утворює характерне температурне «віяло», з координатами збіжності $\alpha_0 = 2\cdot10^{-7}$ см$^{-1}$ і $E_0 =$ 2.55 еВ. З підвищенням температури зразка від 80 до 293 K край власного поглинання зміщується в довгохвильову область спектра (рис. 3.13), що є відображенням зменшення ширини забороненої зони зі збільшенням температури кристала. Температурний коефіцієнт зміни $E_g$ складає $dE_g/dT = -0.9\cdot10^{-3}$ еВ/K. Близьке значення $dE_g/dT = -0.93\cdot10^{-3}$ еВ/K приводять автори [150] для кристалів $Si_2Te_3$, вирощених методом Бріджмена. Температурна залежність нахилу краю поглинання описується виразом [156]:

$$\sigma(T) = \sigma_0 \cdot \left(\frac{2kT}{\hbar\omega_\phi}\right) \cdot th\left(\frac{\hbar\omega_\phi}{2kT}\right), \qquad (3.7)$$

де $\sigma_0$ – параметр, зв’язаний з константою екситон (електрон)-фононної взаємодії (ЕФВ) g співвідношенням $\sigma_0 = (2/3)g^{-1}$; $\hbar\omega_ф$ – характерна енергія фононів, які найбільш ефективно взаємодіють з електронами (екситонами). Для більшості кристалів $\hbar\omega_ф$ близьке до енергії найбільш високоенергетичного LO-фонона [156]. Аналіз критерію Тоядзави [155] вказує на те, що в кристалах $Si_2Te_3$ має місце сильна електрон-фононна взаємодія ($\sigma_0 < 0.61 < 1$), яка проявляється у виникненні локалізованих екситонів.

Із аналізу спектрів крайового поглинання (рис. 3.13) авторами [140] визначено значення ефективної частоти фононів $\hbar\omega_ф$ = 38.8 меВ (313 см$^{-1}$), порівняння якої з реальними значеннями частот коливань кристалічної ґратки $Si_2Te_3$ дозволяє з'ясувати, якого саме типу фонони приймають участь у формуванні краю власного поглинання. Отримане значення параметра $\hbar\omega_ф$ = 38.8 меВ (313 см$^{-1}$) близьке до енергії повздовжніх оптичних LO-фононів 335 см$^{-1}$, які проявляються в КР- та ІЧ-спектрах $Si_2Te_3$ [157]. Це характерно для більшості кристалів, в яких край поглинання описується правилом Урбаха. Таким чином, експоненціальна форма краю власного поглинання кристалів $Si_2Te_3$ визначається не тільки впливом заряджених домішок, але й поздовжніми оптичними LO-фононами.

У зв'язку з тим, що непрямі оптичні переходи в кристалах $Si_2Te_3$ маскуються довгохвильовими урбахівськими «хвостами» поглинання, визначення значення ширини забороненої зони є складним [156].

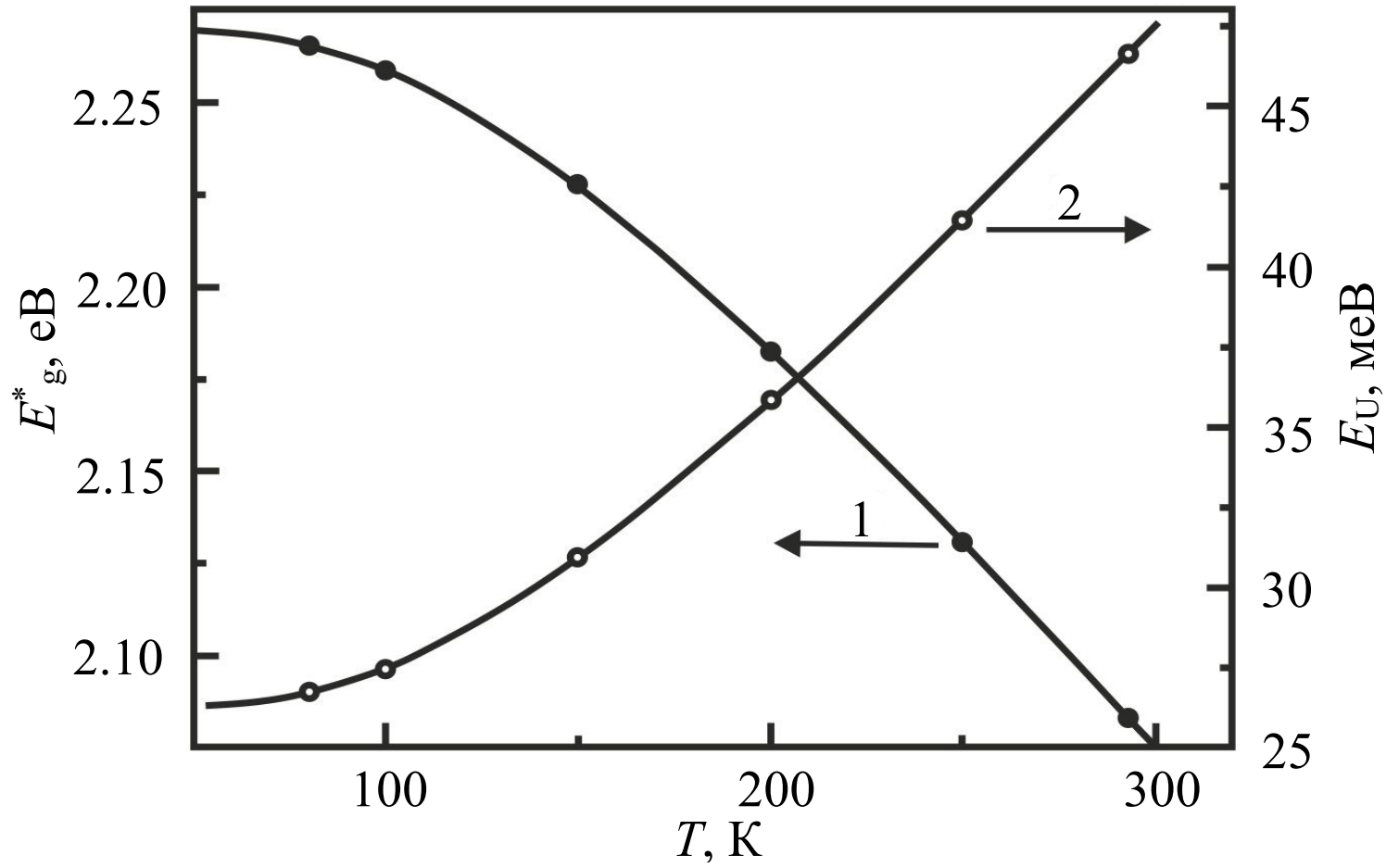

Рис. 3.14. Температурні залежності оптичної псевдощілини $E^*_g$ (1) і урбахівської енергії $E_U$ (2) для кристала $Si_2Te_3$ [140].

У такому випадку часто за значення оптичної псевдощілини $E^*_g$ приймають енергію, яка відповідає енергетичному положенню краю поглинання при фіксованому рівні поглинання α = 10$^3$ см$^{-1}$. Температурна поведінка, визначеної таким чином ширини оптичної псевдощілини $E^*_g$ для кристала $Si_2Te_3$ (рис. 3.14) в рамках моделі Ейнштейна описується співвідношенням [158]:

$$E^*_g(T) = E^*_g(0) - S^*_g k\theta_E \left[\frac{1}{\exp(\theta_E / T) - 1}\right] \quad (3.8)$$

де $E^*_g(0)$ і $S^*_g$ – відповідно ширина оптичної псевдощілини при 0 К і безрозмірна константа взаємодії; $\theta_E$ – температура Ейнштейна, яка відповідає усередненій частоті фононних збуджень системи невзаємодіючих осциляторів. Отримані при описі залежностей $E_g^*(T)$ параметри $E_g^*(0)$, $S_g^*$ і $\theta_E$ для кристала $Si_2Te_3$ наведені в табл. 3.1

Таблиця 3.1. Параметри урбахівського краю поглинання та ЕФВ для кристалів $SiTe_2$ і $Si_2Te_3$

| **Кристал** | **$SiTe_2$** | **$Si_2Te_3$** |
|---|---|---|
| $E_g^*$ (293 К), еВ | | 2.083 |
| $E_U$ (293 К), меВ | | 46.6 |
| $\alpha_0$, см$^{-1}$ | $1.4431\times10^6$ | $9.6\times10^6$ |
| $E_0$, еВ | 2.47 | 2.51 |
| $\sigma_0$ | 0.55563 | 0.61 |
| $\hbar\omega_ф$, меВ | 27.09 | 38.8 |
| $\theta_E$, К | | 381 |
| $(E_U)_0$, меВ | | 26.3 |
| $(E_U)_1$, меВ | | 54.4 |
| $E_g^*(0)$ , еВ | | 2.27 |
| $S_g^\alpha$ | | 15.2 |

Незважаючи на те, що в даний час не існує єдиної універсальної інтерпретації правила Урбаха, не викликає сумнівів те, що експоненціальна форма урбахівського краю поглинання зумовлена розупорядкуванням. У випадку кристалів це динамічне (температурне) розупорядкування, джерелом якого виступає зумовлена ґратковими коливаннями електрон-фононна взаємодія, та статичне (структурне) розупорядкування викликане дрібномасштабними порушеннями періодичного потенціалу кристалічної ґратки із-за наявності у кристалі точкових заряджених дефектів [159–161]. Внесок кожного із цих факторів залежить від концентрації заряджених домішок у конкретному зразку та його температури, яка визначає концентрацію рівноважних фононів. При пониженні температури фонони «виморожуються», але хвости коефіцієнта поглинання не зникають. Їхнє існування пов'язане із неоднорідністю кристала, викликаною саме наявністю власних точкових дефектів. У випадку кристалів $Si_2Te_3$ це, перш за все, велика концентрація стехіометричних катіонних вакансій і дефектів упаковки.

Мірою ступеня розмиття урбахівського краю поглинання, і відповідно, мірою ступеня розупорядкування кристалічної гратки [162], може служити урбахівська енергія $E_U$ (рис. 3.14), яка, як вказувалося вище, визначається динамічним (температурним) і статичним (структурним) розупорядкуванням [159, 163]:

$$E_U = (E_U)_X + (E_U)_T \ , \qquad (3.9)$$

де $(E_U)_X$ і $(E_U)_T$ – відповідно внески структурного (статичного) і температурного (динамічного) розупорядкування в $E_U$, які вважаються незалежними, еквівалентними й адитивними. Для розділення внесків різних типів розупорядкування в $E_U$ застосовувалася методика, запропонована авторами [156]. При цьому використовувалося відоме співвідношення, яке добре описує температурну залежність урбахівської енергії $E_U$ в рамках моделі Ейнштейна [159, 160]:

$$(E_U) = (E_U)_0 + (E_U)_1 \cdot \left[ \frac{1}{\exp(\theta_E / T) - 1} \right], \qquad (3.10)$$

де $(E_U)_0$ й $(E_U)_1$ – постійні величини. Значення параметрів $(E_U)_0$ і $(E_U)_1$, отриманих при описі експериментальних температурних залежностей $E_U$ співвідношенням (3.10), наведені в табл. 3.1. Порівнюючи співвідношення (3.9) і (3.10), знайдено значення $(E_U)_T = 20.3$ меВ (становить 43.6 % від $E_U$), $(E_U)_X = 26.3$ меВ (становить 56.4 % від $E_U$) при $T = 293$ К.

Слід зазначити, що експоненціальна форма краю поглинання може бути зумовлена деформацією ґратки, викликаної рухом фононів або напруженнями, що виникають через присутність дефектів, непружним розсіюванням носіїв фононами, внутрішніми електричними полями в кристалі, а також локалізованими енергетичними рівнями, які формують домішкову зону, що перекривається з дозволеною зоною.

## 3.6. ЕЛЕКТРОННА СТРУКТУРА $SiTe_2$

**3.6.1. Зонна структура і природа електронних станів тригонального $SiTe_2$.** Електронна структура об'ємного $SiTe_2$, розрахована нами методом DFT з використанням гібридного функціоналу HSEO6 в зоні Бріллюена (рис. 3.1) гексагональної ґратки, наведена на рис.

3.15. Початок відліку енергії відповідає енергії верхнього зайнятого рівня.

За результатами розрахунку дителурид кремнію є непрямозонним напівпровідником з вершиною валентної зони в напрямку Г–К і дном зони провідності в напрямку M–L зони Бріллюена і розрахованою шириною забороненої зони $E_{gi}$ = 1.2 еВ.

Шаруватий характер кристала $SiTe_2$ знаходить своє відображення у структурі енергетичного спектра. Спостерігається значна анізотропія закону дисперсії для окремих зон вздовж і поперек тришарових пакетів. Так, у напрямку Г→M та Г→K (вздовж тришарового пакету) дисперсія більшості верхніх валентних зон перевищує у напрямку Г→A (поперек шарів). Слабка дисперсія $E(\boldsymbol{k})$ вздовж осі $c$ кристала $SiTe_2$, тобто вздовж напрямку Г→Δ→A, перпендикулярно до моношарів, утворених атомами Si та Te, свідчить про шаруватий характер даної сполуки та відносно слабкий вплив взаємодії між тришаровими пакетами на електронну структуру. Помітна дисперсія зон вздовж напрямків, паралельних тришаровим пакетам вказує на сильну взаємодію у структурних одиницях [$SiTe_6$], що формують «сендвічі».

У валентній зоні $SiTe_2$, із загальною шириною 11.77 еВ, наявні вісім заповнених енергетичних зон, які утворюють три зв'язки зон. Інформацію про внески атомних орбіталей в кристалічні стани $SiTe_2$ дають розрахунки повної й локальних парціальних густин електронних станів, наведених на рис. 3.16. У валентній зоні дителуриду кремнію переважають парціальні 5*s*- і 5*p*-стани телуру, причому їх енергетичні положення істотно різняться.

Зони з переважаючими внесками станів телуру можна розділити на три типи. Нижня зв'язка з двох низькоенергетичних зон, що формує дно валентної зони в енергетичному інтервалі від –11.77 еВ до –10.27 еВ, утворена переважно 5*s*-станами телуру з незначним домішуванням 3*s*- і 3*p*-станів кремнію. Вище неї на відстані 3.12 еВ розташована сильно дисперсна відокремлена зона, утворена гібридизованими Te5*p*, 5*s*- Si3*s*-станами. Верхня зв'язка з п'яти заповне них зон шириною 4.15 еВ має змішаний характер за участю гібридизованих 5*p*-станів телуру і 3*p*-станів кремнію.

Верхня частина цієї заповненої зони, з шириною 1.5 еВ, має переважно аніонний характер, і $p_z$-стани телуру формують її вершину в напрямку Г–К. Валентний спектр в енергетичному інтервалі від –4.15 до верха валентної зони визначається Te5*p*-станами, що утворюють чотири $\sigma(p_{x,y})$- і дві $\pi(p_z)$-зони. Їх дисперсійні залежності $E$(k)

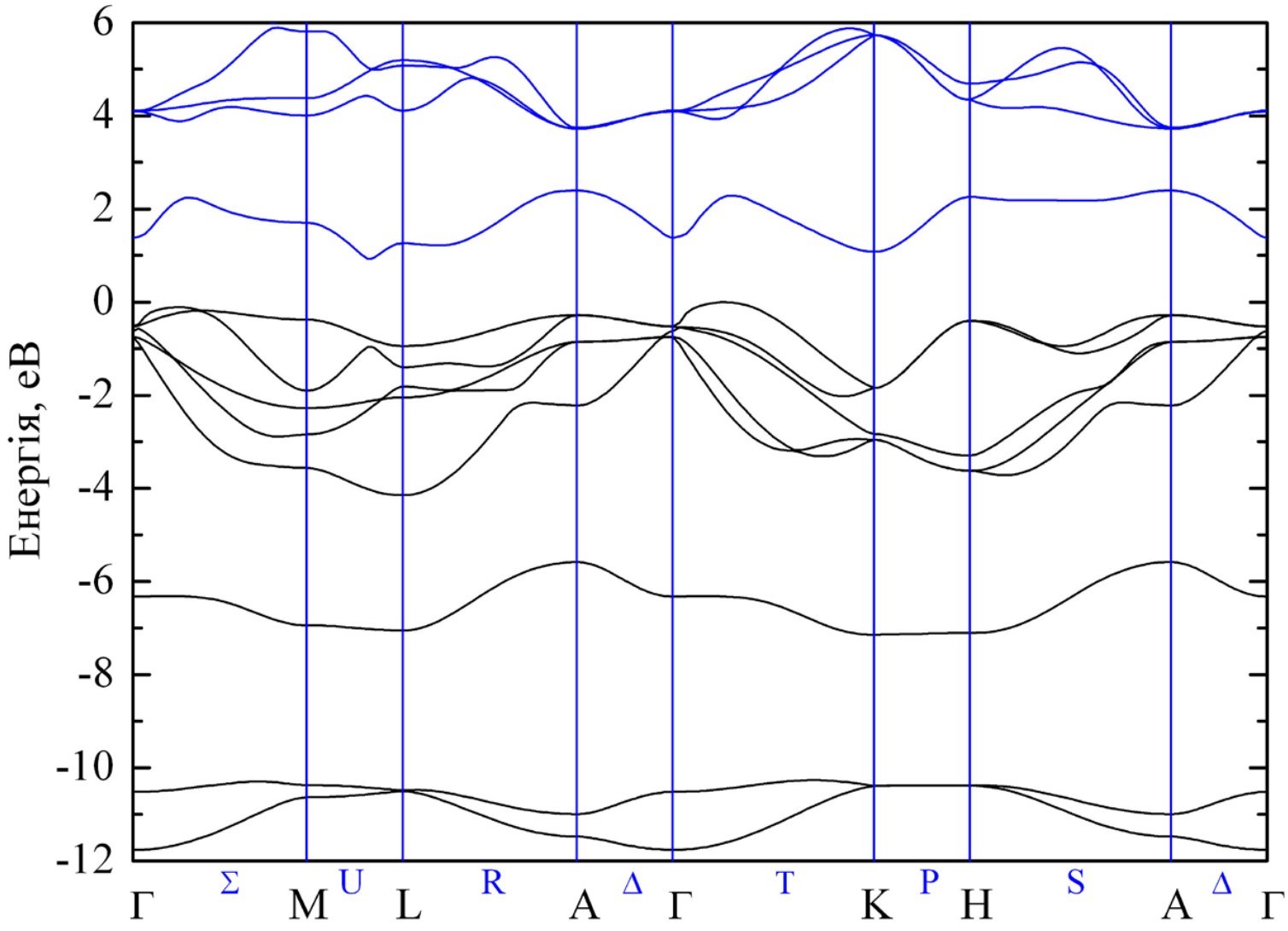

Рис. 3.15. Електронна структура $SiTe_2$.

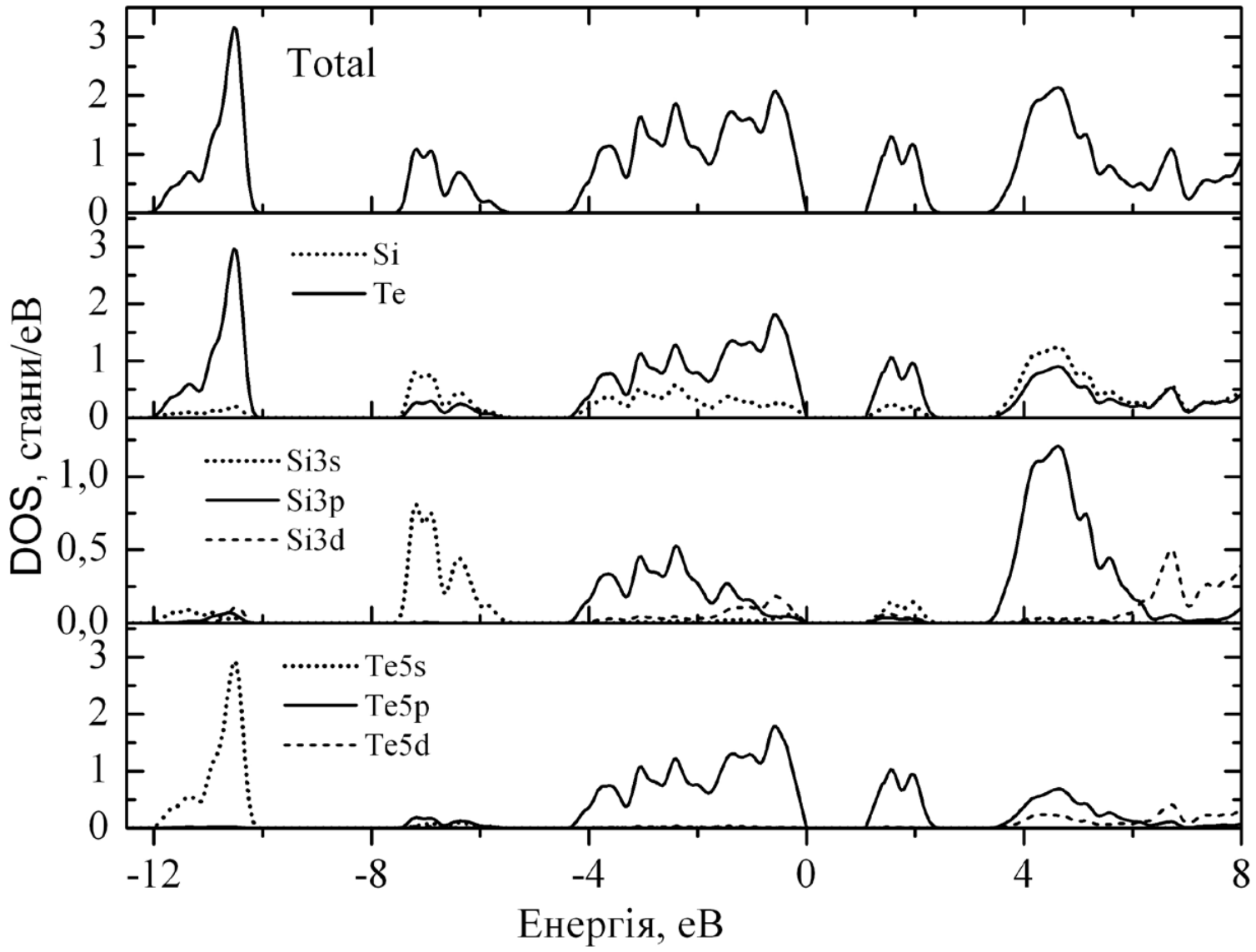

Рис. 3.16. Повна і парціальні густини електронних станів $SiTe_2$.

істотно різняться. Так, для Te5$p_{x,\ y}$-зони дисперсія $E$(k) максимальна в напрямку $k_{x,y}$ (Г–К). Ці зони відображають розподіл станів телуру в атомних шарах телуру, та мають квазідвовимірний (2D) тип, формують близькі до плоских ділянки в напрямку $k_z$ (Г–А). Te 5$p_z$-подібні стани (3D-типу) орієнтовані перпендикулярно атомним шарам телуру і відповідальні за слабкі міжшарові $\pi$-зв'язки. Si $s$-, $p$-стани домішуються до системи Te 5$p$-подібних зон.

Характерною особливістю електронного спектра тригонального $SiTe_2$ є наявність нижньої незайнятої зони, відокремленої забороненим енергетичним інтервалом (1.33 еВ) від наступних незайнятих зон. Ця ізольована незайнята зона має суттєву дисперсію і містить внески вільних Te $p$- і Si $s$-станів.

**3.6.2. Електронна структура прогнозованої триклінної фази $SiTe_2$.** Використовуючи еволюційний алгоритм та першопринципні розрахунки, автори [164] прогнозували нову шарувату структуру дителуриду кремнію (рис. 3.17), яка є більш стабільною, ніж фаза

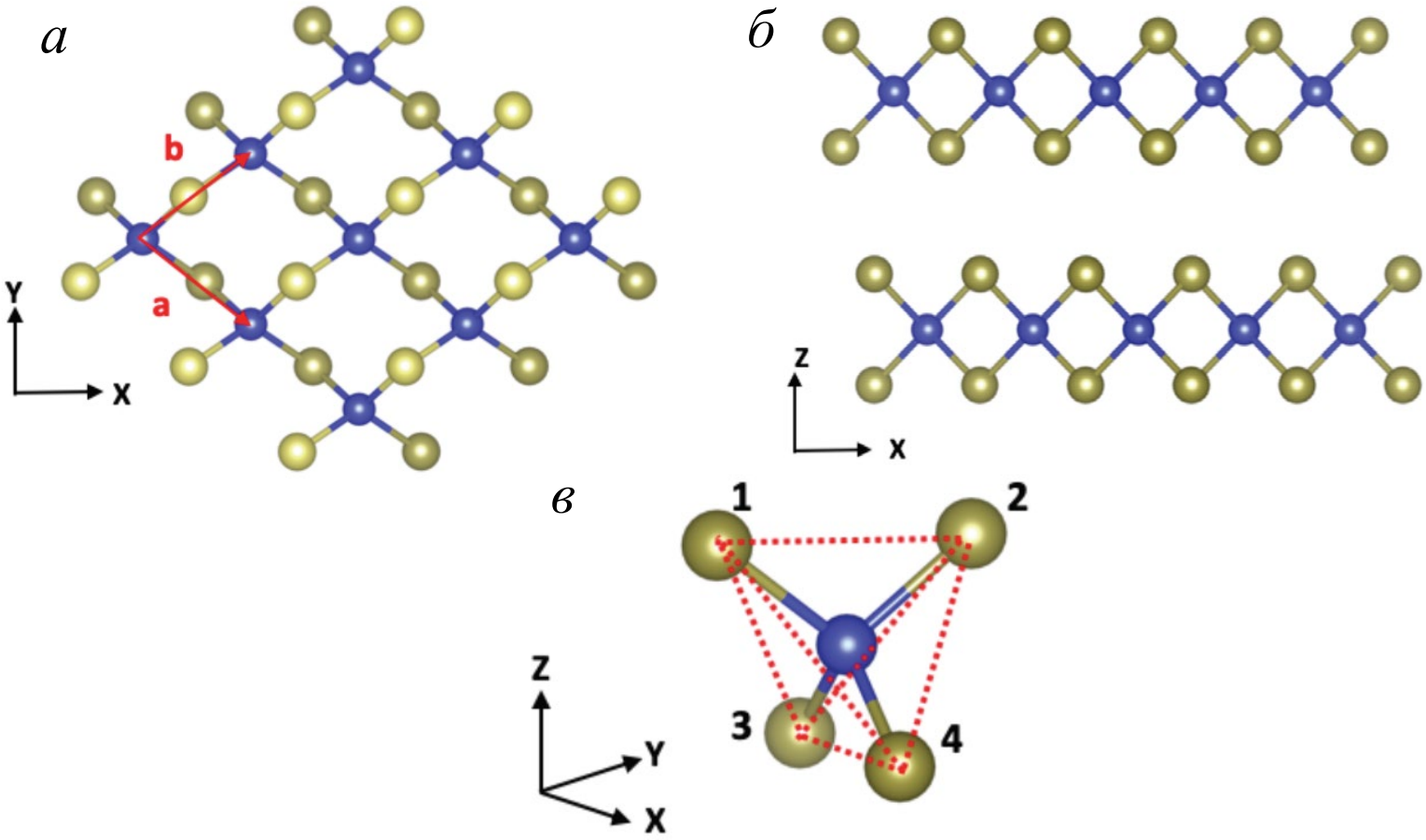

Рис.3.17 Проекції кристалічної структури триклінного $SiTe_2$ на площини XY (*а*) і XZ (*б*); (*в*) тетраедр [$SiTe_4$] [164].

типу $CdI_2$. Прогнозована структура має триклінну кристалічну гратку з просторовою групою P1 і параметрами гратки $a = b = 3.945$ Å, $c = 7.076$ Å, $\alpha = 83.68°$, $\beta = 83.77°$, $\gamma = 75.98°$. Примітивна елементарна комірка складається із трьох атомів: один атом Si і два атоми Te. Прогнозована триклінна структура $SiTe_2$ є більш стабільною, оскільки основною структурною одиницею є тетраедр [$SiTe_4$] (рис. 3.17, *в*),

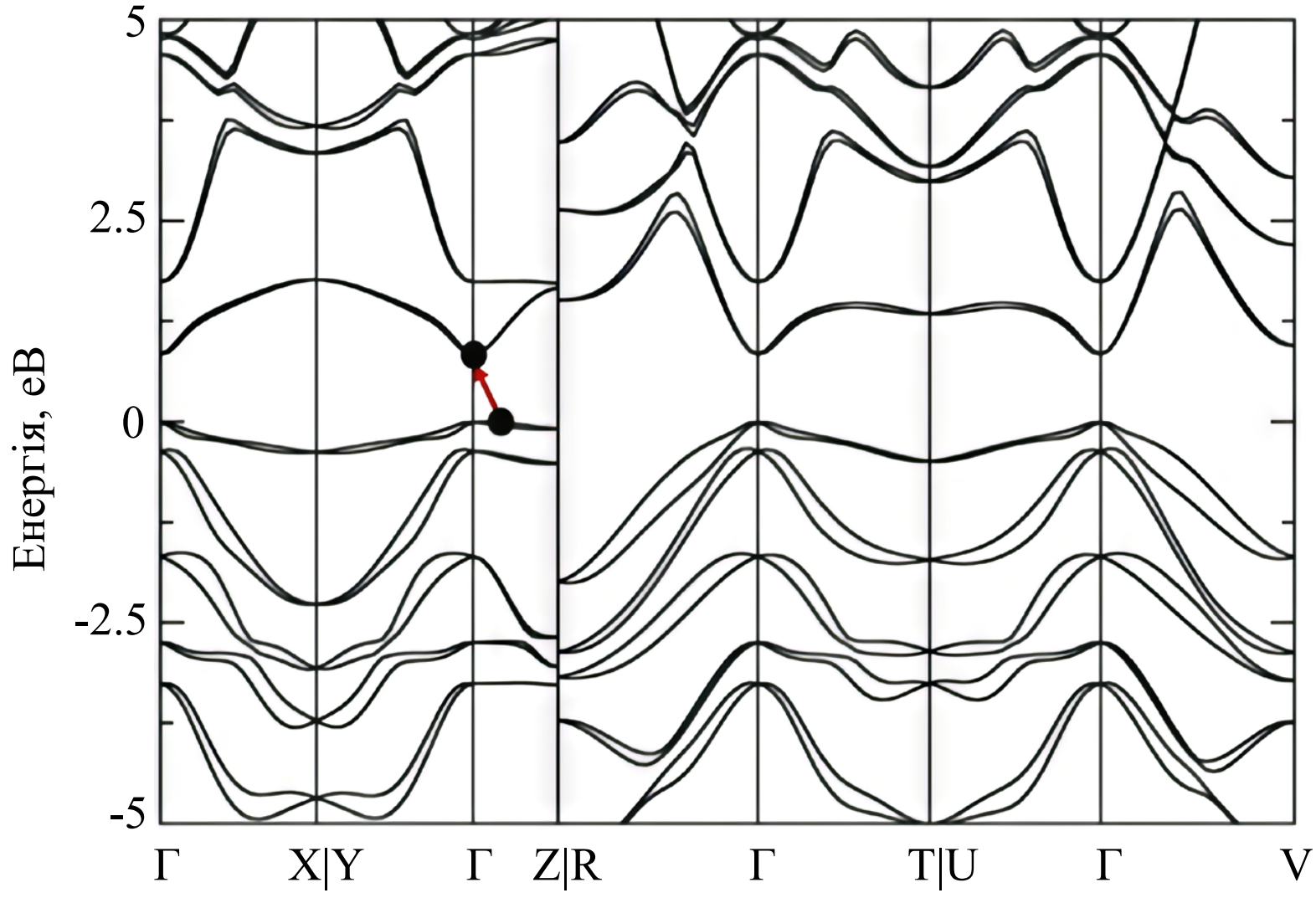

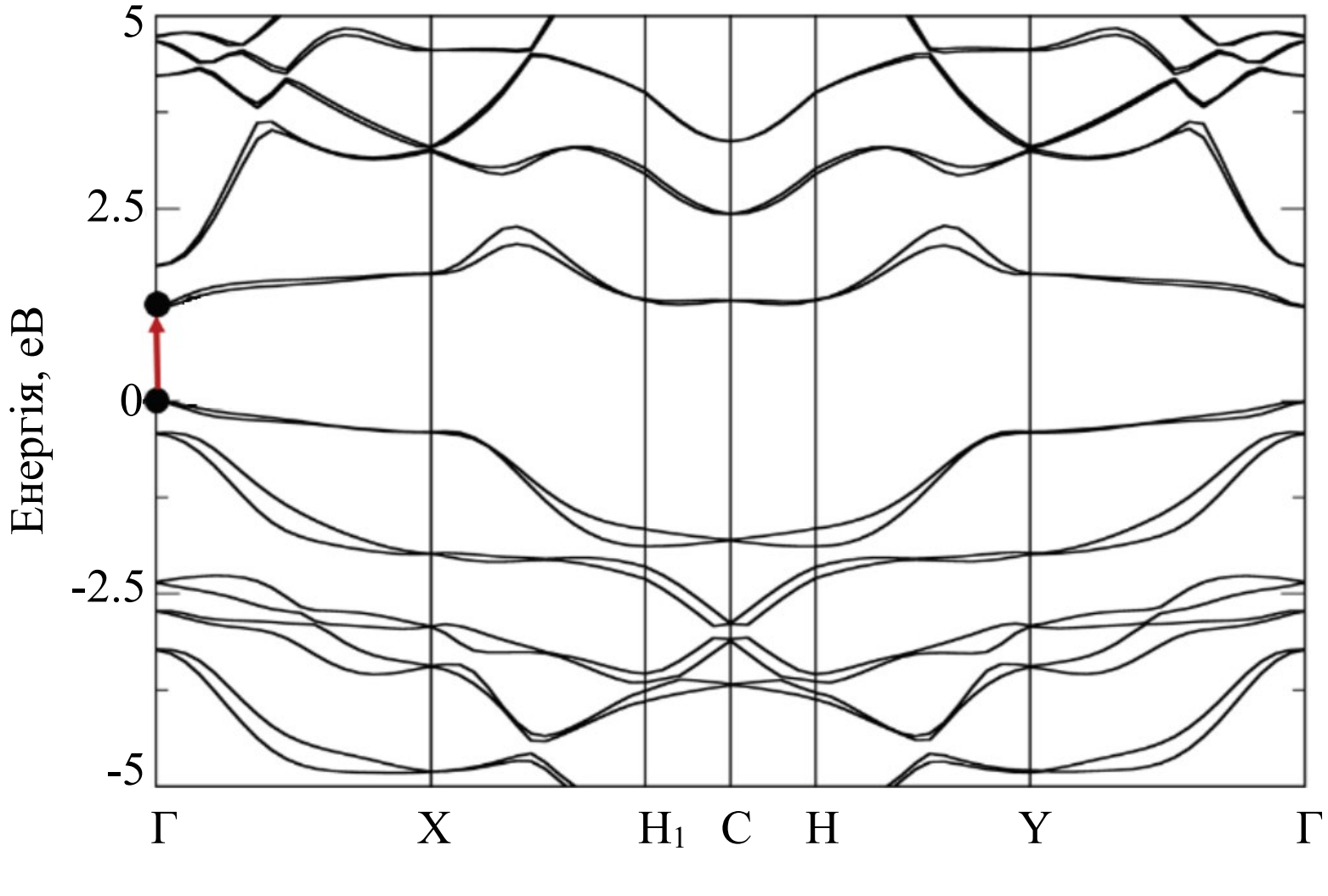

Рис. 3.18. Електронна структура об’ємного (а) і моношару (б) триклінного $SiTe_2$ [164].

в якому атом Si має чіткий тетраедричний зв'язок, що вказує на ковалентний зв'язок за рахунок $sp^3$-гібридизації Si 3*s* та 3*p* орбіталей.

Електронні структури об'ємного і моношару триклінного $SiTe_2$, розраховані з використанням гібридного методу DFT з функціоналом HSE06, наведені на рис. 3.18, *а*, *б* відповідно. Розрахована ширина забороненої зони для об'ємного кристала рівна $E_{gi}$ = 0.83 еВ, а для моношару $E_{gd}$ = 1,222 еВ. Отже, прогнозована об'ємна фаза $SiTe_2$ є непрямозонним напівпровідником і прямою забороненою зоною у моношарі [164].

## 3.7. РОЗПОДІЛ ЕЛЕКТРОННОЇ ГУСТИНИ В ТРИГОНАЛЬНІЙ ФАЗІ $SiTe_2$

Двовимірні карти густини заряду валентних електронів для чотирьох різних кристалографічних площин: двох проведених перпендикулярно тришаровим пакетам (110), (11θ), катіонній площині (001) та аніонній площині (001), зміщеній на 1/4 *c*, наведені на рис. 3.19 відповідно. Суцільні лінії на контурних картах описують поверхні зі сталою електронною густиною, а густина ліній на рисунку характеризує градієнт електронної густини. При формуванні хімічного зв'язку відбувається перерозподіл електронної густини між взаємодіючими атомами, що наглядно видно з наведених карт, де більша густина відображає сильніший міжатомний зв'язок.

Розподіл електронного заряду між різними атомами кремнію (телуру), які належать одному катіонному (аніонному) шару в тришаровому пакеті, передають карти, наведені на рис. 3.19, *а*, *б*. На рис. 3.19, *а* місця розташування іонів Te, які знаходяться в атомному шарі (на відстані 1.78 Å або 1/4 c) вище верхнього атомного шару кремнію позначені точкою, а іони Te, розташовані нижче цієї площини, позначені кружками. Найближчими сусідами іона Te є три іони Si (позначені на рис. 3.19, *б* точками), які знаходяться на рівній відстані. На рис. 3.19, *а* і *б* видно, що не спостерігається спільних ліній рівня $\rho(\mathbf{r})$ для сусідніх катіонів (аніонів) у атомних шарах кремнію (телуру), що свідчить про слабке перекривання їх хвильових функцій.

Карти електронної густини всередині тришарових пакетів, які відображають хімічний зв'язок іонів кремнію з найближчими сусідами – іонами телуру в октаедрі $[SiTe_6]$, наведені на рис. 3.19, *в*, *г*. З цих рисунків видно, що електронна густина всередині тришарових паке-

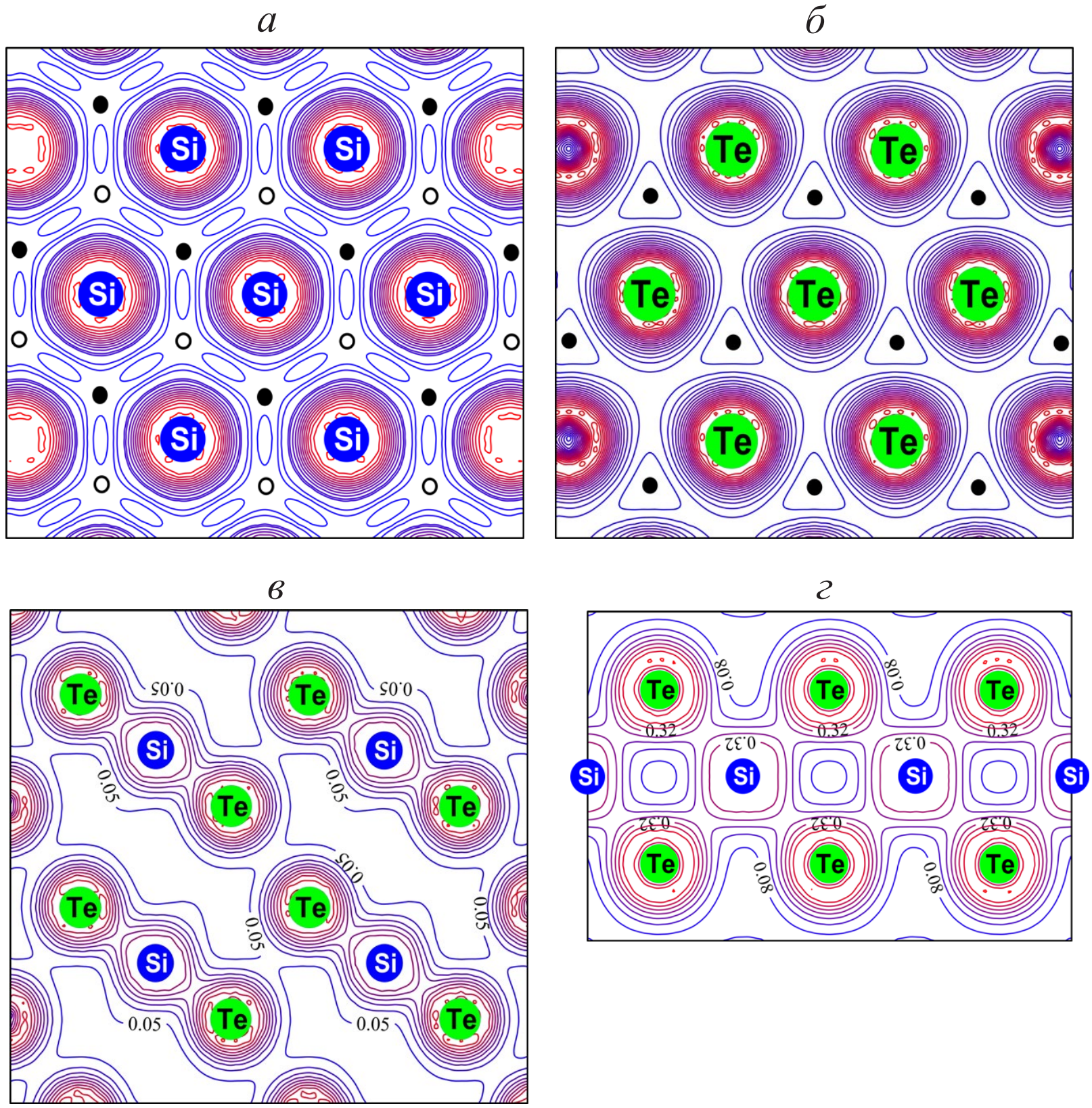

Рис. 3.19. Розподіл валентної еле ктронної густини в сітці атомів кремнію (*а*) (точками і кружками відзначені місця розташування іонів Te, які знаходяться в двох різних аніонних шарах тришарового пакета і телуру (*б*) (точками відзначені місця розташування іонів Si) і в площинах (110) (*в*), $(\bar{1}01)$ (*г*). Значення ρ на ізолініях наведені в $Å^{-3}$.

тів значно вища, ніж на їх границях. Розподіл заряду в одному тришаровому пакеті утворює практичну замкнуту оболонку, що вказує на слабку міжпакетну ван-дер-ваальсову взаємодію, обумовлену $p_z$-станами телуру, які частково входять у міжшаровий простір. Така просторова анізотропія електронної густини і енергетичного розпо-ділу електронних 5*p*-станів телуру є причиною квазідвовимірності кристалів дителуриду кремнію.

Розподіл валентної густини в аніон-катіонній площині, яка проходить уздовж зв'язків Si–Te в октаедрі [$SiTe_6$] (рис. 3.19, *в*), показує наявність спільних контурних ліній між катіоном (Si) і аніоном (Te), які свідчать про гібридизацію Si3*s*- , 3*p*- і Te5*s*-, 5*p*-станів. При цьому простежується поляризація електронної хмари в напрямку від атома Te до Si. Градієнт електронної густини спрямований уздовж зв'язків Si–Te з переважною локалізацією заряду на атомах телуру в силу його більшої електронегативності. Яскраво виражена деформація контурів ρ(**r**) від атомів телуру в бік атомів кремнію уздовж лінії зв'язку Si–Te, і наявність спільних контурів, які охоплюють максимуми електронної густини на катіон-аніонних зв'язках, та відображають ковалентну складову хімічного зв’язку в тришарових пакетах. Наявність ковалентної складової в $SiTe_2$ обумовлена гібридизацією 5*p*-станів телуру і 3*s*-, 3*p*-станів кремнію (рис. 3.16). Саме заряд ко-валентного зв'язку є відповідальним за стабільність октаедричних структурних утворень [$SiTe_6$] у даній сполуці.

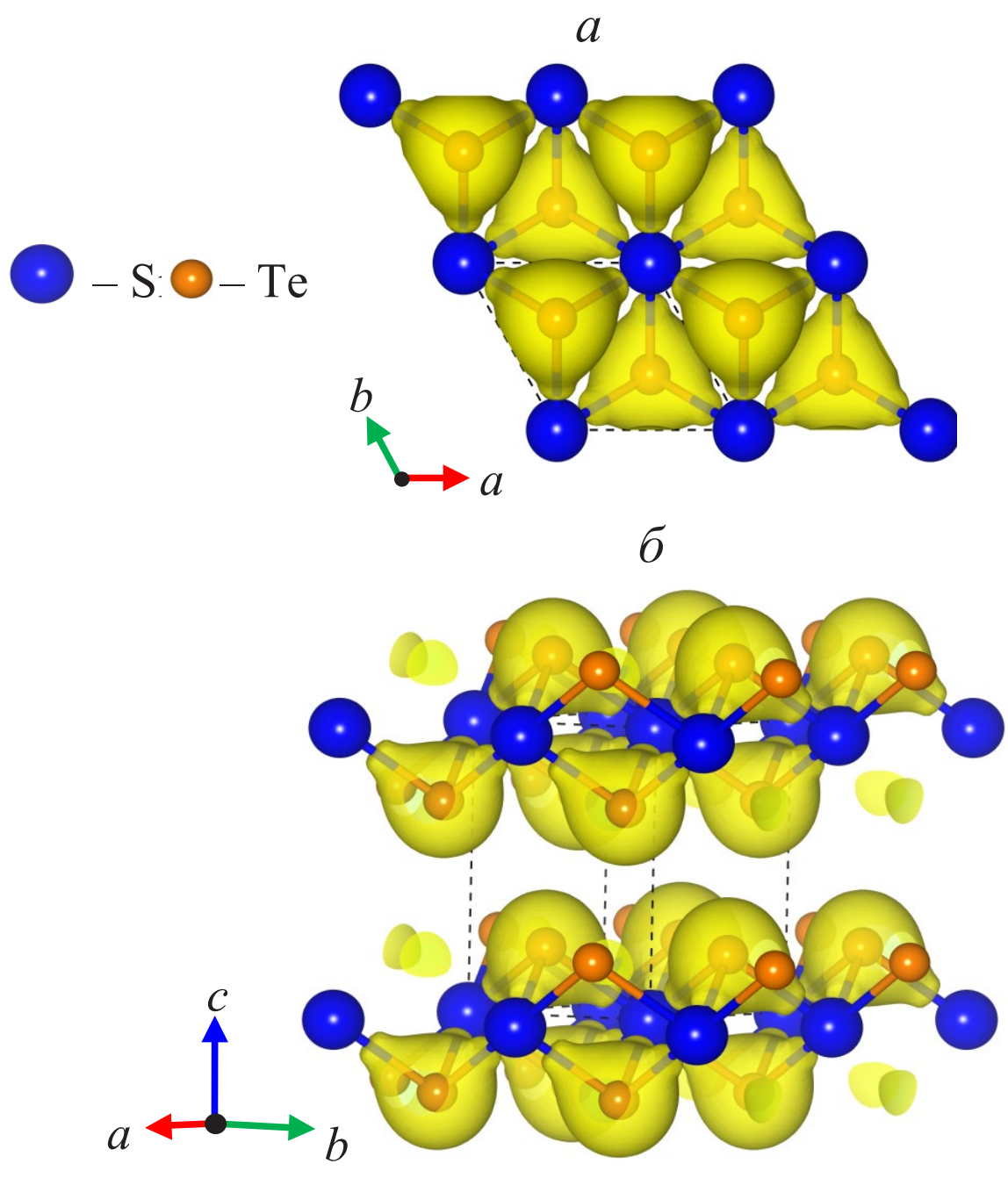

Рис. 3.20. Просторовий розподіл густини валентного заряду в елементарній комірці $SiTe_2$.

Іонна компонента хімічного зв'язку зумовлена частковим перенесенням зарядової густини між атомами кремнію і телуру за рахунок різниці їх електронегативностей ($ЕН^{Si}$ = 1.9, $ЕН^{Te}$ = 2.1). На картах електронної густини це відображається в більшій густині валентних електронів поблизу місць локалізації атомів телуру і скороченням заряду на ковалентному зв'язку між атомами телуру і кремнію. Таким чином, міжатомні взаємодії в $SiTe_2$ мають комбінований характер і включають ковалентну, іонну й ван-дер-ваальсову складові хімічного зв'язку. На рис. 3.20 наведено об'ємні картини розподілу густини заряду в одному тришаровому пакеті (рис. 3.20, *а*) та в елементарній комірці $SiTe_2$ (рис. 3.20, *б*). Видно, що зарядова густина концентрується в основному всередині октаедрів [$SiTe_6$], що утворюють нескінченні тришарові пакети. В октаедрах ізоповерхні $\rho$ сильно деформовані вздовж напрямків зв'язків Si–Te, відображаючи тим самим суттєву ковалентну складову, обумовлену гібридизацією Te 5*p*- і Si 3*p*- станів. Форма розподілу ізоповерхні ρ у тришаровому пакеті –Te–Si–Te– демонструє утворення сильних міжатомних взаємодій між атомами Si і Te і слабкі взаємодії між халькогенами двох сусідніх «сендвічів».

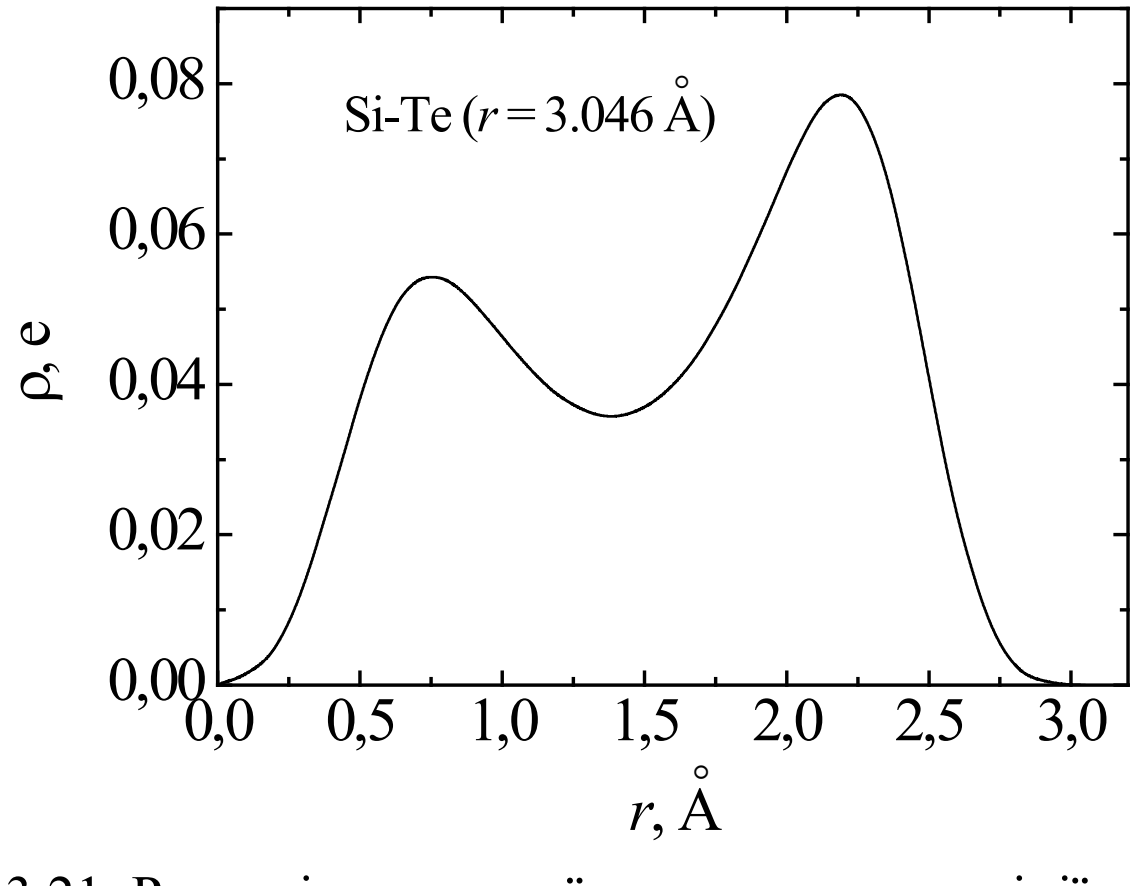

Рис. 3.21. Розподіл зарядової густини вздовж лінії зв'язку Si–Te в октаедрі [$SiTe_6$].

Характер формування міжатомних зв’язків у структурній одиниці [$SiTe_6$], характерній для кристала $SiTe_2$, може бути проілюстрований також за допомогою просторового розподілу радіальної зарядової густини ρ(**r**) (рис. 3.21). Видно, що зарядова густина в октаедрі [$SiTe_6$] уздовж лінії зв'язку Si–Te має два чітко виражених локальних максимуми поблизу ядер аніона (Te) і катіона (Si), причому поблизу іона телуру інтенсивність у максимумі ρ(**r**) більша, ніж поблизу іона кремнію.

### 3.8. ОПТИЧНІ ВЛАСТИВОСТІ КРИСТАЛІВ $SiTe_2$

Вперше оптичні властивості шаруватих кристалів $SiTe_2$, вирощених методом ХТР (транспортер – йод), були досліджені при 300 К в діапазоні довжин хвиль 0.3–10 мкм [20]. Спектральна залежність коефіцієнта поглинання в області власного поглинання наведена на рис. 3.22. Для з'ясування наявності можливих прямих і непрямих переходів у кристалі $SiTe_2$ автори [20] провели аналіз кривої поглинання. У випадку прямих переходів коефіцієнт поглинання описується залежністю $(h\nu – E_g)^n/h\nu$, де $n$ дорівнює 1/2 для дозволених і 3/2 для заборонених переходів за участі фотонів з енергією $h\nu$. Для значень, що перевищують $10^3$ см$^{-1}$, було знайдено відповідність енергетичній залежності для заборонених прямих переходів в інтервалі енеогій 0–6 еВ і показано на рис. 3.23, *а*. Екстрапольоване значення $E_g$ становить 2.18 еВ. Пряма заборонена зона добре узгоджується зі значенням 2.16 еВ, отриманим із вимірювань спектрів фотопровідності (рис. 5.7 розділ 5).

Нижня ділянка кривої поглинання (рис. 3.22) була досліджена для виявлення можливих непрямих переходів з участю фононів. Для таких переходів коефіцієнт поглинання описується залежністю $(h\nu - E_g' \pm E_ф)^m/h\nu$, де $m$ = 2 для дозволених і $m$ = 3 для заборонених переходів; $E_g'$ –мінімальна заборонена енергетична щілина, а $E_ф$ – це енергія для фонона, який поглинається (+) або випромінюється (–) під час переходу. На рис. 3.23, *б*. показано діапазон узгодження для можливої відповідності до енергетичної залежності як для дозволених, так і для заборонених типів переходів. Значення для $E_g'$, отримані шляхом екстраполяції лінійних ділянок, становлять 1.82 еВ і 1.89 еВ для непрямих заборонених й непрямих дозволених переходів відповідно. З огляду на діапазон узгодженості, важко вказати на перевагу однієї із залежностей, наведених на рис. 3.23, *б*. Автори [20] схильні вважати, що $SiTe_2$ є непрямозонним напівпровідником із непрямою забороненою зоною в околі 1.85 еВ.

Таким чином, згідно з результатами дослідження краю власного поглинання, дителурид кремнію є непрямозонним напівпровідником із шириною забороненої зони $E_{gi}$ = 1.85 еВ, а ширина забороненої зони для прямих переходів оцінена як $E_{gd}$ = 2.18 еВ [20].

Більш детально оптичні властивості кристалів $SiTe_2$ в області фундаментального поглинання у широкому інтервалі температур (100–300 К) досліджені авторами [21, 165]. Кристали були вирощені за

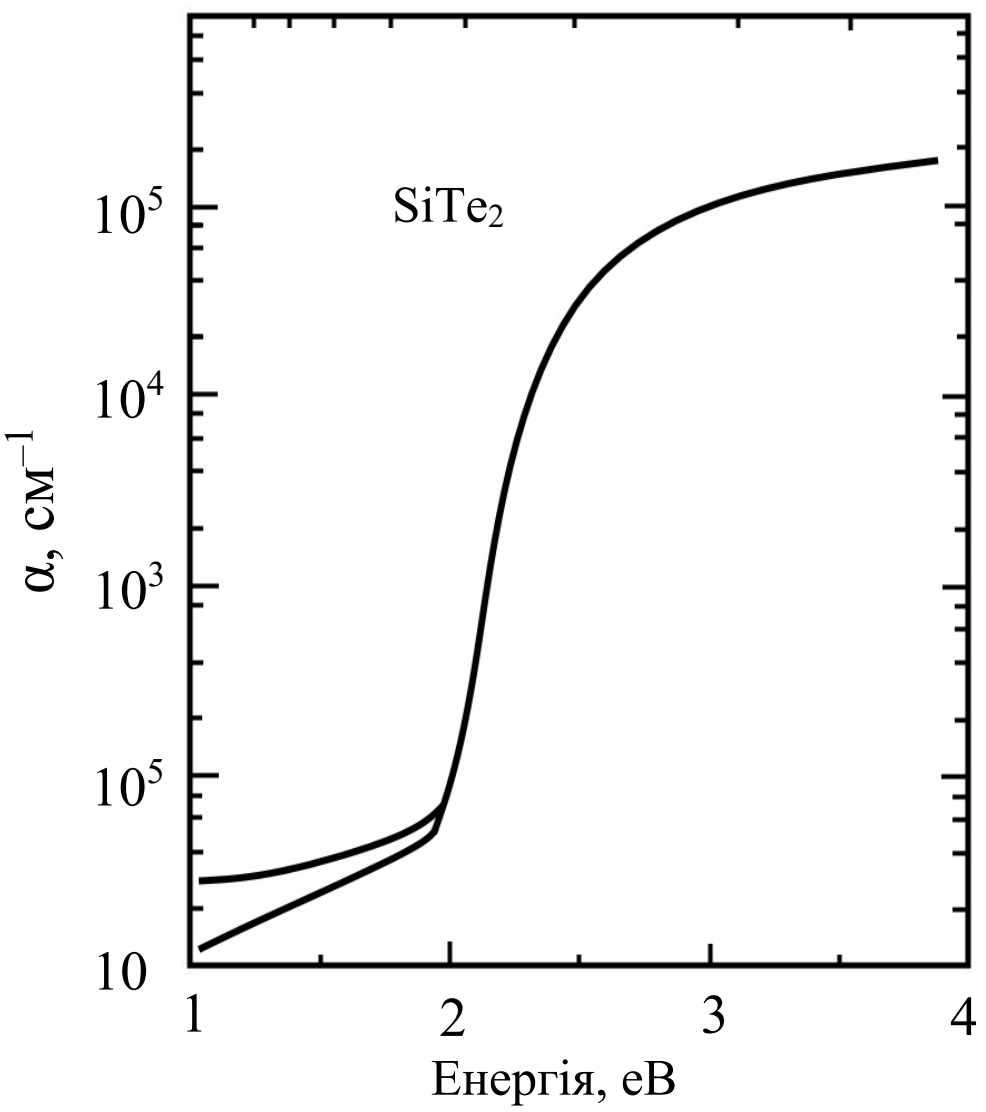

Рис. 3.22. Спектральна залежність коефіцієнта поглинання в області фундаментального поглинання кристала $SiTe_2$ [20].

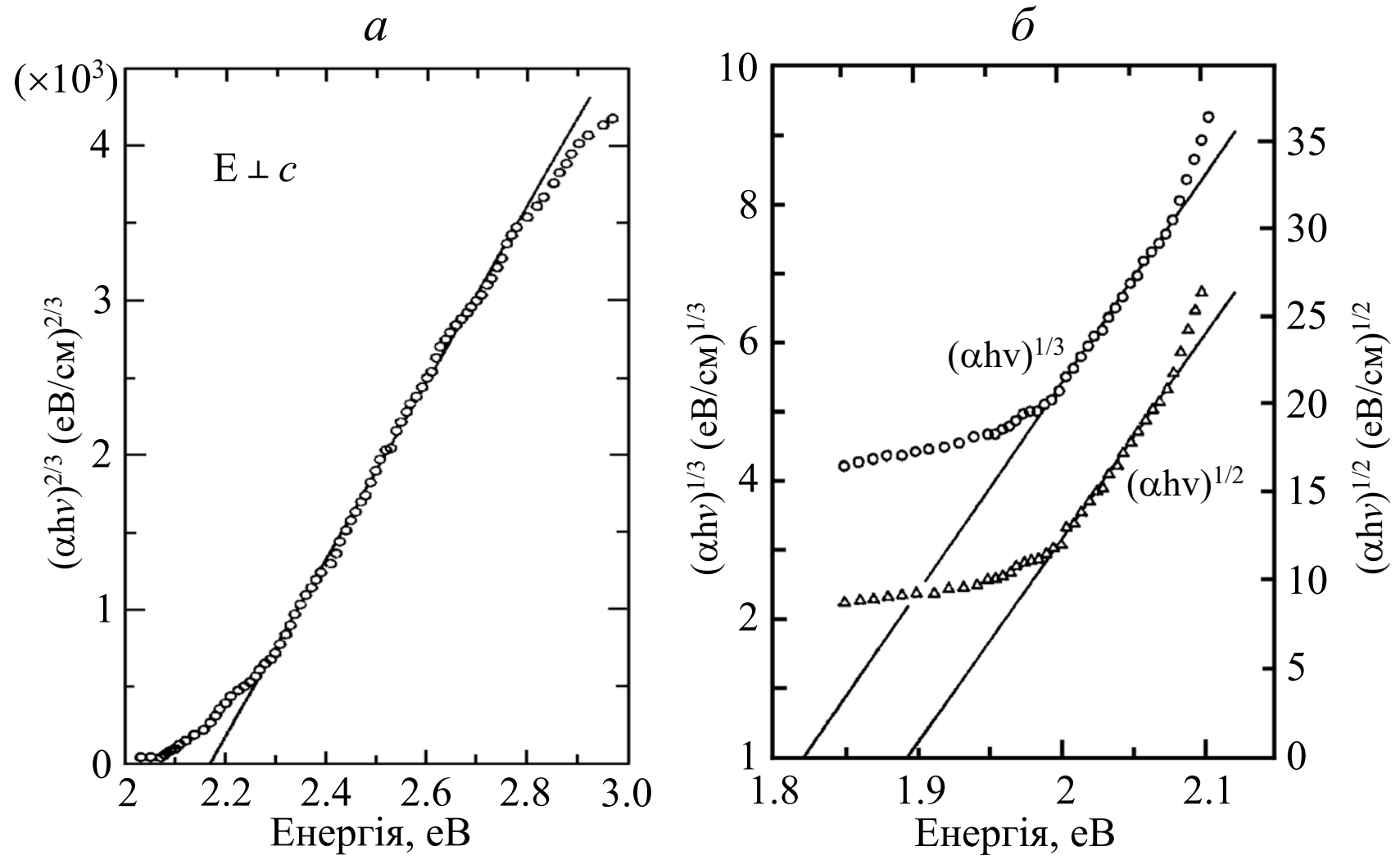

Рис. 3.23. *а*) Залежність $(\alpha h\nu)^{2/3}$ від енергії фотонів, екстрапольована до значення енергії прямої забороненої зони; *б*) залежності $(\alpha h\nu)^{1/3}$ та $(\alpha h\nu)^{1/3}$ від енергії фотонів, екстрапольовані на порогову енергію для непрямих переходів кристала $SiTe_2$ [20].

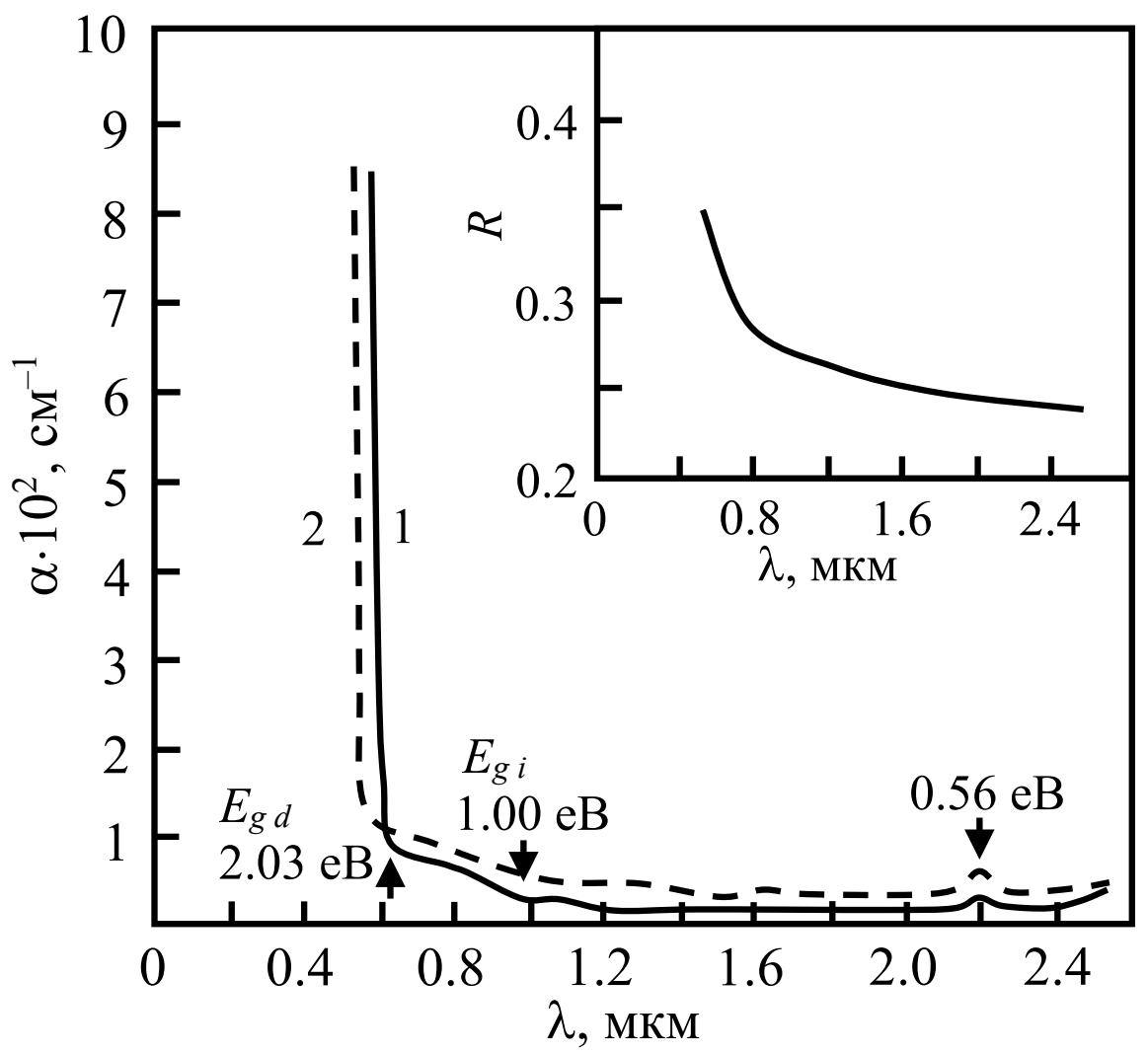

Рис. 3.24. Спектральна залежність коефіцієнта поглинання (α) кристала $SiTe_2$ при 300 К – крива 1, і 100 К – крива 2. На вставці наведений спектр відбивання для цього ж кристала [21].

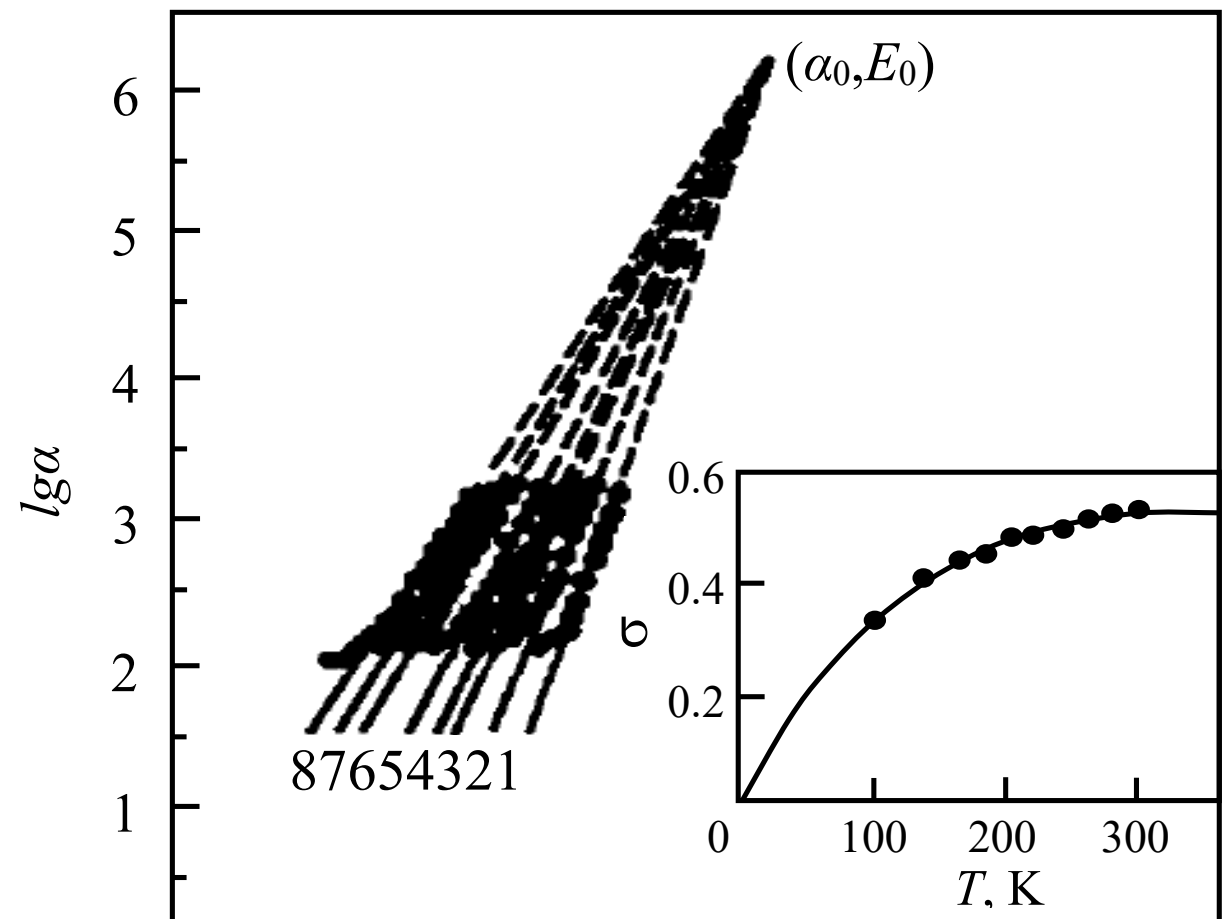

Рис. 3.25. Спектральна залежність крайового поглинання кристала $Si_2Te_3$ при різних температурах. $T$, К: 1 – 100; 2 – 147; 3 – 179; 4 – 200; 5 – 218; 6 – 251; 7 – 267; 8 – 300 [21].

технологією, аналогічною до описаної у роботі [20], і містили велику кількість аніонних вакансій, на що вказував їх хімічний аналіз. Склад кристалів змінювався від $SiTe_{1.996}$ до $SiTe_{1.886}$, тобто кристали виростали дефектними по аніонній підгратці. При $T$ = 300 К в області непрямих переходів край поглинання $SiTe_2$ знаходиться при 1.0 еВ, а в області прямих переходів – при 2.03 еВ (рис. 3.24).

Спектральні залежності коефіцієнта поглинання кристала $SiTe_2$, виміряні при різних температурах в інтервалі 100–300 К, наведені на рис. 3.25 [21]. В інтервалі значень коефіцієнта поглинання 80–$10^3$ $см^{-1}$ спектральна залежність краю власного поглинання описується емпіричним правилом Урбаха (3.6). З рис. 3.25 видно, що високоенергетичні ділянки спектра крайового поглинання кристалічного $SiTe_2$ у досліджуваному інтервалі температур (100–300 К) утворюють характерне температурне «віяло», з координатами збіжності $\alpha_0 = 1.44 \cdot 10^6$ $см^{-1}$ і $E_0$ = 2.48 еВ.

З підвищенням температури зразка від 100 до 300 К край власного поглинання зміщується у довгохвильову область спектра (рис. 3.25), що є відображенням зменшення ширини забороненої зони зі збільшенням температури кристала. Температурна залежність нахилу краю поглинання описується виразом (3.6).

## 3.9. ІНТЕРФЕРЕНЦІЙНІ ФІЛЬТРИ НА БАЗІ ШАРУВАТИХ КРИСТАЛІВ $SiTe_2$

Одним із специфічних застосувань шаруватих кристалів є створення на їх основі інтерференційних фільтрів для різних довжин хвиль. При цьому використовується висока якість поверхні сколу, що дозволяє у досить простий спосіб отримати якісні оптичні фільтри. Авторами [165] запропоновано методику розрахунку параметрів та побудови інтерференційного фільтра з контрольованими параметрами шляхом послідовного розташування шаруватих кристалів, зокрема $SiTe_2$, різної товщини. Використання даних шаруватих кристалів має перевагу завдяки їхній ідеальній кристалічній поверхні та малим коефіцієнтам поглинання. Продемонстровано можливість створення оптичного фільтра із заданими характеристиками на довжині хвилі з ближньої ІЧ області. При цьому фільтр забезпечує високий максимум коефіцієнта пропускання на бажаній довжині хвилі, що залежить від товщин використовуваних зразків, і контрольовану спектральну напівширину.

Існує декілька характеристик продуктивності оптичних фільтрів. Найбільш повною із них є спектральна крива пропускання фільтра. Щоб отримати бажану криву пропускання, об'єднують декілька фільтрів з різними характеристиками. Найбільш поширеним способом комбінування фільтрів є їх послідовне розміщення. У першому наближенні, якщо всі фільтри знаходяться у повітрі, результуючий коефіцієнт пропускання визначається як:

$$T_\lambda = T_{1\lambda} \cdot T_{2\lambda} \ldots T_{i\lambda} \ldots T_{n\lambda}, \qquad (3.11)$$

де $T_{i\lambda}$ – коефіцієнт пропускання $i$-го фільтра на довжині хвилі λ. Рівняння (3.11) справедливе, коли фільтри паралельні, а напрям поширення випромінювання є нормальним до їх поверхні. Ідеальний смуговий фільтр пропускає все падаюче випромінювання в одній спектральній області та відбиває чи поглинає все інше. Такий фільтр повністю описується шириною області пропускання, довжиною хвилі $\lambda_0$, на якій він центрований, і піковим коефіцієнтом пропускання $T_0$. Напівширина $\Delta\lambda_{0.5}$ фільтра – це різниця між довжинами хвиль, на яких коефіцієнт пропускання становить половину $T_0$. Подібним чином визначається базова ширина $\Delta\lambda_{0.01}$. Відношення $\Delta\lambda_{0.01}/\Delta\lambda_{0.5}$ називається коефіцієнттом (фактором) форми, а відношення $T_{min}/T_0$ - коефіцієнтом якості. Нарешті, відстань між двома максимумами пропускання, що знаходяться поруч із основним, називається вільним спектральним діапазоном.

Шаруваті кристали $SiTe_2$ мають дзеркальну поверхню, типу інтерферометра (еталона) Фабрі-Перо, а їх спектри пропускання складаються з інтерференційних смуг. Таким чином можна отримати фільтр з ідеальними параметрами на довжині хвилі $\lambda_0$, комбінуючи низку кристалів різної товщини, спектри пропускання яких дають спільний для всіх кристалів максимум на цій довжині хвилі.

Оптичне пропускання шаруватого напівпровідника з товщиною $d$ описується наступним співвідношенням [165]:

$$T = \frac{(1-R)^2\left(1+\dfrac{k^2}{n^2}\right)}{\left[\exp\left(\dfrac{2\pi k}{\lambda}d\right) - R\cdot\exp\left(-\dfrac{2\pi k}{\lambda}d\right)\right]^2 + 4R\cdot\sin^2(\delta+\psi)}, \qquad 3.12)$$

де $R = (n-1)^2 + k^2 / (n+1)^2 + k^2$ – коефіцієнт відбивання, $\delta = 2\pi\, n\, \mathrm{d} / \lambda$ – фазовий зсув для випадку нормального падіння, $\tan\psi = 2k/n^2 + k^2 - 1$, $n$ і $k$ - дійсна та уявна частини комплексного

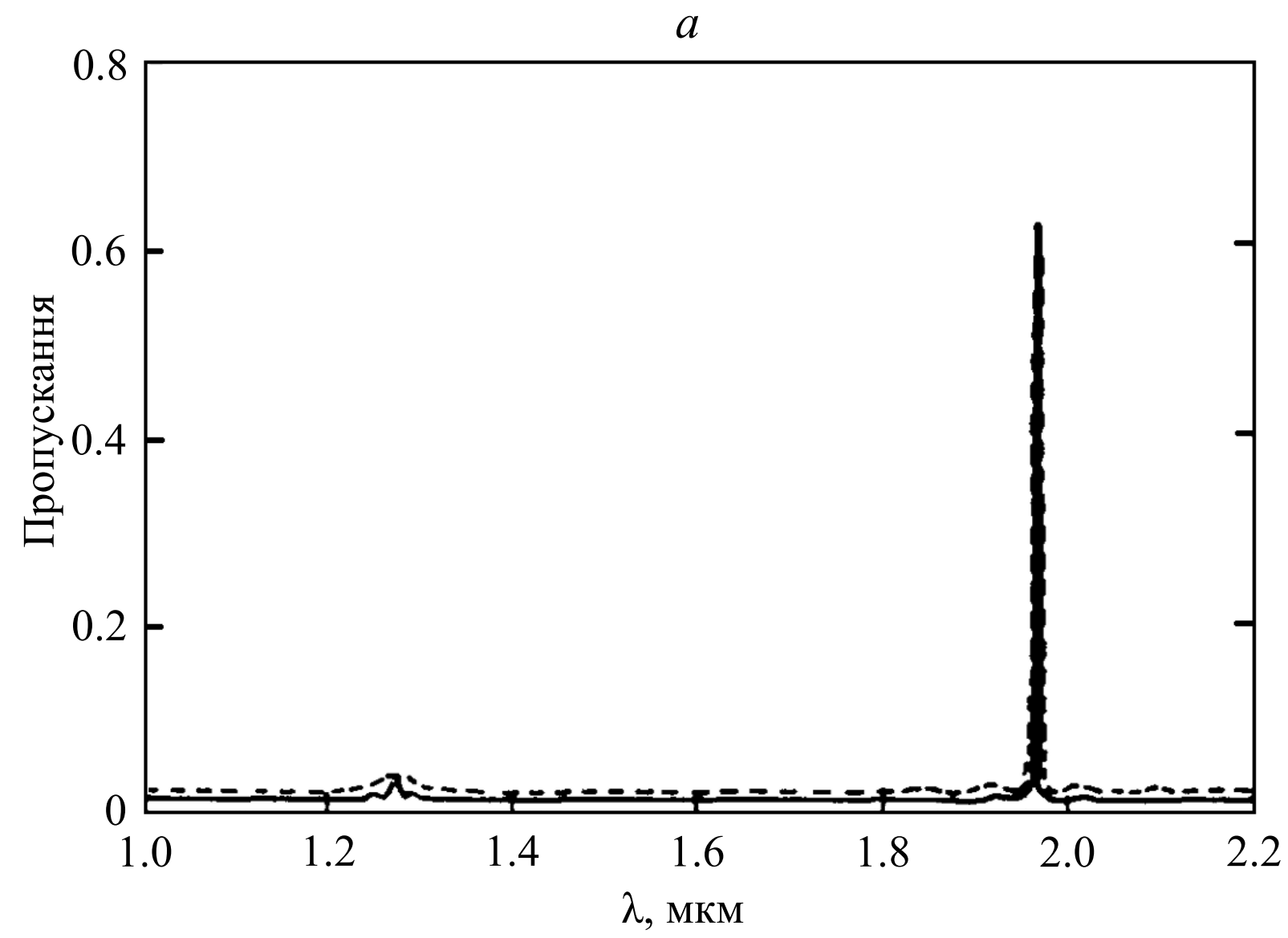

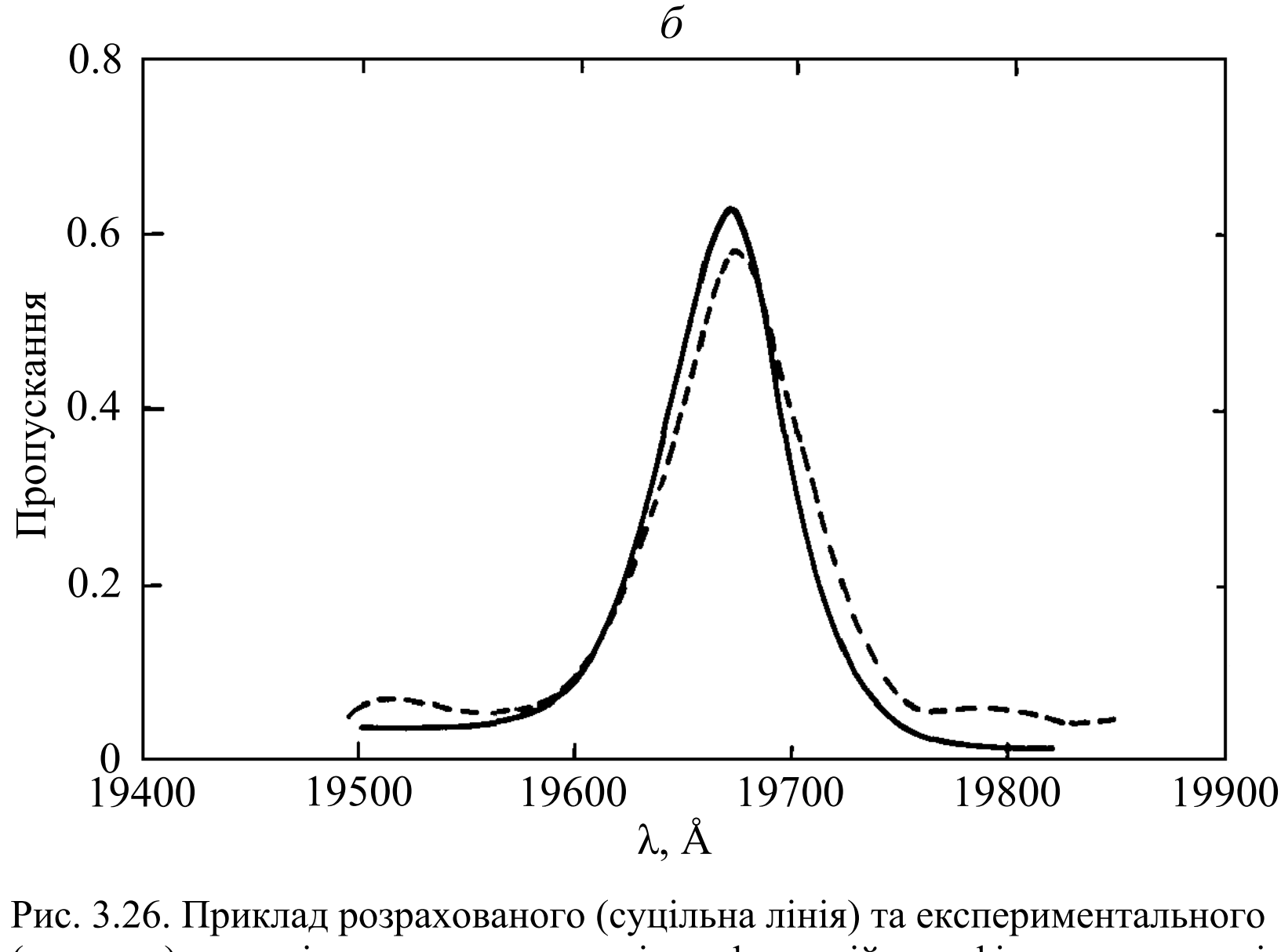

Рис. 3.26. Приклад розрахованого (суцільна лінія) та експериментального (пунктир) спектрів пропускання для інтерференційного фільтра на основі шаруватих кристалів $SiTe_2$ (у двох масштабах спектра). [165].

показника заломлення, відповідно. Умова максимуму для інтерференційної картини :

$$\lambda N = 2\, n\, d, \qquad (3.13)$$

де $N$ – ціле число. Для відомих значень $n$ і $k$ товщини зразків розраховувалися за рівнянням (3.12) таким чином, щоб усі мали максимум пропускання при довжині хвилі $\lambda_0$, після чого вони розташовувалися послідовно. Змінюючи число зразків, можна варіювати спектральну півширину фільтра, максимум пропускання та інші характеристики. Для розрахунку параметрів був розроблений алгоритм, за яким послідовно розраховувались спектри пропускання після кожного з елементів, з використанням формули (3.12).

Приклад спектрів пропускання для отриманого інтерференційного фільтра з повітряними проміжками, що був реалізований з використанням послідовно розташованих п'яти кристалів $SiTe_2$, приведено на рис. 3.26. Максимум пропускання відповідав довжині хвилі 14250 Å. З використанням чотирьох зразків з товщинами від 9.50 до 23.75 мкм був отриманий фільтр з максимумом пропускання більше 97% і водночас низьким значенням півширини $\lambda_{0.5}$. Також були отримані фільтри з різною кількістю зразків, у тому числі більше п’яти.

Розрахункова методика була підтверджена експериментальними вимірюваннями параметрів фільтрів, при цьому розрахунки здійснювалися при наступних оптичних параметрах: $3.14 > n > 2.92$ , $0.00022 < k < 0.00028$, відповідно. Кристали $SiTe_2$ отримували або епітаксіальним вирощуванням, або сколюванням з товстого зразка. Їх товщини варіювалися від 9.19 до 77.43 мкм, а отримані значення піка пропускання $\lambda_0$ становили 19674 і 20025 Å відповідно. При цьому експериментальні та розраховані характеристики фільтра добре відтворюються.

Таким чином, інтерференційні фільтри на основі шаруватих кристалів $SiTe_2$ демонструють відносно високий максимум пропускання $T_0$ на бажаній довжині хвилі $\lambda_0$, яка залежить від товщини використовуваних зразків, а також контрольовану спектральну напівширину. Вони можуть знайти застосування в оптичних приладах і в телекомунікаційних пристроях з використанням волоконної оптики. В аналогічних конструкціях можуть бути використані інші шаруваті напівпровідникові структури. Відзначимо, що перевагою такого використання шаруватих монокристалів є можливість отримання гладких поверхонь без використання будь-якого виду травлення чи полірування, що мінімізує витрати.

## 3.10. ОПТИЧНІ Й ФОТОАКУСТИЧНІ ВЛАСТИВОСТІ СТЕКОЛ $Si_xTe_{100-x}$

**3.10.1. Спектри фундаментального поглинання стекол $Si_xTe_{100-x}$.** Важливими характеристиками спектра пропускання, які визначають комплекс оптичних властивостей халькогенідних склоподібних напівпровідників, є енергетичне положення і форма границі фундаментального поглинання, які у значній мірі залежать від хімічного складу та зовнішніх факторів (температури, тиску, опромінення, тощо). Вперше край власного поглинання стекол $Si_xTe_{100-x}$ дослідили автори [46] при кімнатній температурі. На рис. 3.27 приведені спектри фундаментального поглинання свіжоприготовленого (крива 1) і термічно відпаленого вище температури кристалізації (крива 2) склоподібного $Si_{20}Te_{80}$. Як видно з цього рисунка, край поглинання відпаленого зразка зміщується в область менших енергій, а ширина забороненої зони різко зменшується і є близькою до ширини забороненої зони кристалічного телуру. Рентгеноструктурні дослідження виявили у відпаленому зразку наявність кристалічного телуру і кристалічного $Si_2Te_3$.

Спектри крайового поглинання трьох складів стекол $Si_xTe_{100-x}$, виміряні при кімнатній температурі, проаналізовані в залежності $(\alpha h\nu)^{1/2}$ від $h\nu$ і наведені на рис. 3.28. Перетин прямої лінії таких графіків з віссю енергії дає $E_{opt}$. Як видно з рис. 3.28, збільшення концентрації кремнію в стеклах $Si_xTe_{100-x}$ приводить до збільшення оптичної ширини забороненої зони.

Для склоподібного $Si_{15}Te_{85}$ край власного поглинання при фіксованій температурі в діапазоні 77–300 K описується експоненціальною залежністю коефіцієнта поглинання $\alpha$ від частоти (рис. 3.29), проте кристалічне правило Урбаха (3.6) в цих стеклах не виконується. При нагріванні склоподібного $Si_{15}Te_{85}$ відбувається зміщення краю власного поглинання в довгохвильову область спектра з температурним коефіцієнтом зміщення ~ $7\cdot10^{-4}$ еВ·K$^{-1}$, при цьому крутизна краю $dln\alpha/d\nu$ практично не залежить від температури.

Такий характер температурної поведінки краю фундаментального поглинання характерный для більшості бінарних халькогенідних склоподібних напівпровідників ($As_2S_3$, $As_2Se_3$, $GeS_2$, $GeSe_2$). Для аналітичного опису цих залежностей замість виразу (3.6) авторами [166] запропоновано так зване “склоподібне” правило Урбаха з температурно-незалежним нахилом спектральної характеристики коефіцієнта поглинання:

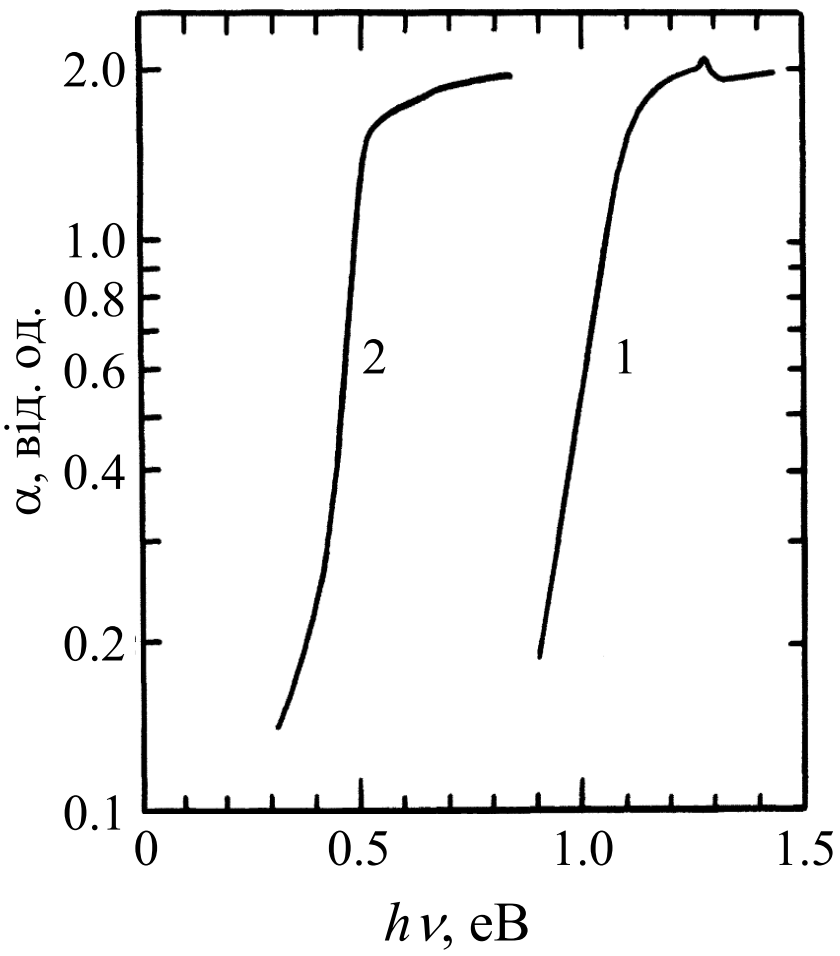

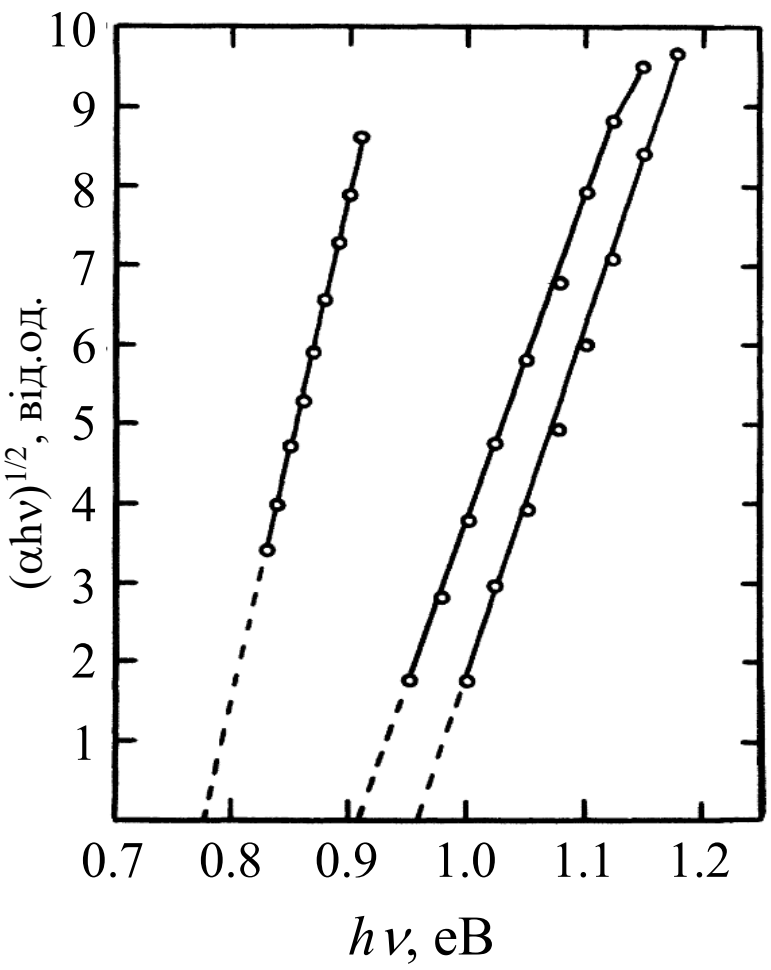

Рис. 3.27. Спектри крайового поглинання невідпаленого (1) і відпаленого вище температури кристалізації (2) скла $Si_{20}Te_{80}$ [46]

Рис. 3.28. Графіки залежності $(\alpha\, h\nu)^{1/2}$ від $h\nu$ для трьох складів стекол $Si_xTe_{100-x}$ . x: 1 – 15; 2 – 20, 3 – 25. [46]

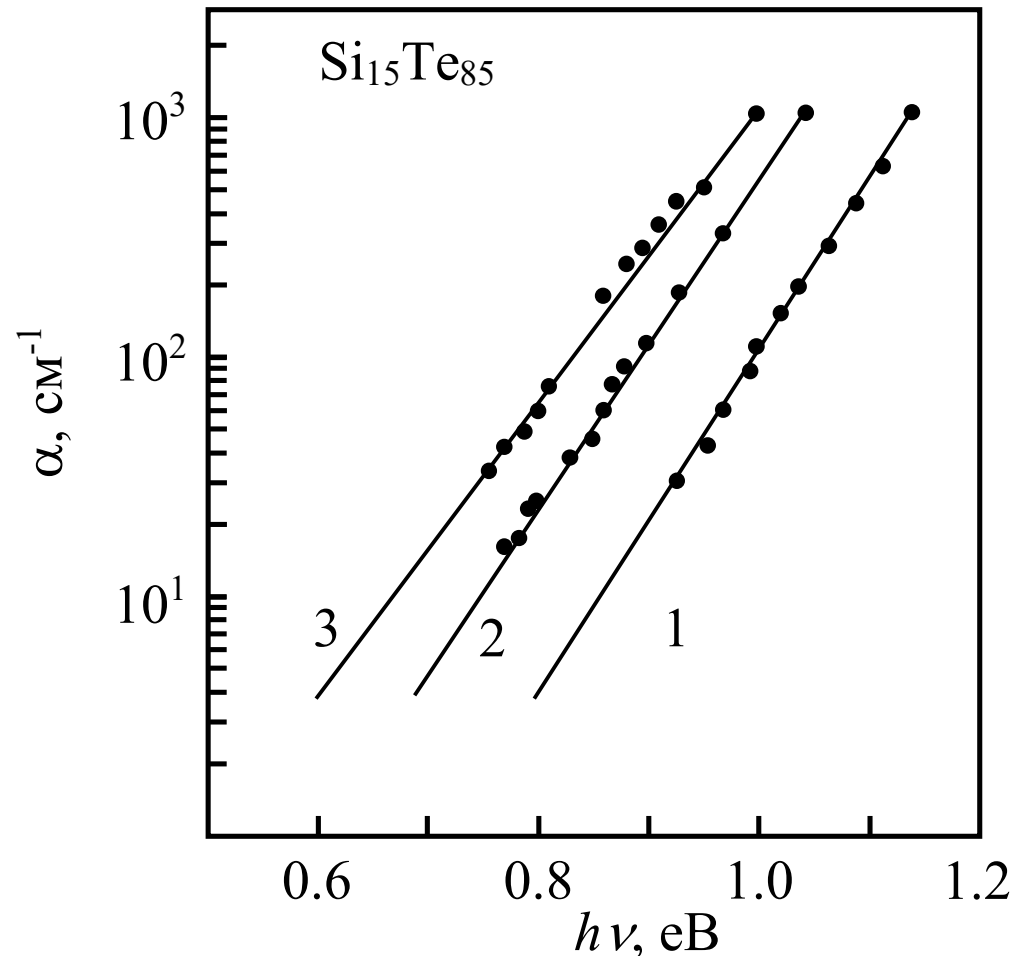

Рис. 3.29. Спектральна залежність крайового поглинання скла $Si_{15}Te_{85}$ при $T$, K: 1 – 100; 2 – 300; 3 – 350.

$$\alpha(h\nu,T)=\alpha_0\exp(\frac{h\nu}{E_0}+\frac{T}{T_0}), \qquad (3.14)$$

де $\alpha_0$ – константа, $T_0$ – деяка характеристична температура, яка пов'язана з параметром статичного розупорядкування $E_0$, $1/E_0 = dln\alpha/dh\nu$ – температурно-залежний логарифмічний нахил спектральної характеристики.

Вираз (3.14) відображує зсув краю поглинання зі збільшенням температури зразка в довгохвильову область без зміни параметра нахилу $E_0$. Другий доданок в показнику експоненти (3.14) відображає лінійну температурну залежність краю поглинання, яка має місце в інтервалі температур від близьких до кімнатної і вище аж до $T_g$.

Відмінність в температурних залежностях спектрів власного поглинання скла $Si_{15}Te_{85}$ і кристалічного $Si_2Te_3$ швидше за все пов'язана з відмінністю в структурі країв дозволених зон впорядкованої і розупорядкованої фаз. Відомо, що в забороненій зоні халькогенідних склоподібних напівпровідників існує значна кількість локалізованих станів на краю дозволених зон. Специфіка температурної залежності краю в склі $Si_{15}Te_{85}$ визначається саме тим, що довгохвильова ділянка спектру власного поглинання скла формується переходами з цих рівнів.

**3.10.2. Фотоакустичні дослідження стекол $Si_xTe_{100-x}$.** Іншим незалежним способом визначення оптичної ширини псевдощілини стекол $Si_xTe_{100-x}$ є фотоакустичний (ФА) метод дослідження оптичного поглинання [167]. На рис. 3.30 *а, б* приведені залежності нормованого ФА-сигналу від довжини хвилі для різних складів стекол $Si_xTe_{100-x}$. Як видно з цього рисунка, в області низьких енергій ФА-сигнал зростає зі збільшенням довжини хвилі і досягає насичення для фотонів з енергією $h\nu > E_g$, де $E_g$ – оптична ширина псевдощілини. Значення оптичної псевдощілини $E_g$ автори [167] визначали графічно по точках перетину дотичних, проведених уздовж ділянки ФА-сигналу в області ФА насичення і області експоненціального збільшення поглинання (рис. 3.30, *а, б*). Визначені в такий спосіб значення $E_g$ для стекол $Si_xTe_{100-x}$ приведені в таблиці 3.2. Як видно

Таблиця 3.2. Значення оптичної псевдощілини стекол $Si_xTe_{1-x}$

| **Склад** | $Si_{28}Te_{72}$ | $Si_{25}Te_{75}$ | $Si_{22}Te_{78}$ | $Si_{20}Te_{80}$ | $Si_{17}Te_{83}$ | $Si_{15}Te_{85}$ | $Si_{10}Te_{90}$ |
|---|---|---|---|---|---|---|---|
| $E_g$, еВ | 1.38 | 1.35 | 1.33 | 1.31 | 1.25 | 1.18 | 1.09 |

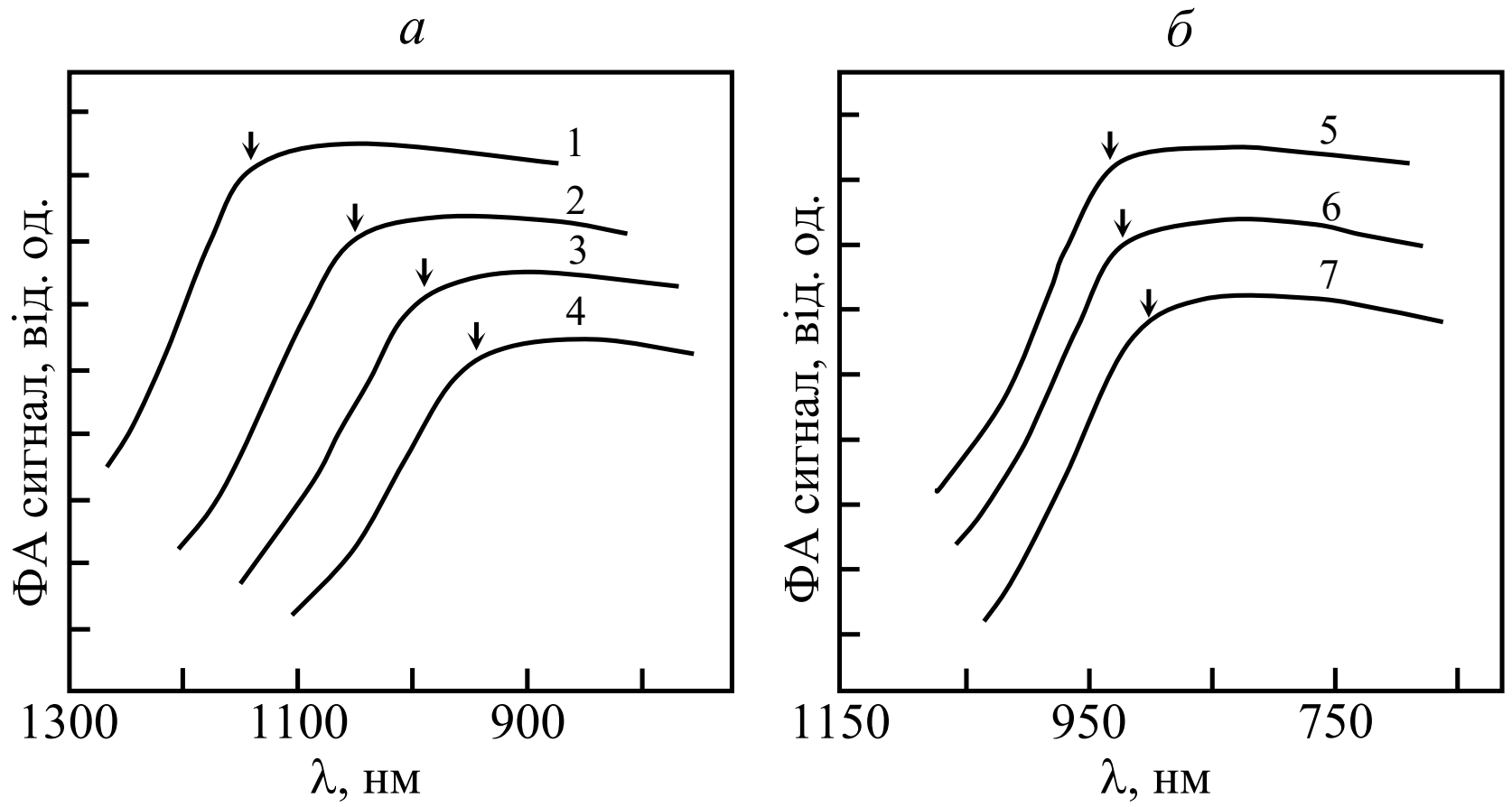

Рис. 3.30. Нормовані фотоакустичні спектри стекол $Si_xTe_{100-x}$.
*x*: 1 – 10; 2 – 15; 3 – 17; 4 – 20, 5 – 22; 6 – 25; 7 – 28 [167].

з даної таблиці, ширина оптичної псевдощілини збільшується з вмістом телуру аж до *x* ~ 20, після чого швидкість росту $E_g$ сповільнюється. Це обумовлено підвищенням ступеня упорядкованості стекол, збільшенням у них концентрації звязків Te–Te і утворення тетраедрів $SiTe_4$.

# РОЗДІЛ 4

# ЕЛЕКТРИЧНІ ВЛАСТИВОСТІ КРИСТАЛІЧНИХ І СКЛОПОДІБНИХ ТЕЛУРИДІВ КРЕМНІЮ

Шаруваті сполуки, в тому числі й сесквітелурид кремнію, є ідеальною системою для дослідження явищ переносу, що мають місце в низькорозмірних системах, таких як від'ємний диференціальний опір, анізотропія провідності, а також для створення різних приладів на їх основі: детекторів, високочутливих сенсорів, генераторів надвисоких частот та інших.

## 4.1. ВОЛЬТ-АМПЕРНІ ХАРАКТЕРИСТИКИ Й ЕФЕКТ ПЕРЕМИКАННЯ У ШАРУВАТИХ КРИСТАЛАХ $Si_2Te_3$

**4.1.1. Вольт-амперні характеристики $Si_2Te_3$.** Вперше вольт-амперні характеристики (ВАХ) шаруватих кристалів $Si_2Te_3$ з вісмутовими контактами, нанесеними термічним випаровуванням у вакуумі, дослідили автори [168] у широкому інтервалі температур 82–370 К (рис. 4.1). При кімнатній температурі кристали мали *p*-тип провідності та питомий опір ~$10^7$ Ом·см, виміряний уздовж кристалографічної осі *c*. ВАХ досліджувались у статичному режимі на постійному струмі вздовж осі *c*. Як видно із рис. 4.1, при низьких напругах ВАХ лінійна, а зі збільшенням напруги з'являється квадратична залежність ($I \sim U^2$). Лінійна омічна ділянка спостерігається в області низьких напруг. З підвищенням температури дана ділянка розширюється, а напруга переходу збільшується. Отже, квадратична ділянка із зростанням температури зменшується.

Наведені на рис. 4.1 ВАХ були проаналізовані авторами [168] на основі теорії інжекційних струмів та активаційної провідності в кристалах [169]. Температурна залежність струму та його енергії активації досліджено окремо для омічної та квадратичної ділянок вольт-амперних характеристик (рис. 4.2). Ці дані були використані для визначення глибини та концентрації локальних центрів, наявних у кристалах $Si_2Te_3$. При цьому автори [168] виходили з уявлень про закономірності зміни енергії активації при переході від омічної до квадратичної ділянки, розвинених раніше в роботі [169], які дозволяють визначити тип провідності (власна або домішкова). У результаті встановлено, що ВАХ монокристалів $Si_2Te_3$ визначаються струмами, обмеженими просторовим зарядом. Як видно з рис. 4.2, при

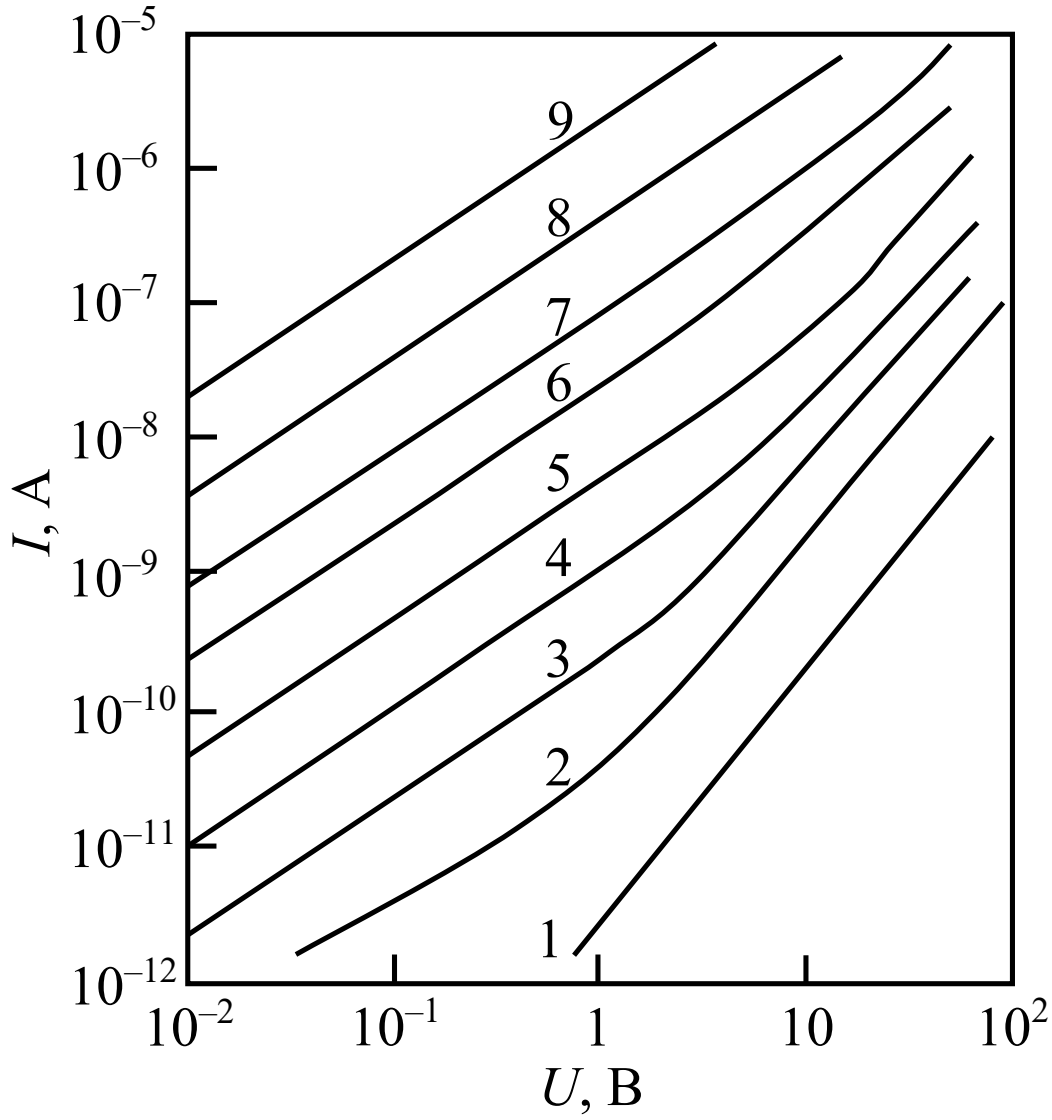

Рис. 4.1. Сімейство вольт-амперних характеристик кристала $Si_2Te_3$, виміряні при прикладанні електричного поля вздовж осі *c*. *T*, K: 1 – 82; 2 – 102; 3 – 122; 4 – 141; 5 – 165; 6 – 202; 7 – 248; 8 – 295; 9 – 370 [168].

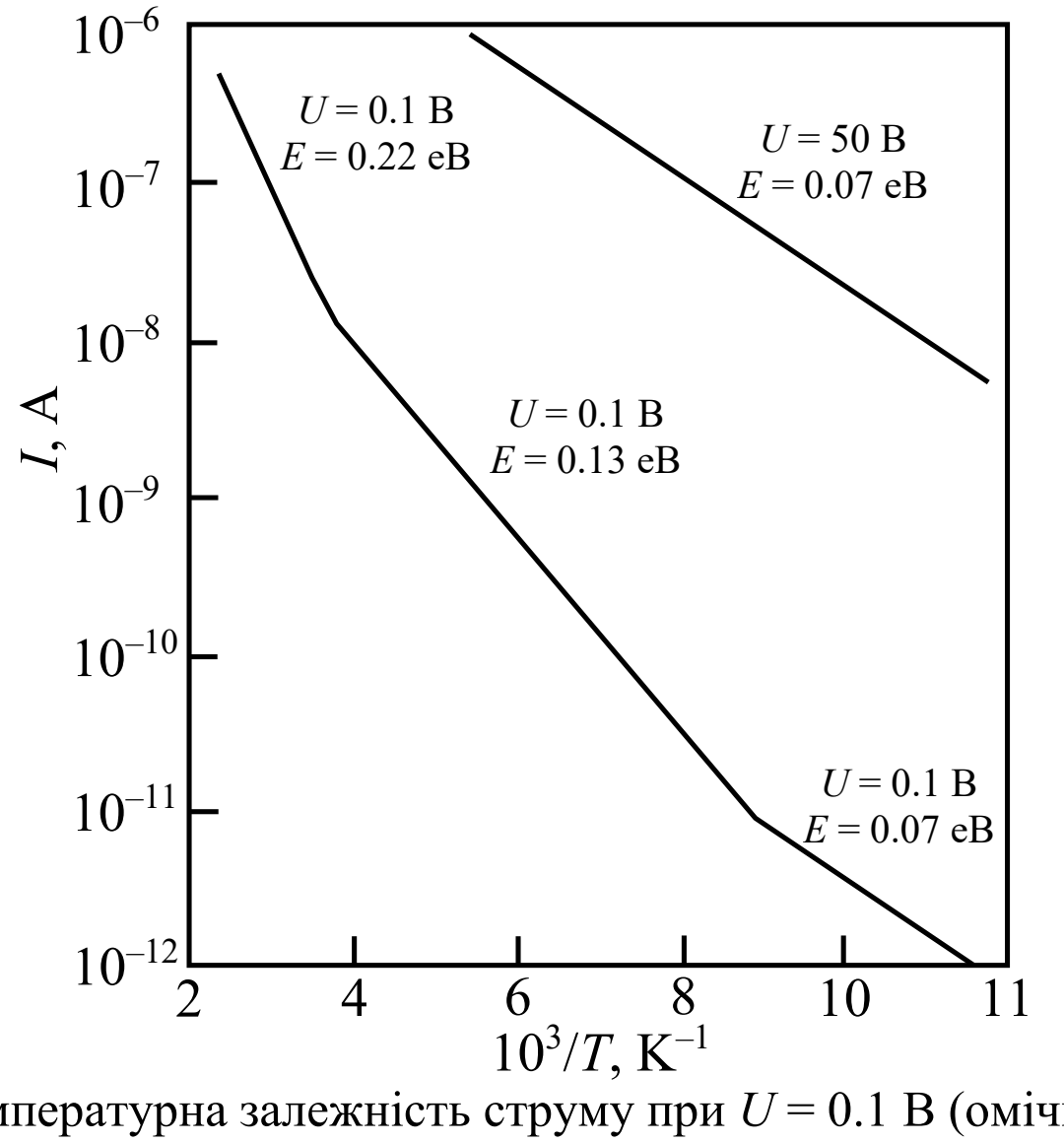

Рис. 4.2. Температурна залежність струму при $U = 0.1$ В (омічна область) та при $U = 50$ В (квадратична область) для кристала $Si_2Te_3$, наведеного на рис. 4.1 [168].

збільшенні температури від 82 до 370 K струм збільшується від $10^{-12}$ до $8\cdot10^{-7}$ A при $U = 0.1$ B, а залежність $\ln I = f(10^3/T)$ має три яскраво виражені ділянки, що вказує на наявність у кристалах $Si_2Te_3$ трьох типів локальних рівнів: одного діркового ($E_V - E_p$ = 0.07 еВ; $N_p = 3\cdot10^{18}$ см$^{-3}$) та двох електронних ($E_V - E_{n1}$ = 0.19 еВ; $N_{n1} = 2\cdot10^7$ см$^{-3}$; $E_V - E_{n2}$ = 0.37 еВ; $N_{n2} = 1\cdot10^9$ см$^{-3}$). Визначено також різницю концентрацій акцепторних та донорних рівнів ($N_A - N_D = 1.5\cdot10^{10}$ см$^{-3}$) і рухливість дірок ($u_p$=14 см$^2$/В·с) при 200 K.

Як вже зазначалось в §1.5 (розділ 1), кристали $Si_2Te_3$ сильно гігроскопічні, тому навіть незначні сліди водяної пари викликають хімічну реакцію, внаслідок чого на поверхні утворюються $SiO_2$ і чистий телур. Для електричних вимірювань необхідно уникати цієї реакції на поверхні, оскільки наявність шару $SiO_2$ і Te на поверхні кристалів $Si_2Te_3$ приводить до різкої зміни їх питомого опору. Тому нанесення контактів та усі вимірювання електропровідності слід проводити у високому вакуумі. Однак у процесі переміщення досліджуваного кристала з ростового контейнера (кварцової ампули) у вакуумну установку для нанесення контактів, а потім у кріостат для вимірювання електропровідності, його поверхня піддається впливу парів води лабораторної атмосфери.

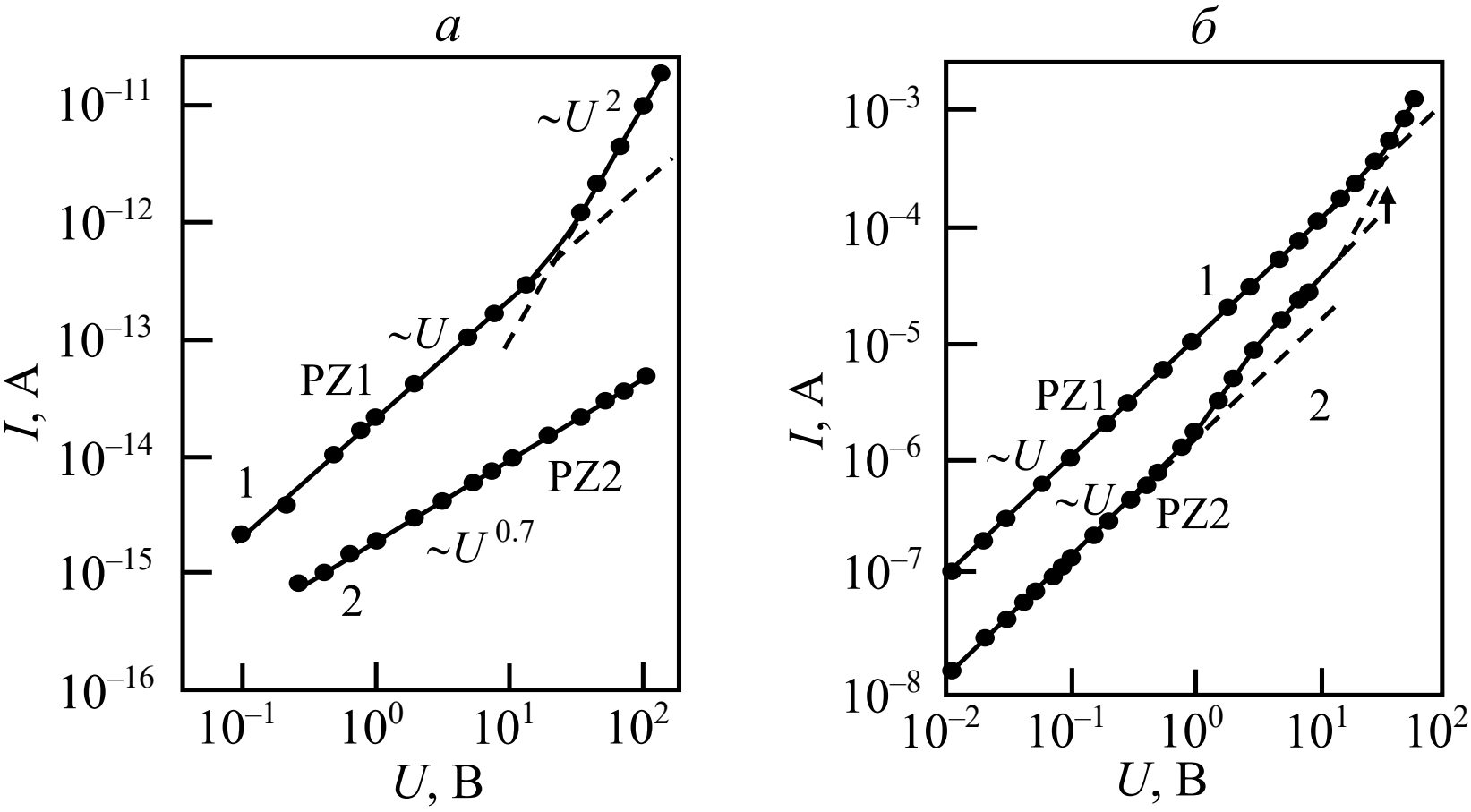

Рис. 4.3. Вольт-амперні характеристики кристалів $Si_2Te_3$ із золотими контактами при *а*) 296 K і *б*) 575 K (паралельно осі *c*). PZ1– омічна поведінка, PZ2 – не омічна поведінка [43].

З метою вивчення впливу водяної пари на ВАХ кристалів $Si_2Te_3$ авторами [43] були приготовлені два зразки (умовно позначені PZ1 і PZ2) шляхом розколу одного монокристала. Для зразка PZ1 золоті контакти були нанесені на чисту природну поверхню (001) сколу. У разі другого PZ2 зразка золоті контакти напилялися на поверхню сколу після попереднього перебування зразка в атмосфері повітря протягом кількох секунд. Вольт-амперні характеристики обох зразків, виміряні при двох температурах 296 і 575 К, наведені на рис. 4.3. Для зразка PZ1 ВАХ лінійна при малих електричних полях та квадратична ($I \sim U^2$) при великих полях (крива 1, рис. 4.3, *а*). Таким чином, якщо золоті контакти напилялися на чисту поверхню (001) кристала $Si_2Te_3$, то характер ВАХ такий самий, як і у випадку з вісмутовими контактами (порівняй рис. 4.3. і рис. 4.1). У разі, якщо золоті контакти нанесені на поверхню кристала, після перебування останнього протягом певного часу в лабораторній атмосфері, то характер ВАХ змінюється і вона стає сублінійною ($I \sim U^{0.7}$) (крива 2, рис. 4.3, *а*) при кімнатній температурі. Зі збільшенням температури зразка PZ2 до 575 К ВАХ стає лінійною в області малих електричних полів, як і у випадку зразка PZ1 (рис. 4.3, *б*).

**4.1.2. Ефект перемикання у шаруватих кристалах $Si_2Te_3$.** При певному значенні електричного поля, прикладеного перпендикулярно до шарів (тобто вздовж осі *с*), у кристалах $Si_2Te_3$ спостерігається ефект перемикання з високоомного стану в низькоомний, тобто наявна S-подібна вольт-амперна характеристика (рис. 4.4, *а*) [43]. Ефект перемикання з високоомного в низькоомний стан спостерігається також і в інших шаруватих кристалах халькогенідів IV групи: $GeSe_2$ і GeS [170, 171]. Для прикладу на рис. 4.4, *б* приведені ВАХ кристалічного і склоподібного $GeSe_2$ [170].

При відносно слабких полях ВАХ описується законом Ома, а починаючи з певної напруги, характерної для конкретного кристала, лінійність ВАХ порушується і переходить у залежність $I \sim U^n$, з показником $n \approx 2$. При подальшому збільшенні прикладеної напруги $n$ збільшується і досягає значення 4–10 у різних кристалах. При напруженості електричного поля $E \sim 10^4$ В/см спостерігається стрибкоподібний перехід зразка з високоомного в низькоомний стан. При цьому опір зразка зменшується на кілька порядків, і в низькоомному стані струм обмежується опором навантаження. Найчастіше ВАХ симетричні. Слід зазначити, що у всіх випадках ефект перемикання запам'ятовуючий і число перемикань залежить від параметрів кристала. Зразки знаходяться тривалий час як у високоомному стані, так і

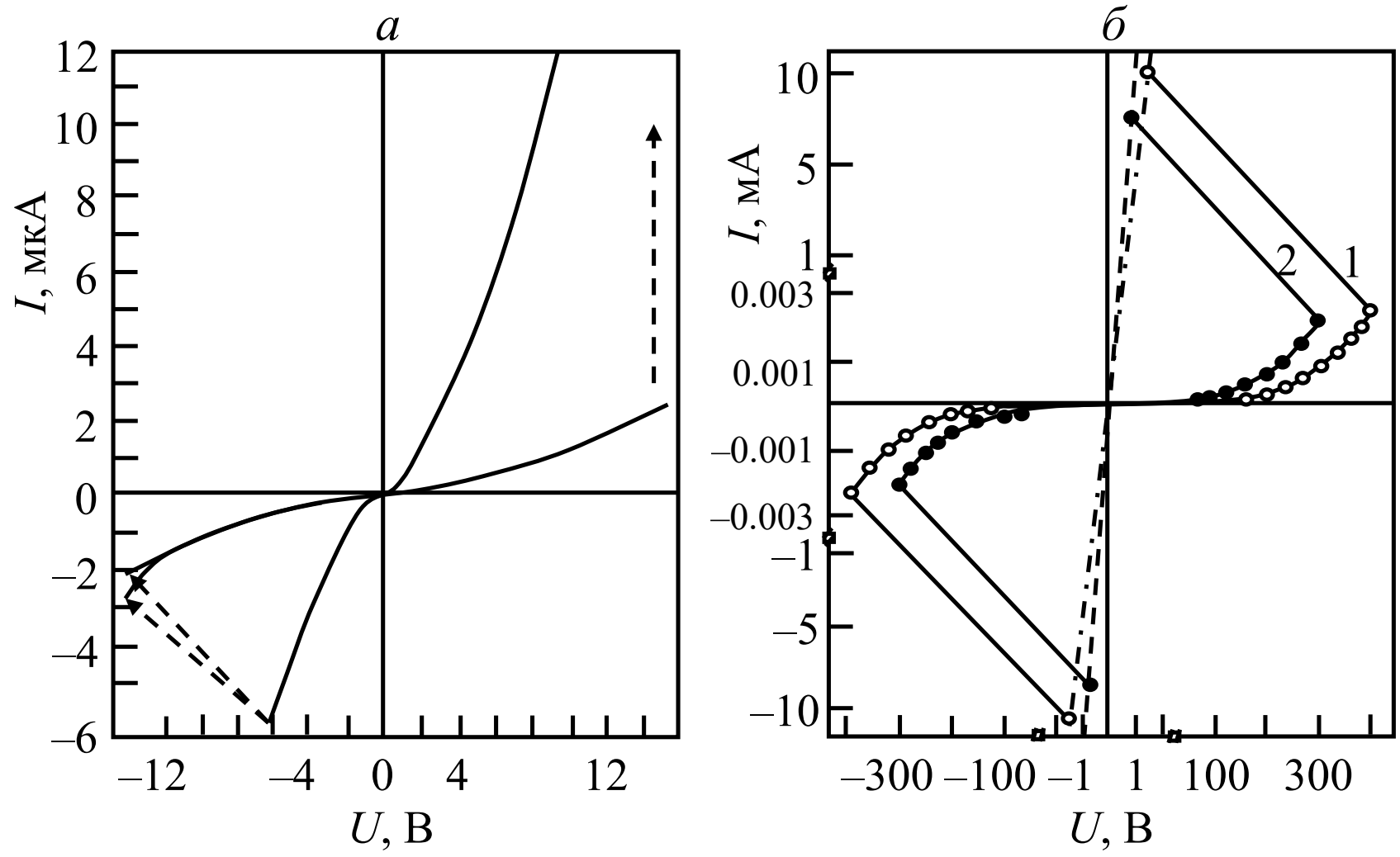

Рис. 4.4. Вольт-амперні характеристики шаруватих кристалів:
*а*) – $Si_2Te_3$ [43]; *б*) – $GeSe_2$ (1) і скла (2) [170].

в стані електричної пам'яті. Стан електричної «пам'яті» можна «стерти» подачею імпульсу напруги будь-якої полярності, амплітуда якого рівна або вища за порогове значення напруги $U_{пер}$.

**4.1.3. Ефект перемикання у нановіскерах $Si_2Te_3$.** Нановіскери (НВ) $Si_2Te_3$, вирощені уздовж напрямку [0001], демонструють унікальну поведінку зворотного перемикання опору з високоомного в низькоомний стан, обумовленого прикладеним електричним полем [59]. Цей перемкнутий стан є стабільним, за умови якщо тільки зворотний потенціал не прикладений для зворотного перемикання опору. Схема установки для дослідження ВАХ нановіскерів $Si_2Te_3$ наведена на вставці на рис. 4.5, *а* [59]. Оскільки НВ є дуже крихкими, важливим є нанесення омічних контактів. Одним контактом служить шар золота, нанесений на підкладку, на якій здійснювався процес зростання НВ. Для забезпечення другого Ga контакту, нановіскер разом з підкладкою, на якій він вирощений, розміщується у вертикальне положення у напрямку до низу. Скляна підкладка, з нанесеним рідким галієм, повільно переміщується вгору до приведення в контакт рідкого Ga з нановіскером.

На рис. 4.5, *а* показано зміну струму ($U$ = 0.1 В) в процесі переміщення неновіскера в напрямку Ga електрода. У відсутності контакту між НВ і Ga електродом початковий струм дорівнює нулю. По мірі наближення Ga електрода до НВ $Si_2Te_3$ струм стрибкоподібно

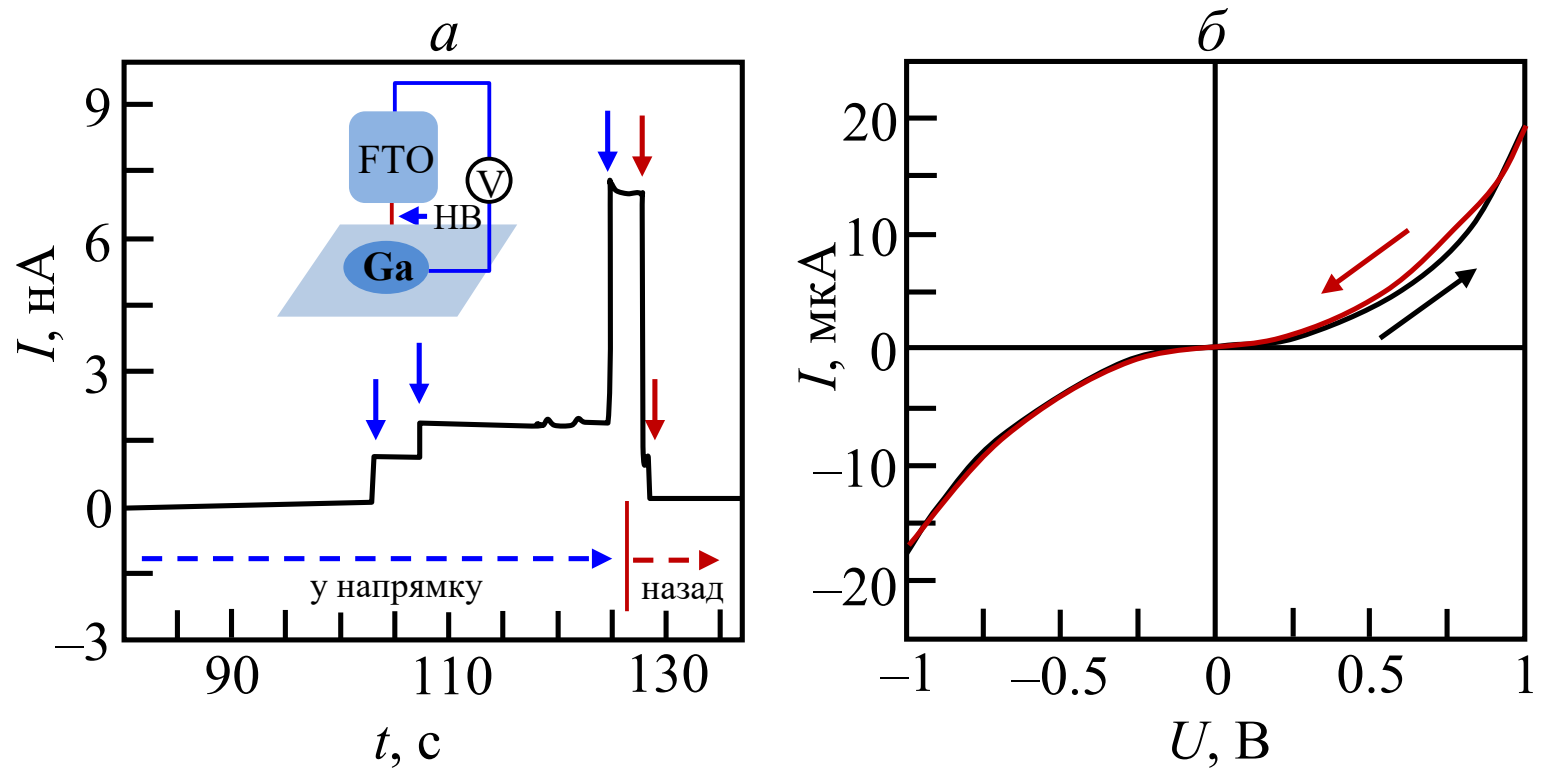

Рис. 4.5. (*а*) Зміни струму для нановіскера $Si_2Te_3$ при напрузі 0.1 В, коли Ga електрод переміщується у напрямку і назад від НВ. На вставці (*а*) зображена схема установки для електричних вимірювань. (*б*) ВАХ нановіскера $Si_2Te_3$ [59].

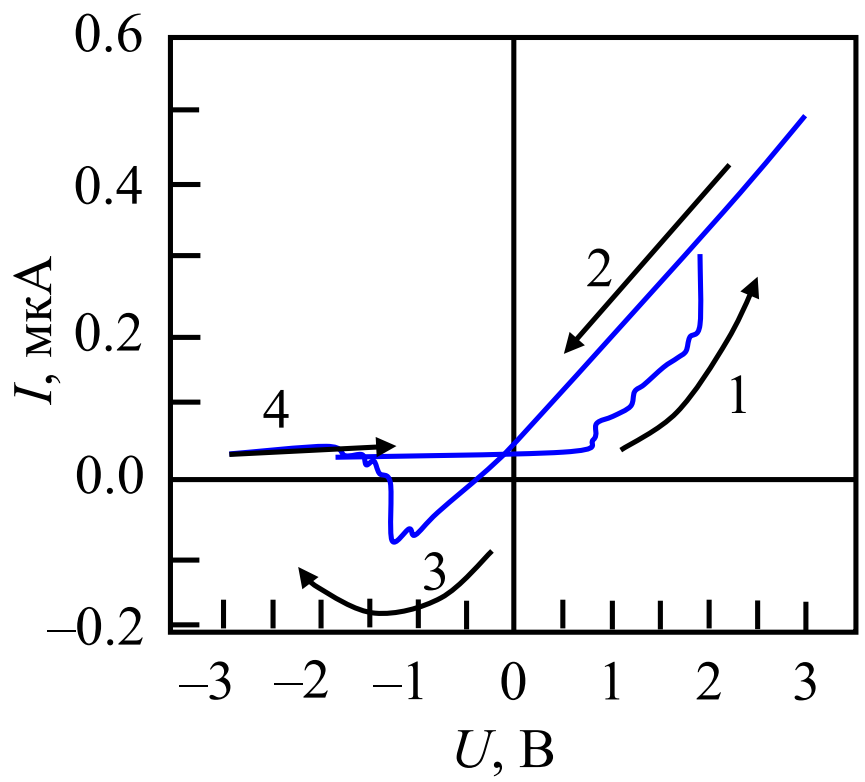

Рис. 4.6. Ефект перемикання в НВ $Si_2Te_3$. Додатня напруга відноситься до потенціалу на електроді з оксиду олова, легованого фтором. Послідовності розгортки напруги позначені стрілками 1, 2, 3 і 4 [59].

змінюється від 0 до ~0.9 нА, демонструючи тим самим наявність контакту НВ з Ga електродом. При наступному переміщенні Ga електрода до зразка НВ, струм спочатку залишається постійним, до поки не відбудеться другий стрибок струму від 0.9 до 1.8 нА із-за додаткового контакту НВ з Ga електродом. Подальше переміщення Ga електрода ближче до зразка НВ приводить до великого стрибка струму до 7 нА, який може бути викликаний великим віскером або групою НВ, що контактують з Ga електродом. По мірі віддалення контакту Ga електрода від НВ струм різко знижується до 0.9 нА і далі до нуля. Цей експеримент яскраво демонструє можливість встановлення електричного контакту з крихкими НВ, використовуючи рідкий електрод, та проводити вимірювання ВАХ для одного НВ $Si_2Te_3$. Як тільки забезпечується контакт НВ із Ga електродом, стає можливим дослідження ВАХ, результати якої приведені на рис. 4.5, *б*. Як видно з цього рисунка, ВАХ є нелінійною, і тим самим демонструє, що свіжовирощені НВ $Si_2Te_3$ є напівпровідниками. У діапазоні низьких напруг від –1 до 1 В крива ВАХ є зворотною і симетричною.

Як показано на рис. 4.6, ефект перемикання з високоомного в низькоомний стан у свіжоприготовленому НВ $Si_2Te_3$ відбувається в кілька етапів. Перший стрибок струму спостерігається при 0.9 В. При подальшому збільшенні напруги до 3 В спостерігається кілька стрибків струму при різних напругах. Коли напруга знижується з +3.0 до –3.0 В, низькоомний стан спочатку зберігається до тих пір, до поки буде прикладена від'ємна напруга – 1.1 В [59].

Ґрунтуючись на даних, отриманих в результаті першопринципних розрахунків , автори [59] дійшли висновку, що резистивне перемикання в НВ $Si_2Te_3$ відбувається внаслідок унікального фазового переходу між напівпровідниковими і металевими сегментами уздовж нанодротини. Під дією зовнішньої напруги, яка прикладається вздовж НВ, димери Si–Si в $Si_2Te_3$ дисоціюють внаслідок джоулевого тепла, в результаті один із двох атомів Si димера під дією електричного поля дифундує через бішар Te, в результаті чого $Si_2Te_3$ перебудовується в метастабільну металеву фазу. Автори [59] провели структурну оптимізацію моделі для реструктурованого $Si_2Te_3$ з використанням теорії функціоналу густини і показали, що ця структура дійсно метастабільна. Розрахунки електронної густини станів для реструктурованої фази, також підтверджують наявність металевої фази.

## 4.2. ЕЛЕКТРИЧНІ ВЛАСТИВОСТІ КРИСТАЛІВ $Si_2Te_3$

Сесквітелурид кремнію за рахунок відхилення від стехіометричного складу характеризується наявністю додатньо заряджених кремнієвих вакансій, що зумовлюють діркову провідність. Отже, для цього матеріалу справедливим є вираз $\sigma = e \cdot p \cdot u_p$. Результати дослідження електропровідності та її температурної залежності шаруватих кристалів *p*-типу $Si_2Te_3$ наведені у роботах [3, 172, 173]. Необхі-дно відзначити істотні відмінності у значеннях питомої електропровідності кристалів $Si_2Te_3$, отриманих різними авторами. Вона змінюється в широких межах від $10^{-1} \div 10^{-4}$ Ом$^{-1}$см$^{-1}$ в [3] до $10^{-13}$ Ом$^{-1}$см$^{-1}$ [172], що може бути пов'язано як із якістю досліджених кристалів, оскільки останні вирощувалися різними методами, так і, в значній мірі, зі станом поверхні кристалів, через їх сильну гігроскопічність.

**4.2.1. Електричні властивості полікристалічного $Si_2Te_3$.** Питомий опір спеціально нелегованого полікристалічного $Si_2Te_3$, отриманого методом загартування-відпалу, становить $10^2$–$10^3$ Ом·см при кімнатній температурі [3]. Відхилення від стехіометричного складу в бік надлишку одного з компонентів у кількості 1 ат. % приводить до збільшення питомого опору до $10^4$–$10^5$ Ом·см.

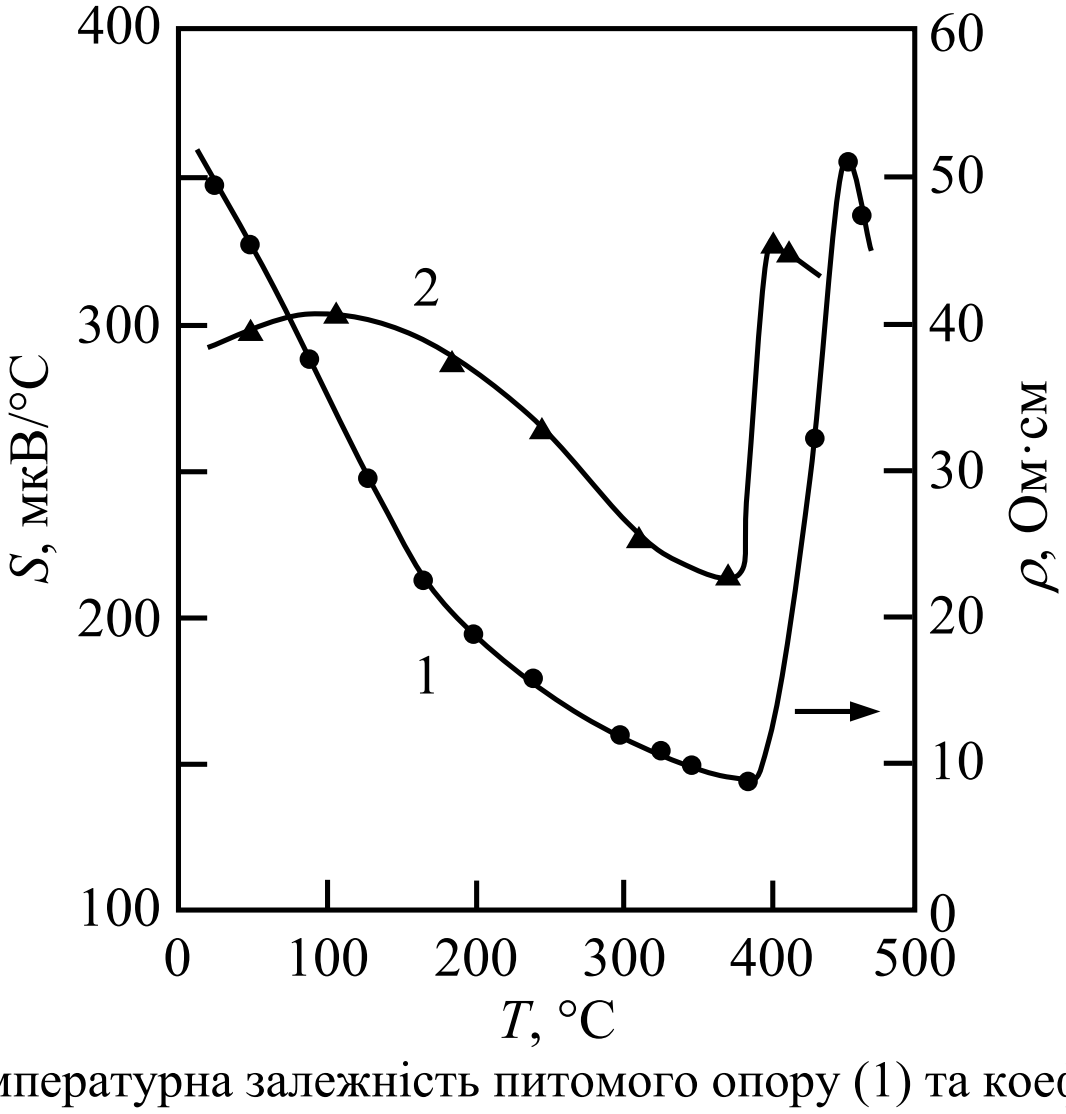

Рис. 4.7. Температурна залежність питомого опору (1) та коефіцієнта Зеебека (2) полікристалічного $Si_2Te_3$ [3].

Автор [3] досліджував також вплив різних домішок, введених у процесі синтезу, на електричні властивості полікристалічного $Si_2Te_3$. Так, введення домішок Cu, Fe, As, Cd і Ga в кількості $1 \cdot 10^{20}$ ат/см$^3$, супроводжується різким збільшенням питомого опору до $10^5$–$10^8$ Ом·см. Введення домішок Sb і Sn у кількості до $1 \cdot 10^{21}$ ат/см$^3$ приводить до різкого зменшення питомого опору до 1 Ом·см. Однак, металографічні дослідження зразків $Si_2Te_3$:Sb(Sn) виявили в них наявність другої фази. Тому, автор [3] вважає, що низький опір цих зразків не є «чистим» ефектом впливу домішок сурми і олова. Спе-ціально нелегований та легований різними домішками полікристалічний $Si_2Te_3$ є напівпровідником *p*-типу провідності.

Теплопровідність полікристалічного $Si_2Te_3$ складає 4 – 5 мВт/°С·см при кімнатній температурі. На рис. 4.7 наведені температурні залежності коефіцієнта Зеебека та питомого опору полікристалічного нелегованого $Si_2Te_3$. На цьому рисунку чітко видно різке збільшення обох параметрів поблизу 400 °С, що пов'язано з фазовим переходом, встановленим за допомогою рентгеноструктурного аналізу [3].

**4.2.2. Вплив атмосфери повітря та подальшого відпалу на електропровідність кристалів $Si_2Te_3$.** Кінетику впливу вологого повітря на процес поверхневого розкладання кристалів $Si_2Te_3$, вирощених методом сублімації, ілюструє рис. 4.8, на якому представлена залежність темнового струму від часу перебування досліджуваного зразка в атмосфері повітря [172]. Як видно із рис. 4.8, при прикладанні постійної напруги до досліджуваного зразка сила струму зростає на кілька порядків протягом перших хвилин після впливу парів води, наявних у повітрі, і має тенденцію до насичення після кількох годин. Для відновлення вихідної провідності кристала необхідно

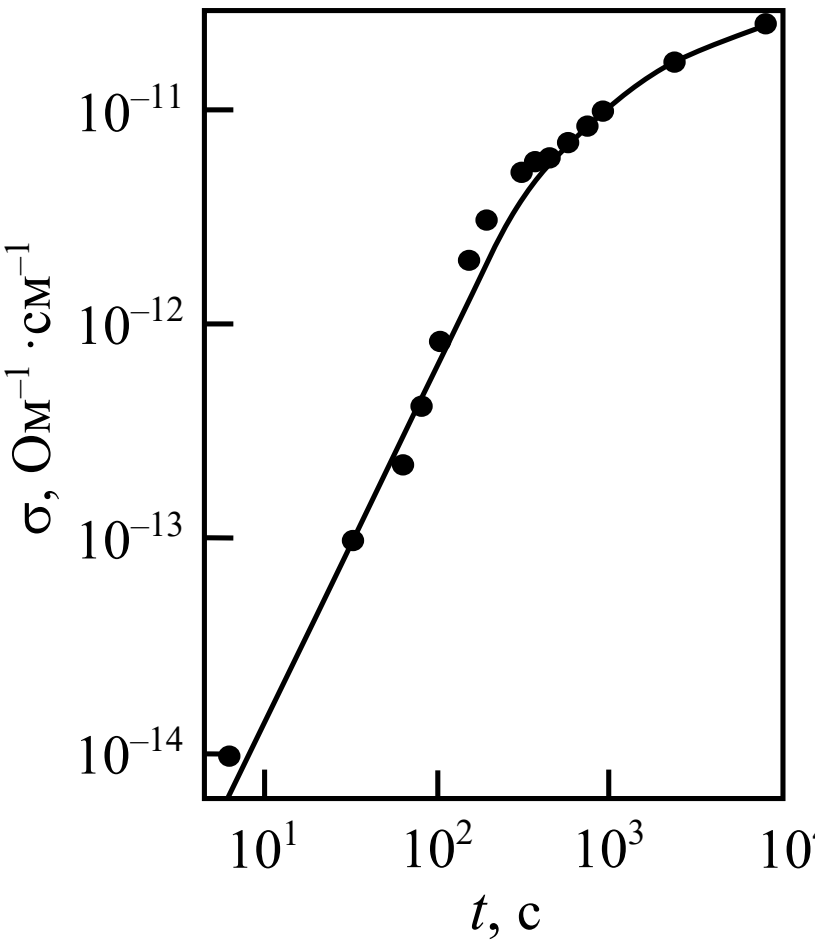

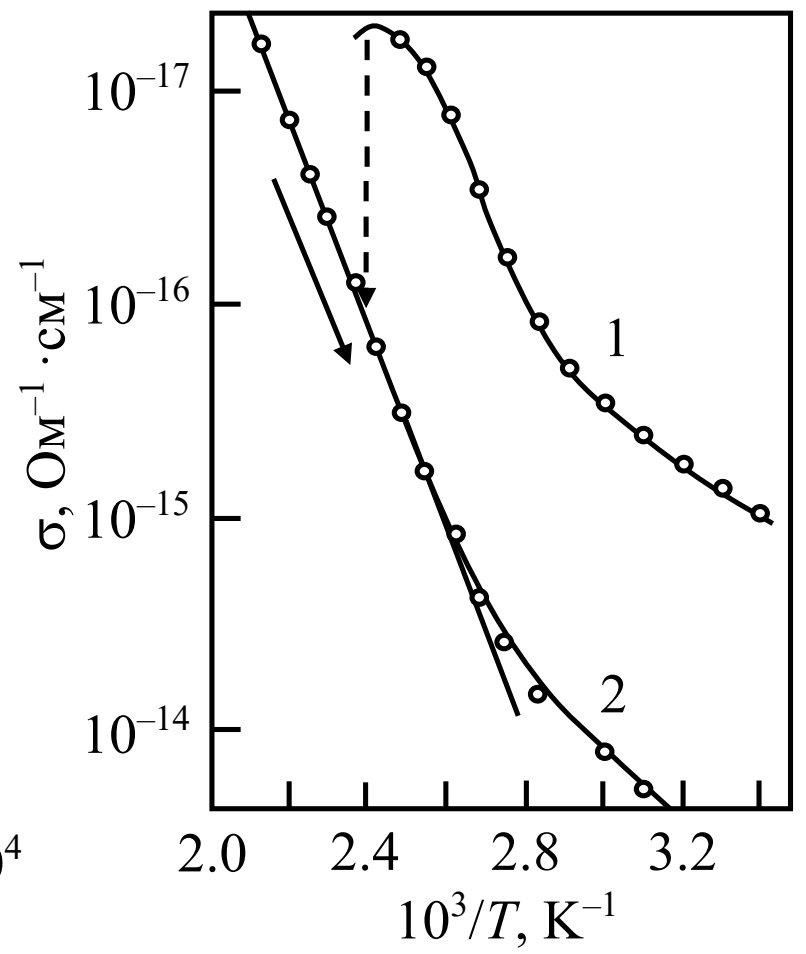

Рис. 4.8. Залежність провідності кристала $Si_2Te_3$ від часу впливу повітря на зразок [172].

Рис. 4.9. Температурна залежність провідності кристала $Si_2Te_3$ до попереднього нагрівання зразка у високому вакуумі (крива 1) та після сублімації Te з поверхні при 400 K (крива 2) [172].

видалити поверхневий шар телуру в процесі нагрівання зразка у вакуумі $10^{-5}$ Торр. Цей процес ілюструє крива 1 на рис. 4.9, що відображає температурну залежність провідності кристала з шаром телуру; при 400 К поверхневий шар телуру починає сублімувати, що супроводжується зменшенням провідності (крива 2, рис. 4.9). Після повного видалення шару телуру з поверхні зразка, гістерезис на температурній залежності провідності зникає і крива нагрівання – охолодження збігається, повторюючи провідність об'ємного кристала [172]. Таким чином, попередній відпал зразків у вакуумі дозволяє стабілізувати стан поверхні, виключити вплив поверхневої провідності й отримати результати, що характеризують об'ємні властивості кристалів $Si_2Te_3$.

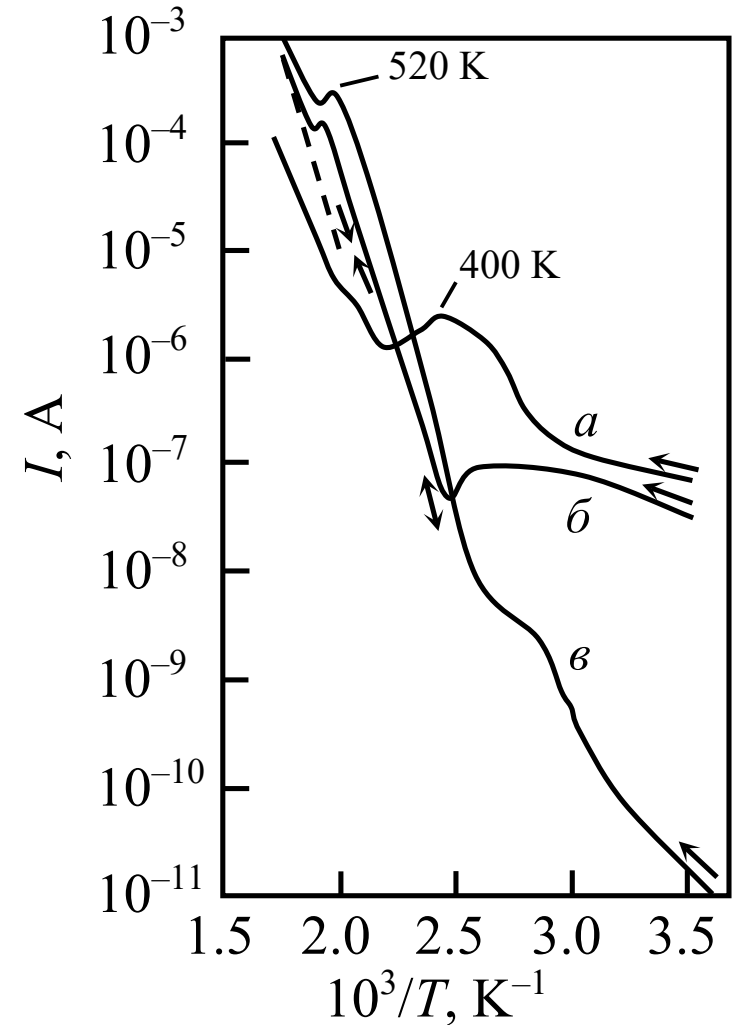

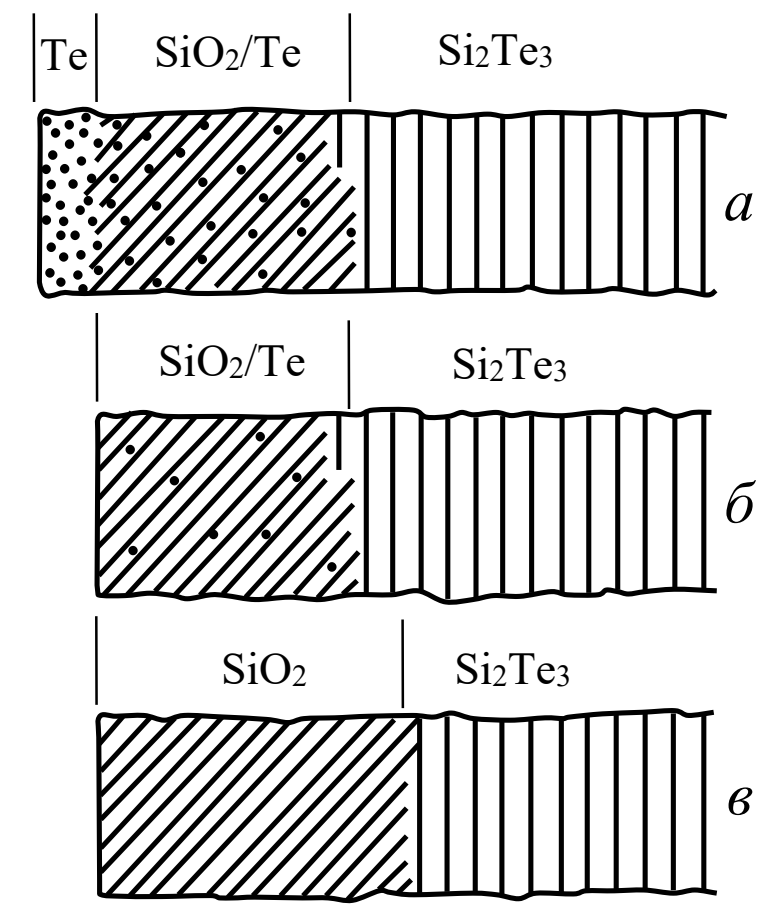

Рис. 4.10. Температурні залежності темнового струму в процесі термічної обробки кристалів $Si_2Te_3$, які знаходились на повітрі протягом різного часу (*а* – 1 хв, *б* – 5 хв, *в* – 1 год). Швидкість нагрівання: *а* – 3; *б* – 0.7; *в* – 0.7 К/хв. [43].

Рис. 4.11. Схема структури поверхневого шару кристала $Si_2Te_3$:
*а*) після експозиції у вологому повітрі протягом кількох тижнів,
*б*) після 2 год. термічної обробки експонованого зразка при 500 К,
*в*) після 4 год. термообробки відкритого зразка при 650 К [43].

Згодом автори [43] докладніше дослідили вплив термовідпалу на електропровідність кристалів $Si_2Te_3$, попередньо витриманих у вологому повітрі. На рис. 4.10 наведено температурні залежності тем-нового струму трьох зразків, що знаходились на повітрі протягом різного часу (*а* – 1 хв, *б* – 5 хв, *в* – 1 год). Для всіх трьох зразків на кривих залежності $I = f(10^3/T)$ при $T$ = 400 К спостерігається особливість у вигляді максимуму, інтенсивність якого залежить від часу перебування кристала на повітрі до початку вимірювань. Чим менший час перебування зразка на повітрі, тим стає інтенсивнішим максимум поблизу $T$ = 400 К. Для зразків (*б* і *в*) з підвищенням температури на залежності $I = f(10^3/T)$ проявляється другий максимум при 520 К. Після того як зразки були відпалені при 520 К їх електропровідність не змінювалася після перебування у вологому повітрі.

Ґрунтуючись на цих результатах, автори [43] зробили висновок, що відпал кристалів, попередньо витриманих на повітрі, викликає пасивацію поверхні. На рис. 4.11 показана схема трансформації поверхневого шару в процесі термовідпалу, запропонована авторами [43]. На першому етапі при 400 К відбувається випаровування верхнього шару атомарного Te, а з підвищенням температури до 520 К відбувається випаровування Te з другого поверхневого шару, що складається з суміші Te і $SiO_2$. В результаті поверхня кристала залишається захищеною шаром $SiO_2$.

**4.2.3. Анізотропія електропровідності шаруватих кристалів $Si_2Te_3$.** Анізотропія кристалічної структури шаруватих кристалів приводить до анізотропії сил зв'язку, що, звісно, зумовлює анізотропію фізичних властивостей, зокрема й електропровідності. Шаруваті кристали $Si_2Te_3$, як і більшість інших шаруватих сполук, відомих у системах $A^{IV}$–$B^{VI}$, є сильно анізотропними. Для вимірювання анізотропії електропровідності та її температурної залежності автори [173] напиляли золоті контакти на протилежні грані кристалів $Si_2Te_3$, вирощених газотранспортним методом, перпендикулярно до осі *c*. Оскільки після напилення золотих контактів провідність зразків збільшувалася на три порядки, тому зразки відпалювали при 570 К протягом 30 годин. Відпал зразків у вакуумі дозволив стабілізувати стан поверхні, виключити вплив поверхневої провідності та отримати результати, що характеризують об'ємні властивості досліджених кристалів.

Температурні залежності електропровідності шаруватих кристалів $Si_2Te_3$, виміряні вздовж шарів ($\sigma_{\perp c}$, тобто перпендикулярно до осі *c*), для свіжосколотого зразка, а також для іонно обробленого та відпаленого зразка в координатах Арреніуса наведені на рис. 4.12 (крива 1). Хороше узгодження результатів для обох зразків вказує, що іонне травлення повністю видаляє плівки телуру та $SiO_2$. Темнова

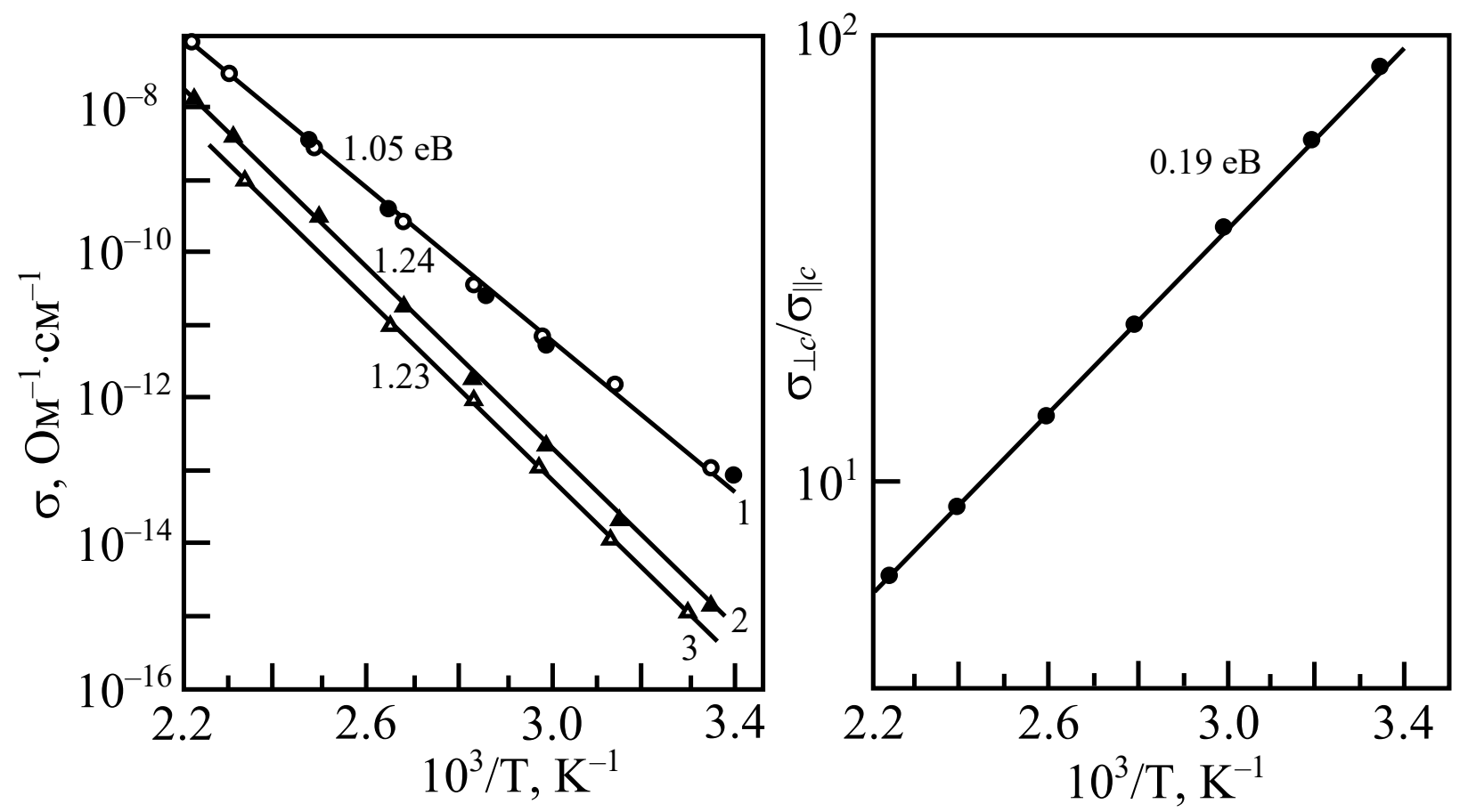

Рис. 4.12. Температурна залежність електропровідності кристалів $Si_2Te_3$ [173] (1) ○ $\sigma_{\perp c}$ – зразок оброблений іонним пучком аргону з наступним відпалом,● $\sigma_{\perp c}$ – свіжосколотий зразок;
(2)▲ $\sigma_{||c}$ – зразок оброблений іонним пучком аргону з послідуючим відпалом, (3) △ $\sigma_{||c}$ – результати, отримані в [43].

Рис. 4.13. Температурна залежність анізотропії електропровідності шаруватих кристалів $Si_2Te_3$ [173].

провідність, виміряна вздовж шарів $\sigma_{\perp c}$, при кімнатній температурі становить $10^{-13}$ Ом$^{-1}$·см$^{-1}$.

Температурна залежність електропровідності іонно обробленого зразка, виміряна перпендикулярно шарам ($\sigma_{||c}$), представлена кривою 2 на рис. 4.12. Для порівняння на рис. 4.12 (крива 3) наведено результати, отримані в роботі [43]. При кімнатній температурі темнова провідність кристалів $Si_2Te_3$ вздовж осі *c* становить $10^{-15}$ Ом$^{-1}$·см$^{-1}$, тобто на два порядки нижча ніж перпендикулярно до осі *c*. Таким чином, шаруваті кристали $Si_2Te_3$ характеризуються сильною анізотропією електропровідності $\Lambda = \sigma_{\perp c} / \sigma_{||c}$, яка становить $10^2$ при кімнатній температурі й зменшується на порядок зі збільшенням температури зразка до 450 К (рис. 4.13) [172]. Термічна енергія активації $E_a$ електропровідності $\sigma_{\perp c}$, виміряної вздовж шарів, рівна 1.05 еВ, а енергія активації провідності, виміряної перпендикулярно до шарів, тобто вздовж осі *c*, рівна 1.24 еВ.

Таким чином, анізотропія провідності на постійному струмі у шаруватих кристалах $Si_2Te_3$ описується виразом:

$$\sigma_{\perp c} / \sigma_{\parallel c} = A \cdot \exp\left(\Delta E_a / kT\right), \tag{4.1}$$

де $A$ рівне відношенню ефективних мас носіїв електричного заряду, а $\Delta E_a$ – енергія активації переносу поперек шарів.

Анізотропію провідності у шаруватих кристалах зазвичай вважають відображенням сильної анізотропії ефективних мас, оскільки $A = m_{\perp c} / m_{\parallel c}$. У цьому випадку передбачається, що взаємодія носіїв заряду з коливаннями ґратки приводить до локалізації їх в окремому тришаровому пакеті. При цьому перенесення носіїв заряду може здійснюватися звичайною зонною провідністю вздовж шарів і стрибкоподібним механізмом поперек шарів.

Розраховані дисперсійні криві (рис. 3.2, розділ 3) були використані нами для оцінки величин ефективних мас носіїв заряду в напрямках вздовж ($m_{\perp c}$) та поперек ($m_{\parallel c}$) шарів у кристалах $Si_2Te_3$. Відношення ефективних мас електронів і дірок у кристалах $Si_2Te_3$ із зазначенням відповідних напрямів у оберненому просторі зони Бріллюена становить $m_e^{Г-K} / m_e^{Г-A} = 0.51$ і $m_e^{Г-L} / m_e^{Г-A} = 0.84$ для електронів, і $m_h^{Г-K} / m_h^{Г-A} = -0.84$ і $m_h^{Г-L} / m_h^{Г-A} = -1.29$ для дірок. Таким чином, ефективні маси електронів і дірок у кристалах $Si_2Te_3$ практично ізотропні.

Враховуючи незначну відмінність величин ефективних мас дірок вздовж і поперек шарів, пояснити сильну анізотропію електропровідності $\sigma_{\perp c} / \sigma_{\parallel c} \sim 10^2 – 10^4$ [172, 173] кристалів $Si_2Te_3$ $p$-типу провідності, виключно анізотропією ефективних мас неможливо. На теперішній час встановлено, що анізотропія провідності, що не відповідає анізотропії ефективних мас носіїв заряду, характерна практично для більшості шаруватих кристалів, наприклад, для шаруватих напівпровідників $SnS_2$ $\sigma_{\perp c} / \sigma_{\parallel c} \sim 10^4$ [174] і InSe $\sigma_{\perp c} / \sigma_{\parallel c} \sim 10^2 – 10^3$ [175], для шаруватого металу $NbSe_2$ $\sigma_{\perp c} / \sigma_{\parallel c} \sim 10^2$ [176] та для шаруватого напівметалу графіту $\sigma_{\perp c} / \sigma_{\parallel c} \sim 10^3 – 10^5$ [177].

Висока анізотропія електропровідності шаруватих кристалів $Si_2Te_3$ не є також наслідком двомірності зонної структури, оскільки виконані теоретичні розрахунки (рис. 3.2, розділ 3) та експериментальні дані дослідження краю оптичного поглинання [140] вказують на тривимірний характер енергетичних зон. Враховуючи, що зонна структура шаруватих кристалів, всупереч очікуваному, далека від двовимірної і нагадує швидше зонну структуру тривимірних кристалів, Еванс і Юнг [178] показали, що в цьому випадку анізотропію електропровідності, обумовлену відношенням ефективних мас, можна очікувати на рівні між 1 і 10. Отже, у досконалих шаруватих кристалах не повинна спостерігатися сильна анізотропія, яка має місце в реальних кристалах $Si_2Te_3$ і досягає $\sigma_{||c} / \sigma_{\perp c} \sim 10^2–10^4$.

Сильна анізотропія електропровідності в реальних шаруватих кристалах, у тому числі і $Si_2Te_3$, обумовлена наявністю в них різного роду протяжних дефектів, що викликають утворення потенціальних бар'єрів для руху носіїв заряду поперек шарів. Виникненню численних плоских дефектів, дефектів упаковки шарів, гвинтових дислокацій сприяє слабкий міжшаровий ван-дер-ваальсовий зв'язок у шаруватих кристалах. Наявність дефектів упаковки в реальних шаруватих кристалах $Si_2Te_3$ [24] приводить до порушення трансляційної інваріантності в напрямку, перпендикулярному до шарів, у той час як вздовж шарів трансляційна інваріантність зберігається. Тому реальні шаруваті кристали слід розглядати як структури з одновимірним розупорядкуванням вздовж кристалографічної осі *c*. Теоретичні викладки впливу одновимірної невпорядкованості шаруватих напівпровідників у напрямку, перпендикулярному до шарів, на величину статичної електропровідності наведені в роботах [179–181]. У цій моделі вважається, що безлад є недіагональним і задається розподілом матричних елементів перескоків електронів з шару в шар, а сам механізм статичної провідності в напрямку, перпендикулярному до шарів, описується одновимірною моделлю перескоків провідності. Такий характер провідності вздовж осі *c* пов'язаний з локалізацією хвильових функцій електронів у кінцеве число шарів, викликаної дефектами упаковки.

За аналогією з аморфними напівпровідниками, феноменологічний опис властивостей перенесення в шаруватих кристалах можна здійснити, ввівши край рухливості ($E_{perc}$) для провідності поперек шарів (рис. 4.14) [179, 180]. Внаслідок поширеного характеру блохівських хвильових функцій вздовж шарів, енергія активації переносу в режимі постійного струму в цьому напрямку дорівнює $E_a = E_0 - E_{perc}$, як у звичайному напівпровіднику, а поперек шарів ефективна енергія активації $E_a = E_{perc} - E_F$ (рис. 4.14 ). Тоді $E_{perc} - E_0 = \Delta E$ є експериментально визначена енергія активації переносу в режимі постійного струму поперек шарів (вираз 4.1). Автори [181] вважають, що енергія $\Delta E$ безпосередньо пов'язана з концентрацією дефектів упаковки, які значною мірою залежать від легування. Необхідно відзначити, що дефекти упаковки в шаруватих структурах не впливають на

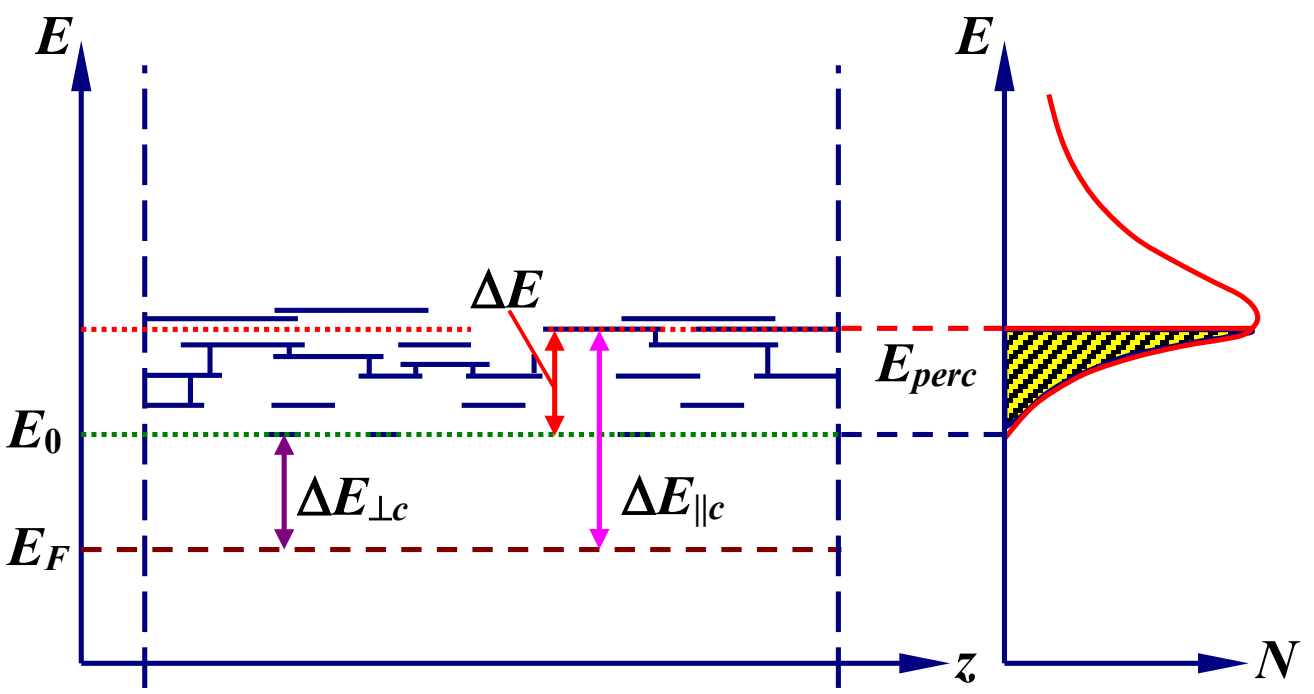

Рис. 4.14. Модель переносу носіїв заряду поперек шарів. Просторовий та енергетичний розподіл електронних станів зображено горизонтальними лініями. Вертикальні лінії вказують напрямки можливих перескоків носіїв заряду [179].

взаємодії між атомами – іншими сусідами, які належать сусіднім шарам. Тому енергія утворення дефектів упаковки досить мала. Ці плоскі дефекти представляють ефективний механізм релаксації напруги зсуву, паралельно площині шарів досконалого (ідеального) кристала. Пластична деформація збільшує концентрацію дефектів упаковки. Другий важливий момент: пластична деформація одночасно збільшує значення $\Delta E$. Отже, високі значення $\Delta E_a$, що спостерігалися в нелегованих шаруватих кристалах $Si_2Te_3$ пов'язані з великою концентрацією дефектів упаковки, експериментально виявлених авторами [24].

## 4.3. ЕЛЕКТРОПРОВІДНІСТЬ НА ПОСТІЙНОМУ СТРУМІ ХАЛЬКОГЕНІДНИХ СКЛОПОДІБНИХ НАПІВПРОВІДНИКІВ

Інтерпретація даних з електричних явищ переносу в аморфних і склоподібних халькогенідних напівпровідниках тісно пов'язана з розподілом густини станів: вузьких хвостів локалізованих станів у країв валентної зони і зони провідності, а також зони локалізованих рівнів поблизу середини забороненої зони (рис. 4.15). Відповідно до моделі Мота–Девіса [182] при інтерпретації експериментальних даних по електричним явищам переносу в аморфних і склоподібних напівпровідниках прийнято розглядати три механізми провідності: провідність по делокалізованим станам вище порогу рухливості, провідність у хвостах густини станів і провідність по локалізованим станам на рівні Фермі.

***Перенесення носіїв заряду, збуджених за край рухливості в делокалізовані стани з енергіями $E_c$ і $E_v$*** (рис. 4.15). У цьому випадку електропровідність описується виразом:

$$\sigma = \sigma_0 \exp(-E_a / kT). \qquad (4.2)$$

У випадку електронної провідності $E_a$ є різницею енергій нижнього краю зони провідності $E_c$ і рівня Фермі $E_F$, $E_a = E_c - E_F$. У випадку діркової провідності $E_a = E_F - E_v$, де $E_v$ – енергія верха валентної зони.

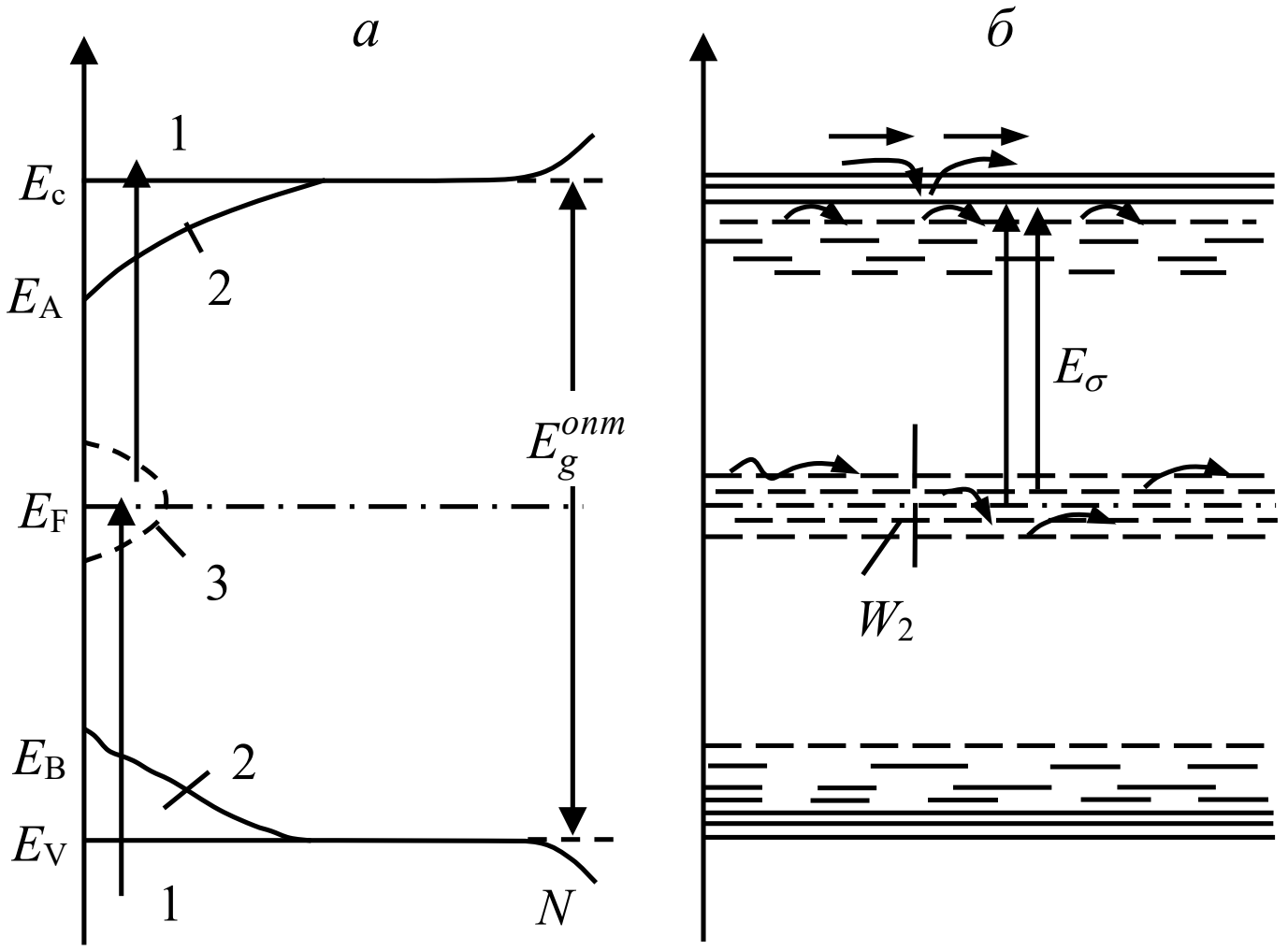

Рис. 4.15. Схема густини станів у ХСН та можливих трьох механізмів перенесення заряду [182]. 1 – збудження у зонні стани; 2 – збудження в локалізовані стани в області ЕЛОК; 3 – стрибки електронів в області енергії Фермі.

Дослідження краю власного поглинання склоподібних та аморфних напівпровідників показали, що заборонена щілина зменшується зі збільшенням температури. Енергетична різниця $E_c - E_F$ також повинна змінюватися, і в припущенні лінійної залежності від температури графік залежностей $ln\,\sigma$ від $1/T$ являє собою пряму лінію. В даному випадку можна записати:

$$E_c - E_F = E(0) - \gamma \cdot T, \qquad (4.3)$$

де $E(0)$ – ширина забороненої зони при $T = 0$ K, а $\gamma$ – її температурний коефіцієнт.

***Провідність у хвостах густини станів.*** Провідність, пов'язана з носіями, які збуджуються в локалізовані стани біля країв зон, тобто, поблизу $E_A$ і $E_B$ (рис. 4.15) здійснюється шляхом перескоків і описується виразом:

$$\sigma = \sigma \exp[-(E_A - E_F + W_1)/kT], \qquad (4.4)$$

де $W_1$ – енергія активації стрибків, яка повинна зменшуватися зі зменшенням температури, оскільки за своєю природою провідність має стрибковий характер зі змінною довжиною стрибка. Але й у цьому випадку слід очікувати лінійну залежність $\ln\sigma$ від $1/T$, оскільки основний внесок у температурну залежність провідності вносить множник, який визначає активацію носіїв. Провідність по локалізованих станах у хвостах дозволених зон характеризується активаційною залежністю рухливості

$$\mu \sim \exp(-\Delta E/kT), \qquad (4.5)$$

де $\Delta E$ – різниця енергій рівнів, між якими здійснюється перехід.

Таким чином, оскільки основний внесок у температурну залежність провідності вносить множник, що визначає активацію носіїв, знову слід очікувати приблизно лінійну залежність $ln\,\sigma$ від $1/T$.

***Провідність по локалізованим станам на рівні Фермі.*** За наявності скінченої густини станів поблизу рівня Фермі $N_F$ у провідність будуть давати внесок носії заряду з енергією поблизу $E_F$. Ці носії можуть здійснювати стрибки між локалізованими станами аналогічно тому, як це має місце у процесі домішкової провідності в сильно легованих і компенсованих напівпровідниках при низьких температурах. Цей внесок у провідность можна записати як

$$\sigma = \sigma_2 \exp(-W_2 / kT), \qquad (4.6)$$

де $\sigma_2 \leq \sigma_1$ і $W_2$ – енергія активації стрибка, величина якого близька до половини ширини зони локалізованих станів.

При зниженні температури число і енергія фононів зменшуються і стрибки з більшою енергією, що стимулюються фононами, стають все менш вигідними. Для носіїв стають вигідними стрибки на великі відстані, що дозволяють потрапити на вузли, що лежать ближче за енергією, ніж найближчі сусіди. Цей механізм називається стрибковим механізмом провідності зі змінною довжиною стрибка. Моттовський розрахунок стрибкової провідності зі змінною довжиною стрибка дає температурну залежність провідності наступного виду:

$$\sigma = \sigma_0 \exp[-(T_0/T)^{1/4}], \qquad (4.7)$$

де $T_0 = 16a^3/[kN_F)]; a = \hbar\sqrt{2m^*}$ – радіус локалізації, $\nu \approx 10^{12}$ – частота фононів.

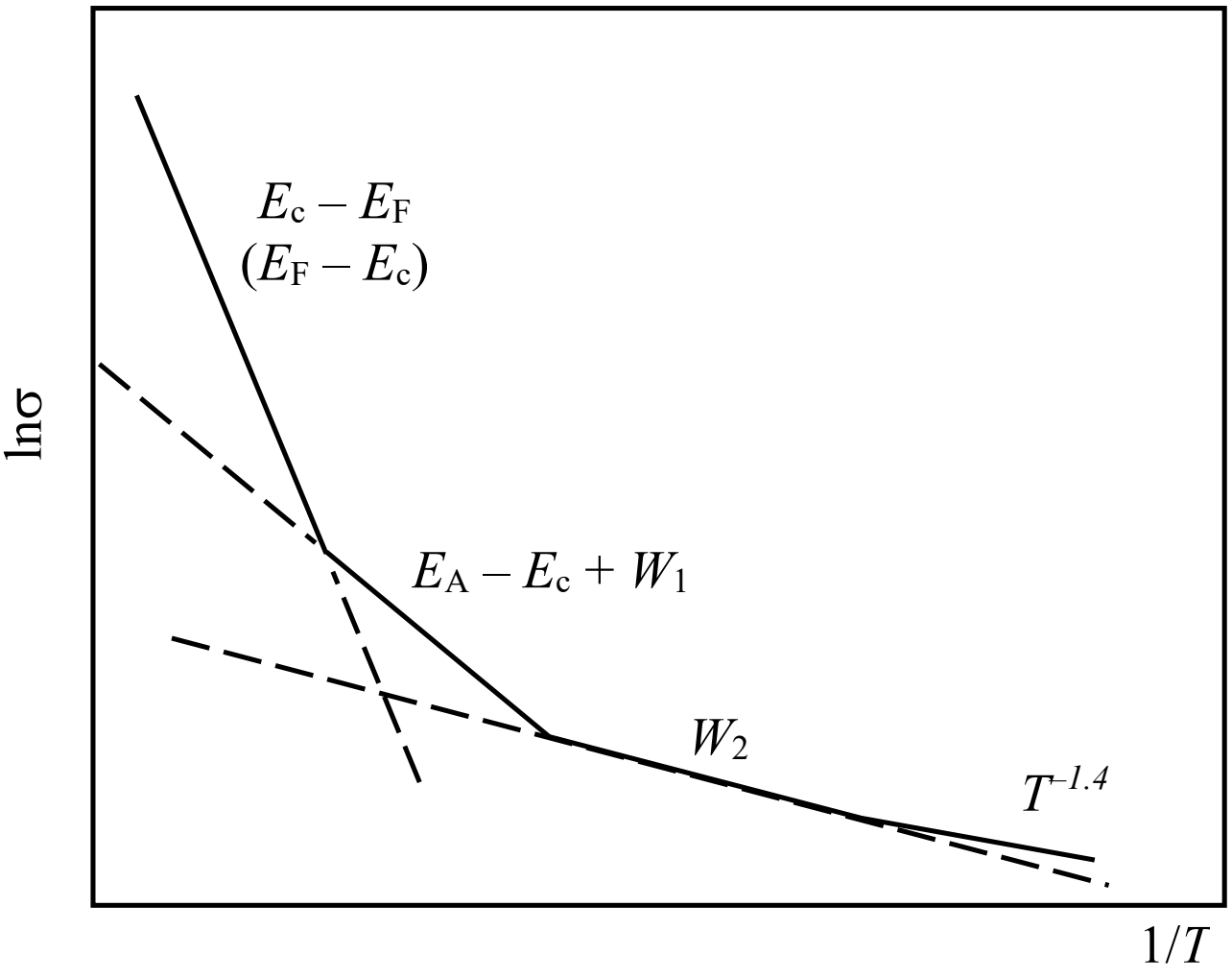

Рис. 4.16. Температурна залежність провідності ХСН [182].

На рис. 4.16 показана температурна залежність провідності, яку слід очікувати в ХСН виходячи з природи локалізованих станів у різних температурних інтервалах. Поблизу прямих ділянок температурної залежності провідності на рис. 4.16 вказані енергії активації, що відповідають розглянутим механізмам провідності.

## 4.4. ЕЛЕКТРОПРОВІДНІСТЬ СТЕКОЛ $Si_xTe_{100-x}$

**4.4.1. Електричні властивості об'ємних стекол $Si_xTe_{100-x}$.** Вперше електропровідність на постійному струмі склоподібних сплавів $Si_xTe_{100-x}$ ($15 \le x \le 25$) у широкому інтервалі температур 200–400 К дослідили автори [46]. Для прикладу на рис. 4.17. наведена температурна залежність питомого опору скла $Si_{20}Te_{80}$, яка описується залежністю $\rho = \rho_0 \exp(E_a/kT)$, з енергією активації $E_a$, яка змінюється від ~ 0.55 еВ при 400 К до 0.44 еВ при 200 К. Концентраційні залежності питомого опору ($\rho$), виміряного при кімнатній температурі та енергії активації ($E_a$), виміряної в температурному інтервалі 323–373 K, стекол $Si_xTe_{100-x}$ наведені на рис. 4.18. Згідно з даними вимірювання ефекту Холла, стекла $Si_xTe_{100-x}$ мають *p*-тип провідності зі значенням рухливості $u$ = 1 см$^2$·В$^{-1}$·с$^{-1}$ незалежної від температури в інтервалі 200–300 K [46].

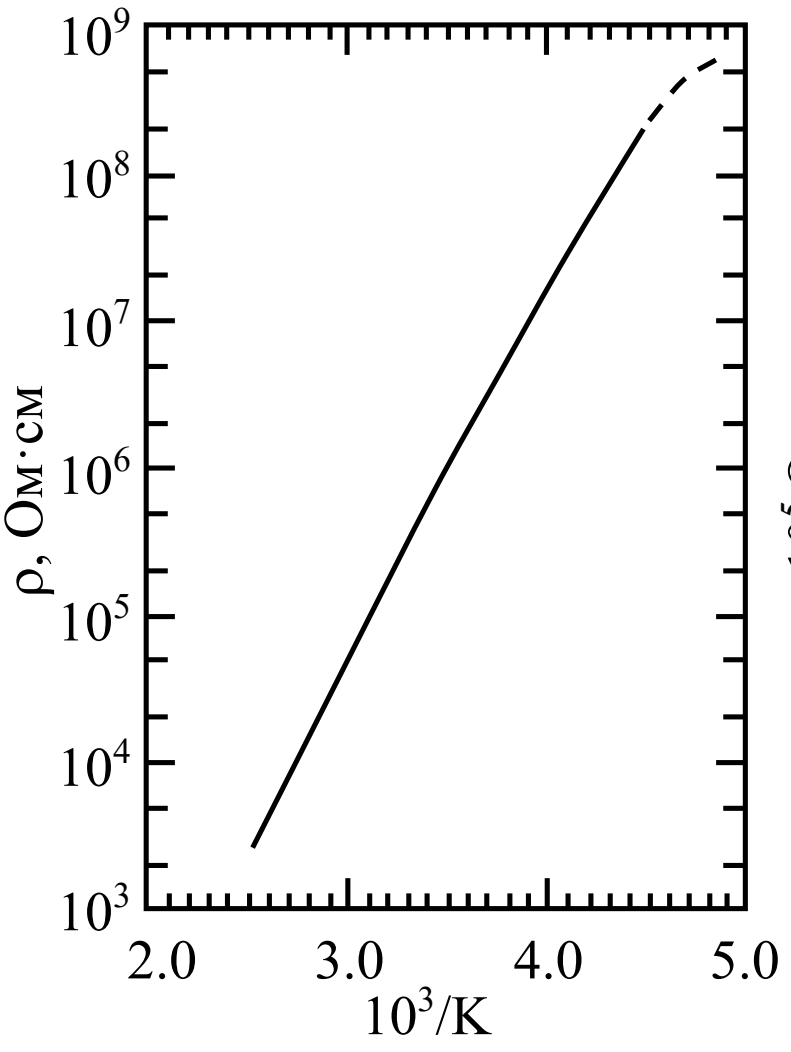

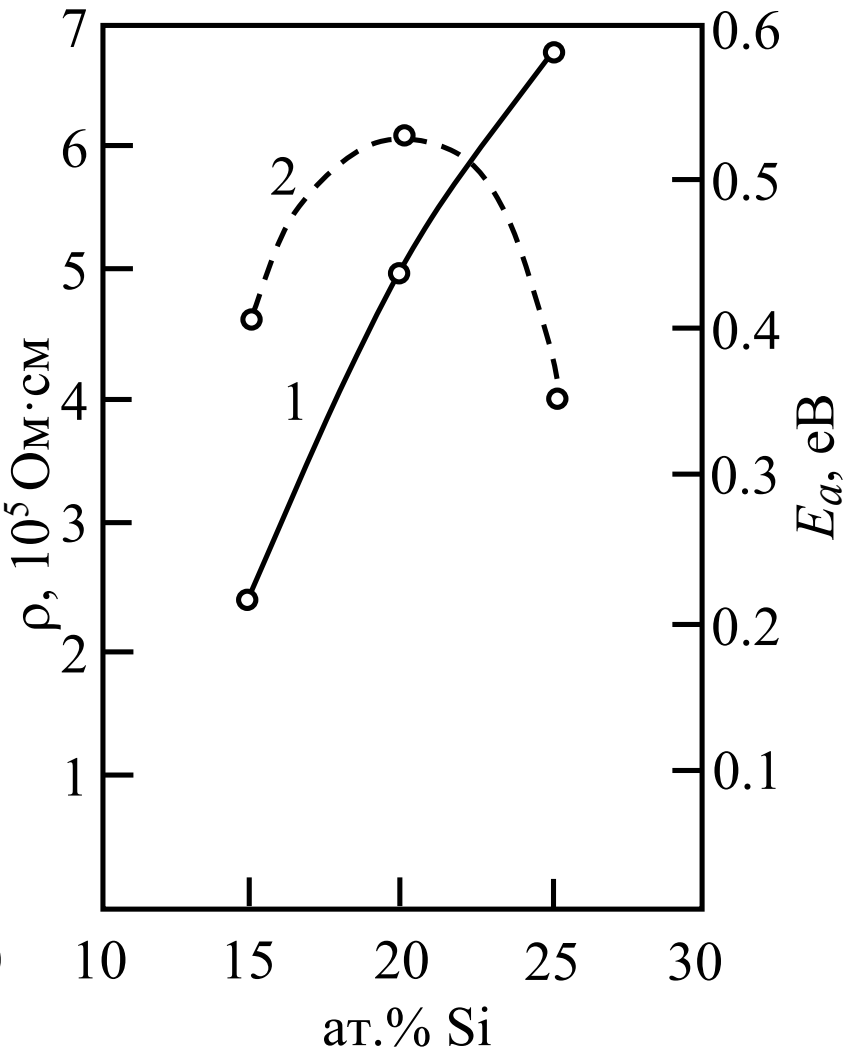

Рис. 4.17. Температурна залежність питомого опору скла $Si_{20}Te_{80}$ [46].
Рис. 4.18. Концентраційні залежності питомого опору (1) та енергії активації (2) стекол $Si_xTe_{100-x}$ [46].

Переважна більшість фізичних властивостей, у тому числі і електропровідність, стекол $Si_xTe_{100-x}$ ($10 \le x \le 27.5$) сильно залежить від умов їх отримання і насамперед від швидкості охолодження розплаву. На прикладі скла $Si_{20}Te_{80}$ автори [69, 183] за допомогою рентгеноструктурного аналізу встановили, що в результаті різних режимів охолодження розплаву його структура зазнає істотних змін: охолодження у крижаній воді дає аморфну структуру (однорідне скло без включень), а при охолодженні на повітрі в сітці скла виникає система нанокристалів телуру. Розмір кристалітів становить ~110 Å.

Використовуючи загартування розплаву від температури на ~523 K вище температури ліквідусу у воду з льодом (швидкість охолодження ≥ 473 K/с) автори [75] встановили для стекол $Si_xTe_{100-x}$ лінійну залежність питомого опору ρ, енергії активації електропровідності $E_a$, оптичної ширини забороненої зони $E_g^{опт}$ та мікротвердості від складу (рис. 4.19, криві 1). При менших швидкостях охолодження

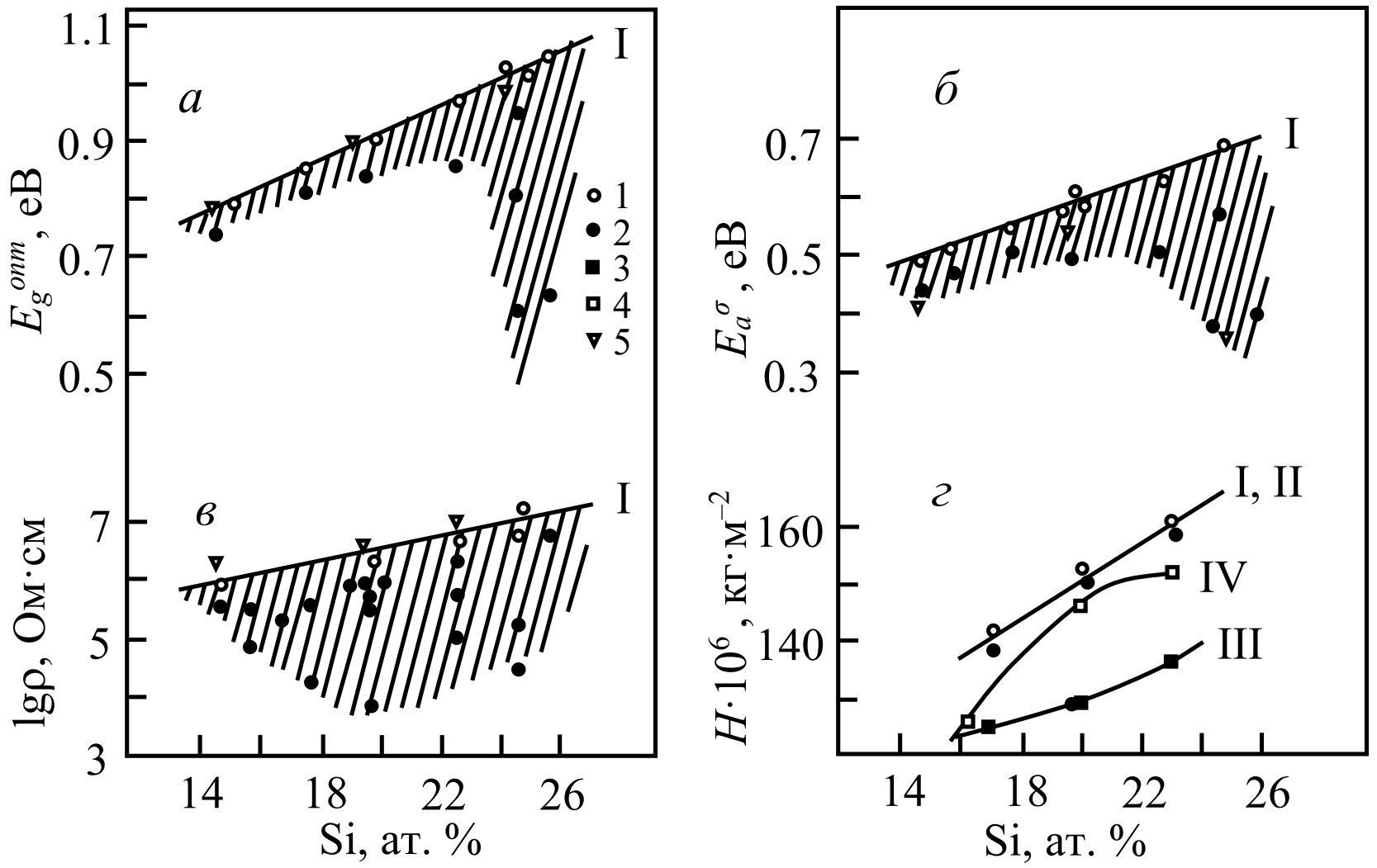

Рис. 4.19. Концентраційні залежності ширини оптичної щілини $E_g^{опт}$ (*а*), енергії активації електропровідності $E_a^\sigma$ (*б*), питомого опору ρ (*в*) і мікротвердості *H* (*г*) стекол $Si_xTe_{100-x}$ при температурах 293 (I, II) і 338 K (III, IV). 1, 3 – різко загартовані зразки, 2, 4 – зразки термооброблені при 373 K протягом 2 год [75], 5 – дані [46].

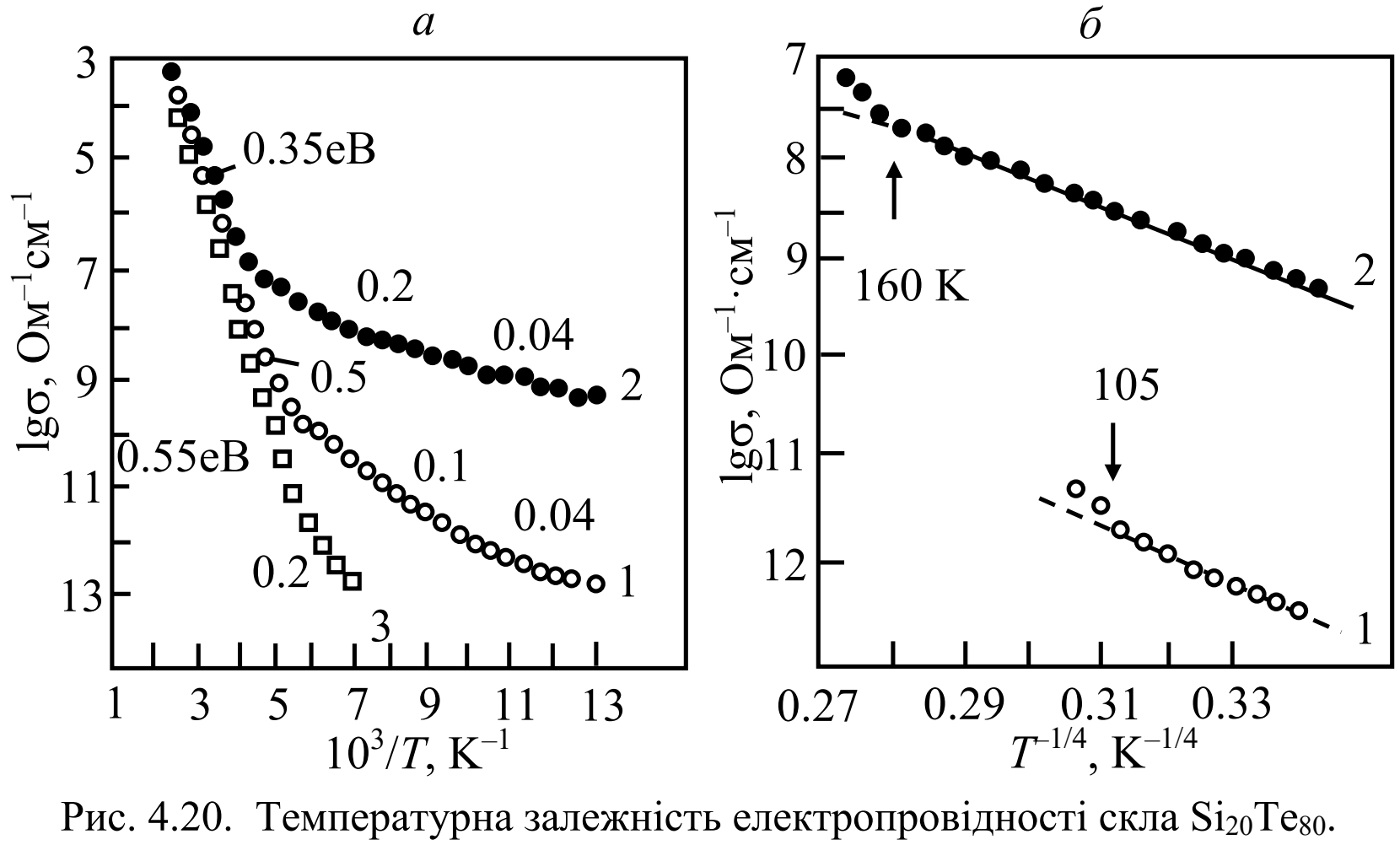

Рис. 4.20. Температурна залежність електропровідності скла $Si_{20}Te_{80}$. *а* – функція $\lg\sigma = f(1/T)$; *б* – функція в координатах Мотта $\lg\sigma = f(T^{-1/4})$ [75].

спостерігається відхилення від лінійної залежності цих параметрів від складу, і проявляється екстремум поблизу складу $Si_{20}Te_{80}$.

Електричні властивості сплаву $Si_{20}Te_{80}$ у кристалічному стані визначаються властивостями компонентів, що співіснують в евтектичній області даного складу. Домінуючим є відносно висока провідність телуру та практично безактиваційна залежність $\sigma(T)$. У склоподібному стані $Si_{20}Te_{80}$ є напівпровідником з $E_g$ = 0.5 еВ [184].

Істотний вплив на зазначені параметри та їх температурну залежність чинять умови термовідпалу стекол: температура і час відпалу, швидкість нагрівання, якість вихідного зразка і т.д. [67, 75, 185]. Вплив термообробки демонструють заштриховані області на рис. 4.19, де замість первісної лінійної залежності виходить область з максимумом або мінімумом поблизу $Si_{20}Te_{80}$ з великим розкидом експериментальних точок, що визначається деталями термообробки.

Вплив термообробки евтектичного скла $Si_{20}Te_{80}$ на характер температурної залежності електропровідності на постійному струмі у діапазоні температур 80–400 К ілюструє рис. 4.20, *а* на якому крива 1 відноситься до вихідного скла, отриманого загартуванням розплаву в крижану воду; крива 2 цього ж скла, відпаленого при 373 К протягом 2 годин; крива 3 для скла витриманого протягом року на повітрі. На кожній з кривих на рис. 4.20, *а* приведені енергії активації електропровідності. З порівняння кривих 1 і 2 на рис. 4.20, *а* видно, що відпал скла $Si_{20}Te_{80}$ при $T < T_g$ приводить до збільшення електропровідності, зменшенню термічної енергії активації $E_a$, і зміни характеру залежності $\sigma = f(1/T)$. Після термообробки скла $Si_{20}Te_{80}$ на кривій $\lg\sigma = f(1/T)$ з'являється низькотемпературна ділянка, яка описується законом Мотта (4.7) для стрибкової провідності зі змінною довжиною стрибка поблизу рівня Фермі [182]. Термообробка скла $Si_{20}Te_{80}$ збільшує протяжність температурного інтервалу стрибкової провідності на рівні Фермі від 77–105 К (крива 1, рис. 4.20, *б*) до 77–161 К (крива 2, рис. 4.20, *б*).

Зміну величини електропровідності та характеру її температурної залежності автори [75] пов'язують з особливостями еволюції структури евтектичного скла в процесі термообробки. При високих температурах структура евтектичного скла прямує до кристалізації через поділ фаз. Згідно [79] кристалізація аморфних евтектичних сплавів кремній-телур є багатостадійним процесом і на початкових стадіях супроводжується формуванням метастабільних фаз, тобто при нагріванні має місце перехід від однорідної просторової сітки до двофазної системи, що супроводжується збільшенням концентрації дефектів, пов'язаних з границями кластерів та фаз, зокрема обірваних зв'язків, дислокацій різного типу тощо. За даними рис. 4.19, *в* можна стверджувати, що для скла $Si_{20}Te_{80}$ границі, які виникають при термообробці, є найбільш електрично активними, оскільки саме для цього складу скла має місце найбільша зміна питомого опору.

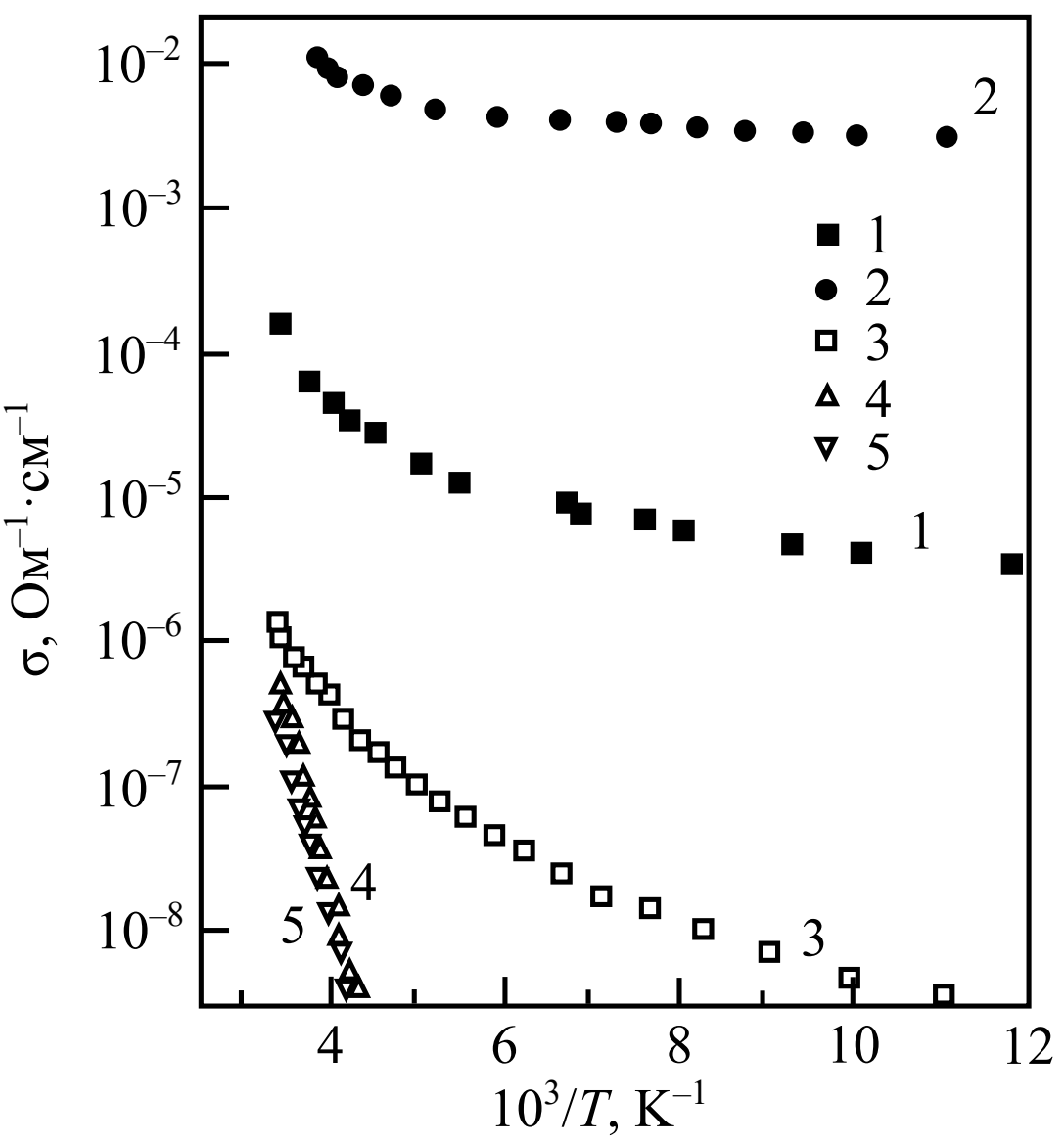

Рис. 4.21. Температурні залежності електропровідності стекол $Si_{20}Te_{80}$, загартованих: 1, 4 – у крижаній воді; 2, 3, 5 – на повітрі. Обробка поверхні: 1, 2, 3 – до травлення; 4, 5 – після травлення [69].

Необхідно відзначити, що при дослідженні електропровідності та її температурної залежності стекол $Si_xTe_{100-x}$ дуже важливим є підготовка зразків до вимірювань. На прикладі склоподібного $Si_{20}Te_{80}$ автори [69] показали (рис. 4.21), що величина і характер температурної залежності електропровідності на постійному струмі істотно відрізняються, якщо вимірювання виконані на полірованих зразках і на зразках, після їх обробки в поліруючому хімічному травнику. Відомо, що в процесі механічної обробки скла на поверхні зразка утворюється порушений шар, який видаляється в процесі травлення. Отже, причиною підвищення електропровідності σ є велика концентрація поверхневих дефектів, що виникають у процесі механічної обробки скла. У шліфованих зразках поверхнева провідність може переважати над об'ємною.

**4.4.2. Електричні властивості скла $Si_{20}Te_{80}$, отриманого в умовах мікрогравітації.** Результати порівняльного дослідження електропровідності склоподібного сплаву $Si_{20}Te_{80}$, отриманого в космосі (к-) в умовах мікрогравітації на ГКС «Мир» та його земного (з-) аналога, викладено у роботах [87–89]. Умови мікрогравітації сприяють отриманню більш однорідного за структурою і менш дефектного скла $Si_{20}Te_{80}$, ніж у земних умовах, що пояснюється зменшенням ймовірності зародження кластерів при затвердінні в умовах мікрогравітації внаслідок «ефекту відриву» розплаву від внутрішніх стінок ампули. Густина обох стекол виявилася близькою $\rho_к$ = 5.033 г/см$^3$ та $\rho_з$ = 5.029 г/см$^3$. Середня мікротвердість космічного скла 136 кг/мм$^2$ менша, ніж наземного скла – 150 кг/мм$^2$, що свідчить про більш високу мікрооднорідність скла, отриманого в умовах мікрогравітації.

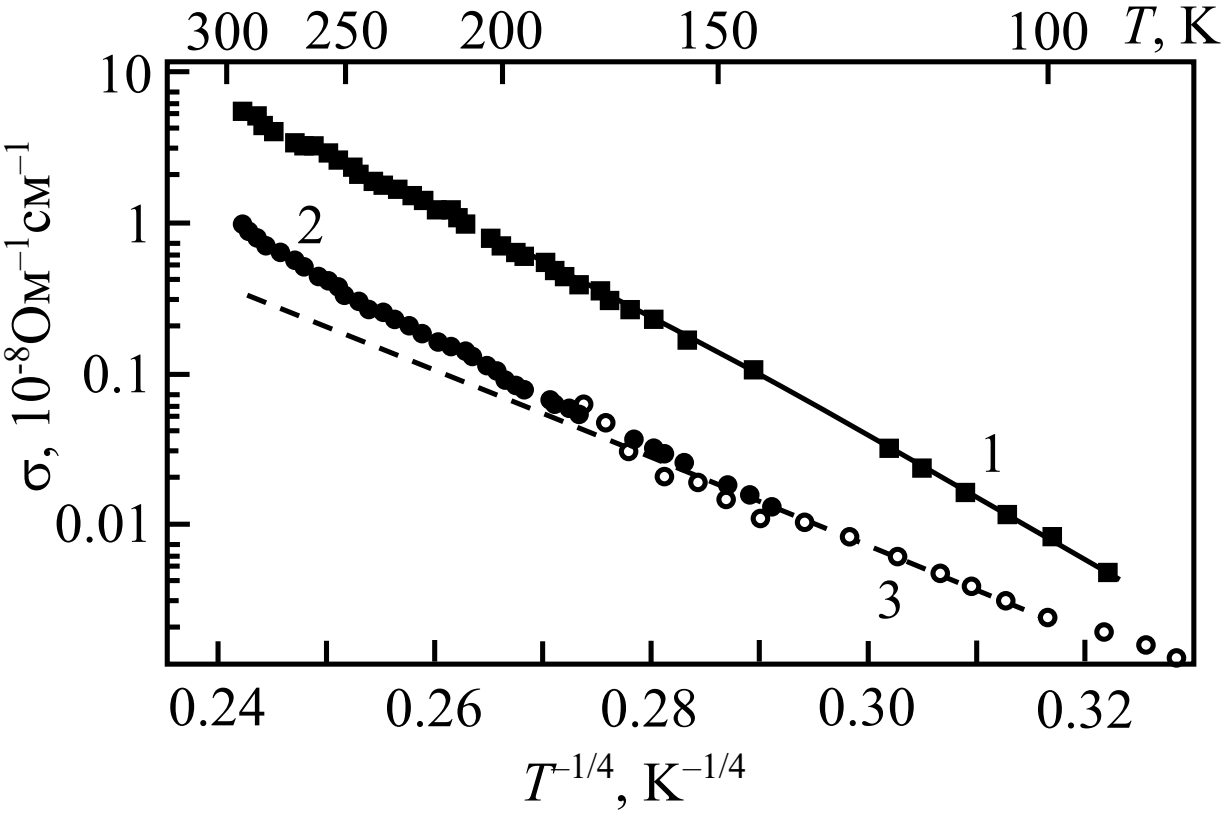

Рис. 4.22. Температурні залежності провідності скла $Si_{0.20}Te_{0.80}$, отриманого в умовах мікрогравітації (1) та наземних умовах (2, 3).
1, 2 – [89]; 3 – [75].

На рис. 4.22 наведено температурні залежності електропровідності для стекол $Si_{20}Te_{80}$, отриманих в умовах мікрогравітації (крива 1) і на землі (криві 2, 3). Низькотемпературна лінійна ділянка залежності $\sigma = f(T^{-1/4})$ добре описується виразом (4.7) з параметрами: $\sigma_0$ = 4.9·10$^{10}$ Ом$^{-1}$см$^{-1}$ і $T_0$ = 7.5·10$^7$ К для космічного зразка; $\sigma_0$ = 4.6·10$^6$ Ом$^{-1}$см$^{-1}$ і $T_0$ = 2.1·10$^7$ К для земного зразка; і відповідає стрибковій провідності зі змінною довжиною стрибка. Таким чином, у зазначеній області температур перенесення заряду в склоподібному $Si_{20}Te_{80}$, незалежно від умов його отримання, здійснюється за допомогою стрибкової провідності носіїв заряду зі змінною довжиною стрибка по локалізованих станах, що лежать у вузькій смузі енергій поблизу рівня Фермі.

Для космічного скла $Si_{20}Te_{80}$ область виконання закону Мотта простягається до більш високих температур, ніж для земного аналога. Густина станів $N_F$ на рівні Фермі, створюваних дефектами, за якими відбувається стрибкова провідність для космічного зразка виявилася меншою, ніж для земного ($2.5\cdot10^{18}$ та $8.8\cdot10^{18}$ $еВ^{-1}см^{-3}$ відповідно). Велика однорідність за складом, менша середня мікротвердість і менша величина $N_F$ для космічного зразка в порівнянні з земним аналогом, дозволила автору [89] зробити висновок, що скло, отримане в космічних умовах внаслідок «ефекту відриву» має меншу кількість дефектів, ніж земний аналог і ближче до ідеального.

**4.4.3. Електричні властивості аморфних $Si_xTe_{100-x}$, отриманих методом спінінгувавння розплаву.** Вимірювання електропровідності аморфних $Si_xTe_{100-x}$, автори [83] проводили двозондовим методом на прямокутних зразках довжиною 10 мм, шириною 4 мм і товщиною 1мм, спресованих із порошка аморфних стрічок, отриманих методом спінінгувавння розплаву. На рис. 4.23 приведена температурна залежність електропровідності аморфного $Si_7Te_{93}$, на якому

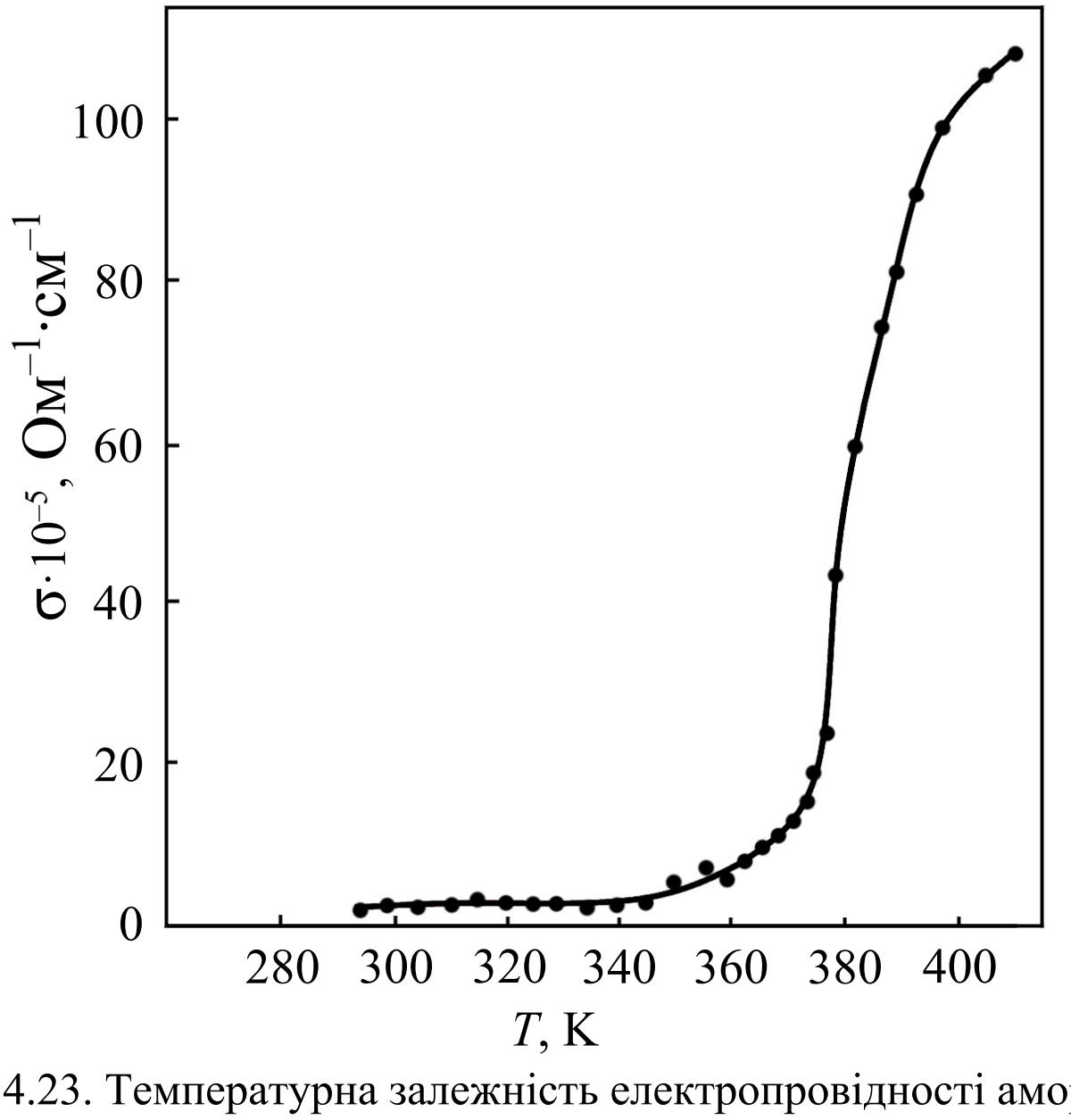

Рис. 4.23. Температурна залежність електропровідності аморфного $Si_7Te_{93}$ [83].

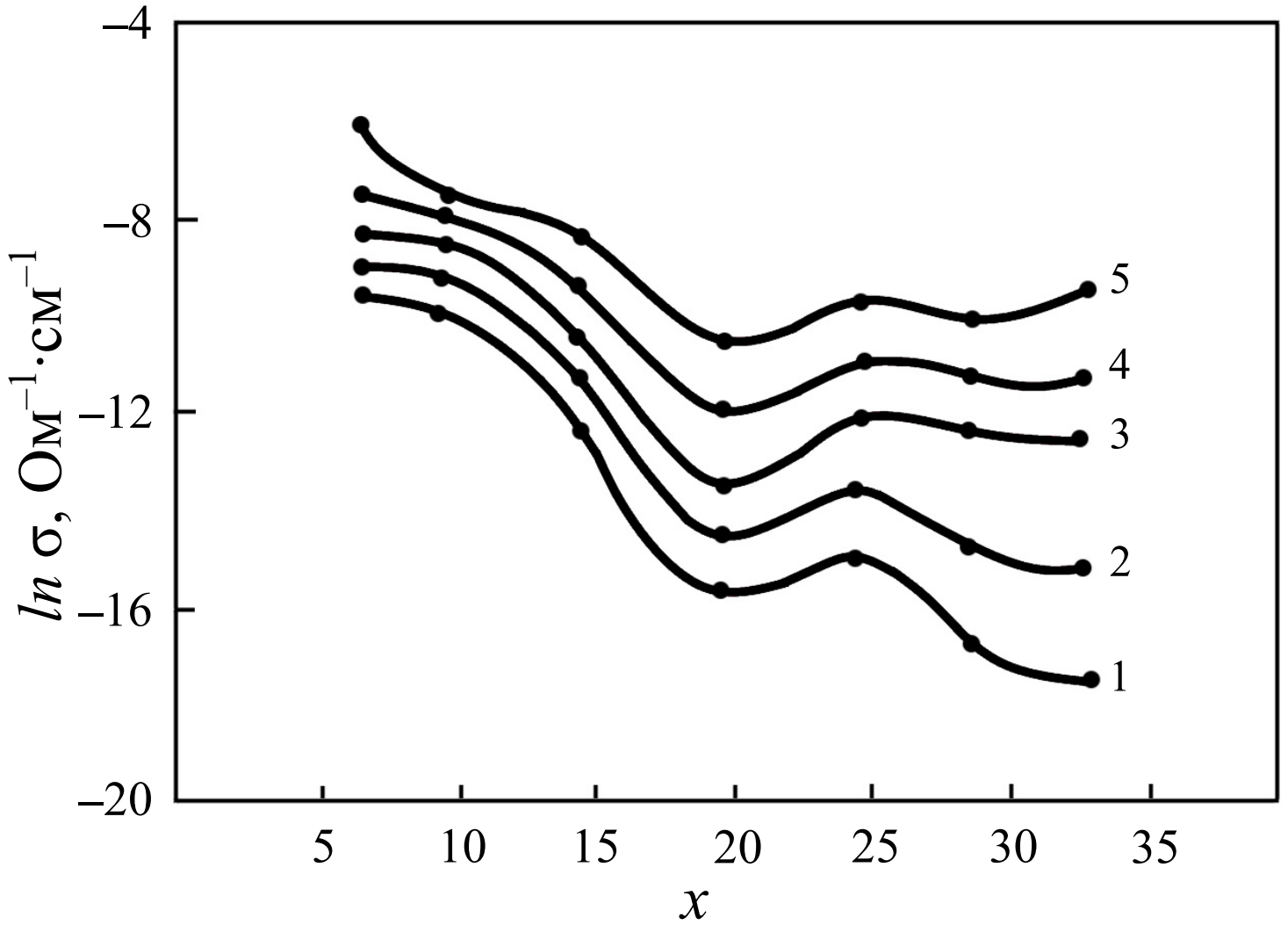

Рис. 4.24. Концентраційні залежності електропровідності аморфних $Si_xTe_{100-x}$, виміряної при різних температурах $T$, K:
1 – 70, 2 – 170, 3 – 293, 4– 313, 5 – 373 [83].

чітко видно різку зміну електропровідності при першій температурі кристалізації $T_{к1}$, яка зв’язана з виділенням кристалічного Te в аморфній фазі. Концентраційні залежності провідності аморфних зразків $Si_xTe_{100-x}$, виміряні при різних температурах, наведені на рис. 4.24. Вимірювання зміни провідності зі складом вказують на наявність одного локального мінімуму на кривих $ln$ σ від $x$ при $x$ = 20, а другого при $x$ = 33, коли температура зменшується.

**4.4.4. Електропровідність аморфних плівок $Si_xTe_{100-x}$.** Електричні властивості тонких аморфних плівок $Si_xTe_{100-x}$ (0 – 82 ат.% Te), отриманих сумісним термічним випаровуванням кристалічного кремнію і телуру у вакуумі $10^{-6}$ Торр на підігріті до 323 K підкладки, вивчені в роботі [93]. Для випаровування Si використовувалася електронна гармата потужністю 4 кВт, а випаровування Te здійснювалося термічно з Mo човника, покритого $Al_2O_3$. Хімічний склад плівок варіювався шляхом зміни відносних швидкостей випаровування між двома джерелами і визначався після завершення процесу осадження методом рентгенофазового мікроаналізу. Для електричних вимірювань на аморфні плівки $Si_xTe_{100-x}$ напилювались Al-контакти.

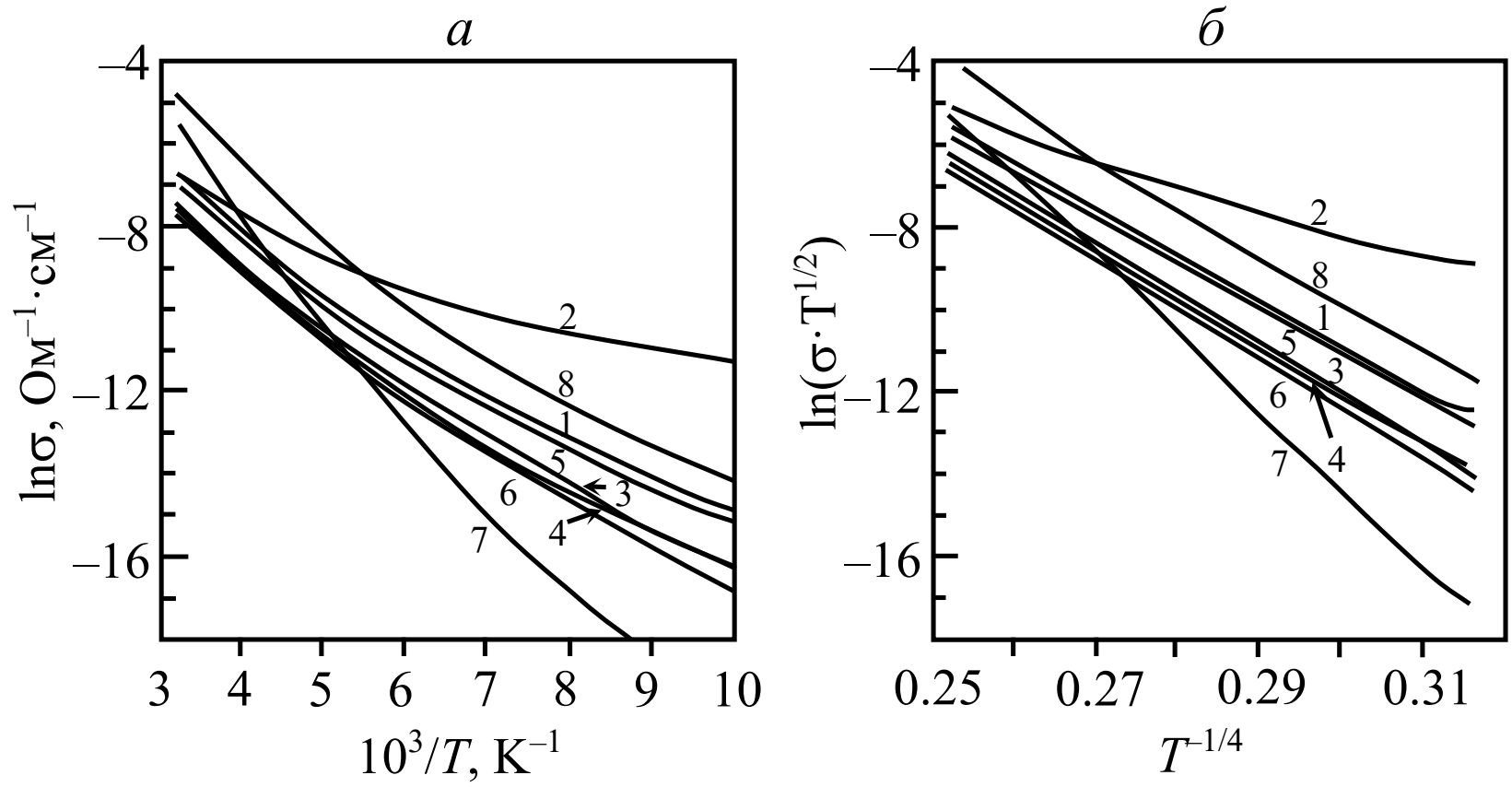

Рис. 4.25. *а* – Температурні залежності електропровідності аморфних плівок $Si_xTe_{100-x}$; *б* – залежність $\ln(\sigma T^{1/2})$ від $T^{-1/4}$. $x$ = ат.%: 1 – 0; 2 – <1; 3 – 9; 4 – 12; 5 – 13; 6 – 41; 7 – 57; 8 – 82 [93].

Температурні залежності темнової провідності аморфних плівок $Si_xTe_{100-x}$, виміряні в області низьких температур 100–300 К, наведені на рис. 4.25 *а*. При кімнатній температурі і вище електронний переніс реалізується за делокалізованими станами, про що свідчить наявність лінійних ділянок на залежності $\ln\sigma = f(1/T)$ вище $T \geq 250$ К. В області низьких температур ($T \leq 250$ К) електронний переніс здійснюється за локалізованими станами поблизу рівня Фермі в режимі стрибкової провідності зі змінною довжиною стрибка. Темнова провідність у цьому режимі описується виразом (4.7). Густина локалізованих станів поблизу рівня Фермі $N_F$, визначена із нахилу лінійної ділянки $\ln(\sigma T^{1/2})$ від $T^{-1/4}$ (рис. 4.25, *б*), зображена на рис. 4.26, *в*.

Концентраційна залежність σ для цих двох режимів провідності (виміряна при 300 і 100 К) наведена на рис. 4.26 *а*. Зі збільшенням вмісту Te провідність при кімнатній температурі збільшується, головним чином, за рахунок зменшення ширини забороненої зони $E_g$ аморфних плівок. Концентраційна залежність енергії активації $E_a$ наведена на рис. 4.26, *б*. Низькі значення $E_a$ відповідають відносно великим значенням провідності.

Висока темнова провідність, мала енергія активації і велика густина станів $N_F$ поблизу рівня Фермі є характерними ознаками аморфних тонких плівок $Si_xTe_{100-x}$, отриманих у процесі випаровування. Проте концентраційна залежність електричних властивостей

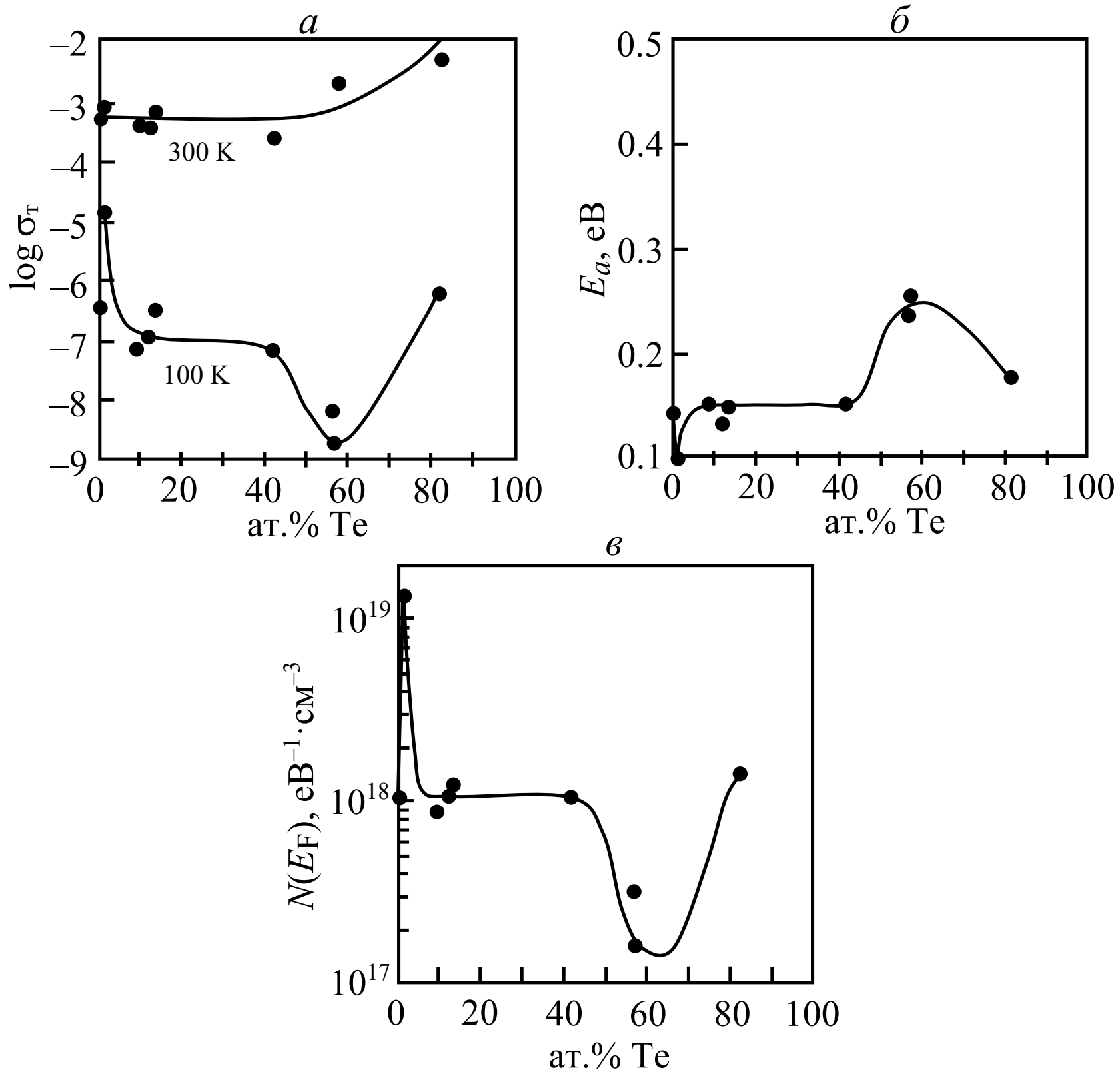

Рис. 4.26. Концентраційні залежності електропровідності σ (*а*), енергії активації $E_a$ (*б*) та густини станів на рівні Фермі $N(E_F)$ (*в*) аморфних плівок $Si_xTe_{100-x}$ [93].

показує деякі цікаві особливості для складів < 1 ат.% Te і для 60 ат.% Te. Малі концентрації Te (< 1 ат.%) збільшують густину станів на рівні Фермі та зменшують енергію активації стрибкової провідності. Це пов'язано з тим, що Te проявляє властивості домішки заміщення – двозарядного донора в *a*-Si. При високих вмістах Te (≥ 60 ат.%) спостерігається зменшення густини локалізованих станів на рівні Фермі та збільшення енергії активації.

Оскільки тонкі аморфні плівки $Si_xTe_{100-x}$ були отримані сумісним випаровуванням, вони містять велику кількість дефектів, таких як дислокаційні пустоти і обірвані зв'язки (як на атомах Si, так і на Te). Ці дефекти викликають високу густину локалізованих станів у псевдощілині, що підтверджується експериментальними даними.

Результати дослідження електричного опору аморфних тонких плівок $Si_xTe_{100-x}$ в інтервалі температур вище кімнатної (273 – 673 К) приведені в роботах [186, 187]. Тонкі плівки $Si_xTe_{100-x}$ товщиною 250 нм отримані спільним напиленням Si і Te на $SiO_2$/Si підкладки. Вимірювання опору проведено двозондовим методом. Температурні залежності опору для трьох складів аморфних тонких плівок $Si_xTe_{100-x}$ ($x$ = 10, 15, 18) наведені на рис.4.27 при нагріванні зі швидкістю 10 К/хв (крива 1) та охолоджені (крива 2) досліджуваного зразка. Як видно із цих рисунків, електричний опір усіх трьох плівок поступово зменшується з підвищенням температури досліджуваного зразка аж до температури 523 К. Вище цієї температури електричний опір усіх плівок трохи збільшується до $T$ = 568 К з наступним зменшенням опору в області вищих температур до 583 К. Крім того, на температурних залежностях опору тонких плівок $Si_xTe_{100-x}$ чітко видно різкий спад опору в околі $T$ = 423 К.

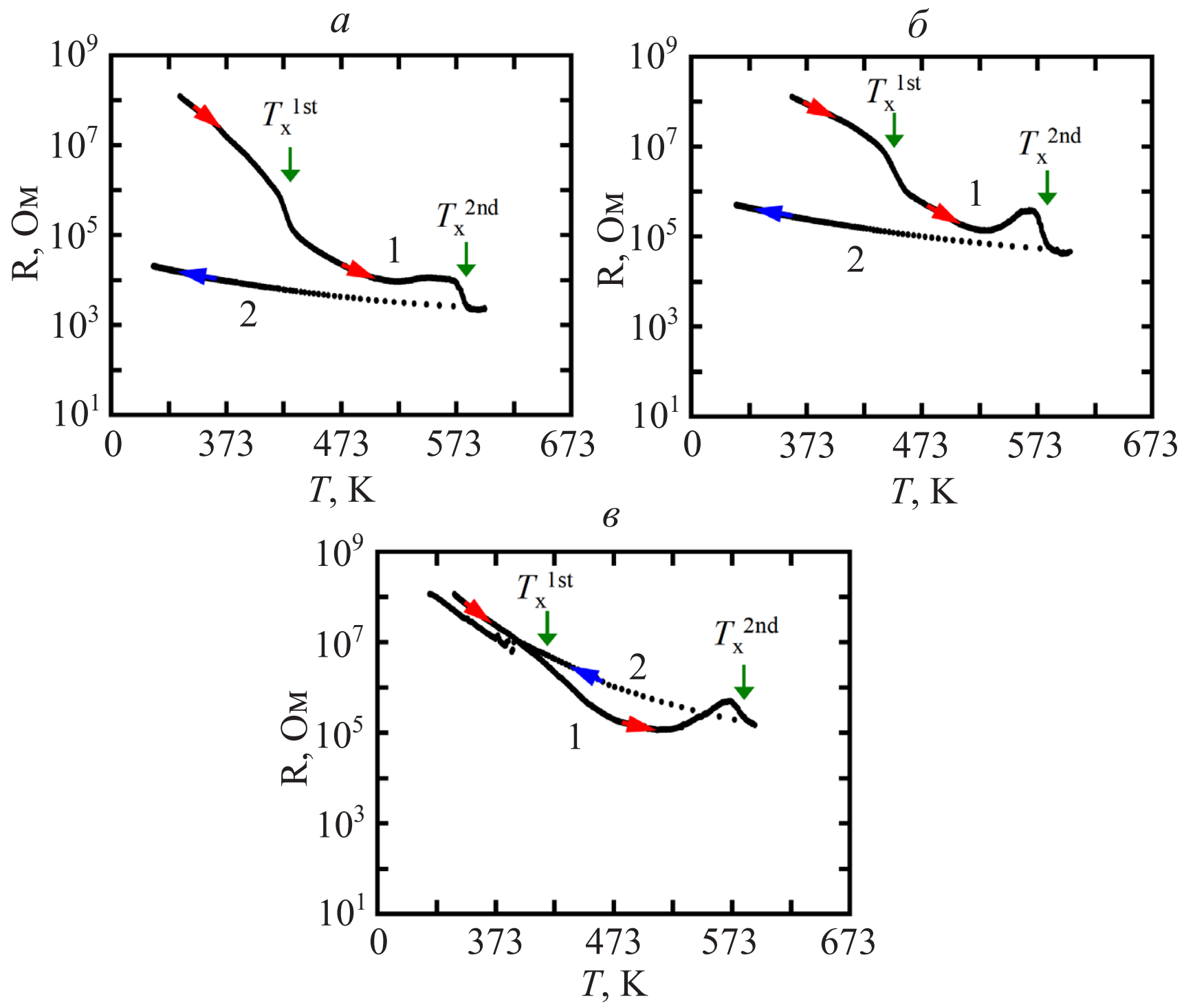

Рис. 4.27 Температурні залежності опору тонких аморфних плівок $Si_{10}Te_{90}$ (*а*), $Si_{15}Te_{85}$ (*б*) і $Si_{18}Te_{82}$ (*в*) [186].

Провівши мікроструктурні дослідження тонких плівок методами ТЕМ і EDS, автори [186, 187] дійшли до висновку, що наявні особливості на температурних залежностях опору аморфних плівок $Si_xTe_{100-x}$ в процесі нагрівання викликані двостадійною кристалізацією. Спочатку в процесі нагрівання аморфних плівок $Si_xTe_{100-x}$ при $T = 453$ К відбувається виділення кристалічних зерен Te розміром 40 – 60 нм, при цьому аморфна фаза знаходиться вздовж меж зерен Te.

Дослідження ПЕМ показали, що незначне збільшення електричного опору в околі $T = 568$ К викликане формуванням високоомної аморфної фази Si, навколо кристалітів Te. Подальше нагрівання приводить до зменшення електричного опору за рахунок кристалізації фази. Таким чином, наявні особливості на температурних залежностях аморфних плівок $Si_2Te_3$ демонструють двостадійний процес кристалізації. Перша температура кристалізації $T_K^{1st}$ трохи зростає зі збільшенням вмісту Si, тоді як друга температура кристалізації $T_K^{2nd}$ не залежить від складу і залишається сталою при 583 К. Після охолодження до кімнатної температури електричний опір кристалічної фази різко зростає зі збільшенням вмісту Si. Електричний опір закристалізованої плівки $Si_{18}Te_{82}$ перевищує $10^8$ Ом при кімнатній температурі. Різке збільшення електричного опору кристалічної фази зі збільшенням Si автори [186, 187] пояснюють збільшенням об'ємної частки фази $Si_2Te_3$, яка має високий питомий опір.

### 4.5. ЕЛЕКТРОПРОВІДНІСТЬ РОЗПЛАВІВ $Si_xTe_{100-x}$.

Як зазначає автор [188], найбільш важливими фізичними характеристиками рідких напівпровідників є температурні залежності електропровідності й термоерс (коефіцієнт Зеебека).

Електропровідність і термоерс розплавів системи Si–Te у широкому інтервалі температур та складів 40–100 ат.%Te вперше досліджені авторами [189]. Для розплаву $Si_{60}Te_{40}$, збагаченого кремнієм, електропровідність різко зростає з температурою від одиниць до десятків $Ом^{-1}см^{-1}$(крива 1, рис. 4.28) і від десятків до сотень $Ом^{-1}см^{-1}$ для складів 50 і 58 ат.% Te в інтервалі температур 1173–1273 К (криві 2 і 3, рис. 4. 28) [189]. У міру збагачення зразків телуром характер політерм має менш різкий нахил і виявляє тенденцію до насичення.

Порядок величин електропровідності, а також характер її температурної залежності свідчить про переважний внесок напівпровідникової компоненти у провідності розплавів системи Si–Te, проте

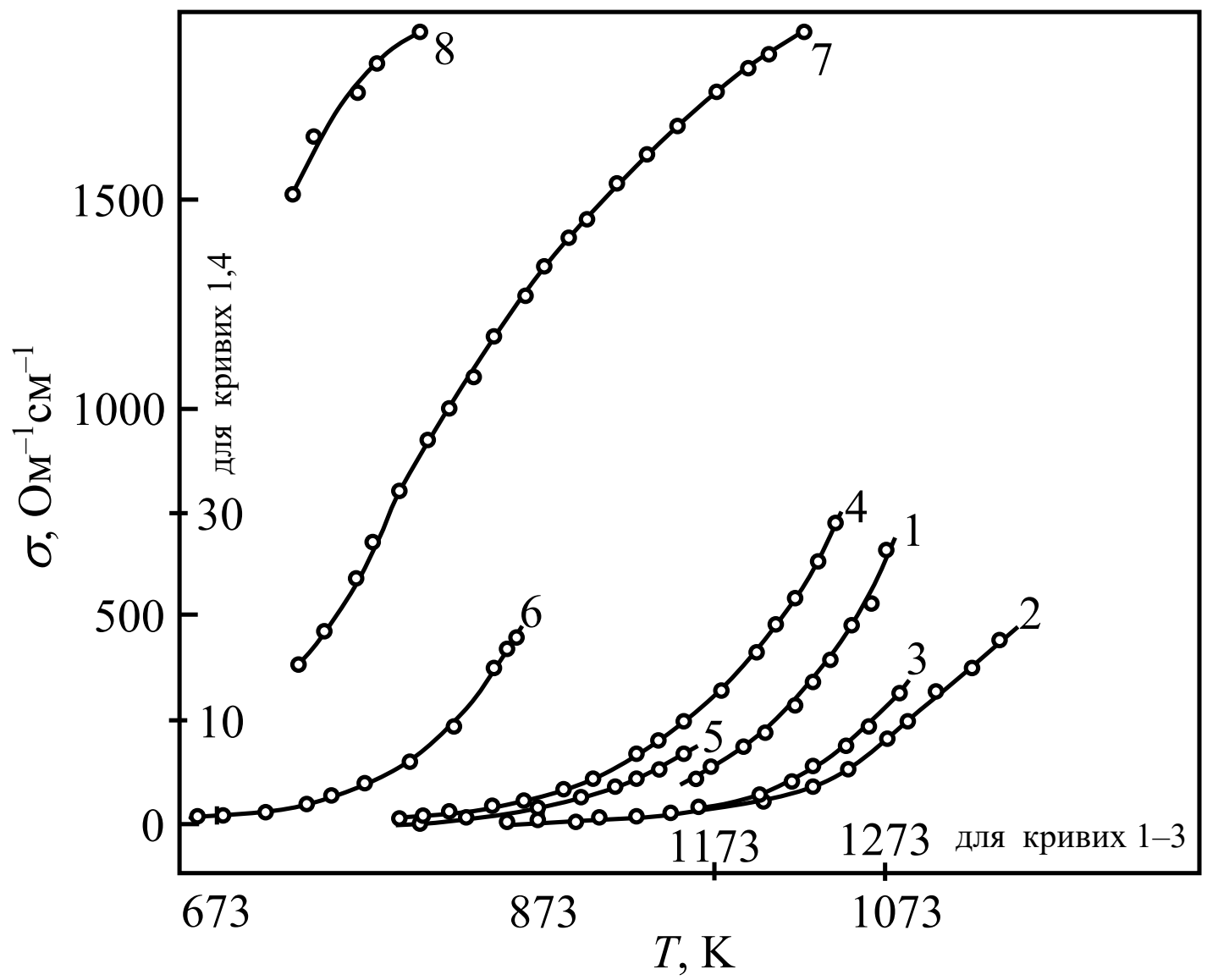

Рис. 4.28. Температурні залежності електропровідності розплавів $Si_xTe_{100-x}$.
*x*, ат. %: 1 – 60; 2 – 50; 3 – 42; 4 – 33,3; 5 – 28;
6 – 20; 7 – 10; 8 –0 [189].

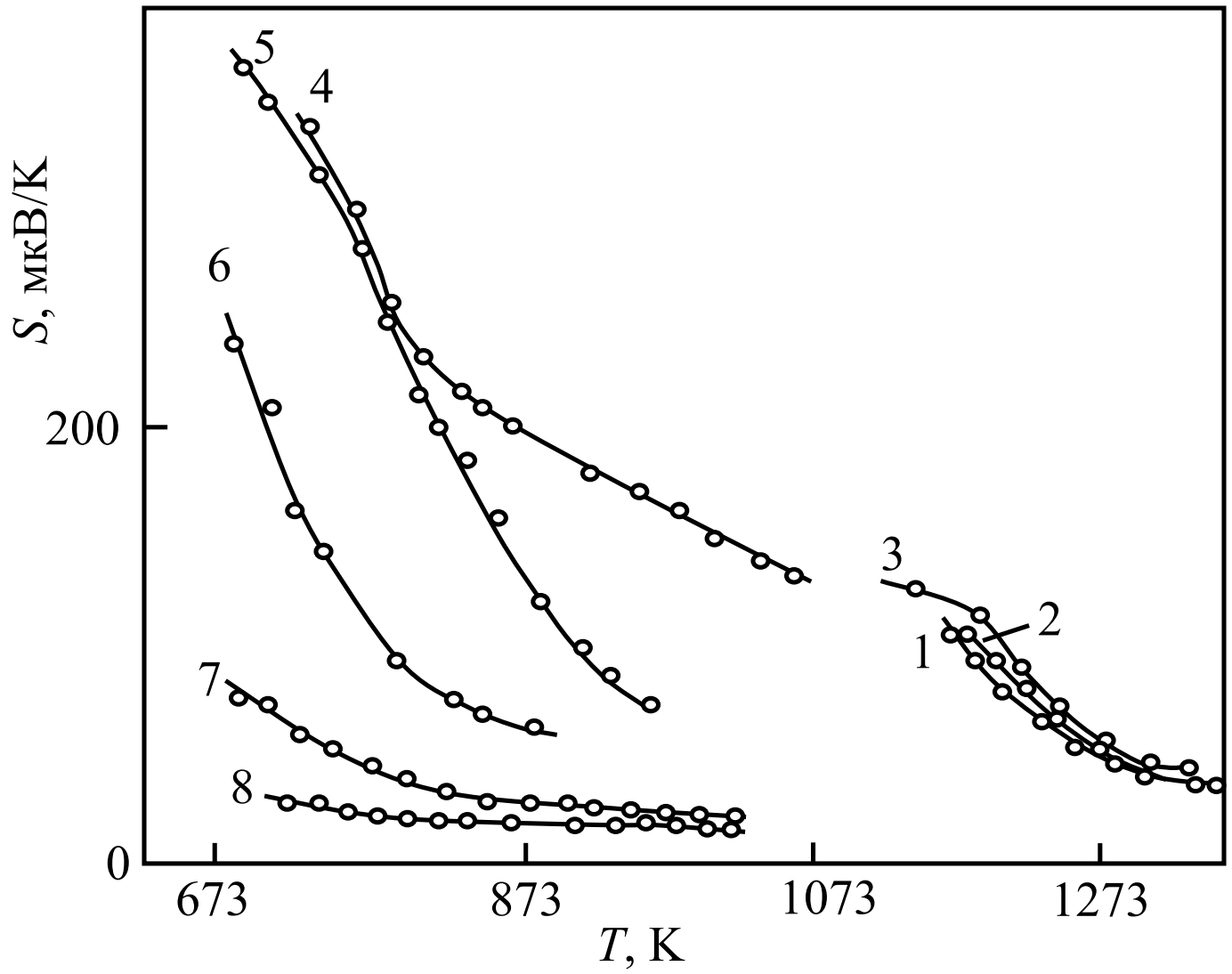

Рис. 4.28. Температурні залежності термоерс розплавів $Si_xTe_{100-x}$.
*x*, ат. %: 1 – 60; 2 – 50; 3 – 42; 4 – 33,3; 5 – 28; 6 – 20; 7 – 10; 8 – 0 [189].

для розплавів, збагачених телуром, за високих температур спостерігається тенденція до виродження напівпровідникового внеску та поступової металізації зв'язків. Для розплавів, збагачених кремнієм, за низьких температур не виключається помітний іонний внесок у загальну провідність. Величина термоерс для зразків із вмістом 40–60 ат.% Te має значення 300–400 мкВ/К, але в міру збільшення концентрації телуру та зміни характеру електропровідності *S* вона зменшується до 40–60 мкВ/К (рис. 4.29). Таким чином, згідно [189] електричні властивості розплавів системи Si–Te укладаються в модель рідкого поліфункціонального провідника, провідність якого включає три компоненти: іонну, напівпровідникову (з переважанням діркових носіїв) і металеву, співвідношення між якими змінюється в залежності від складу та температури.

Електропровідність і термоерс розплавів системи $Si_xTe_{100-x}$ $(0 \le x \le 30)$ також досліджували автори [190]. Вимірювання опору проводились чотирьохзондовим методом з використанням комірки з плавленого кварцу, обладнаного вольфрамовими електродами. На рис. 4.30 наведено температурні залежності електропровідності для різних складів. Видно, що електропровідність змінюється монотонно зі зміною складу від майже металевих значень для чистого телуру до вкрай малих значень при великих *x*. Друга особливість, що заслуговує на увагу, полягає в тому, що електропровідність має великий температурний коефіцієнт, що відображає енергію активації в широкій області складів *x* і *T*.

Вище точки плавлення на залежностях log σ від $10^3/T$ спостерігаються прямі лінії. Перехід від напівпровідникової до металевої поведінки спостерігається для сплавів з високим вмістом Te (> 90 ат.%) при високих температурах, де на залежності $\log\sigma = f(1/T)$ спостерігається відхилення від лінійної залежності. Розплави $Si_xTe_{100-x}$ із вмістом Si від 10 ат.% і вище проявляють напівпровідниковий характер. Для сплаву $Si_{20}Te_{80}$ питомий опір становить $8\cdot10^{-2}$ Ом·см при температурі 723 К. За сталої температури питомий опір збільшується із збільшенням вмісту кремнію у всьому виміряному діапазоні концентрацій (від 0 до 30 ат.% Si).

Додаткову інформацію про властивості рідких напівпровідників дають вимірювання термоерс, особливо якщо вони виконані для тих самих складів, що і вимірювання електропровідності. На рис. 4.31 наведені криві залежності *S(T)* для системи $Si_xTe_{100-x}$ для різних значень *x*. З цього рисунка видно, що для складів $x \ge 10$ термоерс силь-

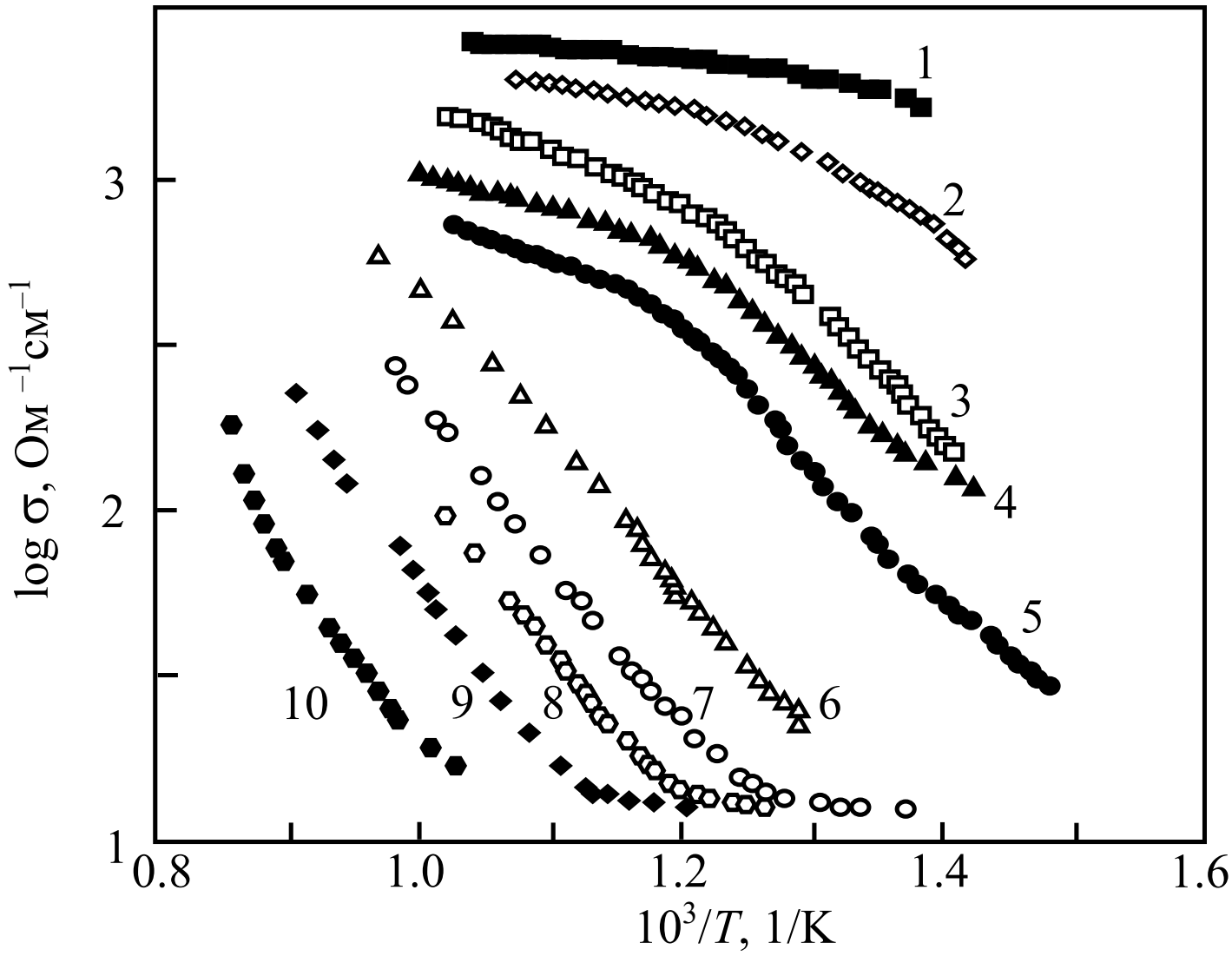

Рис. 4.30. Температурні залежності електропровідності сплавів $Si_xTe_{100-x}$. *x*, ат. %: 1 – 0; 2 – 5; 3 – 10; 4 – 12; 5 – 15; 6 – 18; 7 – 20; 8 – 22; 9 – 25; 10 – 30 [190].

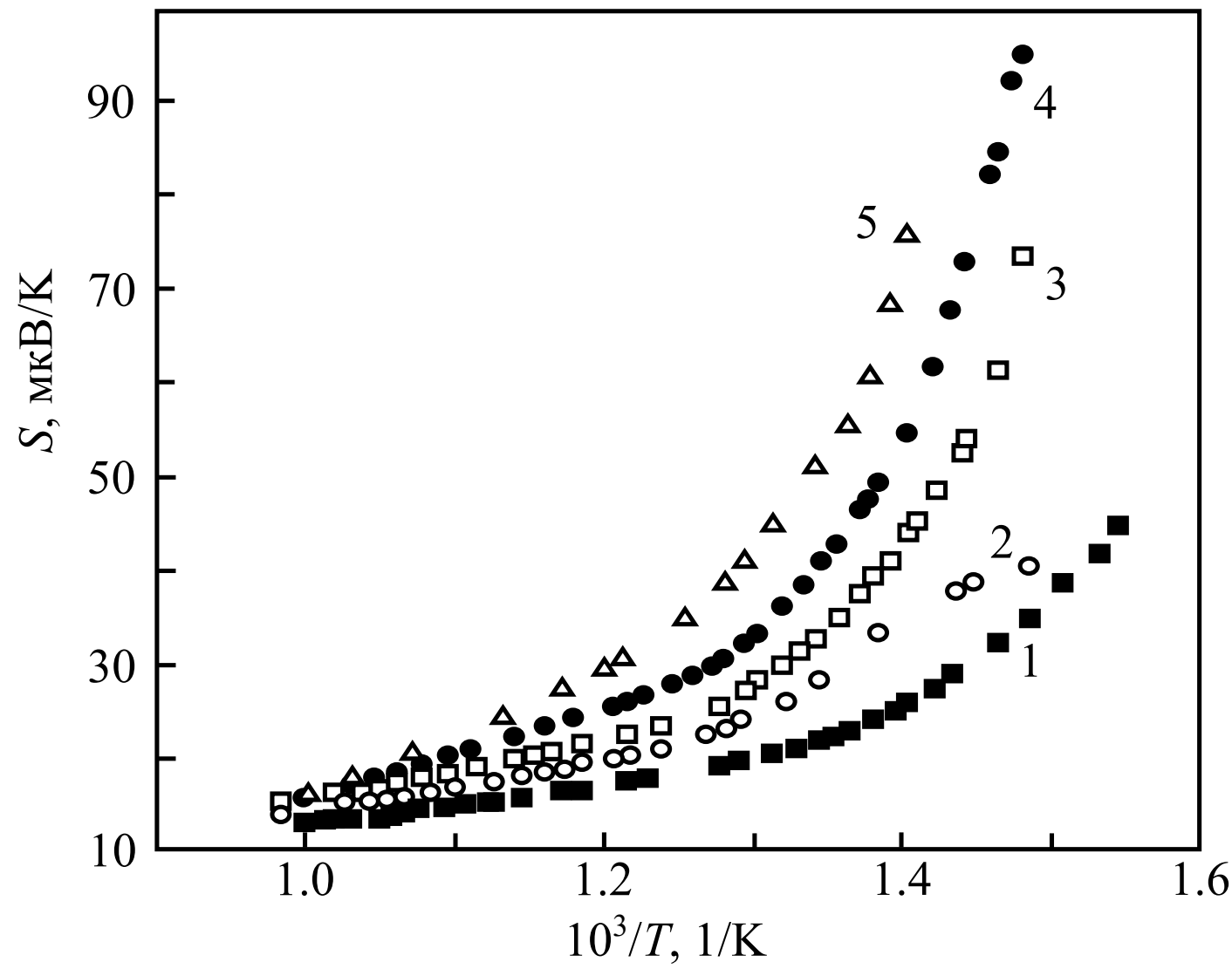

Рис. 4.31. Температурні залежності термоерс сплавів $Si_xTe_{100-x}$. *x*, ат. %: 1 – 0; 2 – 5; 3 – 10; 4 – 15; 5 – 18 [190].

но залежить від температури, а при $x < 10$ ця залежність слабка, як і у випадку $\sigma(T)$ (рис. 4.30). Найбільше значення термоерс (100 мкВ/К) має розплав евтектичного складу ($Si_{15}Te_{85}$) при температурі евтектики. З порівняння рис. 4.30 та 4.31 видно, що між величинами S і $\sigma$ є хороша кореляція. Тобто при сталій температурі зі збільшенням вмісту кремнію спостерігається збільшення питомого опору і термоерс.

## 4.6. ВПЛИВ ТИСКУ НА ЕЛЕКТРИЧНІ ВЛАСТИВОСТІ СТЕКОЛ $Si_xTe_{100-x}$

Дослідження склоподібних телуридів кремнію при високих тисках викликає підвищений інтерес, оскільки вплив тиску приводить до істотних змін їх електричних властивостей і появі фазових переходів. Вплив тиску до 8.5 ГПа на питомий опір стекол $Si_xTe_{100-x}$ ($10 \le x \le 28$), отриманих загартуванням розплаву в крижану воду, досліджено у роботах [80, 104 – 106]. Для генерації тисків використовувалася камера високого тиску з ковадлом типу «закруглений конус-площина» із штучних полікристалічних алмазів «карбонадо» [191]. Ці ковадла мають хорошу провідність і можуть бути використані в якості електродів до зразка. Баричні залежності питомого опору об'ємних стекол $Si_xTe_{100-x}$ наведені на рис. 4.32. З цього рисунка видно, що для всіх складів стекол спостерігається зменшення питомого опору з підвищенням тиску, аж до певного критичного значення $P_к$, характерного для кожного складу, вище якого має місце різке стрибкоподібне зменшення питомого опору приблизно на шість порядків величини. Автори [80, 106] вважають, що при $P_к$ відбуваються фазові переходи скло–напівпровідник–метал, після якого опір зразків практично залишається сталим. Для скла $Si_{20}Te_{80}$, питомий опір якого при атмосферному тиску рівний $\rho = 1.39 \cdot 10^6$ Ом·см, із зростанням тиску до 3 ГПа, $\rho$ спочатку експоненційно зменшується, потім практично не змінюється, аж до критичного тиску $P_к = 7$ ГПа (крива 4, рис. 4.32), який є максимальним для даної системи стекол.

Для складів стекол, як збагачених Te у порівнянні зі складом $Si_{20}Te_{80}$, так із меншим вмістом телуру, характер залежності питомого опору від тиску практично такий самий як і для скла $Si_{20}Te_{80}$ (рис. 4.32). Водночас критичний тиск $P_к$, при якому має місце стрибкоподібне зменшення питомого опору, для всіх складів стекол $Si_xTe_{100-x}$ суттєво знижується порівняно з $x = 20$. Концентраційні залежності

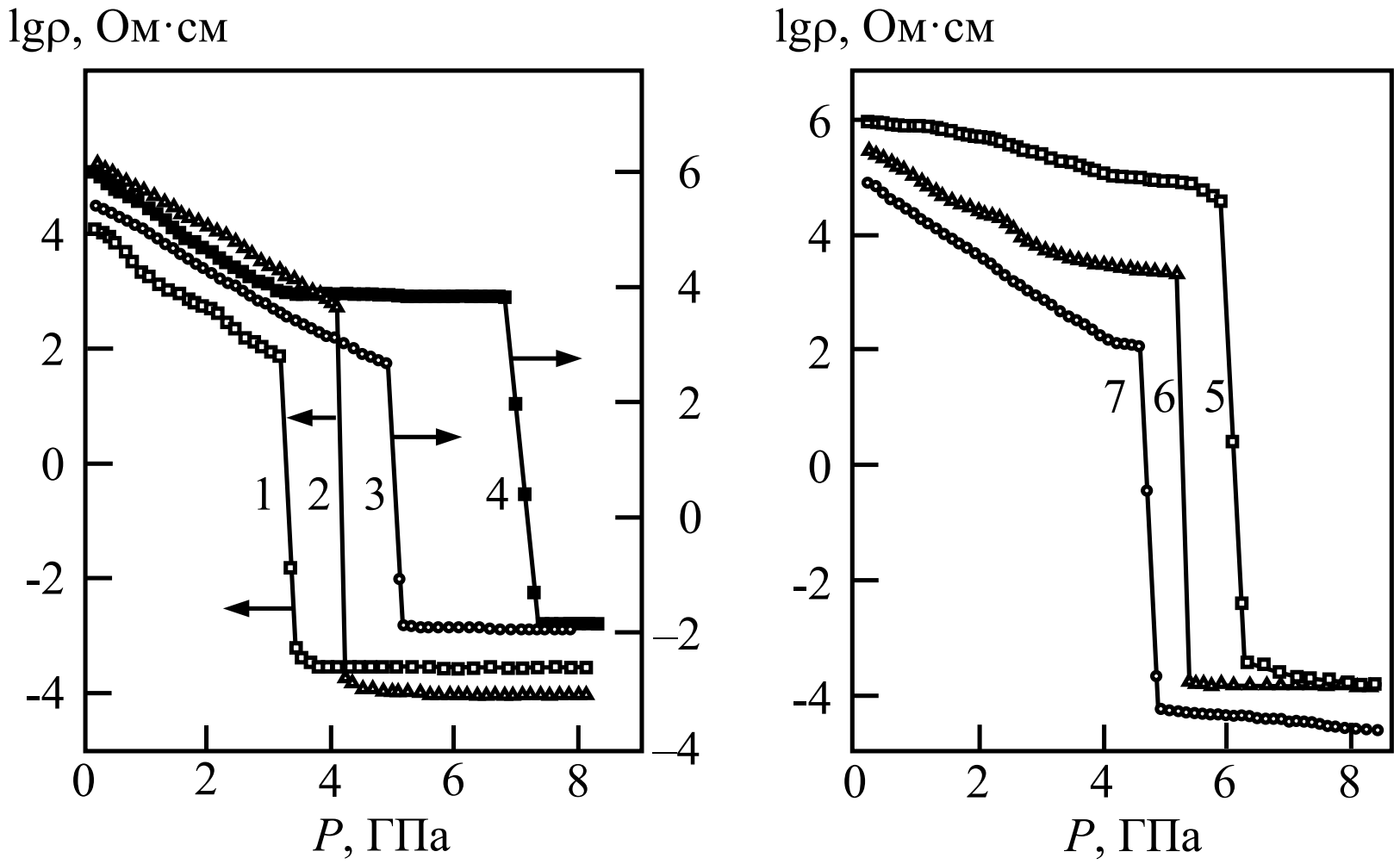

Рис. 4.32. Баричні залежності питомого опору стекол $Si_xTe_{100-x}$.
$x$: 1 – 10; 2 – 15; 3 – 17; 4 – 20; 5 – 22; 6 – 25; 7 – 28 [80].

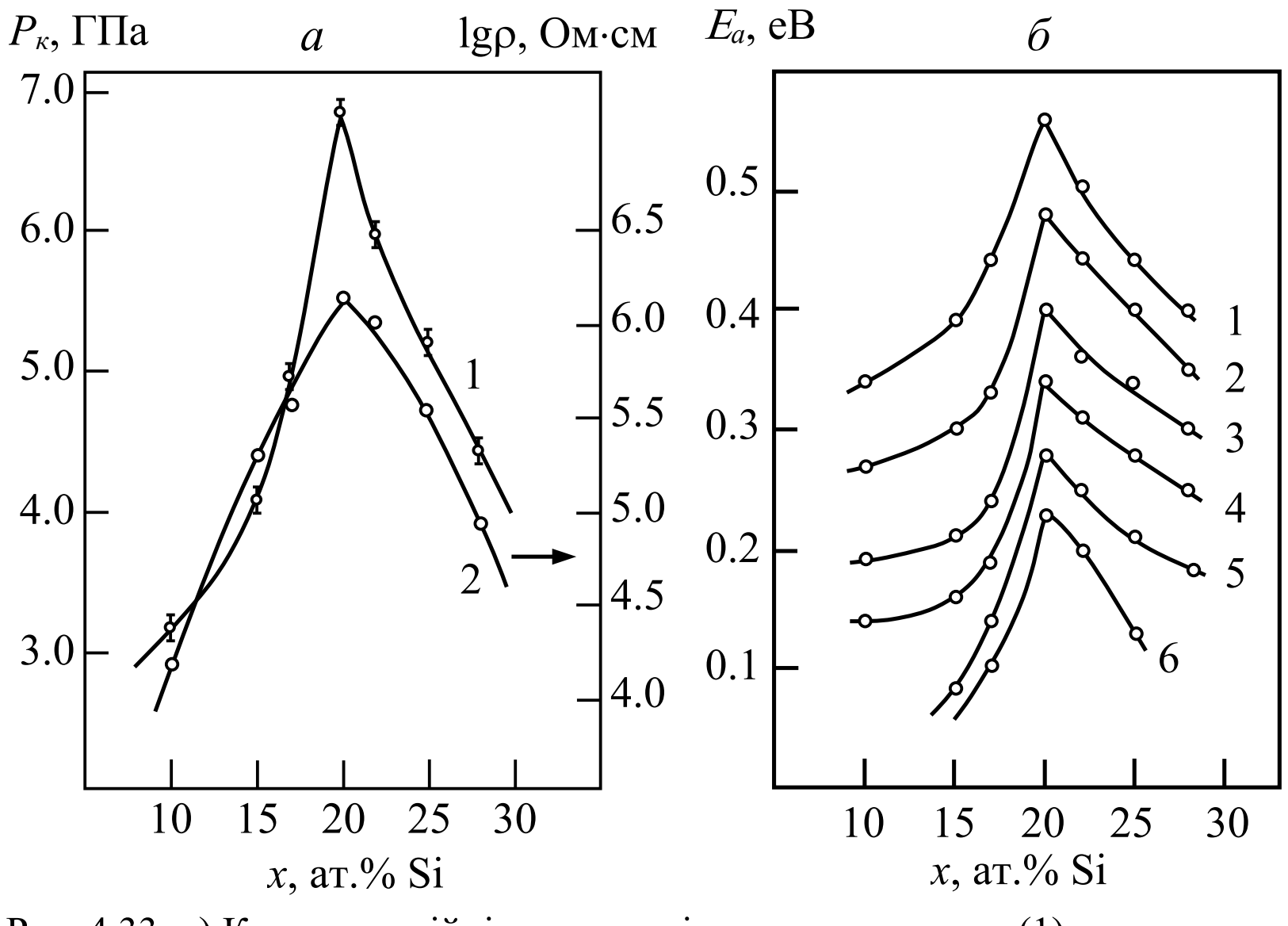

Рис. 4.33. *а*) Концентраційні залежності критичного тиску (1) та питомого опору (2); *б*) концентраційні залежності енергії активації $E_a$ стекол $Si_xTe_{100-x}$, виміряних при різних тисках. $P$, ГПа:
1 – $10^5$ Па; 2 – 1; 3 – 2; 4 – 3; 5 – 4; 6 – 5 [80].

критичного тиску переходу $P_к$ і питомого опору ρ, виміряного при атмосферному тиску та кімнатній температурі, для стекол $Si_xTe_{100-x}$ наведені на рис. 4.33, *а*.

Температурні залежності електропровідності скла $Si_{20}Te_{80}$, виміряні при різних сталих високих тисках до 7 ГПа, представлені в Арреніусових координатах $\lg\sigma = f(T^{-1})$ на рис. 4.34. У цих координатах залежності $\sigma(T)$ добре апроксимуються прямою, характерною для напівпровідникових матеріалів. За нахилом прямих в інтервалі температур 100–300 К визначено енергії активації $E_a$. Зі збільшенням тиску зразка $Si_{20}Te_{80}$ енергія активації зменшується від 0.56 еВ при атмосферному тиску до 0.16 еВ при $P$ = 6.8 ГПа, зменшується при цьому і $\sigma_0$ від $1.89\cdot10^3$ $Ом^{-1}см^{-1}$ до $6.76\cdot10^{-2}$ $Ом^{-1}см^{-1}$ [105]. Подібний характер температурної залежності провідності спостерігається і для інших складів стекол $Si_xTe_{100-x}$. Тобто, для всіх складів стекол $Si_xTe_{100-x}$ електропровідність нижче $P_к$ є термічно активованою з єдиною енергією активації $E_a$ у діапазоні температур 100–300 К. Концентраційні залежності енергії активації $E_a$ при різних тисках нижче $P_К$ приведені на рис. 4.33, *б*.

Як видно із рис. 4.33 *а* і *б*, такі властивості стекол $Si_xTe_{100-x}$ як критичний тиск переходу, питомий опір при атмосферному тиску і кімнатній температурі, а також енергія активації електропровідності при різних тисках демонструють аномальну зміну для складу $x$ = 20. Цю аномальну зміну властивостей при $x$ = 20, автори [80] зв’язують із ідеальністю склоподібної фази для даного хімічного складу.

Для всіх складів стекол $Si_xTe_{100-x}$ виявлено структурний фазовий перехід при критичних тисках, положення якого зсувається у бік низьких тисків як зі збільшенням, так і зі зменшенням процентного вмісту телуру по відношенню до складу $Si_{20}Te_{80}$. Необхідно відзначити, що аналогічну граничну поведінку поблизу складу з $x$ = 20 у бінарних стеклах $Si_xTe_{100-x}$ проявляють градієнт електричного поля, виміряний за ефектом Мессбауера, коливні моди в спектрах КРС і температура кристалізації [96]. Відповідно до [96] ця гранична поведінка зазначених характеристик служить доказом морфологічної структурної зміни, яка може керуватися зв'язністю сітки скла або середнім координаційним числом. При $x \leq 20$ сітка складається переважно з поперечно зв'язаних кремнієм ланцюжкових фрагментів $Te_n$. При $x \geq 20$ ланцюжки перебудовуються в тетраедричні одиниці

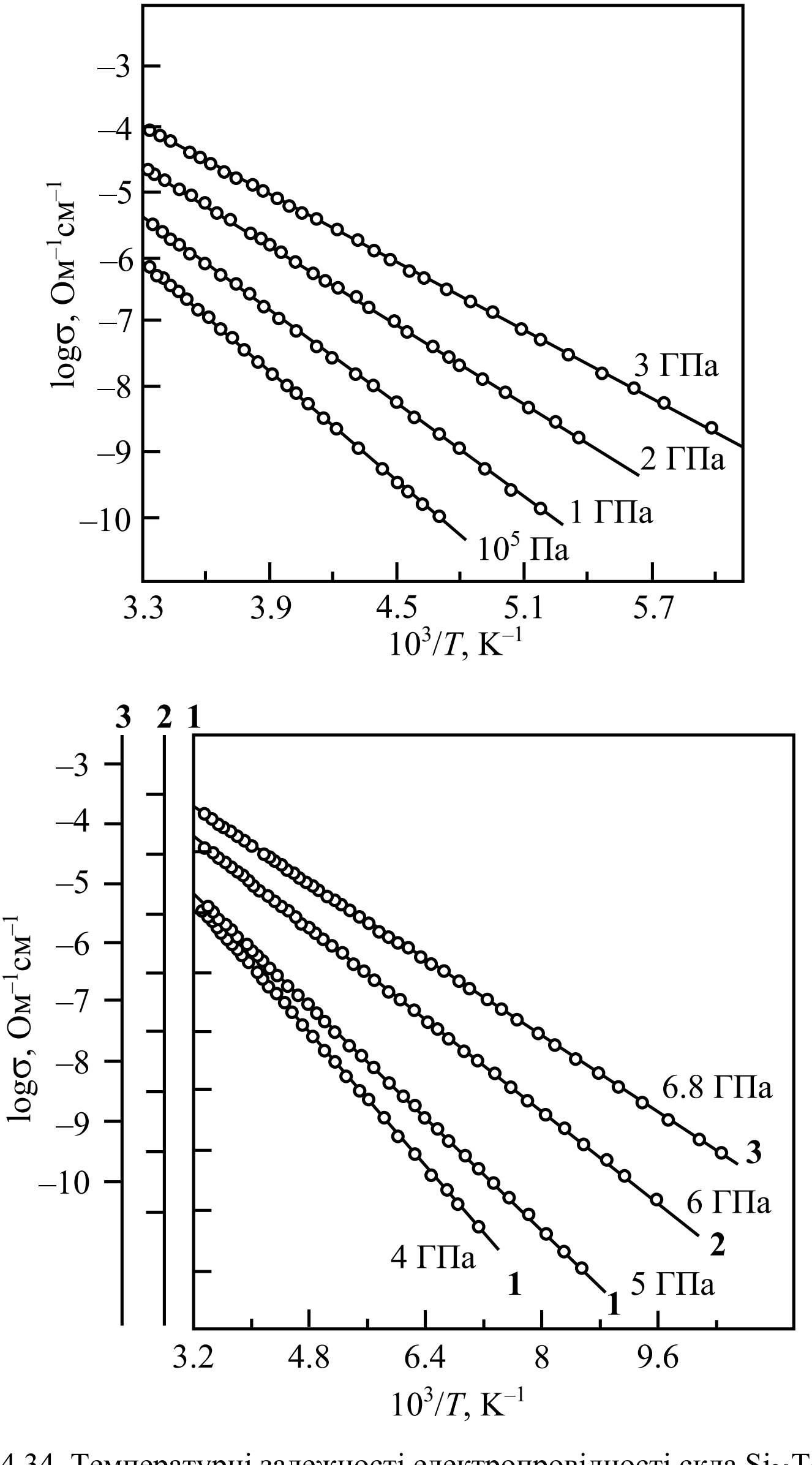

Рис. 4.34. Температурні залежності електропровідності скла $Si_{20}Te_{80}$, виміряні при різних тисках [105].

$Si(Te_{1/2})_4$ і сегрегуються у скріплені дефектом $Si_2Te_3$-подібні молекулярні фрагменти. Ці фрагменти являють собою жорсткі області, які проникають вище за поріг.

Рентгеноструктурні дослідження показали, що стекла $Si_xTe_{100-x}$ при тисках вище $P_к$ кристалізуються [105, 106]. Кристалічні фази високого тиску мають гексагональну структуру з параметрами ґратки $a$ = 6.42 Å, $c$ = 4.74 Å при $10 \le x \le 17$; $a$ = 4.0 Å, $c$ = 6.0 Å при $x$ = 20; $a$ = 6.06 Å, $c$ = 11.58 Å при $22 \le x \le 25$.

Рентгеноструктурні дослідження зразків $Si_xTe_{100-x}$, отриманих при високих тисках вище $P_t$, показали [80], що при цьому тиску має місце поліморфна кристалізація скла, причому цей перехід є також переходом напівпровідник-метал. Підтвердженням цього є той факт, що питомий електричний опір зразків залишається постійним після фазового переходу.

## 4.7. НАДПРОВІДНІСТЬ СТЕКОЛ $Si_{15}Ag_{15}Te_{70}$ І $Si_{15}Ag_5Te_{80}$

Введення до складу стекол системи Si–Te срібла (Ag) приводить до суттєвого впливу на величину питомого опору, характеру зміни ρ із тиском, та появи надпровідності при високих тисках і низьких температурах. Авторами [133] досліджені температурні залежності питомого опору, температури переходу $T_н$ у надпровідний стан і критичні магнітні поля $H_{н2}(T)$ для двох складів стекол $Si_{15}Ag_{15}Te_{70}$ і $Si_{15}Ag_5Te_{80}$.

При кімнатній температурі й атмосферному тиску питомий опір даних потрійних стекол становить $10^5$–$10^6$ Ом·см. На відміну від бінарного скла $Si_{20}Te_{80}$ (крива 1, рис. 4.35), при збільшенні тиску до 150 кбар питомий опір стекол $Si_{15}Ag_{15}Te_{70}$ і $Si_{15}Ag_5Te_{80}$ плавно зменшується на 6–7 порядків (криві 2 і 3, рис. 4.35). При зменшенні тиску від 150 кбар до нуля, має місце невелика незворотна зміна ρ у кілька разів, що характерно для більшості ХСН, які піддаються дії високих тисків. Подальше збільшення тиску до 200 кбар для скла $Si_{15}Ag_{15}Te_{70}$ приводить до незворотних змін питомого опору ρ при зменшенні тиску до нуля. У цьому випадку електропровідність σ після зняття тиску збільшується на 2 порядки в порівнянні з її значенням до початку прикладання тиску.

Згідно даних [133] при тисках нижче 80 кбар питомий опір стекол $Si_{15}Ag_{15}Te_{70}$ і $Si_{15}Ag_5Te_{80}$ зростає при охолодженні (в області температур 300–200 К) за експоненційним законом. На рис. 4.36 приведені

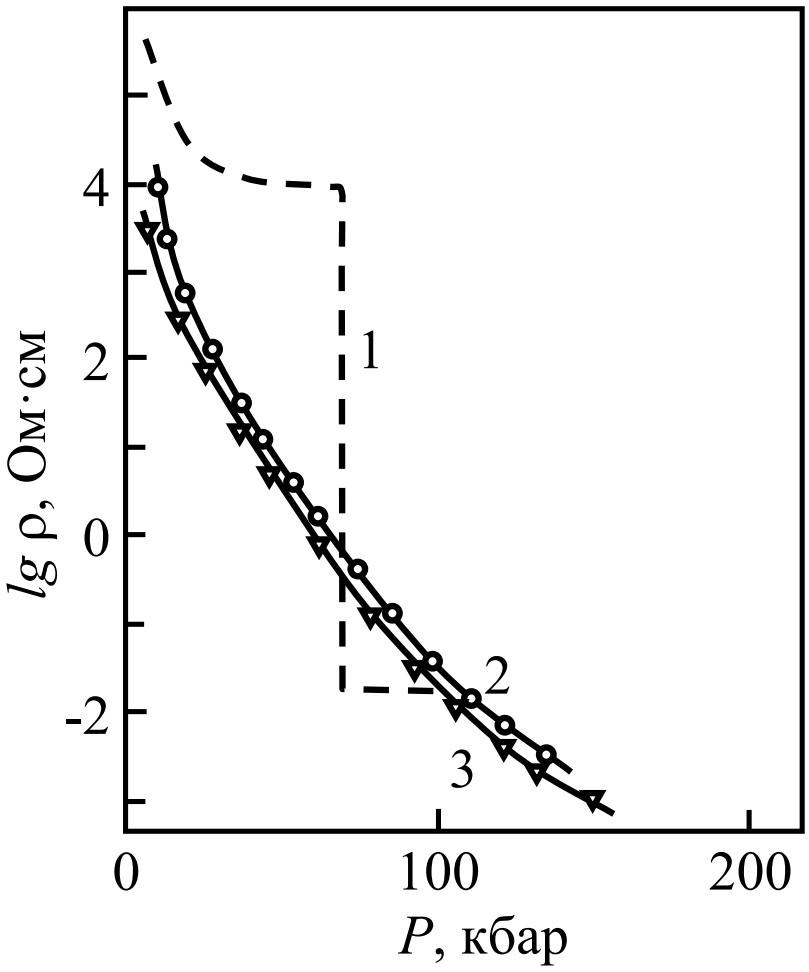

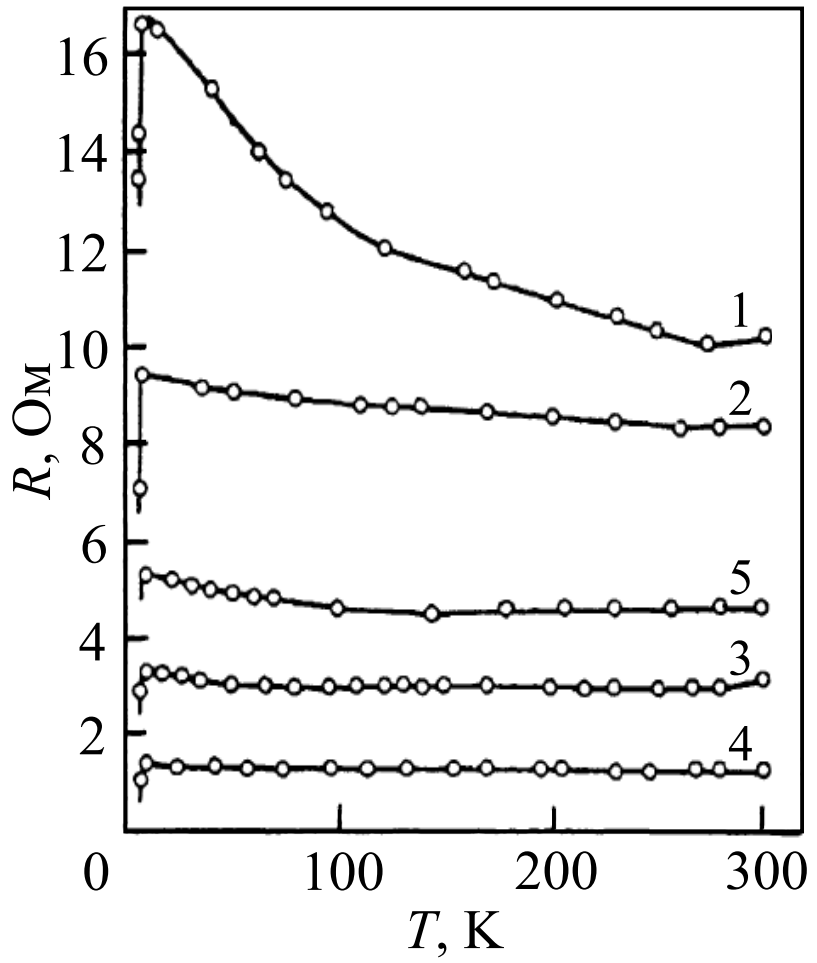

Рис. 4.35. Залежність *lg* ρ від тиску стекол $Si_{20}Te_{80}$(1), $Si_{15}Ag_{15}Te_{70}$(2) і $Si_{15}Ag_{5}Te_{80}$(3) [133].

Рис. 4.36. Температурні залежності опору стекол $Si_{15}Ag_{15}Te_{70}$ (1–4) і $Si_{15}Ag_{5}Te_{80}$ (5), виміряні при різних тисках. *P*, кбар: 1– 90; 2 – 95; 3 – 130; 4 – 150; 5 – 120 [133].

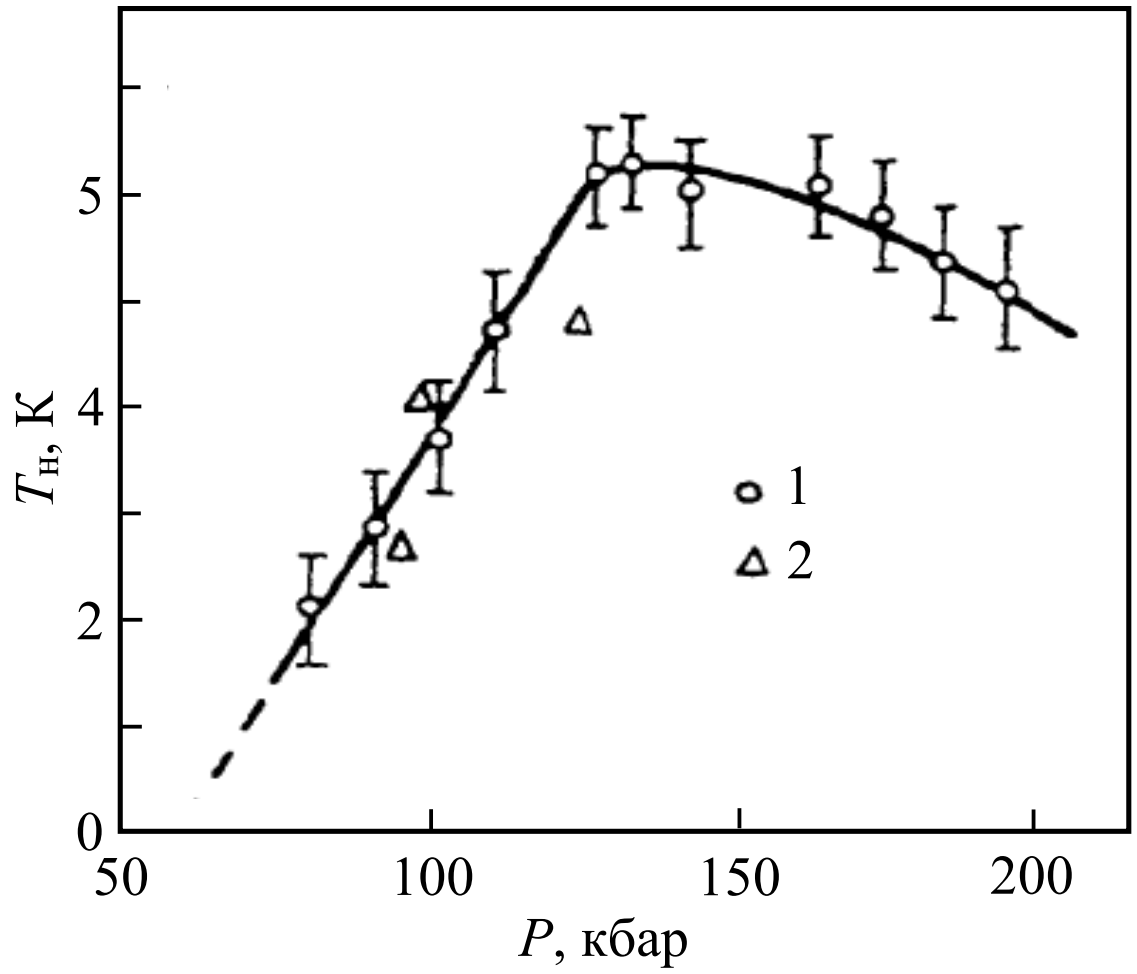

Рис. 4.37. Залежність температури переходу в надпровідний стан ($T_н$) від тиску для стекол $Si_{15}Ag_{15}Te_{70}$(1) і $Si_{15}Ag_{5}Te_{80}$(2) [133].

температурні залежності опору стекол $Si_{15}Ag_{15}Te_{70}$ і $Si_{15}Ag_5Te_{80}$, виміряні при різних тисках. В області тисків вище 80 кбар спостерігається зростання опору при зменшенні температури аж до початку переходу в надпровідний стан. Величина цього зростання поступово зменшується зі збільшенням тиску. При $P > 150$ кбар в $Si_{15}Ag_{15}Te_{70}$ і при $P > 100$ кбар в $Si_{15}Ag_5Te_{80}$ її значення становить кілька відсотків.

Надпровідність у склі $Si_{15}Ag_{15}Te_{70}$ автори [133] виявили при тиску $P \sim 80$ кбар, а з підвищенням тиску температура переходу в надпровідний стан зростає зі швидкістю $dT_н/dP = 0.1$ К·кбар$^{-1}$. Критична температура досягає максимального значення 6.4 К при 130 кбар, а подальше збільшення тиску до 200 кбар приводить до монотонного зменшення $T_н$. За умови, що величина прикладеного тиску не перевищує 140 – 150 кбар, поведінка $T_н$ при прикладанні тиску зворотна. При зменшенні тиску від значення $P = 200$ кбар автори [133] спостерігали незворотню поведінку $T_н$: при $P < 100$ кбар переходи в надпровідний стан стають двоступінчастими. Верхній ступені відповідає практично не залежне від тиску значення $T_н = 5.6$ К, а нижній – $T_н$, що зменшується при зменшенні тиску до $T_н < 1.5$ К. Для скла $Si_{15}Ag_5Te_{80}$ $T_н$ також зростає при стисканні в інтервалі тисків 80 – 120 кбар. Залежності $T_н$ від $P$ наведені на рис. 4.37.

Виникнення надпровідності під впливом тиску непов'язане з радикальною зміною структури зразків при прикладанні тиску. Про це свідчать результати рентгеноструктурних досліджень стекол $Si_{15}Ag_{15}Te_{70}$, підданих дії тиску ~100 кбар, які підтверджують аморфність при прикладанні та подальшому знятті тиску [133].

Результати рентгеноструктурних досліджень та зворотність поведінки електричних та надпровідних характеристик в області тисків $P < 150$ кбар вказують на те, що надпровідність з'являється та існує в склоподібній фазі $Si_{15}Ag_{15}Te_{70}$. Перехід від діелектричного стану до металевого відбувається приблизно при тих самих тисках, при яких з'являється надпровідність, і на думку авторів [133] зв'язаний з перетином порога рухливості $E_c$ рівнем Фермі $E_F$ (перехід діелектрик-метал андерсонівського типу).

### 4.8. ЕФЕКТ ПЕРЕМИКАННЯ І ПАМ'ЯТІ У СТЕКЛАХ $Si_xTe_{100-x}$

Під впливом сильних електричних полів халькогенідні стекла переходять зі свого початкового стану низької провідності (ВИМК) у стан високої провідності (УВІМК) Це електричне перемикання буває двох типів, порогове та з пам'яттю [192]. Стекла порогового перемикання повертаються до свого початкового стану високого опору. Після вимкнення прикладеного електричного поля стекла з пороговим перемиканням повертаються у свій вихідний стан (ВИМК), тоді як стекла з перемиканням пам'яті залишаються у своєму стані високої провідності. У перемикаючих стеклах з пам'яттю із за джоулевого нагрівання відбувається структурний фазовий перехід із аморфного у кристалічний стан (незворотній перехід). Перемикаючі стекла порогового типу не зазнають структурного фазового переходу, і перемикання відбувається за рахунок зворотніх електронних переходів.

Необхідно відмітити, що ефекти електричного перемикання в халькогенідних склоподібних напівпровідниках і, відповідно, ефект пам'яті активно досліджувались починаючи з 1960-х років. Початком цих досліджень було відкриття ефекту перемикання в стеклах Tl–As–Se(Te) авторами [193]. Трохи пізніше, Овшинським [194] на стеклах $Si_{12}Te_{48}As_{30}Ge_{10}$ було виявлено ефект пам'яті зв'язаний з кристалізацією. Електричне перемикання відноситься до керованого електричним полем переходу, який проявляється в аморфних і склоподібних халькогенідах із напівпровідникового вимкнутого (ВИМК) стану в провідний увімкнений (УВІМК) стан, який може бути двох типів, а саме: пам'ять і порогове перемикання. Перемикання памяті – це явище, яке включає структурний фазовий перехід (аморфно-кристалічний); тоді як порогове перемикання є процесом, характерним для аморфної фази, і не включає структурний фазовий перехід. Перемикання памяті відноситься до кристалізації, і це відрізняється від порогового перемикання.

Вольт-амперні характеристики і ефект електричного перемикання в об'ємних стеклах $Si_xTe_{100-x}$ ($15 \le x \le 25$), отриманих загартуванням розплаву від температур 1373 K, у суміші крижаної води і NaOH досліджені авторами [195] в широкому діапазоні складів. Досліджувані зразки у вигляді шліфованих пластинок товщиною 0.2–0.4 мм поміщались у тримач, який складався із плоского нижнього електрода і точкового верхнього електрода з пружинним механізмом для утримання зразка. Ще для двох складів стекол $Si_{15}Te_{85}$ і $Si_{20}Te_{80}$ ефект перемикання дослідили автори [196, 197].

Типові ВАХ для трьох складів стекол $Si_{15}Te_{85}$, $Si_{17}Te_{83}$ і $Si_{20}Te_{80}$ приведені на рис. 4.38, які показують, що досліджувані стекла демонструють вихідну омічну поведінку з високим опором ( стан

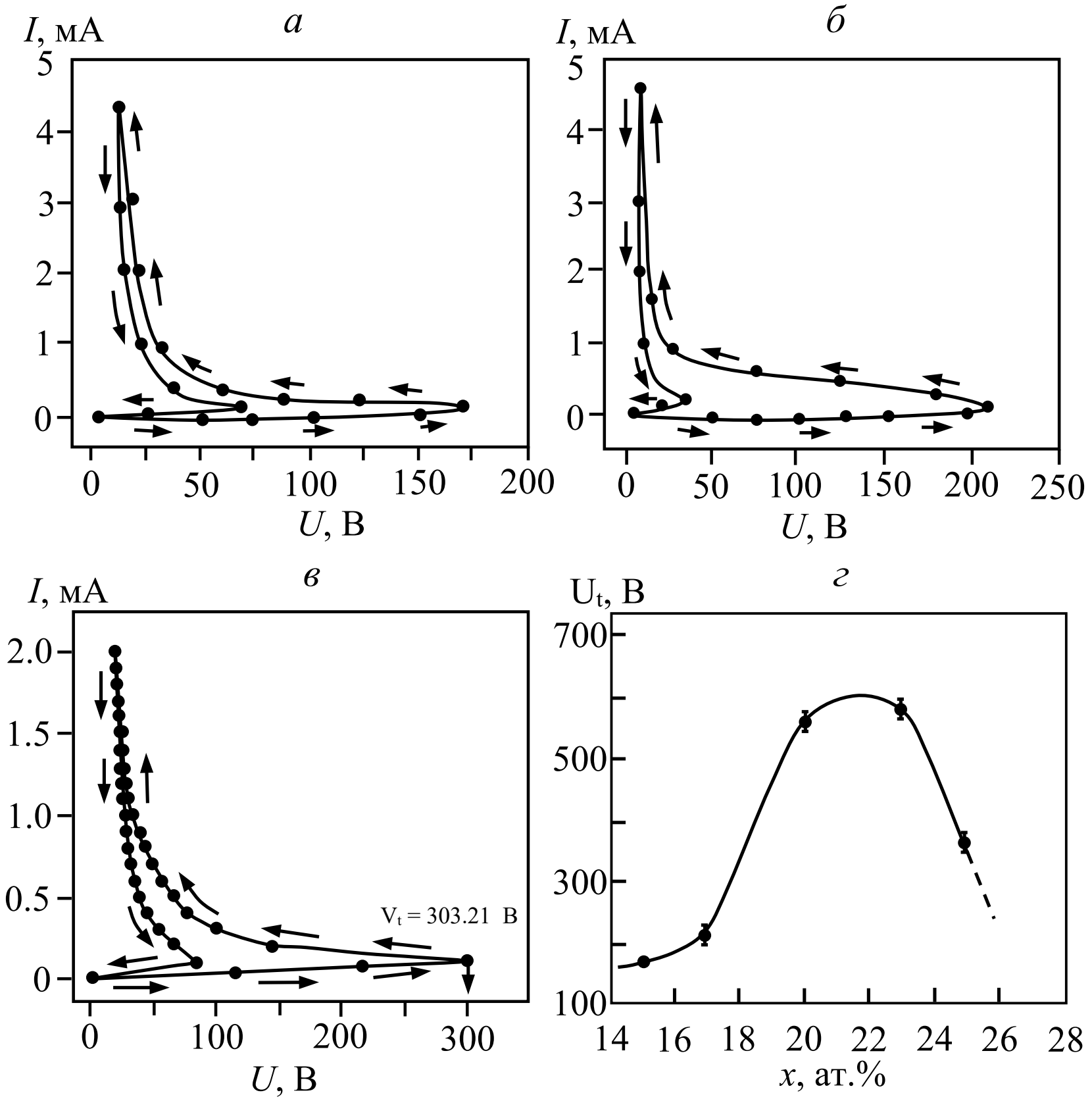

Рис. 4.38. Вольт-амперні характеристики стекол $Si_{15}Te_{85}$(*а*), $Si_{17}Te_{83}$(*б*) [195], $Si_{20}Te_{80}$(*в*) [196]. Стрілки вказують на напрямок руху. *г* – концентраційна залежність напруги перемикання стекол $Si_xTe_{100-x}$ [195].

ВИМК). Поблизу критичної напруги ($U_{th}$) зразки демонструють область від'ємного опору і швидке електричне перемикання із стану високого опору (ВИМК) в стан низького опору (УВІМК). Виявлено, що зразки усіх складів стекол системи Si–Te залишаються у стані УВІМК з високою провідністю і не повертаються у свій початковий стан ВИМК з високим опором навіть після зняття прикладеного електричного поля. Цей факт ясно вказує на те, що стекла $Si_xTe_{100-x}$ демонструють поведінку перемикання пам'яті. Стекла $Si_xTe_{100-x}$ демонструють чисте електричне перемикання без будь-яких коливань ВАХ при переході в стан УВІМК. Поля перемикання для даних бінарних стекол складають 6–25 кВ/см.

На рис. 4.38, *г* приведена залежність напруги порогового перемикання $U_t$ від складу стекол $Si_xTe_{100-x}$ в межах області склоутворення. З цього рисунка видно, що напруга перемикання цих стекол спочатку збільшується зі збільшенням $x$, демонструючи широкий максимум в околі $x = 20$, тобто біля порогу перколяції жорсткості. Через близькість між хімічним і механічним порогами відбувається поворот у $U_t$, який відбувається дуже близько до RPT і призводить до видимого максимуму в $U_t$.

Загально прийнятим є той факт, що ефект перемикання як у порогових, так і в перемикаючих стекол носить електронний характер і виникає, коли пастки заряджених дефектів, які наявні в халькогенідних стеклах, заповнені польовими носіями заряду, збудженими прикладеним електричним полем. Додаткові теплові ефекти вступають в процес у стеклах з перемиканням пам’яті з утворенням кристалічного каналу в області електрода в результаті нагрівання джоулевим теплом.

# РОЗДІЛ 5

## ФОТОЕЛЕКТРИЧНІ ВЛАСТИВОСТІ КРИСТАЛІВ $Si_2Te_3$ і ФОТОДЕТЕКТОРИ НА ЇХ ОСНОВІ

Граничні функціональні параметри напівпровідникових фотоелектричних матеріалів обмежуються їх структурною досконалістю. Значною мірою це стосується і шаруватих кристалів $Si_2Te_3$, яким властива розвинена система власних (природних) точкових і протяжних дефектів [9, 10, 24]. Аналіз впливу точкових і протяжних дефектів на властивості кристалів $Si_2Te_3$ та питання про цілеспрямоване управління їх концентрацією є одними з найважливішими проблемами фізики твердого тіла та технології.

В широкозонних напівпровідниках, до числа яких належить $Si_2Te_3$, фоточутливість визначається головним чином системою енергетичних рівнів, утвореною локальними центрами різної енергетичної природи. Для визначення параметрів локальних центрів і встановлення схеми електронних переходів у широкозонних напівпровідниках використовують комплекс стаціонарних і кінетичних методів дослідження фотопровідності. Найбільш повну інформацію про процеси прилипання і рекомбінації нерівноважних носіїв заряду дають вимірювання температурної залежності фотопровідності, термостимульованої провідності, рухливості і люкс-амперних характеристик.

Проведені на теперішній час дослідження вказують на велику різноманітність фотоелектричних властивостей сесквітелуриду кремнію: можливість отримання фоточутливих кристалів без спеціальних обробок, широкий енергетичний спектр домішкового фотоефекту, його трансформації при зміні умов вирощування кристалів, наявність тонкої структури в спектральному розподілі фоточутливості, сублінійність люкс-амперних характеристик, значна інерційність релаксації фотоносіїв та ін. Подібне багатство явищ, характерне для фотопровідника $Si_2Te_3$, спостерігається у спеціально нелегованих кристалах і вказує на значну фотоелектричну активність власних дефектів ґратки.

Враховуючи той факт, що у разі навіть нетривалого перебування кристалів $Si_2Te_3$ на повітрі, їх поверхня покривається шаром телуру, який утворюється в результаті хімічної реакції з незначними слідами парів води, для вивчення фотопровідності цих кристалів, необхідно регенерувати їх поверхню за допомогою вакуумної сублімації.

## 5.1. СПЕКТРИ ФОТОПРОВІДНОСТІ ШАРУВАТИХ КРИСТАЛІВ $Si_2Te_3$.

Спектральний розподіл фотопровідності визначає залежність фоточутливості напівпровідника від енергії (довжини хвилі) падаючого випромінювання і в загальному випадку відображає наявність двох внутрішніх фотоефектів: домішкового та власного. У першому випадку з домішкових центів відбувається генерація вільних носіїв одного типу – монополярна генерація, у другому – зона-зонна біполярна генерація. По «червоній» границі домішкового фотоефекту визначається енергетичне положення центрів. Величина домішкового фотоефекту зазвичай значно менша власного сигналу через менший коефіцієнт поглинання. Різке зростання сигналу фотопровідності відбувається в області фундаментального поглинання через зростання коефіцієнта поглинання. При певній енергії фотона, близькій до ширини забороненої зони, крива фоточутливості досягає максимуму, а потім спадає в області сильного поглинання. Причиною цього є поверхнева рекомбінація нерівноважних носіїв заряду (ННЗ). При освітленні напівпровідникового кристала світлом із області фундаментального поглинання, процеси генерації та рекомбінації нерівноважних носіїв заряду відбуваються як в об'ємі, так і на його поверхні. Просторовий розподіл швидкості генерації ННЗ задається коефіцієнтом поглинання та його спектральною залежністю $\alpha(h\nu)$, а аналогічна залежність процесу рекомбінації – відповідними швидкостями в об'ємі та на поверхні ($R$ та $R_S$). Як правило, швидкість рекомбінації на поверхні більша, ніж в об'ємі, навіть для ідеальної (атомарно чистої) поверхні через можливе захоплення ННЗ на стани в забороненій зоні, спричинені обривом зв'язків на границі кристалічної ґратки (рівні Тамма). Для реальної поверхні нерівність $R_S > R$ підсилюється внаслідок її великої дефектності як природної, так і тієї, що виникла в результаті обробки та контакту з навколишнім середовищем. Наявність поверхневих дефектів формує додаткові по відношенню до об'єму канали рекомбінації, що зменшують час життя та концентрацію ННЗ біля поверхні. Додатковий до об'єму темп рекомбінації на поверхні характеризується швидкістю рекомбінації, як коефіцієнт пропорційності $R_S = S{\cdot}\Delta n = S\Delta p$.

Питома фотопровідність $\Delta\sigma_ф$ у загальному вигляді описується виразом

$$\Delta\sigma_ф = e(u_n\Delta n + u_p\Delta p)G(\alpha) = eu_n(\tau_n + b^{-1}\tau_p)G(\alpha), \qquad (5.1)$$

де $\Delta n$ і $\Delta p$ – концентрації, $u_n$ і $u_p$ – рухливості ($b = u_n/u_p$), $\tau_n$ і $\tau_p$ – часи життя нерівноважних носіїв заряду – електронів та дірок, $G(\alpha)$ – темп оптичної генерації пар в одиниці об'єму, що залежить від $\alpha$ та швидкості поверхневої рекомбінації $S$. Його величина в кристалі товщиною $d$

$$G(\alpha) = \beta(1 - R^0)\frac{\Phi_0}{d}[1 - \exp(-\alpha d)] \qquad (5.2)$$

визначається спектральною залежністю коефіцієнта поглинання $\alpha(h\nu)$, квантовим виходом внутрішнього фотоефекту $\beta$, коефіцієнтом відбивання $R^0(h\nu)$, густиною потоку фотонів $\Phi_0$.

У широкозонних напівпровідниках фоточутливість визначаються переважно системою локальних центрів, які формують схеми оптичних і термічних переходів (енергетичних рівнів, утворених локальними центрами різної енергетичної природи). Вона фіксує як ступінь компенсації напівпровідника, так і швидкість рекомбінації у ньому нерівноважних носіїв заряду [199–201]. При описі процесу рекомбінації та побудові адекватної йому рекомбінаційної моделі необхідно враховувати наступні фактори [200, 202]:

1. Усі типи центрів у тій чи іншій мірі впливають на фоточутливість (концентрацію ННЗ $\Delta n$, $\Delta p$ та час їхнього життя $\tau$) безпосередньо як центри рекомбінації або опосередковано (за умови електронейтральності) як $t$-центри прилипання основних та неосновних носіїв заряду.
2. Фотоелектрична активність центрів $i$-типу, тобто ступінь їх впливу на $\Delta n$ і $\tau$, залежить від колективних параметрів (концентрації $\Re_i$, темнового $N_{0i}$ та світлового електронного $N_i$ заповнення) та індивідуальних (коефіцієнтів захоплення $C_{ni}$ електронів та $C_{pi}$ дірок, енергетичного положення щодо $c$- або $\upsilon$-зони $E_{ci}$, $E_{\nu i}$, перерізів захоплення фотона $S_\phi$ та ін).
3. Роль центрів $i$-типу в рекомбінації різна в залежності від умов експерименту (температури, рівня збудження): центри рекомбінації можуть стати центрами прилипання або зовсім виключатися з процесу.
4. Електрична природа центрів (донори, акцептори) та їх сумарні концентрації визначають ступінь компенсації напівпровідника і тим самим контролюють час життя ННЗ у разі слабкого рівня збудження.

Незалежно від методу вирощування, кристали $Si_2Te_3$ є фоточут-ливими без спеціальних додаткових обробок. При кімнатній температурі інтегральна фоточутливість кристалів становить $\sigma_ф/\sigma_т = 10^2$–$10^3$, де $\sigma_ф$ – електропровідність при освітленості $10^4$ Лк.

Основною характеристикою будь-якого фотопровідника є спектральний розподіл фоточутливості [201]:

$$I_ф = e(u_n + u_p)\frac{U}{l^2}\Phi_0 \cdot \beta \cdot (1 - R^0)\tau F, \qquad (5.3)$$

де $e$ – заряд електрона, $u_n$, $u_p$ – рухливість електронів та дірок; $U$ – зовнішня різниця потенціалів, прикладена до зразка; $(1–R^0)\Phi_0$ – частка відбитого від поверхні зразка світла, β – квантовий вихід внутрішнього фотоефекту, $R^0$ – коефіцієнт відбивання, $l$ – довжина зразка, $F$ – фактор впливу поверхні, який характеризує відношення виміряного фотоструму до «ідеального», тобто такого, який був би за умови відсутності поверхневої рекомбінації ($S = 0$) і повного поглинання падаючих на поверхню квантів світла $\Phi_0$:

$$F = \frac{1 - \exp(-\alpha d)}{1 + S} + \frac{S}{1 + S}\left[\frac{1}{1 + \alpha L} - \frac{\exp(-\alpha d)}{1 - \alpha L}\right]. \qquad (5.4)$$

Оскільки кристали $Si_2Te_3$, незалежно від методу їх вирощування, мають значну інтегральну фоточутливість ($\sigma_ф/\sigma_т = 10^2$–$10^3$), це дозволило авторам [140] провести дослідження їх спектрів фотопровідності. З цією метою на природні грані кристалічних зразків наносились золоті омічні контакти так, щоб реалізувалась компланарна геометрія, тобто між контактами був зазор в 5–6 мм, в який відбувалося освітлення зразка.

Спектри фотопровідності кристалів $Si_2Te_3$, вирощених методами сублімації та анти-Бріджмена (Піццарелло), досліджені в роботах [26, 140, 172, 198] в інтервалі температур 93 – 373 K. Враховуючи, що сесквітелурид кремнію має двосторонню область гомогенності (див. § 1.1, розділ 1), тому важливим також є вивчення ступеня відхилення складу кристалів від стехіометричного на спектральний розподіл фоточутливості. З цією метою, авторами [140] досліджені спектри фотопровідності кристалів $Si_2Te_3$, вирощених як із стехіометричної шихти, так і шихти, що містила надлишок телуру. Типові неполяризовані спектри ФП кристалів першого типу, виміряні в інтервалі температур 293–440 К на постійному струмі та модульованої освітленості досліджуваного зразка, наведено на рис. 5.1. Як видно з цього рисунка, у спектрах ФП кристалів $Si_2Te_3$, вирощених із стехі-ометричної шихти, спостерігається одна широка смуга, енергетичне положення максимуму якої зсувається в область менших енергій зі збільшенням температури зразка, що відображає зменшення ширини забороненої зони.

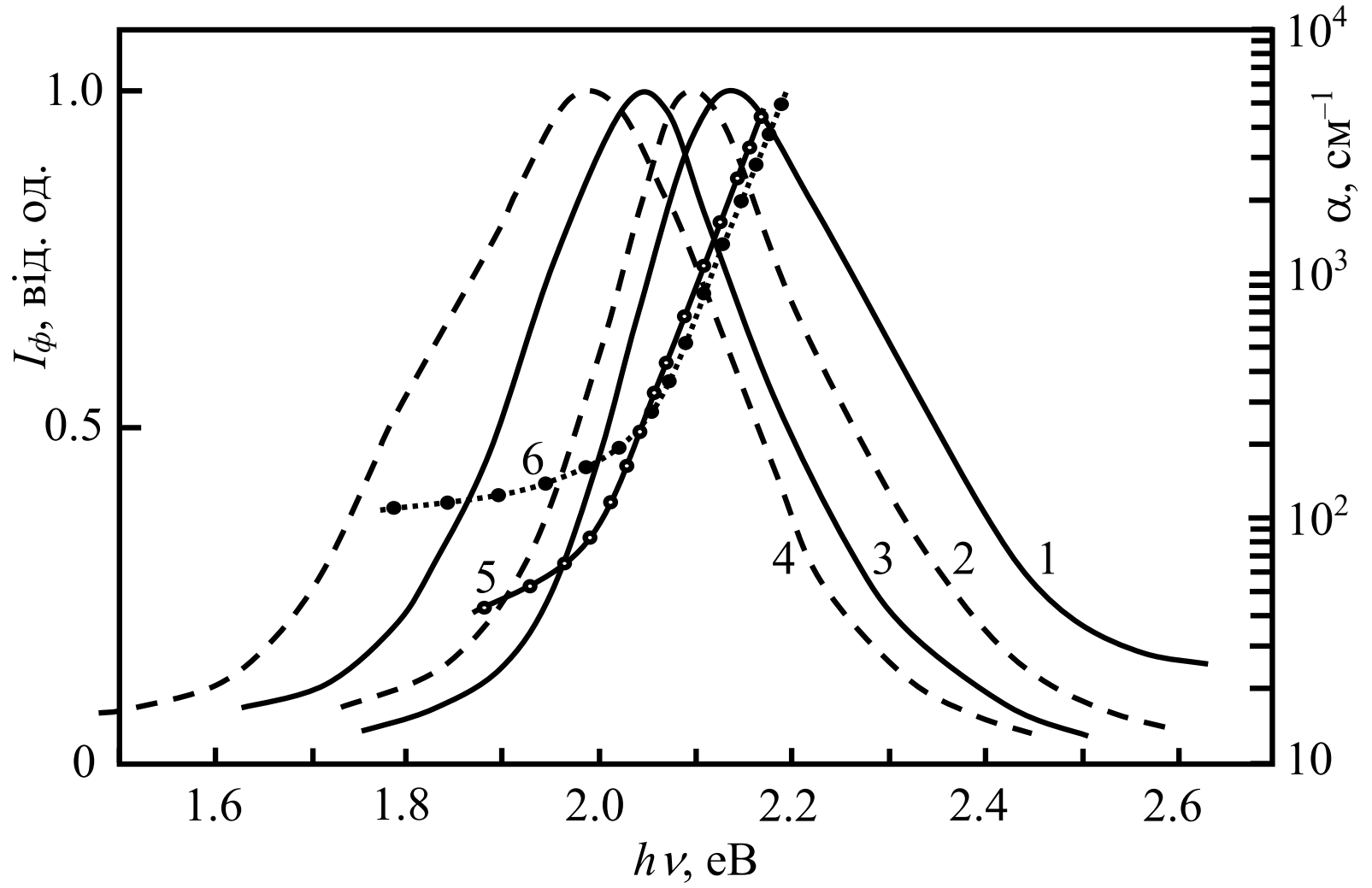

Рис. 5.1. Спектральні залежності стаціонарного фотоструму $I_ф(h\nu)$ (1–4) та крайового поглинання (5, 6) монокристалів $Si_2Te_3$.
*T*, K: 1 – 293; 2 – 350; 3 – 410; 4 – 440. (крива 5 – [140]; крива 6 – [150]).

Для фотопровідності суттєвим є стан поверхні досліджуваного зразка. Поверхнева рекомбінація зменшує ефективний час життя носіїв при поверхневій генерації і приводить, таки чином, до падіння фотопровідності при зменшенні довжини хвилі випромінювання в області власного поглинання. При збільшенні енергії фотона вище краю власного поглинання спостерігається зменшення фотопровідності внаслідок того, що генерація електронно-діркових пар відбувається ближче до поверхні і поверхнева рекомбінація зменшує ефективний час життя. Таким чином, короткохвильовий спад фотопровідності (фотоструму) викликаний поверхневою рекомбінацією носіїв струму. Її роль підвищується у більш короткохвильній частині спектра по мірі зростання коефіцієнта поглинання та зменшення у зв'язку з цим поверхневого шару, в якому генеруються фотоносії.

Для ідентифікації природи максимуму в спектрі ФП на рис. 5.1 наведено спектри крайового поглинання кристала $Si_2Te_3$, взяті з робіт [140, 150] (криві 5 і 6, відповідно). Із зіставлення спектрів ФП і фундаментального поглинання випливає, що при T = 293 K енергетичне положення максимуму $h\nu_{max}$ = 2.13 eB у спектрі фотопровідності знаходиться в області власного поглинання та відповідає значенню $\alpha \approx 2\cdot10^3$ см$^{-1}$. Таким чином, природа цього максимуму зумовлена генерацією нерівноважних носіїв, викликаних оптичними зона-зонними переходами (Г→К) з вершини валентної зони, сформованої 5*p*-станами неподіленої електронної пари (*lone pair*) телуру на дно зони провідності, сформованої змішуванням вільних *p*-станів телуру та кремнію (§ 3.1, рис. 3.3, розділ 3).

При цьому природно виникає питання, яким саме способом правильно визначити значення ширини забороненої зони зі спектрів ФП кристала $Si_2Te_3$? При дослідженні фотопровідності гомополярних напівпровідників (Si, Ge та інш.), край смуги фундаментального поглинання яких різко виражений, $E_g$ визначають за порогом фотопровідності (за правилом Мосса). Однак, у випадку кристалів $Si_2Te_3$ частотна залежність коефіцієнта поглинання $\alpha(h\nu)$ в області краю фундаментального поглинання при $\alpha \leq 10^3$ см$^{-3}$ (рис. 5.1) не коренева, як у випадку прямозонних переходів в ідеальних напівпровідниках.

Крім того, провести оцінку $E_g$ за енергетичним положенням напівспаду довгохвильової ділянки основного (власного) максимуму

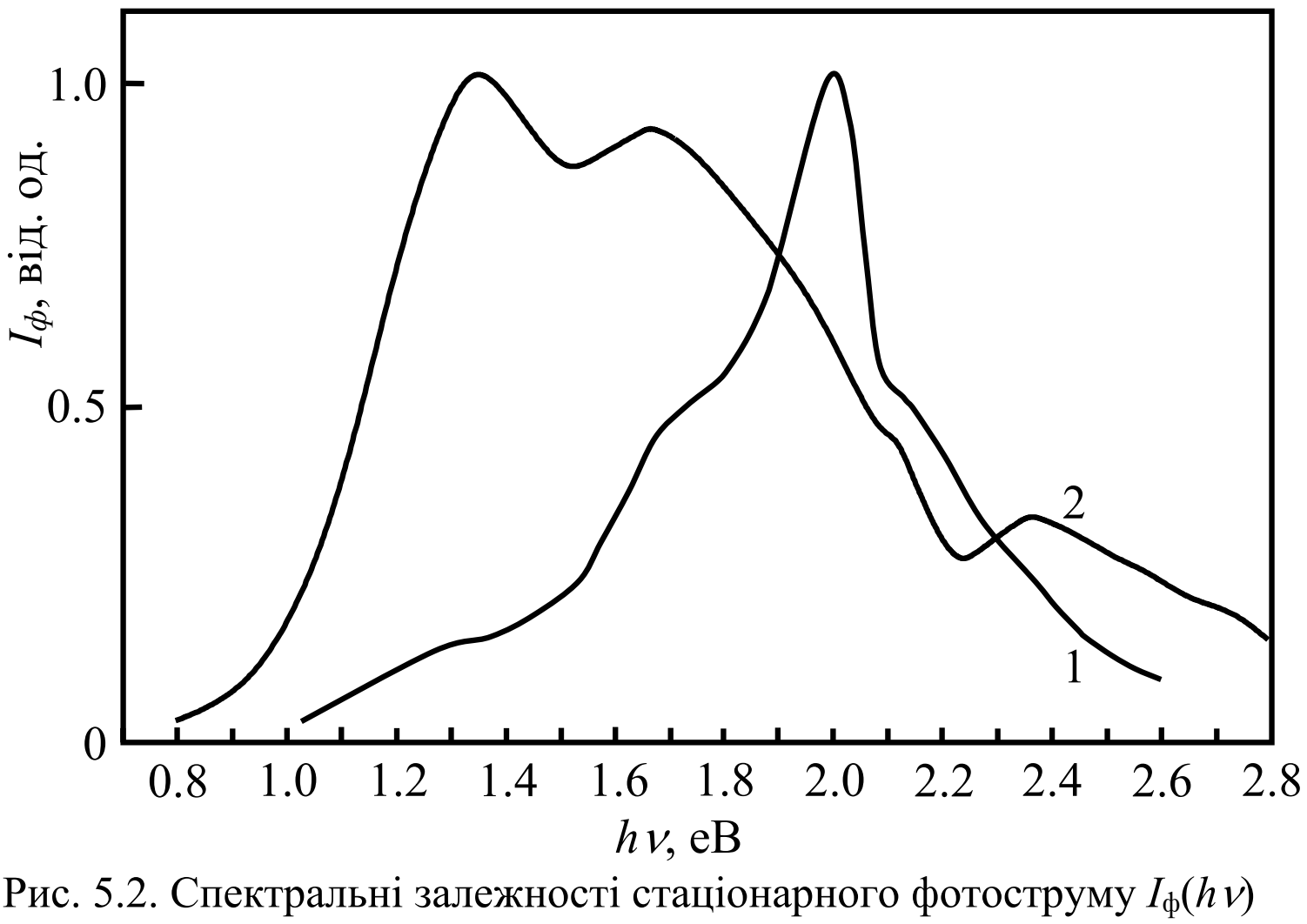

Рис. 5.2. Спектральні залежності стаціонарного фотоструму $I_ф(h\nu)$ кристалів $Si_2Te_3$, вирощених методом сублімації з стехіометричної шихти (1) і шихти, збагаченої телуром (2). $T$ = 293 K [140].

фотопровідності ускладнено ще й тим, що в кристалах $Si_2Te_3$ сильно виражена домішкова смуга, яка до того ж перекривається з власною смугою фотопровідності (рис. 5.2, крива 2).

У такому випадку, як показали автори [203], в широкозонних кристалах з експоненційною залежністю довгохвильового краю власного поглинання в широких межах товщин зразків і швидкостей поверхневої рекомбінації ефективна ширина забороненої зони може бути з великою точністю визначена за енергетичним положенням власного максимуму спектральної характеристики фотопровідності. Таким чином, якщо енергію міжзонних переходів оцінювати за спектральним положенням власного максимуму фотопровідності, то з наведених на рис. 5.1 спектрів ФП випливає, що ширина забороненої зони кристала $Si_2Te_3$ рівна $E_g$ = 2.13 еВ при кімнатній температурі.

Як зазначають автори [140], навіть при вирощуванні кристалів $Si_2Te_3$ методом сублімації із стехіометричної шихти в одній і тій самій ампулі виростають кристали, спектр ФП яких (крива 1, рис. 5.2) відрізняється від описаного вище. Як видно з рис. 5.2, фоточутливість таких кристалів проявляється у більш ширшому спектральному діапазоні 1.0÷2.5 еВ, а сам спектр ФП є складним і містить яскраво виражений інтенсивний пік при 2.02 еВ, одну особливість у вигляді напливу при 2.12 еВ на високоенергетичному спаді основного піка та дві особливості при 1.7 і 1.32 еВ на довгохвильовому спаді.

Спектр фотопровідності нестехіометричних кристалів $Si_2Te_3$, вирощених методом сублімації з вихідної шихти, яка містила надлишок телуру, зазнає ще більших змін (крива 2, рис. 5.2). На цьому рисунку видно, що в спектрі ФП домінуючими є домішкові смуги з максимумами при 1.65 еВ та 1.33 еВ, а інтенсивність власного максимуму різко зменшується і він проявляється у вигляді перегину при ~2.1 еВ. Враховуючи, що нестехіометричні кристали $Si_2Te_3$ містять одночасно вакансії кремнію (за природою самої речовини) та надлишкові атоми телуру для встановлення природи домішкових смуг у спектрах ФП необхідні додаткові комплексні дослідження стаціонарних та кінетичних характеристик фотопровідності.

Спектри ФП кристалів $Si_2Te_3$, вирощених методом Піццарелло, досліджені авторами [26, 172], приведені на рис. 5.3. При $T$ = 93 К в спектрах ФП наявна інтенсивна смуга з максимумом при 2.2 еВ в області фундаментального поглинання, особливість у вигляді перегину при 1.9 еВ на довгохвильовому спаді основної смуги і широка довгохвильова смуга з максимумом при ~1 еВ (рис. 5.3, крива 4). ІЧ максимум (~1 еВ) спостерігається тільки після освітлення досліджу-

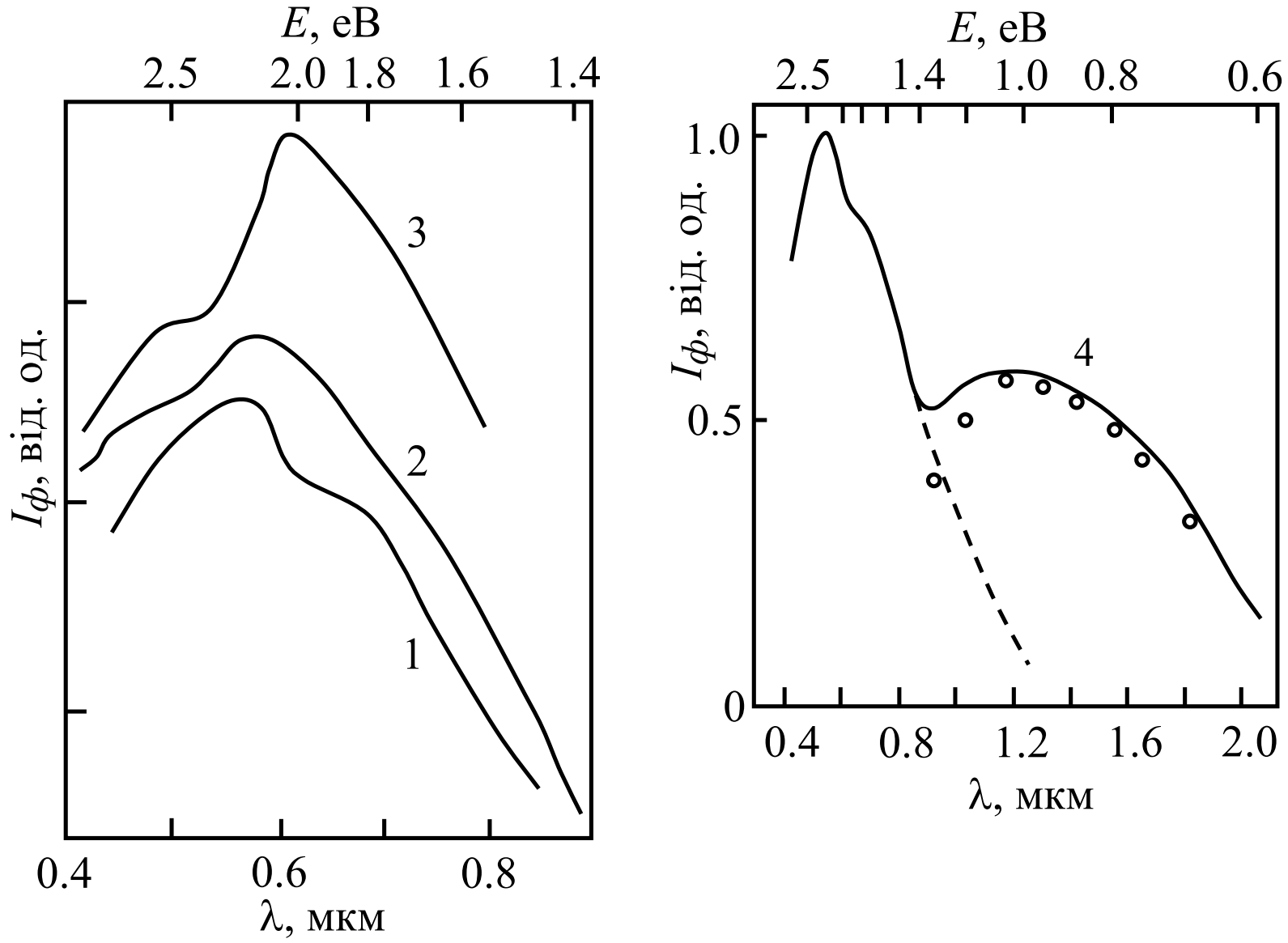

Рис. 5.3. Спектральний розподіл фотоструму в кристалі $Si_2Te_3$, вирощеного методом Піццарелло. *T*, К: 1, 4 – 93, 2 – 292, 3 – 373 [26, 172].

ваного зразка світлом із області власного поглинання. Автори [26, 172] вважають, що цей ІЧ максимум обумовлений пастками, які заповнюються дірками при освітленні власним світлом, а в процесі вимірювання опустошуються.

## 5.2. ТЕМПЕРАТУРНА ЗАЛЕЖНІСТЬ СТАЦІОНАРНОЇ ФОТОПРОВІДНОСТІ КРИСТАЛІВ $Si_2Te_3$.

Вивчення стаціонарних залежностей фотопровідності (температурної залежності фотоструму $I_ф$ та люкс-амперних характеристик) дає певні відомості про рекомбінаційні процеси у напівпровідниках. Хід залежностей дозволяє в загальноному встановити схему рекомбінації (кількість фотоелектрично активних центів, кількість та інтенсивність фотоелектричних переходів у дослідженому інтервалі температур при даному рівні збудження) ,а також визначити ряд параметрів цих центрів.

Поряд із температурним гасінням фотоструму, у більшості фоточутливих напівпровідників спостерігається зростання його (активація) зі збільшенням температури зразка. Апріорі можна вказати на кілька можливих причин такої залежності. Активація фотоструму може бути зумовлена температурними змінами рухливості основних носіїв (у разі домішкового механізму розсіювання), перерізів захоплення їх на центри рекомбінації, квантового виходу фотоефекту при складному механізмі генерації (наприклад, екситонному) і, нарешті, прилипанням основних носіїв струму [199, 202].

За наявності у фундаментальній щілині напівпровідника домішкових рівнів, зайнятих електронами (або дірками), генерація нерівноважних носіїв заряду можлива при оптичному збудженні квантами з енергією $h\nu < E_g$. Як зазначалося в § 1.4.3, розділ 1, особливістю кристалічної структури сесквітеллуриду кремнію є статистичне розміщення 8 атомів кремнію у двох позиціях 12і та одній 4е, причому зазначені позиції заповнені з дефіцитом 71%, оскільки в них замість 28 атомів кремнію розміщуються лише 8. Саме наявність цих катіонних вакансій (власних точкових дефектів) у ґратці кристалічного $Si_2Te_3$ приводить до утворення локалізованих станів у забороненій зоні, які є фотоактивними та відповідають за появу довгохвильових особливостей при 1.7 та 1.32 еВ у спектрі фотопровідності. Фоточутливість нескомпенсованих кристалів зумовлена наявністю у них так званих *r*-центрів повільної рекомбінації. Глибина залягання *r*-центрів може бути визначена за межею спектра домішкового фотоефекту (рис. 5.2). З цими ж центрами зв'язана домішкова фотолюмінесценція з $h\nu_{max}$ = 1.1 еВ. Фотолюмінесценція виникає у тому випадку, якщо всі або частина нерівноважних дірок локалізується на *r*-центрах, де вони можуть випромінювально рекомбінувати з вільними електронами.

Стаціонарне значення фотопровідності для випадку мономолекулярної кінетики рекомбінації за умови, що все падаюче світло поглинається зразком, рівне:

$$\sigma_{\Phi}(T) = e\cdot\beta(T)N\cdot\tau(T)\cdot u(T), \qquad (5.5)$$

де $\beta$ – квантовий вихід, $N$ – кількість падаючих квантів, $u$ – рухливість, $\tau$ – час життя носіїв.

Температурний хід фотопровідності визначається температурними залежностями рухливості, квантового виходу та часу життя нерівноважних носіїв. При поглинанні світла у власній смузі залежність ФП від температури визначається переважно множником $u(T)\cdot\tau(T)$, оскільки квантовий вихід має бути близьким до одиниці і слабо залежати від температури.

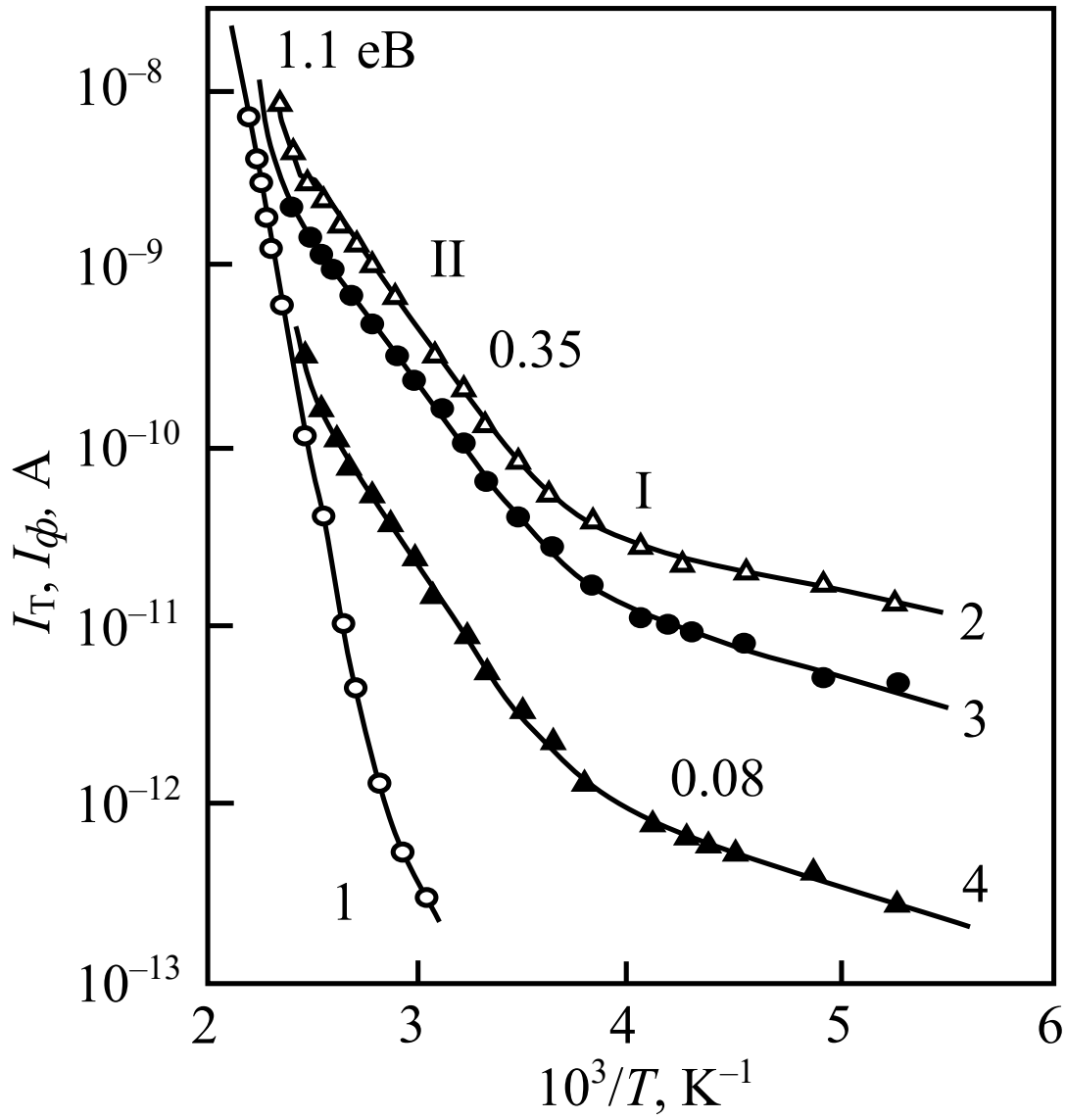

Рис. 5.4. Температурні залежності темнового струму (1) і фотоструму (2–4) кристала $Si_2Te_3$, виміряні для трьох рівнів збудження: 2 – $5.2 \cdot 10^{14}$; 3 – $7.3 \cdot 10^{13}$; 4 – $1.6 \cdot 10^{13}$ фотон/см$^2$·с [26].

Температурні залежності фотоструму кристала $Si_2Te_3$, виміряні при різних рівнях освітлення, наведено на рис. 5.4 [26, 172]. На кривих температурної залежності фотоструму спостерігаються дві області (I та II) температурної активації (ТА) фотоструму з енергіями процесу $E_I$ = 0.08 еВ і $E_{II}$ = 0.35 еВ. Температура, що відповідає межі області I, зі зростанням інтенсивності збудження *L* зміщується в область більш високих температур.

## 5.3 ТЕРМОСТИМУЛЬОВАНА ПРОВІДНІСТЬ КРИСТАЛІВ $Si_2Te_3$.

Аналіз температурної залежності фотоструму в ряді напівпровідників (CdS, CdSe, ZnS, GeS та ін.) показав [200], що в переважній більшості випадків активація фотоструму зумовлена центрами прилипання основних носіїв (*t*-центрів). Центри прилипання визначають заповнення центрів рекомбінації (через умову електричної нейтральності), що за умов несталого заповнення *r*-центрів приводить до температурної та концентраційної залежності часу життя основних носіїв [200].

У переважній більшості широкозонних фотопровідників активація фотоструму викликана термічною перезарядкою між центрами прилипання основних нерівноважних носіїв заряду ($t$) та центрами їхньої рекомбінації ($r$) [200]. Для виявлення центрів прилипання та визначення їх основних параметрів авторами [26] проведено дослідження термостимульованого струму (ТСС) у кристалах $Si_2Te_3$. На рис. 5.5 наведені криві ТСС, які відповідають трьом різним швидкостям нагрівання зразка. Як видно з цього рисунка, в інтервалі температур 210 ÷ 310 K спостерігається один максимум струму при $T = 280$ K, зумовлений наявністю одного рівня прилипання. Зі збільшенням швидкості нагрівання зразка максимум зміщується в область більших температур з одночасним зростанням інтенсивності піка.

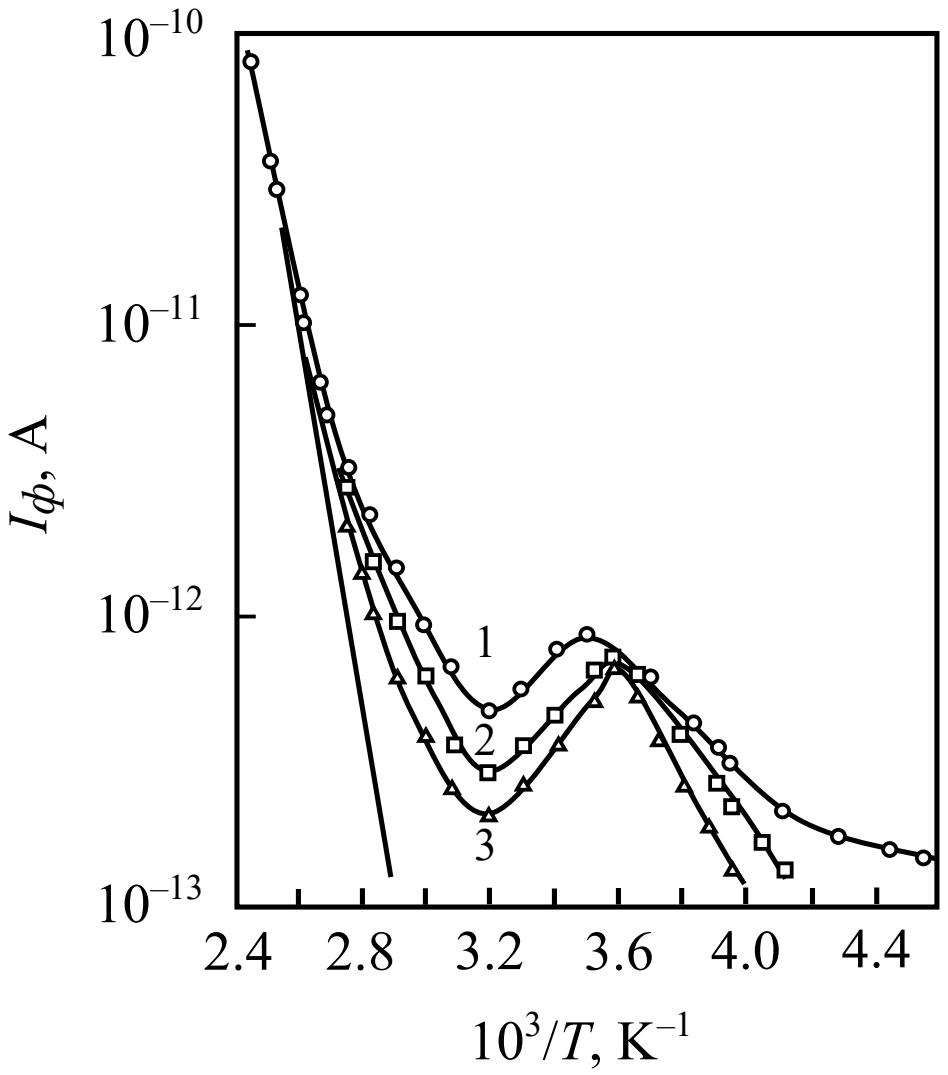

Рис. 5.5. Криві термостимульованого струму, виміряні при різних швидкостях нагріву зразка V, K/c: 1 – 0.1; 2 – 0.03; 3 – 0.01 [26].

Для визначення основного параметра рівнів прилипання – глибини залягання – за кривими ТСС, автори [26] застосували методи аналізу, незалежні від типу рекомбінації (мономолекулярна або бімолекулярна) та носіїв заряду в кристалах (метод початкового підйому [204] і Адамса – Хаєрінга [205]). Початкова ділянка (до максимуму) кривих ТСС незалежно від типу рівнів прилипання (швидкі або повільні) описується виразом $I_{TCC} = \text{const}\cdot\exp(-E_t/kT)$. З нахилу залежностей $lnI_{TCC} = f\left(T^{-1}\right)$ і залежностей $lnI_{TCC}^{макс} = f\left(T_m^{-1}\right)$, не пов'язаних із швидкістю нагрівання та частотним фактором, визначено енергетичну глибину залягання рівня прилипання: $E_t$ = 0.45 еВ та концентрацію пасток $10^{17}$ $см^{-3}$.

На рис 5.6 приведені залежності фотоструму $I_ф$ від швидкості генерації $G$ кристала $Si_2Te_3$ виміряні при двох різних температурах: 293 К (крива 1) і 373 К (крива 2). При кімнатній температурі (крива 1) на залежності $I_ф \sim G$ чітко видно наявність двох ділянок: при малих значеннях $G$ залежність $I_ф \sim G$ сублінійна, а зі збільшенням швидкості генерації спостерігається лінійна ділянка. До сублінійної залежності приводить заповнення центрів нерівноважними носіями, генерованими при освітленні зразка.

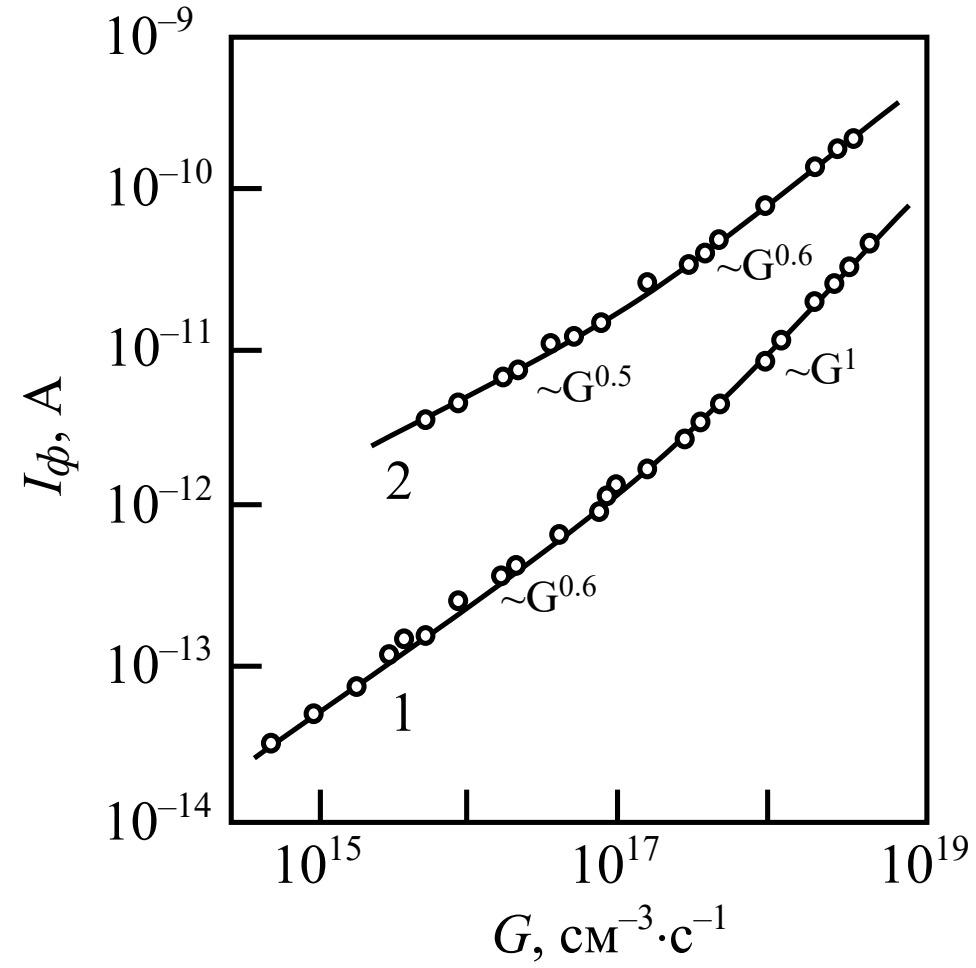

Рис. 5.6. Залежності фотоструму від швидкості генерації $G$ кристала $Si_2Te_3$, виміряні при різних $T$, К: 1 – 293, 2 – 373 [26].

Дослідження кінетики наростання та спаду фотоструму в кристалах $Si_2Te_3$ показало, що при освітленні прямокутними імпульсами світла з часом відсічення ~ $10^{-5}$ с, спостерігається короткохвильова складова фотоструму із $\tau \leq 10^{-5}$ с. На кінетичних кривих релаксації фотоструму при кімнатній температурі фіксується цілий набір ділянок із $\tau$ від $10^{-3}$ до кількох секунд, зумовлених прилипанням нерівноважних носіїв.

## 5.4. СПЕКТРИ ФОТОПРОВІДНОСТІ КРИСТАЛІВ $SiTe_2$

Шаруваті кристали $SiTe_2$, вирощені методами сублімації та ХТР, є високоомними ($\sigma_T = 10^{-9}$ Ом$^{-1}$ см$^{-1}$ при 300 К) та фоточутливими без додаткових спеціальних обробок [20].

Спектри фотопровідності свіжоприготовленого кристала $SiTe_2$ та кристала, який зберігався 6 місяців у вакуумному боксі (ексикаторі), виміряні при кімнатній температурі, наведено на рис. 5.7. Як видно із цього рисунка в спектрі свіжоприготованого кристала (крива 1) спостерігається чітко виражений максимум при 2.16 еВ. Енергетичне положення цього максимуму відповідає прямим оптичним переходам, що слідує з аналізу функціональної залежності краю фундаментального поглинання кристалів $SiTe_2$ [20].

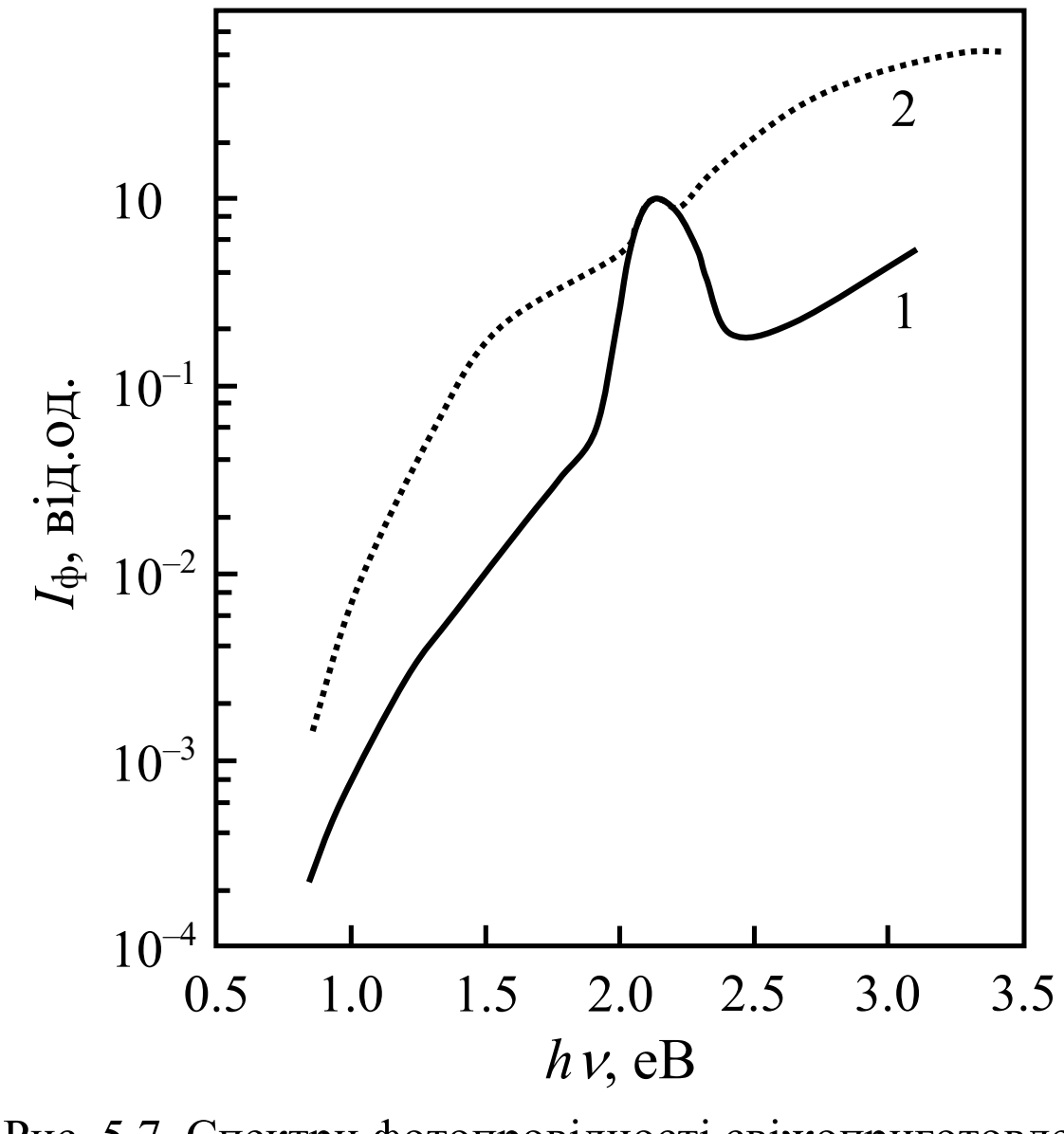

Рис. 5.7. Спектри фотопровідності свіжоприготовленого кристала (1) та кристала $SiTe_2$, який зберігався 6 місяців у вакуумному боксі (2). T = 300 K [20].

Довгохвильова особливість у вигляді напливу в інфрачервоній області та подальше збільшення відгуку для енергій фотонів, які перевищують 2.5 еВ, автори [20] зв'язують із впливом поверхневих ефектів на фотовідповідь. Фотовідповідь, що спостерігається поза областю прямих переходів, пов'язана з явищем швидкого хімічного розкладання, яке, як відомо, відбувається на поверхні кристалів $SiTe_2$. Підтвердженням цього висновку є збільшення цих двох областей в спектрі ФП, оскільки відомо, що невелике поверхневе розкладання відбулося в кристалах, які зберігалися протягом тривалих періодів часу, навіть у вакуумних ексікаторах.

## 5.5. ФОТОЕЛЕКТРИЧНІ ВЛАСТИВОСТІ СТЕКОЛ $Si_{15}Te_{85}$

Стекла $Si_{15}Te_{85}$ є фоточутливими без спеціальних додаткових обробок. При 293 К й освітленості $L = 10^4$ лк білим світлом кратність відношення фотопровідності ($\sigma_ф$) до темнової провідності ($\sigma_т$) $к = \sigma_ф/\sigma_т$ складає $10^2$–$10^3$. Типові спектри фотопровідності склоподібного $Si_{15}Te_{85}$, виміряні при різних температурах, наведені на рис. 5.8. При кімнатній температурі у спектрі ФП спостерігаються два максимуми $h\nu_{max1} = 0.88$ еВ і $h\nu_{max2} = 1.57$ еВ. Для ідентифікації природи максимумів, наявних у спектрах фотопровідності, авторами [198] було проведено вимірювання крайового поглинання на тих самих зразках, на яких вимірювалась фотопровідність. Із співставлення кривих 2 і 5, рис. 5.8 видно, що низькоенергетичний максимум у спектрах ФП скла $Si_{15}Te_{85}$ знаходиться в області фундаментального поглинання, тобто він є власним.

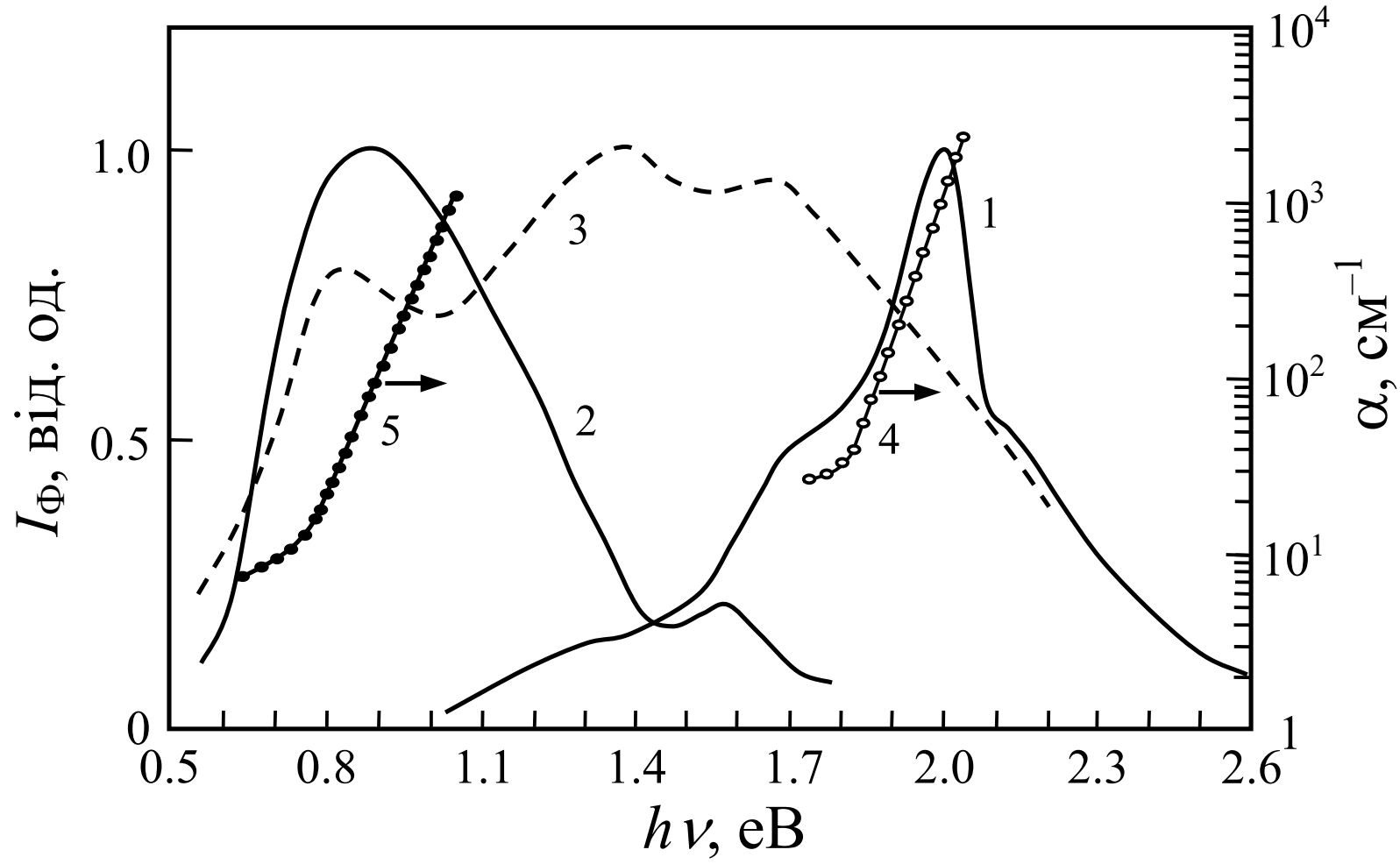

Рис. 5.8. Спектри фотопровідності (1–3) та крайового поглинання (4, 5) кристала $Si_2Te_3$ (1, 4) і скла $Si_{15}Te_{85}$ (2, 3, 5), виміряні при різних температурах *T*, К: 1, 2, 4, 5 – 293; 3 – 100 [198].

У випадку стехіометричних халькогенідних стекол $GeSe_2$ на кривих спектрального розподілу фоточутливості спостерігається тільки один максимум поблизу краю власного поглинання [206]. Для цих стекол характерним є те, що при переході від низькоенергетичної (довгохвильової) до високоенергетичної (короткохвильової) області краю поглинання фотовідгук змінюється майже від нуля, потім швидко зростає при досягненні краю поглинання і знову спадає, тоді як коефіцієнт поглинання продовжує зростати, і за звичай, наближається до деякого нульового асимптотичного значення. У цій останній спектральній області генерація носіїв обмежується вузьким приповерхневим шаром зразка і струм визначається поверхневою рекомбінацією.

На відміну від стехіометричних стекол $GeSe_2$, загальний вигляд спектрів ФП нестехіометричних стекол $Si_{15}Te_{85}$ є більш складним і містить два максимуми ($h\nu_{max1}$ = 0.88 еВ і $h\nu_{max2}$ = 1.57 еВ) при кімнатній температурі і три максимуми $h\nu_{max1}$ = 0.84 еВ, $h\nu_{max2}$ = 1.38 еВ і $h\nu_{max3}$ = 1.67 еВ при $T$ = 100 К. Зіставляючи криву 2 спектра ФП з кривою 5 (рис. 5.8) спектра фундаментального поглинання, бачимо що низькоенергетичний максимум $h\nu_{max1}$ = 0.88 еВ у спектрі ФП знаходиться в області власного поглинання ($\alpha = 10^2$ см$^{-1}$) і обумовлений генерацією електронно-діркових пар. Враховуючи той факт, що досліджувані зразки готувались шліфуванням і поліруванням, кількість поверхневих рівнів при цьому в них значна і тому в високоенергетичній області повинен мати різкий спад ФП, що і спостерігається експериментально (крива 2, рис. 5.8). Проте цей спад ФП не відбувається до нуля, а навпаки у високоенергетичній області спостерігається ще один максимум $h\nu_{max2}$ = 1.57 еВ. Цікавою є перебудова спектра ФП склоподібного $Si_{15}Te_{85}$ з пониженням температури досліджуваного зразка. Має місце як перерозподіл інтенсивностей в максимумах ФП, так і їх енергетичне зміщення. Інтенсивність власного зона-зонного максимуму дещо зменшується, а його енергетичне положення зміщується у довгохвильову область ($h\nu_{max1}$ = 0.82 еВ при $T$ = 100 К). У високоенергетичній області при $T$ = 100 К чітко проявляються два максимуми $h\nu_{max2}$ = 1.38 еВ і $h\nu_{max3}$ = 1.67 еВ, інтенсивність яких перевищує інтенсивність власного максимуму. Для порівняння на рис. 5.8, крива 1 приведено неполяризований спектр ФП кристала $Si_2Te_3$, вирощеного з газової фази методом сублімації. Цей спектр є досить складним, в якому яскраво виражений інтенсивний пік при $h\nu_{max}$ = 2.0 еВ та особливість у вигляді плеча при $h\nu = 2.12$ еВ на короткохвильовому спаді основного максимуму знаходяться в області фундаментального поглинання $Si_2Te_3$. Ще дві особливості, які спостерігаються на довгохвильовому спаді основного максимуму в спектрі ФП кристала $Si_2Te_3$, проявляються також у високоенергетичній області спектра ФП склоподібного $Si_{15}Te_{85}$.

При низьких температурах має місце перерозподіл інтенсивностей у максимумах ФП і поява третього – більш високоенергетичного максимуму $h\nu_{max3} = 1.67$ еВ (крива 3, рис. 5.8). Для інтерпретації складної будови спектрів ФП стекол $Si_{15}Te_{85}$ необхідно залучити дані з вивчення їх структури. Як показали прямі дифракційні методи дослідження структури, стекла $Si_xTe_{1-x}$ характеризуються мікронеоднорідною будовою, тобто у них одночасно наявні дві групи структурних одиниць: тетраедри [$SiTe_4$] і ланцюжки Te–Te. Евтектичні стекла $Si_{17}Te_{83}$, в залежності від умов склування, містять у собі нанокристаліти Te, довжиною до 100 Å.

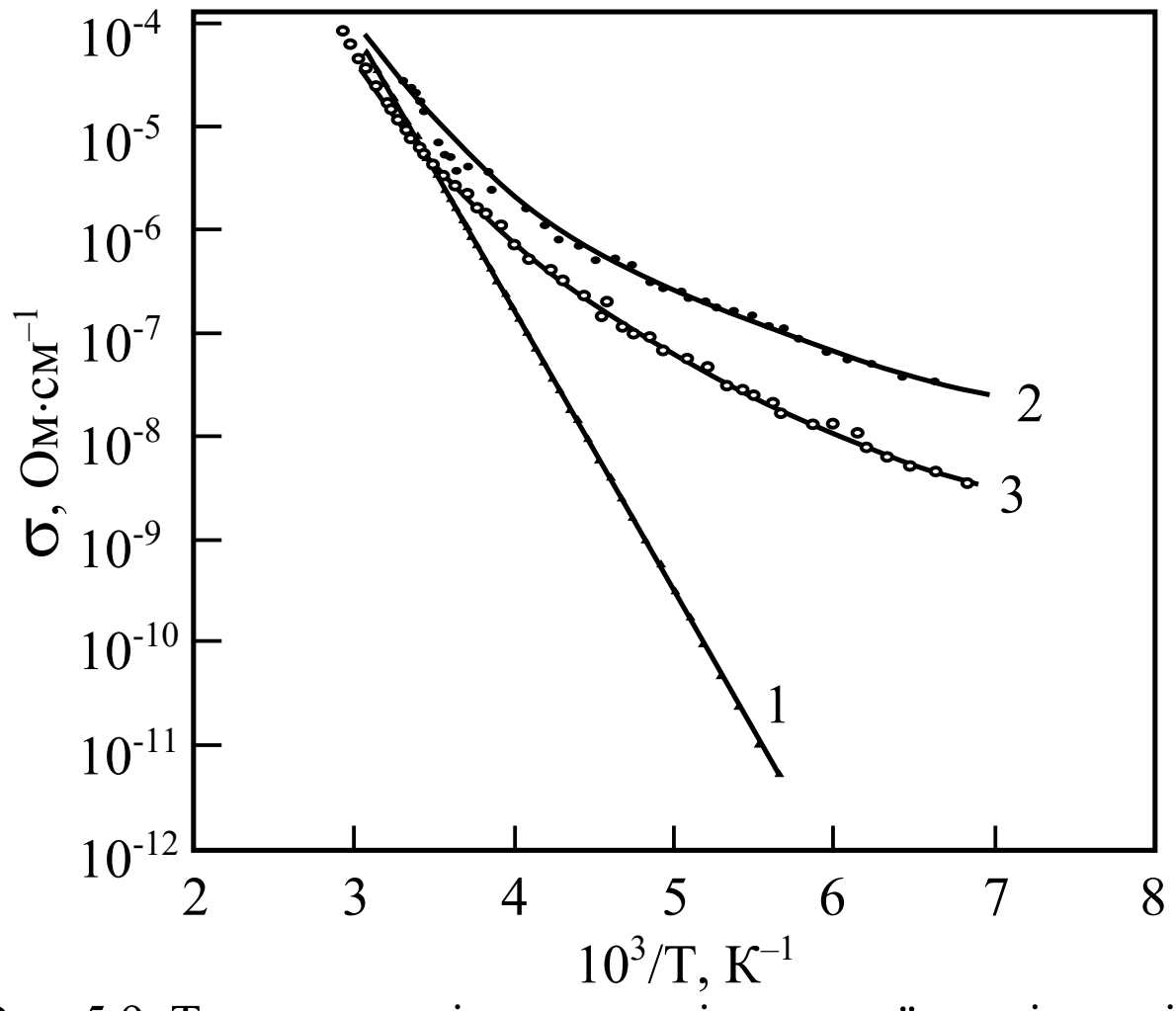

Рис. 5.9. Температурні залежності темнової провідності (1) і фотопровідності (2, 3) склоподібного $Si_{15}Te_{85}$, виміряні при різних освітленостях L, лк: 2 – $10^4$; 3 – $10^2$ [198].

Температурні залежності фотопровідності склоподібного $Si_{15}Te_{85}$, виміряні при двох різних інтенсивностях сталого освітлення $L$ зразка, наведені на рис. 5.9, криві 2, 3. Зміна інтенсивності падаючого на зразок випромінювання не приводить до істотних змін характеру температурної залежності фотопровідності. Із рис. 5.9 видно, що у широкій області температур має місце термічна активація фотопро-відності (ТАФ). Зауважимо, що на відміну від стехіометричних халькогенідних стекол $GeSe_2$, у досліджуваних стеклах $Si_{15}Te_{85}$ процес ТАФ не може бути описаний експоненціальним законом зі сталою енергією активації. Існує велике число моделей, які пояснюють процес ТАФ у широкозонних фотопровідниках. Проте жодна із відомих моделей не може бути прямо використана для випадку стекол $Si_{15}Te_{85}$, оскільки останні є мікронеоднорідними.

В інтервалі температур, де має місце температурна активація фотопровідності, люкс-амперні характеристики стекол $Si_{15}Te_{85}$ описуються залежністю $I_ф \sim L^n$, де $0.5 < n < 1$, тобто є сублінійними. Сублінійне зростання фотопровідності, а також термоактивація ФП зі змінним нахилом залежності $\sigma_ф = f(1/T)$, характерні для широкозонних напівпровідників з наявністю в них великої концентрації розподілених за енергією локальних станів, які виконують роль центрів прилипання. Не виключена можливість внеску в термоактивацію ФП й активації температурної залежності рухливості носіїв. Однозначна відповідь на це може бути дана тільки після вивчення температурної залежності дрейфової рухливості у даних стеклах.

## 5.6 ФОТОДЕТЕКТОРИ НА ОСНОВІ ШАРУВАТИХ КРИСТАЛІВ $Si_2Te_3$

Враховуючи що шаруваті кристали $Si_2Te_3$, незалежно від методу і умов їх одержання, є фоточутливими в широкій спектральній області (рис. 5.1–5.3), авторами [207] були виготовлені фотодетектори на основі об’ємного сесквітелуриду кремнію на підкладці пластини $SiO_2/Si$ 300 нм (рис. 5.10, *а*). Замість стандартної електронно-променевої літографії була застосована технологія переносу електродів, що дало можливість запобігти пошкодженню зразків $Si_2Te_3$ електронним променем. Для уникнення хімічного розкладу та виникнення станів розриву, спричинених дефектами, золоті (Au) електроди були нанесені безпосередньо на зразки $Si_2Te_3$.

З типових напівлогарифмічних графіків струм-напруга ($I_{ds}$–$V_{ds}$) пристрою $Si_2Te_3$ при різних температурах (рис. 5.10, *б*) можна побачити, що струм зменшується на два порядки при зниженні температури від 300 до 80 К. Хороша симетрія кривих $I_{ds}$–$V_{ds}$ вказує на якісні контакти електрод/канал. Температурна залежність опору пристрою показує, що зі зниженням температури опір зростає, виявляючи типову напівпровідникову поведінку (рис. 5.10, *в*).

Залежності фотоструму $I_ф$ від густини потужності $P$ падаючого

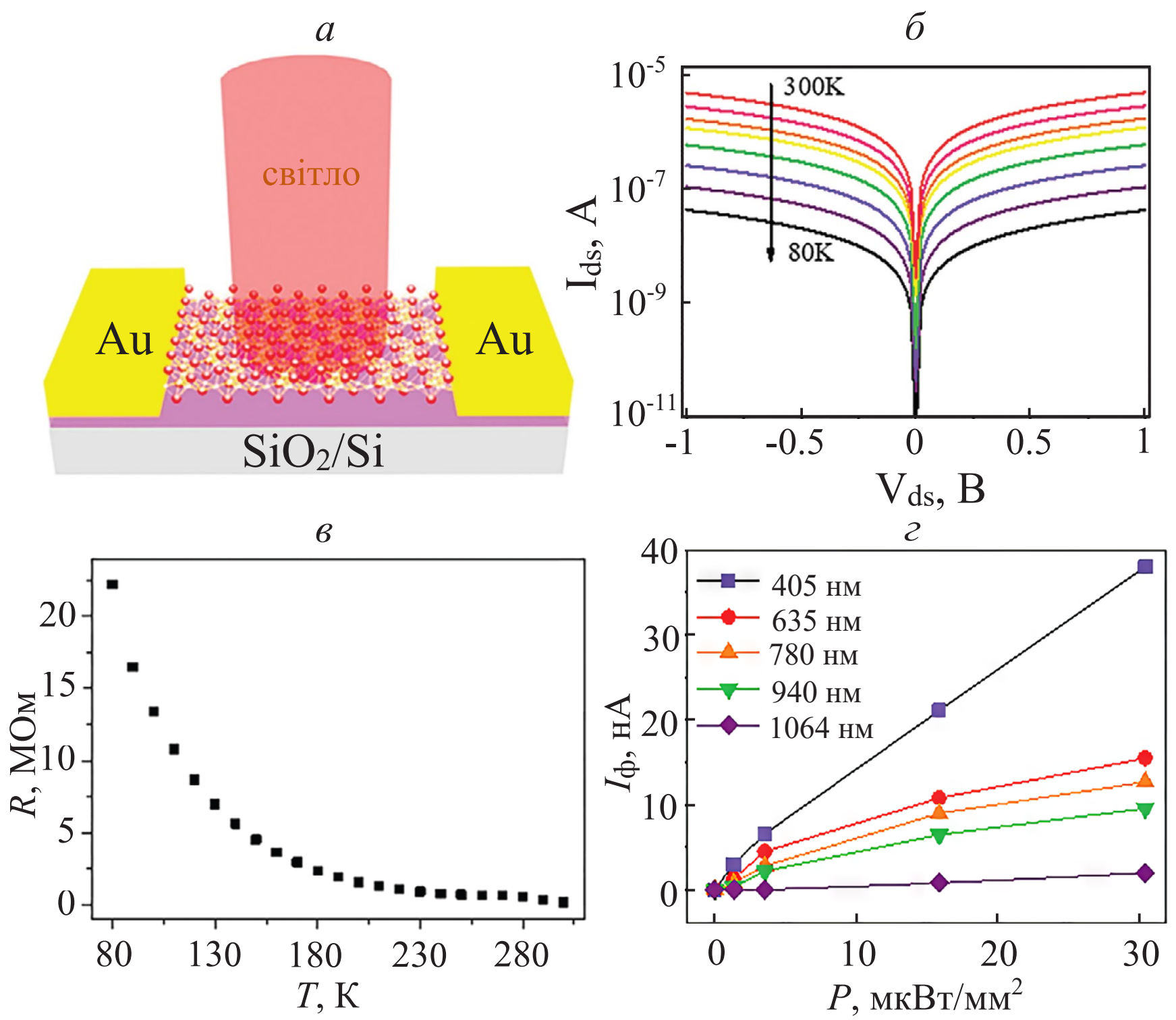

Рис. 5.10. *а*) Схематичне зображення фотоприймача $Si_2Te_3$.
*б*) Типові напівлогарифмічні ВАХ при різних температурах.
*в*) Температурна залежність опору об'ємного $Si_2Te_3$.
*г*) Фотострум як функція падаючої густини потужності [207].

світла, наведені на рис. 5.10, *г*, чітко вказують на збільшення $I_ф$ зі збільшенням $P$ на всіх довжинах хвиль 405 – 1064 нм. На відміну від лінійної світлової ВАХ, $I_ф$ демонструє сублінійну залежність від $P$.

Світлові вольт-амперні характеристики ($I_ф$–$V_{ds}$) фотодетектора $Si_2Te_3$, освітленого різними довжинами хвиль (405 – 1064 нм) при густині потужності 15,93 мкВт/мм$^{-2}$, демонструють що спектральний відгук знаходиться в межах від видимої до ІЧ області (рис. 5.11, *а*).

Чутливість фотодетектора $R_\lambda = I_ф/P{\cdot}S$, де $I_ф$ – фотострум, $P$ - густина потужності падаючого світла, $S$ – ефективна площа пристрою, в залежності від довжини хвилі λ (рис. 5.11, *б*) вказує на те, що максимальне значення $R_\lambda$ досягає при довжині хвилі падаючого світла 405 нм. Зі збільшенням довжини падаючої хвилі $R_\lambda$ зменшується і

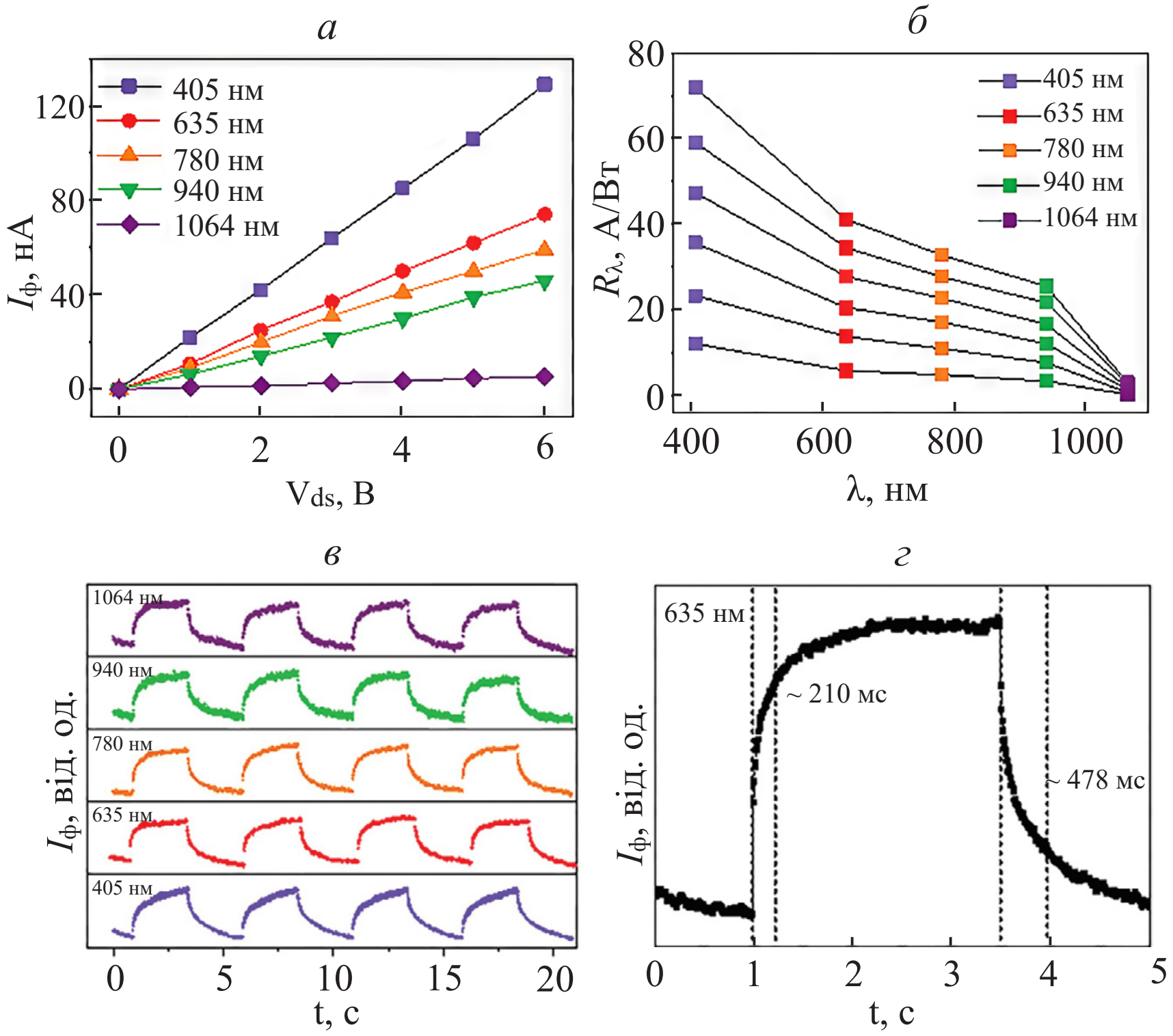

Рис. 5.11. *а*) Світлові ВАХ;
*б*) залежність чутливості $R_\lambda$ фотоприймача $Si_2Te_3$ від довжини хвилі при освітленні різними довжинами хвиль;
*в*) кінетика наростання та спаду фотоструму при вмиканні та вимиканні падаючого світла;
*г*) час наростання і спаду фотоструму при освітленні 635 нм [207].

при 1064 нм це зменшення досягає 28 разів у порівнянні $R_\lambda$ при 405 нм. Це вказує на те, що фотодетектори на основі шаруватих кристалів $Si_2Te_3$ є високоспектрально селективними та сильно залежать від довжини хвилі падаючого світла.

Фотовідповідь, як функція часу під впливом модульованого світла, чітко демонструє хорошу світлочутливість на різних довжинах хвиль (рис. 5.11, *в*), і, навіть після 250 циклів $I_ф$ (>26 хв), фотодетектор все ще показує стабільні та повторювані характеристики струму увімкнення/вимкнення. Розрахований час наростання та час спаду при освітленні 635 нм становить приблизно 210 і 478 мс (рис. 5.11, *г*), відповідно, що краще, ніж у $MoS_2$, InSe та багатьох інших традиційних фотодетекторів.

Завдяки ефекту електростатичного екранування здатність модуляції нижнього затвора зменшується зі збільшенням товщини зразка. Для подальшого дослідження двовимірної природи та кореляційних властивостей заднього затвора пластівців $Si_2Te_3$ авторами [207] був додатково виготовлений польовий транзистор (FET) із товщиною 13.5 нм, довжиною каналу 3 мкм і шириною 8.5 мкм шляхом електродного перенесення. Зображення пристрою на атомно-силовій мікрофотографії (АСМ) показано на вставці на рис. 5.12, *а*. Вихідні криві $I$–$V_{ds}$ при різних напругах затвора ($V_g$) в діапазоні від -30 до 30 В, що свідчить про якісні контакти, а також сильний регулюючий вплив $V_g$ на $I_{ds}$ (рис. 5.12, *а*).

З передавальних кривих при різних значеннях $V_{ds}$ в інтервалі від 0.1 до 5 В (рис. 5.12, *б*) можна побачити, що $I_{ds}$ зменшуються зі збільшенням $V_g$, що вказує на поведінку провідності *p*-типу $Si_2Te_3$.

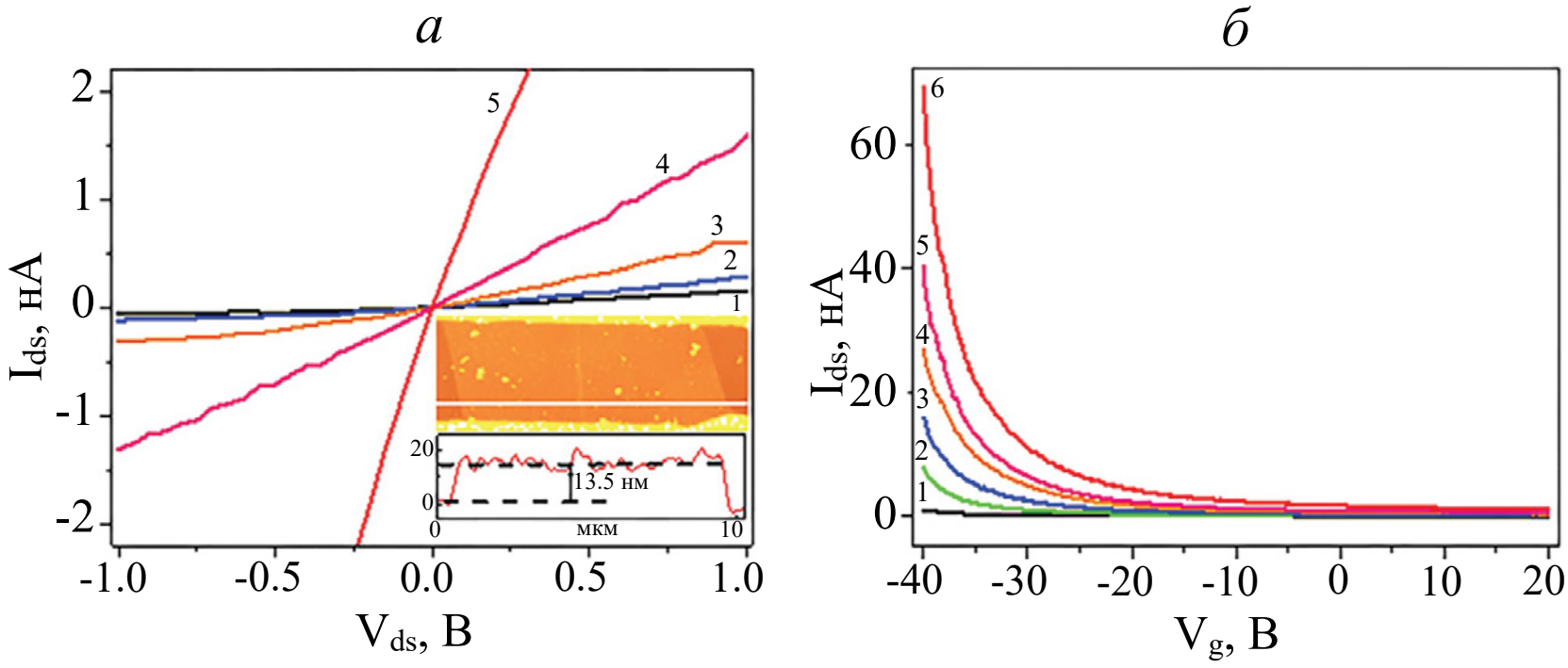

Рис. 5.12 *а*) Вихідні криві з різними значеннями $V_g$:
1 – 30В, 2 – 20В, 3 – 10В, 4 – 0В, 5 – -30В,.
Вставка: АСМ зображення типового пристрою, його товщина 13.5 нм. *б*) Криві передачі при різних значеннях $V_{ds}$:
1 – 0.1В, 2 – 0.5В, 3 – 1В, 4 – 2В, 5 – 3В, 6 – 5В [207].

Розрахована рухливість носіїв заряду польового транзистора становить 1.35 $см^2В^{-1}с^{-1}$, що є відносно загальним у порівнянні з багатьма зареєстрованими 2D FET. Це пов'язано з впливом внутрішніх дефектів вакансій у кристалічній структурі та адсорбентів, які призводять до розсіювання носіїв і приводять до низької рухливості. Більшої рухливість можна очікувати за рахунок покращення контактів або при заміні ізоляційного шару на $HfO_2$ з високим показником *k*, в якості підзатворного шару.

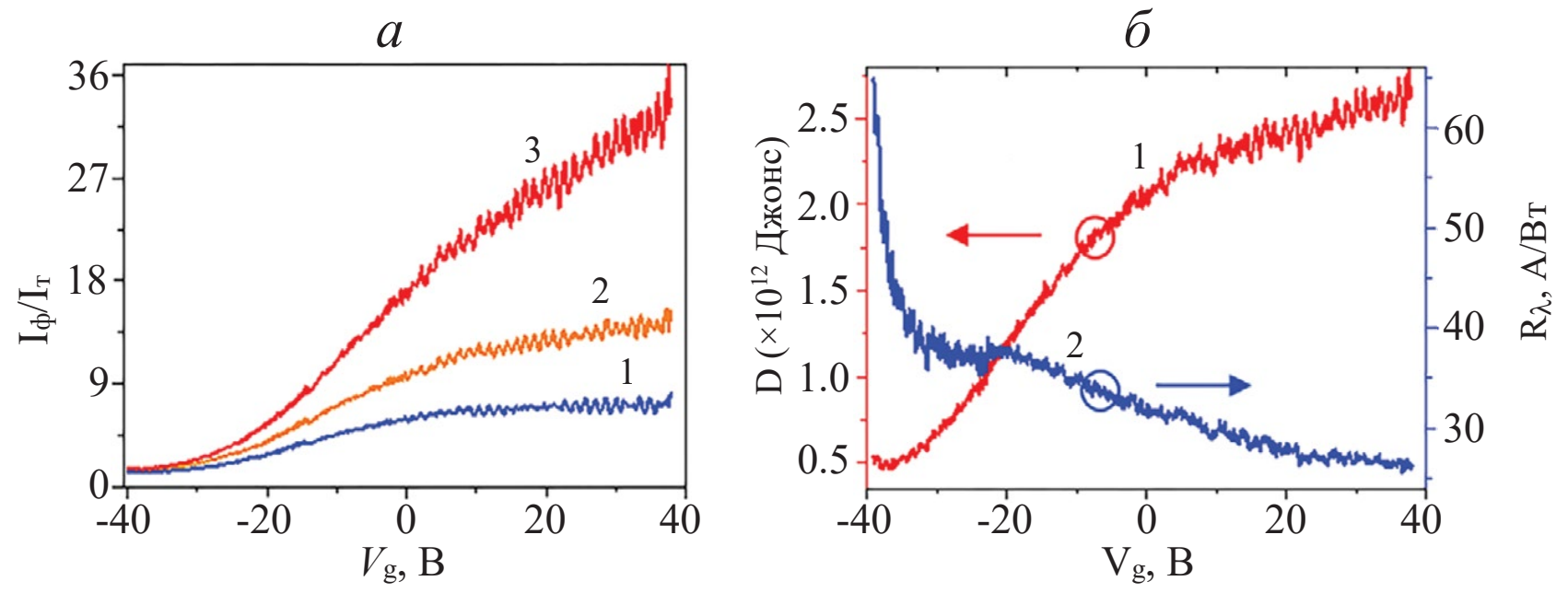

Рис. 5.13 *а*) Криві $I_ф/I_т$–$V_g$ фотодетектора $Si_2Te_3$ при різних рівнях густини падаючої потужності: 1 – 1.34 мкВт/мм$^2$, 2 – 1.87 мкВт/мм$^2$, 3 – 3.55 мкВт/мм$^2$, при $V_g$ від -40 до 40 В та $V_{ds}$ = 1 В.
*б*) *D* (крива 1) і $R_\lambda$ (крива 2) як функція $V_g$ [207].

Коефіцієнт перемикання може сягати приблизно 35, коли $V_g$ становить 40 В при $P$ = 3.55 мкВт мм$^{-2}$ (рис. 5.13, *а*) і може надалі збільшуватися до понад сто разів шляхом подальшого збільшення інтенсивності освітлення. Цей результат [207] кращий, ніж деякі з зареєстрованих двовимірних матеріалів для фотодетекторів.

Чутливість (визначається як $D^* = S^{1/2}R_\lambda/(2eI_т)^{1/2}$), один із дуже важливих параметрів, яку можна використовувати для порівняння ефективної чутливості різних фотодетекторів. Із залежностей розрахованих $R_\lambda$ і $D^*$ від $V_g$ (рис. 5.13, *б*) можна побачити, що $R_\lambda$ зменшується, тоді як $D^*$ збільшується зі збільшенням $V_g$. Максимальний $D^*$ (рис. 5.13, *б,* крива 1) може досягати 2.81×10$^{12}$ Джонса при зміні $V_g$ до 40 В. Тим часом $R_\lambda$ досягає 65 AW$^{-1}$ при зміні $V_g$ до -40 В. У роботі [207] продемонстровано значне покращення фотоелектричних характеристик двовимірного фотодетектора $Si_2Te_3$ порівняно зі звичайними фотодетекторами $MoS_2$, $WS_2$, $SnS_2$ та InSe.

Таким чином, представлені в роботі [207] результати дослідження польового (FET) транзистора на основі $Si_2Te_3$ і фотоприймача на його основі, демонструють чудову широкосмугову спектральну характеристику в діапазоні 405–1064 нм. Фоточутливість даного фотоприймача з товщиною 13.5 нм при освітленні 405 нм може досягати 65 А Вт$^{-1}$ і 2.81×10$^{12}$ Джонса, відповідно, що перевищує багато традиційних широкосмугових фотоприймачів.

# РОЗДІЛ 6

# ФОТОЛЮМІНЕСЦЕНЦІЯ ОБ'ЄМНОГО ТА НАНОСТРУКТУРОВАНОГО $Si_2Te_3$

У загальному комплексі досліджень електронних властивостей шаруватих кристалів $Si_2Te_3$ важливе місце займає вивчення фотолюмінесценції (ФЛ). Люмінесценція, або випромінювальна рекомбінація, у напівпровідниках може спостерігатись або не спостерігатись, що пов'язано з переважанням одного з двох конкуруючих процесів рекомбінації: випромінювальної і безвипромінювальної. Випромінювальна рекомбінація в широкозонних кристалічних напівпровідниках здебільшого пов'язана з наявністю локалізованих станів у забороненій зоні. Тому дослідження стаціонарних і кінетичних характеристик фотолюмінесценції, які дають багато інформації про нерівноважні процеси в кристалах, набувають винятково важливого значення при вивченні властивостей об’ємного та наноструктурованого сесквітелуриду кремнію. Дослідження характеристик фотолюмінесценції дозволяє отримати інформацію як про локалізовані електронні стани в забороненій зоні $Si_2Te_3$, так і про механізми випромінювальних та безвипромінювальних процесів у цьому матеріалі.

Однією з проблем при дослідженні фізичних властивостей наноструктурованого $Si_2Te_3$ є стабільність матеріалу в умовах навколишнього середовища. Це пов'язано з тим, що велике відношення площі поверхні до об'єму на нанорівні приводить до поверхневої реакції з водяною парою в атмосфері, внаслідок чого утворюється тонкий шар Te. Крім того, ускладнення структурних характеристик $Si_2Te_3$ при низьких розмірах через орієнтацію димерів кремнію при різних температурах і деформаціях також може привести до разюче різних оптичних або електронних властивостей.

Дослідження випромінювальної рекомбінації нерівноважних носіїв заряду в тривимірних (об’ємних), двовимірних (нанопластинах) і одновимірних (нанодротинах) $Si_2Te_3$ мотивоване тим, що люмінесценція є одним із найбільш чутливих методів визначення енергетичних характеристик носіїв заряду, екситонних станів, домішкових центрів, власних дефектів, електрон-фононної взаємодії тощо. Відомо, що той чи інший механізм випромінювальної рекомбінації визначається зонною структурою напівпровідника, концентрацією та природою центрів рекомбінації, температурою кристала та рівнем збудження.

## 6.1. СПЕКТРИ ВИПРОМІНЮВАЛЬНОЇ РЕКОМБІНАЦІЇ ОБ'ЄМНИХ КРИСТАЛІВ $Si_2Te_3$.

Вперше дослідження фотолюмінесценції монокристалів $Si_2Te_3$, вирощених методом ХТР, були виконані авторами [46]. При 77 К в спектрі фотолюмінесценції наявна широка смуга з максимумом при енергії 1.3 еВ (рис. 6.1, крива 1), природа якої пов'язується з існуванням власних дефектів. Широку смугу фотолюмінесценції з максимумом при 1.1 еВ (рис. 6.1, крива 3) спостерігали також автори [26] при лазерному збудженні (514 нм, 0.1 Вт) і температурі 93 К кристала $Si_2Te_3$, вирощеного методом Бріджмена. На рис. 6.1 (крива 3) приведено також спектр фотолюмінесценції при лазерному збудженні 488 нм нанопластини $Si_2Te_3$, вирощеної методом CVD [208]. Оскільки, незалежно від методу вирощування і габітусу, кристали $Si_2Te_3$ мають велику концентрацію власних точкових дефектів до $10^{17}$ см$^{-3}$, це сприяє наявності широкого спектра випромінювальної рекомбінації від видимого до ближнього ІЧ діапазону (рис. 6.1).

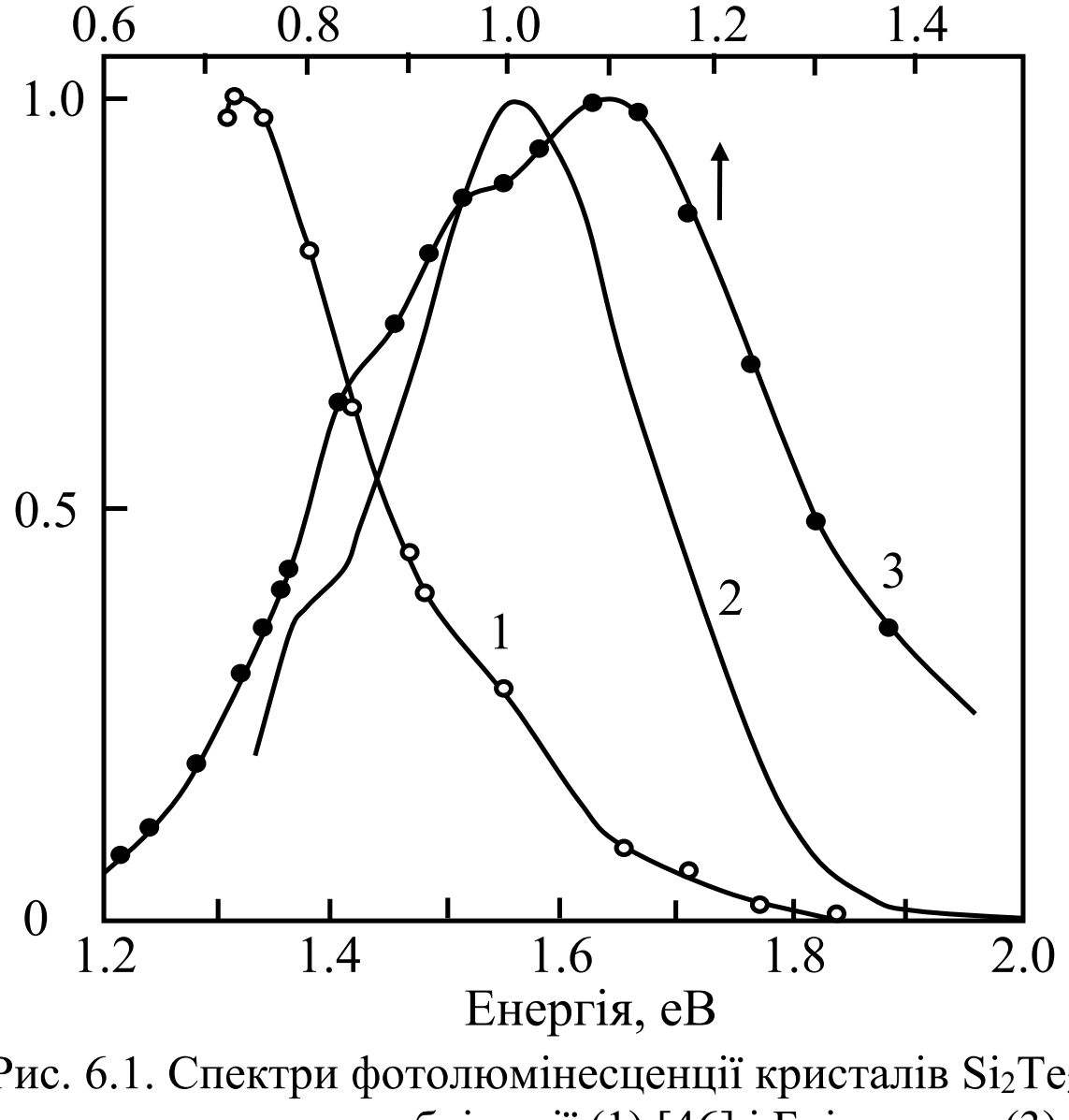

Рис. 6.1. Спектри фотолюмінесценції кристалів $Si_2Te_3$, вирощених методами сублімації (1) [46] і Бріджмена (3) [26], та нанопластини (2) [208]. *T*, K: 1 – 77; 3 – 93, 2 – 293.

Докладніше фотолюмінесценцію монокристалів $Si_2Te_3$, вирощених методом хімічного осадження із парової фази, в широкому інте-рвалі температур (80–290 K) дослідили автори [207]. У спектрі ФЛ фіксуються дві широкі домішкові смуги випромінювальної рекомбінації з максимумами при 1.59 еВ (780 нм) і 1.31 еВ (947 нм), позначених А і В відповідно (рис. 6.2). Внаслідок сильної електрон-фононної взаємодії «домішково-дефектні» смуги ФЛ в кристалах $Si_2Te_3$ при 80 К є досить широкі і частково перекриваються. Із підвищенням температури досліджуваного зразка енергетичне положення піка В у спектрі ФЛ практично не змінюється, тоді як для піка А чітко спостерігається «червоне» зміщення в інтервалі температур 80 – 140 K, і «синє» зміщення зі збільшенням температури від 140 до 290 K (рис. 6.2).

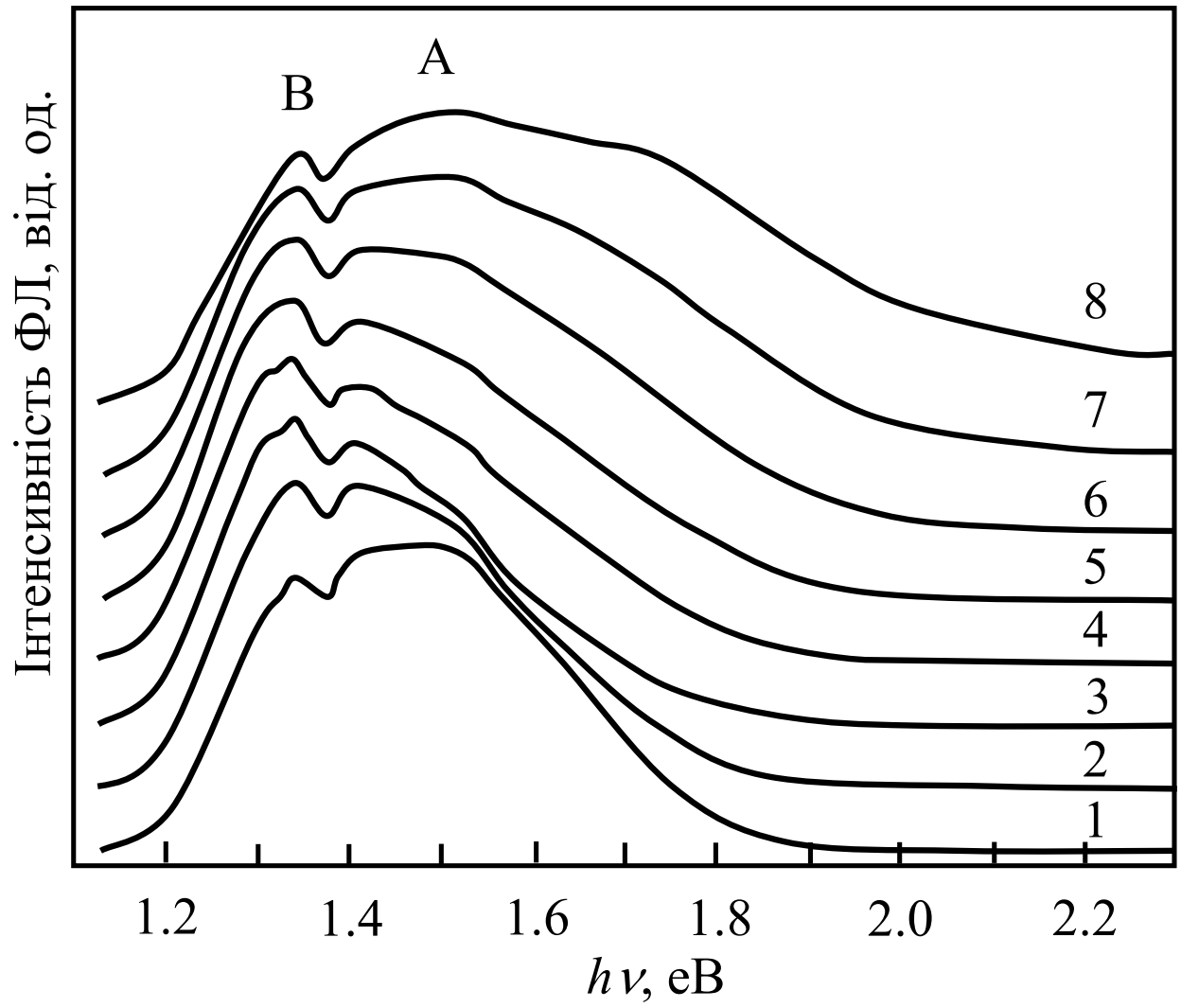

Рис. 6.2. Нормовані спектри ФЛ кристала $Si_2Te_3$, вирощеного методом хімічного осадження із парової фази (CVD метод), виміряні при різних температурах. *T*, K: 1 – 80; 2 – 110; 3 – 140; 4 – 170; 5 – 200; 6 – 230; 7 – 260; 8 – 290 [207].

Крім двох домішкових смуг у спектрі ФЛ об'ємного кристала $Si_2Te_3$ при низьких температурах (80 К) автори [207] виявили прикрайову смугу з максимумом при 594.3 нм (2.09 еВ). Із підвищенням температури інтенсивність домішкової і прикрайової фотолюмінесценції кристалів $Si_2Te_3$ різко зменшується. Термічне гасіння люмінесценції спостерігається у тому випадку, коли частина енергії збудження розсіюється при безвипромінювальних переходах. У реальних кристалах $Si_2Te_3$ поряд із випромінювальними центрами існу-ють центри безвипромінювальних переходів, які створюють у забороненій зоні енергетичні рівні, через які з великою імовірністю може відбуватися рекомбінація вільних електронів і дірок. Якщо в кристалах наявні центри люмінесценції, то буде відбуватися конкуренція між випромінювальними і безвипромінювальними переходами.

Отже, спектральний склад домішкової ФЛ об'ємного кристалічного сесквітелуриду кремнію значною мірою визначається умовами та методом росту, ступенем відхилення від стехіометричного складу та чистотою вихідних компонентів, які використовуються для синтезу бінарної речовини.

Необхідно відзначити, що фотолюмінесцентні властивості шаруватого 2D-напівпровідника $Si_2Te_3$ істотно залежать від товщини досліджуваного зразка. Дослідження ФЛ відлущених нанопластин та об'ємного кристала $Si_2Te_3$ дозволили авторам [207] виявити зміни в спектрах власного та домішкового випромінювання. Спектри ФЛ об'ємного кристала і відлущених нанопластин різної товщини наведені на рис. 6.3. Як видно з цього рисунка, широкосмугова фотолюмінесценція $Si_2Te_3$ від 550 до 1050 нм різко зменшується для пластівців товщиною 20 нм і повністю зникає при товщині 9 нм. Поряд зі зменшенням інтенсивності фотолюмінесценції має місце зсув смуги A в короткохвильову область по мірі зменшення товщини відлущеного шару.

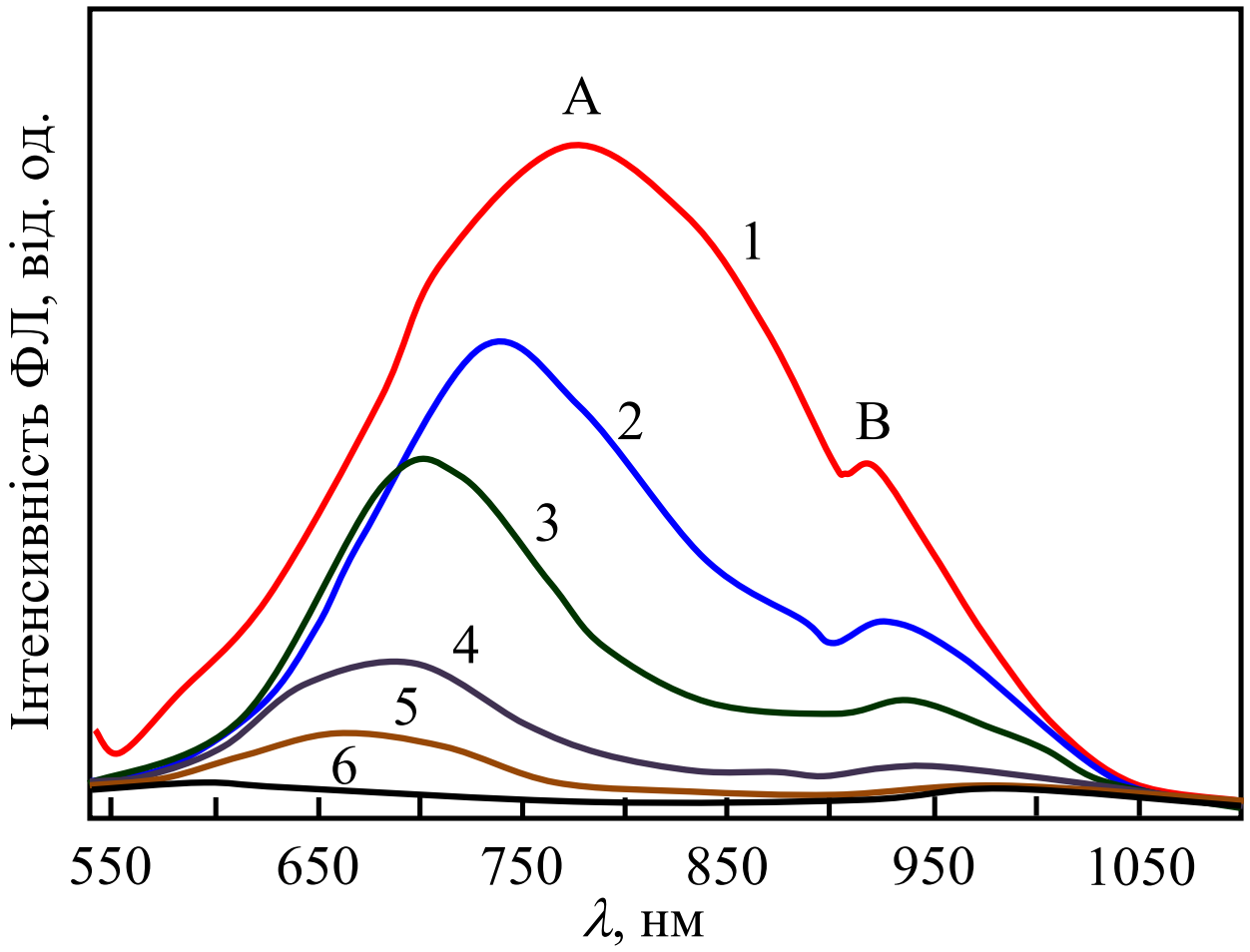

Рис. 6.3. Спектри ФЛ об'ємного (1) і відшарованих пластівців $Si_2Te_3$ різної товщини $d$, нм: 2 – 78.4; 3 – 36.8; 4 – 24.4; 5 – 20.2; 6 – 9.3 [207].

Зменшення квантового виходу при зменшенні товщини відлущеної пластини викликане підсиленням безвипромінювальної рекомбінації із-за виходу носіїв заряду через поверхневі стани і/або впровадження міжшарових дефектів різного типу. Необхідно також відмітити, що механічне відлущення супроводжується розривом ван-дерваальсових зв'язків і помітними напругами ґратки, які можуть створювати сприятливі умови для формування протяжних дефектів. Безвипромінювальний центр походить із поверхневих станів $Si_2Te_3$, у той час як випромінювальна ФЛ виникає внаслідок рекомбінації електронів і дірок в об'ємних шарах. Зменшення інтенсивності ФЛ більш тонкого зразка може бути зв'язано з ростом безвипромінювального внеску у поверхневі стани за рахунок збільшення поверхні по відношенню до об'єму при утоншенні шаруватого $Si_2Te_3$.

## 6.2. ПРИКРАЙОВА ТА ДОМІШКОВА ФОТОЛЮМІНЕСЦЕНЦІЯ НАНОПЛАСТИН $Si_2Te_3$

Результати досліджень фотолюмінесценції 2D-нанопластин $Si_2Te_3$, вирощених методом хімічного осадження з парової фази, у широкому інтервалі температур 10–300 K приведені у роботі [50]. Типові спектри ФЛ нанопластин $Si_2Te_3$, при збудженні світлом He-Cd лазера (441 нм), виміряні при різних температурах, наведені на рис. 6.4. При низьких температурах (10 – 60 K), як і у випадку об'ємних кристалів $Si_2Te_3$, у спектрах ФЛ нанопластин також спостерігається інтенсивна широка домішкова смуга A з максимумом при 1.606 еВ, на довгохвильовому спаді якої проявляється особливість у вигляді «плеча» (B) при 1.441 еВ (рис. 6.4, крива 1).

Енергетичні положення максимумів домішкових смуг A і B у спектрі ФЛ нанопластин $Si_2Te_3$ залежать від температури. Так, зі збільшенням температури від 10 до 180 K спостерігається зміщення енергетичного положення максимумів домішкових смуг ФЛ в область менших енергій внаслідок зменшення ширини забороненої зони. Подальше підвищення температури вище 180 K супроводжується інверсією характеру енергетичного положення максимумів домішкових смуг A і B у спектрі ФЛ, тобто зсув відбувається у високоенергетичну область.

Дослідження температурної залежності інтенсивності рекомбінаційного випромінювання домішкових смуг A і B виявили подібну тенденцію (рис. 6.5). При збільшенні температури від 10 до 60 K інтенсивність обох смуг практично не змінюється, а при подальшому збільшенні температури від 60 до 300 K спостерігається температурне гасіння фотолюмінесценції. При цьому температурна залежність інтенсивності обох домішкових смуг ФЛ є експоненціальною.

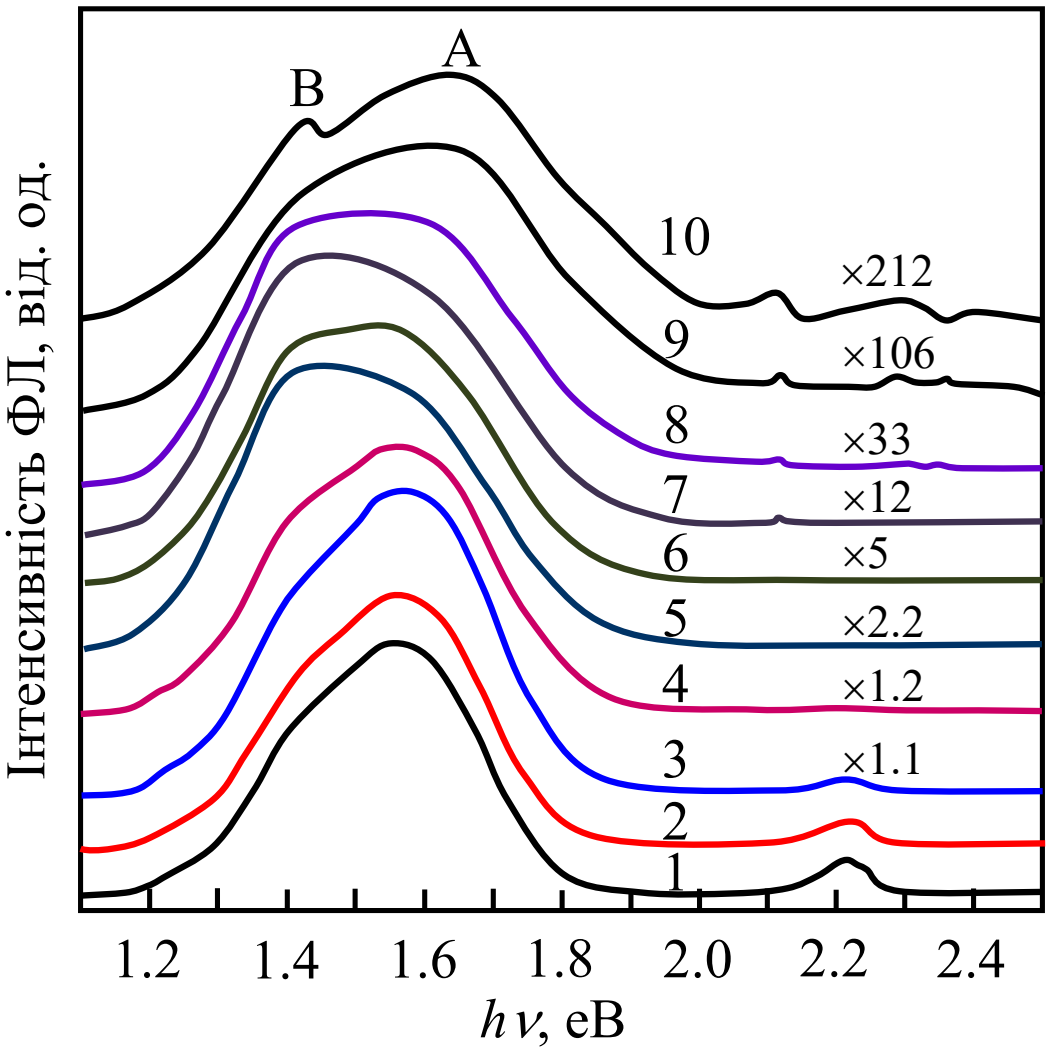

Рис. 6.4. Спектри фотолюмінесценції нанопластини $Si_2Te_3$, виміряні при різних температурах. $T$, K: 1 – 10; 2 – 30; 3 – 60; 4 – 90; 5 – 120; 6 – 150; 7 – 180; 8 – 210; 9 – 250; 10 – 300 [50].

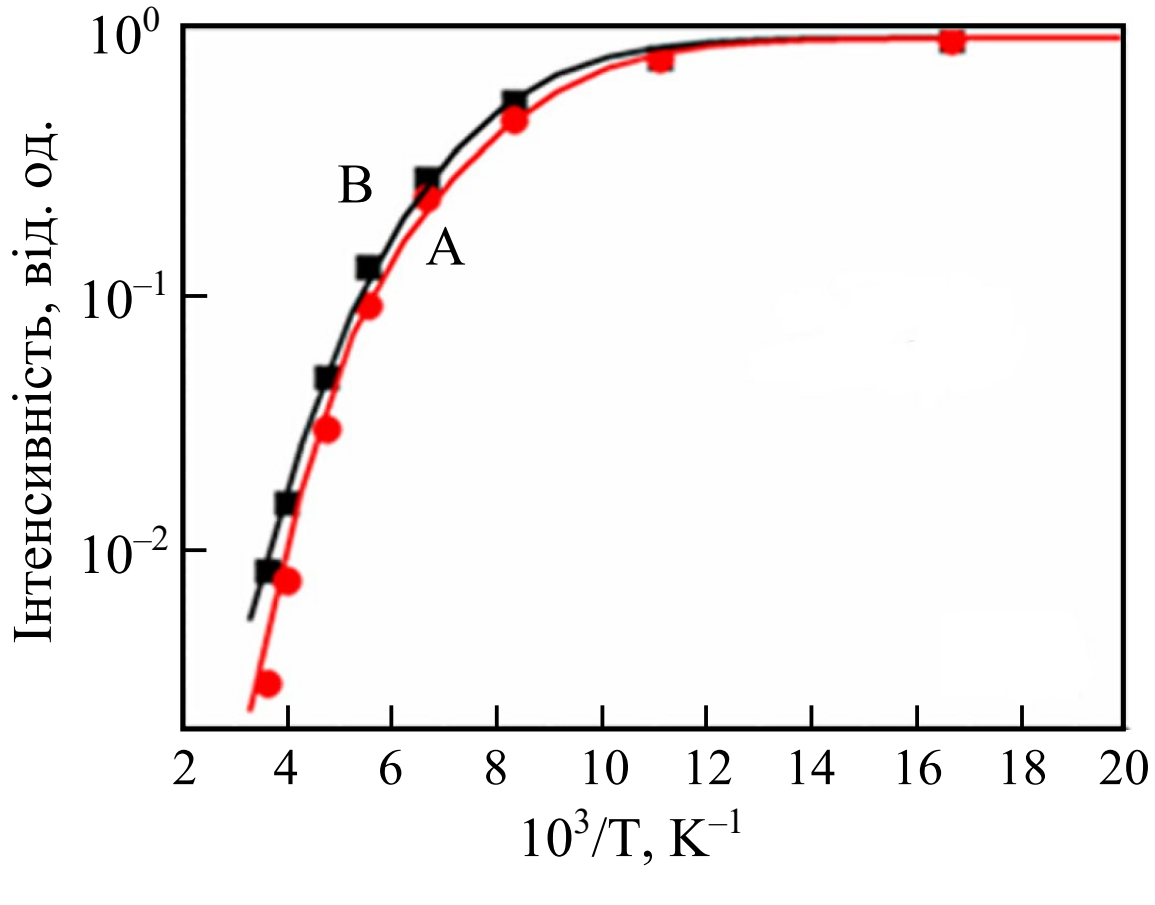

Рис. 6.5. Температурні залежності інтенсивностей випромінювальних смуг A і B [50]

Відмінною особливістю нанопластин $Si_2Te_3$ є наявність у спектрі ФЛ, крім інтенсивних домішкових смуг, слабкого прикрайового випромінювання з максимумом при 2.24 еВ (рис. 6.4, крива 1). Термін «прикрайове випромінювання» використовується в широкому розумінні для позначення процесів випромінювальної рекомбінації, які приводять до появи квантів світла з енергією, яка відрізняється на кілька десятих долей електрон-вольт від ширини забороненої зони. Такі процеси стають суттєвими при низьких температурах, і у випадку нанопластин $Si_2Te_3$ спостерігаються в інтервалі 10 – 90 K (рис. 6.6).

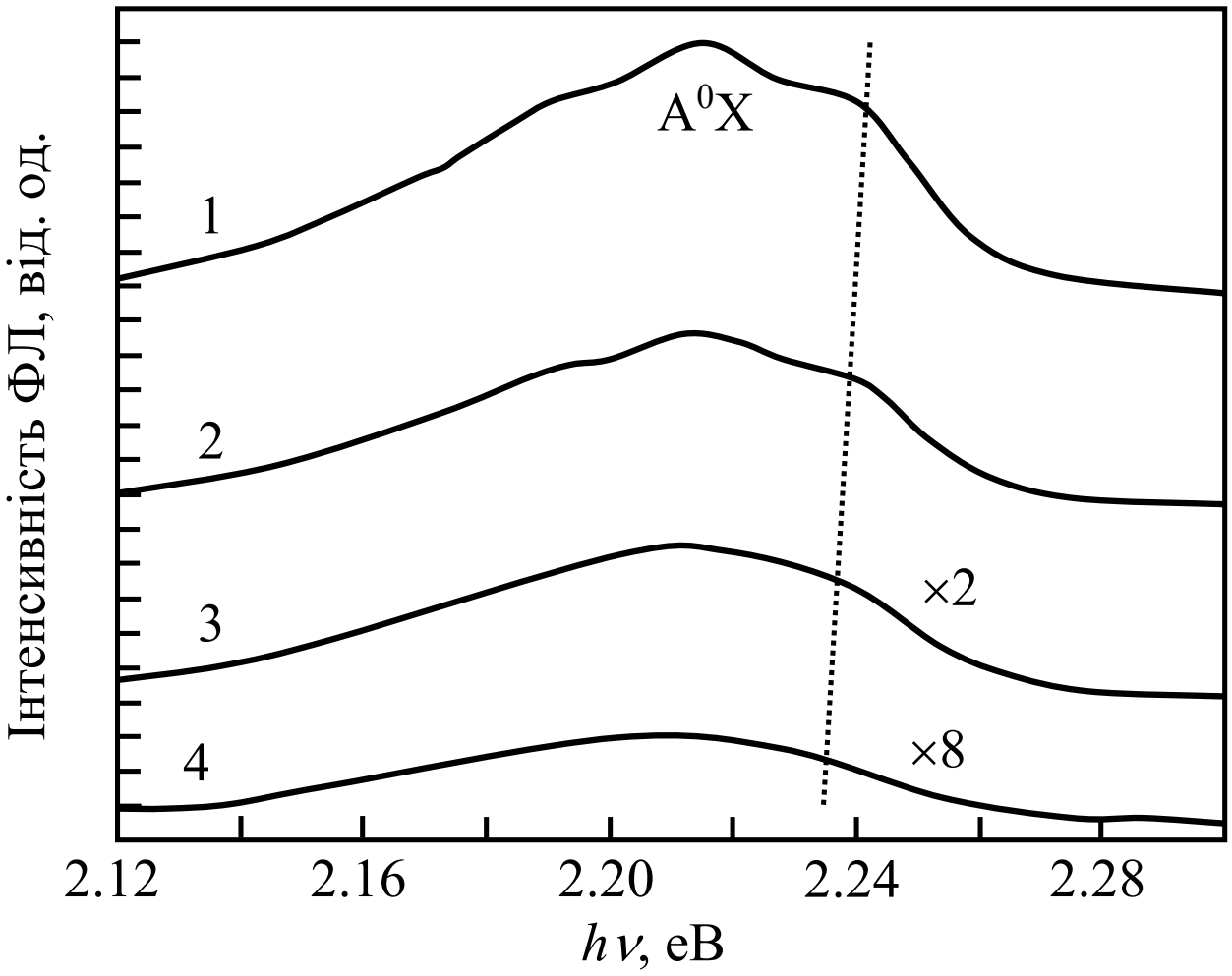

Рис. 6.6. Низькотемпературні спектри прикрайової ФЛ нанопластини $Si_2Te_3$. *T*, K: 1 – 10; 2 –30; 3 – 60; 4 – 90 [50].

Із рис. 6.6 видно, що спектр прикрайової фотолюмінесценції при $T$ = 10 K є більш складним і може бути апроксимований чотирма смугами з максимумами при 2.240, 2.216, 2.190 і 2.161 еВ (рис. 6.7). Природу піка 2.240 еВ автори [50] зв'язують із анігіляцією вільного екситону. З довгохвильової сторони від цієї смуги наявна більш інтенсивна смуга 2 із максимумом при 2.216 еВ, зумовлена рекомбінацією акцепторно-зв'язаних екситонів ($A^0X$). Збуджені стани, які відповідають за виникнення цієї смуги прикрайового випромінювання нанорозмірних кристалів $Si_2Te_3$ можна представити та напівкількісно описати за допомогою моделі електронно-діркових пар, локалізованих поблизу іонізованих або нейтральних дефектів кристалічної ґратки. Тому термін «зв'язані екситонні комплекси» став загально-прийнятим. Два наступних піки (2.190 і 2.161 еВ) є фононними повтореннями LO фононів. Оптичні переходи в області краю смуги поглинання відбуваються за участю коливань ґратки. Еквідистантні лінії у серіях крайового випромінювання пояснюється одночасним випромінюванням фотона та 0, 1, 2 … поздовжніх (LO) фононів.

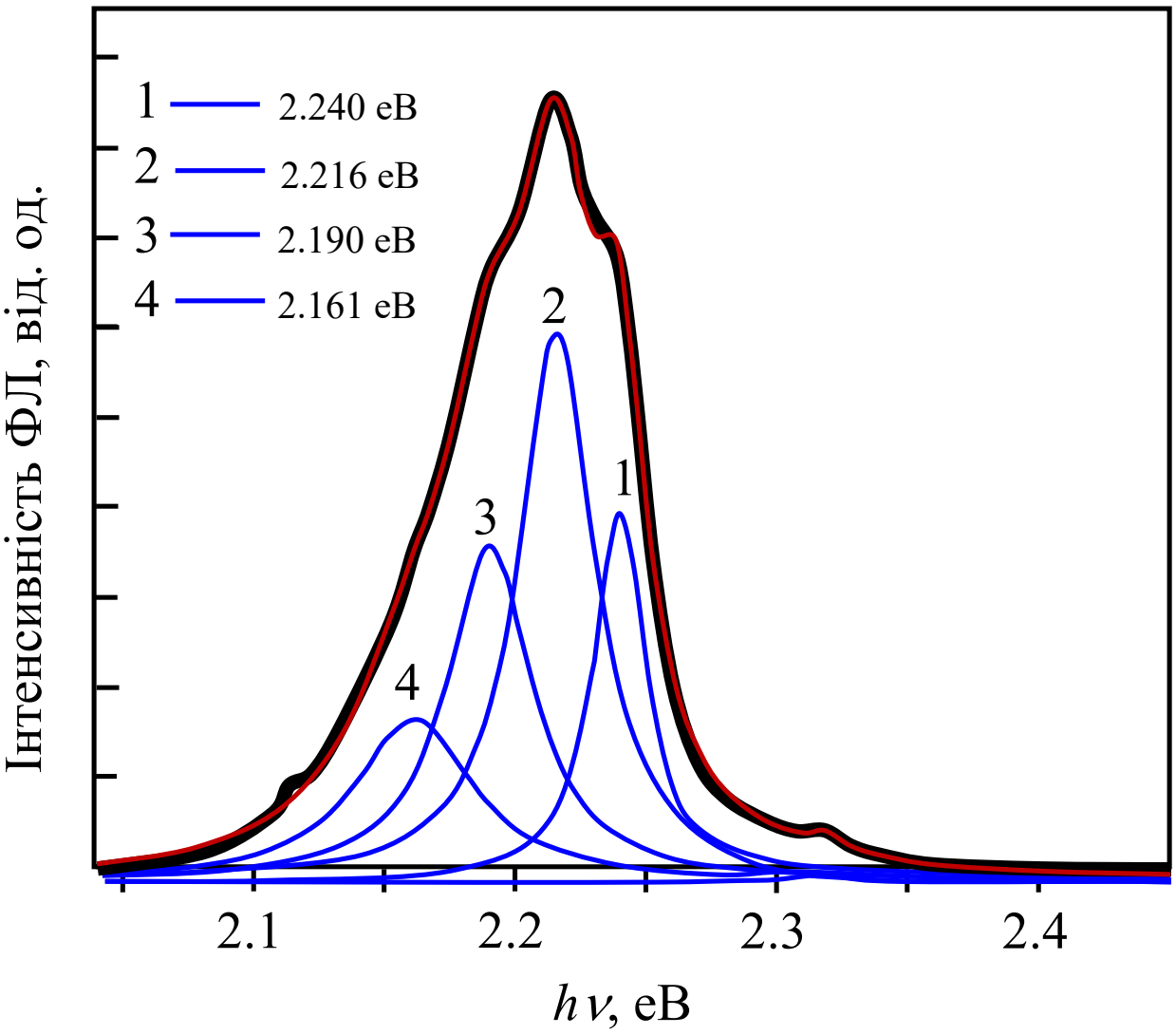

Рис. 6.7. Спектр прикрайової ФЛ нанопластини $Si_2Te_3$ та його розкладання на гаусіани: смуги вільного екситону (1), зв'язаного екситону (2) і смуги фононного повторення (3, 4), при 10 K [50].

При збільшенні температури зразка (нанопластини) у спектрах ФЛ спостерігається ряд змін (рис. 6.6). Напівширина ліній при низьких температурах значно перевищує *kT*, з підвищенням температури ще збільшується, так що структура, пов'язана з LO-фононом, зникає.

## 6.3. ВПЛИВ ДОМІШОК НА ФОТОЛЮМІНЕСЦЕНЦІЮ НАНОПЛАСТИН $Si_2Te_3$

Дослідження впливу різних домішок на основні характеристики фотолюмінесценції наноструктурованого $Si_2Te_3$ спрямовані на з'ясування ролі домішок у формуванні енергетичного спектра локалізованих станів у забороненій зоні даного матеріалу. Вплив легування та інтеркалювання на спектри фотолюмінесценції нанопла-стин $Si_2Te_3$ досліджено у роботі [51]. Нанопластини $Si_{2-x}Ge_xTe_3$ із різною концентрацією ізоелектронної домішки германію ($x$ = 0.04 – 0.22) були вирощені із парової фази. Легування сесквітелуриду кремнію ізоелектронною домішкою германію приводить до збільшення інтенсивності випромінювання та зміщення смуги фотолюмінесценції в довгохвильову область (рис. 6.8). Вже при малих концентраціях введеної домішки германію (0.04 ат. %) спектр ФЛ зазнає істотних змін, що проявляється в повному гасінні піка 750 нм (1.65 еВ) та появі яскраво вираженого піка при ~860 нм (~1.44 еВ), енергетичне положення якого практично вже не залежить від подальшого збільшення концентрації домішки аж до $x$ = 0.22. Червоний зсув піка фотолюмінесценції в легованих германієм нанопластин $Si_2Te_3$ автори [51] пов'язують зі зменшенням енергії прямої забороненої зони.

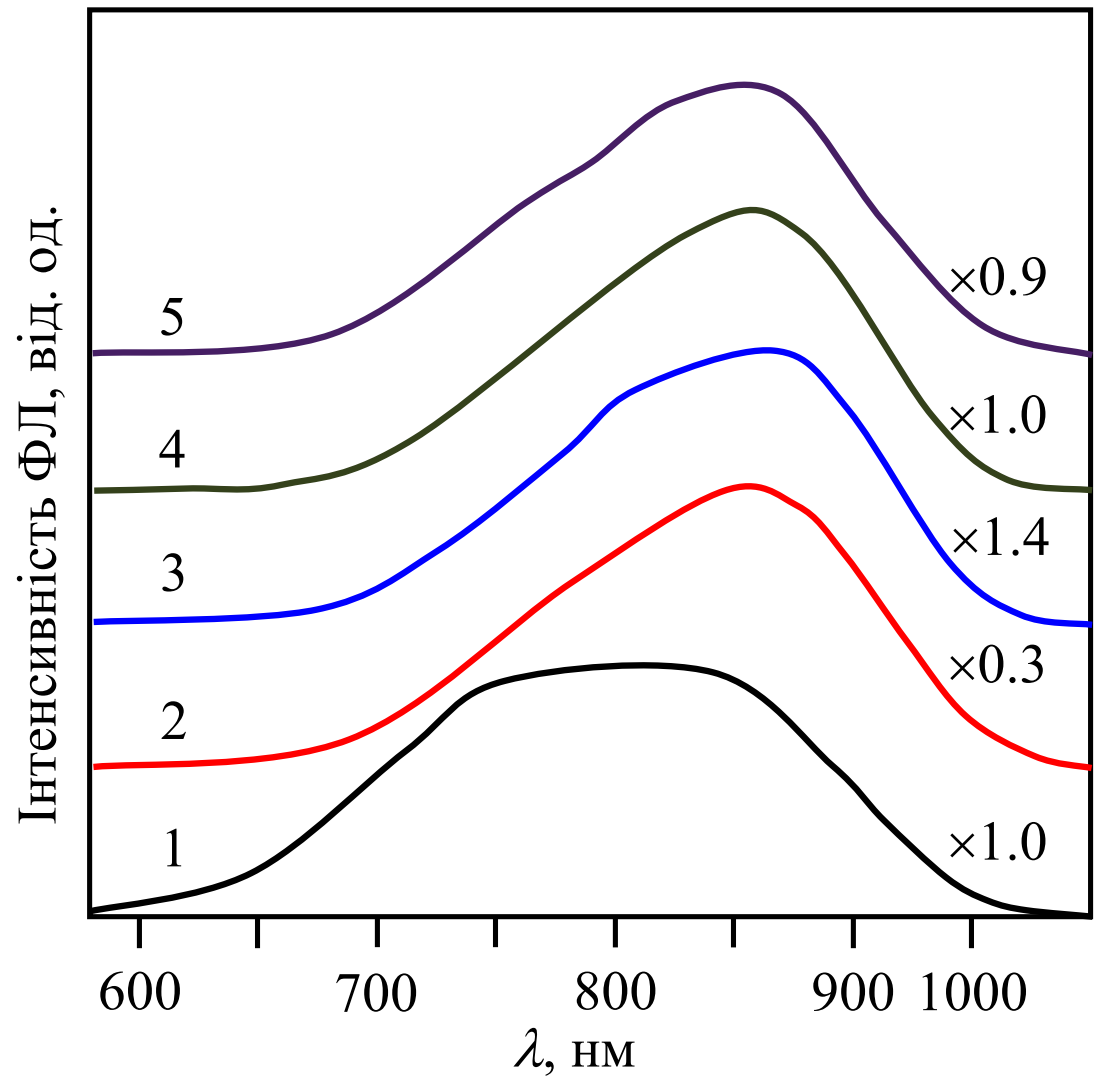

Рис. 6.8. Спектри ФЛ нанопластин $Si_{2-x}Ge_xTe_3$.
$x$: 1 – 0; 2 – 0.04; 3 – 0.10; 4 – 0.12; 5 – 0.22 [51].

Нуль-валентну мідь і германій автори [6] інтеркалювали в нанопластини $Si_2Te_3$ з використанням окисно-відновної реакції диспропорціювання $Cu^{+1}$ до $Cu^0$ в ацетоні або броміді германію $GeBr_4$ до $Ge^0$ при 423 К в октадецені ($C_{18}H_{36}$) відповідно. На рис. 6.9 наведені спектри ФЛ спеціально не активованої та інтеркальованих Ge і Cu нанопластин $Si_2Te_3$. У разі інтеркалювання нанопластин $Si_2Te_3$ германієм або міддю низькотемпературний спектр фотолюмінесценції не зазнає істотних змін (рис. 6.9, криві 2, 3), у порівняні зі спектром ФЛ спеціально нелегованої нанопластини $Si_2Te_3$ (рис. 6.9, крива 1), що є додатковим доказом того, що інтеркаляція суттєво не змінює матрицю. Інтеркаляція Cu і Ge впливає тільки на квантовий вихід випромінювальної рекомбінації. Так, інтеркаляція германію приводить до зменшення інтенсивності ФЛ, що ймовірно викликано утворенням додаткових безвипромінювальних центрів у матриці в процесі її нагрівання (для забезпечення процесу інтеркаляції).

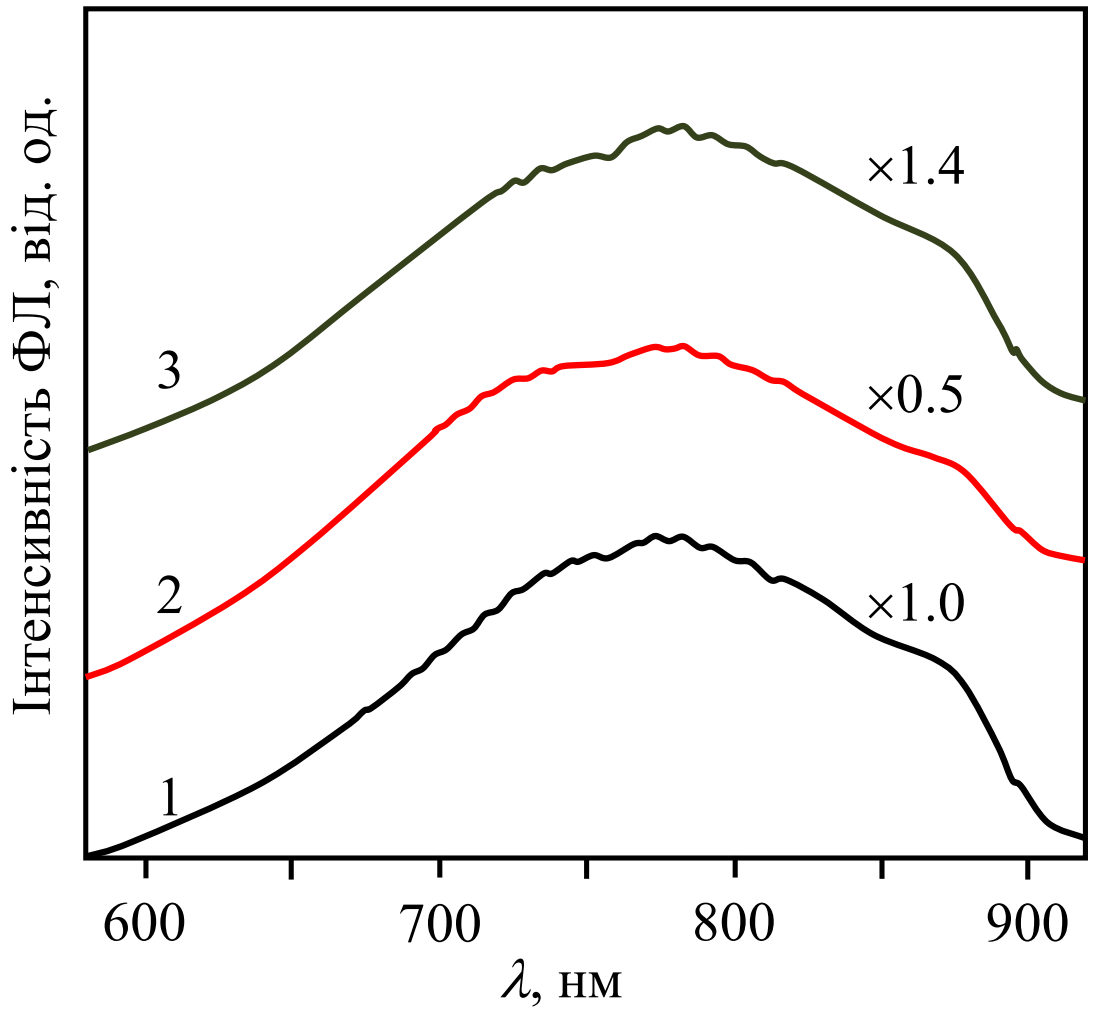

Рис. 6.9. Спектри ФЛ спеціально нелегованої (1) та інтеркальованих міддю (2) і германієм (3) нанопластин $Si_2Te_3$ [51].

Інтеркаляція міді в міжшаровий ван-дер-ваальсовий простір нанопластин $Si_2Te_3$ навпаки приводить до збільшення інтенсивності ФЛ. Однією з можливих причин цього є той факт, що при інтеркаляції Cu в шаруваті кристали мідь не тільки локалізується у міжшаровому просторі, але й може дифундувати в шари самої кристалічної ґратки, створюючи таким чином додаткові дефекти в матриці.

Оскільки фотолюмінесценція є результатом випромінювальної рекомбінації за участю глибоких локальних рівнів у забороненій зоні, і матриця залишається незмінною, смуга ФЛ залишається широкою. Враховуючи, що мідь створює більше дефектів усередині матриці, загальний квантовий вихід фотолюмінесценції збільшується, оскільки через інтеркаляцію міді може створюватися більше станів у забороненій зоні. Інтеркаляція германію навпаки приводить до зменшення квантового виходу ФЛ, можливо, через високі температури, необхідні для інтеркаляції, які можуть заліковувати деякі дефекти матриці $Si_2Te_3$. Інтеркаляція металу може зробити матеріал більш провідним, це означає, що інтеркалювання може бути використане для збільшення провідності $Si_2Te_3$ без послаблення бажаних властивостей фотолюмінесценції.

## 6.4. ФОТОЛЮМІНЕСЦЕНЦІЯ ПОЛІКРИСТАЛІЧНИХ ТОНКИХ ПЛІВОК $Si_2Te_3$

Полікристалічні плівки $Si_2Te_3$ площею ~ 8×2 см$^2$ автори [62] отримували методом хімічного осадження з парової фази (CVD метод) на підкладки $SiO_2$/Si. Ріст плівок відбувався за механізмом пара-рідина-тверде тіло, що дозволяло контролювати товщину плівки та її кристалічну структуру в залежності від температури підкладки.

На рис. 6.10 приведені спектри ФЛ (збуджені випромінюванням лазера 532 нм) тонких плівок $Si_2Te_3$, вирощених при різних температурах підкладок. Квантовий вихід ФЛ був відносно низьким для всіх досліджуваних зразків, разом з тим було виявлено, що товщина плівки впливає на спектри ФЛ. У спектрах ФЛ тонких плівок наявна

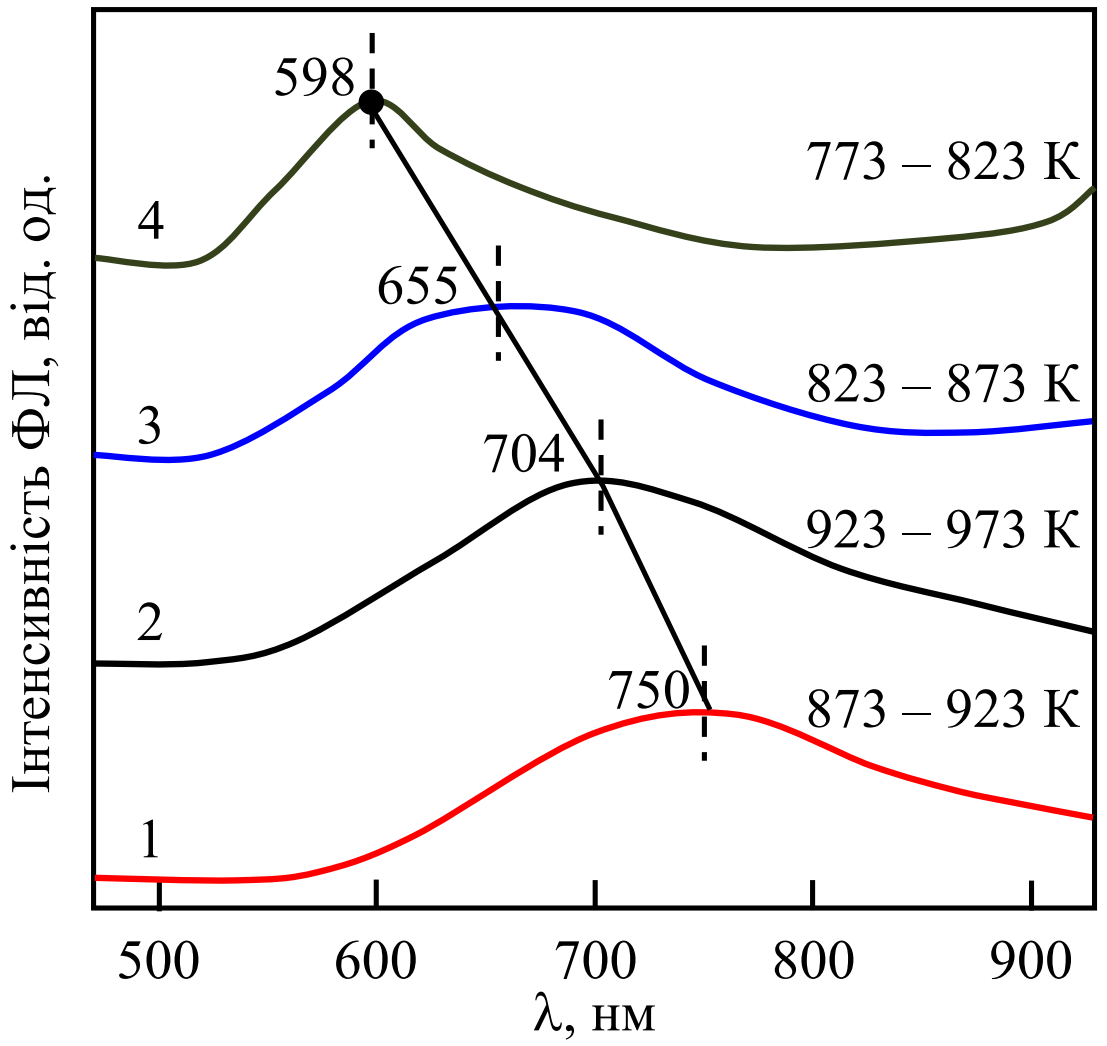

Рис. 6. 10. Спектри ФЛ полікристалічних тонких плівок $Si_2Te_3$ різної товщини, осаджених на підігріті до 773–973 К підкладки $SiO_2$/Si [62]. Товщина, нм: 1 – 8; 2 – 6; 3 – 4.5; 4 – 3.

широка смуга з максимумом у діапазоні від 598 до 750 нм, який демонструє чітке червоне зміщення зі збільшенням товщини плівки. Червоний зсув ФЛ свідчить про те, що крім внеску зміни структури в заборонену зону, міжшаровий зв'язок може також відігравати ключову роль у $Si_2Te_3$, так що ширина забороненої зони тонких плівок $Si_2Te_3$ збільшується від 1.65 до 2.07 еВ зі зменшенням товщини плівки. У спектрі ФЛ тонкої плівки $Si_2Te_3$ товщиною 8 нм наявний пік випромінювання при ~750 нм, що є близьким до піка випромінювання об'ємного кристала (рис. 6.2) і нанопластини (рис. 6.3) $Si_2Te_3$, вирощеної за допомогою методу CVD. Інший пік при меншій довжині хвилі ~600 нм демонстрували більш тонкі плівки $Si_2Te_3$ (~3 нм), що відповідає товщині 4 – 5 моношарів. Ці результати свідчать про те, що $Si_2Te_3$ може бути перспективним оптоелектронним матеріалом, випромінюванням якого можна модулювати в широкому діапазоні довжин хвиль у всій області видимого світла.

Фотолюмінесценцію однозернистих тонких плівок $Si_2Te_3$, епітаксіально вирощених на підкладках $SiO_2$/Si, досліджено в роботі [63]. У спектрі ФЛ, виміряного при кімнатній температурі, зафіксовано смугу з максимумом $h\nu_{max}$ = 1.58 еВ

## 6.5. ФОТОЛЮМІНЕСЦЕНЦІЯ НАНОДРОТІВ $Si_2Te_3$

Спектри фотолюмінесценції нанодротів (нановіскерів, НВ) $Si_2Te_3$, синтезованих з використанням золота в якості каталізатора на кремнієвій підкладці методом хімічного осадження з парової фази, досліджені авторами [60, 209] в широкому інтервалі температур 9–290 К. Нанодроти діаметром 300 нм і довжиною 10 мкм мали довільну орієнтацію на кремнієвій підкладці. Незважаючи на те, що переважна більшість НВ були прямими і відносно однорідними по діаметру, виявилося, що нановіскери мають шорсткі поверхні, які можуть впливати на фотолюмінесценцію. Через довільну орієнтацію ансамбля нановіскерів для збудження фотолюмінесценції автори [60, 209] використовували циркулярно поляризоване лазерне збудження (475 нм), задля того, щоб усі НВ були однаково збуджені з точки зору поляризації.

Спектри фотолюмінісценції ансамбля нановіскерів $Si_2Te_3$ на кремнієвій підкладці, виміряні при різних температурах і потужностях збуджуючого світла, наведені на рис. 6.11 і 6.12, відповідно. При температурах нижче 100 К спектр ФЛ нановіскерів характеризується

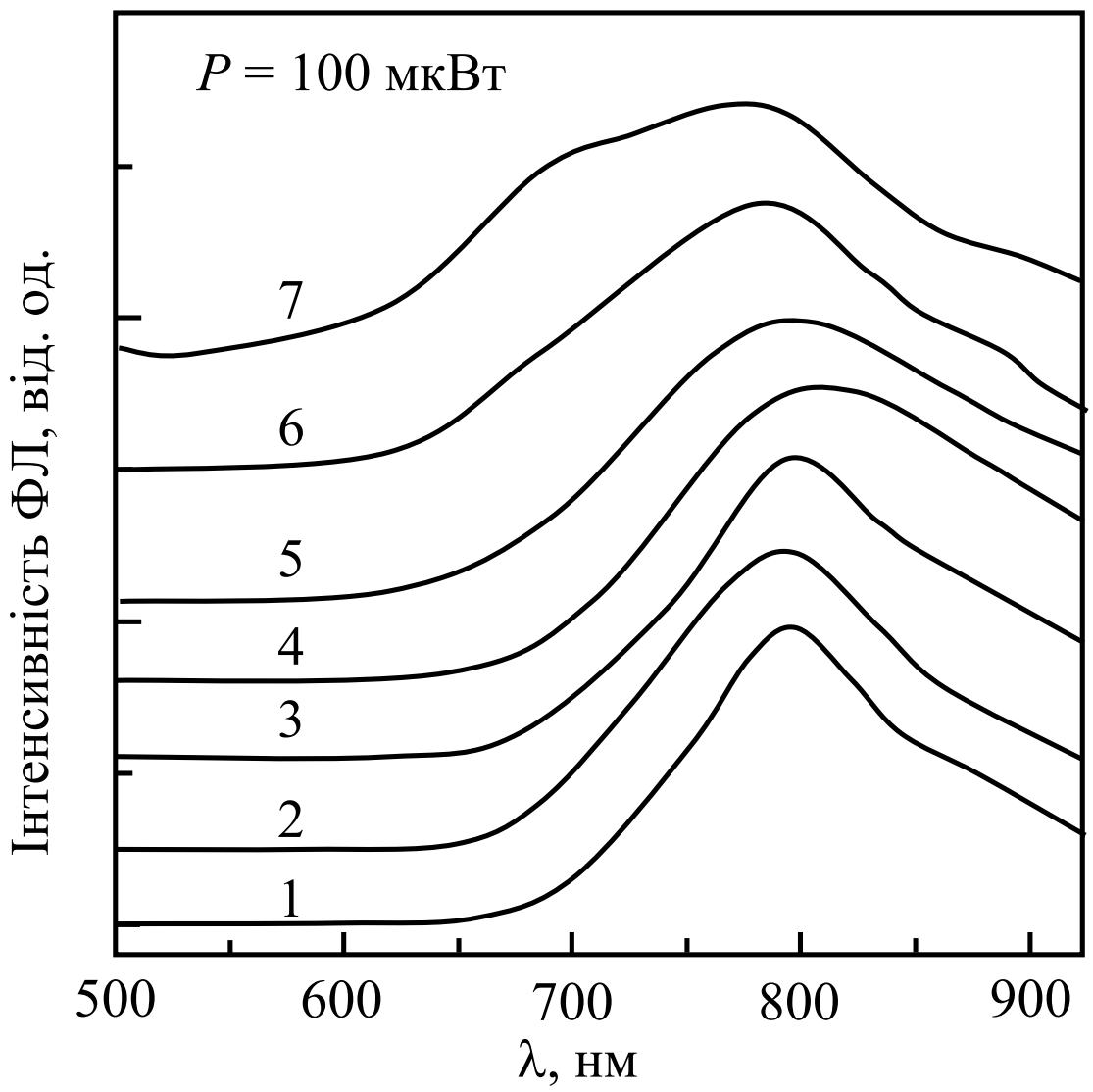

Рис. 6.11. Спектри фотолюмінесценції ансамбля нанодротів $Si_2Te_3$. *T*, K: 1 – 9; 2 – 50; 3 – 100; 4 – 150; 5 – 200; 6 – 250; 7 – 290 [60].

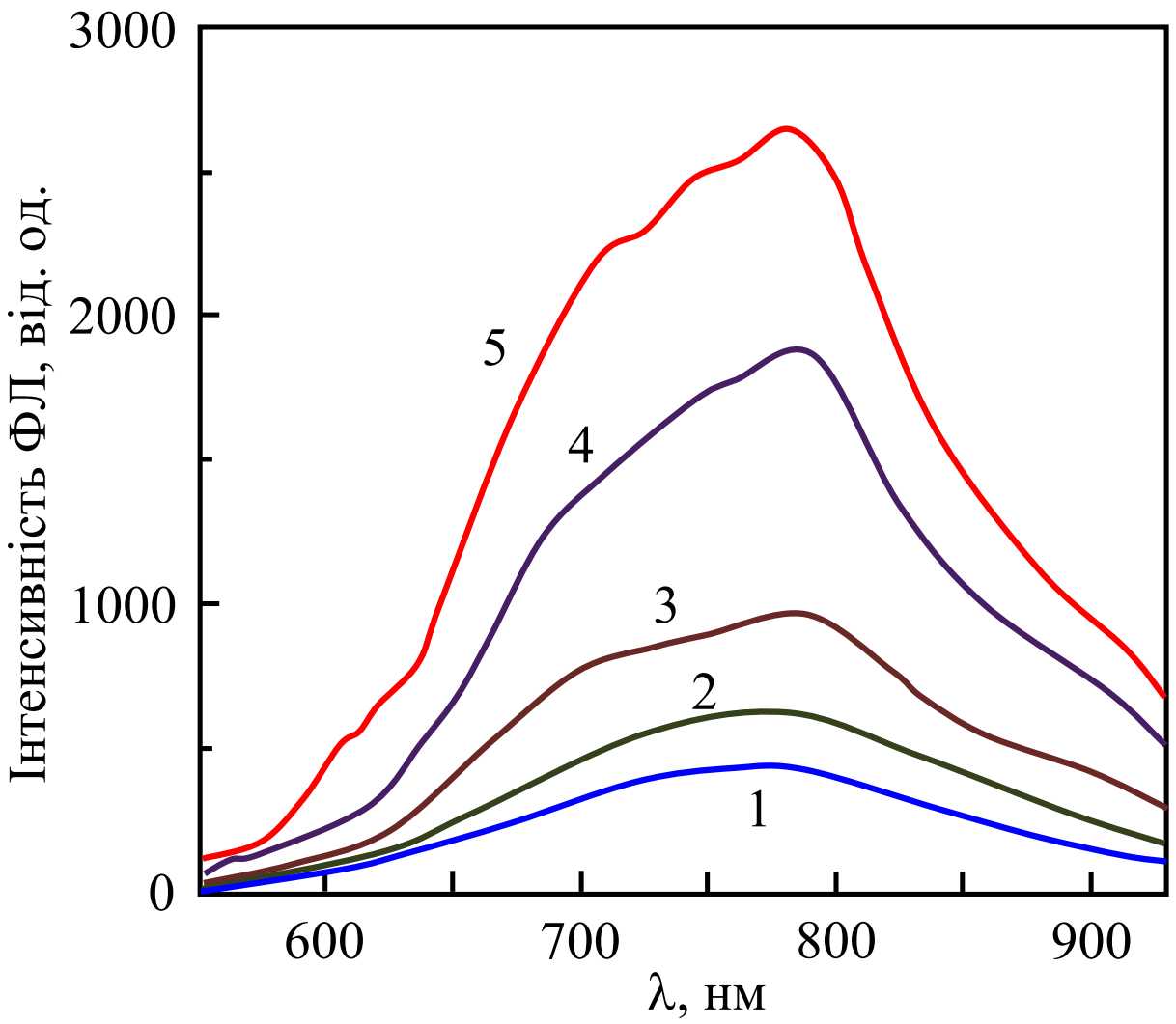

Рис. 6.12. Спектри ФЛ ансамбля нанодротів $Si_2Te_3$, виміряні при $T = 290$ К і різних потужностях збуджуючого світла *P*, мкВт: 1 – 20; 2 – 50; 3 – 100; 4 – 400; 5 – 750 [59].

широкою смугою з максимумом близько 790 нм і напівшириною ~90 нм та напливом на довгохвильовому спаді основної смуги ~890 нм. Подібні спектри ФЛ отримали автори [50] для нанопластин $Si_2Te_3$ (рис. 6.4). Із збільшенням температури зразка (нанодротів) у спектрі ФЛ з'являється ще одна особливість також у вигляді напливу, але вже на короткохвильовому спаді основної смуги ~700 нм. На відміну від нанопластин, у низькотемпературних спектрах ФЛ нановіскерів прикрайову смугу не зафіксовано.

Люмінесцентне випромінювання НВ $Si_2Te_3$ автори [60] аналізували в широкому діапазоні потужності збуджуючого лазерного випромінювання (20 – 750 мкВт). Із рис. 6.12 видно, що при збільшенні потужності збудження у спектрі ФЛ проявляється особливість у вигляді напливу в околі 680 нм.

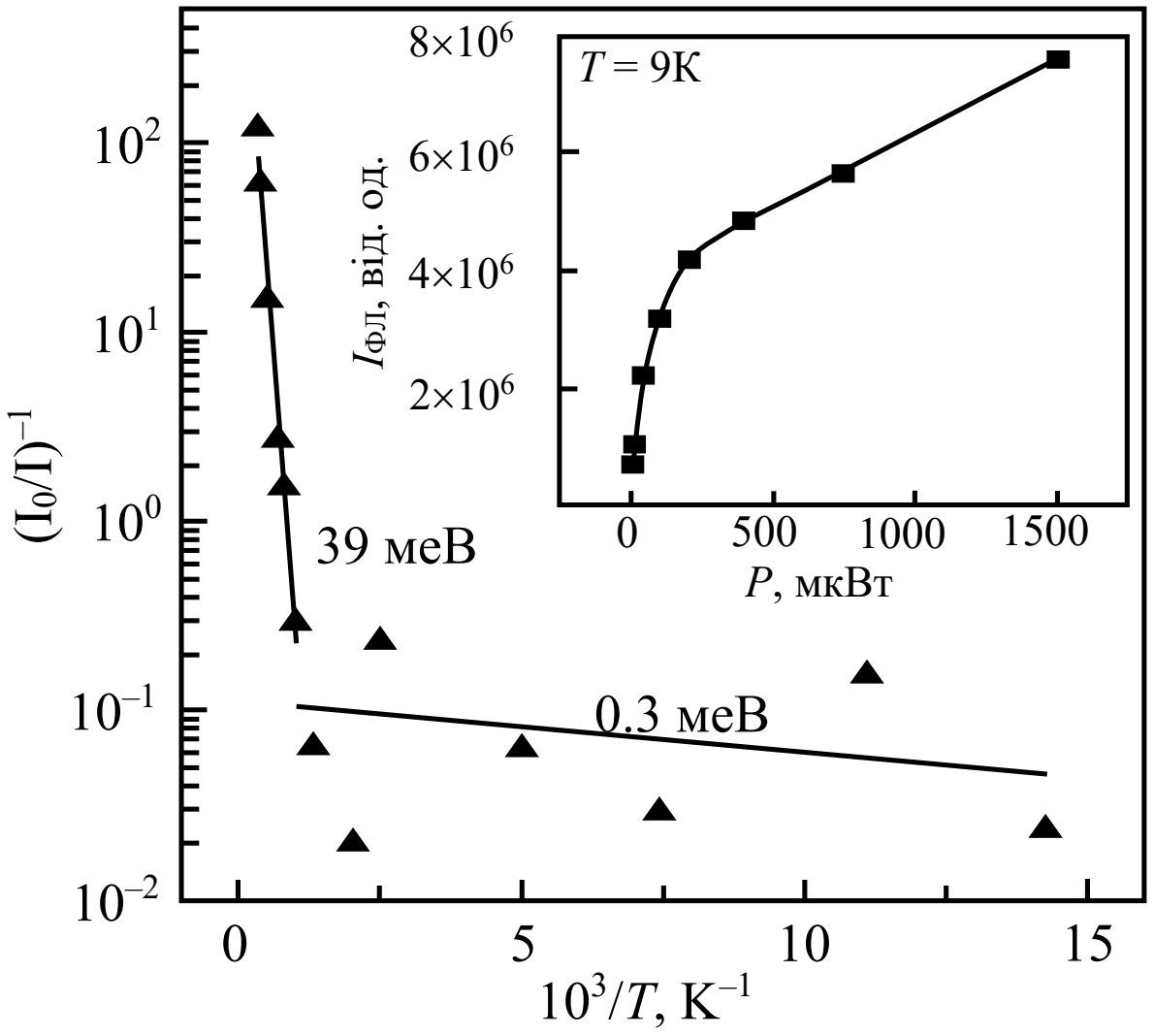

Рис. 6.13. Температурна залежність інтегральної інтенсивності ФЛ нанодротів $Si_2Te_3$ при фіксованій середній потужності збудження $P$ = 100 мкВт. $I$ та $I_0$ – інтенсивності ФЛ при даних температурах $T$ і 0 К відповідно. На вставці наведено залежність інтенсивності випромінювання від потужності збудження при T = 9 К [60].

Температурна залежність інтегральної інтенсивності ФЛ нанодротів наведена на рис. 6.13. При низьких температурах інтегральна інтенсивність випромінювання є досить високою і не змінюється до 75 К, а вище цієї температури спостерігається різке гасіння ФЛ з енергією активації $E_a$ = 39 меВ. Наявність енергії активації 39 меВ, що перевищує значення теплової енергії $kT$ при кімнатній температурі (~25.7 меВ), свідчить про те, що енергетичний перехід фотозбуджених носіїв був викликаний не просто тепловою енергією, а тепловою модифікацією рівнів енергії стану, які пов'язані з дефектами, зокрема з неконтрольованими домішками або поверхневими дефектами, спричиненими шорсткістю поверхні нановіскерів.

На вставці рис. 6.13 наведено залежність інтегральної ФЛ від потужності лазерного збудження при $T$ = 9 K. Нижче певного значення збудження (250 мкВт) інтенсивність ФЛ лінійно зростає зі збільшенням потужності збудження. Вище 250 мкВт спостерігається сублінійна люкс-яркісна характеристика з показником менше одиниці. Автори [60, 209] припускають, що при високих потужностях збудження велика середня кількість фотозбуджених носіїв приводить до безвипромінювальної рекомбінації з поверхневими станами, спричиненими шорсткістю поверхні НВ.

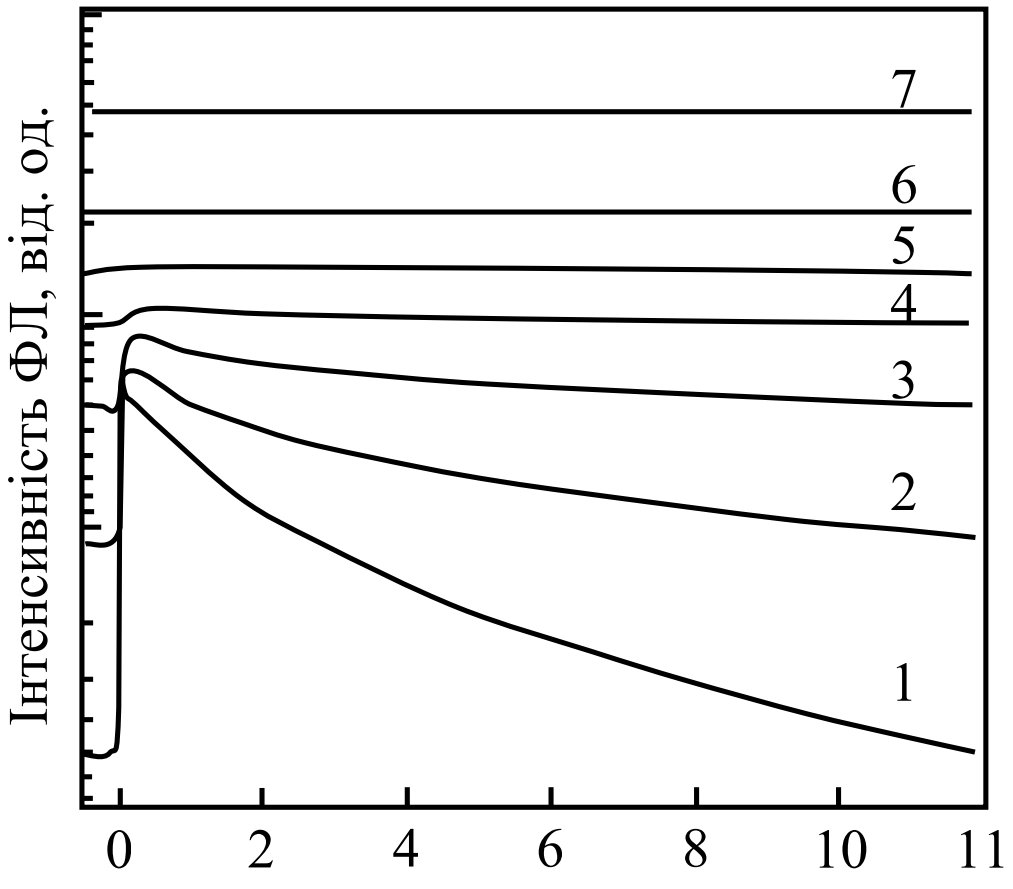

Рис. 6.14. Кінетика загасання ФЛ нанодротин $Si_2Te_3$ при різних температурах та фіксованій середній потужності збудження $P$ = 100 мкВт. $T$, K: 1 – 290; 2 – 250; 3 – 200; 4 – 150; 5 – 100; 6 – 50; 7 – 9 [60].

Значне скорочення часу загасання у разі підвищення температури вказує на кілька ефектів. Спочатку відбувається процес термалізації нейтральних донорів та термічне гасіння фотозбуджених носіїв. Це узгоджується із спостереженням значного зниження загальної інтег-ральної інтенсивності випромінювання ФЛ (рис. 6.13). Теплові процеси приводять до збільшення швидкості безвипромінювальної рекомбінації і в кінцевому підсумку до зменшення часу розпаду фотонів, які випускаються. По-друге, згідно даних [60], може відбутись перебудова структури ґратки $Si_2Te_3$ за різних температур, що приведе до зміни структури зон і динаміки несущих.

У спектрах ФЛ нанодротів $Si_2Te_3$ переважає випромінювання, пов'язане з дефектами та поверхневими станами як при низьких, так і при кімнатній температурах. Час загасання фотозбуджених носіїв був досить тривалим (> 10 нс) при низьких температурах і зменшився (< 2 нс) при $T_{кімн}$ (рис. 6.14). При високих швидкостях збудження час розпаду носіїв зменшується. Прискорення швидкості розпаду фотозбуджених носіїв вказує на термічне гасіння разом із безвипромінювальною рекомбінацією при високій температурі та значній потужності збудження.

Однією з основних проблем люмінесценції сесквітелуриду кремнію залишається проблема ідентифікації центрів, які відповідають за ту чи іншу смугу випромінювання. Люмінесцентні властивості спеціально нелегованих кристалів різного габітусу, визначаються складом і концентраційним співвідношенням власних дефектів. Тому ідентифікація центрів люмінесценції може бути однозначною тільки за умови, коли комплексне дослідження люмінесценції (спектри збудження, поглинання, температурне гасіння, кінетика) буде проведено.

# РОЗДІЛ 7

# ПОТРІЙНІ СПОЛУКИ НА БАЗІ ТЕЛУРИДІВ КРЕМНІЮ В СИСТЕМАХ M–Si–Te (M = Na, K, Cu, Ag, Al, In)

Вивчення характеру взаємодії між телуридами кремнію ($SiTe_2$, $Si_2Te_3$) і телуридами $M_2Te$, виявлення нових потрійних сполук та дослідження їх властивостей має як наукове, так і практичне значення. Однак дослідженню потрійних систем M–Si–Te присв'ячена обмежена кількість робіт. На теперішній час побудовано кілька діаграм стану цих потрійних систем, але не достатньо досліджено властивостей потрійних сполук, які існують у цих системах.

## 7.1. СИСТЕМА $Si_2Te_3$–$Na_2Te$

У системі $Si_2Te_3$–$Na_2Te$ встановлено існування двох потрійних сполук $Na_6Si_2Te_6$ та $Na_8Si_4Te_{10}$ [210, 211]. Що стосується третьої сполуки $Na_2SiTe_3$, то відомо тільки про її використання як активного матеріалу негативного електрода перезаряджаючих акумуляторних літієвих батарей із неводним електролітом [212].

**7.1.1. Одержання та кристалічна структура $Na_6Si_2Te_6$ і $Na_8Si_4Te_{10}$**. Синтез телурогіподисилікату ($Na_6Si_2Te_6$) автори [210] здійснювали з високочистих елементарних компонент (Na, Si, Te) у вакуумованих кварцових ампулах. Максимальна температура синтезу становила 1123 K. Стехіометричні кількості елементарних компонентів автори [210] зважували в атмосфері аргону і завантажували у кварцову ампулу. З метою компенсації втрати Na внаслідок реакції із стінками ампули, його кількість повинна бути більшою на ~10% у порівняні з стехіометричним складом. Свіжоприготовлений полікристал прозорий і має червонуватий відтінок. При потраплянні вологого повітря він миттєво розкладається, в результаті чого на стінках посудин та поверхні кристалів утворюються шари Te в результаті окислення повітрям первісно утвореного $H_2Te$, так що речовина залишається металево-сірою після короткотривалої витримки.

Рентгеноструктурні дослідження показали, що $Na_6Si_2Te_6$ кристалізується в моноклінній структурі з просторовою групою $P2_1c$ і параметрами ґратки: $a$ = 8.786 (3) Å, $b$ = 12.780 (4) Å, $c$ = 8.864 (3) Å, β = 119.71 (5)°, $Z$ = 2, $\rho_{роз}$ = 3.686 г/см$^3$ та $\rho_{екс}$ = 3.70 г/см$^3$ [210]. Кристалічна структура $Na_6Si_2Te_6$ наведена на рис. 7.1. Базовими структурними блоками є октаедри [$Si_2Te_6$], утворені двома тригона-

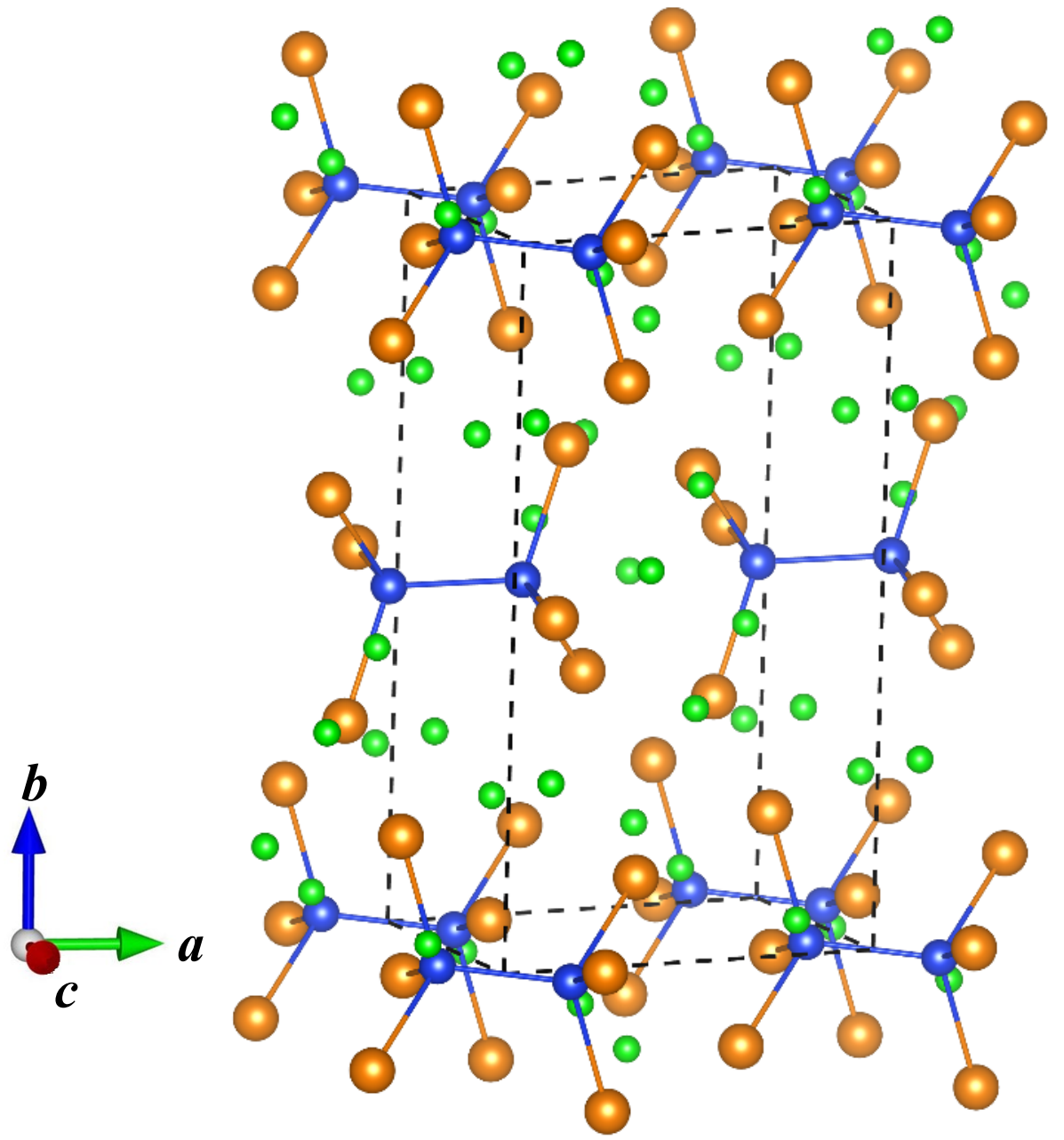

Рис. 7.1. Кристалічна структура $Na_6Si_2Te_6$ [210].

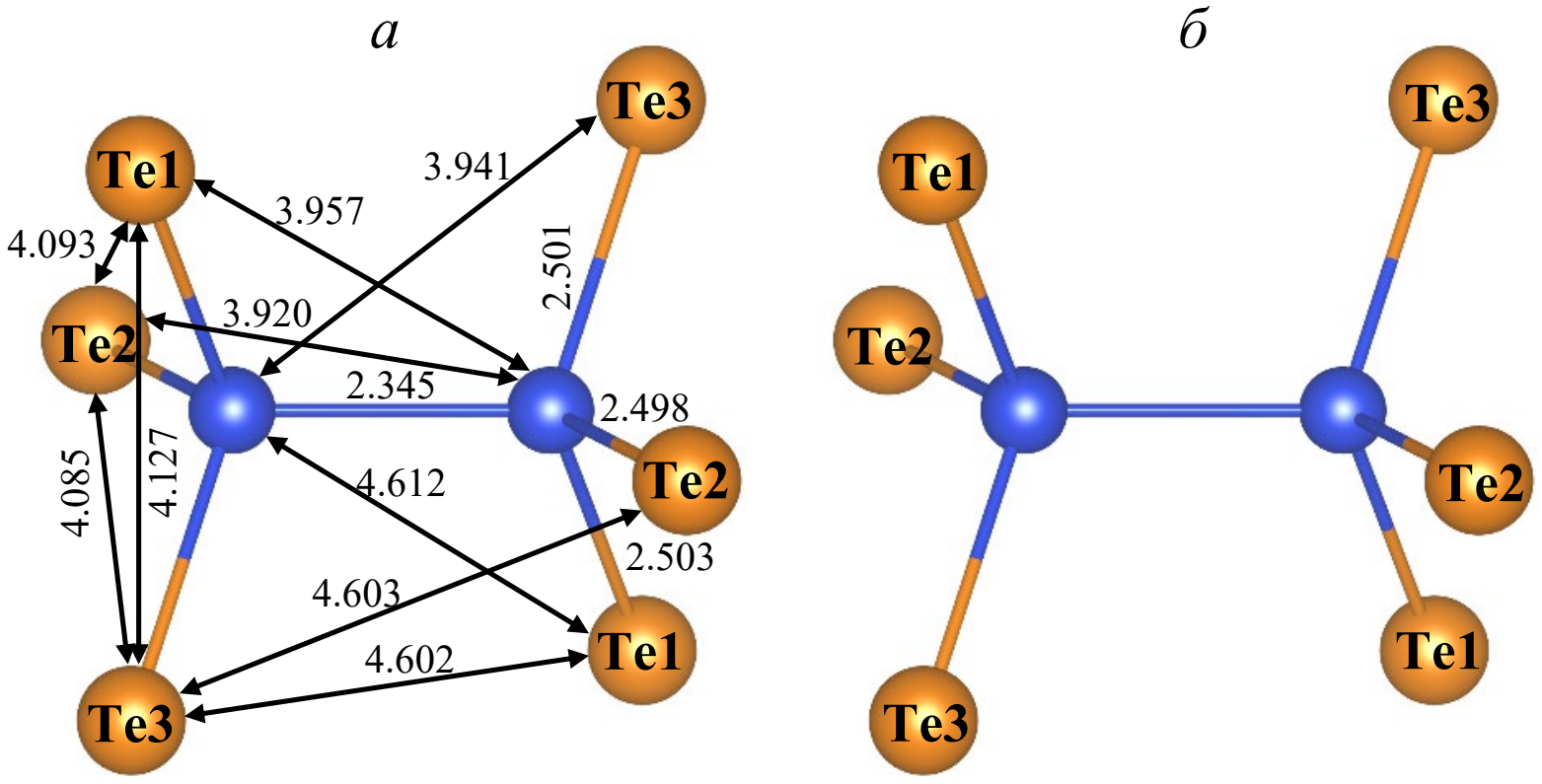

Рис. 7.2. Міжатомна відстань в Å (*а*) і кути в град. (*б*) в аніоні $[Si_2Te_6]$ [210].

льними пірамідами [$SiTe_3$], з'єднаними вершинами з атомів кремнію (рис. 7.2). Дві трикутні грані основи $Te_3$ розгорнуті майже на 60° і утворюють «ступінчасту» конфігурацію. Кожен атом Si знаходиться у злегка спотвореному тетраедричному оточенні трьох атомів Te і одного сусіднього атома Si. Відхилення кута зв'язку від ідеального тетраедричного кута у центрального атома Si незначне, незважаючи на відмінності в розмірах лігандів (рис. 7.2, *б*). Октаедри [$Si_2Te_6$]$^{6-}$ укладені один відносно одного так, що між ними є досить великі спотворені октаедричні порожнечі заповнені іонами Na.

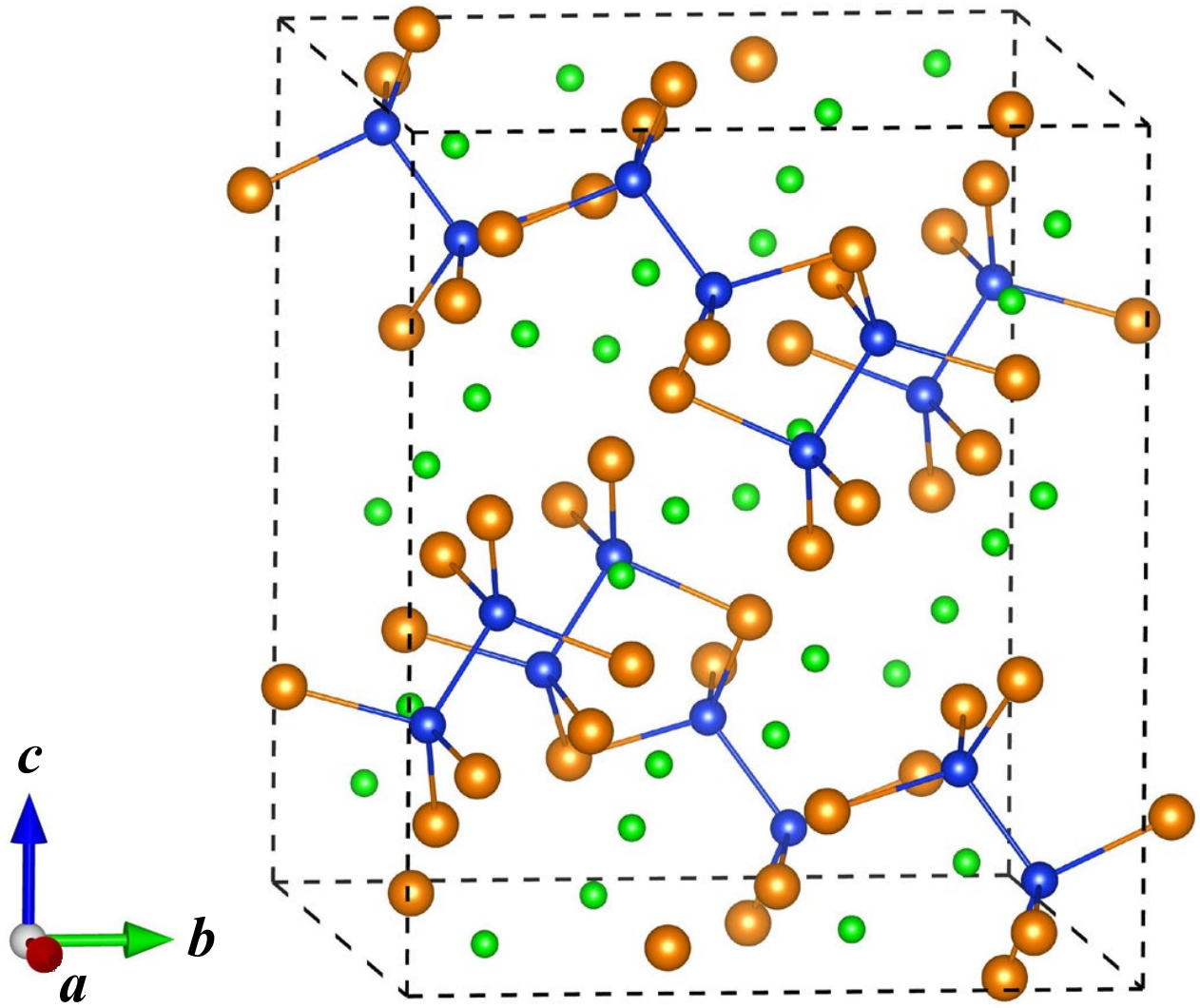

Рис. 7.3. Кристалічна стр уктура $Na_8Si_4Te_{10}$ [211].

Сполуку $Na_8Si_4Te_{10}$ автори [211] також отримували взаємодією елементарних компонентів в атмосфері аргону при 823 K та гартуванням протягом двох годин. Дана сполука кристалізується в моноклінній структурі (ПГ $P2_1/c$, Z = 4) з параметрами ґратки: $a$ = 14.073 Å, $b$ = 12.842 Å, $c$ = 14.882 Å, β = 92.22°; $\rho_{екс}$ = 3.90 г/см$^3$, $\rho_{теор}$ = 3.885 г/см$^3$. Кристалічна структура $Na_8Si_4Te_{10}$ наведена на рис. 7.3. Структура містить дискретні аніони [$Si_4Te_{10}$]$^{4-}$, які утворюються з двох ок-таедрів [$Si_2Te_6$], з’єднаних між собою по ребру містковими атомами телуру. Гантелі $Si_2$ оточені 6 атомами Te у спотвореному октаедрі. Залежно від стехіометрії, більш висока полімеризація цих одиниць може привести до більших аніонних асоціацій (при з’єднанні більше двох структурних одиниць [$Si_2Te_6$]).

**7.1.2. Електронна структура $Na_6Si_2Te_6$.** Електронна структура $Na_6Si_2Te_6$ розрахована авторами [213] методом функціоналу електронної густини (DFT) в LDA наближенні приведена на рис. 7.4, *а*. За початок відліку шкали енергії вибрано положення верха валентної зони, яке знаходиться в точці Y. Дно зони провідності розташовано в центрі ЗБ у точці Г, отже, телурогіподисилікат є непрямозонним напівпровідником з розрахованою шириною забороненої зони $E_{gi}$ = 1.88 еВ.

Валентні зони мають слабку дисперсію і утворюють чотири енергетично відокремлені зв'язки зон. Інформацію про внески атомних орбіталей у кристалічні стани $Na_6Si_2Te_6$ дають розрахунки повної та локальних парціальних густин станів. Профілі розподілу повної густини станів, а також внески від окремих станів різних атомів для $Na_6Si_2Te_6$ наведені на рис. 7.4, *б*. З аналізу енергетичного розподілу локальних парціальних густин станів натрію, кремнію і телуру випливає, що в кожну з чотирьох зв'язок заповнених зон *s*-і *p*-стани дають неоднакові внески, що відрізняються один від одного величиною. У валентній зоні $Na_6Si_2Te_6$ переважають 5*s*- і 5*p*-стани телуру, причому їхнє енергетичне положення істотно відрізняється. Найбільш низькоенергетична зв'язка з 12 заповнених зон, розташована в енергетичному інтервалі від –13.53 до –9.89 еВ, формується переважно 5*s*-станами атомів телуру. Незважаючи на переважний характер 5*s*-станів телуру, для даної валентної зв'язки зон істотними є ефекти гібридизації станів атомів Si і Te, які приводять до появи внесків 3*s*-станів атомів кремнію, які виявляються в основному локалізованими в області енергій нижніх двох зон, внесків Si 3*s*- 3*p*-станів у середню частину з двох зон і внесків 3*p*-станів кремнію у верхню частину цієї підзони (8 зон), що мають слабку дисперсію. Середня зв'язка з чотирьох розділених (2+2) зон, що слідує за *s*-зонами телуру, відокремлена від них інтервалом енергій у 2.35 еВ, не перекривається з верхньою валентною підзоною, формуючи таким чином дві ізольовані підзони. Ці заповнені підзони мають гібридизований характер і утворені внаслідок перекриття Si3*s*- та Te5*p*-станів.

Верхню зв'язку заповнених зон умовно можна розділити на дві частини: нижню (–3.67 ÷ –2.32 еВ) з 10 зон, яка має змішаний характер за участю гібридизованих 5*p*-станів телуру та 3*p*-станів кремнію з дуже незначною домішкою *s*- і *p*-станів натрію; верхню (–2.20 ÷ 0 еВ), що містить 24 зони, утворену переважно не містковими 5*p*-орбіталями телуру, заповнені двома електронами (неподіленою парою), з домішуванням *p*-, *d*-станів кремнію та *s*-, *p*-станів

*а*

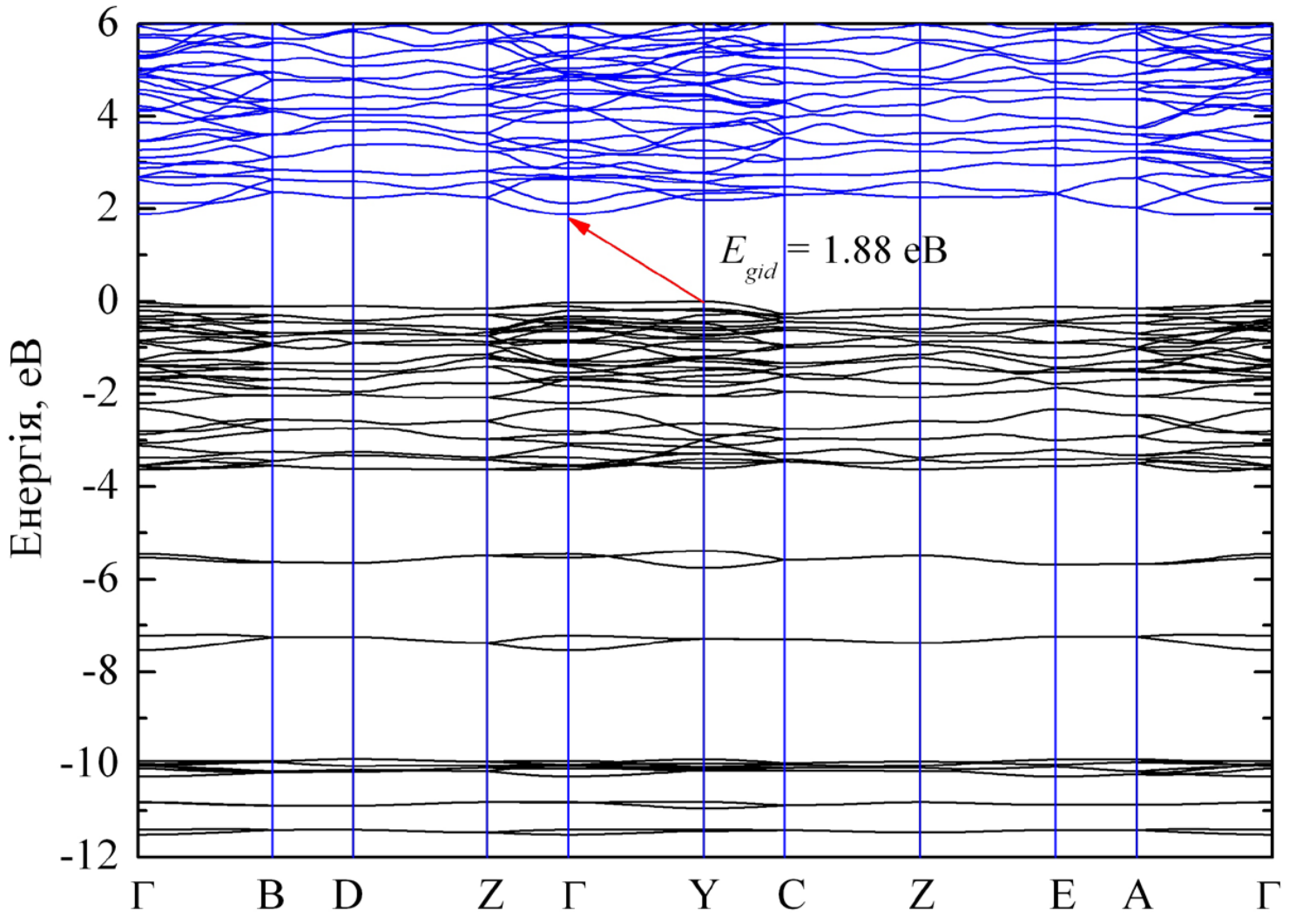

*б*

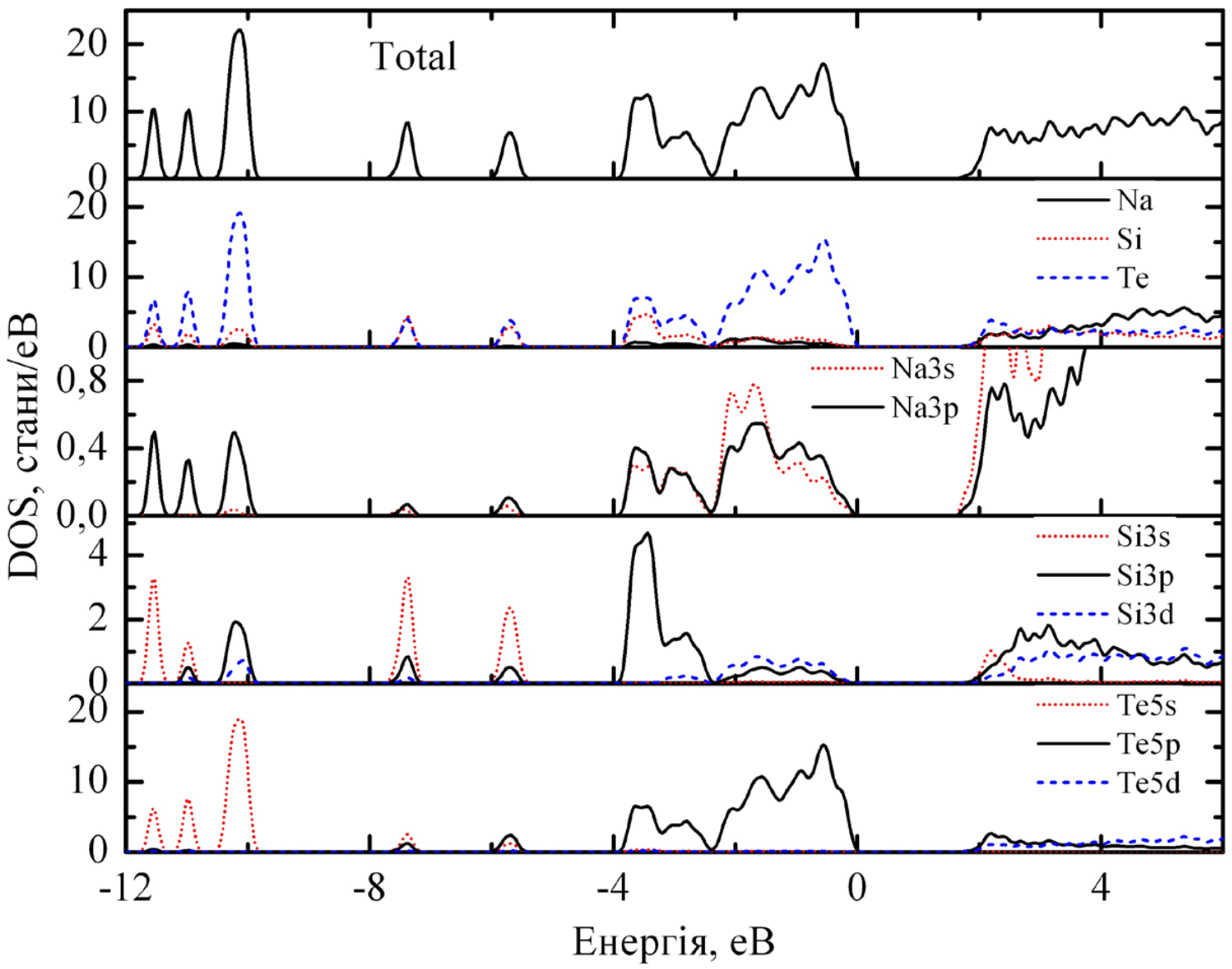

Рис. 7.4. Електронна структура (*а*), повна та локальні парціальні густини електронних станів (*б*) $Na_6Si_2Te_6$ [213].

лужного металу. Вершину валентної зони в точці Г формують виключно незв'язуючі $5p_{\pi}$-стани телуру.

Електронна низькоенергетична структура незаповнених електронних станів телурогіподисилікату натрію утворена в основному «замішуванням» вільних *p*-станів телуру з *s*- та *p*-станами кремнію та натрію.

## 7.2. СИСТЕМА $Si_2Te_3$ – $K_2Te$

У цій системі встановлено наявність трьох сполук $K_4Si_4Te_{10}$ ($K_2Si_2Te_5$), $K_6Si_2Te_6$ і $K_2SiTe_3$. Найменш вивченою є сполука $K_2SiTe_3$. Відомо тільки, що її так само як і $Na_2SiTe_3$, використовують в якості активного матеріалу негативного електрода перезаряджаючих акумуляторних літієвих батарей [212].

**7.2.1. Одержання і кристалічна структура $K_4Si_4Te_{10}$.** Синтез сполуки $K_4Si_4Te_{10}$ автори [214] здійснювали сплавленням стехіометричних кількостей елементарних компонент (K, Si, Te) у вакуумованих кварцових ампулах. Зважаючи на неминучу реакцію калію зі стінками кварцової ампули, його додавали з надлишком ~10 ваг.% порівняно із стехіометричним складом. Процес синтезу здійснювали у два етапи. На першій стадії вихідну суміш повільно нагрівали до 673 К протягом 6 годин і витримували при цій температурі протягом двох годин. Потім температуру підвищували до 923 К з подальшою витримкою протягом 30 хвилин. Після цього ампулу з розплавом знижували до 623 К і отриманий продукт відпалювали при цій температурі протягом 14 годин. У результаті отримані прозорі, блідо-жовті листоподібні кристали. Кристали $K_4Si_4Te_{10}$ є гігроскопічними. Перебування на повітрі супроводжується утворенням шару телуру на поверхні кристала.

Сполука $K_4Si_4Te_{10}$ кристалізується в ромбічній структурі (просторова група *Pnma*) з параметрами ґратки: $a$ = 21.258 (8) Å, $b$ = 12.005 (7) Å, $c$ = 10.608 (7) Å [214]. Пікнометрична густина складає $\rho_{вим}$ = 3.85 г/см$^3$. Структура сполуки $K_4Si_4Te_{10}$ побудована з катіонів $K^+$ та аніонів $[Si_4Te_{10}]^{4-}$. Аніони побудовані з чотирьох тетраедрів $[SiTe_4]$ і мають конформацію адамантового скелету з атомів телуру, в якому атоми Si зв'язані з чотирма вузловими атомами Te (рис. 7.5). Подібну будову має аніон $[Si_4Te_{10}]^{4-}$ у структурі $Na_4Si_4S_{10}$ [215].

Упаковка аніонів $[Si_4Te_{10}]^{4-}$ один щодо одного вимагає різну координацію для іонів калію. Іони K1 і K3 знаходяться в спотвореному октаедричному оточенні з 6 атомів Te (K1–Te 3.518 – 4.056 Å;

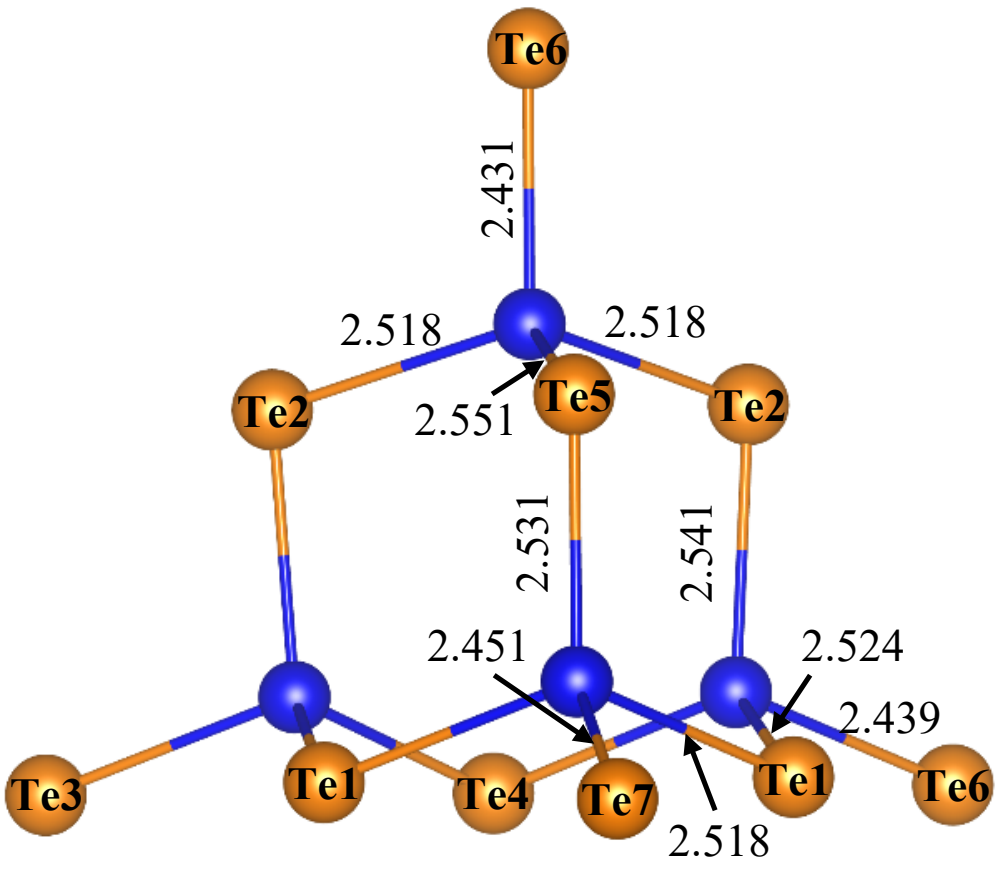

Рис. 7.5. Супертетраедр $[Si_4Te_{10}]^{4-}$ в $K_4Si_4Te_{10}$. (Довжини зв'язків у Å) [214].

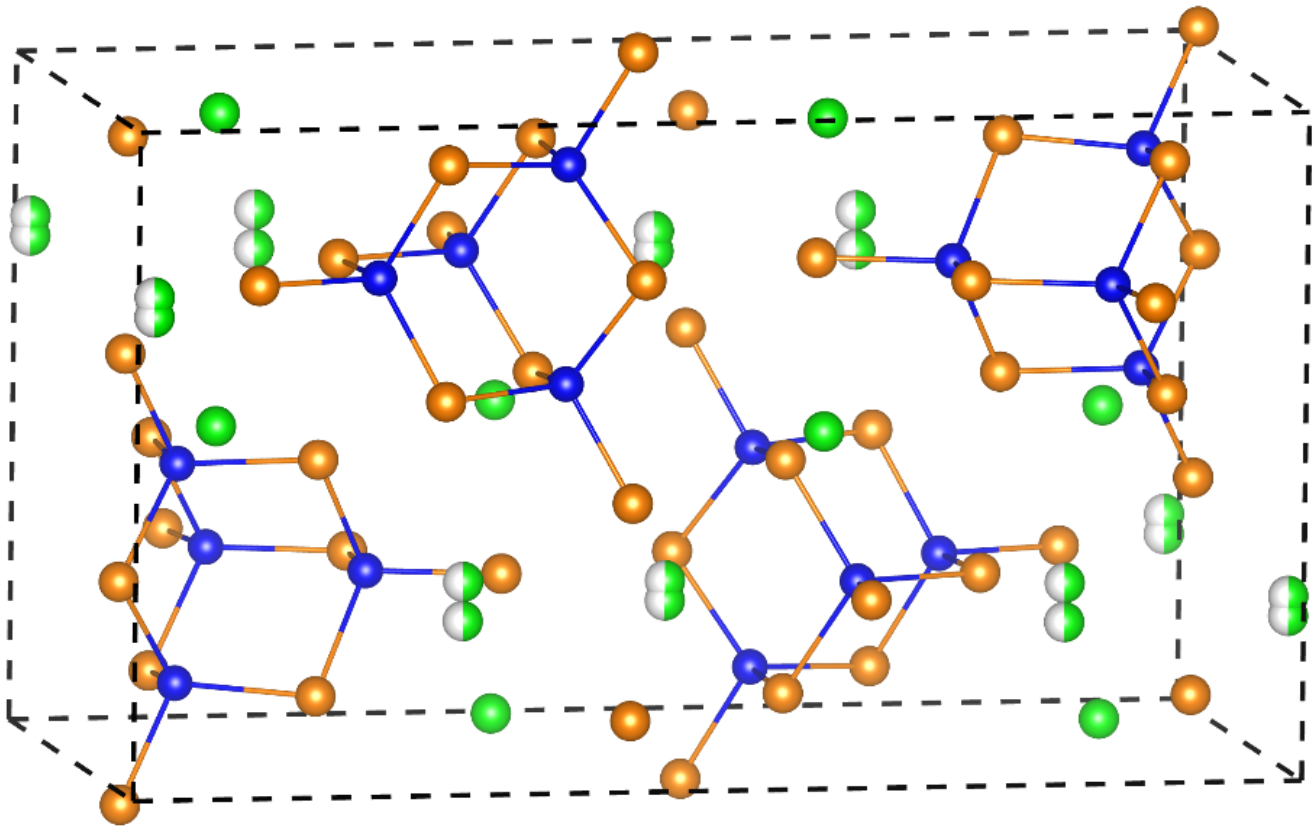

Рис. 7.6. Перспективна проекція кристалічної структури $K_4Si_4Te_{10}$ [214].

K3–Te 3.428–3.802 Å). Іони K2, з іншого боку, початково координовані 8 атомами Te (K2–Te 3.450–4.211 Å) у спотвореному та однобічно відкритому багатограннику. На цій відкритій поверхні розташовані ще 3 атоми Te, які знаходяться на відносно великих відстанях Si–Te (4.553 і 4.755 Å).

Встановлена структура $K_4Si_4Te_{10}$, зокрема будова аніону, зумовлює формулу $K_4Si_4Te_{10}$, а не $K_2Si_2Te_5$.

**7.2.2. Одержання і кристалічна структура $K_6Si_2Te_6$.** Синтез потрійної сполуки $K_6Si_2Te_6$ автори [216, 217] проводили нагріванням стехіометричних кількостей елементарних Si, K і Te у вакуумованих кварцових ампулах до температури 903 K, з наступною витримкою на протязі 1 години, після чого ампулу з речовиною повільно охолоджували до кімнатної температури.

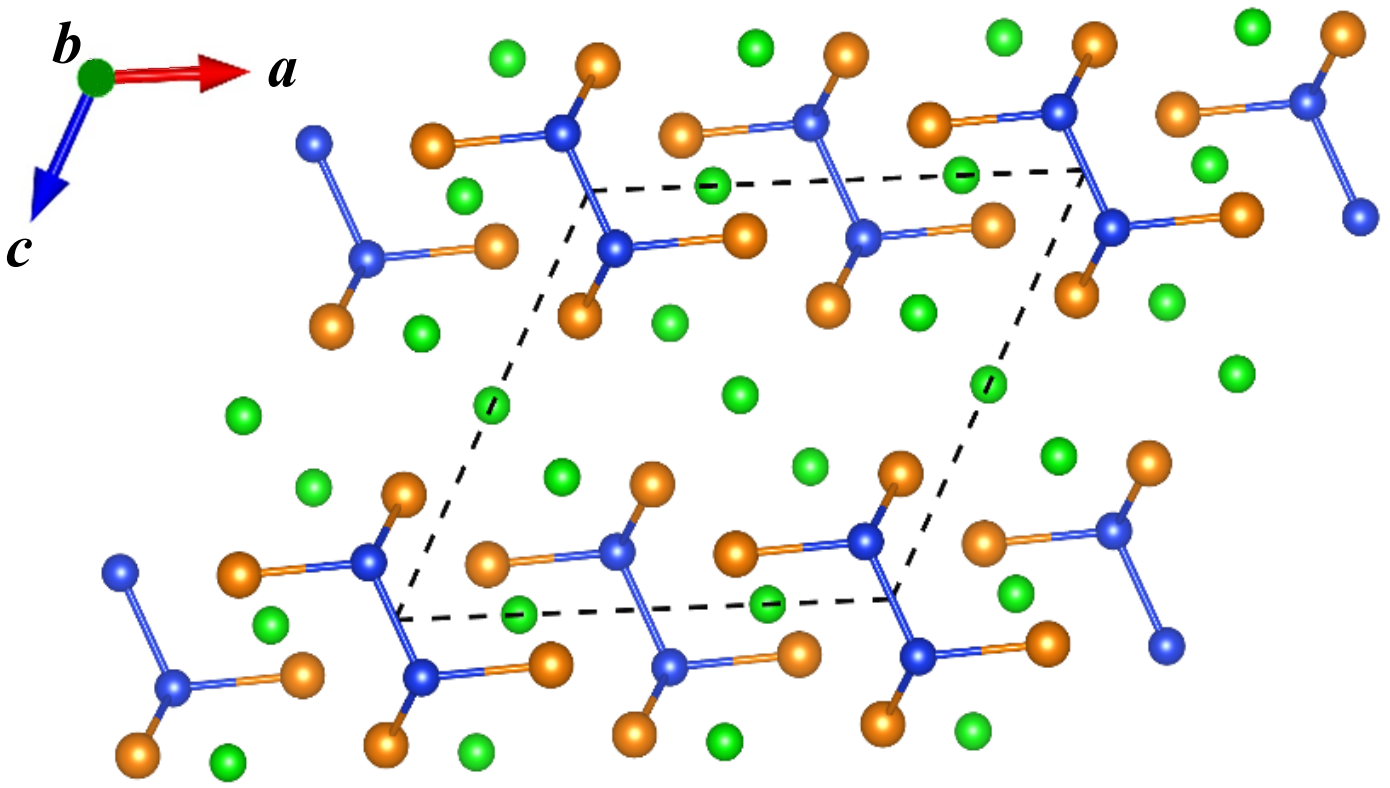

Рис. 7.7. Проекція кристалічної структури $K_6Si_2Te_6$ на площину (010) [217].

$K_6Si_2Te_6$ кристалізується в моноклінній структурі, симетрія якої описується просторовою групою *C2/m* з параметрами ґратки: $a$ = 9.652 (5), $b$ = 13.621 (8), $c$ = 8.902 (5) Å, β = 117.34 та $Z$ = 2 [216–218]. Основним структурним елементом $K_6Si_2Te_6$ є аніон $[Si_2Te_6]^{6-}$. На відміну від $Si_2Te_3$, де структурні одиниці $[Si_2Te_6]$ двомірно зв'язані між собою, утворюючи нескінченні подвійні шари (рис. 1.17, розділ 1), в $K_6Si_2Te_6$ ці дисилікатні групи ізольовані (рис. 7.7). Таким чином, кристалічна структура $K_6Si_2Te_6$ містить ізольовані октаедри $[Si_2Te_6]$ у ступінчастій (шаховій) конформації, з’єднані атомами K у спотворених октаедричних або тригонально-призматичних конфігураціях. Середня відстань Si–Te становить 2.510 Å, а відстань Si–Si 2.40 Å. Координація атомів Si є майже ідеально тетраедричною.

## 7.3. СИСТЕМА Cu–Si–Te

Фазова діаграма системи Cu–Si–Te досліджена авторами [219] методами диференціально-термічного, рентгенівського фазового та металографічного аналізів. Вихідні зразки отримували шляхом сплавлення суміші елементарних компонент (Cu, Si і Te) у відкачаних кварцових ампулах. Ця система має п’ять потрійних евтектик і дві зони розшарування в рідкому стані, одна з яких повністю розташована в потрійній. У системі Cu–Si–Te встановлено наявність тільки однієї потрійної сполуки $Cu_2SiTe_3$, яка плавиться інконгруентно при 851 К. За даними [220–222] ця потрійна сполука кристалізується в гранецентрованій кубічній ґратці (ПГ $F\bar{4}3m$ ) з параметром $a$ = 5.93 Å, Z = 1 (надструктура типу цинкової обманки). Пікнометрично визначена густина $\rho_{екс}$ = 5.47 г·см$^{-3}$, $\rho_{раз}$ = 5.69 г·см$^{-3}$ [220–222]. У роботі [223] вказується на існування моноклінної фази (ПГ $Cc$) $Cu_2SiTe_3$ з параметрами гратки $a$ = 12.86, $b$ = 6.07, $c$ = 9.05 Å, β = 99°, $\rho_{екс}$ = 5.90 г·см$^{-3}$, $\rho_{раз}$ = 5.96 г·см$^{-3}$.

## 7.4. СИСТЕМА $SiTe_2$ – $Ag_2Te$

**7.4.1 Діаграма стану системи $SiTe_2$–$Ag_2Te$**. Діаграма стану системи повністю не побудована. За результатами диференціально-термічного, мікроструктурного та рентгенофазового аналізів автори [224, 225] дослідили і побудували діаграму стану системи $Ag_2Te$–$SiTe_2$ тільки зі сторони збагаченого $Ag_2Te$ (рис. 7.8). Вихідні зразки сплавів отримували прямим синтезом з елементів Si, Te і Ag у відкачаних кварцових ампулах при 1073–1273 К. У системі встановлено утворення однієї тернарної сполуки $Ag_8SiTe_6$, яка плавиться конгруентно при 1143 К, і евтектику з $Ag_2Te$ при 1103 К та 90 мол.% $Ag_2Te$.

Сполука $Ag_8SiTe_6$ є триморфною і характеризується наявністю трьох поліморфних модифікацій γ, β і α. Стабільною при кімнатній температурі є гранецентрована кубічна γ-фаза. Сполука $Ag_8SiTe_6$ має два зворотних поліморфних перетворення. Температура фазового переходу гранецентрованої кубічної ґратки у низькотемпературну кубічну рівна 263 К (β ⇔ γ), а другий низькотемпературний фазовий перехід спостерігається при 195 К (α ⇔ β) [224, 225].

Результати дослідження впливу високого тиску ($P$ = 0–30 кбар) на температуру β ⇔ γ фазового переходу у $Ag_8SiTe_6$ приведені в роботі [226]. Високий тиск створювався в апараті типу поршень–циліндр. Залежність температури фазового переходу β ⇔ γ для $Ag_8SiTe_6$ від тиску ($P$ = 0 – 30 кбар) наведена на рис. 7.9 [226]. Як видно з даного рисунка, температура фазового перетворення β ⇔ γ в $Ag_8SiTe_6$ монотонно збільшується з підвищенням тиску.

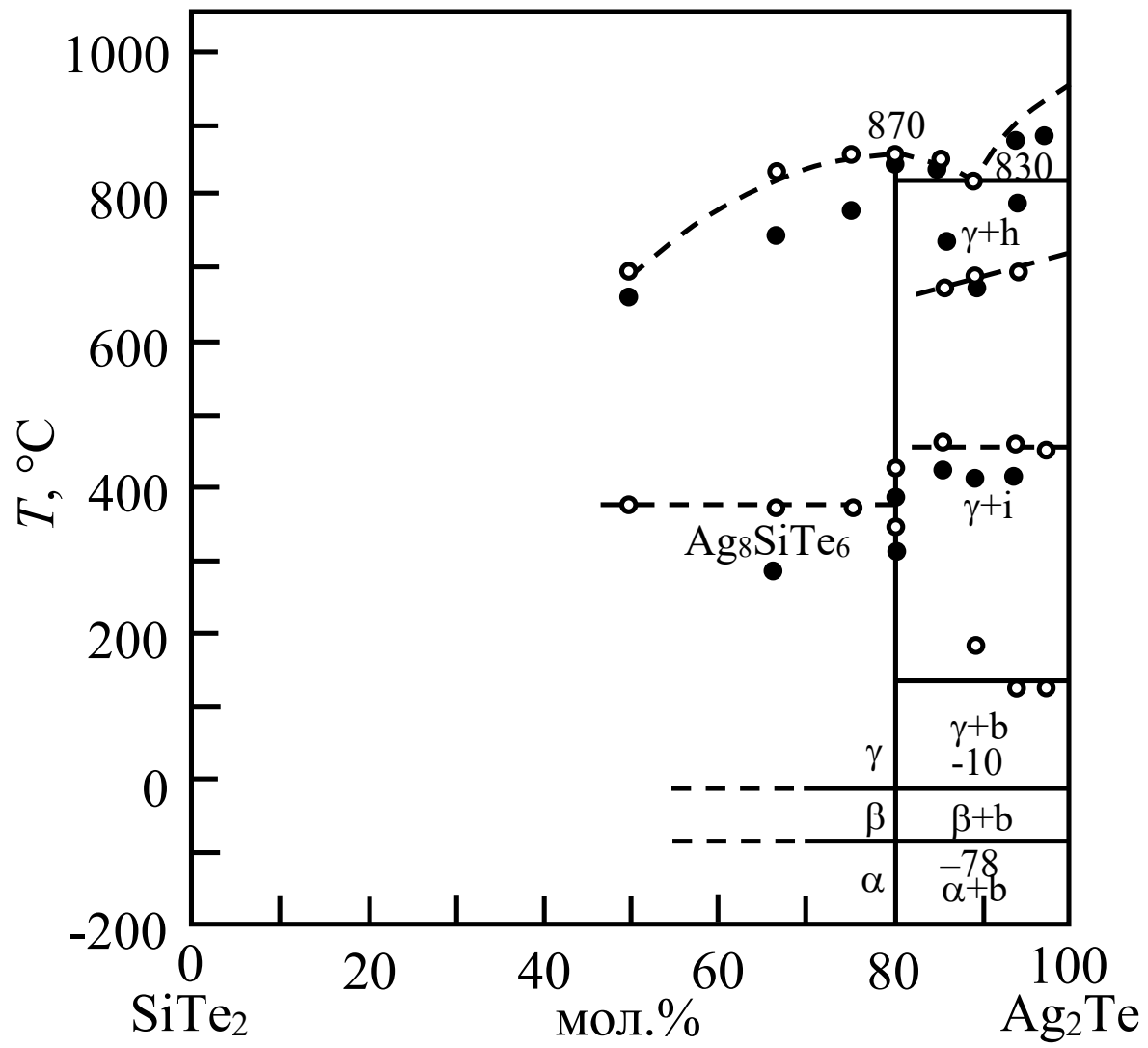

Рис. 7.8. Діаграма стану системи $SiTe_2$– $Ag_2Te$ в області 50–100 мол% $Ag_2Te$ [224].

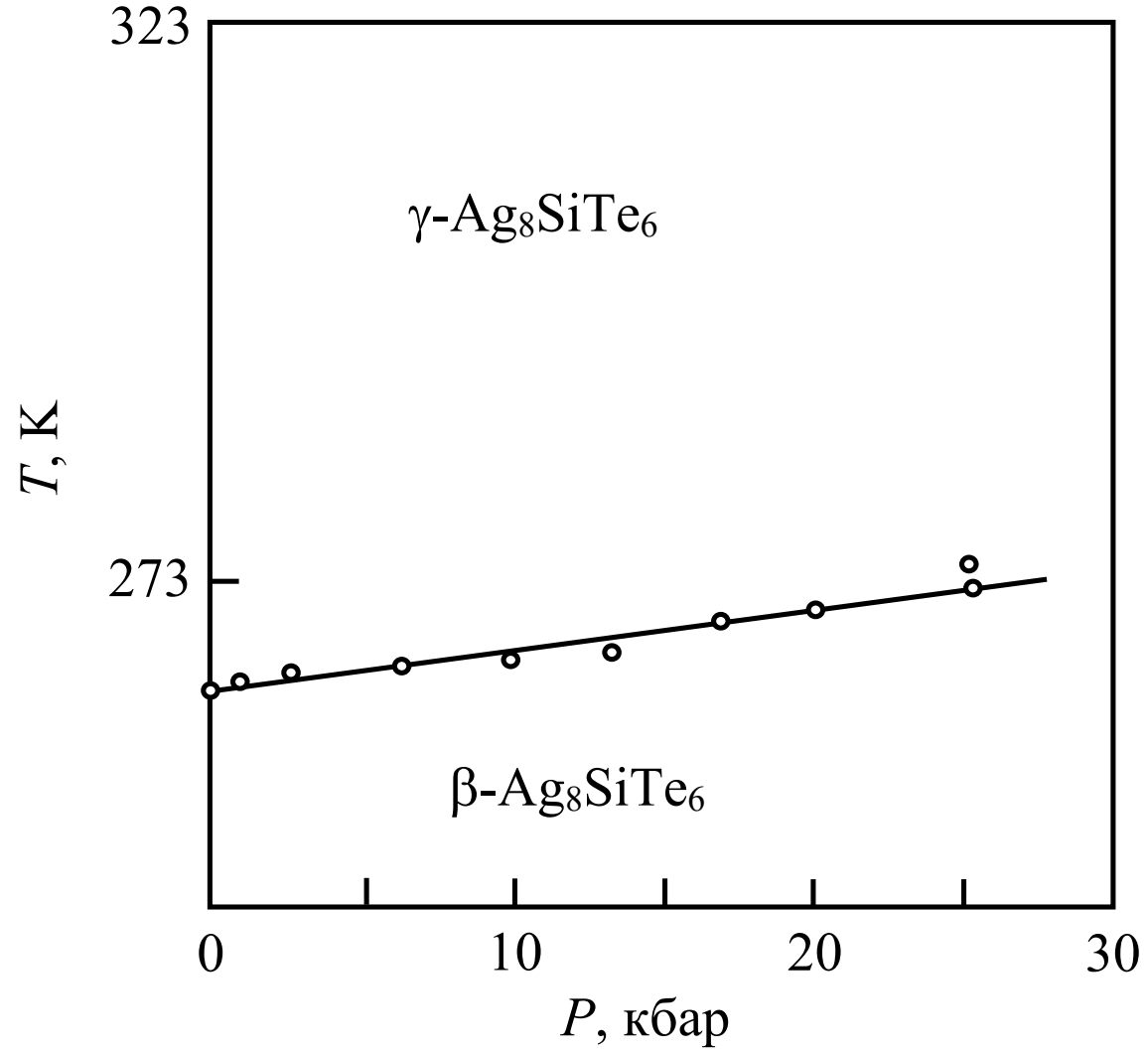

Ри. 7.9. Барична залежність температури β ⇔ γ фазового переходу в $Ag_8SiTe_6$ [226].

**7.4.2. Одержання і кристалічна структура γ-$Ag_8SiTe_6$.** Найбільш простим і поширеним способом синтезу бінарних та тернарних телуридів елементів першої групи та кремнію є сплавлення вихідних компонентів у відкачаних кварцових ампулах. У разі синтезу сполуки γ-$Ag_8SiTe_6$ такий спосіб є непридатним, оскільки при сплавленні стехіометричних кількостей вихідних компонентів (тобто Ag:Si:Te – 8:1:6), отримується неоднорідна фаза [227, 228]. Тому було запропоновано синтез сполуки γ-$Ag_8SiTe_6$ проводити шляхом прямого сплавлення нестехіометричного співвідношення 5:1:4 елементарних компонентів у вакуумованій кварцовій ампулі [228].

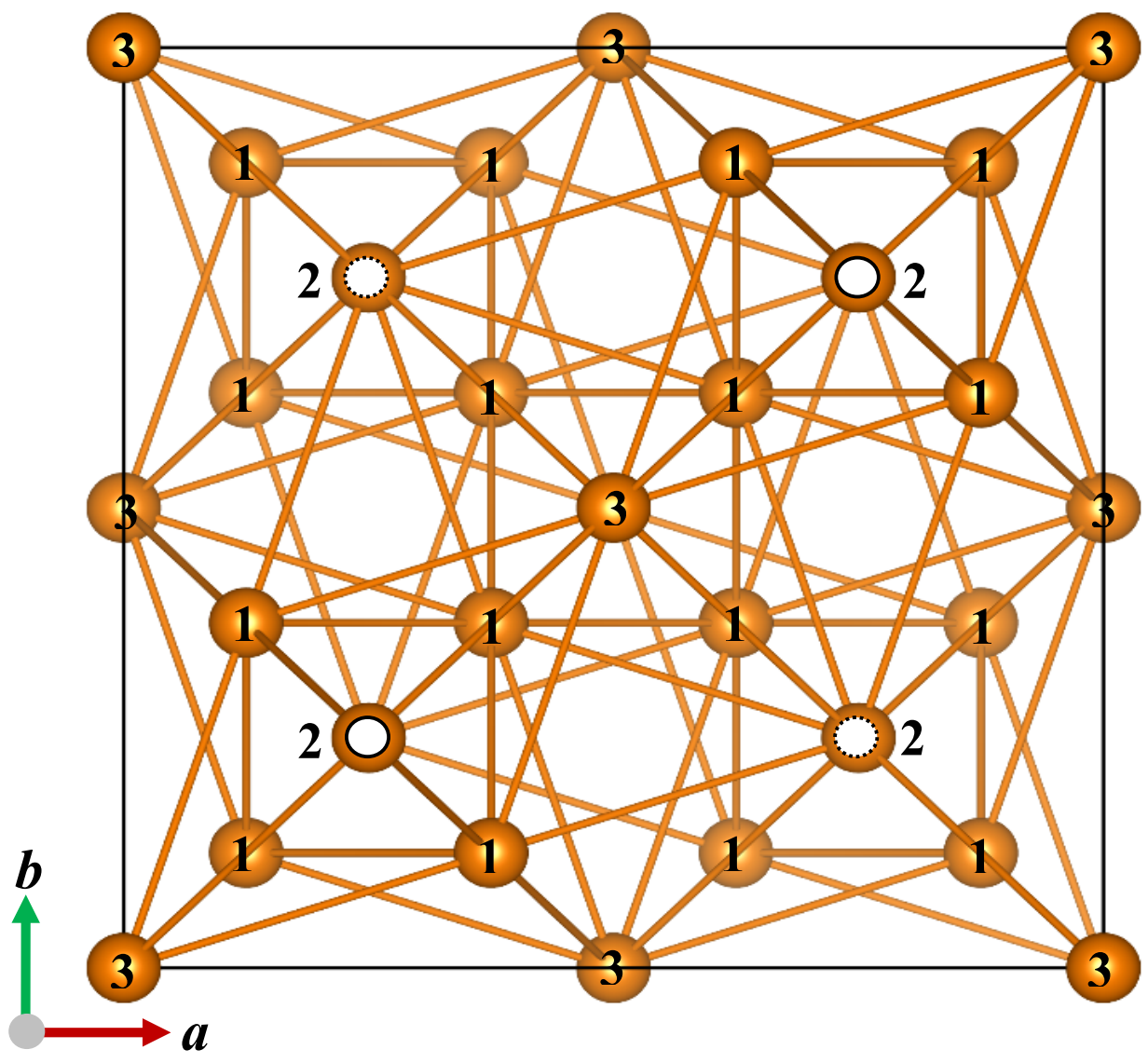

Рис. 7.10. Проекція каркасу атомів телуру у площині (110). Атоми кремнію (не зафарбовані кружки) також наведено [230].

Використовувалися два режими синтезу. Автори [226] нагрівали зразки до температури 1073 К протягом 7 днів, а потім охолоджували до 773 К протягом шести діб, із наступним охолодженням до кімнатної температури на протязі доби. Для отримання однофазного $Ag_8SiTe_6$ автори [227] також використовували нестехіометричний склад вихідних компонентів Ag:Si:Te – 5:1:4 і сплавляли їх при 1273 К на протязі 24 годин із наступним відпалом при 873 К протягом одного тижня.

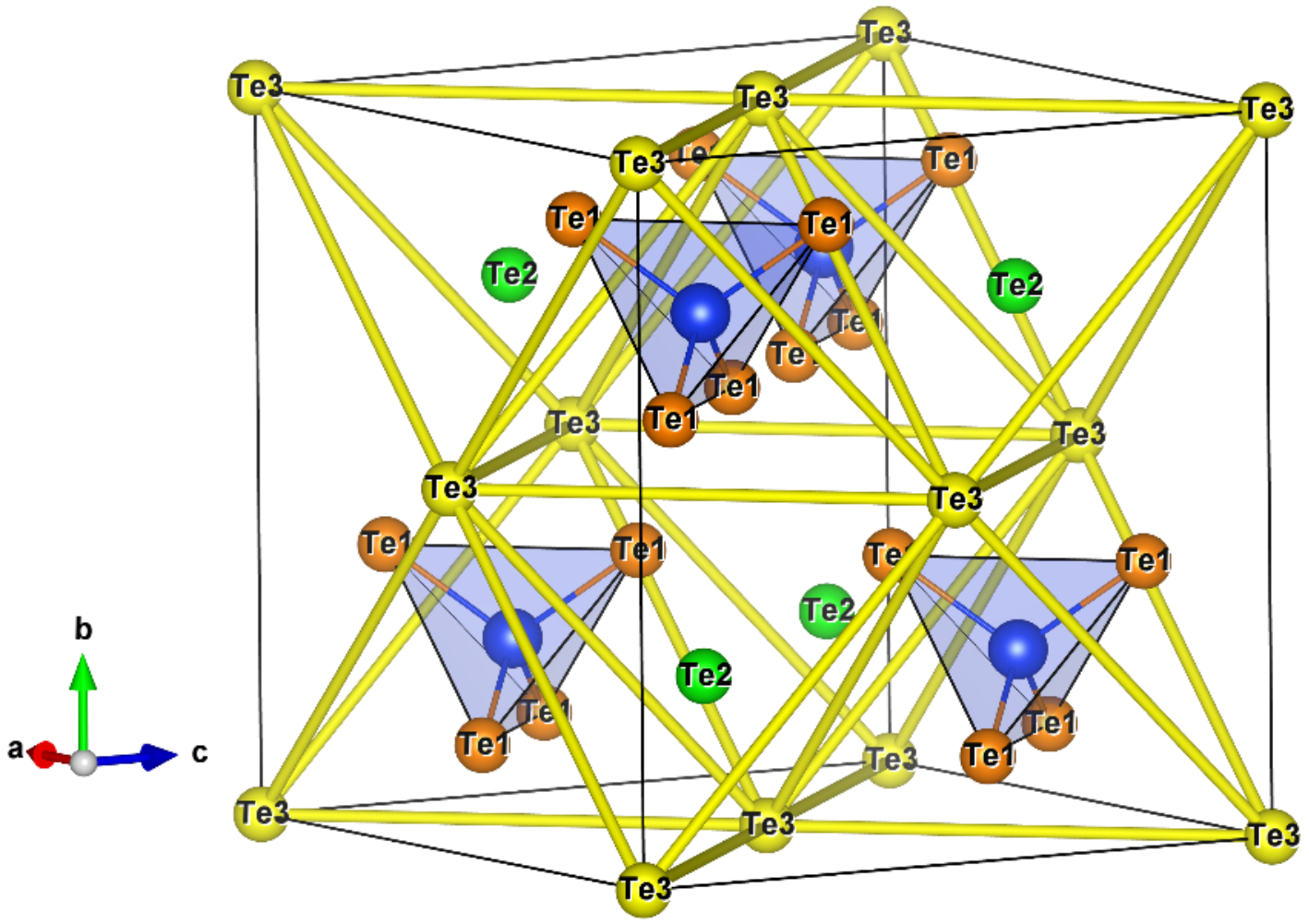

Рис. 7.11. Стереоскопічне зображення структури [$SiTe_6$] в γ-$Ag_8SiTe_6$ [227].

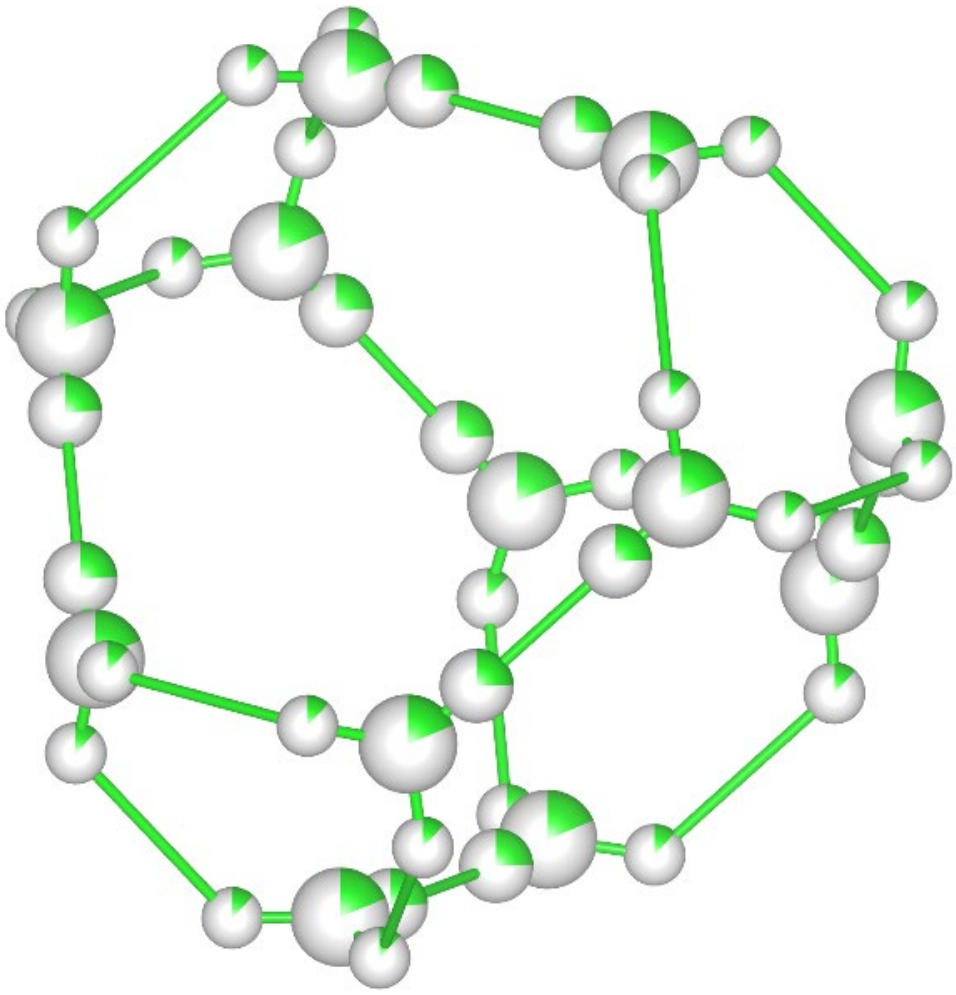

Рис. 7.12. Стереоскопічне зображення кластера срібла в γ-$Ag_8SiTe_6$ [227].

Структура γ-$Ag_8SiTe_6$ є похідною від структурного типу аргіродиту $Ag_8GeS_6$. Фаза γ-$Ag_8SiTe_6$, ізоструктурна $Ag_8GeTe_6$ [229, 230], кристалізується в гранецентрованій кубічній ґратці з параметром $a$ = 11.515 Å [225] і $a$ = 11.5225 Å [227], просторова група $F\bar{4}3m$ . Елементарна комірка містить чотири формульні одиниці (Z = 4), тобто 32 атоми срібла, 4 атоми кремнію та 24 атоми телуру. В елементарні комірці є три нееквівалентні атоми телуру в позиціях 16e (Te1), 4c (Te2) і 4a (Te3) (рис. 7.11). Атоми Te1 та атоми Si утворюють структурну одиницю – тетраедр [$SiTe_4$]. Аніонна ґратка структури утворена тетраедрами [$SiTe_4$] та атомами телуру. Атоми Te3 розташовані у вершинах примітивної елементарної комірки, тобто в кутах і посередині граней *F*-центрованої комірки, утворюючи таким чином 8 тетраедричних вузлів, з яких половина зайнята тетраедрами [$SiTe_4$], а друга половина – атомами Te2 (рис. 7.11). Атоми Ag розташовуються в тетраедричних пустотах і розподілені по двох 48-кратних (Ag1 і Ag2), однією 16-кратною (Ag3) та однією 4-кратною позицією з різним ступенем заповнення (рис. 7.12). Атоми Ag зміщені з центрів тетраедрів у бік зчленування граней. У структурі $Ag_8SiTe_6$ є велика кількість шляхів переміщення іонів Ag через структуру, що забезпечує високу іонну складову провідності.

Отже, як і у випадку $Ag_8SiS_6$ [231], структура $Ag_8SiTe_6$ побудована з тетраедрів [$SiTe_4$], зв'язаних через атоми срібла та телуру (Te2 і Te3). Цю фазу умовно можна записати як $Ag_8(SiTe_4)Te_2$.

Для сполуки $Ag_8SiTe_6$ характерним є поліморфізм: при високих температурах вона кристалізується у розупорядкованій відносно розподілу катіонів Ag кубічній модифікації, при низьких температур за рахунок упорядкування катіонів симетрія знижується до ромбічної або моноклінної.

**7.4.3. Електричні й термоелектричні властивості полікристалічного $Ag_8SiTe_6$.** Результати дослідження питомого опору (ρ), коефіцієнта Зеебека (S), коефіцієнта потужності ($S^2$/ρ) і теплопровідності (*κ*) полікристалічного $Ag_8SiTe_6$ в широкому інтервалі температур 300 – 850 K приведені в роботі [228] і представлені на рис. 7.13.

Питомий опір полікристалічного $Ag_8SiTe_6$ складає $10^{-3}$ – $10^{-4}$ Ом·м, що на один порядок більший ніж у типічних термоелектричних матеріалів, таких як $Bi_2Te_3$. Великі значення питомого опору $Ag_8SiTe_6$ обумовлені низькою концентрацією носіїв. Як видно з рис 7.13, *б* коефіцієнт Зеебека є додатнім, що вказує на те, що основними типами носіїв заряду є дірки. Питомий опір ρ і коефіцієнт Зеебека S з підвищенням температури зменшуються (рис. 7.13, *а*, *б*).

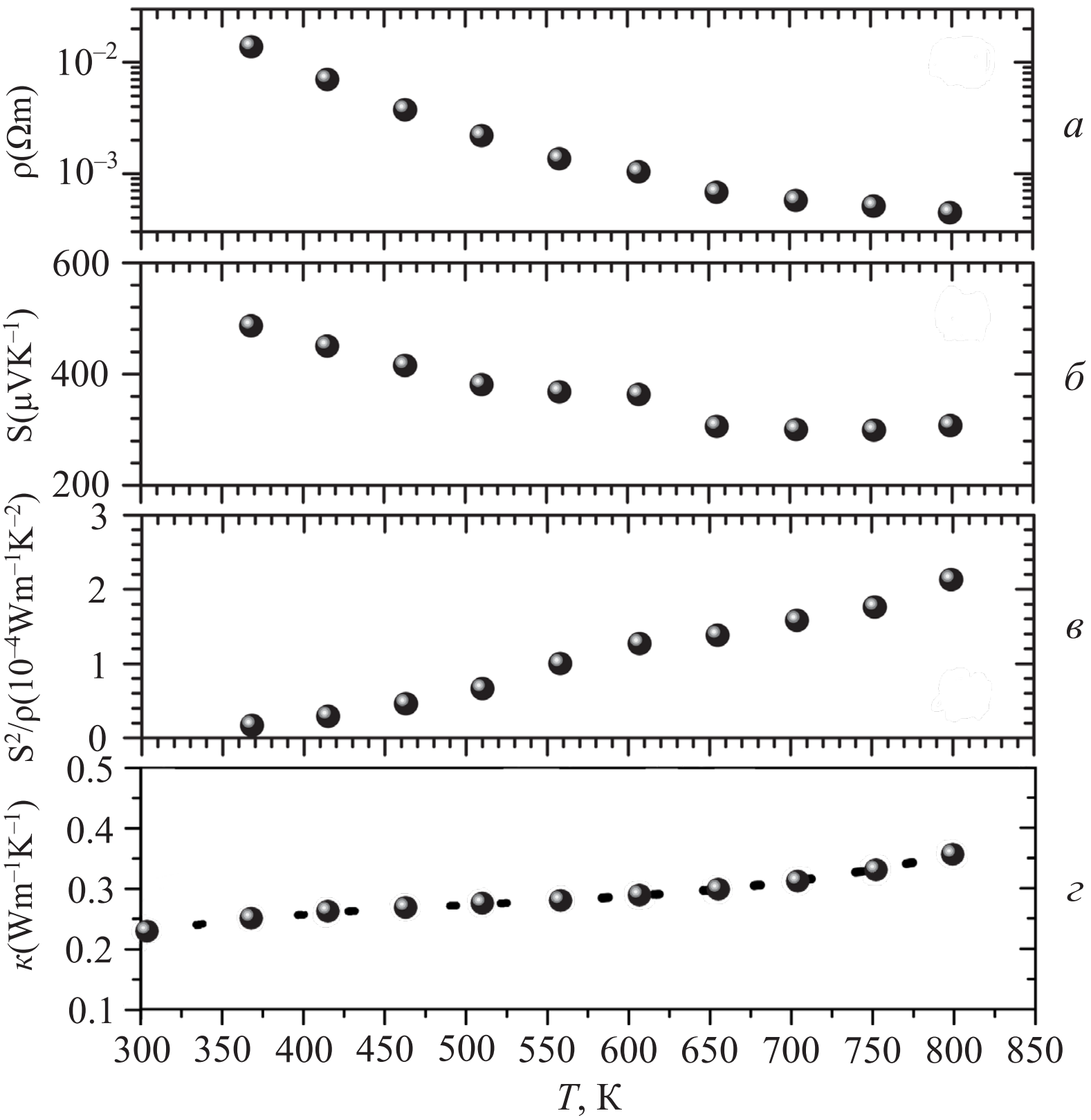

Рис. 7.13. Температурні залежності питомого опору (*а*), коефіцієнта Зеебека (*б*), коефіцієнта потужності (*в*) і теплопровідності (*г*) $Ag_8SiTe_6$ [228].

$Ag_8SiTe_6$ є напівпровідником *p*-типу з досить низькою теплопровідністю (*κ*). Максимальне значення коефіцієнта потужності досягає $0.21\cdot10^{-3}$ Вт·м$^{-1}$К$^{-2}$ при 800 К. $Ag_8SiTe_6$ демонструє відносно високі значення *ZT*, 0,48 при 800 К, завдяки своїй надзвичайно низькій теплопровідності ≈ 0.25 Вт·м$^{-1}$ К$^{-1}$ при кімнатній температурі. Низьке значення теплопровідності обумовлене складною кристалічною структурою $Ag_8SiTe_6$, в якій реалізується велика ступінь свободи в положенні атомів Ag. Отже, $Ag_8SiTe_6$ є високоефективним термоелектричним матеріалом з низькою теплопровідністю.

### 7.5. СИСТЕМИ M–Si–Te (M = Al, In)

У системах M–Si–Te (M = Al, In) встановлено наявність потрійних сполук $AlSiTe_3$ і $InSiTe_3$. Телуросилікати алюмінію ($AlSiTe_3$) та індію ($InSiTe_3$) належать до родини шаруватих телуридів кремнію $MSiTe_3$ (M = Al, In, Cr, Cs) з наявними кремнієвими димерами (Si–Si). Підвищений інтерес до шаруватих кристалів $InSiTe_3$ викликаний створенням на їх основі широкосмугових фотодетекторів [232]. Крім того, спеціально нелегований і легований фосфором $InSiTe_3$ є перспективним термоелектричним матеріалом [233]. Тому дослідження фізичних властивостей цих сполук є актуальною задачею. Особливе місце займають дослідження електронних і фононних станів як експериментально, так і теоретично.

**7.5.1. Кристалічна структура.** Телуросилікати алюмінію ($AlSiTe_3$) та індію ($InSiTe_3$), автори [234, 235] отримували безпосереднім сплавленням елементарних компонентів, взятих у стехіометричному співвідношенні, у відкачаних кварцових ампулах при 923 К на протязі двох тижнів. Подібним способом автори [232] також синтезували $InSiTe_3$. Вихідні порошки In, Si, Te у молярному співвідношенні 1:1:3 запаювали у вакуумовану кварцову ампулу, яку розміщували в піч. Спочатку температуру повільно підвищували до 1123 К і витримували при цій температурі протягом 1 доби після чого охолоджували до кімнатної температури в режимі вимкнутої печі. Отримані кристали $InSiTe_3$ мали блочну структуру з типовим розміром 4×4×2 мм$^3$. Структура синтезованих кристалів $InSiTe_3$ була визначена за допомогою рентгенівської дифракції.

Ізоструктурні сполуки $AlSiTe_3$ і $InSiTe_3$ кристалізуються у тригональній ґратці, симетрія якої описується просторовою групою $P\overline{3}m1$. Параметри гексагональної комірки рівні: $a$ = 6.834 Å, $c$ = 6.995 Å для $AlSiTe_3$ [234] та $a$ = 7.0411 Å, $c$ = 7.1001 Å для $InSiTe_3$ [235]. Число формульних одиниць для обох сполук рівне двом (Z = 2). Кристалічна структура потрійних сполук $MSiTe_3$ є двовимірною і побудована з тришарових пакетів-сендвічів, кожен з яких складається з двох моноатомних шарів телуру, між якими знаходяться димери (гантелі) кремнію (Si–Si) та катіони металу (Al, In), які займають октаедричні позиції [$Si_2Te_6$] та [$MTe_6$] у співвідношенні 1:2. Димери кремнію розташовані однин над одним уздовж напрямку осі $c$ (рис. 7.14). Октаедри [$Si_2Te_6$] утворені двома тригональними пірамідами [$SiTe_3$], які зв’язані вершинами з атомів кремнію (рис. 7.14, *а*).

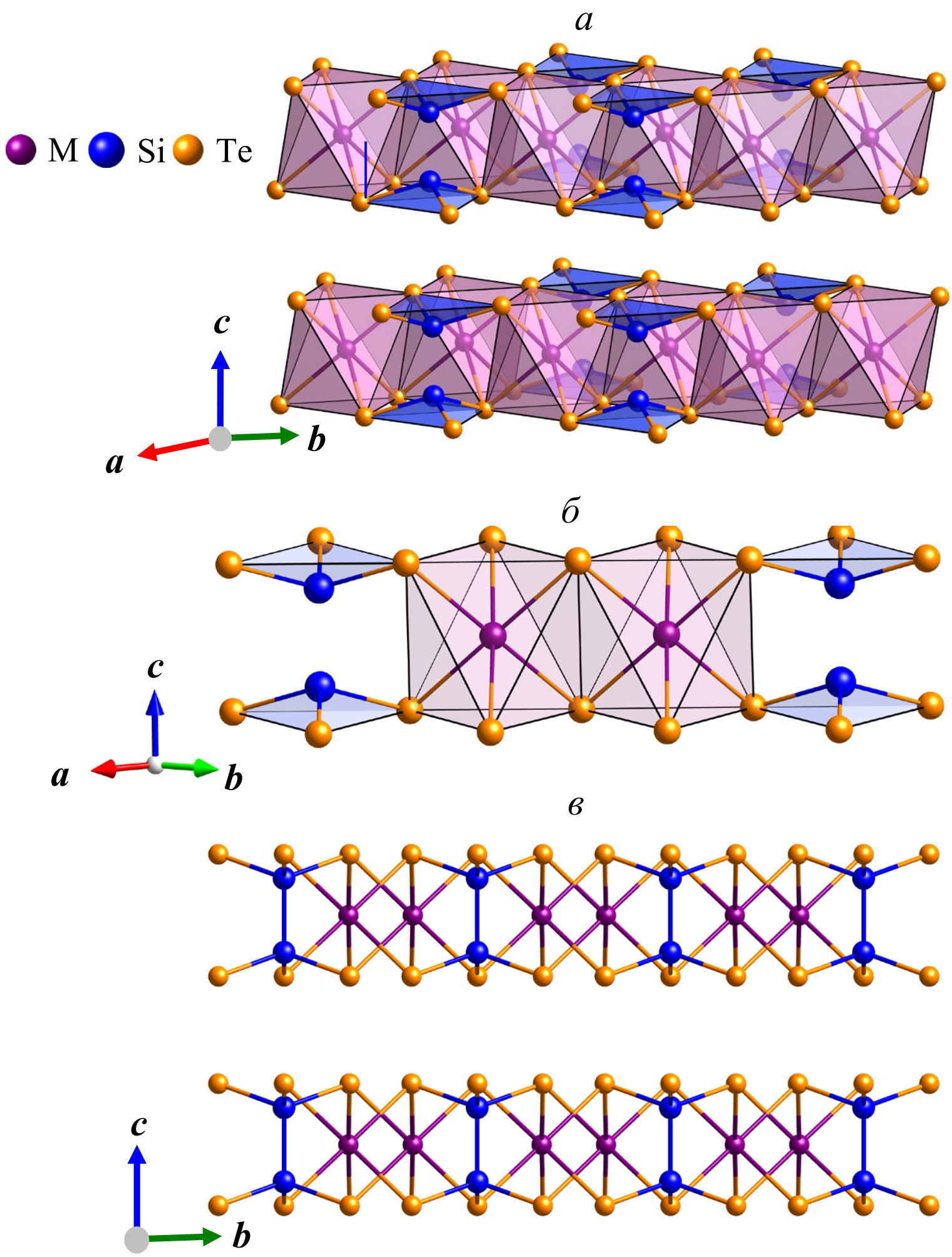

Рис. 7.14. Кристалічна структура (*а*) з виділеними двома октаедрами $[Si_2Te_6]$ та $[MTe_6]$ (*б*) та проекція структури на площину YZ (*в*) $MSiTe_3$.

Гратка моношарів $AlSiTe_3$ і $InSiTe_3$ є гексагональною, кожен атом кремнію з'єднаний із одним атомом кремнію та трьома атома ми телуру, тоді як атоми Al(In) знаходяться в центрі октаедрів $[MTe_6]$, утворених атомами телуру (рис. 7.14).

**7.5.2. Електронна структура $AlSiTe_3$ та $InSiTe_3$.** Електронні структури кристалів $AlSiTe_3$ та $InSiTe_3$, розраховані методом теорії функціоналу густини з використанням гібридного функціонала HSE06 без урахування спін-орбітальної взаємодії у точках високої симетрії та вздовж симетричних напрямків у незвідній частині зони Бріллюена (рис. 7.15), наведені на рис. 7.16 та 7.17 відповідно [236, 237]. За початок відліку енергії прийнято останній заповнений стан.

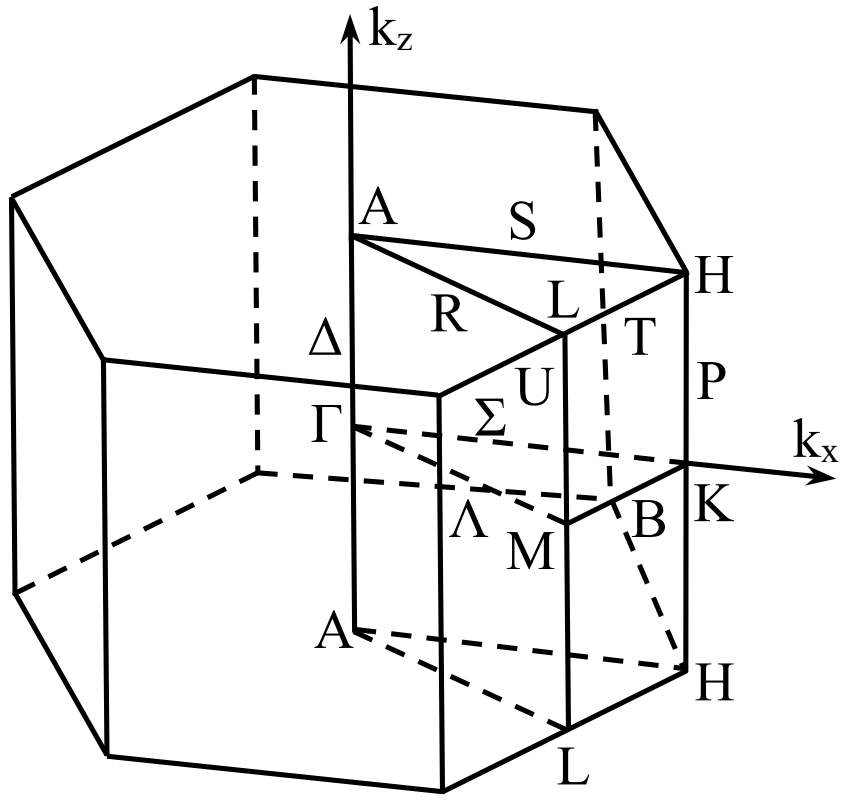

Рис. 7.15. Зона Бріллюена.

Як видно з рис.7.16 і 7.17, зонні структури ізоструктурних сполук $AlSiTe_3$ і $InSiTe_3$ демонструють кількісну і якісну подібність, яка виражена в топології та кількості заповнених енергетичних зон, що пояснюється в рамках моделі жорстких зон [238]. Згідно цієї моделі, для ізоструктурних сполук характерна подібна структура енергетичних зон, зумовлена однаковою атомною будовою даних сполук. Важливими кваліфікаційними принципами при аналізі електронної структури валетної зони кристалів є число валентних електронів, яке дозволяє встановити кількість дисперсійних віток, характер партнерів по хімічному зв'язку (визначається взаємне енергетичне розташування валетних підзон) і кристалічна структура речовини (показує на розщеплення станів, особливо верхніх підзон валетної зони) [143]. Оскільки, елементарні комірки $AlSiTe_3$ та $InSiTe_3$ містять 6 шестивалентних аніонів (Te), 2 тривалентних катіони (Al, In) і 2 чотиривалентних катіони (Si), то число валентних електронів в ЗБ рівне 50 і відповідно енергетичний спектр валетної зони формується з 25 енергетичних зон, об'єднаних у чотири зв'язки заповнених зон, розділених по енергії забороненими проміжками.

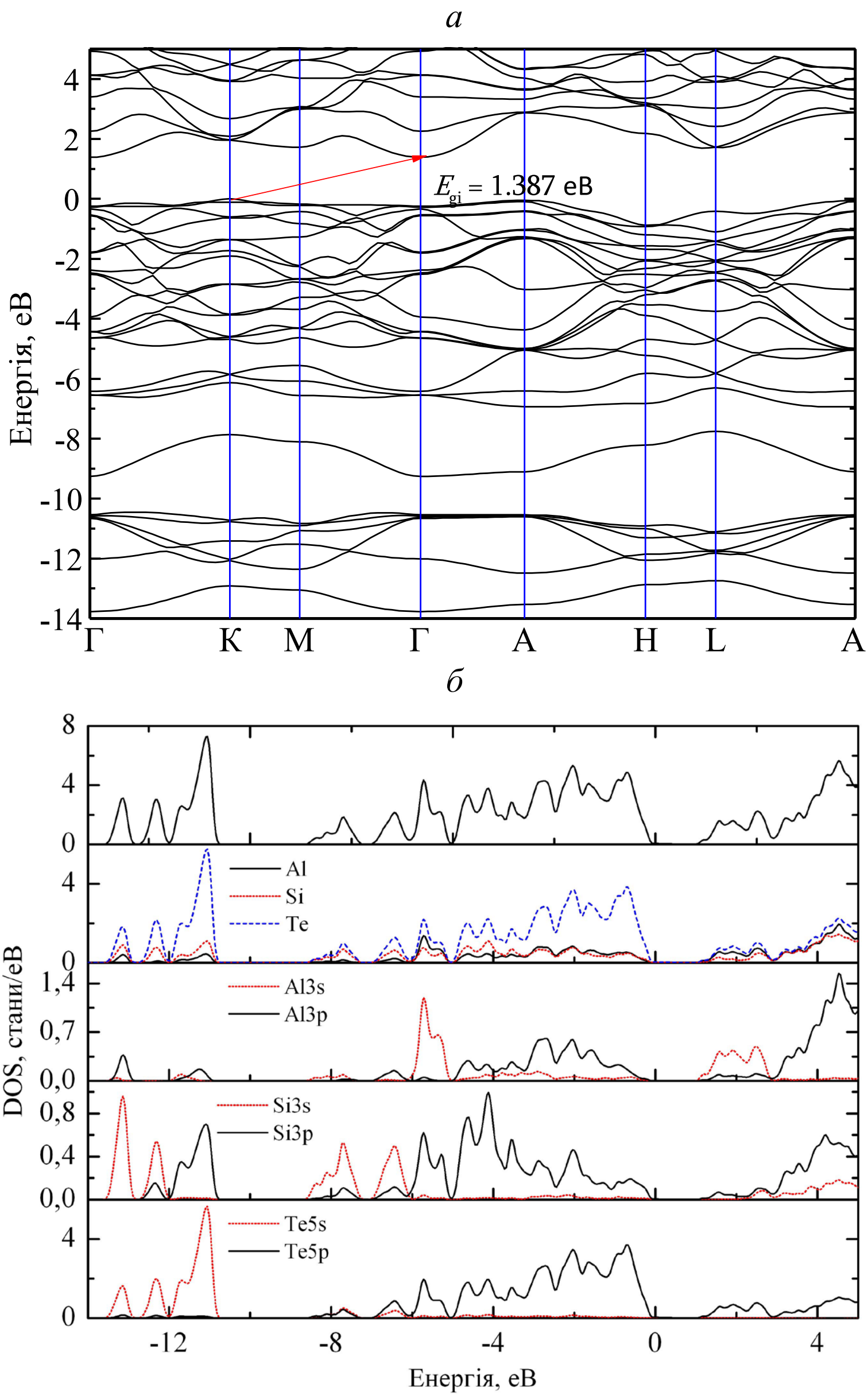

Рис. 7.16. Електрона структура (*а*), повна та локальні парціальні густини станів (*б*) $AlSiTe_3$ розраховані гібридним функціоналом HSE06 [236].

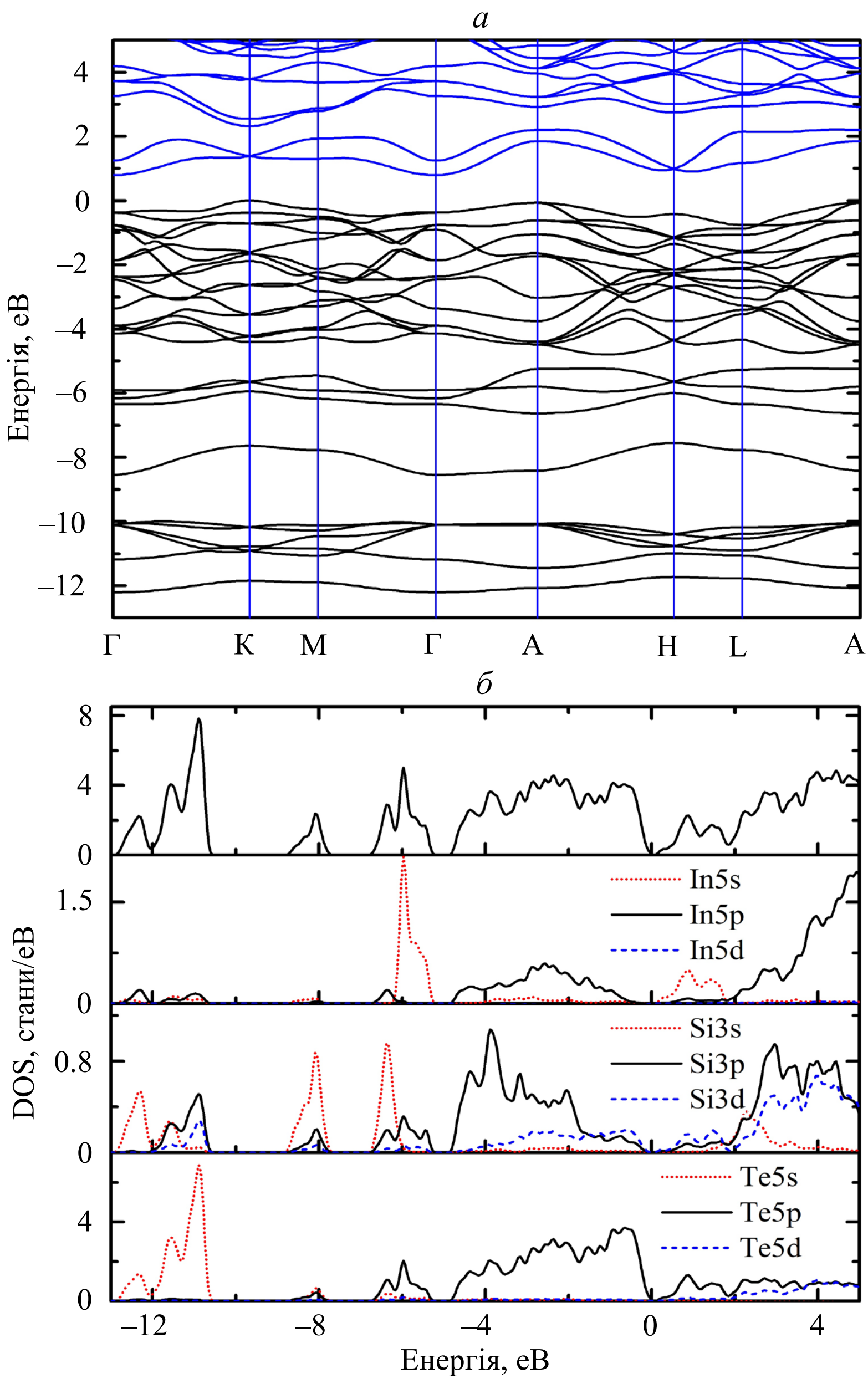

Рис. 7.17. Електрона структура (*а*), повна та локальні парціальні густини станів (*б*) $InSiTe_3$ розраховані гібридним функціоналом HSE06 [237].

Будова країв енергетичних зон – максимумів валетної зони і мінімумів зони провідності – визначає фундаментальні фізичні властивості напівпровідників і є досить чутливою до наближень, в яких проводяться розрахунки, особливо при наявності конкуруючих екстремумів. Як видно із рис. 7.16 і 7.17 краї енергетичних зон $AlSiTe_3$ та $InSiTe_3$ в околі забороненої зони характеризуються наявністю близьких максимумів і мінімумів. Згідно [236, 237] для обох об'ємних кристалаїв вершина валентної зони знаходиться в точці K, а дно зони провідності локалізовано в центрі зони Бріллюена. Таким чином, $AlSiTe_3$ та $InSiTe_3$ є непрямозонними напівпровідниками з розрахованими в HSE06 наближенні ширинами забороненої зони $E_{gi}$ = 1.39 еВ і $E_{gi}$ = 0.8 еВ, відповідно. Експериментально визначене значення $E_{gi}^{opt}$ з аналізу краю власного поглинання є тільки для об'ємного $InSiTe_3$ $E_{gi}^{opt}$ = 0.79 еВ [232]. Результати розрахунків ширин заборонених зон об'ємних і моношарів $AlSiTe_3$ та $InSiTe_3$ з використанням різних наближень приведені в табл. 7.1.

Таблиця 7.1. Значення ширин заборонених
зон об'ємних імоношарів $AlSiTe_3$ і $InSiTe_3$

| **Сполука** | **Форма кристала** | **Наближення** | **$E_{gi}$, еВ** | **Тип переходу** | **Література** |
|---|---|---|---|---|---|
| $AlSiTe_3$ | об'ємний | GGA + PBE | $E_{gd}$=1.24 | | [242] |
| $AlSiTe_3$ | об'ємний | HSE06 | 1.387 | K→Г | [236] |
| $AlSiTe_3$ | моношар | GGA + PBE | $E_{gd}$=1.41 | | [242] |
| $AlSiTe_3$ | моношар | HSE06 | 2.05 | M-Г→K-M | [239] |
| $AlSiTe_3$ | моношар | PBE | 1.34 | | [239] |
| $InSiTe_3$ | об'ємний | | 0.545 | K→Г | [241] |
| $InSiTe_3$ | об'ємний | HSE06 | 0.8 | K→Г | [237] |
| $InSiTe_3$ | об'ємний | HSE06 | 0.78 | A→Г | [232] |
| $InSiTe_3$ | моношар | PBE | 0.89 | K→ K-M | [239] |
| $InSiTe_3$ | моношар | HSE06 | 1.54 | K→ K-M | [239] |
| $InSiTe_3$ | моношар | HSE06 | 1.30 | Г→ K | [232] |

Зіставлення зонних спектрів ізоструктурних сполук $AlSiTe_3$ та $InSiTe_3$ дозволяє прослідкувати за змінами, які відбуваються при заміщенні атомів в катіонній підгратці. Згідно розрахунків повна ширина валентної зони даних сполук складає 13.8 еВ і 12.21 еВ для $AlSiTe_3$ і $InSiTe_3$ відповідно. Таким чином, заміщення Al→In в сполуках $MSiTe_3$ приводить до зменшення ширини забороненої зони, повної ширини валентної зони та верхньої і нижньої зв'язок валент-них зон. Істотно змінюється топологія краю зони провідності $InSiTe_3$, що проявляється в існуванні відокремленої підзони в низькоенергетичній частині спектра з двох незаповнених зон відокремлених забороненою енергетичною щілиною.

Для встановлення генезису кристалічних орбіталей із атомних станів елементів, які входять до складу кристала, були проведені розрахунки повної $N(E)$ і локальних парціальних густин станів з використанням результатів зонного розрахунку власних функцій $\psi_{i,k}(r)$ та власних значень енергій $E(k)$. Профілі розподілу повних густин, а також внески від окремих станів різних атомів для $AlSiTe_3$ та $InSiTe_3$ приведені на рис. 7.16, *б* і 7.17, *б* відповідно. Основні закономірності розподілу густин електронних станів однакові для обох сполук. Аналіз парціальних внесків у повну густину станів $N(E)$ дозволяє ідентифікувати генетичне походження різних підзон валентної зони і зони провідності $AlSiTe_3$ та $InSiTe_3$. Співвідношення між інтенсивностями максимумів у парціальних густинах станів різного типу симетрії різні. У глибині валентної зони обох сполук в повній густині електронних станів $N(E)$ домінує внесок 5*s*-станів телуру, тоді як у верхній частині валентної зони домінуючим є внесок 5*p*-станів атомів Te. Таким чином, нижня валентна підзона обох кристалів, сформована переважно 5*s*-станами телуру. Незважаючи на домінуючий характер Te5*s*-станів, для даної підзони істотними є ефекти гібридизації станів атомів кремнію (алюмінію, індію) і телуру, що приводить до появи внесків 3*s*-станів атомів кремнію, які виявляються в основному локалізованими в нижній частині цієї підзони і *s*-, *p*-, *d*-станів Si у її верхній частині.

Середня зв’язка з чотирьох розділених (1+3) зон, що слідує за 5*s*-станами телуру, відокремлена від них забороненим інтервалом енергій, не перекривається з самою верхньою валентною підзоною, формуючи таким чином дві ізольовані підзони. У формуванні відокремленої валентної зони в околі –8 еВ основний внесок дають Si 3*s*-стани. Наступна підзона з трьох зон в околі –6 еВ сформована Te 5*p*-, Si 3*s*-, 3*p* - і Al 3*s*-, In 5*s*-станами.

Найбільш складною є верхня підзона зайнятих станів, що складається з 15 дисперсійних віток. Самий верх цієї підзони, розташований безпосередньо поблизу вершини валентної зони, сформований переважно 5*p*-станами телуру з незначним домішуванням 3*p*-, 3*d*-станів кремнію. Нижня частина цієї підзони сформована гібридизованими 5*p*-станами телуру, *p*-станами кремнію та алюмінію (індію).

Електронна низькоенергетична структура незаповнених елек-тронних станів у цих сполуках формується в основному замішуванням вільних Te *p*-, *d*-, Si *s*-, *p*-, *d*- і Al(In) *s*-, *p*-станів, з основним внеском *p*-станів усіх атомів. Отже, аналіз повних та парціальних густин станів вказує на значну гібридизацію *s*- та *p*-станів атомів Si та Te, що свідчить про сильно іонно-ковалентний характер хімічного зв'язку Si–Te у координаційних октаедрах [$Si_2Te_6$] – структурних одиницях $AlSiTe_3$ та $InSiTe_3$, а основну роль в оптичних міжзонних переходах має відігравати перенесення заряду між Te 5*p*- зайнятими станами і Te *p*- + Si *s*-, *p*- незайнятими станами у зоні провідності.

**7.5.3. Просторовий розподіл валентного заряду.** Для опису природи хімічного зв'язку в кристалах дуже важливим є вивчення електронної густини, інформацію про яку можна отримати, побудувавши контури постійної густини. Оскільки електронна густини ρ(**r**) є функцією в тривимірному просторі, то зазвичай контури будують в найбільш актуальних площинах. Враховуючи, що основними структурними одиницями кристалів $AlSiTe_3$ і $InSiTe_3$ є октаедри [$Si_2Te_6$], утворені двома тригональными пірамідами [$SiTe_3$], з’єднаними вершинами з атомів кремнію (рис. 7.14, *а*) і октаедри [$MTe_6$], у цьому випадку найбільш зручно представити контурні карти ρ(**r**) у площинах, які проходять через два атоми телуру і один атом кремнію (рис. 7.18, *а*), вздовж димера $Si_2$ і двома атомами телуру (рис. 7.18, *б, в*), а також в площині, яка проходить через основу тригональної піраміди [$SiTe_3$] (моношар телуру) (рис. 7.18, *г*) в октаедрі [$Si_2Te_6$]; у площині, що проходить через шість атомів телуру і два атоми M в двох реберно-увязаних октаедрах [$MTe_6$] (рис. 7.18, *д*).

Із порівняння карт розподілу електронної густини заряду ізоструктурних сполук $AlSiTe_3$ і $InSiTe_3$ (рис. 7.18 і 7.19) видно, що вони якісно ідентичні по топології контурів густини ρ(**r**), а їх кількісні відмінності обумовлені різними параметрами гратки цих сполук. В октаедрах [$Si_2Te_6$] заряд валентних електронів розподілений переважно на атомах телуру з вираженою деформацією контурів у бік атомів кремнію. Яскраво виражена деформація контурів ρ(**r**) від атомів телуру в бік атомів кремнію уздовж лінії зв’язку Te–Si і наявність спільних контурів ρ(r), які охоплюють максимуми електронної густини на катіон-аніонних зв’язках (рис. 7.18, *а*, 7.19, *а*), відображають ковалентну складову хімічного зв’язку в октаедрах [$Si_2Te_6$], за формування якої відповідає гібридизація Si 3*s*-, 3*p*- і Te 5*s*-, 5*p*-станів. Поляризація зарядової густини у напрямку Si→Te вказує на наявність крім ковалентної ще й іонної складової зв’язку. Іонна

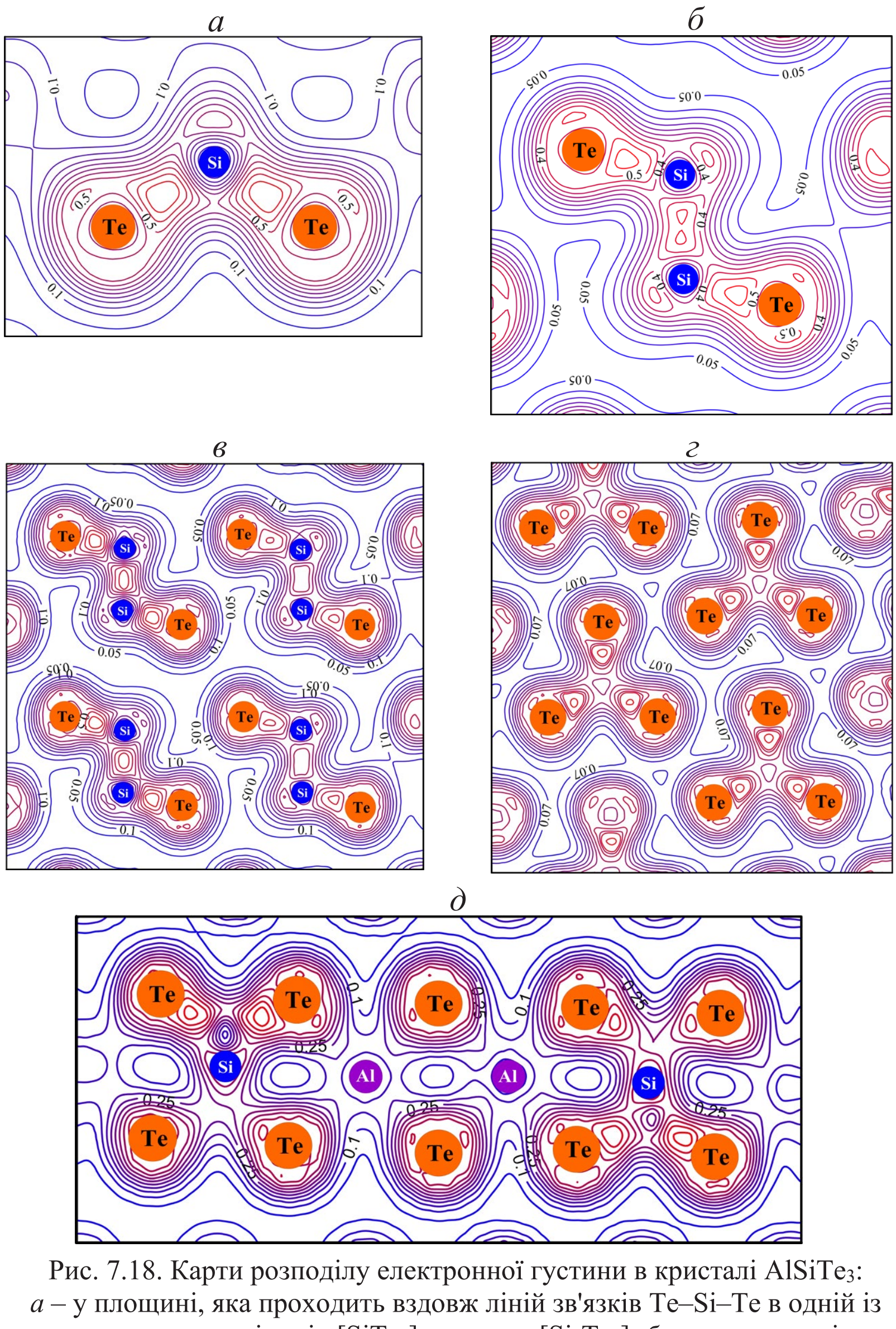

Рис. 7.18. Карти розподілу електронної густини в кристалі $AlSiTe_3$: *а* – у площині, яка проходить вздовж ліній зв'язків Te–Si–Te в одній із тригональних пірамід $[SiTe_3]$ октаедра $[Si_2Te_6]$; *б*, *в* – у площині, перпендикулярній тришаровим пакетам Te–Si–Si–Te, яка проходить через «вертикальні» (*б*, *в*) димери $Si_2$; *г* – у площині моношару телуру; *д* – в площині, яка проходить через грані ув’язаних спільними ребрами октаедрів $[Si_2Te_6]$ і $[AlTe_6]$.

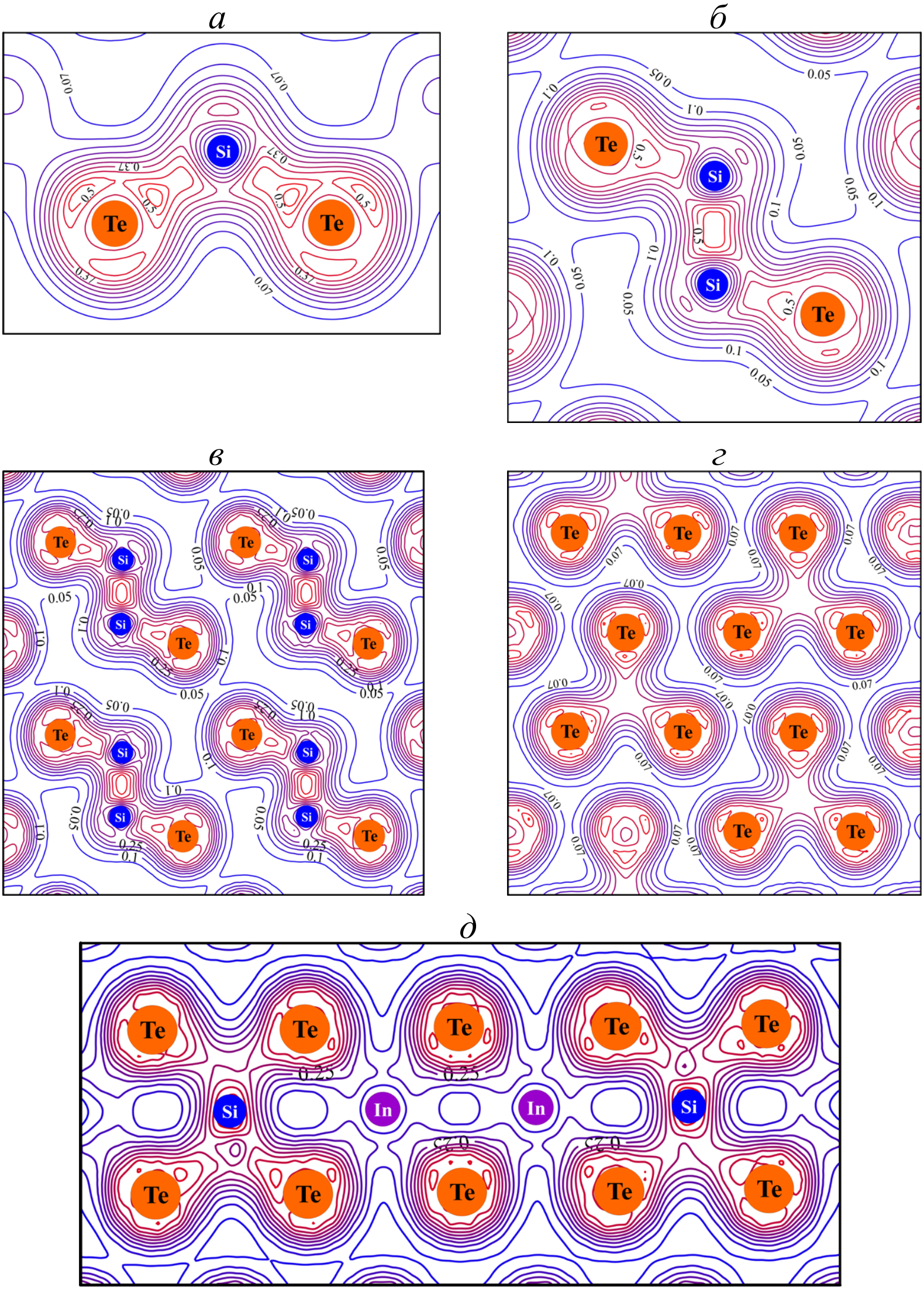

Рис. 7.19. Карти розподілу електронної густини в кристалі $InSiTe_3$:
*а* – у площині, яка проходить вздовж ліній зв'язків Te–Si–Te в одній із тригональних пірамід $[SiTe_3]$ октаедра $[Si_2Te_6]$; *б*, *в* – у площині, перпендикулярній тришаровим пакетам Te–Si–Si–Te, яка проходить через «вертикальні» (*б*, *в*) димери $Si_2$; *г* – у площині моношару телуру; *д* – в площині, яка проходить через грані ув’язаних спільними ребрами октаедрів $[Si_2Te_6]$ і $[InTe_6]$.

складова хімічного зв’язку характеризується: зарядами, локалізованими на самих атомах; асиметрією (поляризацією) ковалентного зв’язку, що проявляється у зміщенні зарядів на зв’язку і деформації контурів сталої (постійної) густини. На картах розподілу заряду валентних електронів чітко видно наявність максимумів заряду у вигляді замкнутих контурів на зв’язках Si–Si димерів Si2 (рис. 7.18, *б*, 7.19, *б*), які за формою такі ж самі, як і у випадку кристалічного кремнію (рис. 8.16, розділ 8).

У структурних одиницях, утворених за участю атомів Al(In), тобто в октаедрах [$AlTe_6$] і [$InTe_6$], основний заряд також зосереджений на атомах телуру (рис. 7.18, *д* і 7.19, *д*). Спільні контури, що охоплюють атоми катіона Al(In) та аніона Te, також характеризують ковалентну складову хімічного зв'язку в цих потрійних сполуках.

Розподіл електронного заряду між різними іонами телуру, які належать одному аніонному атомарному шару у тришаровому пакеті (сендвічі) передають карти, приведені на рис. 7.18, *г,* і 7.19, *г.* Із цих рисунків видно спільні лінії рівня ρ(**r**) між трьома атомами телуру в атомарному шарі телуру, що належать окремому октаедру [$Si_2Te_6$], що не властиве іншим шаруватим кристалам, які кристалізуються в структурі $CdI_2$, наприклад, $SnSe_2$ [147].

Крім того, валентна електронна густина має спільні контури для різних структурних одиниць, зв’язаних між собою через атоми телуру (рис. 7.18, *д*, 7.19, *д*). Проте характер деформації контурів на лініях зв'язку аніон–катіон навколо спільних атомів телуру, які з’єднують октаедр [$Si_2Te_6$] з [$AlTe_6$] або [$InTe_6$], істотно відрізняється. Так, уздовж лінії зв'язку контури навколо халькогену більш деформовані у напрямку атома кремнію, а ніж уздовж лінії зв'язку Te–In.

Аналіз контурних карт ρ(**r**) для даних потрійних сполук показує, що основна частина електронної густини локалізована на атомах телуру. Відмінність хімічної природи атомів Al, In, Si і Te визначає відмінність хімічних зв’язків Al–Te, In–Te і Si–Te. Зв’язки Al–Te та In–Te є більш іонними, а ніж зв’язок Si–Te. Усі три зв’язки мають ковалентну природу і формуються завдяки обмінному і донорно-акцепторному механізмів, що супроводжується $sp^3$-гібридизацією атомних станів. При цьому зв’язки Al–Te, In–Te є більш слабкими, а ніж Si–Te, ковалентний характер якого простежується по картам розподілу заряду наведених на рис. 7.18, *д* і 7.19, *д*.

Електронна густина всередині тришарових пакетів, що відображає хімічний зв'язок атомів кремнію (алюмінію, індію) з найближчими сусідами (атомами телуру) в октаедрах [$Si_2Te_6$] і [$AlTe_6$], [$InTe_6$], значно вища, ніж на їх межах. Не спостерігається спільних ліній рівня ρ(**r**) для сусідніх атомів телуру, що належать двом різним тришаровим пакетам, що свідчить про слабке перекривання їх хвильових функцій. Така просторова анізотропія електронної густини і енергетичного розподілу електронних 5*p*-станів телуру є причиною квазідвовимірності цих потрійних сполук.

**7.5.4. Фононні спектри $AlSiTe_3$ і $InSiTe_3$.** Фононний спектр є фундаментальною характеристикою кристала, яка визначає термодинамічні властивості матеріалу, кінетичні властивості носіїв заряду та оптичні властивості в інфрачервоній області. Такі характеристики фононного спектра, як функція густини фононних станів та фононні дисперсійні криві відображають специфічні особливості кристалічної структури та міжатомних взаємодій і дають важливу інформацію про динаміку кристалічної гратки. Як правило, відомості про фононні дисперсійні криві та функції густини фононних станів отримують з експериментів з розсіювання нейтронів.

Розраховані криві дисперсії фононів уздовж основних напрямків зони Бріллюена та густини фононних станів для кристалів $AlSiTe_3$ і $InSiTe_3$ приведені на рис. 7.20 і 7.21 відповідно [236]. Розрахунки проводились методом функціонала густини з використанням псевдопотенціалів і розкладу хвильових функцій по плоским хвилям. Обмінно-кореляційна взаємодія описується в наближенні локальної густини. В якості псевдопотенціалів використовувались нелокальні нормозберегаючі псевдопотенціали для атомів Al, In, Si і Te. Для обох сполук частоти фононів додатні в усій зоні Бриллюена, що свідчить про статичну і динамічну стабільність цих кристалів. Характер фононних спектрів для обох сполук у значній мірі подібний, що можна було очікувати виходячи з близькості їх кристалохімічної будови. Великі енергетичні щілини у фононних спектрах потрійних сполук $AlSiTe_3$ і $InSiTe_3$ є результатом великої різниці мас між атомами катіонів (Al, In) Si та аніона Te.

Із рис. 7.20 і 7.21 видно, що оптичні фонони в напрямку Г–А мають незначну дисперсію, що вказує на слабку фонон-фононну взаємодію в напрямку осі *c*. Однак у напрямку Г–М–К–Г і A–L–H–A має місце пересікання низькочастотних оптичних фононів з акустичними вітками фононів. Отже, у площині *xy* має місце значна фононфононна взаємодія. Попарне розташування фононних віток паралельно шарам є наслідком шаруватості кристала. Із-за шаруватої структури кристалів відбувається формування низькочастотних оптич-

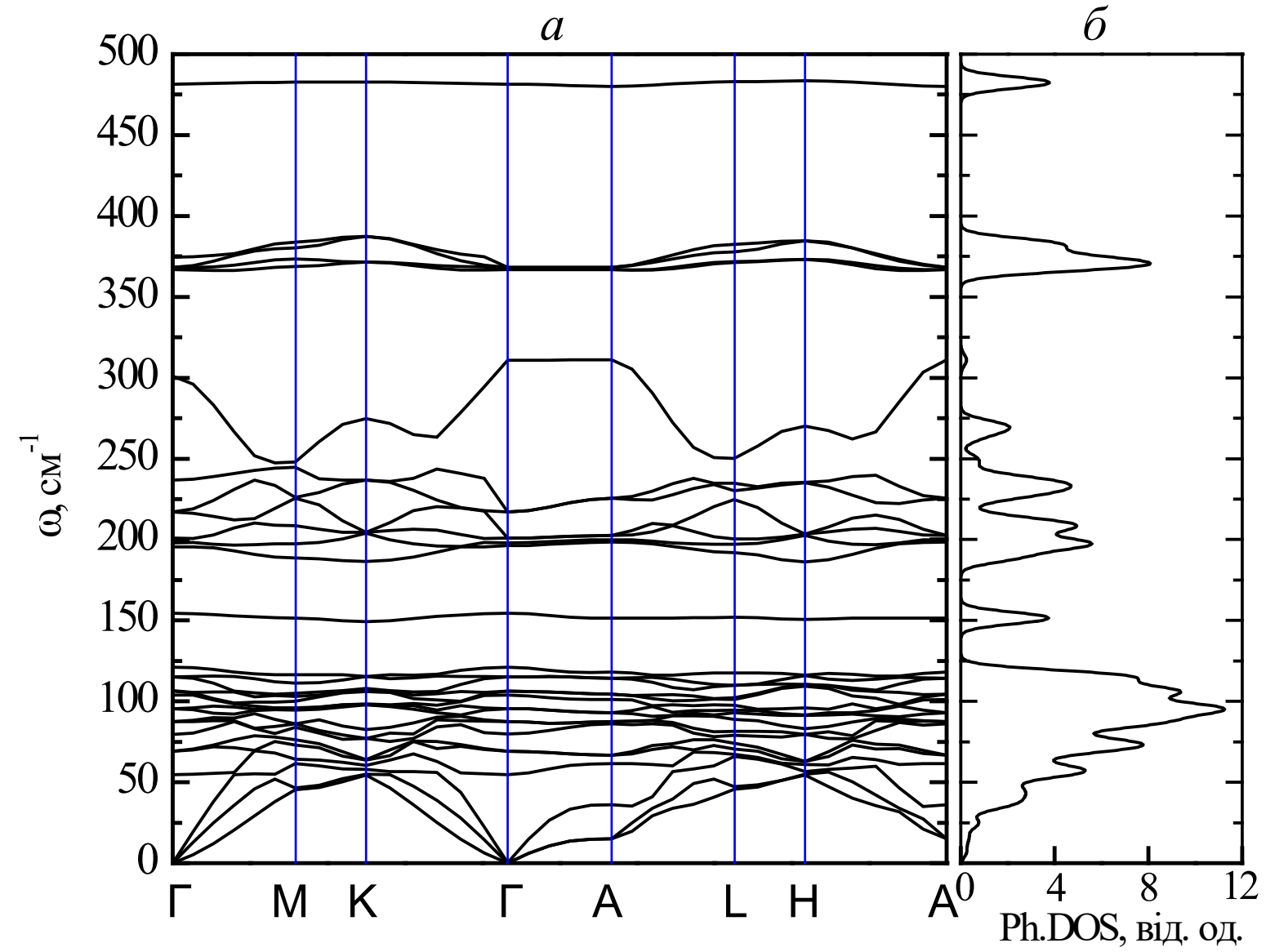

Рис. 7.20 Фононний спектр (*а*) та повна густина фононних станів (*б*) кристала $AlSiTe_3$.

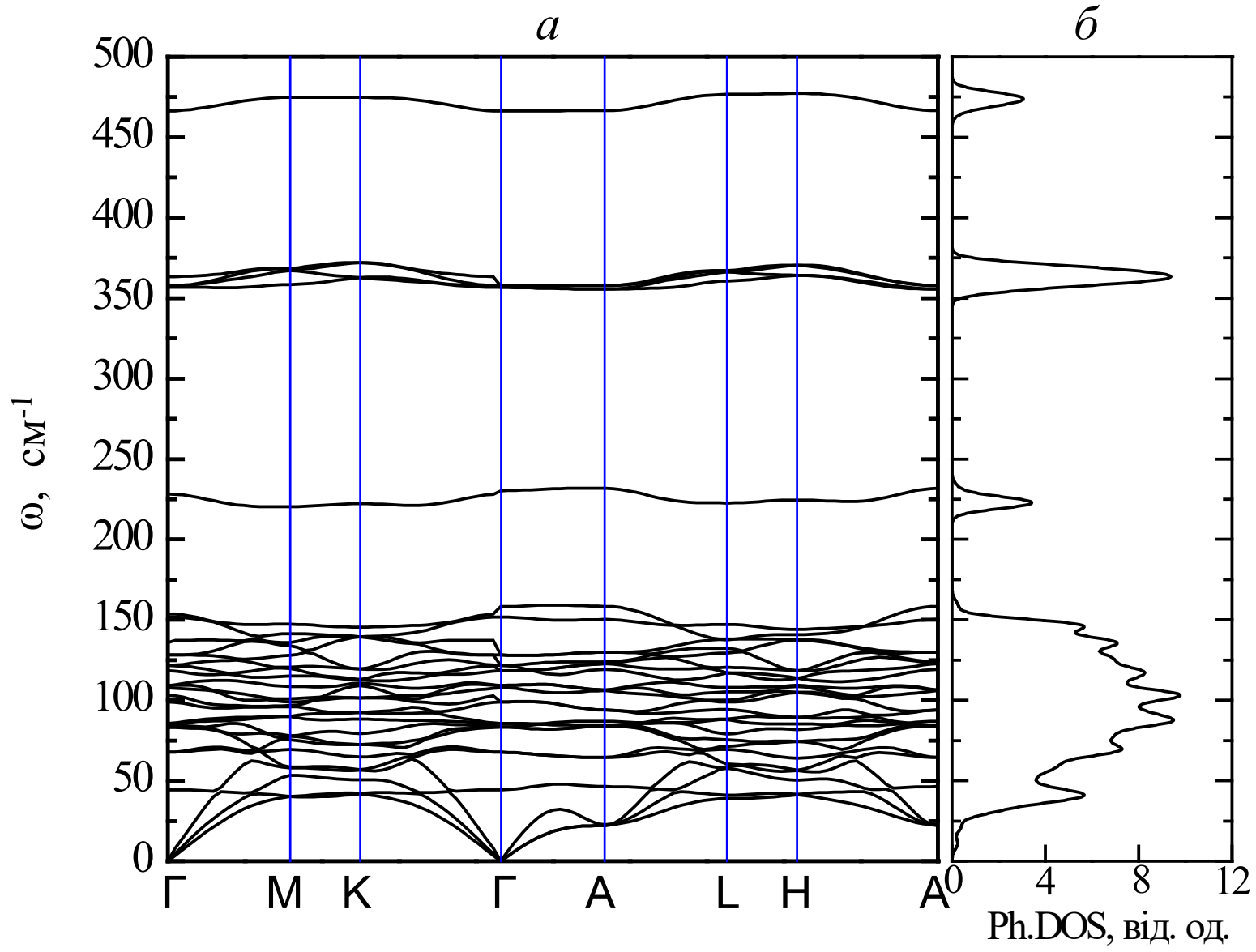

Рис. 7.21. Фононний спектр (*а*) та повна густина фононних станів (*б*) кристала $InSiTe_3$.

них мод, які відповідають коливанням шарів один відносно одного. Крім того, як випливає з розрахунків коливного спектра, має місце суттєва анізотропія для низькочастотних віток коливань. Вздовж напрямку сильного зв'язку Г–М нахил акустичних віток більш різкий, ніж для напрямку слабого зв'язку Г–А. Також спостерігається наявність низькочастотних оптичних віток, яким відповідають зміщення шарів один відносного іншого. Причому поздовжні акустичні вітки взаємодіють з цими низькочастотними вітками коливань. Високочастотні оптичні моди відокремлені в окремі групи, між якими наявні досить великі зонні щілини. При переході від Al-вмісної сполуки до In-вмісної відбувається незначне зниження частот верхніх коливних мод, що пов'язано зі збільшенням маси катіона. При цьому вид дисперсійних кривих змінюється тільки в центральній частині.

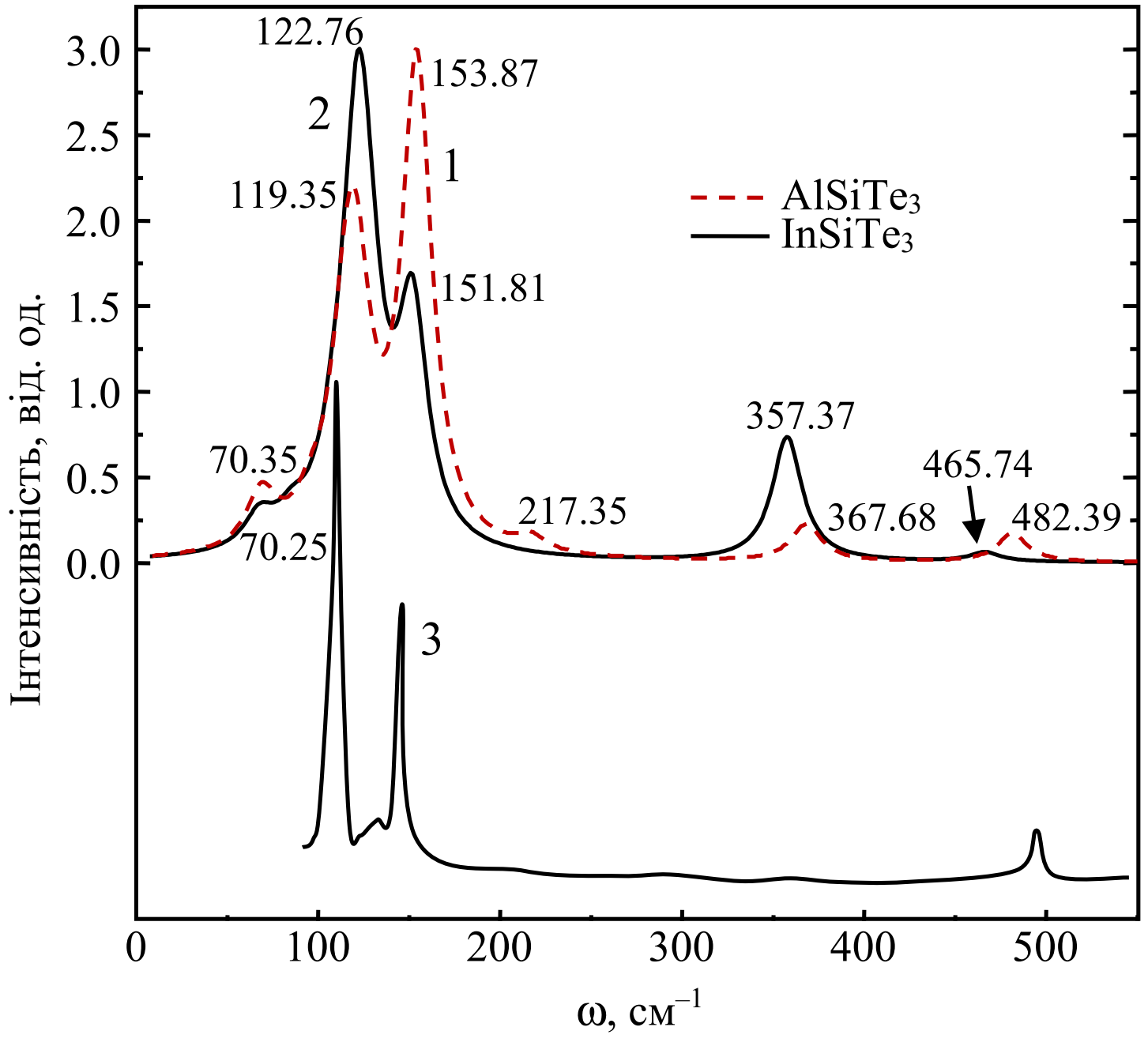

Рис. 7.22. Розраховані спектри КРС кристалів $AlSiTe_3$ (1) і $InSiTe_3$(2) [236] та експериментальний спектр КРС $InSiTe_3$(3) [232].

Як зазначалося вище, елементарна комірка $MSiTe_3$ містить 10 атомів, відповідно коливний спектр складається з 30 коливних мод і описується наступними незвідними зображеннями:

$$\Gamma = 5A_g+5A_u+5^1E_g+5^1E_u+5^2E_g+5^2E_u\,.$$

В спектрах комбінаційного розсіювання світла (КРС) повинно спостерігатись 15 оптично-активних мод симетрії $A_g$, $^1E_g$ і $^2E_g$, в ІЧ-спектрі – 12 мод симетрії $A_u$, $^1E_u$ і $^2E_u$. Кожна мода $E_g$ і $E_u$ двократно вироджена, а моди симетрії $A_g$ і $A_u$ – не вироджені. Одна мода симетрії $A_u$ та одна двократно вироджена мода $E_u$ є акустичними. Внаслідок наявності центра симетрії діє правило альтернативної заборони і в спектрах КРС не можуть проявляється ІЧ активні моди.

Розраховані спектри КРС кристалів $MSiTe_3$ наведені на рис. 7.22, а для кристала $InSiTe_3$ зіставлений з експериментальним спектром. У розрахованих спектрах КРС лінії розширено лоуренцевим контуром шириною 3 см$^{-1}$. Розраховані частоти (70.25, 122.76, 151.81, 357.37, 465.74) для $InSiTe_3$ у центрі зони Бріллюена та КРС спектри добре узгоджуються з експериментом.

## 7.6. ВИСОКОШВИДКІСНИЙ ШИРОКОСМУГОВИЙ ФОТОДЕТЕКТОР НА ОСНОВІ ШАРУВАТИХ КРИСТАЛІВ $InSiTe_3$.

Фотодетектори на основі 2D-матеріалів (2DM) із широкосмуговим фотовідгуком мають істотне значення для великої кількості застосувань, таких як багатохвильове фотодетектування, зображення та нічне бачення. Однак порівняно з традиційними фотодетекторами на основі об’ємного матеріалу, відносно низька швидкість фотодетекторів на основі 2D-матеріалів перешкоджає їх практичному застосуванню. Автори [232] розробили субмікросекундний фотодетектор на основі потрійного телуриду $InSiTe_3$ з тригональною симетрією та шаруватою структурою. Фотодетектори на основі $InSiTe_3$ демонструють надшвидкий фотовідгук (545–576 нс) і широкосмугові можливості виявлення від ультрафіолетового (УФ) до ближнього інфрачервоного (ІЧ) діапазону оптичного зв'язку (365–1310 нм). Крім того, фотодетектор демонструє видатний оборотний та стабільний фотовідгук, при якому характеристики відгуку залишаються незмінними протягом 200 000 циклів роботи перемикача. Ці значення свідчать про те, що $InSiTe_3$ є перспективним кандидатом для створення широкосмугових оптоелектронних пристроїв на основі 2D-матеріалів.

Фотодетектори були виготовлені нанесенням пластинчастих кристалів $InSiTe_3$ товщиною 300 нм на підкладки $SiO_2/Si$ за допомо-

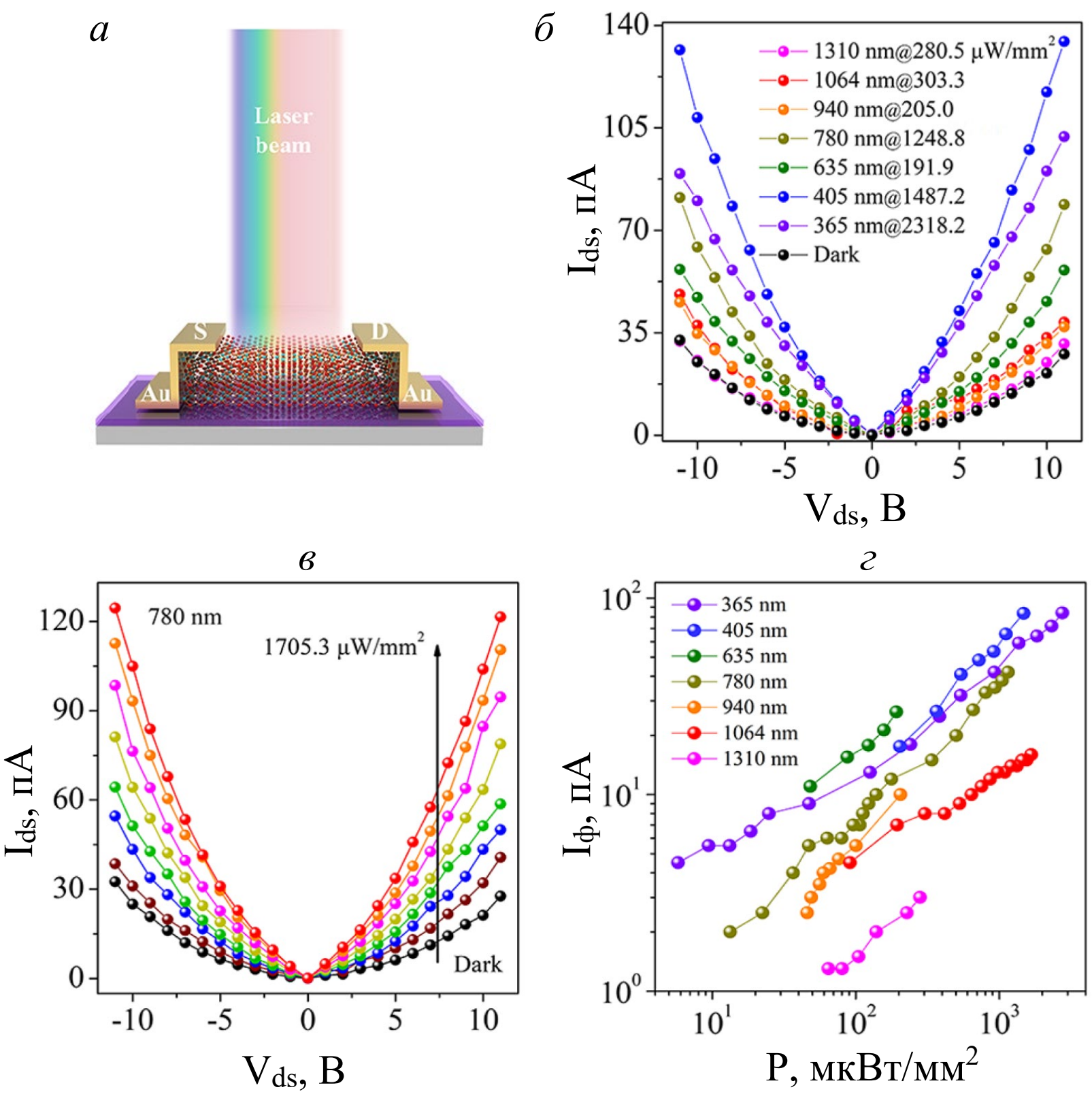

Рис. 7.23 *а* – Принципова схема фотодетектора на основі $InSiTe_3$ [232];
*б* – світлові ВАХ при різних довжинах хвиль від 365 до 1310 нм;
*в* – ВАХ при опроміненні світлом 780 нм різної інтенсивності;
*г* – залежність фотоструму від густини потужності падаючого світла.

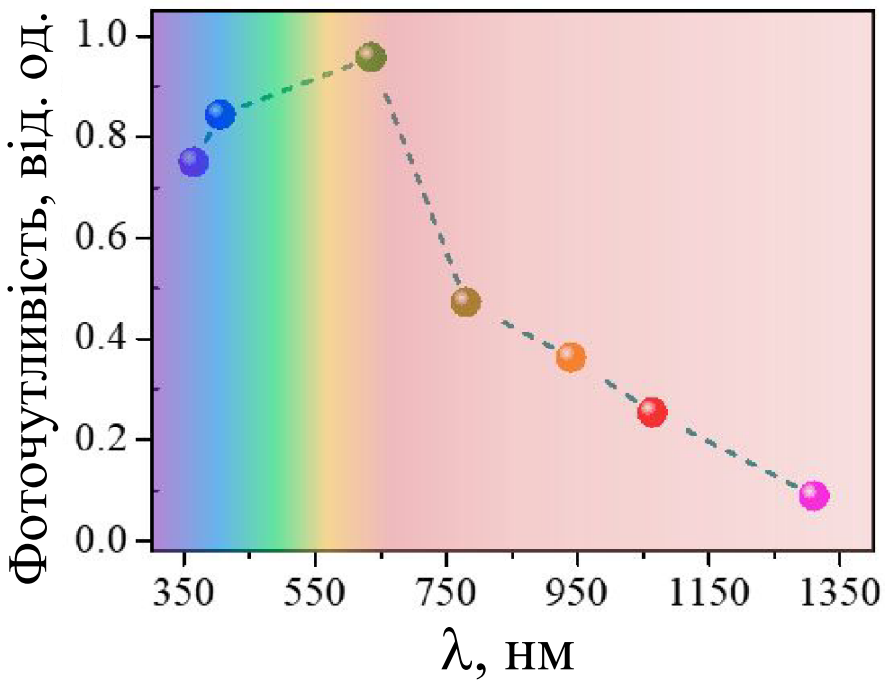

Рис. 7.24. Спектральна чутливість фотодетектора на основі $InSiTe_3$ [232].

гою технології перенесення електродів (рис. 7.23, *а*). Світлові вольт-амперні характеристики $I_{ds}$–$V_{ds}$ фотодетектора, виміряні при різних довжинах хвиль падаючого світла (365, 405, 635, 780, 940, 1064 і 1310 нм) наведені на рис. 7.23, *б*, демонструють широкий спектральний діапазон від УФ до ближнього інфрачервоного діапазону.

ВАХ $I_{ds}$–$V_{ds}$ фотодетектора, виміряні при різних густинах потужності падаючого світла (780 нм) показані на рис.7.23, *в*. Зі збільшенням густини потужності $P$ падаючого світла від 63,4 до 1705,3 мкВт/мм$^{-2}$, фотострум зростає від 13,0 до 93,8 пА, що викликано генерацією нерівноважних електронно-діркових пар в кристалі $InSiTe_3$ під час інтенсивного освітлення. Фотодетектори на основі шаруватих кристалів $InSiTe_3$ демонструють сублінійну степеневу залежність $I_ф \sim P^{\alpha}$ ($\alpha$ = 0.8), яку можна пояснити складними процесами генерації, захоплення та рекомбінації носіїв у таких детекторах.

Залежності фотоструму $I_ф = I_{освітл} - I_{темн}$ фотодетектора від потужності падаючого світла, виміряні при різних довжинах хвиль (365 – 1310 нм), наведені на рис. 7.23, *г*. Із даного рисунка чітко видно, що 2D фотодетектор на основі $InSiTe_3$ виявився спектрально селективним до падаючого світла (рис. 7.24).

# РОЗДІЛ 8

## КРЕМНІЄВІ КЛАТРАТИ: $Si_{46}$, $Si_{136}$, $Na_8Si_{46}$, $Na_{24}Si_{136}$, $Te_{16}Si_{38}$ і $Te_{7+x}Si_{20-x}$ ($x$~2.5). СИНТЕЗ, СТРУКТУРА ТА ВЛАСТИВОСТІ

**Блецкан Д. І., Гапак А. І.**

Клатрати – це сполуки з тривимірною каркасною структурою («хазяїна»), у порожнинах якої розташовані атоми або іони («гостей»), які не утворюють ковалентних зв'язків з каркасом [243]. Кристалічна гратка «хазяїна» не може існувати у відсутності «гостя». Для стабілізації каркаса необхідно заповнити хоча б невелику частину порожнин, тому клатрати отримують кристалізацією «хазяїна» і «гостя». Частинки, які заключні у тривимірну матрицю, як правило, не можуть її покинути без руйнування всієї архітектури. Звідси і пішла назва «клатрат» (від латинського clathratus – захищений решіткою).

Клатратні кристали кремнію відомі з 1965 року, коли вперше був синтезований клатрат $Na_8Si_{46}$ [244]. Однак тільки останніми роками вони стали об'єктом інтенсивних експериментальних та теоретичних досліджень. Такий інтерес викликаний насамперед активним пошуком нових напівпровідникових матеріалів. Оскільки кремній є найпоширенішим і доступним напівпровідником, багато досліджень спрямовано на отримання нових наноформ саме кремнію. При синтезі кремнієвих клатратних сполук існувала проблема стабілізації кремнієвої клатратної структури. Вирішити її вдалося шляхом впровадження в порожнини напівпровідникових грат атомів лужних і (або) лужноземельних металів [245, 246]. При цьому виявилося, що властивості кремнієвих клатратів визначаються атомом металу, який заповнює порожнини кремнієвої гратки. Так, клатрати $M_8Si_{46}$ (M = Na, K) є термоелектричними матеріалами, володіють гранично низькою теплопровідністю [247]. Часткова заміна атомів Na на атоми Ba призводить до появи надпровідності. Клатрат $Na_xBa_{8-x}Si_{46}$ стає надпровідником при температурі 2–4 K, залежно від співвідношення концентрацій атомів Na і Ba [248]. При заміні всіх атомів Na на атоми Ba ($Ba_8Si_{46}$) температура надпровідного переходу підвищується до 8 K [246]. Таким чином, змінюючи сорт легуючих атомів металу, можна синтезувати кремнієві клатрати із заданими властивостями. Однак вивчення впливу різних домішок на властивості кремнієвих клатратних сполук неможливе без детального дослідження електронної структури кремнієвих та кремній-металевих клатратів.

## 8.1. ДІАГРАМА СТАНУ СИСТЕМИ Na–Si. СИНТЕЗ І КРИСТАЛІЧНА СТРУКТУРА БІНАРНОЇ СПОЛУКИ NaSi

**8.1.1 Діаграма стану системи Na–Si.** Методами диференціального термічного аналізу і рентгенофазового аналізу авторами [249] вивчено *T–x* діаграму стану системи Na–Si. На рис.8.1 наведена *T–x* проекція діаграми стану системи Na–Si. Установлено, що в системі Na–Si існує тільки одна сполука – селіцид натрію (NaSi, $Na_4Si_4$), яка плавиться інконгруентно при 1071 K. На кривій ДТА для зразка з 50 мол.% Si спостерігаються два ендотермічні піки при нагріванні, один при 885 K, а інший при 1071 K. Нагрітий до 973 K зразок зберігає свою форму, а нагрітий до 1173 K повністю розплавляється. Екзотермічний пік також спостерігався при охолодженні приблизно до 885 K для всіх зразків з концентрацією 25 – 80 мол.% Si. На підставі цього автори [249] дійшли до висновку, що зворотний фазовий перехід NaSi відбувається при 885 K. Фазова діаграма Na–Si чітко показує, що розплав Na–Si існує вище 953 K у складі, збагаченому Na, і вище 1023 K у складі, збагаченому Si.

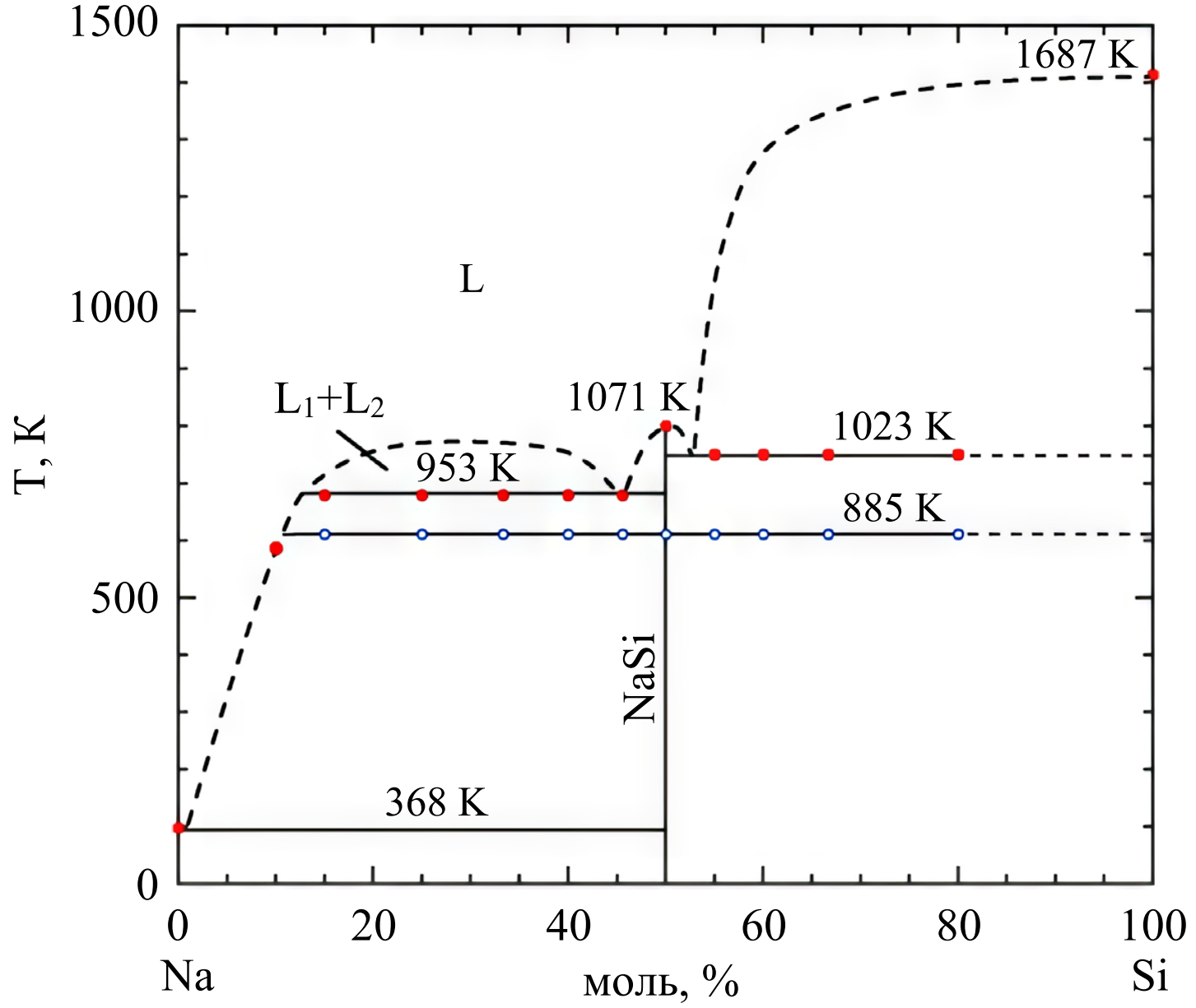

Рис. 8.1 Діаграма стану системи Na–Si [249]

**8.1.2. Синтез і кристалічна структура $Na_4Si_4$.** Силіцид натрію отримують шляхом прямої реакції елементарних натрію і кремнію [250–252]. Оскільки кремній має досить високу температуру плавлення 1685 K і, відповідно, низьку реакційну здатність, тому синтез стехіометричного $Na_4Si_4$ може бути проведений тільки при високій температурі. У зв'язку з високою реакційною здатністю натрію, синтез $Na_4Si_4$ проводять при температурах 923 – 1023 K, в атмосфері інертного газу (аргон, гелій) у герметично закритих ніобієвих, вольфрамових або танталових тиглях, які додатково розміщують у середині вакуумованих кварцових ампулах. Процес синтезу, як правило, триває від кількох днів до одного тижня, коли потрібен фазово чистий матеріал. Покращена кристалічна морфологія була отримана з невеликим надлишком натрію у співвідношенні Na : Si 1.1 : 1 [251], або 1.2 : 1. [252]

Силіцид натрію автори [253] синтезували в атмосфері аргону із суміші порошків Si і NaH у співвідношенні 1 : 2.1 моль при температурі 693 K протягом 90 год. Пізніше так само синтез $Na_4Si_4$ проводили автори [254] згідно реакції:

$$4NaH_{(s)} + 4Si_{(s)} \rightarrow Na_4Si_{4(s)} + 2H_{2(g)}$$

у потоці аргону 55 мл $хв^{-1}$ протягом 24 годин при температурі 668 K. За цих оптимальних умов було отримано $Na_4Si_4$ з оптимізованою чистотою. Атомне співвідношення між Na та Si було оцінено як Na : Si = 1 : 1.02 за допомогою аналізу з використанням енергодисперсійної рентгенівської спектроскопії.

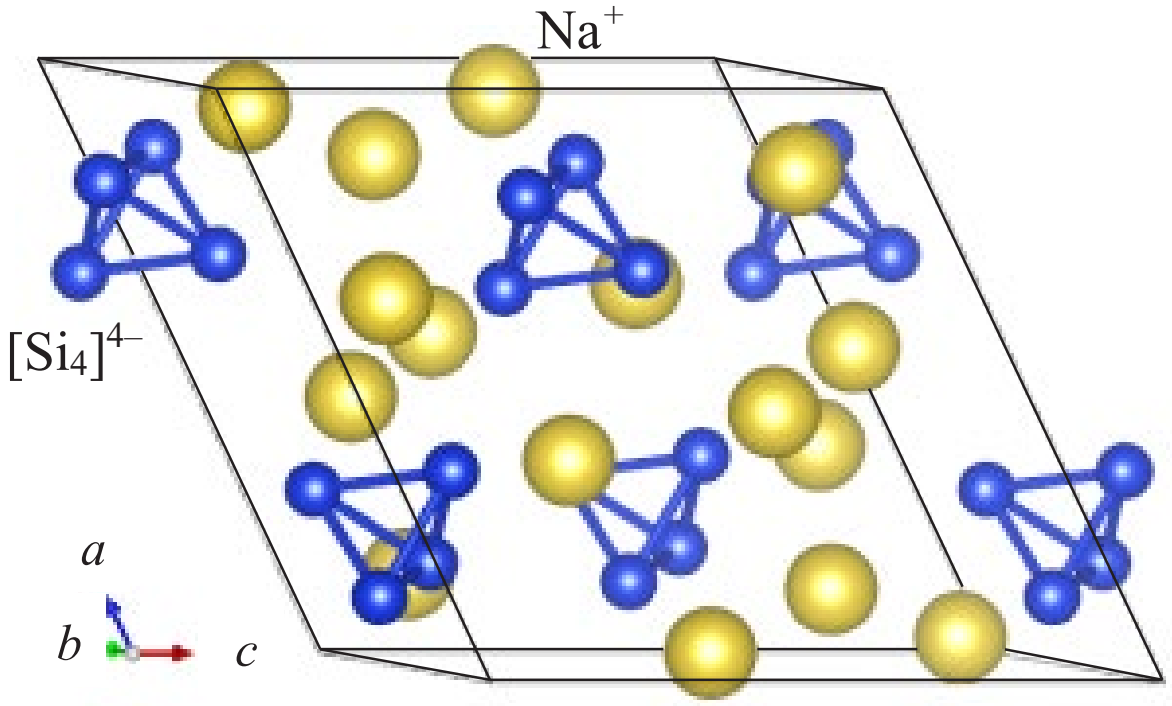

Рис. 8.2. Кристалічна структура $Na_4Si_4$.

Силіцид натрію $Na_4Si_4$ є Zintl фазою, кристалізується в моноклінній структурі з двома кристалографічно нееквівалентними позиціями для Si та Na: Si1 та Si2, Na1 та Na2 з просторовою групою $C2/c$ і параметрами гратки $a$ = 12.1561 Å, $b$ = 6.5465 Å, $c$ = 11.1320 Å, $\alpha$ = 90.000°, $\beta$ = 118.923°, $\gamma$ = 90.000°, Z = 4 [255]; $a$ = 12.1536 Å, $b$ = 6.5452 Å, $c$ = 11.1323 Å, α = 90.000°, β = 118.9°, γ = 90.000°, Z = 4 [256]. Кристалічна структура $Na_4Si_4$ побудована з ізольованих деформованих тетраедричних аніонів $[Si_4]^{4-}$, оточених катіонами $Na^+$ на кожній грані тетраедра (рис. 8 .2).

### 8.2 МЕТОДИ СИНТЕЗУ КЛАТРАТІВ $Si_{46}$, $Si_{136}$, $Na_8Si_{46}$ і $Na_{24}Si_{136}$

**8.2.1 Термічне розкладання Zintl фази $Na_4Si_4$.** Масивні зразки клатратів Na–Si спіканням порошків приготувати складно, так як клатрати мають сильний ковалентний зв'язок Si–Si і при високій температурі розкладаються на кремній алмазного типу. Вперше клатратні сполуки $Na_8Si_{46}$ і $Na_xSi_{136}$ були синтезовані шляхом термічного розкладання прекурсора $Na_4Si_4$ [244]. Під час термічного розкладання прекурсора $Na_4Si_4$, коли іони Na сублімуються, виникає дисбаланс зарядів, який змушує оточуючі атоми Si утворювати чотирикоординовані клітинні структури клатрату. Ці катіони вважаються необхідними для формування матриці клітки. Тому ця методика вимагає спочатку синтезувати фазу прекурсора $Na_4Si_4$ і подальшого його нагрівання в контрольованому середовищі для формування клатратної структури.

Клатрат $Na_8Si_{46}$ типу-I автори [245, 258] отримували шляхом видалення частини атомів Na з $Na_4Si_4$ в атмосфері Ar при 683 К. Клатрат $Na_xSi_{136}$ типу-II також був отриманий шляхом видалення атомів Na з прекурсора $Na_4Si_4$ при температурах 623 – 715 К, але під високим вакуумом [250, 258, 259]. Виділення фаз у клатратах кремнію, утворених шляхом термічного розкладання $Na_4Si_4$, можна досягти шляхом контролю тиску парів Na [257]. Повністю заповнені клатрати типу- I віддають перевагу високому локальному тиску пари Na (наприклад атмосфера Ar) під час термічного розкладання $Na_4Si_4$.

Який вплив чинить атмосфера Ar на утворення клатратної сполуки типу-I $Na_8Si_{46}$? Відповідь на це питання дано в роботі [257]. В умовах вакууму пари Na, що виділяються в результаті розкладання прекурсора $Na_4Si_4$, швидко видаляються із системи, тоді як за наявності атмосфери Ar пари натрію деякий час залишаються в атмосфері Ar, що оточує розкладений $Na_4Si_4$, через коротку довжину вільного пробігу атомів Na в атмосфері Ar. Таким чином наявність парів Na при розкладанні $Na_4Si_4$ сприяє утворенню клатратної сполуки $Na_8Si_{46}$.

Як показали автори [261], натрій повільно видаляється з прекурсора $Na_4Si_4$ шляхом реакції парової фази з просторово розділеним графітом у ефективно закритому об'ємі під одновісним тиском. Дослідивши вплив температури і часу на продукти реакції встановлено, що селективність росту кристалів $Na_8Si_{46}$ або $Na_{24}Si_{136}$ досягається простою зміною температури реакції. В інтервалі температур 853–863 К виростають виключно кристали клатрату $Na_8Si_{46}$ типу-I, тоді як в інтервалі 933–943 К спостерігається ріст кристалів клатрату $Na_{24}Si_{136}$ типу-II (рис. 8.3). Рентгеноструктурні дослідження, підтвердили ідентичність та фазову чистоту монокристалів $Na_8Si_{46}$ та $Na_{24}Si_{136}$. В обох фазах спостерігається повне заповнення всіх вузлів каркаса Si, а обидві клітинки в обох клатратних структурах повністю зайняті «гостями» Na, що підтверджує стехіометричний склад.

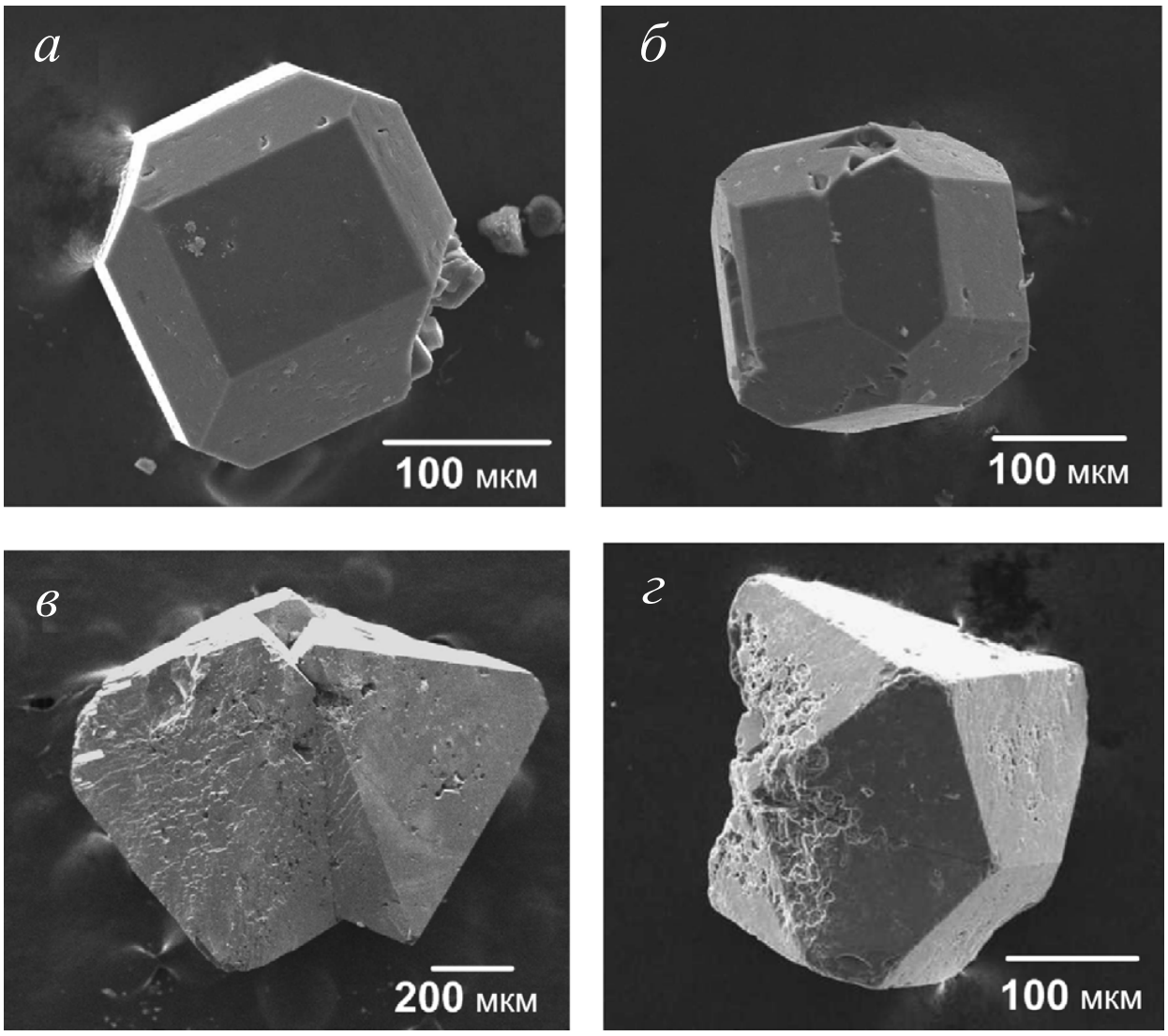

Рис.8.3 Електронні мікрофотографії монокристалів $Na_8Si_{46}$ (*а* і *б*), вирощених при 858 К, та монокристалів $Na_{24}Si_{136}$ (*в* і *г*), вирощених при 938 К. Час реакції для обох фаз становив 8 год [261].

**8.2.2. Синтез при високому тиску і високій температурі.** Авторами [262, 263] синтезовані клатрати $Na_8Si_{46}$, $Na_{24}Si_{136}$ при високих тисках і температурах, а також нова структура $NaSi_{46}$. Синтез кла-тратів проводився в стандартних Париж-Единбургських комірках при тисках і температурах до 6 ГПа і 1500 К відповідно. Експериментальні і теоретичні результати однозначно вказують на те, що клатрати інтеркальовані Na, термодинамічно стабільні тільки в умовах високого тиску. Хімічна взаємодія в системі Na–Si і переходи між двома структурами клатратів відбувається при температурах нижче температури плавлення кремнію. Клатрат $Na_8Si_{46}$ автори [262, 263] синтезували безпосередньо з елементарних компонентів при тисках від 2 до 6 ГПа в діапазоні температур 900–1100 К. За цих умов клатрат $Na_{24}Si_{136}$ утворюється лише як проміжна сполука перед кристалізацією $Na_8Si_{46}$. При більш високих тисках виявлено утворення нової інтеркальованої сполуки – металічного $NaSi_6$, який кристалізується в орторомбічній структурі типу $Eu_4Ga_8Ge_{16}$. Встановлено велику чутливість продуктів кристалізації до концентрації натрію. Клатрат $Na_{24-x}Si_{136}$ типу-II є стабільним при більш низьких температурах у порівнянні з $Na_8Si_{46}$, принаймні до 6 ГПа.

**8.2.3. Метод окислення прекурсора.** Для синтезу клатратів $Na_8Si_{46}$ і $Na_{24}Si_{136}$ автори [251] використали низькотемпературний метод синтезу шляхом окислення прекурсора $N_4Si_4$ термічним розкладанням іонної рідини *n*-додецилтриметаламоній хлориду (DTAC) у поєднанні з $AlCl_3$. Автори [251] повідомляють про одностадійний селективний синтез стехіометричних клатратів $Na_8Si_{46}$ і $Na_{24}Si_{136}$ шляхом реакції окислення $Na_4Si_4$, в якій DTAC/$AlCl_3$ забезпечує джерело газу. Кислі протони, які поставляє DTAC, та іони хлору, які поставляє $AlCl_3$, утворюють пари HCl, які керують реакцією. Наявність NaCl у попередньо промитому продукті підтверджує, що клатратні сполуки утворюються в результаті окислення $Na_4Si_4$.

Для утворення відповідних клатратних фаз температура реакції повинна бути досить високою, щоб відбувалося перенесення маси для забезпечення структурного перегрупування, водночас і достатньо низькою, щоб швидкість окислення була не дуже високою, щоб забезпечити суттєве видалення натрію із прекурсора. Низький парціальний тиск іонної рідини забезпечує такий баланс швидкості окислення. При підвищенні температури збільшується і парціальний тиск для конкретної реакційної посудини, що забезпечує певний контроль швидкості окислення і, таким чином, зменшує вміст Na в прекурсорі коли прекурсор $Na_4Si_4$ реагував на протязі 24 *год*. при температурі 483–513 К.

**8.2.4 Синтез клатрату кремнію $Si_{136}$ типу-II без «гостя».** Синтез кремнієвого клатрату $Si_{136}$ типу-II, що не містить «гостя», описаний у роботі [264]. Спочатку, з метою утворення клатрату складу $Na_xSi_{136}$ з $x > 10$ проведено термічну обробку прекурсора $Na_4Si_4$ під високим вакуумом в інтервалі температур 623–643 К. Значення $x$ отриманого таким чином клатрату згодом було знижено до $x \leq 4$ шляхом тривалої обробки під вакуумом, а потім промито концентрованою соляною кислотою. Повторюючи кілька разів такі обробки було отримано зразок вагою 100 мг, що містив не більше 600 частин на мільйон натрію. Однак у цьому дослідженні остаточний вміст натрію залишився невідомим, і крім того зразок містить до 5% кремнію алмазного типу. Таким чином, залишалось відкритим питання, чи можливо, або ні, отримати справжню, вільну від «гостя», клатратну форму кремнію $Si_{136}$.

Для синтезу клатратів $Na_xSi_{136}$ з дуже низьким вмістом натрію автори [259] на першому етапі використовували класичний метод термічної обробки прекурсора $Na_4Si_4$ у високому вакуумі в інтервалі температур 613–693 К протягом 60 год. Концентрація натрію в отриманому таким чином зразку залежала як від температури, так і від тривалості відпалу. Отриманий зразок очищали шляхом промивання у водному розчині $HNO_3$ для видалення можливих слідів гідрооксиду натрію, а потім сушили у вакуумі. Щоб отримати зразок із найменшим можливим значенням $x$, зразок піддавали повторним термічним обробкам у високому вакуумі при тій самій температурі, до тих пір, до поки випаровування натрію не припинялось. У результаті автори [259] отримали зразок з вмістом натрію в діапазоні $x = 0.5 – 1$. Задля зменшення залишкового вмісту натрію був використаний хімічний процес, який полягав у кількаразовій реакції зразка з невеликими кількостями елементарного йоду в діапазоні температут 573 – 673 К у герметичних скляних пробірках.

Дуже низький вміст залишкового натрію 37 частин на мільйон, що виявлено у даному дослідженні, вказує на те що клатратна форма кремнію $Si_{136}$ може бути стабільною без атомів-«гостів», що відкриває нові надії на синтез даного кремнієвого алотропу як масивних, так і тонкоплівкових зразків. Таким чином при термічному відпалі у вакуумі та послідучій обробці в парах йоду кількість атомів натрію в структурі $Na_xSi_{136}$ може бути зменшено до нуля [259]. Це вигідно відрізняє даний клатрат від клатратів типу-I в структурі якого не вдається повністю видалити атоми «гостя».

**8.2.5. Вирощування кристалів клатратів $Na_8Si_{46}$ типу–I і $Na_{24}Si_{136}$ типу–II шляхом випаровування Na з розчину Na–Si–Sn.** Авторам [265] вперше вдалося виростити монокристали клатрату $Na_8Si_{46}$ типу-I розміром ~ 1.5 мм шляхом випаровування Na з розчину Na–Si–Sn (молярне співвідношення Na:Si:Sn = 10:2:1), отриманого шляхом нагрівання суміші Na, $Na_4Si_4$ та $Na_{15}Sn_4$ при 723 K протягом 24 год. У даному методі $Na_4Si_4$ використовувався як прекурсор для джерела Si, а розплав $Na_{15}Sn_4$ використовувався як флюс для розчинення прекурсора. Вирощування кристалів проводилось при зниженому тиску в атмосфері аргону $10^4$ Па. Умова пересичення потоку росту монокристалів реалізувалась шляхом зменшення вмісту Na в розчині. Суміш полікристалічних клатратів I-, II-типу і Si була отримана випаровуванням Na при 773 K протягом 12 год, але єдину фазу клатрату II-типу авторам [262] отримати не вдалося.

В наступному дослідженні [266] було зроблено спробу виростити більші за розміром кристали клатрату $Na_8Si_{46}$ шляхом зменшення швидкості випаровування Na з розчину Na–Si–Sn (молярне співвідношення Na:Si:Sn = 6:2:1). Щоб зменшити швидкість випаровування Na автори [266] нагрівали розчин при 723 – 873 K в атмосфері аргону при тиску $10^5$ Па, що є близьким до атмосферного тиску та вищим, ніж у попередньому дослідженні ($10^4$ Па). Більш високий тиск сприяв росту монокристалів клатрату $Na_{24}Si_{136}$ розміром грані {111} ~ 2мм, шляхом нагрівання при 873 K протягом 9 год.

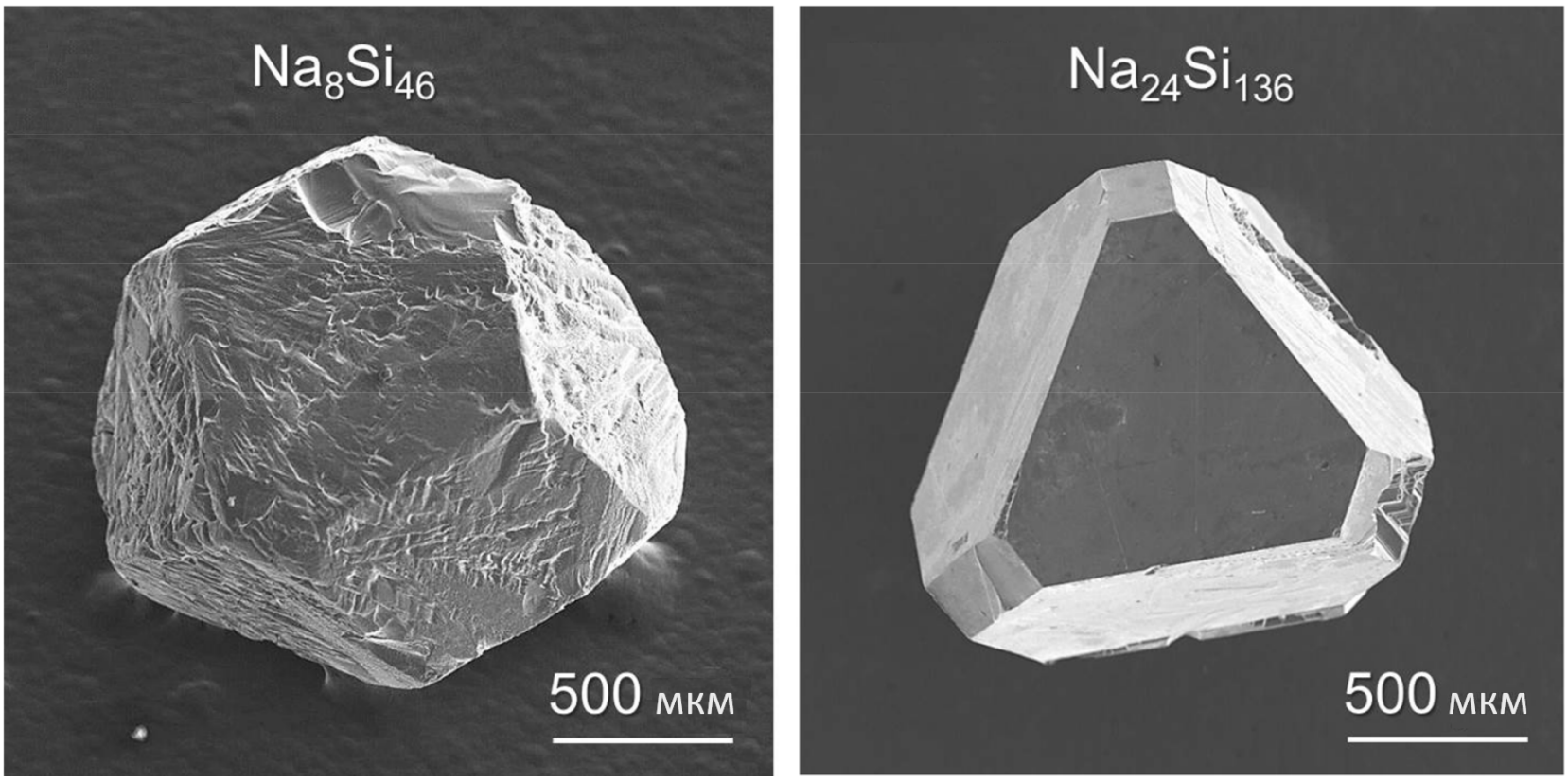

Рис.8.4. Монокристали типу-I клатрату $Na_8Si_{46}$ і типу-II клатрату $Na_{24}Si_{136}$, отриманих випаровуванням Na з розчину Na-Si-Sn [266].

При вирощуванні кристалів клатрату тип-I $Na_8Si_{46}$, під час випаровування Na з розчину Na-Si-Sn при 723 K склад потоку Na-Sn у вихідному матеріалі сильно впливає на морфологію та розміри утворених кристалів клатрату [268].

## 8.3. КРИСТАЛІЧНА СТРУКТУРА КРЕМНІЮ ТА КРЕМНІЄВИХ КЛАТРАТІВ $Si_{46}$, $Si_{136}$ і $Na_8Si_{46}$

**8.3.1. Структура кристалічного кремнію c-Si.** Кремній кристалізується у структурі алмаза, в елементарній комірці якого наявні 8 атомів. Гратка алмазного типу є кубічною гранецентрованою (рис. 8.5). Половина атомів займають вершини гранецентрованого куба, друга половина – центри чотирьох малих октантів із восьми.

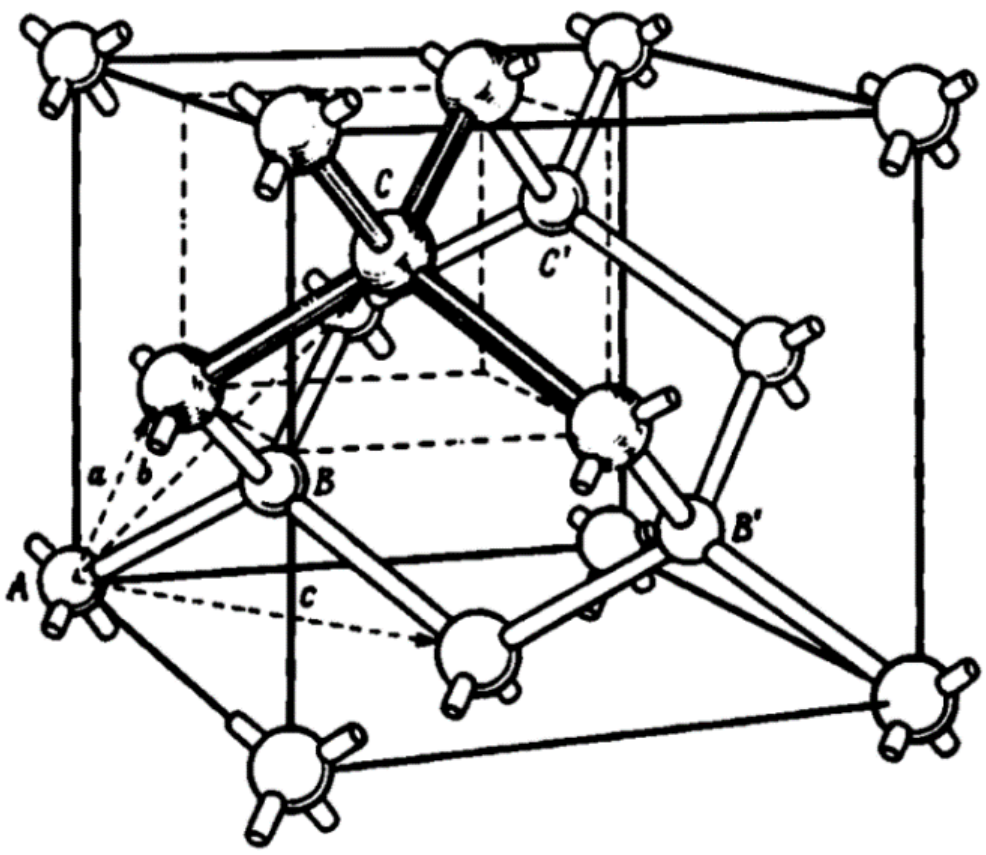

Рис. 8.5. Кристалічна структура кремнію.

Кожний атом тетраедрично зв'язаний з чотирма ближніми сусідніми атомами кремнію парноелектронним зв'язком, в утворенні якого приймають участь чотири його валентні електрони. Тому елементарна комірка кристала кремнію має вісім валентних електронів. Стала гратки (ребро куба елементарної комірки) рівна 5,43075 Å [270], на кожну елементарну кубічну комірку припадає по 8 атомів, пр. гр. $Fd\bar{3}m$. Густина цієї фази становить $\rho = 2.3294$ г/см$^3$.

**8.3.2. Кристалічна структура клатрату $Si_{46}$ і $Si_{136}$.** Напівпровідникові клатрати кристалізуються в трьох структурних типах, їх позначають римськими цифрами: клатрат типу-I, типу-II і типу-III. Опис будови клатратів базується на виділенні координаційних поліедрів, які оточують атоми «гостя», і називаються клатратоутворюючими поліедрами. Кількість вершин таких поліедрів не менше 20, а грані цих поліедрів представляють собою правильні п'яти- і шестикутники. У напівпровідникових клатратах наявні чотири типи клатратоутворюючих поліедрів: пентагональний додека-

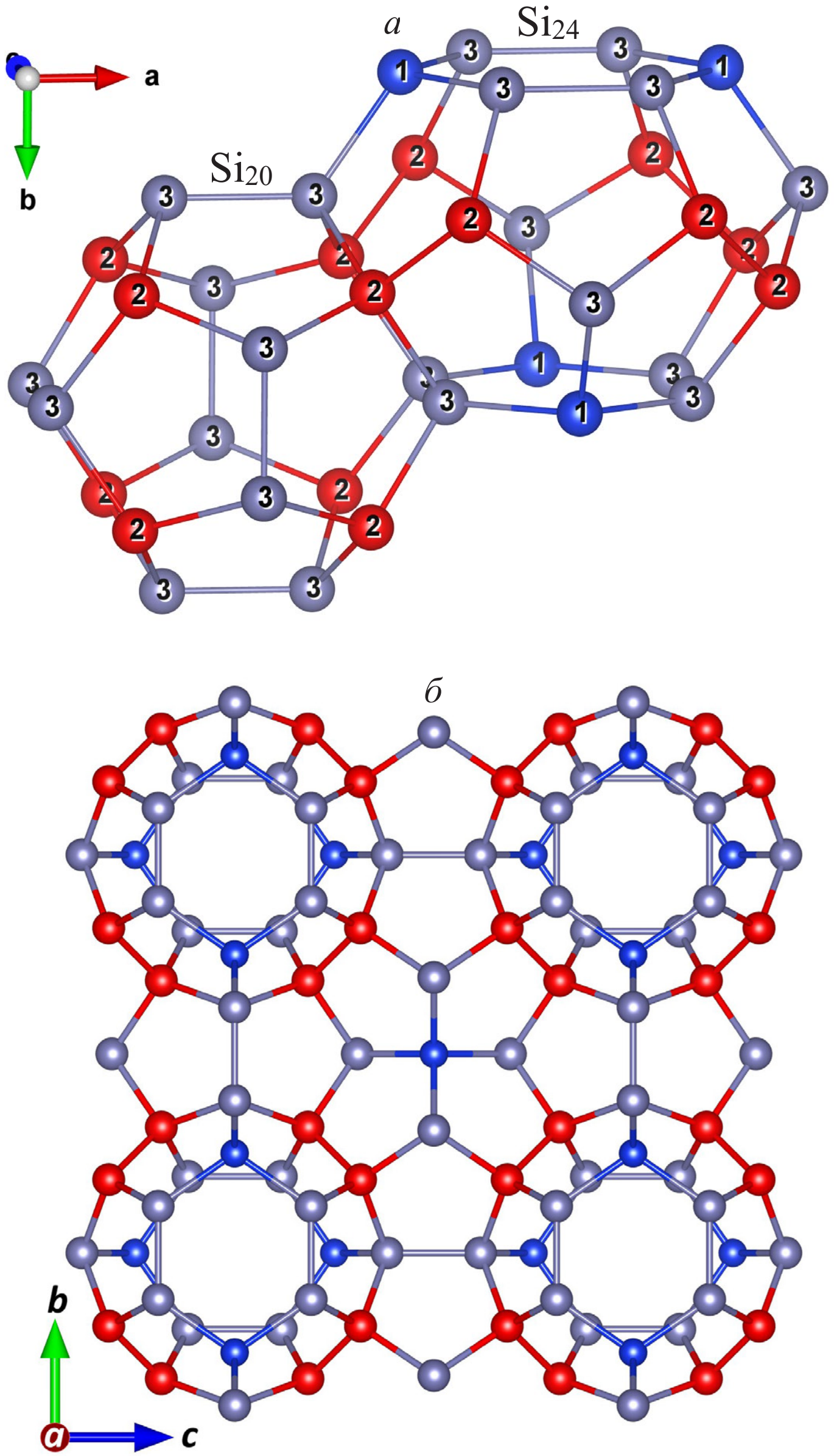

Рис. 8.6. *а* – Ув'язування двох правильних поліедрів $Si_{20}$ і $Si_{24}$;
*б* – проекція кристалічної структури клатрату $Si_{46}$ типу-I на площину (100).

едр, тетракайдекаедр, пентагондекаедр, гексакайдекаедр. Пентагональний додекаедр – це правильний платон поліедр, тобто в якому всі вершини еквівалентні, а всі грані однотипні. Тільки пентагональний додекаедр зустрічається у структурі всіх типів клатратів.

Клітини Si в клатратних структурах мають спільні грані, щоб забезпечити $sp^3$ зв'язок (рис. 8.6). Усі атоми кремнію в клатратах тетраедрично скоординовані з атомами кремнію, які знаходяться у центрах трохи деформованих тетраедрів. Кути зв'язку та довжини зв'язку мають кілька значень замість одного значення, як у випадку кристалічного кремнію.

Кремнієвий клатрат $Si_{46}$ відноситься до клатрату типу-I, і кристалізується в кубічній сингонії в центросиметричній просторовій групі $Pm\overline{3}n$ ($a$ = 10.2 Å). Кристалічна ґратка клатрату $Si_{46}$ типу-I побудована із двох типів поліедрів: пентагонального додекаедра $Si_{20}$ і тетракайдекаедра $Si_{24}$. Пентагональний додекаедр $Si_{20}$ має 20 вершин, 30 ребер й 12 п'ятикутних граней і позначається $[5^{12}]$, а тетракайдекаедр $Si_{24}$ має 24 вершини, 36 ребер, 12 п'ятикутних і дві шестикутні гексагональні грані і позначається як $[5^{12}6^2]$ (рис. 8.6, *а*). В елементарній комірці наявні два додекаедри $Si_{20}$ і шість тетракайдекаедрів $Si_{24}$. Атоми Si2 і Si3 в позиціях 16i і 24k наявні в обох типах поліедрів $Si_{20}$ і $Si_{24}$, а атоми Si1 в позиції 6c – тільки в більш об'ємних тетракайдекаедрах $Si_{24}$ (рис. 8.6, *а*). Тетракайдекаедри з'єднані один з одним через спільні шестикутні грані, а пентагональні додекаедри заключні всередині цієї системи каналів та ізольовані один від одного. Проекція кристалічної структури клатрату $Si_{46}$ типу-I на площину (100) приведена на рис. 8.6, *б*.

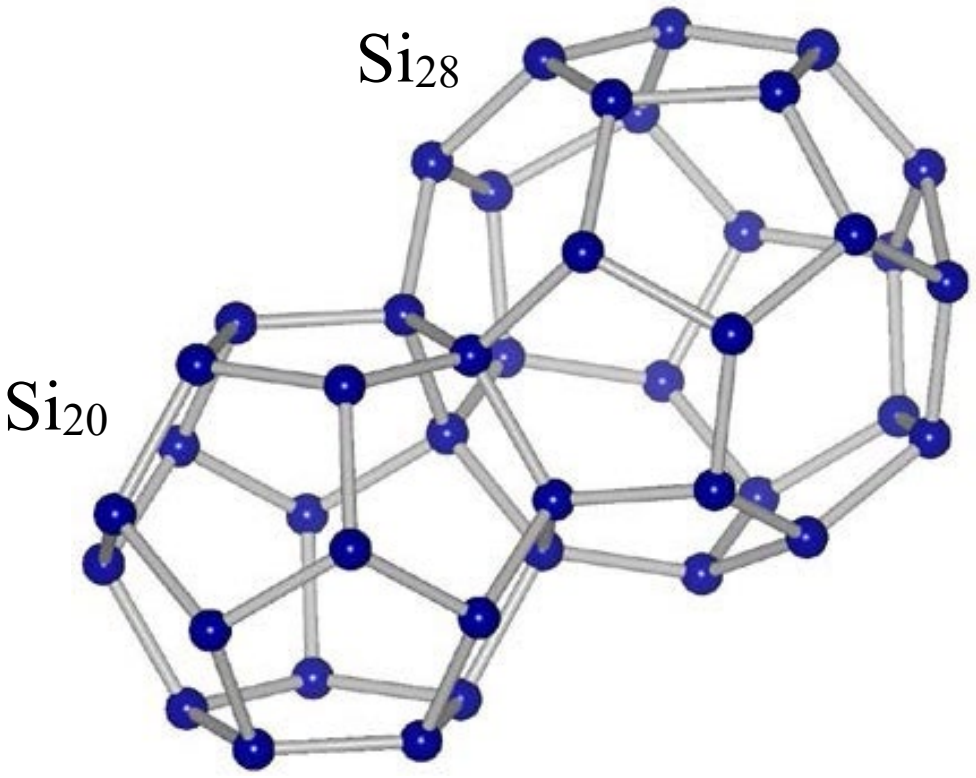

Рис. 8.7. Ув'язування двох правильних поліедрів $Si_{28}$ і $Si_{20}$ .

Кремнієвий клатрат $Si_{136}$ відноситься до клатрату типу-II, і кристалізується в гранецентрованій кубічній сингонії з центросиметричною просторовою групою $Fd\overline{3}m$ ($a$ = 14.62601 Å) [264]. Структура клатрату $Si_{136}$ типу-II подібна до типу-I, але відрізняється від неї. У клатрату $Si_{136}$ типу-II 136 атомів кремнію утворюють гранецентровану кубічну структуру, яка складається з 16 менших п'ятикутних додекаедрів $Si_{20}$ і восьми більших за розміром гексакайдекаедрів $Si_{28}$ [$5^{12}6^{4}$] в елементарній комірці. На відміну від клатрату $Si_{46}$ типу-I, де додекаедри не зв'язані один з одним, у структурі клатрату $Si_{136}$ типу-II (рис 8.7) вони з'єднані спільними гранями і укладені в шари. Шари пентагональних додекаедрів об'єднані в тривимірний каркас через гексакайдекаедри, які також з'єднані між собою спільними гексагональними гранями і розташовуються у кубічній структурі по алмазоподібному мотиву.

У структурі обох клатратів $Si_{46}$ і $Si_{136}$ кожен атом кремнію є тетраедрично координованим, з міжатомними відстанями трохи більшими ніж у кремнію алмазного типу, і кутами зв'язку в діапазоні від 108° до 124°, із середнім значенням, близьким до 109.47°.

**8.3.3. Кристалічна структура клатрату $Na_8Si_{46}$.** $Na_8Si_{46}$ відноситься до структурного клатрату типу-I, кристалізується в примітивній кубічній гратці ($a$ = 10.1983 Å) просторової групи $Pm\overline{3}n$ [244, 258] (рис. 8.8). Розрахована густина $\rho_{роз}$ = 2.316 г/см$^3$, експериментальна – $\rho_{експ}$ = 2.244 г/см$^3$ [244]. За даними [250] $a$ = 10.1964 Å , $\rho_{експ}$ = 2.271 г/см$^3$, а $\rho_{роз}$ = 2.292 г/см$^3$. В елементарній комірці наявні два п'ятикутні додекаедри $Si_{20}$ і шість тетракайдекаедрів (12 п'ятикутних і 2 шестикутні грані) $Si_{24}$ (рис. 8.8, *а*). Середня міжатомна відстань Si–Si становить 2.369 Å і близька до значення в кремнію алмазного типу (2.352 Å). Кути зв'язку Si–Si–Si варіюються від 105° до 125°, а середнє значення близьке до 109.54°, що характерно для $sp^3$-гібридизації [258]. Центри багатогранників $Si_{20}$, і $Si_{24}$ зайняті гостьовими атомами натрію, приводячи до формули $Na_8Si_{46}$. Атоми «гостя» Na1 і Na2 заповнюють порожнечі в кремнієвих поліедрах, займаючи 2*a*- і 6*c*-кратну позиції і характеризуються координаційними числами 20 і 24 відповідно (рис. 8.8, *а* і *б*).

**8.3.4 Кристалічна структура клатрату $Na_{24}Si_{136}$.** Структура клатрату $Na_{24}Si_{136}$ типу-II подібна до типу-I, але відрізняється від неї. У типі-II 136 атомів утворюють гранецентровану кубічну структуру, яка складається з 16 менших п'ятикутних додекаедрів $Si_{20}$ і восьми більших гексакайдекаедрів $Si_{28}$ в елементарній комірці (рис. 8.9). Елементарна комірка також є кубічною ($a$ = 14.62 Å) з просторовою

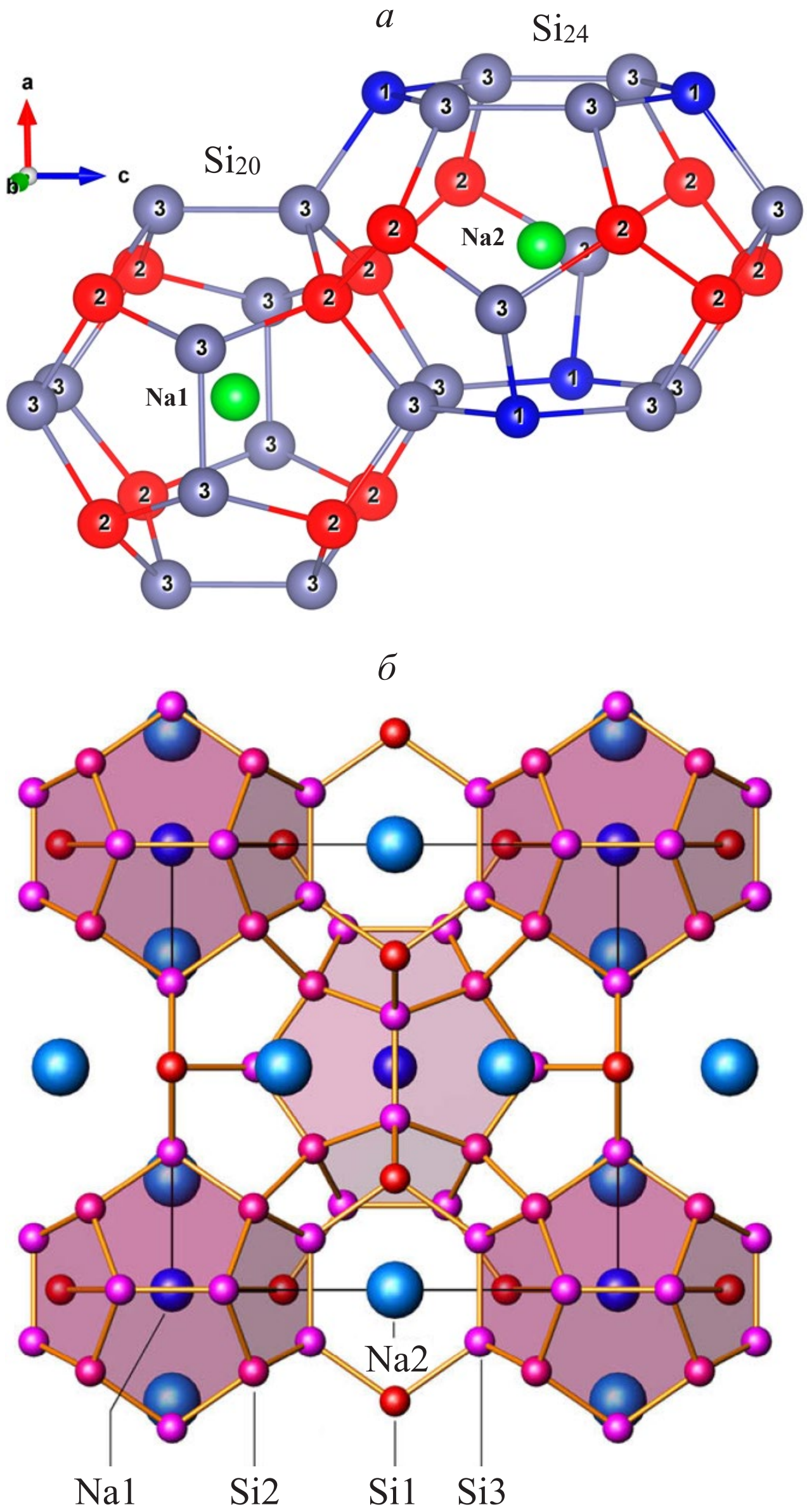

Рис. 8.8. *а* – Ув'язування двох правильних поліедрів $Si_{20}$ і $Si_{24}$, заповнених атомами Na1 і Na2 відповідно; *б* - проекція кристалічної структури клатрату $Na_8Si_{46}$ на площину [100] [269].

групою $Fd\overline{3}m$ [269]. Однак на відміну від клатрату типу-I, структура типу-II показала високу схильність до нестехіометрії залежно від умов синтезу.

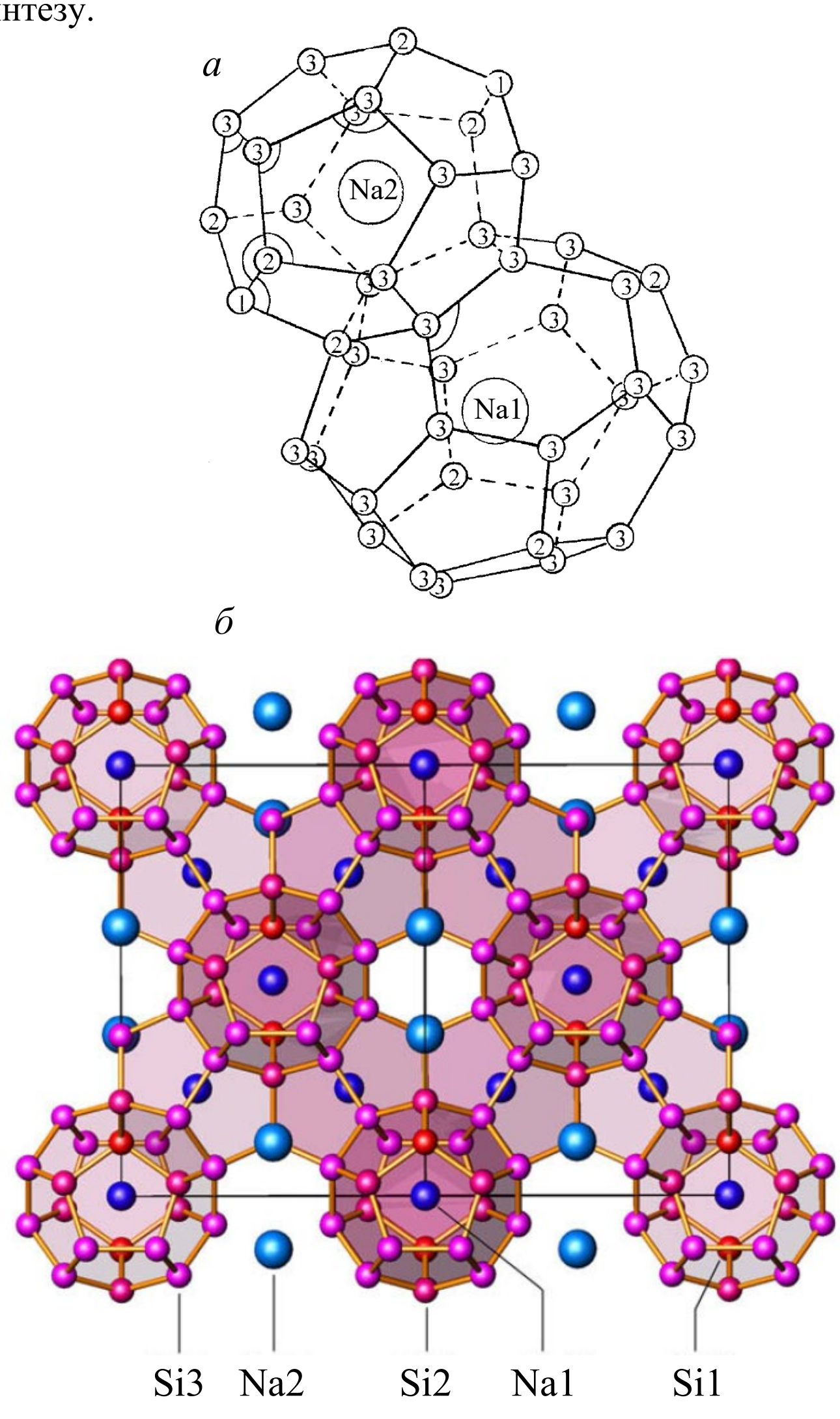

Рис. 8.9. *а* – Увязування двох правильних поліедрів $Si_{20}$ і $Si_{28}$ інкапсульованих атомами Na1 і Na2 [258]; *б* – проекція кристалічної структури клатрату $Na_{24}Si_{136}$ типу-II на площину [110] [269].

## 8.4. ЕЛЕКТРОННА СТРУКТУРА КРИСТАЛІЧНОГО КРЕМНІЮ ТА КЛАТРАТІВ ТИПУ-I $Si_{46}$ і $Na_8Si_{46}$

**8.4.1. Електронна структура кристалічного кремнію.** Електронна зонна структура кристала кремнію, розрахована методом функціонала електронної густини з використанням гібридного функціоналу HSEO6 без врахування спін-орбітальної взаємодії у точках високої симетрії й симетричних напрямках зони Брилюена (рис. 8.10) наведена на рис. 8.11, *а*. Початок відліку шкали енергії поєднано з абсолютним максимумом валентної зони, розташованим у центрі ЗБ (точка з симетрією $\Gamma'_{25}$). Атоми кремнію мають чотири валентні електрони. Оскільки кожна зона Бриллюена має число місць для електронів рівне подвоєному числу елементарних комірок у кристалі, тому в кристалі кремнію наявні чотири валентні зони.

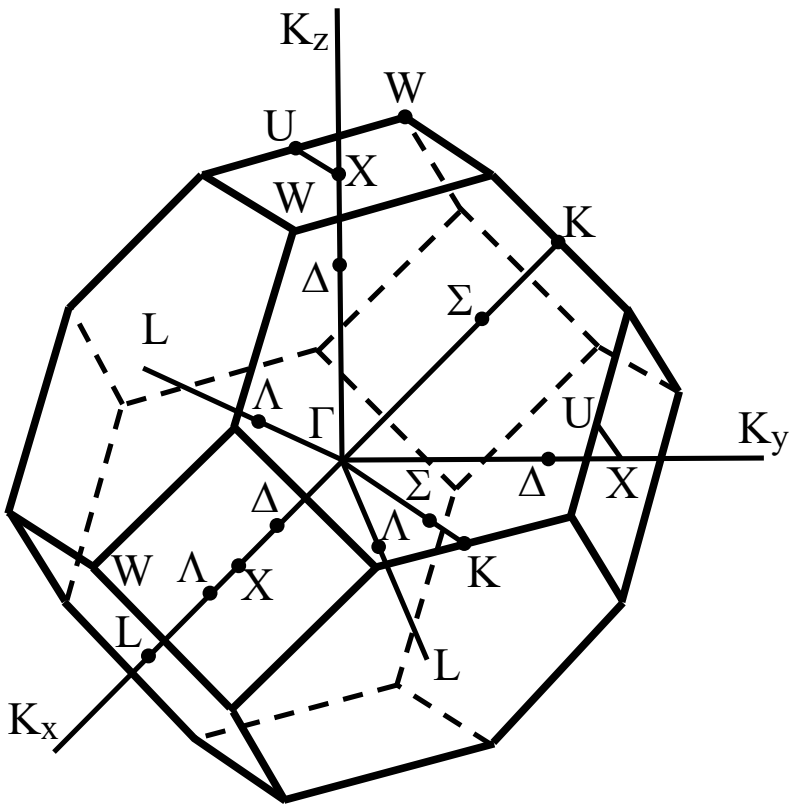

Рис. 8.10. Перша зона Бриллюена гранецентрованої кубічної гратки кремнію.

Згідно наших розрахунків кристал c-Si є непрямозонним напівпровідником, оскільки вершина валентної зони знаходиться у точці Г, а абсолютний мінімум зони провідності знаходиться на границі ЗБ в Δ-точках, – приблизно на відстані в 1/6 частини відрізка Г–X до X-точки. Таким чином, кристалічний кремній є непрямозонним напівпровідником з розрахованою шириною забороненої зони $E_{gi}$ = 0.67 еВ в LDA наближенні та $E_{gi}$ = 1.12 еВ в HSE06 наближенні, що добре узгоджується з експериментальним значенням $E_{gi}$ = 1.123 еВ [271].

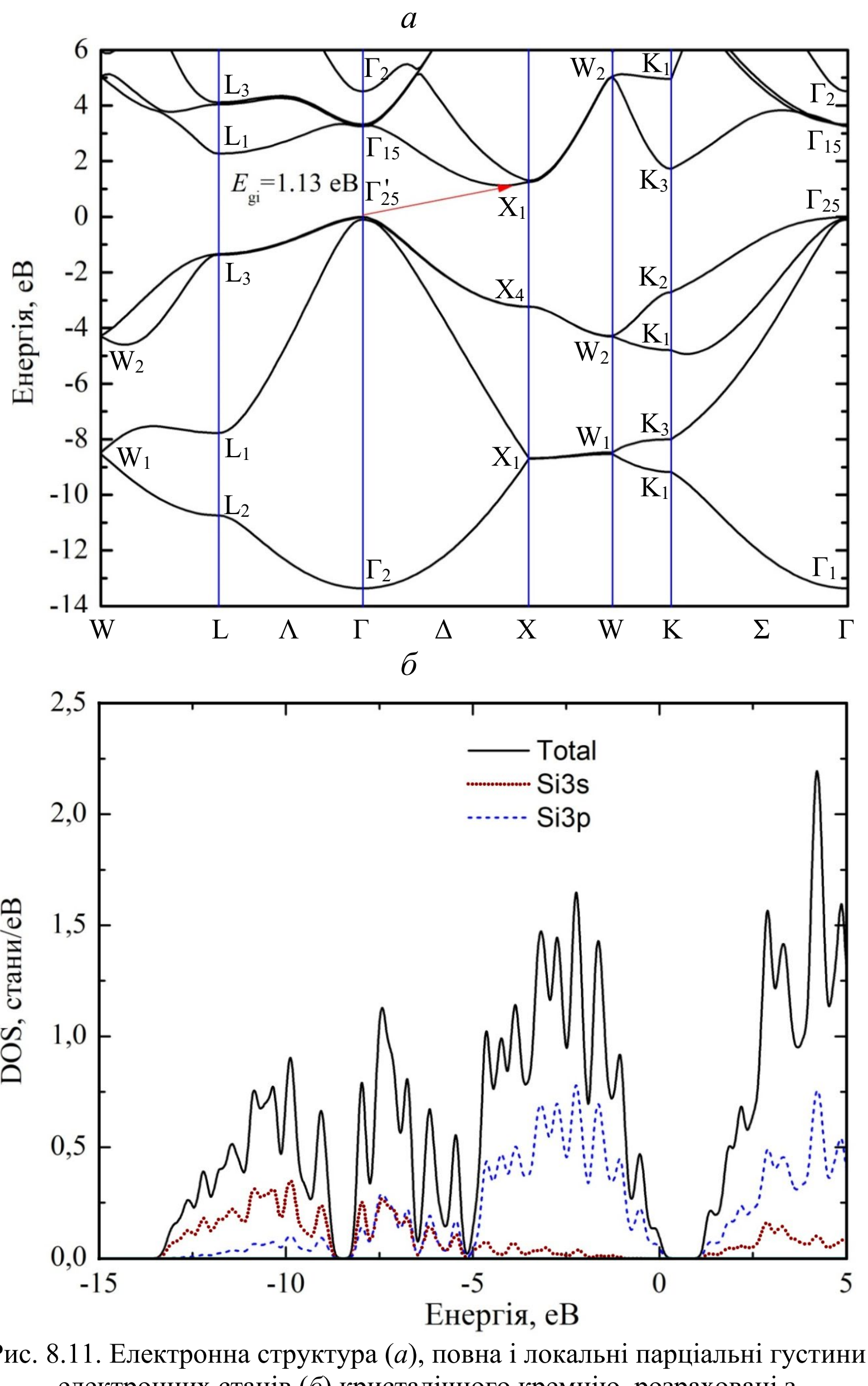

Рис. 8.11. Електронна структура (*а*), повна і локальні парціальні густини електронних станів (*б*) кристалічного кремнію, розраховані з використанням гібридного функціоналу HSEO6.

**8.4.2. Електронна структура клатратів $Si_{46}$ і $Na_8Si_{46}$.** Розраховані методом функціонала електронної густини в LDA-наближенні без врахування спін-орбітальної взаємодії електронні зонні структури клатратів $Si_{46}$ і $Na_8Si_{46}$, у точках високої симетрії й симетричних напрямках зони Брилюена (рис.8.12), наведені на рис. 8.13 і 8.14 відповідно, де за нуль енергії прийнято останній заповнений стан.

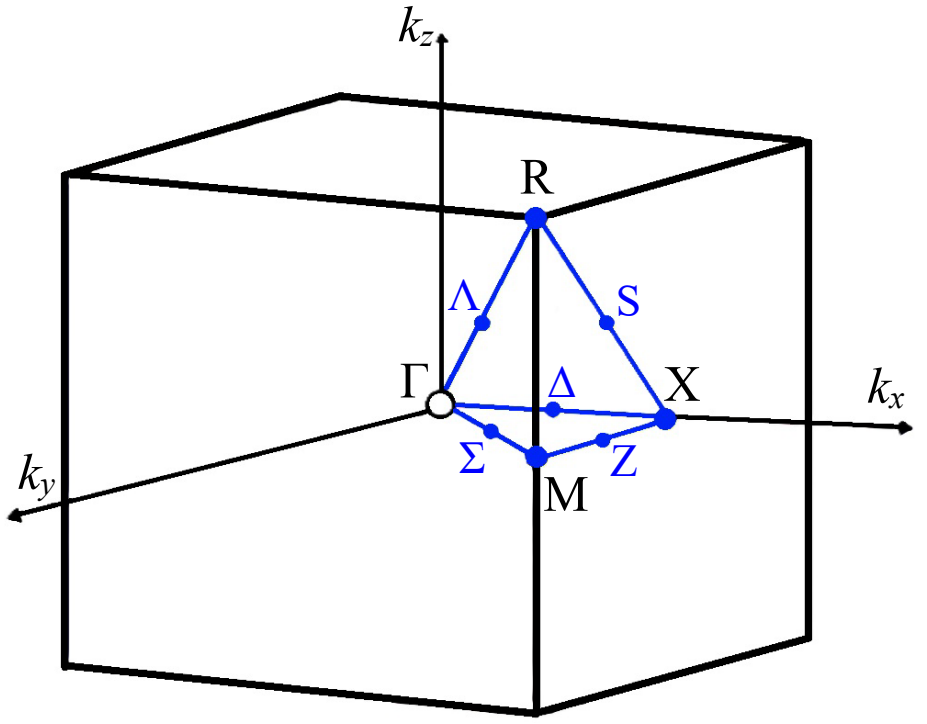

Рис. 8.12. Перша зона Бріллюена кубічних клатратів $Si_{46}$ і $Na_8Si_{46}$.

Оскільки всі атоми Si у клатратах тетраедрично координовані з атомами Si, як і у випадку гратки кремнію, клатрат $Si_{46}$ також є напівпровідником. Разом з тим електронна структура клатрату $Si_{46}$ значно відрізняється від електронної структури кристала кремнію (c-Si), структура якого також складається з тетраедричних зв'язків атомів Si. Згідно наших [272] розрахунків і даних, приведених в роботах [273, 274], у клатрату $Si_{46}$ верх валентної зони і дно зони провідності розташовані на лінії Γ– X і дуже близькі в k просторі, хоча прямий перехід між найвищою валентною зоною та найнижчою зоною провідності оптично заборонений.

В клатрату $Si_{46}$ типу-I рівень Фермі знаходиться у забороненій зоні, отже даний клатрат є непрямозонним напівпровідником з розрахованою в LDA наближенні шириною забороненої зони $E_{gi}$ = 1,22 еВ, що набагато більше, ніж для кристалічного кремнію (0.67 еВ), розрахованого також в наближенні LDA. Близьке значення ширини забороненої зони $E_{gi}$ = 1.26 еВ для клатрату $Si_{46}$, розраховане в LDA наближенні, приводять автори [273]. Хоча LDA зазвичай недооцінює значення $E_g$, різниця між двома структурами свідчить про те, що реальне значення ширини забороненої зони клатрату $Si_{46}$ складає приблизно 1.9 еВ, враховуючи експериментальне значення ши-

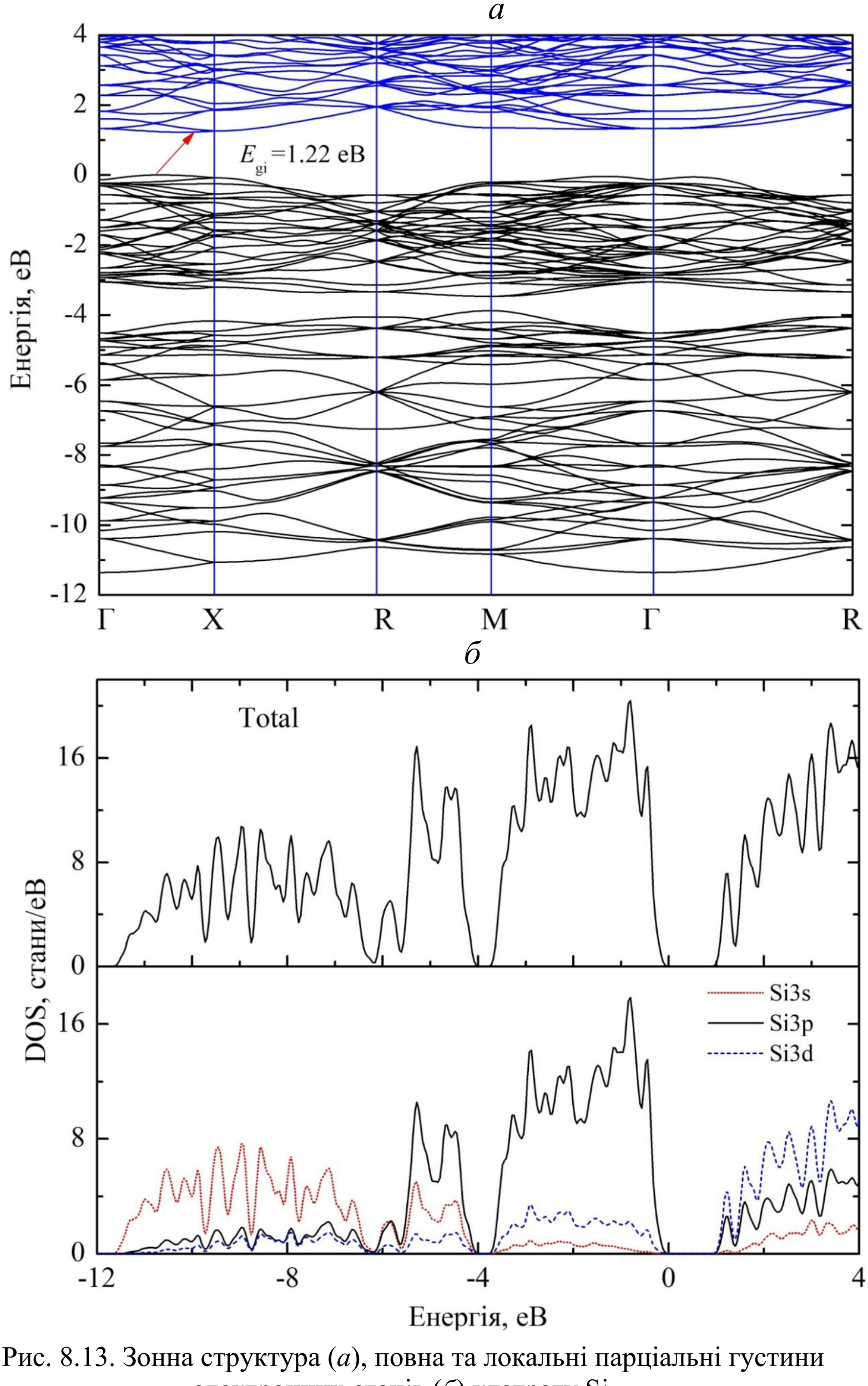

Рис. 8.13. Зонна структура (*а*), повна та локальні парціальні густини електронних станів (*б*) клатрату $Si_{46}$.

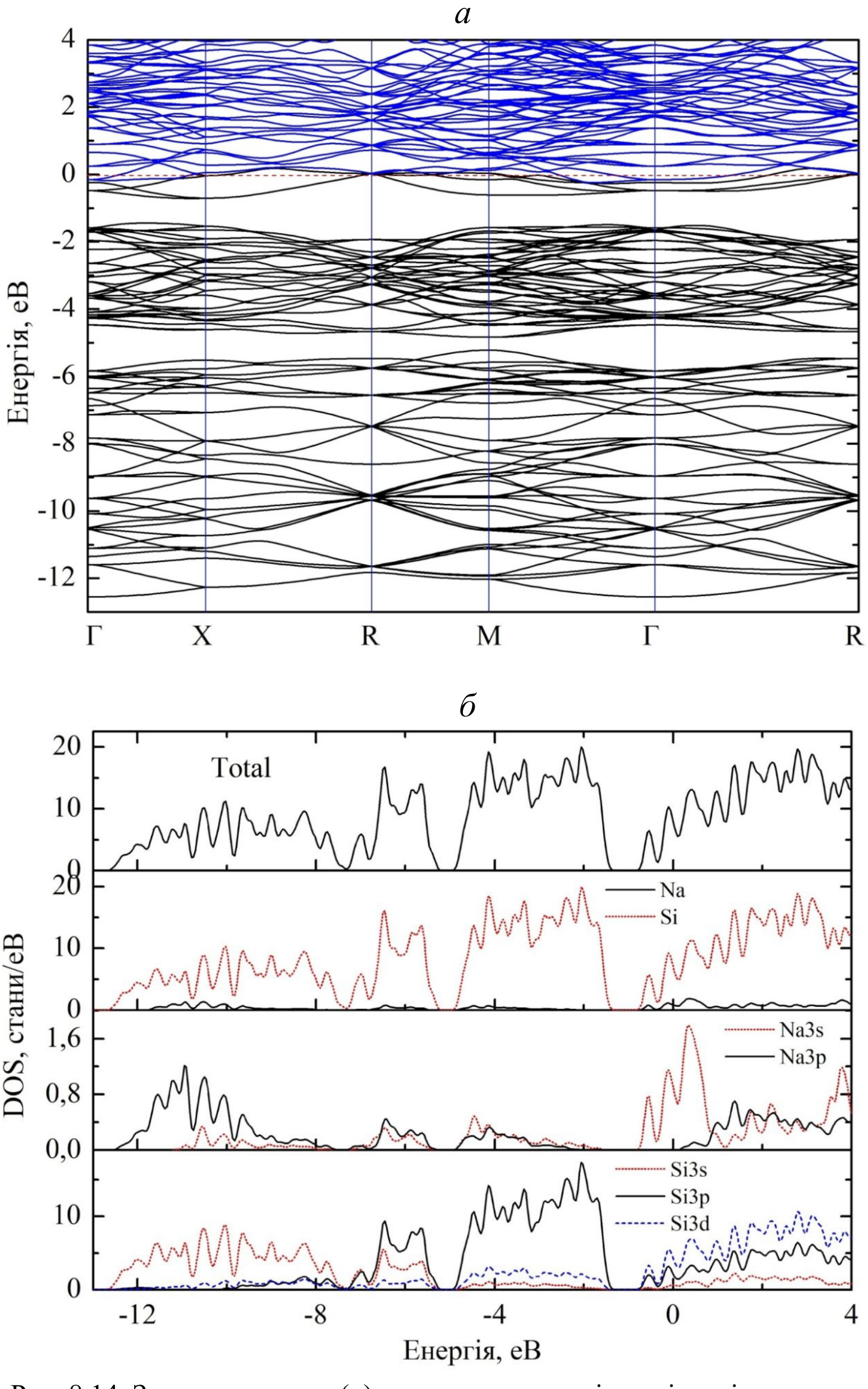

Рис. 8.14. Зонна структура (*а*), повна та локальні парціальні густини електронних станів (*б*) клатрату $Na_8Si_{46}$ [272].

рини забороненої зони c-Si ($E_g$ = 1.123 еВ) [271]. Дійсно, при розрахунках електронної структури клатрату $Si_{46}$ з використанням гібридного функціоналу GGA-PPE (BLYP), автори [274] отримали значення $E_g$ = 1.618 еВ, що є значно ближчим до прогнозованого.

Повна ширина валентної зони клатрату $Si_{46}$ (рис. 8.13) складає 11.36 еВ і є вужчою, ніж для кристала c-Si (13.34 еВ). Крім того у валентній зоні клатрату $Si_{46}$ наявна щілина шириною 0.4 еВ, натомість валентна зона кремнію є неперервною. Наявність щілини у валентній зоні кремнієвого і кремній-натрієвого клатратів відрізняє ці сполуки від кристалічного кремнію з алмазною структурою, валентна зона якого неперервна. Ці унікальні властивості зонної структури клатрату $Si_{46}$ автори [275] пояснюють наявністю в його гратці п'ятичленних кілець, оскільки Si 3*s*-орбіталі не можуть утворювати повний антизв'язуючий стан, на відміну від шестичленних кілець в c-Si.

Порівняння зонних структур клатратів $Si_{46}$ (рис. 8.13) і $Na_8Si_{46}$ (рис.8.14) вказує на їх ідентичність, яка виражена в топології й числі дозволених валентних зон. Так як елементарні комірки клатрату $Si_{46}$ і $Na_8Si_{46}$ містять 46 чотири валентних атомів Si, то число валентних електронів у ЗБ рівне 184 і відповідно енергетичний спектр E(k) валентної зони містить 92 дисперсійні вітки, які згруповані в три дозволені підзони (VBI, VBII і VBIII), загальною шириною 11.36 еВ для клатрату $Si_{46}$ та 12.55 еВ для клатрату $Na_8Si_{46}$.

Відмінність полягає лише у відносному положенні рівня Фермі та величині забороненої зони, на які впливає тип атома «гостя». Пояснення цього ефекту можна провести у моделі жорстких зон [238]. Згідно з цією моделлю, для ізоструктурних сполук, якими і є клатрати $Si_{46}$ і $Na_8Si_{46}$, характерна подібність структур енергетичних зон. Однак після інкапсуляції атомів Na у вісім порожнин кремнієвих кліток край зони провідності буде утворений внеском восьми валентних електронів атомів натрію, що приводить до зсуву рівня Фермі у бік вищих енергій і, отже, рівень Фермі знаходиться в зоні провідності. Вісім валентних електронів від атомів Na в $Na_8Si_{46}$ приймають участь у формуванні краю зони провідності. Це сильно впливає на звуження фундаментальної енергетичної щілини клатрату $Na_8Si_{46}$ у порівнянні з чистим $Si_{46}$ із-за підсилення екрануючих ефектів.

В клатрату $Na_8Si_{46}$ рівень Фермі розташований у зоні провідності, що вказує на його металевий характер. Рівень Фермі відсікає від зони провідності енергетичний інтервал, який має максимальну шири-ну в точці X, рівну 0,6 еВ. Цей інтервал відокремлений від основної частини валентної зони щілиною шириною в 0,7 еВ. Ширина області, розташованої біля дна валентної зони, становить 7.4 еВ в чисто кремнієвому клатраті і зменшується до 7.2 еВ $Na_8Si_{46}$. Область, розташована ближче до рівня Фермі, хоч і включає також 46 зон, але значно вужча: 2.3 еВ в клатраті $Si_{46}$ і 2.5 еВ $Na_8Si_{46}$.

## 8.5. ПОВНА ТА ПАРЦІАЛЬНІ ГУСТИНИ ЕЛЕКТРОННИХ СТАНІВ c-Si ТА КЛАТРАТІВ $Si_{46}$ і $Na_8Si_{46}$

Отримані в результаті зонного розрахунку власні функції $\psi_{i,k}(r)$ та власні значення енергії $E$(k) використовувалися нами для розрахунку повних та парціальних густин електронних станів кристала c-Si та клатратів $Si_{46}$ і $Na_8Si_{46}$. Розраховані повні $N(E)$ та локальні парціальні густини електронних станів кристалічного кремнію та клатратів $Si_{46}$ і $Na_8Si_{46}$ наведені на рис. 8.11, *б*, 8.13, *б* і 8,14, *б* відповідно. Із цих розрахунків випливає, що внески *s*-, *p*-, *d*-станів кремнію в різні підзони валентної зони сильно різняться.

Аналіз парціальних внесків у повну $N(E)$ густину електронних станів дозволяє ідентифікувати генетичне походження різних підзон валентної зони і зони провідності. Як показує розрахунок парціальних густин електронних станів (рис. 8.11, *б*) в низькоенергетичній частині валентної зони кристалічного кремнію переважає внесок 3*s*-станів Si з незначним домішуванням 3*p*-станів. У високоенергетичній області валентної зони домінуючий внесок дають 3*p*-стани Si. В енергетичному інтервалі від –7.5 до –4.5 еВ має місце гібридизація валентних 3*s*- і 3*p*-станів Si. Дно зони провідності c-Si містить внески *s*-, *p*- і *d*-станів Si.

Повні густини $N(E)$ електронних станів у клатратних сполуках $Si_{24}$ і $Na_8Si_{46}$ мають подібні профілі, натомість при легуванні кремнієвого клатрату атомами натрію відбувається зміщення рівня Фермі в область більш високих енергій, і він потрапляє в зону провідності. Як уже зазначалось, при переході від кремнієвого клатрату до кремній-натрієвого клатрату структура енергетичних зон залишається практично незмінною, а змінюється тільки положення рівня Фермі. Отже не повинно бути істотних відмінностей і в профілях повних густин електронних станів.

Основні закономірності розподілу повної та парціальних густин електронних станів подібні для обох клатратів. Відмінність проявляється головним чином у ступені заповнення енергетичних зон ва-лентними електронами і енергетичним положенням відносно рівня Фермі. Для клатрату $Si_{46}$ валентна зона повністю заповнена, а зона провідності пуста і вони розділені інтервалом енергії забороненої зони $E_{gi}$ = 1.22 еВ, що вказує на належність даного клатрату до власного напівпровідника. У випадку $Na_8Si_{46}$ додаткові електрони натрію заповнюють наступну енергетичну зону провідності, забезпечуючи металічні властивості цього легованого клатрату.

Співвідношення між інтенсивностями максимумів у парціальних густинах електронних станів різного типу симетрії різні. Для клатрату $Si_{46}$ сама низькоенергетична зв'язка із 36 зон, розташована в енергетичному інтервалі (–11,3 ÷ –5,5 еВ), сформована переважно Si 3*s*-станами. У валентній зоні клатрату $Si_{46}$ наявна область перекриття *s*- і *p*-станів кремнію. Ця область розташована біля верха першої звязки заповнених зон Низькоенергетична підзона з домінуючим внеском Si *s*-станів перекривається з середньою підзоною (–11,3 ÷ –5,5 еВ), сформованою гібридизованими 3*s*- і 3*p*-станами кремнію. У зонній картині даного клатрату, у цій області наявні 10 енергетичних зон. Сама верхня зв'язка із 46 зон (–4,5 ÷ 0 еВ), відокремлена від другої (середньої) підзони енергетичним інтервалом 0.4 еВ і сформована з домінуючим внеском Si 3*p*-станів. Важливо відмітити, що у формування верхньої валентної підзони і нижньої зони провідності обох клатратів $Si_{46}$ і $Na_8Si_{46}$ вносять також 3*d*-стани кремнію. Цей факт є характерним не тільки для клатратів кремнію, але й для кристалічного кремнію (рис. 8.11, *б*).

Для незаміщених кремнієвих клатратів, у тому числі й для клатрату $Na_8Si_{46}$, характерним є розділення валентної зони на три групи енергетичних підзон (рис. 8.14). В клатрату $Na_8Si_{46}$ внесок від станів атомів натрію в повну густину електронних станів є незначним і переважає в нижній частині валентної зони. У всьому розглядуваному енергетичному інтервалі переважає внесок від електронних станів атомів кремнію. Область з домінуючим внеском Si *s*-станів відокремлена від області, в якій переважає внесок Si *p*-станів, щілиною ширина якої складає 1,4 еВ в клатраті $Si_{46}$ і 1,8 еВ в $Na_8Si_{46}$.

Зазначимо, що валентна зона кристалічного кремнію є неперервною, тобто області утворені переважно Si *s*-станами і переважно Si *p*-станами, перекриваються (рис. 8.11, *б*). Появу енергетичної щілини в повній густині клатратів $Si_{46}$ і $Na_8Si_{46}$ автори [273, 275] пояснюють виходячи з особливостей кристалічної будови даних сполук. У клатратній кремнієвій гратнці 87% складають п'ятичленні кільця і тільки 13% шестикутні, тоді як у структурі кристалічного кремнію c-Si усі кремнієві кільця шестикутні. В алмазній гратці кремнію *s*-орбіталі утворюють насичені зв'язки, тому Si 3*s*-стани розподілені при досить високих енергіях і перекриваються з Si 3*p*-станами, за рахунок чого утворюється неперервна валентна зона. У комірці з пятичленними кільцями зв'язок не є насиченим. Тому в клатратній кремнієвій структурі Si 3*s*-стани розташовані при значно нижчих енергіях, а ніж в алмазній структурі і не перекриваються зі Si 3*p*-станами. Як результат у валентній зоні кремнієвих і кремній-натрієвих клатратів зявляється щілина між областями з домінуючим внеском Si *s*-станів та Si *p*-станів.

Оскільки внесок Na *s*-, *p*-станів у повну густину *N*(*E*) в зоні провідності значно менший половини внесків Si *s*-, *p*-, *d*-станів, тому густина електронних станів у зоні провідності слабо змінюється при включенні атомів Na у порівнянні з первинним $Si_{46}$, що підтверджує слабку гібридизацію між станами зони провідності $Si_{46}$ та станами Na. В кремній-металічному клатрату $Na_8Si_{46}$, як і в клатрату $Si_{46}$, в області незаповнених електронних станів домінує внесок *p*- і *d*-станів кремнію. Помітним є також внесок *s*-станів Na в околі рівня Фермі.

Аналіз парціальних густин електронних станів та карт розподілу заряду валентних електронів вказує на слабку взаємодію між інкапсульованими атомами Na в клітках та атомами кремнію, розташованими в каркасах.

## 8.6. КАРТИ РОЗПОДІЛУ ЗАРЯДУ ВАЛЕНТНИХ ЕЛЕКТРОНІВ

Для точного опису хімічного зв'язку необхідно знати загальну картину просторового розподілу електронної густини. Метод функціонала густини дозволяє провести розрахунки розподілу електронної густини в кристалі кремнію (c-Si) та клатратах $Si_{46}$ і $Na_8Si_{46}$. Розподіл електронної густини заряду дає детальну інформацію про особливості формування хімічного зв'язку в кристалах. Теоретично розрахована і експериментальна карти розподілу густини валентного заряду ρ(r) для кристала кремнію наведені на рис. 8.15, *а*, *б* відповідно. Прецизійні рентгенодифракційні дослідження кристала кремнію, проведені авторами [276], дозволили експериментально вивчити розподіл електронної густини у міжатомному просторі, тобто знайти деформацію електронної густини, обумовлену наявністю хімічного зв'язку між атомами Si (рис. 8.15, б). Як видно з рис. 8.15, розраховані нами карти розподілу густини валентного заряду добре узгоджуються з експериментально отриманими для кристала Si, де чітко видно, що в обох випадках максимум густини валентного заряду припадає на середину зв'язку Si–Si, які об'єднані між собою спільними контурами. Ковалентний зв'язок на картах розподілу електронної густини характеризується зарядом, локалізованим на зв'язку, про що свідчить наявність замкнутих контурів сталої густини ρ(r) на лінії зв'язку. Наявність контурів сталої густини, які охоплю-

*а*

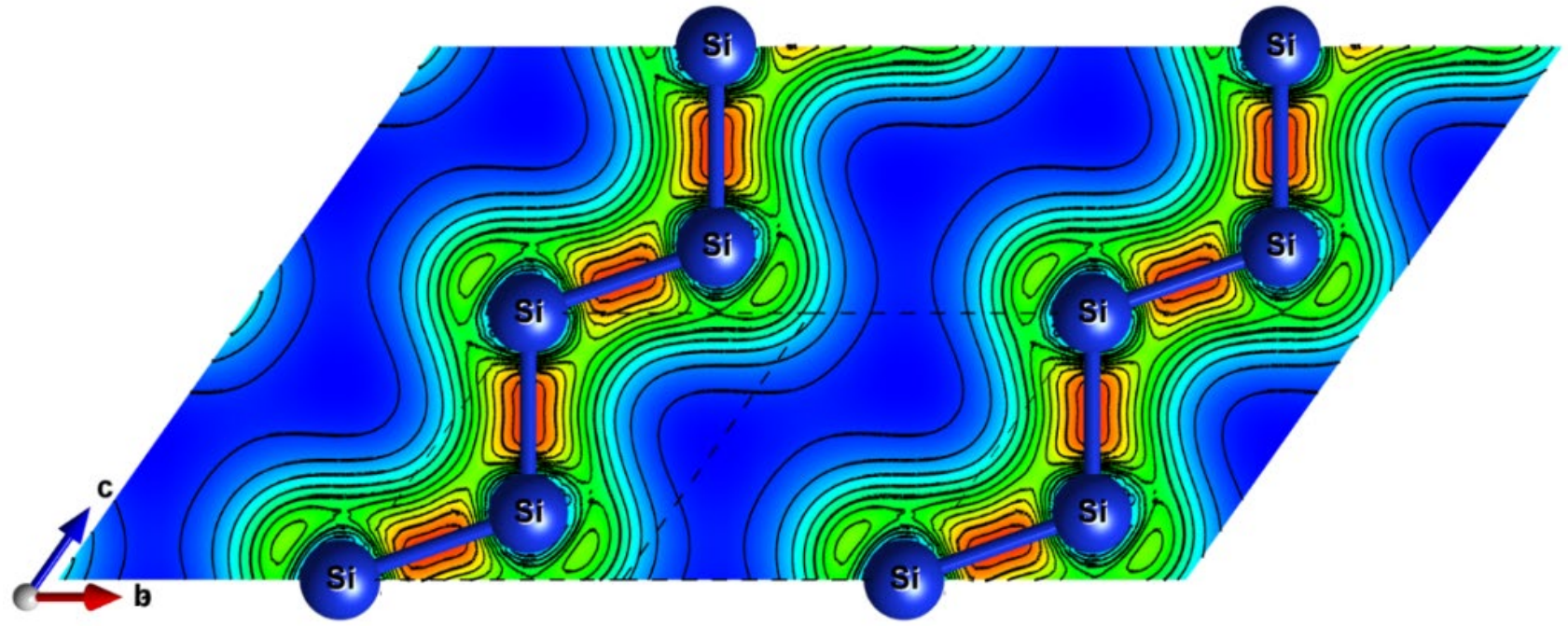

*б*

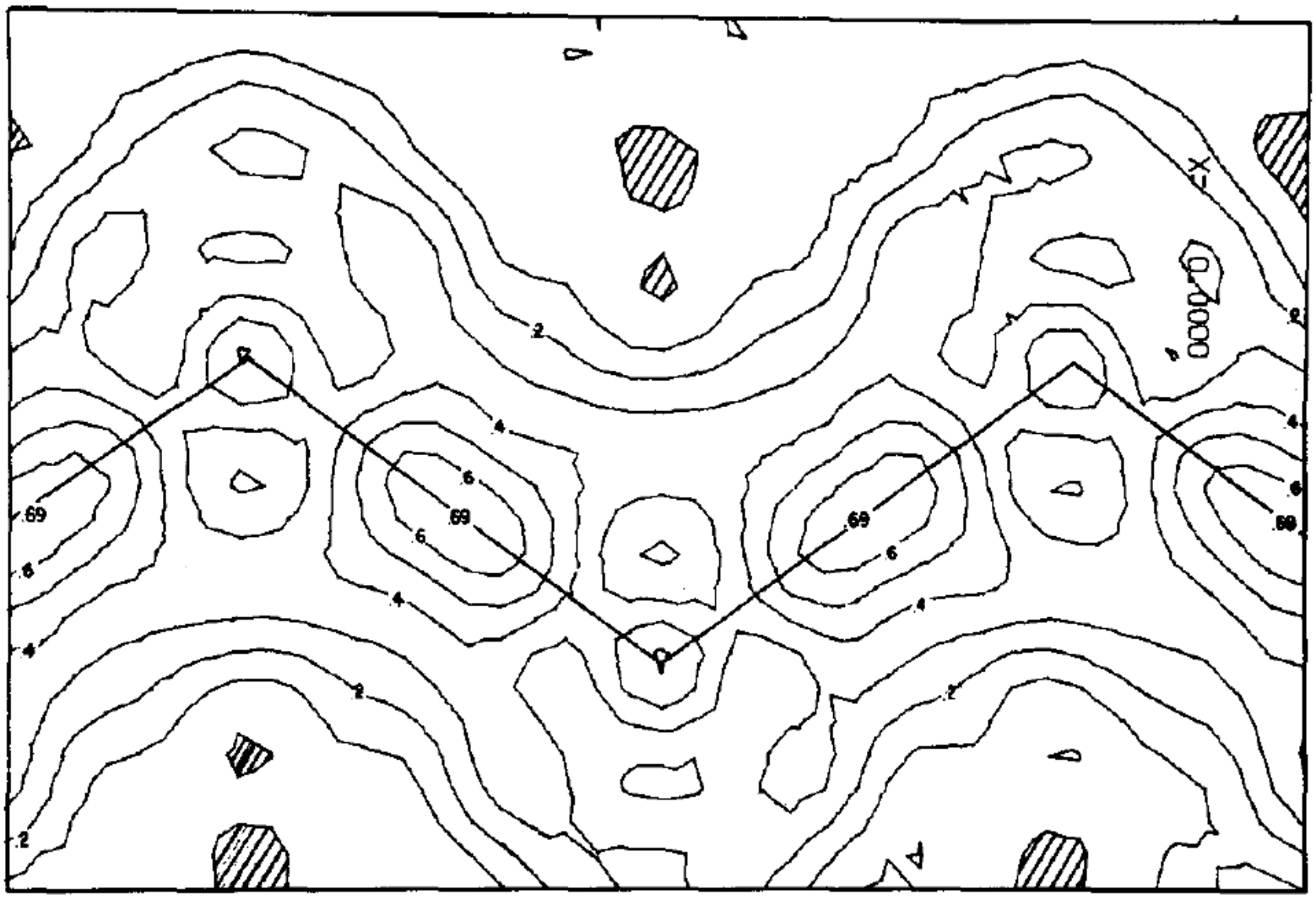

Рис. 8.15. Розрахована (*а*) і експерементальна (*б*) [276] карти розподілу повної густини валентного заряду для кристала Si в площині ($1\bar{1}0$):

ють взаємодіючі атоми Si–Si є ознакою *sp*$^3$-гібридизованих атомних станів. Ці максимуми характеризують ковалентний тип хімічного зв’язку.

На рис. 8.16, *а* наведена карта розподілу густини валентного заряду в тетраедрі $[Si]^{4-}$ структурному елементі кристалічного кремнію і кремнієвих клатратів. На цьому рисунку видно наявність спільних контурів, які охоплюють атоми Si.

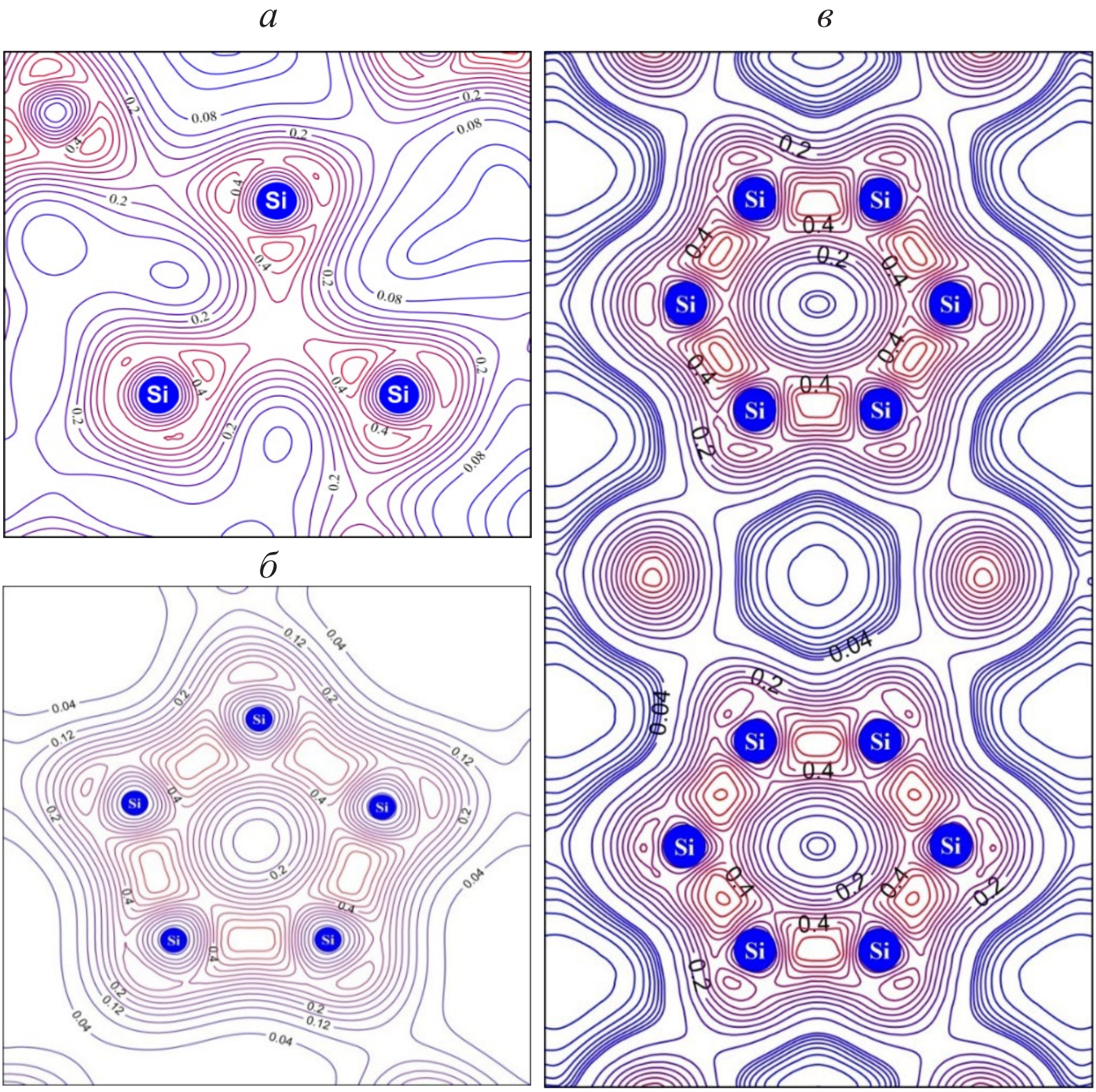

Рис. 8.16. Карти електронної густини клатрату $Si_{46}$: *а*) в площині, що проходить через одну з граней тетраедра $[Si]^{4-}$ ; *б*) через грань Si5 багатогранника $Si_{20}$; *в*) через грань Si6 багатогранника $Si_{24}$ в площині (010).

*а* *б*

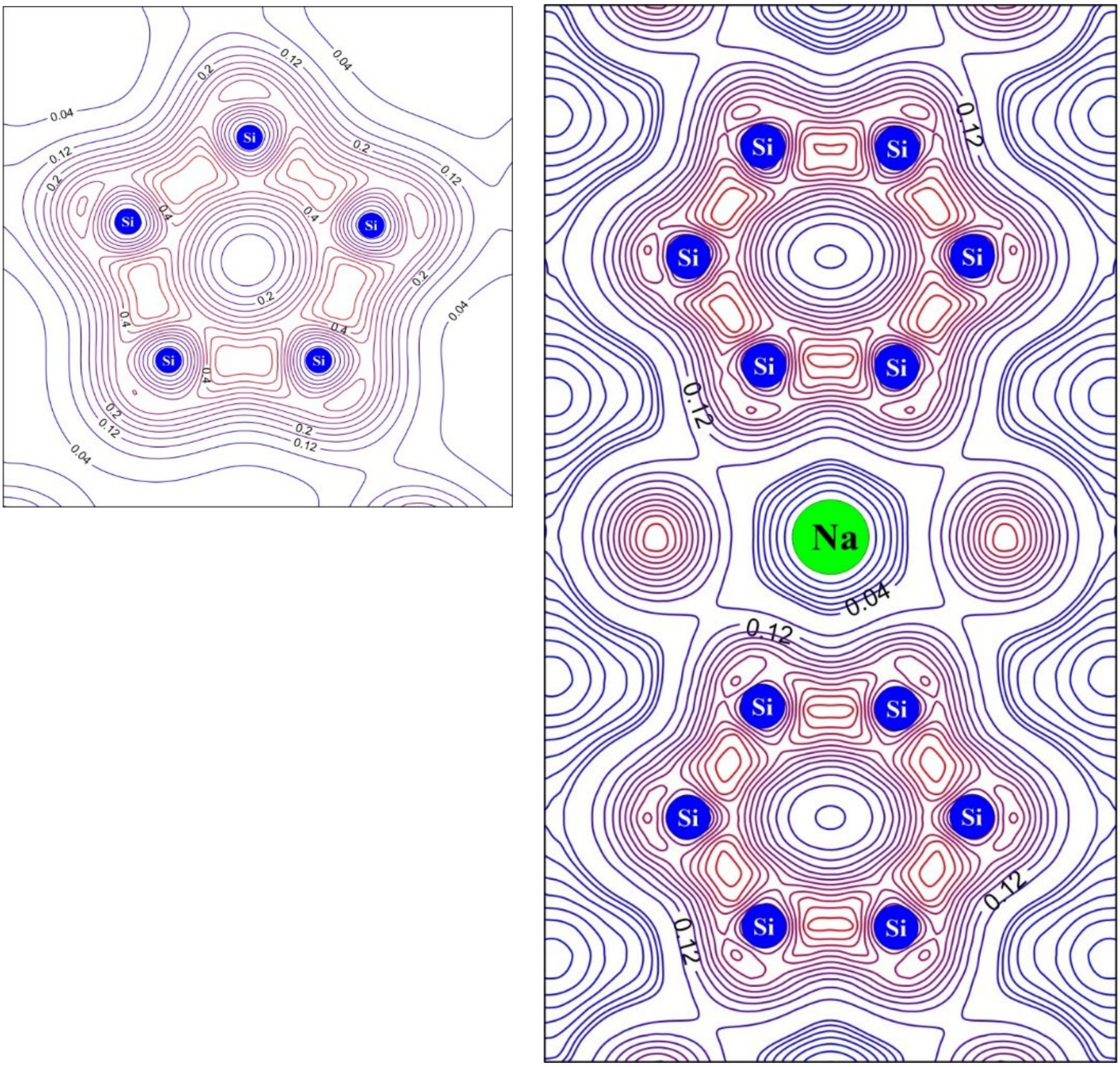

Рис. 8.17. Карти електронної густини клатрату $Na_8Si_{46}$: *а*) через грань Si5 багатогранника $Si_{20}$ (додекаедра); *б*) через грань Si6 багатогранника $Si_{24}$ (тетракайдекаедра) в площині (010).

Результати обчислень густини валентного заряду для клатратів $Si_{46}$ та $Na_8Si_{46}$ у площині [001] приведені на рис. 8.16 і 8.17 відповідно. Карти розподілу електронної густини в обох клатратах якісно ідентичні і наглядно демонструють наявність сильного зв’язку між атомами кремнію. На рис. 8.17 видно, що розподіл густини валентного заряду в додекаедрі $Si_{20}$ і в тетракайдекаедрі $Si_{24}$ клатрату $Na_8Si_{46}$ носить подібний характер і засвідчує наявність сильного зв’язку між атомами Si в каркасі «хазяїна» і відсутність зв’язку між атомами Si та інкапсульованими атомами «гостя» – Na.

Наявність ковалентної складової в кремнієвих клатратах обумовлена гібридизацією 3*s*- і 3*p*-станів Si, як і в кристалічному кремнію. Іонна складова хімічного звязку характеризується: зарядами, локалізованими на самих атомах; асиметрією (поляризацією) ковалентного звязку, що проявляється у зміщенні зарядів на звязку і деформації контурів сталої густини.

## 8.7. ЗІСТАВЛЕННЯ ТЕОРІЇ З ЕКСПЕРИМЕНТОМ

Основну експериментальну інформацію про електронну структуру кристалів отримують взаємодоповнюючими спектроскопічними методами: ультрафіолетова (Ultraviolet Photoelectron Spectra – UPS) та рентгенівська фотоелектронна спектроскопія (X-ray Photoelectron Spectra – XPS), м'яка рентгенівська емісійна спектроскопія внутрішніх і валентних рівнів (X-ray Emission Spectroscopy – XES), рентгенівська абсорбційна спектроскопія (X-ray Absorption Spectroscopy – XAS), оптичне відбивання та поглинання.

Валентні стани зазвичай вивчають за допомогою ультрафіолетової та рентгенівської фотоелектронної спектроскопії. Зі збільшенням роздільної здатності рентгенівських фотоелектронних спектрометрів стало можливим успішно вивчати як валентні, так і остовні стани методами рентгенівської фотоелектронної спектроскопії.

У методах класичної фотоелектронної спектроскопії вимірюють енергетичний розподіл електронів, вибитих із зразка ультрафіолетовим або рентгенівським випромінюванням. Під дією падаючого на зразок випромінювання електрони збуджуються з валентних станів у зону провідності й далі після ряду складних процесів залишають зразок і реєструються у надвисокому вакуумі. Спектр цих електронів залежить від густини електронних станів валентної зони і зони провідності, їхньої симетрії, ймовірності переходів між зонами, точки зони Бріллюена переходу, а також стану поверхні досліджуваного зразка. В ультрафіолетовій фотоелектронній спектроскопії на розподіл густини електронних станів у валентній зоні кристалів накладається розподіл станів у зоні провідності. Збільшення енергії збудження до значень, характерних для рентгенівської фотоелектронної спектроскопії, дозволяє ефективно виключити вплив структури зони провідності.

На теперішній час наявні результати дослідження РФС спектра тільки кристала кремнію c-Si [277]. Оскільки рентгенівські фотоелектронні спектри кристала c-Si відображають розподіл повної густини електронних станів у валентній зоні, тому на рис. 8.18 в єдиній енергетичній шкалі зіставлені експерементальний РФС спектр, взятий з роботи [277], з розрахованою нами повною густиною електронних станів *N*(*E*). При зіставленні результатів теоретичних обчислень густин станів з експериментальними даними необхідно враховувати, що експеримент завжди виконується з певним енергетичним розділенням. Так, у рентгенівській емісійній спектроскопії енергетичне розділення визначається переважно шириною внутрішнього рівня, а в рентгенівській фотоелектронній спектроскопії – шириною лінії рентгенівського випромінювання, що використовується для збудження фотоемісії [278].

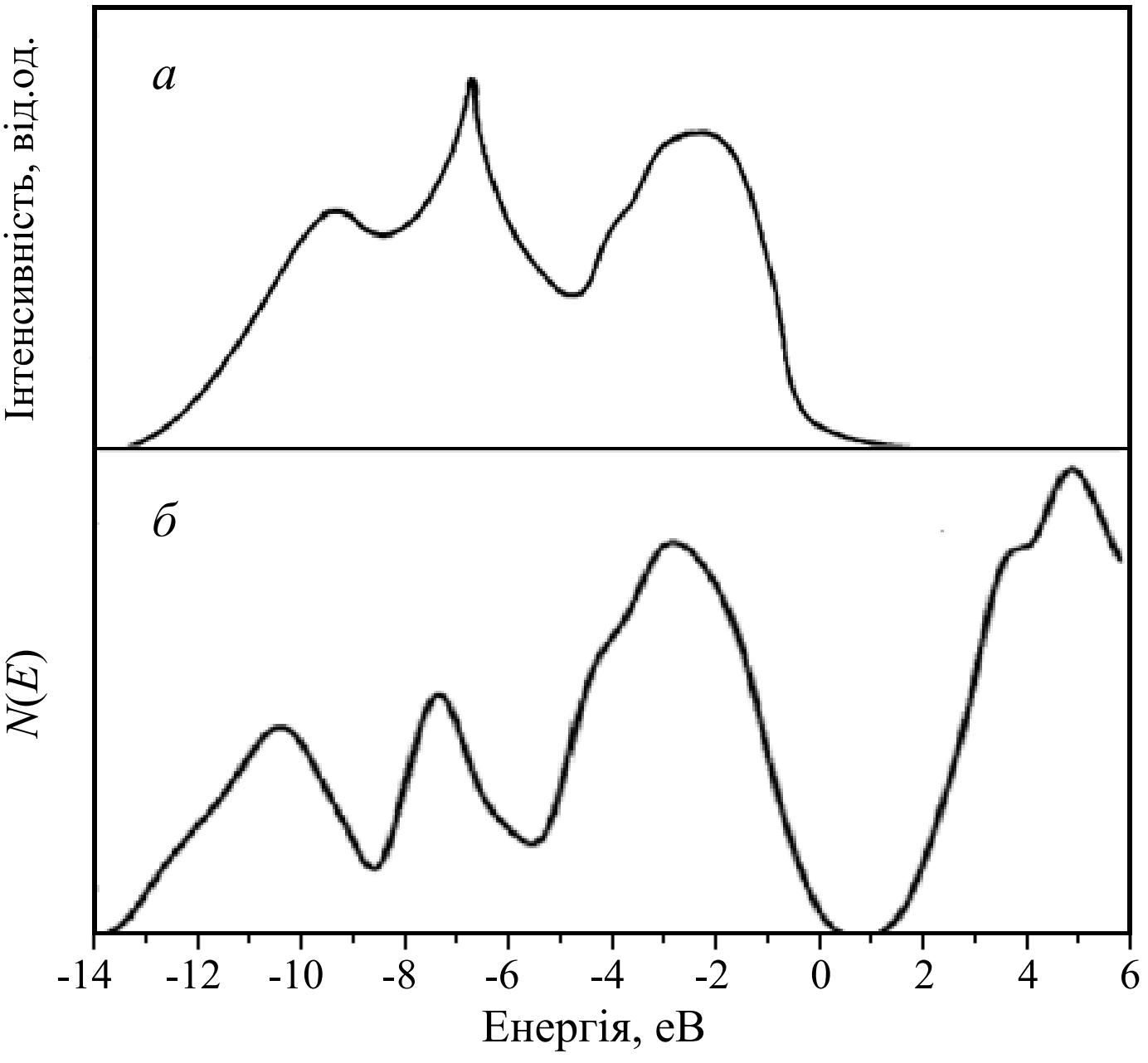

Рис. 8.18. Порівняння розрахованої згладженої повної густини електронних станів *N*(*E*) у валентній зоні (*б*) з експериментальним рентгенівським фотоемісійним спектром (*а*) кристала c-Si [277].

При встановленні відповідності між особливостями на експериментальних кривих і особливими точками енергетичної зони *E*(k) і густини електронних станів *N*(*E*) необхідно враховувати величину роздільної здатності спектрометра. Тому обчислені густини електронних станів розмивають на криву апаратних спотворень спек-трометра. Розмиття кривої $N(E)$ здійснюється гаусовою кривою з напівшириною на половині максимуму 0.8 еВ для станів, розташованих біля верха валентної зони.

Одним із добре апробованих методів дослідження електронної структури кристалів є метод рентгенівської емісійної спектроскопії (РЕС), який дозволяє отримати інформацію про характер розподілу парціальних густин електронних станів окремих компонент [278]. Суть методу РЕС полягає у тому, що в результаті опромінення твердого тіла (наприклад, c-Si, $Na_8Si_{46}$) на остовних атомних рівнях Si 1*s*-, Si 2*s*- утворюються вакантні стани (дірки). На ці стани переходять електрони із валентної зони. Такі переходи супроводжуються емісією рентгенівського випромінювання. Інтенсивність випромінювання з точністю до залежності матричного елемента переходу від енергії пропорційна густині електронних станів у валентній зоні та імовірності переходу. Короткохвильові лінії кожної серії характеристичного рентгенівського спектра (емісійні смуги), що виникають під час заповнення вакансій на тому чи іншому внутрішньому рівні електронами валентної зони, містять у собі інформацію про стани валентних електронів у твердому тілі. У разі хімічної сполуки є можливість роздільного вивчення рентгенівських емісійних спектрів різних серій від кожної компоненти сполуки, що в сукупності дає повну інформацію про електронну структуру кристала. Спільний розгляд рентгенівських спектрів різних серій і компонент зазвичай проводиться шляхом їх розташування в єдиній шкалі енергій. Рентгенівські емісійні $K\beta_{1,3}$- і $L_{2,3}$- – спектри відображають розподіл по енергії зайнятих *p*- і *s*-станів кремнію у валентній смузі кристалічного кремнію і клатрату $Na_8Si_{46}$. Такі переходи супроводжуються емісією рентгенівського випромінювання. Інтенсивність випромінювання пропорційна густині електронних станів у валентній зоні та ймовірності переходу. Оскільки повна ширина валентної зони кристалічного кремнію і клатрату $Na_8Si_{46}$ складає 13.3 еВ і 12.55 еВ відповідно, то вважається, що матричний елемент переходу слабо залежить від енергії. У цьому випадку рентгенівські емісійні спектри відображають розподіл парціальних густин електронних станів у валентній зоні.

З метою перевірки коректності виконаних розрахунків електронної структури кристала c-Si і клатрату $Na_8Si_{46}$, необхідно провести зіставлення в єдиній енергетичній шкалі розрахованих парціальних густин електронних станів з наявними в літературі [279–281] рентгенівськими $K\beta_{1,3}$- і $L_{2,3}$- емісійними спектрами, які несуть інфор-мацію про розподіл валентних електронів різної симетрії в заповненій частині валентної зони.

Розраховані згладжені парціальні густини Si 3*s*- і 3*p*-станів у кристалічного кремнію і клатрату $Na_8Si_{46}$, а також експериментальні Si $K\beta_{1,3}$- і Si $L_{2,3}$ – емісійні спектри, взяті з робіт [281] і [283] відповідно, суміщені в єдиній енергетичній шкалі, наведені на рис. 8.19 і 8.20. За початок відліку енергії прийнято положення верха валентної зони.

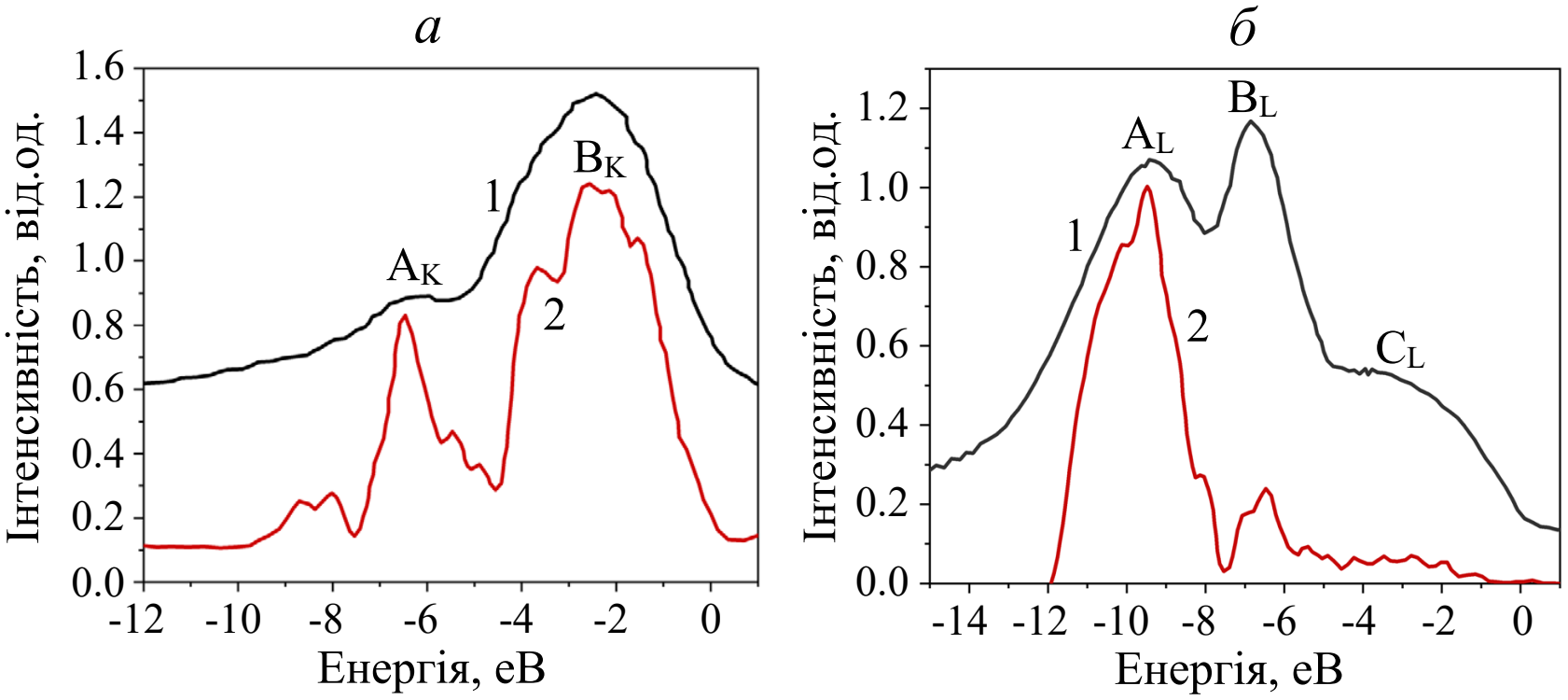

Рис. 8.19. Розраховані згладжені 3*p*- і 3*s*- парціальні густини електронних станів (криві 2) у валентній зоні та експериментальні рентгенівські спектри емісії Si $K_{\beta1,3}$ (*а*, крива 1) і Si $L_{2,3}$ (*б*, крива 1) кристала Si [281].

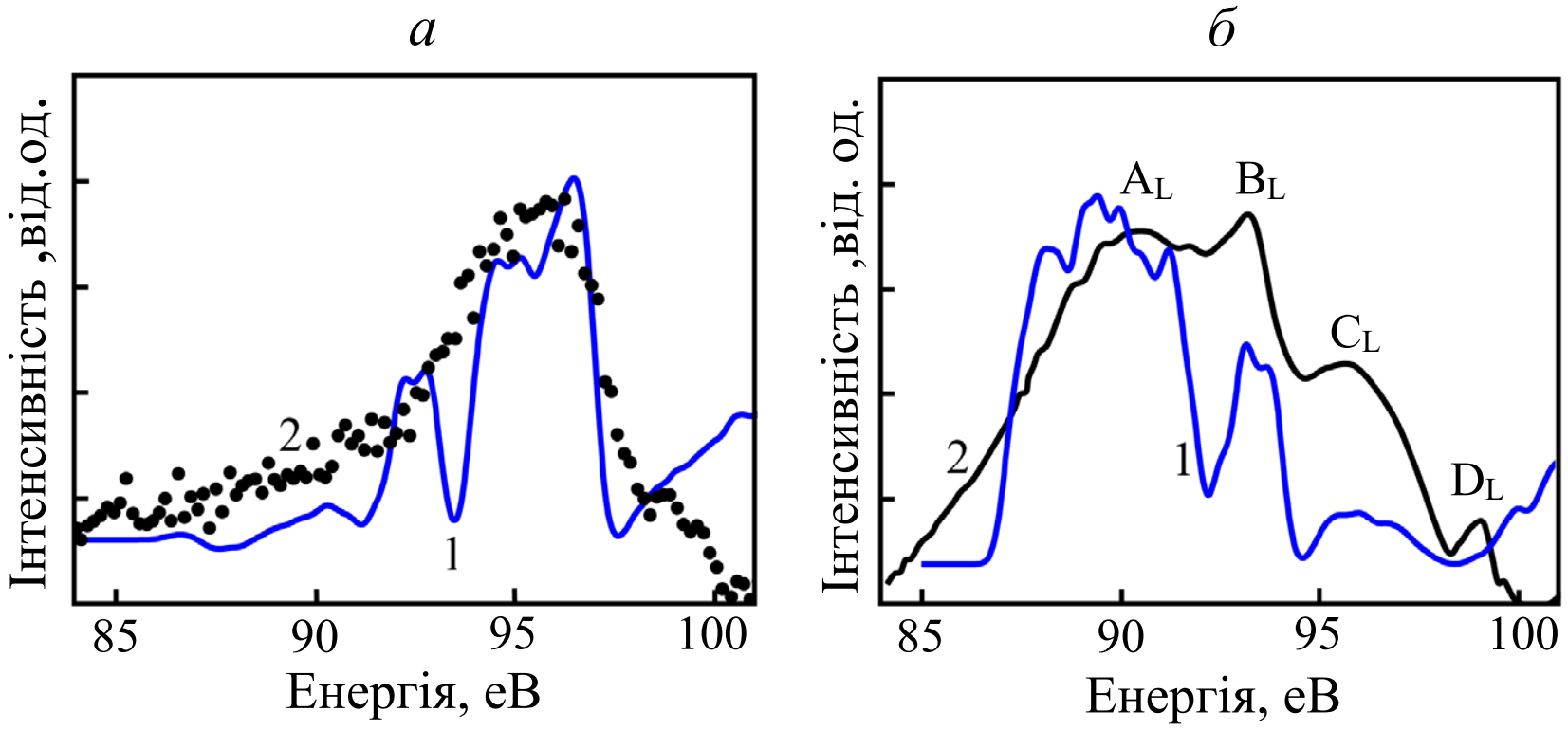

Рис. 8.20. Розраховані згладжені 3*p*- і 3*s*- парціальні густини електронних станів кремнію (криві 1) у валентній зоні та експериментальні Si $K_{\beta1,3}$ – (*а*, крива 2) і Si $L_{2,3}$ – емісійні спектри (*б*, крива 2) клатрату $Na_8Si_{46}$ [281].

У дипольному наближенні рентгенівський емісійний Si $L_{2,3}$ – спектр виникає завдяки переходу електронів з 3*s*- і 3*d*-рівнів на вакансії Si 2*p*-рівня. Накладення рентгенівських смуг на криві густини електронних станів проводилося шляхом поєднання головного максимуму Si $K\beta_{1,3}$- і Si $L_{2,3}$- смуг із максимумом середньої смуги густини електронних станів. В Si$K\beta_{1,3}$- спектрах реєструються переходи з Si 3p-станів валентної зони на Si 1*s*- атомні стани. Їхній внесок максимальний поблизу середини верхньої валентної зони. Принципово зазначити, що в цю саму верхню підзону, в якій домінують 3*p*-стани, дають також внесок Si 3*d*-стани. У відповідності з дипольними правилами відбору $L_{2,3}$- спектри атомів Si відображають розподіл Si 3*s*- і Si 3*d*- станів по валентній зоні. Як випливає з розрахунку, максимуми при енергіях ~8.2 і ~2.1 eВ нижче верха валентної зони в $L_{2,3}$ – спектрах c-Si обумовлені *s*-станами. *d*-стани проявляються тільки у вигляді незначного напливу в області 3.5 eВ в $L_{2,3}$ – спектрі c-Si. Відносна інтенсивність останнього невелика, оскільки матричний елемент ймовірності переходу для *s*-станів приблизно на порядок більша нід для *d*-станів. Характерно, що цей слабкий за інтенсивністю пік в $L_{2,3}$ – спектрі c-Si проявляється і в клатрату $Na_8Si_{46}$, який має таку ж саму кристалічну гратку, як і клатрат $Si_{46}$. Таким чином, перенормований Si $L_{2,3}$ – спектр клатрату $Na_8Si_{46}$ являє собою суперпозицію Si 3*s*- і Si 3*d*- парціальних густин електронних станів. Загалом енергетичні положення основних особливостей розрахованих кривих pDOS узгоджуються з особливостями, спостережуваними в PEC спектрах. Форма рентгенівської Si $L_{2,3}$ – смуги чутлива не тільки до хімічного зв’язку з іншими елементами, але й до структурного стану кремнію, трохи відрізняючись для клатрату $Na_8Si_{46}$ у порівнянні з кристалічним c-Si.

Порівняння теоретично розрахованих парціальних густин електронних станів і експериментальних Si $K\beta_{1,3}$- і Si $L_{2,3}$-спектрів показує, що проведені нами розрахунки в цілому відтворюють як енергетичну структуру Si $K\beta_{1,3}$- і Si $L_{2,3}$-спектрів кристалічного кремнію і клатрату $Na_8Si_{46}$ , так і їх форму. Разом з тим, відмітимо, що особливості наявні в розрахованих pDOS *s*- і *p*- електронів Si не проявляються в експериментальних спектрах PEC кристала c-Si і клатрату $Na_8Si_{46}$ із за обмеженої роздільної здатності спектрометра, вони представлені в емісійних спектрах піками A, B і напливом C в Si $L_{2,3}$ – спектрах.

У Si $L_{2,3}$-спектрі клатрату $Na_8Si_{46}$ зя’вляється додатковий максимум у прифермієвській області (рис. 8.20, *б*). Появу додаткової особливості в спектрі клатрату $Na_8Si_{46}$ (кремнію інтеркальованого атомом лужного металу Na) можна пояснити наступним чином. При заповненні порожнин кремнієвої гратки атомами Na збільшується число валентних електронів на елементарну комірку клатрату, що приводить до зміщення рівня Фермі у зону провідності. У Si $K\beta_{1,3}$- і Si$L_{2,3}$-спектрах це приводить до появи в околі $E_F$ додаткового максимуму. Зіставлення розрахованих нами парціальних густин електронних станів з рентгенівськими $K\beta_{1,3}$ – а також емісійними $L_{2,3}$-спектрами атомів Si для кристалів c-Si і клатрату $Na_8Si_{46}$ шляхом накладання експериментальних [279–281] і розрахованих кривих дозволило отримати хорошу якісну і кількісну узгодженість теорії з експериментом, а також ідентифікувати природу цих максимумів.

Метод DFT–LDA використаний нами для розрахунків електронних структур клатратів $Si_{46}$ і $Na_8Si_{46}$, хоча і занижує ширину забороненої зони, але достовірно відтворює енергетичну структуру валентних зон і зони провідності.

## 8.8. СИНТЕЗ І КРИСТАЛІЧНА СТРУКТУРА КЛАТРАТІВ У СИСТЕМІ Si–Te

**8.8.1. Синтез клатратів $Te_{7+x}Si_{20-x}$ ($x$~2.5) і $Te_{16}Si_{38}$.** Дослідивши бінарну систему Si–Te в умовах високого тиску до 5 ГПа і високих температур автори [284–286] виявили ще дві нові сполуки, структура яких близька до структури класичного клатрату типу-I. Перша з них відповідає формулі $Te_{7+x}Si_{20-x}$ з $x \sim 2.5$ (молярне співвідношення Te/Si = 0.543). Другою виявилась сполука $Te_{16}Si_{38}$ (Te/Si = 0.421), яка існує у двох фазах – кубічній і ромбоедричній. Зміна в структурах цих легованих телуром кремнієвих клатратів, у порівнянні з вихідними клатратами $Si_{46}$ і $G_8Si_{46}$ типу-I, є наслідком особливого характеру атома телуру, який не лише поводиться як «гість», але й приймає участь у формуванні каркасу.

Клатратні сполуки $Te_{7+x}Si_{20-x}$ ($x$~2.5) і $Te_{16}Si_{38}$ автори [284–286] синтезували при високих тисках 3–7 ГПа та високих температурах 973–1123 К. Синтез клатратної сполуки $Te_{7+x}Si_{20-x}$ ($x$~2.5) автори [284] проводили в апараті високого тиску стрічкового типу з діаметром отвору 12 мм, здатним працювати до 9 ГПа в діапазоні температур від 293 до 1773 К. Подрібнені Si та Te у відповідних кількостях були завантажені в платинові тиглі ізольовані від графітової мікропечі з пірофілітом. Ці комірки тиску потім були поміщені в камеру високого тиску. Зразки спочатку піддавали дії тиску, а потім нагрівали до необхідної температури зі швидкістю 150 К·*хв*$^{-1}$. Після завершення реакції зразки охолоджували до кімнатної температури, після чого декомпресували. Оптимальними умовами синтезу клатрату $Te_{7+x}Si_{20-x}$ ($x$~2.5) є тиск 5 ГПа, температура 1073 К і тривалість процесу 30 хв [284].

Синтез клатратів $Te_{16}Si_{38}$ також здійснювали в камері високого тиску з використанням стрічкового преса, здатного створювати тиск на зразок до 5 ГПа в діапазоні температур від 293 до 1673 К [285]. Суміш вихідних елементів була спресована у вигляді циліндричної таблетки відповідного розміру і завантажена в тигель із нітриду бору, який був поміщений у графітову піч. Електричні контакти між піччю та джерелом живлення були виконані за допомогою сталевих штампів. Температуру контролювали за допомогою термопари, розміщеної в отворі, що проходить через камеру тиску до стінки тигля з нітриду бору. Пірофіліт був використаний як середовище передачі тиску. Тиск повільно підвищували до 5 ГПа, після чого зразок нагрівали до температури 1098 або 1473 К зі швидкістю 10 – 18 К·хв$^{-1}$. При заданому тиску зразок витримували протягом 60 хв, а потім гартували до кімнатної температури. Після цього тиск повільно зменшували протягом однієї години до атмосферного тиску, і комірку високого тиску виймали із стрічки. Циліндричний зразок автори [285] вилучали та аналізували методом порошкової рентгенівської дифракції.

**8.8.2. Кристалічна структура кубічної й ромбоедричної фаз $Te_{16}Si_{38}$.** Результати структурного аналізу показали, що обидві фази $Te_{16}Si_{38}$ відноситься до родинного класичного клатрату типу-I, $G_8Si_{46}$ (G = тип атома «гостя»), в яких наявні два типи клатратоутворюючих поліедрів – пентагональний додекаедр $Si_{20}$ і тетракайдекаедр $Si_{24}$ (рис. 8.21). Атоми «гостя» Te у структурі кубічної фази $Te_{16}Si_{38}$ клатрату типу-I знаходяться в позиціях Te1(2*a*) і Te2(6*c*) у центрах додекаедра і тетракайдекаедра відповідно. У кубічному клатраті атоми $Te_3$ заміщують атоми кремнію в позиціях 8*c*. Симетрія кубічної фази описується просторовою групою $P\overline{4}3n$ і параметром гратки $a$ = 10.457 Å, ромбоедричної – ПГ *R3c* і параметром гратки $a$ = 10.465 Å, $\alpha$ = 89.88° [286].

Обидві просторові групи, визначені для кубічної та ромбоедричної фаз $Te_{16}Si_{38}$, відповідають родинній підгрупі, а їх параметри ґратки трохи більше, ніж у класичного клатрату типу-I, що має чисту кремнієву ґратку-носій (табл. 8.1). Цей факт разом із тим, що молярне відношення Si/Te у кубічній і ромбоедричній фазах $Te_{16}Si_{38}$ мен-ше 46/8, тобто 5.75, вказує на те, що частина атомів телуру заміщує атоми кремнію в ґратці-«хазяїна». Уточнення структур ясно показало, що дійсно має місце часткове заміщення атомів Si на атоми Te і що це заміщення відбувається тільки в 16*i* Si2 позиціях родинної структури, які тепер розщеплені або на дві (кубічна фаза) або чотири (ромбоедрична фаза). Одним із наслідків цього заміщення є зміна складу, яку можна представити як $Te_8@(Si_{38}Te_8)$. Атомне відношення Si/Te становить 2.375 і відповідає експериментальному значенню. Це значення є проміжним між значенням незаміщеного клатрату типу-I (5.75) і більш заміщеного $Te_{7+x}Si_{20-x}$ ($x$ ~ 2.5), для якого Si/Te становить 1.84.

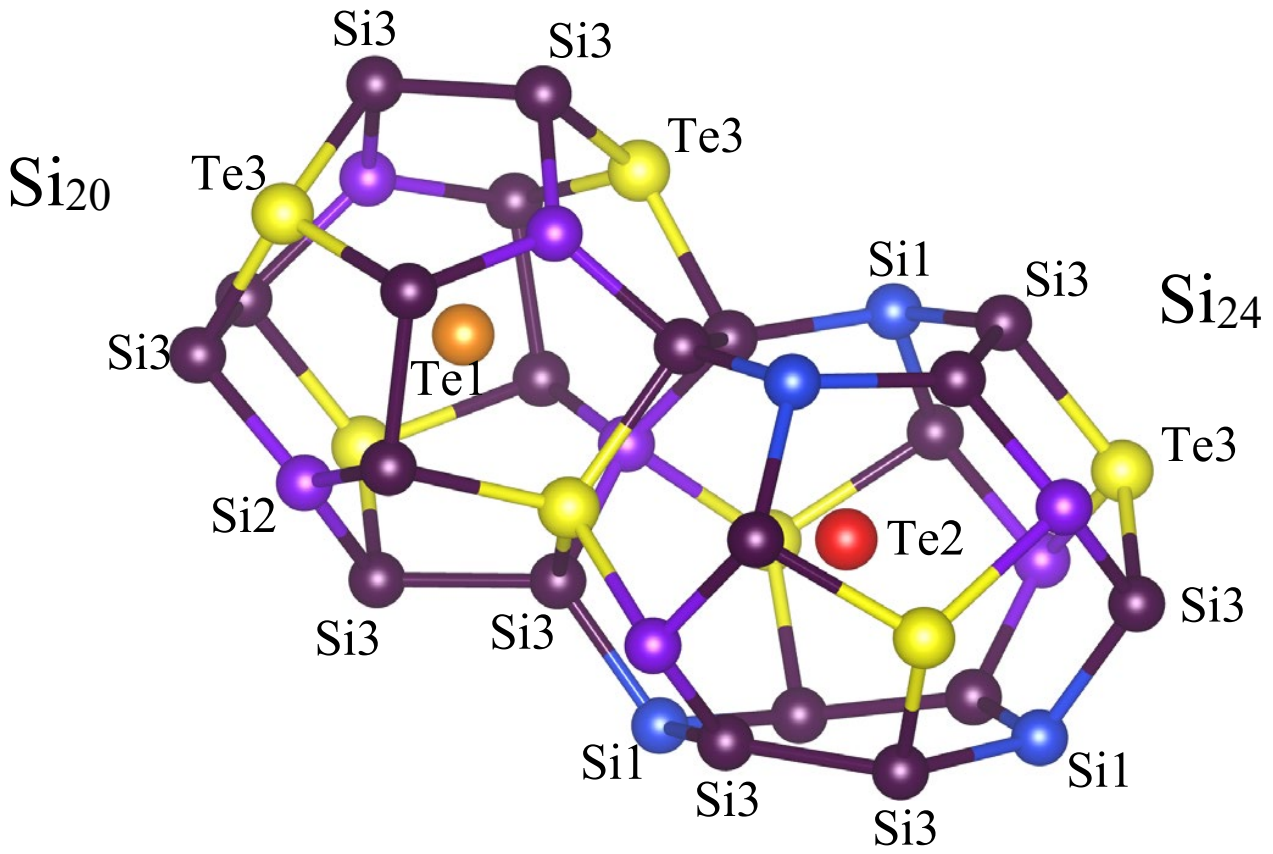

Рис. 8.21. Увязування двох поліедрів $Si_{20}$ і $Si_{24}$, легованих телуром.

Проекція кубічної фази $Te_{16}Si_{38}$ вздовж одного з напрямків (100) наведена на рис. 8.22, *а*. Структура виглядає дуже подібною до структури класичного клатрату типу-I, за винятком того, що п'ятикутні та гексагональні грані клітинок тут деформовані внаслідок заміщення 8 кремнієвих атомів у позиції 16*i* родинної структури на 8 більших атомів Te3. Наявність атомів Si32 не має великого впливу на додекаедричні клітинки, за винятком того, що вони сприяють їх спотворенню, але це викликає більш важливі зміни в тетракайдекаедричних клітинках. Близький вигляд двох із цих тетракайдекаедричних клітинок, з'єднаних спільною деформованою гексагональною поверхнею, представлений на рис. 8.22, *б*.

У кубічній фазі $Te_{16}Si_{38}$, яка є найближчою до вихідної структури з ПГ $Pm\bar{3}n$, найбільш важливою зміною є наявність міцних зв'язків

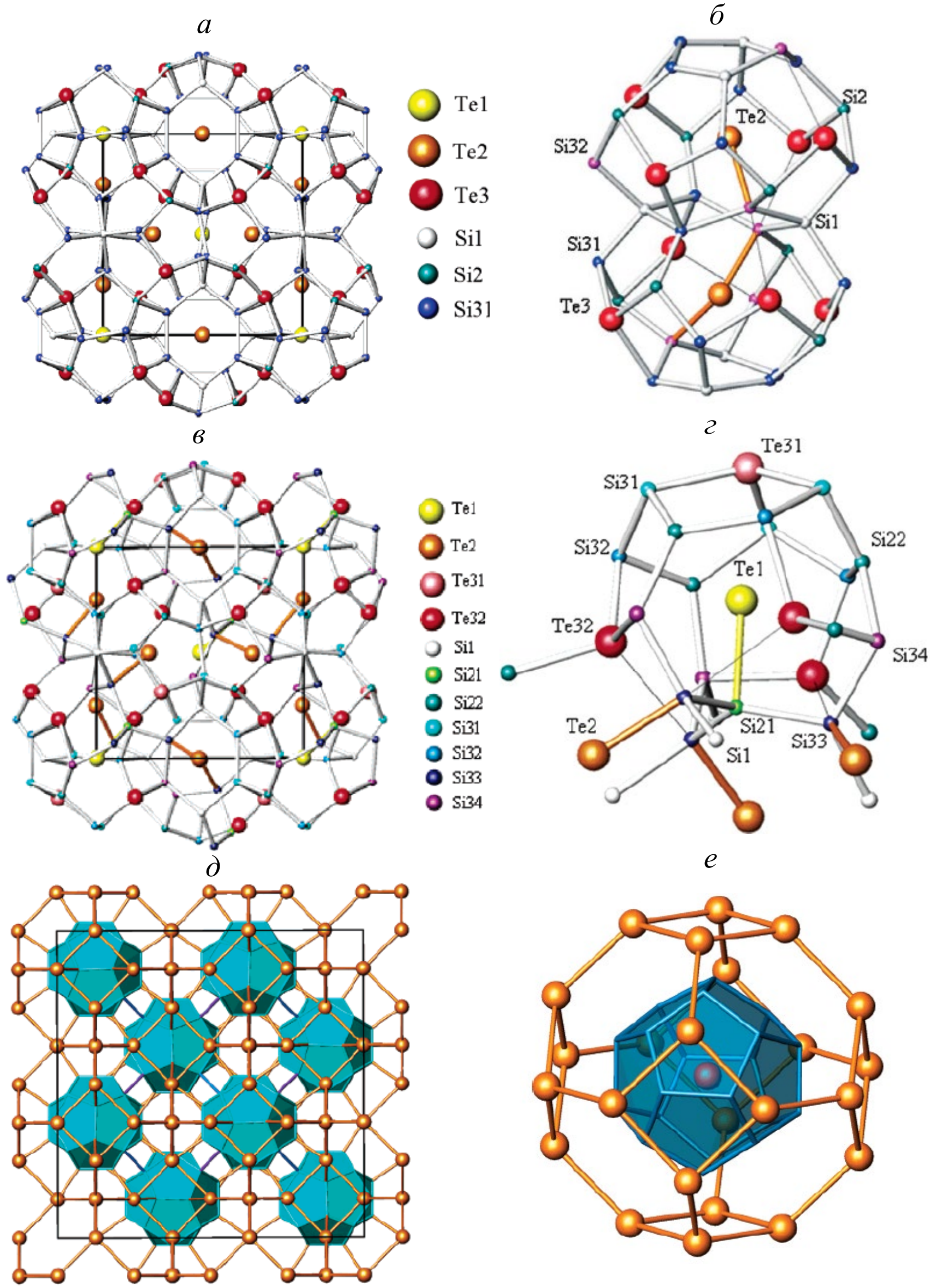

Рис. 8.22. Структура кубічного (*а*) і ромбоедричного (*б*, *г*) клатрату $Te_{16}Si_{38}$; *в*, *д* – кубічного клатрату $Te_{7+x}Si_{20-x}$ [286].

між атомами Te2, розташованими в центрі тетракайдекаедричних клітинок, і одним або двома атомами кремнію його найближчої клітинки. Ці зв'язки виникають у результаті зміщення деяких атомів кремнію в положенні 24k до центру клітинки, внаслідок чого зникають антагоністичні зв'язки між цими атомами кремнію та атомами Te3 у положенні заміщення на клітинці (8e позиції, випущені з родинної структури 16i). Наявність таких трикоординованих атомів телуру в кремнієвих кристалічних ґратках має тенденцію до розкриття клітинок. Однак атоми Te1 у центрі додекаедричних клітинок залишаються ізольованими, оскільки відстань Te1–Te3, що рівна 3.353 Å, передбачає лише слабкі взаємодії 7-го типу [285].

Проекція структури ромбоедричної фази $Te_{16}Si_{38}$ уздовж напрямків (100) зображена на рис. 8.22, *в* яка демонструє зв'язок з кубічною фазою (рис. 8.22, *а*). Основні відмінності пов'язані з подвійним розщепленням кристалографічних позицій Te3, Si2, Si31 та Si32 (табл. 8.1).

Таблиця 8.1. Атомні позиції, зайняті «гостями» (G, Te) та атомами господаря (Si) [286].

<table>
<tr><td>$G_2Si_{46}$</td><td colspan="2">$Te_{16}Si_{38}(Te_8(Te_8Si_{38}))$</td><td>$Te_{7+x}Si_{20-x}$<br>($x$~2.5)</td></tr>
<tr><td>Кубічна;<br>$a$ = 10.3Å</td><td>Кубічна;<br>$a$ = 10.457 Å</td><td>Ромбоедр.;<br>$a$ = 10.465 Å</td><td>Кубічна;<br>$a$ = 21.136 Å</td></tr>
<tr><td>$Pm\bar{3}n$ (223)</td><td>$P\bar{4}3n$ (218)</td><td>α=89.88°; $R3c$ (161)</td><td>$Fd\bar{3}c$ (228)</td></tr>
<tr><td>G1(2a)</td><td>Te1(2a)</td><td>Te1(2a)</td><td>Te1</td></tr>
<tr><td>G2(6d)</td><td>Te2(6c)</td><td>Te2(6b)</td><td rowspan="2">Te4(96g)</td></tr>
<tr><td>Si1(6c)</td><td>Si1(6d)</td><td>Si1(6c)</td></tr>
<tr><td rowspan="4">Si2(16i)</td><td rowspan="2">Te3(8e)</td><td>Te31(2a)</td><td></td></tr>
<tr><td>Te32(6b)</td><td>X 2(64e)</td></tr>
<tr><td rowspan="2">Si2(8e)</td><td>Si21(2a)</td><td>X 3(64e)</td></tr>
<tr><td>Si22(6b)</td><td></td></tr>
<tr><td rowspan="4">Si3(24k)</td><td rowspan="2">2/3 Si31(24i)</td><td>Si31(6b)</td><td rowspan="4">Si1(192h)</td></tr>
<tr><td>Si32(6b)</td></tr>
<tr><td rowspan="2">1/3 Si32(24i)</td><td>Si33(6b)</td></tr>
<tr><td>Si34(6b)</td></tr>
</table>

На відміну від кубічної фази, де атоми Te2 в центрі тетракайдекаедричної клітинки з’єднані з одним або двома атомами Si32 (в середньому 1.33), що займають статистично 33% позицій, у ромбоедричній фазі вони з’єднані з одним атомом кремнію (Si33), який займає одну і чітко визначену позицію, що має однакову множинність (рис. 8.22, *в*). Кристалографічні центри, які займають атоми Te3, розщеплені в позиціях Te32 (6*b*) і Te31 (2*a*) (табл. 8.1).

Основна відмінність між ромбоедричною та кубічною фазами полягає в розщепленні атомів Si2 у позиції 8*e* кубічної фази на дві позиції 2*a* (Si21) та 6*b* (Si22) у спотвореній ромбоедричній фазі. Атом Si21 зміщений до центру додекаедричної клітинки та утворює міцний зв'язок з атомом Te1 (Si21–Te1 = 2.716 Å). Атом Te1 має формальний заряд 2– ($5s^25p^6$) і, отже, має три незв'язані пари, спрямовані на протилежний бік клітинки, до трьох п'ятикутників, для яких Te31 є спільною вершиною (рис. 8.22, *г*). Атом Te31, що виходить із положення Te3 кубічної структури (позиція 8e), займає тут позицію 2*a*, що має таку ж кратність, що і об'єкт Te1. Він зв'язаний з трьома атомами $Si31^+$ при 2.667 Å, що відповідає чверті з 24k позицій родинної структури $Pm\overline{3}n$ типу I.

У ромбоедричній фазі $Te_{16}Si_{38}$ атоми Te2 у центрі тетракайдекаедричних каркасів (позиція 6*b*) міцно зв'язані з одним типом кремнієвих каркасів (Si33), які займають чітко визначене місце з однаковою множинністю, причому ці атоми Si33 походять із розщеплення 24k позицій родинної структури $Pm\overline{3}n$. На відміну від ситуації в кубічній фазі $P\overline{4}3n$, атом Te1 у центрі додекаедричних вузлів каркасу (Te1:2*a*) утворює тут міцні зв'язки з атомами кремнію оточуючого каркаса, причому ці атоми знаходяться у вузлі, що має однакову кратність (Si21: 2*a*) та отриманий з розщеплення 16*i* позиції родинної структури $Pm\overline{3}n$. Зв'язки між гостьовими атомами Te в центрі обох клітинок і кремнієвими атомами Si21 і Si33 ґратки-господаря мають координаційне число лише 3 для атомів Te31 і Te32 у положенні заміщення на кремнієвих клітинках.

**8.8.3. Кристалічна структура клатрату $Te_{7+x}Si_{20-x}$.** Кристалічна структура клатрату $Te_{7+x}Si_{20-x}$ ($x$ ~ 2.5) тісно зв'язана зі структурою клатрату типу-I, але з іншим параметром елементарної комірки (~ 2 × $a_0$) та просторовою групою ($Fd\overline{3}c$ замість $Pm\overline{3}n$) [284]. Його головна особливість полягає в тому, що він відповідає подвійній клатрації атома Te, вкладеного в частково заміщений Te (12,5%) п'ятикутний додекаедр $Si_{20}$, який сам вкладений у великий поліедр Te24 у вигляді усіченого октаедра. Ця клатратна структура кремнію є першою, в якій представлені п'ятикутні додекаедри $Si_{20}$, які пов'язані тільки міжкластерними зв'язками.

У структурі клатрату $Te_{7+x}Si_{20-x}$ ($x \sim 2.5$) є ізольовані пентагональні додекаедри $Te@Te_{2.5}Si_{17.5}$, простір між якими заповнений атомами телуру. Дана сполука цікава ще й тим, що у ній спостерігається інверсія заряду – атоми телуру, які не утворюють поліедр, несуть негативний заряд, тобто, формально є діаніонами $Te2^{-}$.

Проекція кристалічної структури клатрату $Te_{7+x}Si_{20-x}$ на площину (100) приведена на рис. 8.21, *д*. Розподіл п'ятикутних додекаедрів $Si_{20}$ такий самий, як у клатраті родинного типу-I, за винятком того, що два сусідні п'ятикутні додекаедри тут зв'язані тільки міжкластерними зв'язками. Причиною збільшення майже вдвічі параметра елементарної комірки в клатраті $Te_{7+x}Si_{20-x}$ порівняно з клатратом родинного типу-I, зумовлено різницею у розташуванні п'ятикутних додекаедрів у двох структурах [284].

Вісім із двадцяти атомів Si кожного додекаедра частково заміщені атомами Te, що приводить до двох різних кристалографічних позицій (64e) для так званих атомів X2 та X3 (табл. 8.1). Слід зазначити, що ці частково заміщені Te атоми Si (X2 та X3), як було виявлено, мають однаковий коефіцієнт заміщення (заміни) (уточнене середнє значення 2.5/8), і вони відповідають атомам, що утворюють зв'язки між сусідніми п'ятикутними додекаедрами.

Центр усіх кластерів $Si_{20}$ зайнятий гостьовим атомом телуру (Te(1)), утворюючи першу стадію клатрації. Кластери $Si_{24}$ (багатогранники, що мають 14 граней), які заповнюють простір між п'ятикутними додекаедрами в структурі типу-I, тут відсутні через злиття 6*c* і 6*d* кристалографічних позицій просторової групи $Pm\overline{3}n$ клатрату типу-I в одну (96g) для нової групи $96 = (6 + 6) \times 8$). Ці позиції зайняті лише атомами Te(Te4). Це приводить до утворення нового типу поліедральної мережі, що складається з великих усічених октаедрів, розділяючи їх гексагональні та квадратні грані, і вершини яких зайняті 24 атомами Te. Кожен усічений октаедр включає п'ятикутний додекаедр з частково Te-заміщеного Si, центр якого зайнятий атомом телуру, утворюючи тим самим подвійну клатрацію.

## 8.9. ЕЛЕКТРОННА СТРУКТУРА КУБІЧНОЇ ТА РОМБОЕДРИЧНОЇ ФАЗ $Te_{16}Si_{38}$

Електронні структури кубічної та ромбоедричної фаз клатрату $Te_{16}Si_{38}$, розраховані методом функціонала електронної густини в LDA-наближенні без врахування спін-орбітальної взаємодії у точках високої симетрії й симетричних напрямках зони Брилюена (рис. 8.12), наведені на рис. 8.23 і 8.24 відповідно.

У результаті квантово-хімічних розрахунків встановлено подібність енергетичних структур валетної зони і зони провідності кубічної та ромбоедричної фаз $Te_{16}Si_{38}$. Подібність електронних структур кубічної й ромбоедричної фаз вказує на те, що основні особливості електронної структури визначаються у першу чергу сортом атомів, які приймають участь в утворенні хімічних зв'язків.

Оскільки, елементарна комірка клатрату $Si_{38}Te_{16}$ містить 16 шестивалентних атомів Te і 38 чотиривалентних атомів Si, то число валентних електронів у зоні Бріллюена рівне 248 і відповідно у валентній зоні Te-заміщеного клатрату $Te_{16}Si_{38}$ наявні 124 енергетичні зони. Тобто у валентній зоні кубічної і ромбоедричної фаз клатрату $Te_{16}Si_{38}$ міститься однакова кількість зон, а саме 124 енергетичні зони, об'єднані у чотири зв'язки, розділені енергетичними щілинами. Причина полягає в тому, що атоми гратки-«хазяїна» прагнуть до утворення тетраедричних зв'язків Si–Te.

Повна ширина валентних зон кубічної та ромбічної фаз становить 13.67 еВ і 13.14 еВ відповідно. Для кубічної фази верх валентної зони локалізований у напрямку Γ→X, а дно зони провідності знаходиться в напрямку Γ→M. Для ромбоедричної фази верх валентної зони локалізований у напрямку Γ→X, а дно зони провідності знаходиться в центрі зони Брилюена. Таким чином, згідно виконаних розрахунків електронної зонної структури кубічного та ромбоедричного клатрату $Te_{16}Si_{38}$, обидві фази є непрямозонними напівпровідниками з розрахованими ширинами забороненої зони $E_{gi}$ = 0.91 еВ і $E_{gi}$ = 0.94 еВ відповідно.

Розраховані енергетичні зонні структури та хвильові функції були використані для обчислення густин електронних станів і зарядового розподілу. Розраховані повні та локальні парціальні густини електронних станів кубічної й ромбоедричної фаз клатрату $Te_{16}Si_{38}$ наведені на рис. 8.25, *а* і 8.25, *б* відповідно.

Аналіз парціальних внесків у повну густину електронних станів $N(E)$ (рис. 8.25 *а*, *б*) дозволяє ідентифікувати генетичне походження

*а*

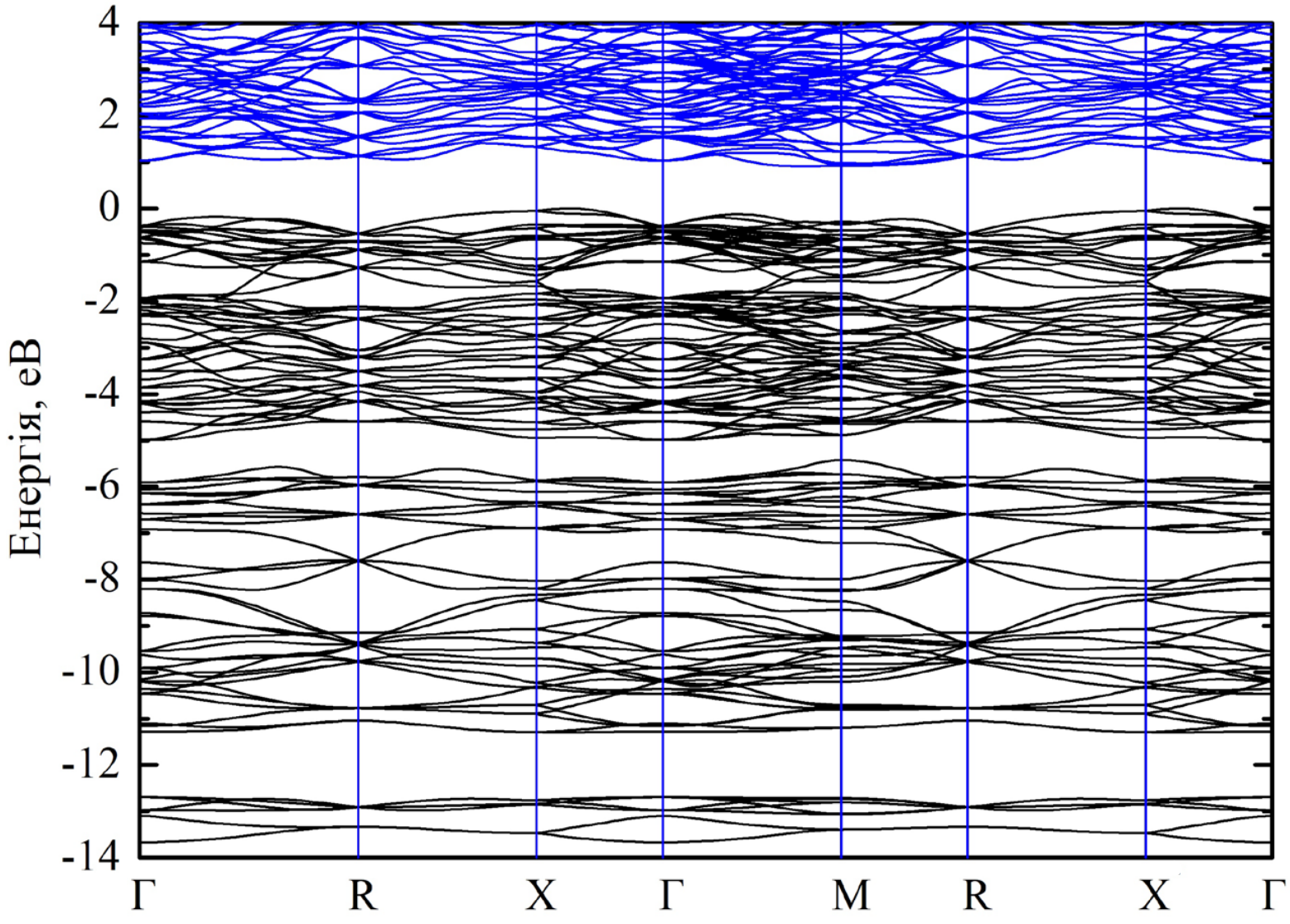

*б*

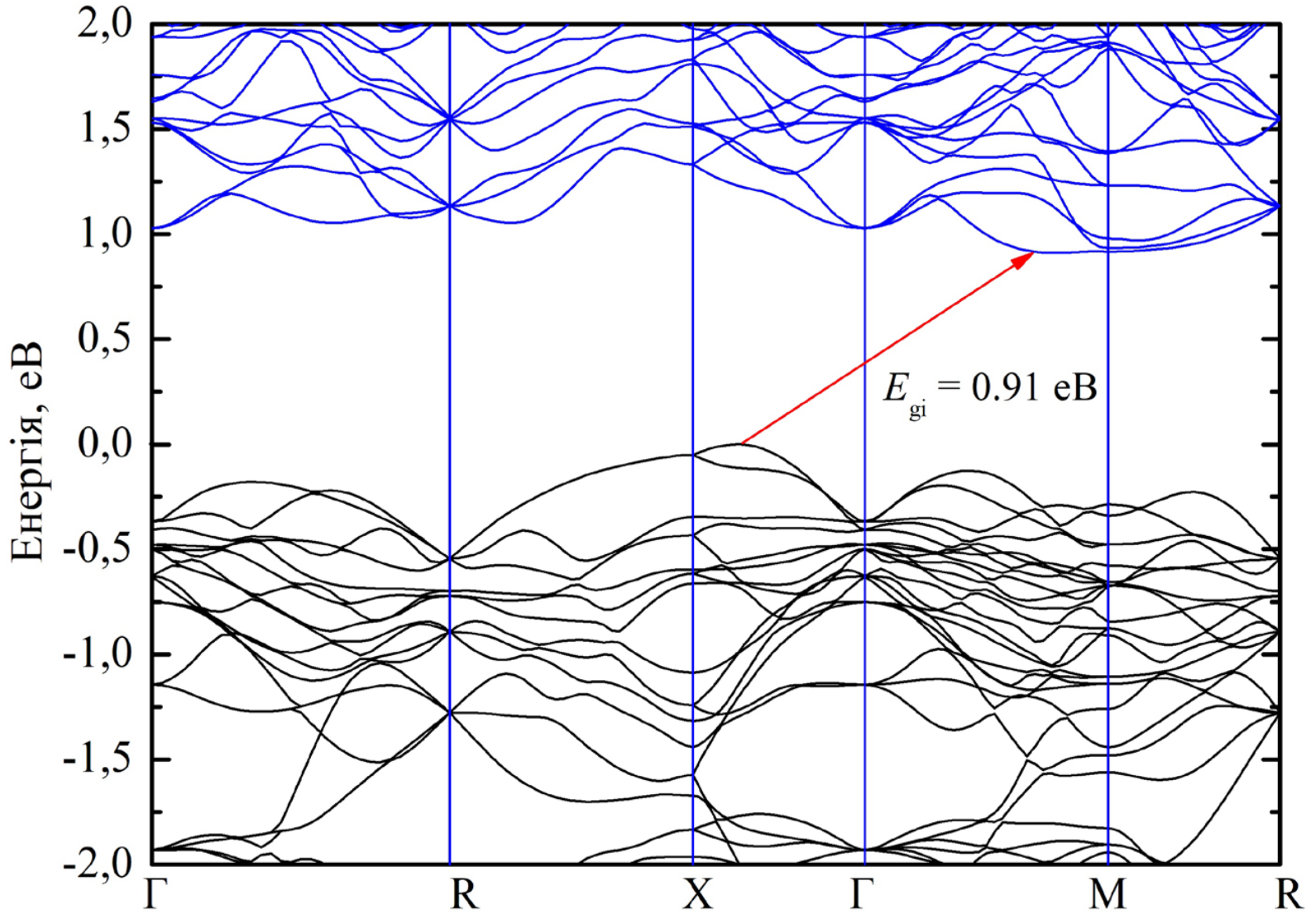

Рис. 8.23. Електронна структура кубічного клатрату $Te_{16}Si_{38}$.

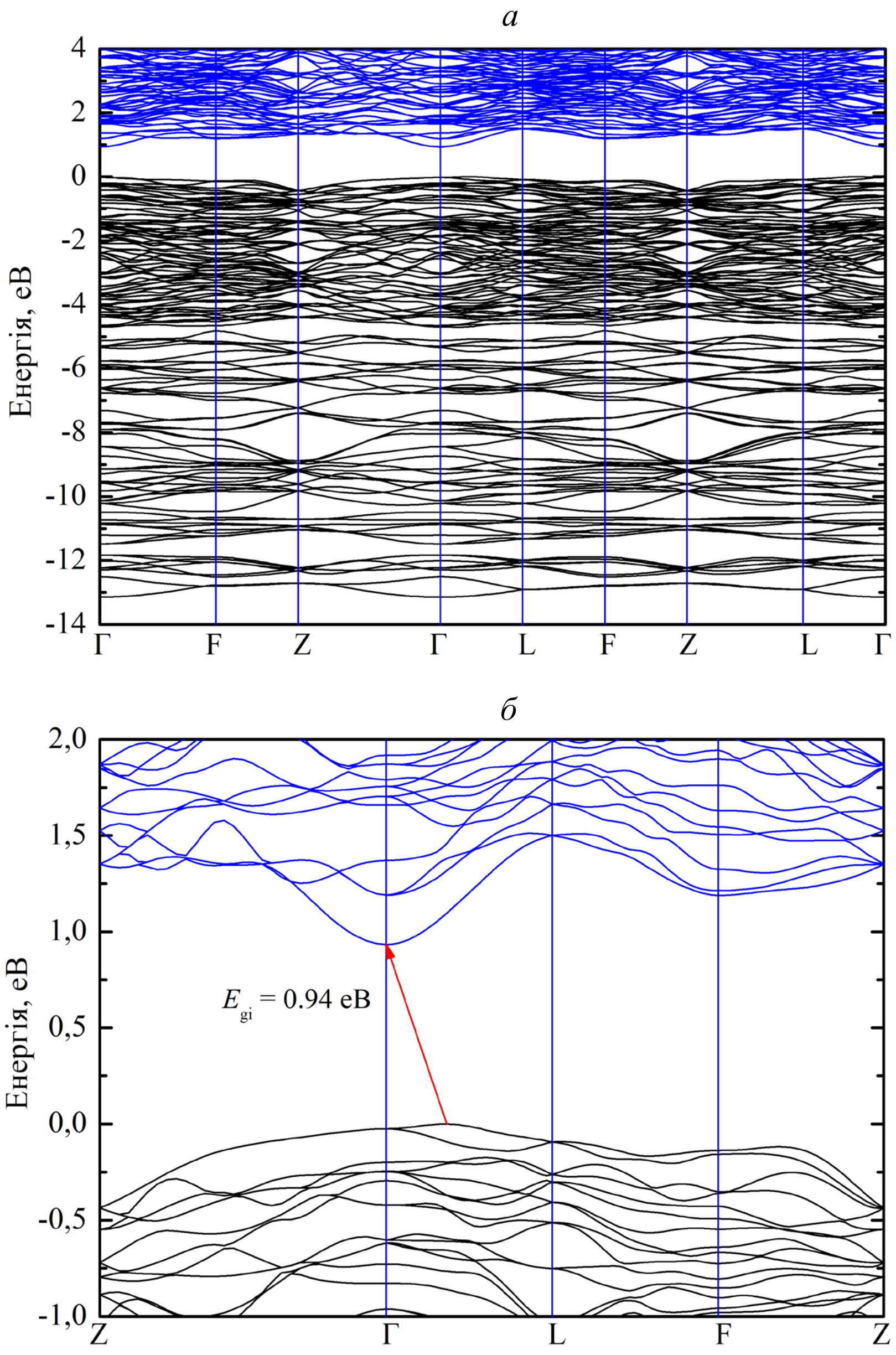

Рис. 8.24. Електронна структура ромбоедричного клатрату $Te_{16}Si_{38}$.

*а*

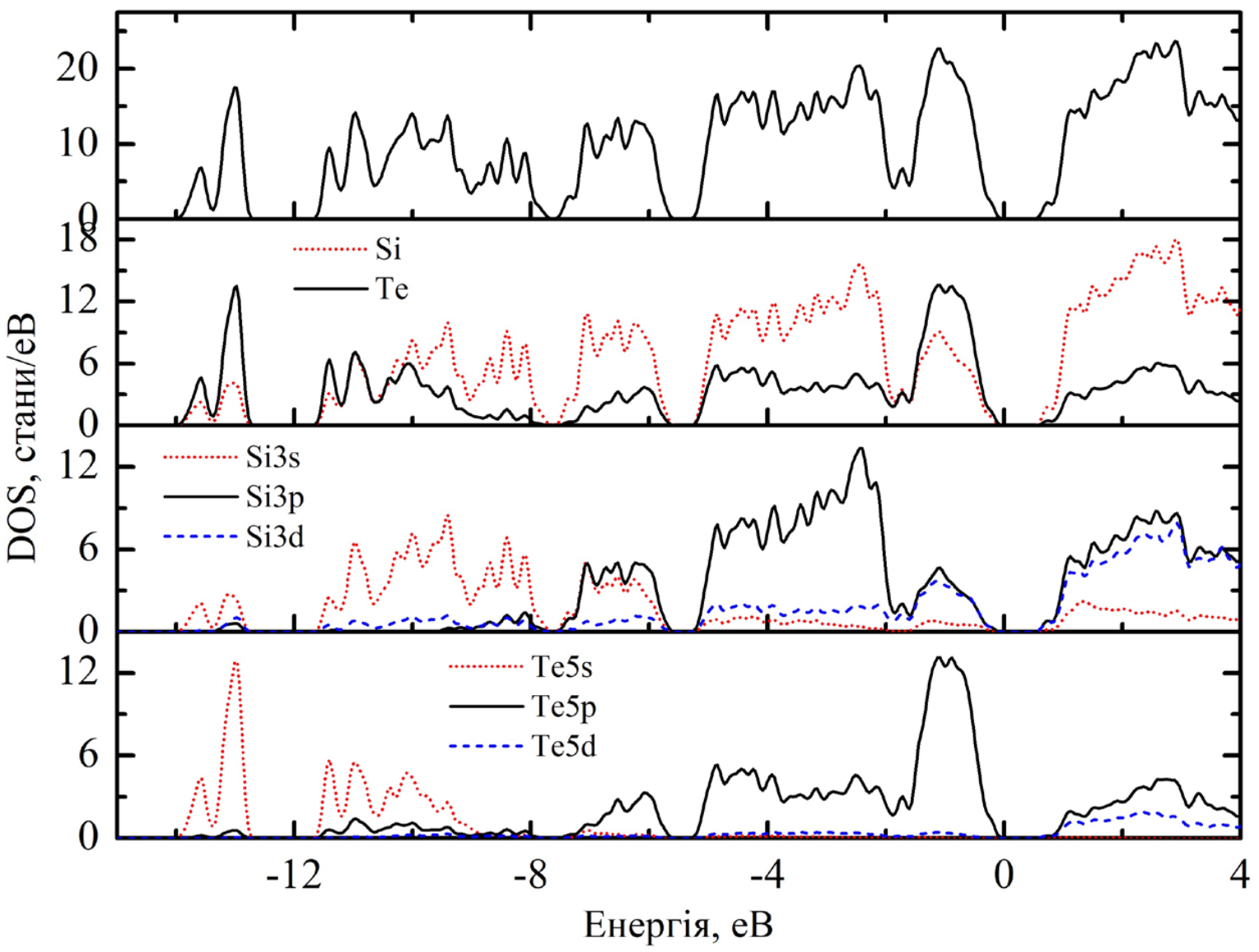

*б*

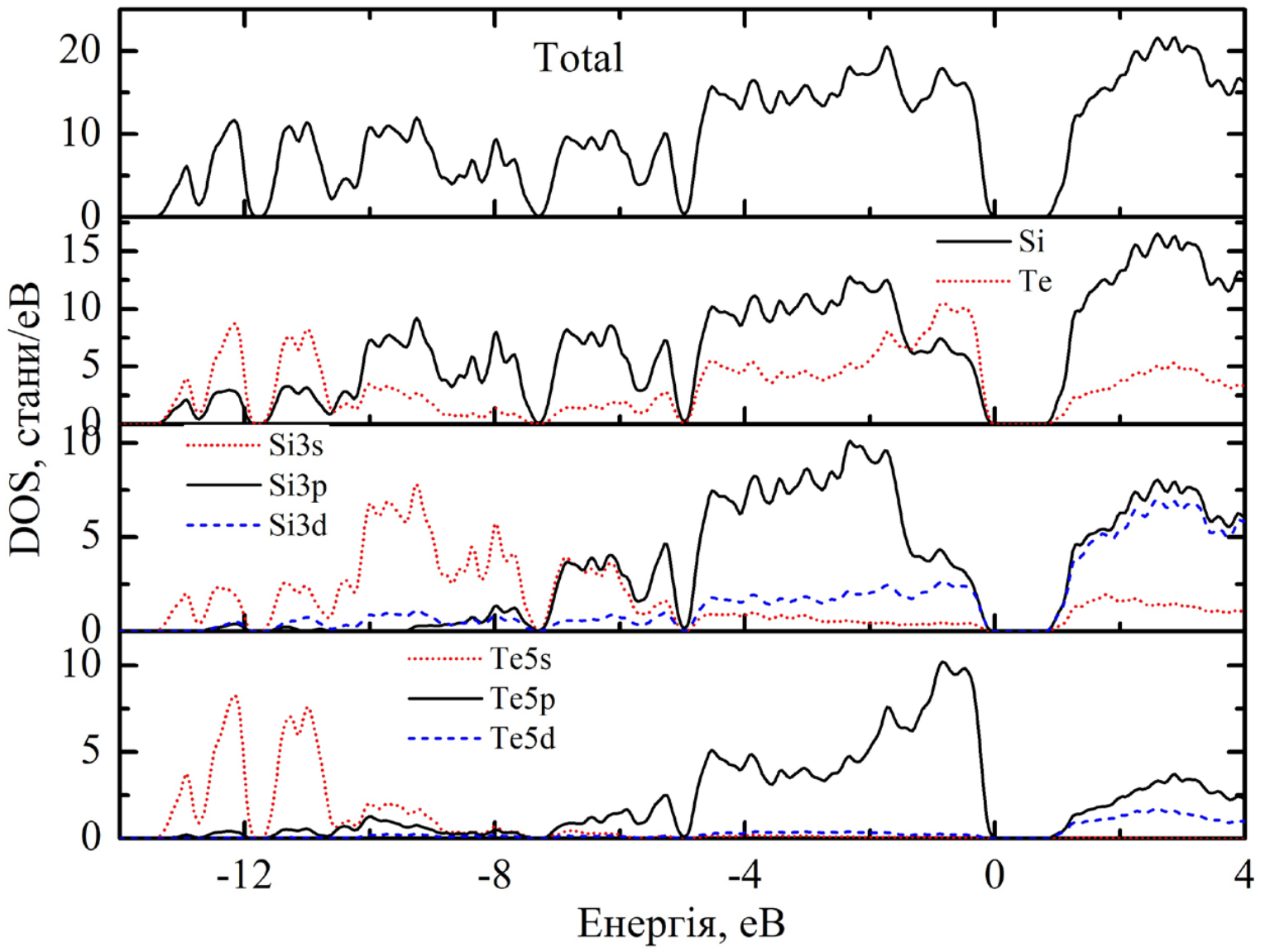

Рис. 8.25. Повна та локальні парціальні густини електронних станів кубічного (*а*) і ромбоедричного (*б*) клатрату $Te_{16}Si_{38}$.

різних підзон валентної зони та зони провідності кубічної й ромбоедричної фаз клатрату $Te_{16}Si_{38}$. Співвідношення між інтенсивностями максимумів у парціальних густинах електронних станів різного типу симетрії різні. У глибині валентної зони обох фаз у повній густині електронних станів $N(E)$ домінує внесок 5*s*-станів телуру, тоді як у верхній частині валентної зони переважаючим є внесок 5*p*-станів атомів Te. Дві найнижчі валентні підзони із 8 і 24 енергетичних зон, сформовані переважно 5*s*-станами телуру. Незважаючи на переважаючий характер Te 5*s*-станів, для цих підзон суттєвими є ефекти гібридизації станів атомів кремнію і телуру, що приводить до появи внесків 3*s*-станів атомів кремнію.

Третя зв’язка із 22 заповнених зон сформована гібридизованими 3*s*- і 3*p*-станами кремнію з незначним домішуванням 5*p*-станів телуру. Саму верхню підзону зайнятих станів, яка містить 70 енергетичних зон, умовно можна розділити на дві частини. Нижня частина цієї підзони сформована гібридизованими 3*p*-станами кремнію і 5*p*-станами телуру, які забезпечують сильний ковалентний зв'язок в тетраедрах $Si_3Te$. Самий верх даної підзони розташований поблизу верха валентної зони, сформований переважно 5*p*-станами неподіленої електронної пари телуру з незначною домішкою 3*p*-, 3*d*-станів кремнію.

Електронна низькоенергетична структура незаповнених електронних станів, яка примикає до забороненої зони, сформована в основному із *p*-, *d*-станів Si з незначною домішкою *p*-, *d*-станів Te.

**8.9.1 Карти розподілу заряду валентних електронів**. Карти просторового розподілу електронної густини дозволяють порівняти характер розподілу повного заряду в одних і тих самих кристалографічних площинах багатогранників $S_{20}$ і $S_{24}$ кубічної й ромбоедричної фаз клатрату $Te_{16}Si_{38}$ (рис. 8.26 і 8.27).

Оскільки усі атоми Si в кремнієвих клатратах, у тому числі й у $Te_{16}Si_{38}$, тетраедрично координовані з атомами Si, які займають центри трохи деформованих тетраедрів $[Si]^{4-}$, тому доцільно провести контурні карти в площині даного тетраедра. На картах розподілу заряду валентних електронів $\rho(r)$ у тетраедрах $[Si]^{4-}$ (рис. 8.26, *а* і 8.27, *а*) чітко видно наявність локалізованих максимумів заряду у вигляді замкнутих контурів на зв’язках Si–Si, як це має місце в кристалічному кремнію (рис. 8.15). Яскраво виражена деформація контурів $\rho(r)$ у напрямках атомів кремнію уздовж ліній зв’язку Si–Si і наявність спільних контурів, що охоплюють максимуми електронної густини на Si–Si зв’язках у тетраедрах $[Si]^{4-}$ (рис. 8.26, *а* і 8.27, *а*),

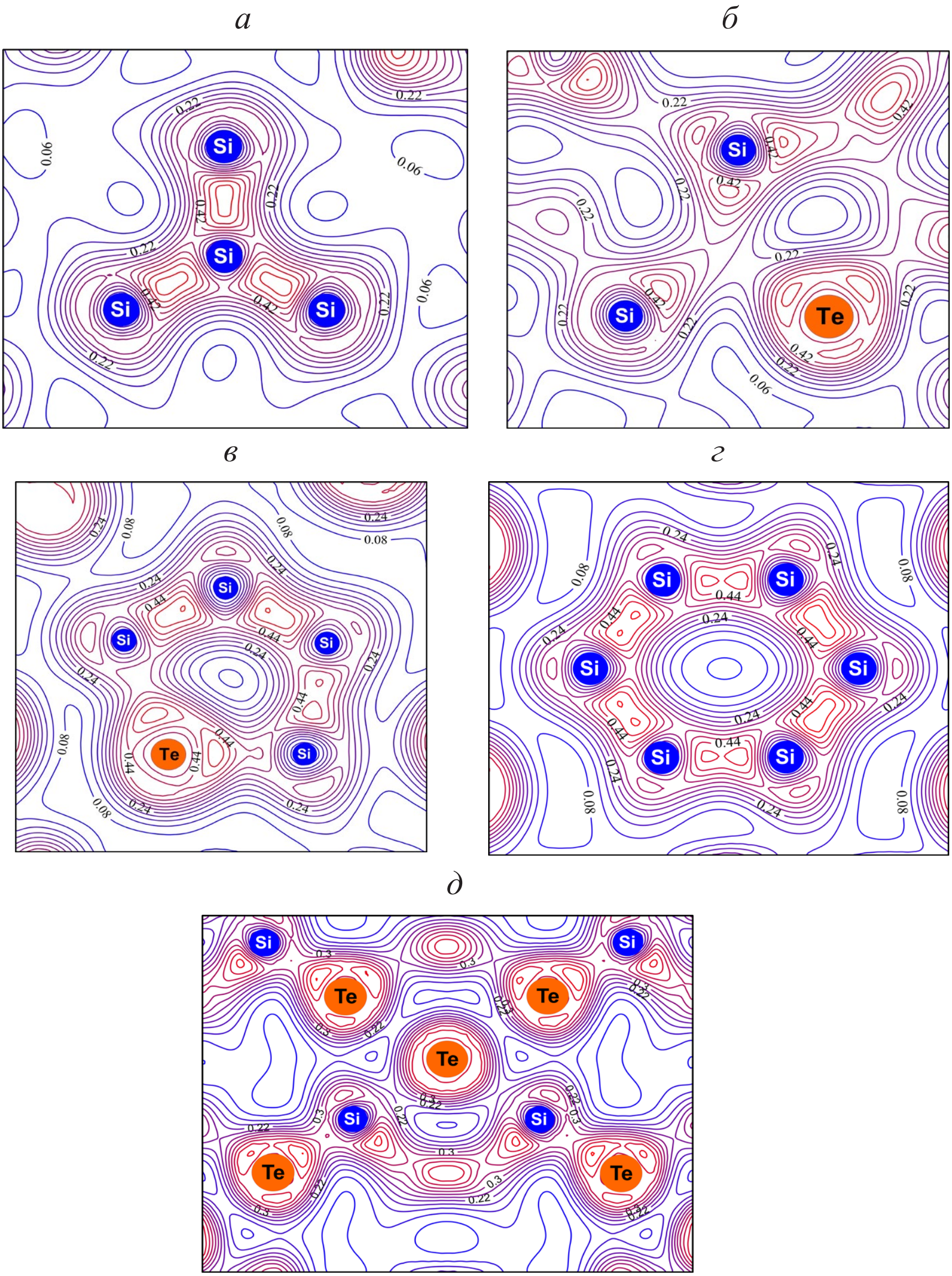

Рис. 8.26. Карти розподілу електронної густини кубічного клатрату $Te_{16}Si_{38}$ в площинах, що проходять: *а* – через грань (утворену із атомів Si) тетраедра $[Si]^{4-}$; *б* – через грань (утворену із 2 атомів Si та 1 атома Te) тетраедра $[Si_3Te]$; *в* – через пятикутну грань додекаедра $Si_{20}$, *г* – через шестикутну грань тетракайдекаедра $Si_{24}$; *д* – через центральний атом Te додекаедра $Si_{20}$.

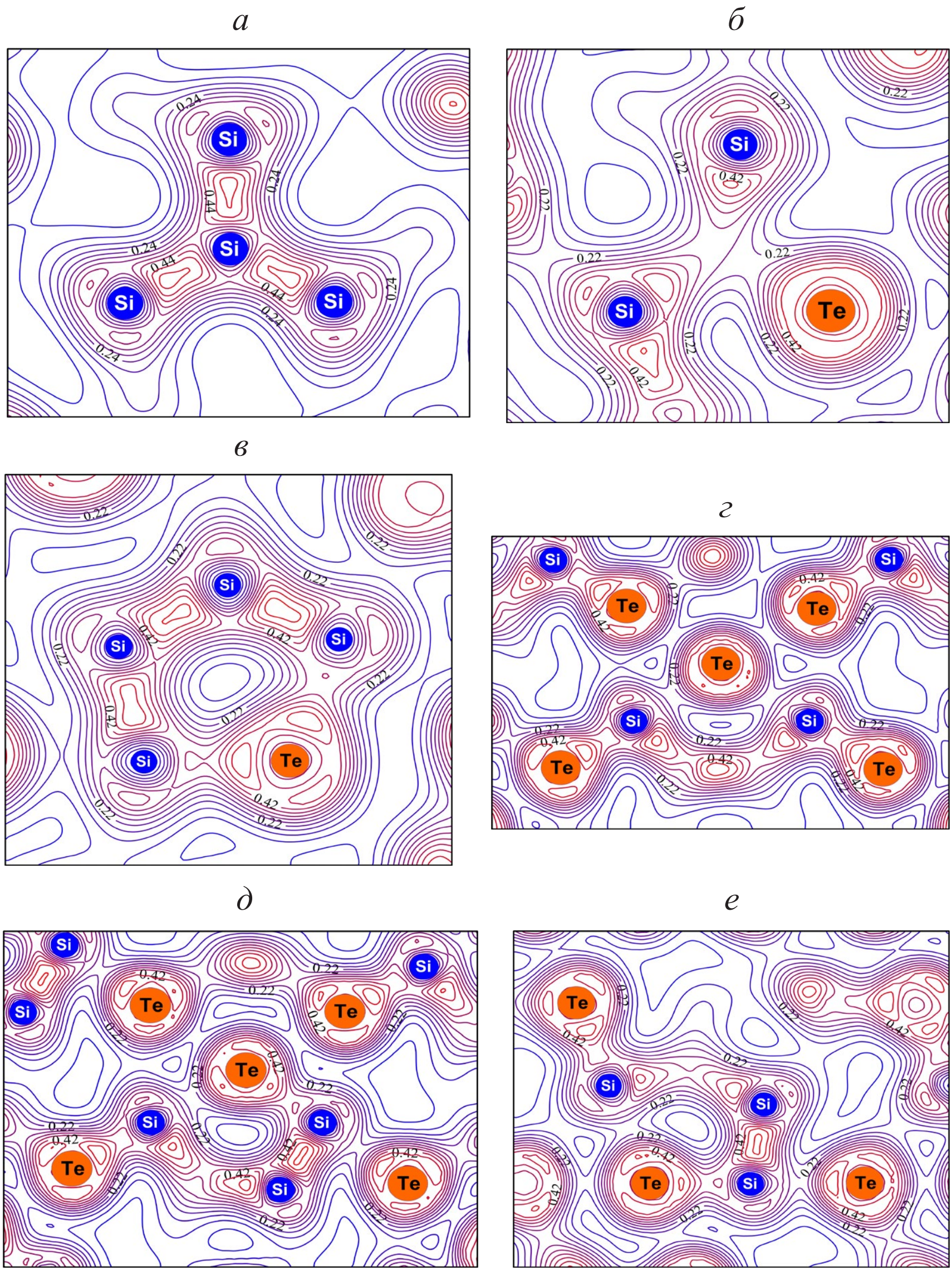

Рис. 8.27. Карти розподілу електронної густини ромбоедричного клатрату $Te_{16}Si_{38}$ в площинах, що проходять: *а* – через грань (утворену із атомів Si) тетраедра $[Si]^{4-}$; *б* – через грань (утворену із 2 атомів Si та 1 атома Te) тетраедра $[Si_3Te]$; *в* – через пятикутну грань додекаедра $Si_{20}$; *г, д, е* – через центральний атом Te додекаедра $Si_{20}$.

відображають ковалентну складову хімічного зв'язку, обумовлену гібридизацією 3*s*- і 3*p*- станів Si.

На рис. 8.26, *в*, *г* і 8.27, *в*, *г*, наведені заглньні картини зарядової густини ρ(r) у п'ятикутних і шестикутних гранях поліедрів $Si_{20}$ і $Si_{24}$ кубічного і ромбоедричного клатрату $Te_{16}Si_{38}$, які ілюструють картини міжатомних взаємодій. На цих рисунках чітко видно, що тільки в п'ятикутних гранях наявні заміщуючі атоми Te3 в позиціях 16*i*. Загальні контури ρ(r), що охоплюють атоми кремнію й телуру в п'ятикутних гранях додекаедрів і тетракайдекаедрів, вказують на наявність ковалентної складової хімічного зв'язку. Просторовий розподіл електронної густини засвідчує виражений ковалентний тип зв'язку Si–Te за рахунок перекриття Si 3*s*-, 3*p*- і Te 5*s*-, 5*p*–орбіталей (рис. 8.25). Поляризація зарядової густини у напрямку Si→Te вказує на наявність крім ковалентної, ще й іонної складової зв'язку за рахунок часткового перенесення зарядової густини від атомів кремнію до більш електронегативних атомів телуру. Відмінність у розмірах атомів телуру і кремнію приводить до виникнення локальних пружних деформацій в околі вузлів кристалічної ґратки, зайнятих атомами Te, що також наглядно ілюструють карти розподілу електронної густини (рис. 8.26 і 8.27).

# СПИСОК ВИКОРИСТАНИХ ДЖЕРЕЛ

1. Блецкан Д. И. Кристаллические и стеклообразные халькогениды Si, Ge, Sn и сплавы на их основе: Монография. – Ужгород: ВАТ «Видавництво „Закарпаття"», 2004, 292 с.
2. Bletskan D.I. Phase equilibrium in the binary systems $A^{IV}B^{VI}$. Part. I. The systems Silicon –Chalcogen // J. Ovonic Res. – 2005. – V.1, № 5. – P. 45–50.
3. Bailey L.G. Preparation and properties of silicon telluride // J. Phys. Chem. Solids. – 1966. – V. 27, № 10. – P. 1593–1598.
4. Legendre B., Souleau C., Hancheng C., Rodier N. The ternary system gold–silicon–tellurium; a contribution to the study of the binary systems silicon–tellurium and gold–silicon, and the structure of $Si_2Te_3$ // J. Chem. Res. Synop. – 1978. – № 5. – P. 165–169.
5. Davey T. G., Baker E. H. A note on the Si–Te phase diagram // J. Mater. Science. – 1980. – V. 15, № 6. – P. 1601–1602.
6. Odin I. N., Ivanov V. A. Ptot-T-x-diagram of the state of the Si–Te system // J. Inorgan. Chem. – 1991. – V. 36, № 5. – P. 1314–1319.
7. Mishra R., Mishra P.K., Phapale S., Babu P.D., Sastry P.U., Ravikumar G., Yadav A.K. Evidences of the existence of $SiTe_2$ crystalline phase and a proposed new Si–Te phase diagram // J. Solid State Chem. – 2016. – V. 237. – P. 234–241.
8. Phapale S., Samui P., Mishra R. Thermodynamic stability of $Si_2Te_3$(s) and $SiTe_2$(s) compounds // J. Alloys Compd. – 2017. – V. 726 – P. 1101–1105.
9. Ploog K., Stetter W., Nowitzki A., Schönherr E. Crystal growth and structure detrmination of silicon telluride $Si_2Te_3$. – Mater. Res. Bull. – 1976. – V.11, № 9. – P. 1147–1154.
10. Gregoriades P. E., Bleris G. L., Stoemenos J. Electron diffraction study of the $Si_2Te_3$ structural transformation // Acta Cryst. B.– 1983. – V. 39, № 4. – P. 421–426.
11. Brebrick R.F. Si–Te system: partial pressures of $Te_2$ and SiTe and thermodynamic properties from optical density of the vapor phase // J. Chem. Phys. – 1968. – V. 49, № 6. – P. 2584–2592.
12. Gauer M.K., Dézsi I., Gonser U., Langouche G., Ruppersberg H. The crystallization of amorphous $Si_xTe_{1-x}$ // J. Non-Cryst. Solids. –1989. – V. 109, № 2-3. – P. 247–254.
13. Cornet J. The eutectic law for binary Te-based systems: a correlation between glass formation and the eutectic composition // Тр. 6-й Междунар. конф. по аморфн. и жидк. полупроводн. 1975.

Структура и свойства некристал. полупроводн. – Л.: Наука, 1976. – С. 72–77.
14. Barrow B. F. Ultra-violet band system of silicon monotelluride // Nature. – 1938. – V. 142. – P. 536.
15. Smirous K., Stourac L., Bednar J. Die halbleitende Verbindung SiTe // Czech. J. Phys. – 1957. V. 7, № 1. – P. 120–122.
16. Насиров Я.Н., Ахмедова Г.М., Зейналова А.А. Термоэлектрические свойства монокристаллов монотеллурида кремния // Неорган. материалы. – 1974. – Т. 10, № 6. – С. 1129.
17. Chen Y., Sun Q., Jena P. SiTe monolayers: Si-based analogues of phosphorene // J. Mater. Chem. C. – 2016. – V. 4, № 26. – P. 6353–6361.
18. Weiss Alarich, Weiss Armin Zur Kenntnis von Siliciumditellurid (II. Mitt. über Siliciumchalkogenide) // Z. Naturforsch. – 1953. Bd. 8b, № 1. – S. 104.
19. Weiss Alarich, Weiss Armin Zur Kenntnis von Siliciumditellurid // Z. Anorg. Allg. Chem // 1953. – V. 273, № 3-5. – P. 124–128.
20. Rau J.W., Kannewurf C.R. Intrinsic absorption and photoconductivity in single crystal $SiTe_2$ // J. Phys. Chem. Solids. – 1966. – V 27, № 6–7. – P. 1097–1101.
21. Lambros A.P., Economou N.A. The optical properties of silicon ditelluride // Phys. Stat. Solidi (b). – 1973. – V. 57, № 2. – P. 793–799.
22. Taketoshi K., Andoh F. Structural studies on silicon ditelluride ($SiTe_2$) // Jap. J. Appl. Phys. A – 1995. – V. 34, № 6. – P. 3192–3197.
23. Pizzarello F. Vapor phase crystal growth of lead sulfide crystals // J. Appl. Phys. – 1954 – V. 25, № 6 – P.804–805.
24. Grigoriadis P. , Stoemenos J. Dislocations and stacking fault energy in silicon ditelluride // J. Mater. Science. – 1978. – V. 13, № 3. –P. 483–491.
25. Блецкан Д.И., Кабаций В.Н., Сакал Т.А. Получение и структура слоистых кристаллов $Si_2Te_3$ // Поверхность. Рентгеновские, синхротронные и нейтронные исследования. – 2004. – № 9. – С. 22–25.
26. Ziegler K., Berkholz P. Photoelectric properties of $Si_2Te_3$ single crystals // Phys. Stat. Solidi. A. – 1977. – V. 39, № 2. – P. 467–475.
27. Haneveld A.J., Van der Veer W., Jellinek F. On silicon tritelluride $Si_2Te_3$, and alleged phosphorus tritelluride // Rec. Tvar. Chim. Pays-Bas – 1968. – V. 87, № 3. – P. 255–256.

28. Зигбан К., Норман К., Фальман А. и др. Электронная спектроскопия М., 1971. 493 с.
29. Кочубей Д. И., Канажевский В. В. Рентгеновская спектроскопия поглощения – инструмент для исследования и создания новых материалов // Химия в интересах устойчивого развития. – 2013. – Т. 21, № 1. – С. 21 – 36.
30. Wu K., Sun W.m, Jiang V., Chen J., Li Li, Cou C., Shi S., Shen X., Cui I. Structure and photoluminescence study of silicon based two-dimensional $Si_2Te_3$ nanostructures // J. Appl. Phys. – 2017. – V. 122. – P. 075701-1–075701-8.
31. Мазалов Л. Н. Рентгеновские спектры и электронная структура молекул // Соросов. Образов. Ж. – 1997. – № 6. – С. 77.
32. Adenis C., Langer V., Lindqvist O. Reinvestigation of the structure of tellurium // Acta Cryst. – 1989. – V. 45. – P. 941–942.
33. Cherin P., Unger P. Two-dimensional refinement of the crystal structure of tellurium // Acta Cryst. – 1967. – V. 23. – P. 670–671.
34. Bhattarai R., Shen X. Predicting a novel phase of 2D $SiTe_2$ // ACS Omega. – 2020. – V. 5, № 27. – P. 16848–16855.
35. Johnson V. L., Anilao A., Koski K. J. Pressure-dependent phase transition of 2D layered silicon telluride ($Si_2Te_3$) and manganese intercalated silicon telluride // Nano Res. − 2019. − V.12, № 9. − P. 2373−2377.
36. Bhattarai R., Shen X. Pressure-induced insulator-metal transition in silicon telluride from first-principles calculations // J. Phys. Chem. C. − 2021.− V. 125.− P. 11532−11539.
37. Bhattarai R., Shen X. Ultra-high mechanical flexibility of 2D silicon telluride// ACS Omega. – 2020. – V. 116. – P. 023101-1–023101-5.
38. Grzechnik A., Crichton W.A., Druzhbin D., Fečík M., Stoffel R. P., Brauksiepe S., Steinberg S., Dronskowski R., Hakala B.V., Friese K. Chemical reactions and phase stabilities in the Si–Te system at high pressures and high temperatures // Inorg. Chem. – 2022. – V. 61, № 19. – P. 7349–7357.
39. Анализ поверхности методами оже- и рентгеновской фотоэлектронной спектроскопии / Под ред. Д. Бриггса, М.П. Сиха. – М.: Мир, 1987. – 600с.
40. Углов В.В., Черенда Н.Н., Анищик В.М. Методы анализа элементного состава поверхностных слоев. – Минск : БГУ, 2007. – 167 с.

41. McGuire G. Auger electron spectroscopy reference manual: a book of standard spectra for identification and interpretation of auger electron spectroscopy data. – 1979. – 144 p.
42. Thomas S. Electron-irradiation effect in the Auger analysis of $SiO_2$ // J. Appl. Phys. – 1974. – V 45, № 1. – P. 161–166.
43. Bauer H.P., Birkholz U. Electrical conductivity of passivated $Si_2Te_3$ single crystals // Phys. Stat. Sol. A. – 1978. – V. 49, № 1. – P. 127–131.
44. Madden H.H. Chemical information from Auger electron spectroscopy // J. Vac. Sci. Technol. – 1981. – V. 18, № 3. – P. 677–689.
45. Галлон Т. Актуальные вопросы электронной оже-спектроскопии // Электронная и ионная спектроскопия твердых тел. Под. ред. Фирмэнса. – М.: Мир, 1981. – С. 236–280.
46. Petersen K.E., Birkholz U., Adler D. Properties of crystalline and amorphous silicon telluride // Phys. Rev. B. – 1973. – V.8. № 4. – P. 1453–1460.
47. Erlandsson R., Birkholz U., Karlsso S.-E. Surface investigation of $Si_2Te_3$ with Auger spectroscopy // Phys. Stat. Solidi. A. – 1977. – V. 41, № 2. – P. K163–K165.
48. Erlandsson R., Birkholz U., Karlsso S.-E. Study of $Si_2Te_3$ Surface Reactions with Auger Electron Spectroscopy // Phys. Stat. Solidi. A. – 1978. – V. 47, № 1. – P. 85–90.
49. Keuleyan S., Wang M., Chung F.R., Commons J., Koski K.J., A silicon-based two-dimensional chalcogenide: growth of $Si_2Te_3$ nanoribbons and nanoplates // Nano Lett. – 2015. – V. 15. – P. 2285–2290.
50. Wu K., Sun W., Jiang Y., Chen J., Li L., Cao C., Shi S., Shen X., Cui J. Structure and photoluminescence study of silicon based two-dimensional $Si_2Te_3$ nanostructures // J. Appl. Phys. – 2017. – V. 122. – P. 075701-1–075701-8.
51. Wang M., Lahti G., Williams D., Koski K.J. Chemically tunable full spectrum optical properties of 2D silicon telluride nanoplates // ACS Nano. – 2018. – V. 12, № 6. – P. 6163–6169.
52. Wu K., Cui J. Morphology control of $Si_2Te_3$ nanostructures synthesized by CVD // J. Mater. Sci. Mater. Electron. – 2018. – V. 29, №18. – P. 15643–15648.
53. Wagner R.S., Ellis W.C., Jackson K.A., Arnold S.M. Study of the filamentary growth of silicon crystals from the vapor // J. Appl. Phys. – 1964. – V. 35. – P. 2993–3000.
54. Okamoto H., Massalski T.B. The Au–Si (Gold-Silicon) system // Bulletin of Alloy Phase Diagrams. – 1983. – V. 4. – P. 190–198.

55. Дружинін А.О. Ниткоподібні кристали кремнію і твердого розчину кремній-германій в мікро- та наноелектроніці : монографія / А.О. Дружинін, І.П. Островський, Ю.М. Ховерко, С.І. Нічкало. Львів : "Тріада плюс", 2016. – 264 с.
56. Дружинін А.О., Островський І.П., Ховерко Ю.М., Нічкало С.І. Вирощування нанорозмірних кристалів Si методом газофазової епітаксії // Вісник Національного університету "Львівська політехніка": Електроніка –2009 – №.646. – с.11–16.
57. Большакова І.А., Заячук Д.М. та ін.. Віскери напівпровідникових матеріалів як результат конкуруючого росту нановіскерів. // Вісник Національного університету "Львівська політехніка", Електроніка, –2013– № 764, с. 107–111.
58. Sen S., Bhatta U. M., Kumar V., Muthe K. P., Bhattacharya S., Gupta S. K., Yakhmi J.V. Synthesis of tellurium nanostructures by physical vapor deposition and their growth mechanism // Cryst. Growth Des. – 2008. – V. 8, № 1. – P. 238–242.
59. Wu K., Chen J., Shen X., Cui J. Resistive switching in $Si_2Te_3$ nanowires // AIP Advances. – 2018. – V. 8, № 12. – P. 125008-1–125008-7.
60. Chen J., Wu K., Shen X., Hoang T. B., Cui J. Probing the dynamics of photoexcited carriers in $Si_2Te_3$ nanowires // J. Appl. Phys. – 2019. – V. 125, № 2. – P. 024306-1–024306-6.
61. Wu K., Chen J., Shen X., Cui J. Growth of $Si_2Te_3$/Si heterostructured nanowire and its photoresponse property // Optic. – 2020. – V. – P. 163475-1–163475-8.
62. Song X., Ke Y., Chen X., Liu J., Hao Q., Wei D., Zhang W. Synthesis of large-area uniform $Si_2Te_3$ thin films for p-type electronic device // Nanoscale. − 2013.− V. 12, № 20.− P. 11242−11250.
63. Giri A., Kumar M., Kim J., Pal M., Banerjee W., Nikam R.D., Kwak J., Kong M., Kim S.H., Thiyagarajan K., Kim G., Hwang H., Lee H. H., Lee D., Jeong U. Surface diffusion and epitaxial self-planarization for wafer-scale single-grain metal chalcogenide thin films // Adv. Mater. – 2021. – V. 33, № 35. – P. 2102252-1–2102252-8.
64. Hilton A. R., Jones C. E., Dobrott R. D., Klein H. M., Bryant A. M., George T. D. Non-oxide IVA–VA–VIA chalcogenide glasses. Part 3. Structural studies // Phys. Chem. Glasses. – 1966. – V. 7, № 4. – P. 116–126.
65. Cornet J. The eutectic law for binary Te-based systems: a correlation between glass formation and the eutectic composition // Тр. 6-й

Междунар. конф. по аморфн. и жидк. полупроводн. 1975. Структура и свойства некристал. полупроводн. – Л.: Наука, 1976. – С. 72–77.
66. Andreev A. A., Ablova M. S., Malek B. T., Nasredinov F. S., Seregin P. P., Turaev E. Eutectic glassy semiconductors in the $A^{III}$–Te and $A^{IV}$–Te systems // In Amorphous and Liquid Semiconductors; 7th Intern. Conf. Edinburgh, 1977. – P. 44–47.
67. Мелех Б. Т., Аблова М. С., Берман И. В., Жукова Т. Б., Маслова З. В. Эвтектические теллуридные стекла, особенности стеклования и проявления физико-химических и физических свойств // Матер. конф. "Некристаллические полупроводники-89". – Ужгород, 1989. – Т. 1. – С. 136–138.
68. Школьников Е. В. О стеклообразующей способности расплавов вблизи эвтектических составов // Физ. и хим. стекла. – 1987. – Т. 13. № 1. – С. 145–149.
69. Kulakova L. A., Melech B. T., Yakhkind E. Z., Kartenko N. F., Bakharev V. I., Yakovlev Y. P. Physical properties of $Si_{20}Te_{80}$ glasses with various structures and their use in acoustooptic devices // Semiconductors – 2001 – V.35, №6 – P.630–636.
70. Алтунян С. А., Минаев В. С., Минаждинов М. С., Скачков Б. К. Стеклообразование в системах Si–Te и диодные переключающие структуры с «памятью» на основе полупроводникового стекла в этой системе // ФТП. – 1970. – Т. 4, № 11. – С. 2214–2215.
71. Petersen K. E., Birkholz U., Adler D. Properties of crystalline and amorphous silicon telluride // Phys. Rev. B. – 1973. – V. 8, № 4. – P. 1453–1461.
72. Feltz A., Maul W., Schönfeld I. Über glasbildung und eigenschaften von chalkogenidsystemen. II. zur glasbildung in den systemen As–Ge–Si–Te und Ge–Si–Te // Z. anorg. allg. Chem. – 1973. – Bd. 396, № 1. – S. 103–107.
73. Минаев В. С., Шипатов В. Т. Исследование структуры стекол в системе кремний–теллур с помощью эффекта Мёссбауэра // Неорган. материалы. – 1980. – Т. 16, № 9. – С. 1526–1529.
74. Bartsch G. E. A., Bromme H., Just T. Radial distribution studies of glassy tellurium–silicon alloys // J. Non-Cryst. Solids. – 1975. – V. 18. № 1. – P. 65–75.
75. Аблова М. С., Андреев А. А., Мелех Б. Т., Маслова З. В., Идрисова Р. М., Жукова Т. Б. Электрофизические свойства эвтекти-

ческих стекол системы Si–Te // Физ. и хим. стекла. – 1988. – Т. 14. № 3. – С. 413–417.

76. Мелех Б. Т., Маслова З. В., Бульченко В. П., Зуев С. Н., Жукова Т. Б., Дедегкаев Т. Т., Насрединов Ф. С., Подхалюзин В. П., Андреев А. А. Получение, физико-химические свойства и структура стекол в двойных и тройных системах $A^{III}(A^{IV})$–Te и $A^{III}$ –$A^{IV}$ –Te // В кн.: Стеклообразные полупроводники. – Л. – 1985. – С. 145–146.
77. Андреев А. А., Аблова М. С., Подхалюзин В. П., Маслова З. В., Мелех Б. Т. Свойства стекол $A^{III}$ –Te, $A^{IV}$ –Te в сильном электрическом поле // Физ. и хим. стекла. – 1979. – Т. 5, № 3. – С. 375–378.
78. Asokan S., Parthasarathy G., Gopal E.S.R. Double glass transition and double stage crystallization of bulk $Si_{20}Te_{80}$ glass // J. Mater. Sci Lett. – 1985, V. 4, № 5. – P. 502–504.
79. Asokan S., Parthasarathy G., Gopal E. S. R. Crystallization studies on bulk $Si_xTe_{100-x}$ glasses // J. Non-Cryst. Solids. – 1986. – V. 86, № 1–2. – P. 48–64.
80. Asokan S., Parthasarathy G., Gopal E. S. R. Evidence for a cristical composition in group-IV–VI chalcogenide glasses // Phys. Rev. B. – 1987. – V. 35, № 15. – P. 8269–8272.
81. Asokan S., Gopal E. S. R. Double glass transition and double stage crystallization in tellurium based chalcogenide glasses // Reviews of Solid State Science. – 1989. – V.3, № 314. – P. 273 – 289.
82. Gauer M. K., Dézsi I., Gonser U., Langouche G., Ruppersberg H. The crystallization of amorphous $Si_xTe_{1-x}$ // J. Non-Cryst. Solids. –1989. – V. 109, № 2–3. – P. 247–254.
83. Vengrenovitch R. D., Podolyanchuk S. V., Lopatniuk I. A., Stasik M. O., Tkachova S.D. Preparation of amorphous $Si_xTe_{1-x}$ alloys and their crystallization // J. Non-Cryst. Solids. – 1994. – V. 171. – P. 243–248.
84. Венгренович Р.Д., Лопатнюк И.А., Подолянчук С.В., Стасик М.О., Цалый В.З. Метастабильная кристаллизация сплавов $Si_xTe_{1-x}$ // Неорган. материалы. – 1996. – Т. 32, № 9. – С. 1087–1091.
85. Лопатнюк І.О. Застосування методу спінінгування розплаву для отримання аморфних сплавів системи Si–Te // Науковий вісник Чернівецького університету: Збірник наук. праць. – Вип 23: Інженерно-технічні науки. – Чернівці:УДУ. – 1998. – С. 55–61.

86. Минаев В.С., Стеклообразные полупроводниковые сплавы. М. Металлургия – 1991.
87. Parfeniev R.V., Regel L.L. Gravity application to anisotropic semiconductor materials: from high- to microgravity conditions // Acta Astronaut. – 2001. – V. 48, Issues 2–3. – P. 163–168.
88. Парфеньев Р.В., Фарбштейн И.И., Якимов С.В., Мелех Б.Т. Особенности условий затвердевания стеклообразного сплава $Te_{80}Si_{20}$ в невесомости и электрофизические свойства полученного образца. Сборник трудов IV международная конференция, Санкт Петербург, 5-8 июля 2004, С. 288.
89. Якимов С.В. Структура и электрофизические свойства кристаллов теллура и сплава $Te_{80}Si_{20}$, полученных при разных условиях выращивания. : Автореф. канд. физ.-мат. наук : 01.04.10 : СПб., 2004, 23 с.
90. Фарбштейн И.И., Мелех Б.Т., Шалимов В.П., Шулька Н.К., Якимов С.В. Образование пузырей в расплаве телур-кремний в условиях микрогравитации и их динамика // Изв. РАН Механика жидкости и газа. – 1994. – Т. 5. – С. 135.
91. Балявичюс С., Декснис А., Лисаускас В., Пошкус А., Шикторов Н. Условия стабильности наносекундного переключения в аморфных пленках теллуридов In, Ga, Ge, Si // Литовский физический сборник. – 1984. – Т. 24, № 2. – С 95–101.
92. Балявичюс С., Декснис А., Пошкус А., Шикторов Н. Эффект «памяти» в случае наносекундного переключения в неупорядоченных пленках теллуридов In, Ge, Si // ФТП. – 1984. – Т. 18, № 8. – С. 1513–1516.
93. Shufflebotham P. K., Card H. C., Kao K. C., Thanailakis A. Amorphous silicon–tellurium alloys // J. Appl. Phys. – 1986. – V. 60, № 6. – P. 2036–2040.
94. Tsunetomo K., Shimizu R., Imura T., Osaka Y. EXAFS and X-ray diffraction studies on the local structure of sputterdeposited amorphous $Si_xTe_{1-x}$ alloy // J. Non-Cryst. Solids. – 1990. – V. 116, № 2–3. – P. 262–268.
95. Lasocka M., Matyja H. Thermal stability of chalcogenide glasses Te – $A^{IV}$ in relation to the atomic number of the $A^{IV}$ element // Phys. Stat. Solidi. A. – 1974. – V. 26, № 2. – P. 671–680.
96. Norban B., Pershing D., Enzweiler R. N., Boolchand P., Griffiths J. E., Phillips J. C. Coordination-number-induced morphological structural transition in a network glass // Phys. Rev. B. – 1987. – V. 36, №15. – P. 8109–8114.

97. Boolchand P., Norban B., Enzweiler R., Griffiths J.E., Phillips J.C. Molecular structure of $Si_xTe_{1-x}$ glasses and oxygen alloying effects // SPIE V. 822 – Raman and Luminescence Spectroscopy in Technology. – 1987. – P. 114–121.
98. Madhusoodanan K.N., Philip J., Asokan S., Parthasarathy G., Gopal E.S.R. Photoacoustic study of the glass transition and crystallization in bulk $Ge_xTe_{1-x}$ and $Si_xTe_{1-x}$ glasses // Photoacoustic and Photothermal Phenomena II. – 1990. – V. 62. – P. 183–185.
99. Zhang, S. N., Zhu, T. J., Zhao, X. B. Crystallization kinetics of $Si_{15}Te_{85}$ and $Si_{20}Te_{80}$ chalcogenide glasses // Physica B. – 2008. – V.403 – P. 3459–3463.
100. Abu El-Oyoun M. DSC studies on the transformation kinetics of two separated crystallization peaks of $Si_{12.5}Te_{87.5}$ chalcogenide glass: An application of the theoretical method developed and isoconversional method // Mater. Chem. Phys. – 2011. – V 131, № 1–2. – P. 495–506.
101. Moharram A.H., Abu El-Oyoun M. Glass transition kinetics of the binary $Si_{12.5}Te_{87.5}$ alloy // Appl. Phys. A. – 2014. – V. 116, № 1. – P. 311–317.
102. Цалий В. З., Вергренович Р.Д., Юречко Р.Я., Цалий З.П. Кристаллізація та структура аморфних сплавів системи $Si_xTe_{100-x}$ // Науковий вісник Чернівецького університету. Фізика. – 1998. – Вип. 40. – С. 93–94.
103. Цалий В. З. Термічна стабільність аморфних сплавів системи $Si_xTe_{100-x}$ // Науковий вісник Чернівецького університету. Фізика, електроніка. – 2005. – Вип. 268. – С. 93–94.
104. Asokan S., Gopal E. S. R., Parthasarathy G. Pressure-induced polymorphous crystallization in bulk $Si_{20}Te_{80}$ glass // J. Mater. Sci. –1986. – V. 21, № 2. – P. 625–629.
105. Asokan S., Parthasarathy G., Subbanna G. N., Gopal E. S. R. Electrical transport and crystallization studies of glassy semiconducting $Si_{20}Te_{80}$ alloy at high pressure // J. Phys. Chem. Solids. – 1986. – V. 47, № 4. – P. 341–348.
106. Asokan S., Gopal E. S. R., Parthasarathy G. Si–Te glasses: relation between the structure and the physical properties // Key Enginieering Materials. – 1987. – V. 13-15. – P. 119–130.
107. Malyj M., Espinosa G. P., Griffiths J. E. Structure and delocalized vibrational modes in vitreous $Si_x(Se_{1-y}Te_y)_{1-x}$ // Phys. Rev. B. –1985. – V. 31, № 6. – P. 3672–3679.

108. Kulakova L.A., Kudoyarova V. Kh., Melekh B.T., Bakharev V. I. Si(Ge)-Se-Te glasses: electrical and acoustic properties // J. Optoelectron. Adv. Mater. – 2006. – V. 8, № 2. – P. 800–804.
109. Kulakova L.A., Kudoyarova V.Kh., Melekh B.T., Bakharev V. I. Synthesis and physical properties of Si(Ge)–Se–Te glasses // Journal of Non-Crystalline Solids. – 2006. – V. 352, № 9–20. – P. 1555–1559.
110. Feltz A., Büttner H. J., Lippmann F. J., Maul W. About the vitreous systems GeTeI and GeTeSi and the influence of microphase separation on the semiconductor behaviour of Ge–Se glasses // J. Non-Cryst. Solids. – 1972. – V. 8–10. – P. 64–71.
111. Feltz A., Maul W., Schönfeld I. Über Glasbildung und Eigenschaften von Chalkogenidsystemen. II. Zur Glasbildung in den Systemen As–Ge–Si–Te und Ge–Si–Te // Z. anorg. allg. Chem. – 1973. – Bd. 396. – № 1. – S. 103–107.
112. Feltz A., Amorphe und Glasartig Anorganische Festkorper // Academie, Berlin 1983.
113. Anbarasu M., Asokan S. The influence of network rigidity on the electrical switching behaviour of Ge–Te–Si glasses suitable for phase change memory applications // J. Phys. D: Appl. Phys. – 2007. – V. 40, № 23. – P. 7515–7518.
114. Anbarasu M., Singh K., Asokan S. Evidence for a thermally reversing window in bulk Ge – Te – Si glasses revealed by alternating differential scanning calorimetry// Phil. Mag. – 2008. – V. 88, № 5. – P. 599–605.
115. Gunti S. R., Asokan S. Observation of high pressure o-GeTe phase at ambient pressure in Si-Te-Ge glasses // AIP Advances. – 2012. – V. 2, № 1. – P. 012172-1–012172-6.
116. Gunasekera K., Boolchand P., Micoulaut M. Effect of mixed Ge/Si cross-linking on the physical properties of amorphous Ge–Si–Te networks // J. Appl. Phys. – 2014. – V. 115, № 5. – P. 164905-1–164905-15.
117. Andreev A. A., Ablova M. S., Manukian A. L. et al. Physico-chemical, electrical and optical properties of glass semiconductors Si –Ge–Te and As–Se over a wide range of temperature // Proc. In-tern. Conf. amorphous semiconductors. 1976. – Balatonfüred, 1976. – P. 429–435.
118. Jagannatha K. B., Roy D., Varma S. G., Asokan S., Das C. Effect of Sn addition on glassy Si-Te bulk sample // AIP Conference Proceedings. – 2018. – V. 1966, № 1. – P. 020034-1 – 020034-4.

119. Jagannatha K. B., Roy D., Das C. Electrical switching and crystalline peak studies on $Si_{20}Te_{80-x}Sn_x$ ($1 \leq x \leq 7$) chalcogenide bulk glasses // J. Non-Cryst. Solids. – 2020. – V. 544. – P. 120196-1 – 120196-7.
120. Jagannatha K. B., Tanujit B., Roy D., Asokan S., Das C. A composition- dependent thermal behavior of $Si_{20}Te_{80-x}Sn_x$ glasses: Observation of Boolchand intermediate phase // J. Non-Cryst. Solids. – 2022. – V. 577, №11 – P. 121311-1 – 121311-6.
121. Jagannatha K.B., Das C. Switching, Raman and morphological studies on $Si_{20}Te_{74}Sn_6$ chalcogenide glass // Materials Today: Proceedings. – 2022. – V. 56, № 6. P. 3755–3759.
122. Leonowicz M., Lasocka M. Thermally induced glass-to-crystal transition in Te–Si–Pb system // Mater. Chem. – 1980. – V. 5, № 2. – P. 109–116.
123. Leonowicz M., Lasocka M. The presence of double Tg and phase separation in $Te_{80}Si_{20-x}Pb_x$ glasses // J. Mater. Science. – 1980. – V. 15, № 6. – P. 1586–1588.
124. Leonowicz M., Lasocka M. Crystallization of Te-Si-Pb glasses exhibiting double Tg effect // J. Mater. Science. – 1981. – V. 16, № 8. – P.2290–2296.
125. Załuski L., Leonowicz M., Trykozko R. Thermal and electrical properties of $Te_{80}Si_{20-x}Pb_x$ glasses // Solid State Commun. – 1981. – V. 39, № 9. – P. 997–999.
126. Leonowicz M., Lasocka M. Effect of preheating on crystallization kinetics of Te–Si–Pb glasses // J. Mater. Sci. Lett. – 1982. – V. 1, № 5. – P. 207–210.
127. Борисова З. У. Халькогенидные полупроводниковые стекла. Ленинград. Изд. Ленинградского ун-та. 1983.
128. Минаев В. С., Шипатов В. Т., Киселев В. Н., Куприянова Р. М., Крупышев Р.С. Стекла в системе Cu–Si–Te // Неорган. материалы. – 1980. – Т. 16, № 8. – С. 1481–1485.
129. Минаев В.С. Новые стекла и некоторые особенности стеклообразования в тройных теллуридных системах // Физ. и хим. стекла. – 1983. – Т. 9, № 4. – С. 432–436.
130. Roy D., Nadig C.H.S., Krishnan A., Karanam A., Abhilash R., Jagannatha K.B., Das C. Electrical switching studies on $Si_{15}Te_{85-x}Cu_x$ bulk ($1 \leq x \leq 5$) glasses // AIP Conference Proceedings. – 2018. – V. 1966. – P. 020033-1–020033-5.

131. Roy D., Tanujit B., Jagannatha K. B., Varma G.S., Asokan S., Das C. Manifestation of intermediate phase in Cu doped Si–Te glasses // J. Non-Crystal. Sol. – 2020. – V. 531. – P. 119863-1–119863-6.
132. Зуев С. Н., Жукова Т. Б., Мелех Б. Т., Андреев А. А. Стеклообразование в системе Si–Ag–Te // Физ. и хим. стекла. – 1987. – Т. 13. № 1. – С. 618–619.
133. Андреев А. А., Берман И. В., Кистаубаев Т. З., Мелех Б. Т. Сверхпроводящие свойства нового теллуридного стекла в системе Si–Ag–Te //ФТТ. – 1988. – Т. 30. № 7. – С. 2177–2181.
134. Gunti S. R., Asokan S. Thermodynamic, Raman and electrical switching studies on $Si_{15}Te_{85-x}Ag_x$ ($4 \leq x \leq 20$) glasses // J. Appl. Phys. – 2012. – V. 111, № 3. – P. 033518-1–033518-5.
135. Pumlianmunga P., Ramesh K. Electrical switching, local structure and thermal crystallization in Al–Te glasses // Mater. Res. Bull – 2017 – V.86 – P. 88–94.
136. Anbarasu M., Singh K. K., Asokan S., The presence of thermallyreversing window in Al–Te–Si glasses revealed by alternating differentialscanning calorimetry and electrical switching studies. // J.Non-Cryst. Solids – 2008. – V. 354, № 28 – P.3369–3374.
137. Wilson P. T., Ramanna R., Chahal S., Shekhawat R., Kumar M. M., Ramesh K. Local structure and electrical switching in $Al_{20}Te_{75}X_5$(X = Si, Ge, As, Sb) glasses // Appl. Phys. A – 2020. – P. 289-1–289-9
138. Gunti S.R., Ayiriveetil A., Asokan S. Thermodynamic, kinetic and electrical switching studies on $Si_{15}Te_{85-x}In_x$ glasses: Observation of Boolchand intermediate phase // J. Solid State Chem. – 2011. – V.184, № 12. – P. 3345–3352.
139. Bletskan D.I., Studenyak I.P., Vakulchak V.V., Bletskan M.M. Electronic structure, optical and photoelectrical properties of $Si_2Te_3$ crystal // 22nd International Conference on Electronic Properties of Two Dimensional Systems // 18th International Conference on Modulated Semiconductor Structures. – Pennsylvania State University, 31 july – 4 august, 2017, MS-12.
140. Bletskan D.I., Vakulchak V.V., Studenyak I.P. Electronic structure, optical and photoelectrical properties of crystalline $Si_2Te_3$ // Semiconductor Physics, Quantum Electronics and Optoelectronics. –2019. – V. 22, № 3. – P. 76–82.
141. Juneja R., Pandey T., Singh A. K. High thermoelectric performance in *n*-doped silicon-based chalcogenide $Si_2Te_3$ // Chem. Mater. –2017 – V. 29 – P.3723–3730.

142. Bhattarai R. Computational study of optical, mechanical, and phase transition properties of silicon telluride ($Si_2Te_3$) // Electronic Theses Dissertation – The University of Memphis – 2021.
143. Meisel A., Leonhardt G, Szargan R. Röntgenspektren und chemische bindung // Publisher: Leipzig : Akademische Verlagsgesellschaft Geest u. Portig – 1977.
144. Shen X., Puzyrev Y.S., Combs C., Pantelides S.T. Variability of structural and electronic properties of bulk and monolayer $Si_2Te_3$ // Appl. Phys. Lett. – 2016. – V. 109, № 11. – P. 113104-1–113104-5.
145. Xian X., Yu N., Zhao J., Wang J. Tellurium vacancy in two-dimensional $Si_2Te_3$ for resistive random-access memory // J. Solid State Chem. – 2021. – V. 303 – P.122448.
146. Cohen M.L. Electronic change densities in semiconductors // Science – 1973. – V. 179, № 4079. – P.1189–1195.
147. Bletskan D.I., Glukhov K.E., Frolova V.V. Electronic structure of 2H-$SnSe_2$: ab initio modeling and comparison with experiment // Semiconductor Physics, Quantum Electronics and Optoelectronics. – 2016. – V. 19, № 1. – P. 98–108.
148. Соболев В.В., Немошкаленко В.В. Электронная структура твердых тел в области фундаментального края поглощения. – Киев: Наук. думка, 1992. – 568 с.
149. Vennik J., Callaerts R. Sur les properietes optique du tellurure de silicium $Si_2Te_3$ // C. R. Acad. Sc. Paris. – 1965. – V. 260. – P. 496–499.
150. Brückel B., Birkholz U., Ziegler K. Fundamental absorption and Franz-Keldysh effect in silicon telluride // Phys. Stat. Solidi (b). – 1976 – V. 78, № 1. – P. K23–K25.
151. Уханов Ю.И. Оптические свойства полупроводников. – М.: Наука, 1977. – 368 с.
152. Цебуля Г.Г., Лисица М.П., Малинко В.Н. Новая интерпретация красного поглощения на Ge и CdTe // УФЖ. – 1967. – Т. 12, № 7. – С. 1144–1150.
153. Froza A., Selloni A. Tetragedrally-bonded amorphous semiconductors. – N.Y. : London, 1985. – P. 271–285.
154. Urbach F. The long-wavelenth edge of photographic sensitivity and of the electronic absorption of solids // Phys. Rev. – 1953. – V. 92, № 5. – P. 1324–1331.
155. Toyozawa Y., Sumi H. Urbach-Martiensen rule and exciton trapped momentaliry by lattice vibration // Jap. J. Phys. Soc. – 1971. – V. 31, № 2. – P. 342–358.

156. Студеняк І. П., Краньчец М., Курик М. В. Оптика розупорядкованих середовищ. Ужгород: Гражда, 2008. – 220 с.
157. Zwick U., Rider K. H. Infrared and Raman study of $Si_2Te_3$ // Z. Physic B. – 1976 – V. 25 – P. 319–322.
158. Beaudoin M., DeVries A. J. G., Johnson S. R., Laman H., Tiedje T. Optical absorption edge of semi-insulating GaAs and InP at high temperatures // Appl. Phys. Lett. – 1997. – V. 70, № 26. – P. 3540–3542.
159. Cody G. D., Tiedje T., Abeles B. [et al.] Disorder and the optical-absorption edge of hydrogenated amorphous silicon // Phys. Rev. Lett. – 1981. – V. 47, № 20. – P. 1480–1483.
160. Johnson S. R., Tiedje T. Temperature dependence of the Urbach edge in GaAs // J. Appl. Phys. – 1995. –V. 78, № 9. – P. 5609–5613.
161. Sa-Yakanit V., Glyde H. R. Urbach tails and disorder // Comm. Cond. Matt. Phys. – 1987. – V. 13, № 1. – P. 35–48.
162. Pistoulet B., Robert J. L., Dusseau J. M., Ensuque L. Conduction mechanisms in amorphous and disordered semiconductors explained by a model of medium-range disorder of composition // J. Non-Cryst. Solids. – 1978. – V. 29, № 1. – P. 29–40.
163. Tauc J. Absorption edge and internal electric fields in amorphous semiconductors // Mater. Res. Bull. – 1970. – V. 5. – P. 721–729.
164. Bhattarai, R., Shen, X. Predicting a novel phase of 2D $SiTe_2$. // ACS Omega – 2020 – V. 5 – P.16848–16855.
165. Doni-Caranicola E. G., Lambros A. P. Use of single $SiTe_2$ crystals with a layered structure in optical filter design // J. Opt. Soc. Am. – 1983. – V. 73, № 3. – P. 383–386.
166. Вайнштейн И. А., Зацепин А. Ф., Кортов В. С., Щапова Ю. В. Правило Урбаха в стеклах PbO–$SiO_2$ // ФТТ – 2000 – Т. 42, №. 2 – С. 224–229.
167. Madhusoodanan K. N., Philip Jacob, Asokan S., Parthasarathy G., Gopal E. S. R. Photoacoustic investigation of the optical absorption and thermal diffusivity in $Si_xTe_{100-x}$ glasses // J. Non-Cryst. Solids – 1989 – V.109, № 2–3. – P.255–261.
168. Roberts G. G., Lind E. L. Space charge conduction in single crystal $Si_2Te_3$ // Phys. Letters. A. – 1970. – V. 33, № 6. – P. 365–366.
169. Roberts G. G., Schmidlin F. W. Study of Localized Levels in Semi-Insulators by Combined Measurements of Thermally Activated Ohmic and Space-Charge-Limited Conduction // Phys. Rev. – 1969. - V. 180, № 3. – P. 785–794.
170. Блецкан Д. И., Таран В. И., Сичка М.Ю. Эффект переключения в

слоистых кристаллах $A^{IV}B^{VI}$ // УФЖ. – 1976. – Т. 21, № 9. – С. 1436–1441.
171. Мадатов Р. С., Алекперов А. С., Гасанов О. М. Эффект переключения и памяти в слоистых кристаллах GeS // Прикладная физика. – 2015. – № 4. С. 11–15.
172. Ziegler K., Junker H.-D., Birkholz U. Electrical conductivity and Seebeck coefficient of $Si_2Te_3$ single crystals // Phys. Stat. Solidi. A. – 1976. – V. 37, № 1. – P. K97–K99.
173. Rick M., Rosenzweig J., Birkholz U. Anisotropy of electrical conductivity in $Si_2Te_3$ // Phys. Stat. Solidi. A. – 1984. – V. 83, № 2. – P. K183–K186.
174. Блецкан Д. І., Фролова В. В. Вплив методу та умов вирощування на електричні властивості кристалів $SnS_2$ // Науковий вісник Ужгородського університету. Серія Фізика. – 2015. – Вип. 37. – С. 36–50.
175. Belen'kii G. L., Abdullaev N. A., Zverev V. N., Shteinshraiber V.Ya., Nature of the conductivity anisotropy and distinctive features in the localization of electrons in layered indium selenide // JETP Letters. – 1988. – V. 47, № 10. – С. 584–586.
176. Edwards J., Frindt R. F. Anisotropy in the resistivity of $NbSe_2$ // J. Phys. Chem. Sol. – 1971. – V. 32, № 9. – P. 2217–2221.
177. Uher C., Sander L. M. Unusual temperature dependence of the resistivity of exfoliated graphites // Phys. Rev. B. – 1983. – V. 27, № 2. – P. 1326–1333.
178. Evans B. L., Young P. A. Delocalized excitons in thin anisotropic crystals // Phys. Stat. Sol. (b). – 1968. – V. 25, № 1. – P. 417–425.
179. Maschke K., Overhof H. Influence of stacking disorder on the dc conductivity of layered semiconductors // Phys. Rev. B. – 1977. – V. 15, № 4. – P. 2058–2061.
180. Maschke K., Schmid Ph. Influence of stacking disorder on the electronic properties of layered semiconductors // Phys. Rev. B. – 1975. – V. 12, № 10. – P. 4312–4315.
181. Fivaz R. C., Schmid Ph. Transport properties of layered semiconductors / Optical and Electrical Properties / Ed. P.A. Lee. – Dordrecht: D. Reidel Publ. Co, – 1976. – P. 343–384.
182. Mott N. F., Davis, E. A. Electronic processes in non-crystalline materials // Clarendon Press: Oxford. 1979. 590 p.
183. Кулакова Л. А., Мелех Б. Т., Яхкинд Э. З., Картенко Н. Ф., Бахарев В.И. Влияние термообработки на структуру и физические свойста стекол состава $Si_{20}Te_{80}$ // Физ. и хим. стекла. – 2001. – Т. 27, № 3. – С. 353–364.

184. Мелех Б. Т., Кулакова Л. А., Бахарев В. И., Кудоярова В. Х., Грудинкин С.А. Эвтектические теллуридные стекла–ХСП с изменением структуры ближнего порядка при переходе «стекло-кристалл»: получение, электрические и оптические свойства, возможные приложения в акустооптике // Сборник трудов VI Международной конференции «Аморфные и микрокристаллические полупроводники», Россия, Санкт-Петербург, 7–9 июля 2008 г. С. 185–186.
185. Аблова М. С., Певцов А. Б., Илисавский Ю. В., Кулакова Л.А., Кулиев Р., Яхкинд Э.З., Мелех Б.Т., Шубников М.Л., Андреев А.А. Электрические, оптические и акустические свойства стекол системы Si–Te // Тез. докл. всес. конф. «Стеклообразные полупроводники». Ленинград, 1985, С. 109–110.
186. Saito Y., Sutou Y., Koike J. Electrical resistance change with crystallization in Si-Te amorphous thin films // Mater. Res. Soc. Symp. Proc. – 2010 – V. 1251.
187. Saito Y., Sutou Y., Koike J. Crystallization behavior and resistance change in eutectic $Si_{15}Te_{85}$ amorphous films // Thin Solid Films – 2012 – V. 520 – P. 2128–2131.
188. Cuttler M. Liquid semiconductors // Academic: Press. New York and London, 1977, 266 p.
189. Зинченко В. Ф., Великанов А. А., Шевчук П. П. Природа проводимости расплавов системы Si–Te–As // Тр. I Укр. респ. конф. по электрохимии. Киев. 1973. Ч. II. С. 147–153.
190. Gasser J.-G., Halim H., Wax J.-F., Vinckel J. Transport effects in a liquid IV–VI alloy: the Te–Si system // J. Non.-Cryst. Solids. – 1996. – V. 205–207, Part 1. – P. 120–125.
191. Bandyopadhyay A. K., Nalini A. V., Gopal E. S. R., Subramanyam S. V. High pressure clamp for electrical measurements up to 8 GPa and temperature down to 77 K // Rev. Sci. Instrum. – 1980. – V. 51, № 1. – P. 136–139.
192. Bogoslovskiy N. A., Tsendin K. D. Phisics of switching and memory effects in chalcogenide glassy semiconductors // Semiconductors – 2012 – V.46 – P.559–590.
193. Kolomiets B. T., Lebedev E. A., Current-voltage characteristics of a point contact with vitreous semiconductors // Radiotekh. Elektron. – 1963. – V.8, № 12 – P. 2097–2098.
194. Ovshinsky S.R. Reversible electrical switching phenomena in disordered structures// Phys. Rev. Lett. – 1968. – V. 21, № 20. – P. 1450.

195. Murthy C.N. , Ganesan V., Asokan S. Electrical switching and topological thresholds in Ge–Te and Si–Te glasses // Appl. Phys. A. – 2005. – V. 81, № 5. – P. 939–942.
196. Fernandes J., Ramesh K., Udayashankar N.K. Electrical switching in $Si_{20}Te_{80-x}Bi_x$ ($0 \le x \le 3$) chalcogenide glassy alloys // J. Non-Cristal. Solids. – 2018. – V. 483. – P. 86–93.
197. Fernandes B. J., Munga P., Ramesh K., Udayashankar N. K. Electrical switching studies of ternary $Si_{15}Te_{85-x}Bi_x$ ($0 \le x \le 2$) chalcogenide glasses // Mater. Today: Proceedings – 2018. – V. 5 – P.21292–21298.
198. Блецкан Д. И., Вакульчак В. В. Электрические и фотоэлектрические свойства кристаллического $Si_2Te_3$ и стеклообразно-го $Si_{15}Te_{85}$ // IX Международная конференции «Аморфные и микрокристаллические полупроводники». – Россия, Санкт-Петербург, 2014 г., 7–10 июля. С. 24–25.
199. Bube R. Photoelectronic properties of semiconductors. – Cambridge University Press; 1st edition – 1962. – 340 p.
200. Лашкарев В. Е., Любченко А. В., Шейнкман М. К. Неравновесные процессы в фотопроводниках. – Киев: Наук. думка, 1981. – 264 с.
201. Любченко А. В., Сальков Е. А., Сизов Ф. Ф. Физические основы полупроводниковой инфракрасной фотоэлектроники. Киев: Наук. Думка, 1984. – 256 с.
202. Гарягдыев Г., Городецкий И. Я., Любченко А. В., Нурягдыев О., Султанмурадов С. Температурная активация примесного фототока в кристаллах $Mg_xCd_{1-x}Se$ // УФЖ. – 1987. – Т. 32, № 1. – С. 137–141.
203. Власенко А. И., Власенко З. К., Любченко А. В. Спектральные характеристики фотопроводимости полупроводников с экспоненциальным краем фундаментального поглощения // ФТП. – 1999. Т. 33, № 11. – С. 1295–1299.
204. Garlick G. F. J., Gibson A. F. The electron trap mechanism of luminescence in sulfide and silicate phosphors // Proc. Phys. Soc. – 1948. – V. 60, № 6. – P. 574–590.
205. Haering R. R., Adams E. N. Theory and application of thermally stimulated currents in photoconductors // Phys. Rev. – 1960. – V. 117, № 2. – P. 451–454.
206. Блецкан Д. И., Полажинец Н. В., Чепур Д. В. Фотоэлектрические свойства кристаллического и стеклообразного $GeSe_2$ // ФТП. – 1984. – Т. 18, № 2. – С. 223–228.

207. Chen J. W., Tan C. Y., Li G. Chen L.J., Zhang H.L., Yin S.Q., Li M., Li L., Li G.H. 2D Silicon-based semiconductor $Si_2Te_3$ toward broadband photodetection // Small . – 2021. –V. 17, № 13. – P. 2006496-1–2006496-9.
208. Chen J., Bhattarai R., Cui J., Shen X., T. Hoang Anisotropic optical properties of single $Si_2Te_3$ nanoplates // Scientific Reports – 2020 – V.10, №11 – P.1–9.
209. Chen J., Wu K., Shen X., Hoang T.B., Cui J. Temperature dependent dynamics of photoexcited carriers of $Si_2Te_3$ nanowires // Mesoscale and Nanoscale Physics. – 2018. – P. 1–12.
210. Eisenmann B., Schäfer H. $Na_6Si_2Te_6$ – A new tellurohypodisilicate // Z. Naturforsch. B. –1981. – V. 36, № 12. – P. 1538–1541.
211. Eisenmann B., Schwerer H., Schafer H. Neuartige $Si_4Te_{10}(^{8-})$ und $Ge_4Te_{10}(^{8-})$ anionen im $Na_8Si_4Te_{10}$ bzw. $Na_8Ge_4Te_{10}$ // Rev. Chim. Miner. – 1983. – V. 20, № 1. – P. 78–81.
212. Sato Toshitada, Bito Yasuhiko, Murata Toshihide, Ito Shuji, Matsu-da Hiromu, Toyoguchi Yoshinori Non-aqueous electrolyte seconda-ry battery // European patent specification № EP 0 880 187 B1. 25.11.1998 Bulletin 1998/48.
213. Bletskan D., Vakulchak V., Studenyak I. Electronic structure of $Na_6Si_2Te_6$ // Proc. VIII Int. Seminar "Properties of ferroelectric and superionic systems". – Ukraine, Uzhgorod, 29–30 october, 2019. – P. 41–44.
214. Eisenmann B., Schäfer H. $K_4Si_4Te_{10}$, das erste tellurosilicat mit adamantan-analogen $Si_4Te_{10}^{4-}$-anionen // Z. Anorg. Allg. Chem. – 1982. – V. 491, № 1. – P. 67–72.
215. Ribes M., Olivier-Fourcade J., Philippot E., Maurin M. Etude structuralle de thiocomposes a groupement anionique de type tetrane $Na_4X_4S_{10}$ (X = Ge, Si) et $Ba_2Ge_4S_{10}$ // J. Solid State Chem. – 1973. – V. 8, № 3. – P. 195–205.
216. Dittmar G. $K_6[Si_2Te_6]$–Synthesis and structure of the first telluro-disilicate // Angew. Chem. – 1977. – V.16, № 8. – P. 554–554.
217. Dittmar G. Die Kristallstruktur des Hexakaliumhexatellurodisilicats, $K_6[Si_2Te_6]$ // Acta Cryst. – 1978. – B. 34. – P. 2390–2393.
218. Dittmar G. Die Kristallstrukturen von $K_6[Ge_2Te_6]$ und $K_6[Sn_2Te_6]$ und ihre kristall-chemische Beziehung zum $K_6[Si_2Te_6]$-Typ // Z. anorg. allg. Chem. – 1979. – V. 453, №1. – P. 68–78.
219. Dogguy M., Rivet J., Flahaut J. Description du systeme ternaire Cu–

Si–Te // J. Less-Com. Metals. – 1979. – V. 63, № 2. – P. 129–145.
220. Rivet J., Flahaut J., Laruelle P. Sur un groupe de composes ternaires a structure tetraedrique // C. R. Acad. Sci.– 1963.– V. 257, № 1.– P. 161–164.
221. Rivet J., Gorochov O. , Flahaut J. Etude des properties electriques des composes de formule generale $A_2^{I}B^{IV}X_3^{VI}$ dans laquelle $A^{I}$ = Cu, $B^{IV}$ = Si, Ge ou Sn et $X^{VI}$ = S, Se ou Te // C. R. Acad. Sci. Paris –1965.– V. 260.– P. 178–181.
222. Rivet J. Contribution a l'etude de quelques combinaisons ternaires sulfurees, seleniees ou tellurees du cuivreavec les elements du groupe IV // Ann. Chim.– 1965.– V. 10, № 5–6. – P.243–270.
223. Hahn H., Klingen W., Ness P., Schulze H. Ternäre Chalkogenide mit Silicium, Germanium und Zinn // Naturwissenschaften – 1966. – V.53, №1 – P. 18.
224. Gorochov O. Les composes $Ag_8MX_6$ (M – Si, Ge, Sn et X – S, Se, Te) // Bull. Soc. Chim. France. –1968. – № 6, – P. 2263–2268.
225. Gorochov O., Flahaut J. Les composes $Ag_8MX_6$ avec M = Si, Ge, Sn et X – S, Se, Te // C. R. Acad. Sci. Paris. – 1967. – V. 264, № 26. – P. 2153–2155.
226. Pistorius C.W.F.T., Gorochov O. Polymorphism and stability of the semiconducting series $Ag_8MX_6$ (M = Si, Ge, Sn, and X = S, Se, Te) to high pressures // High Temp. – High Pressures. – 1970. – V. 2, № 1. P. 31–42.
227. Boucher F., Evain M., Brec R. Single-crystal structure determination of $\gamma$-$Ag_8SiTe_6$ and powder X-ray study of low-temperature $\alpha$ and $\beta$ phases // J. Solid State Chem. – 1992. – V. 100, № 2. – P. 341–355.
228. Charoenphakdee A., Kurosaki K., Muta H., Uno M., Yamanaka S. $Ag_8SiTe_6$: A New Thermoelectric Material with Low Thermal Conductivity // Jap. J. Appl. Phys. – 2009. – V. 48, №1 – P. 011603-1–011603-3.
229. Rysanek N., Laruelle P., Katty A. Structure cristalline de $Ag_8GeTe_6(\gamma)$ // Acta Cryst. B. – 1976. – V. 32, № 3. – P. 692–696.
230. Geller S. The crystal structure of $\gamma$-$Ag_8GeTe_6$, a potential mixed electronic-ionic conductor // Z. Kristallogr. - Crystalline Materials. – 1979. – V. 149, № 1-4. – P. 31–48.
231. Krebs B., Mandt J. Zur Kenntnis des Argyrodit-Strukturtyps: Die Kristallstruktur von $Ag_8SiS_6$ // Z. Naturforsch. B. – 1977. – V. 32, № 4. – P. 373–379.
232. Chen J., Li L., Gong P. et al. A submicrosecond-response ultraviolet –visible–near-infrared broadband photodetector based on 2D tel-

lurosilicate $InSiTe_3$ // ACS Nano. – 2022. – V. 16. – P. 7745–7754.
233. Suriwong, T., Kurosaki, K., Thongtem, S. Thermoelectric properties of phosphorus-doped indium tellurosilicate: $InSiTe_3$ // J. Alloys Compd. – 2018. – V. 735. – P. 75–80.
234. Sandre, E., Carteaux, V., Marie, A. M., & Ouvrard, G.. Structural determination of a new lamellar tellurosilicate, $AlSiTe_3$ // J. Alloys Compd. – 1994. – V. 204, № 1-2. – P. 145–149.
235. Sandre E., Carteaux V., Ouvrard, G. Un nouveau tellurosilicate lamellaire $InSiTe_3$ // C. R. Acad. Sci. Paris. – 1992. – V. 314. P. 1151–1156.
236. Блецкан Д. І., Вакульчак В. В., Малець О. О. Електронна структура і оптичні функції $AlSiTe_3$ // The 8th International scientific and practical conference "Modern research in world science", October 29-31, 2022, Lviv, Ukraine. P. 442–448.
237. Блецкан Д.І., Вакульчак В.В., Гапак А.І., Кабацій В.М. Електронна структура $InSiTe_3$ // IX Українська наукова конференція з фізики напівпровідників. – Україна, Ужгород – 2023 – С. 286–287.
238. Зейман Дж. Принципы теории твердого тела. М.: Мир. 1974, 478с.
239. Debbichi L., Kim H., Björkman T., Eriksson O., Lebègue S. First-principles investigation of two-dimensional trichalcogenide and sesquichalcogenide monolayers // Phys. Rev. B. – 2016. – V. 93. – P. 245307-1–245307-6.
240. B. Legendre, C. Souleau, C. Hancheng, N. Rodier, The ternary system gold-silicon-tellurium; a contribution to the study of the binary system silicon-tellurium and gold-silicon, and the structure of $Si_2Te_3$, J. Chem. Res. Synop. – 1978 – №.5 – P.165 – 169.
241. Korkmaz M. A., Deligoz E., Ozisik H. Strong Elastic Anisotropy of Low-Dimensional Ternary Compounds: $InXTe_3$ (X = Si, Ge) // J. Electron. Mater. – 2021. – V. 50. – P. 2779–2788.
242. Güler E., Ugur S., Güler M., Ugur G. Unveiling the electronic, optical, and thermoelectrical properties of bulk and monolayer $AlSiTe_3$ by first principles // Chem. Phys. – 2023. – V.575. – P.112068-1–112068-7.
243. Уэллс А. Структурная неорганическая химия. М.:Мир. 1987. – 408 с.
244. Kasper, J.S., Hagenmul.P, Pouchard, M., Cros, C. Clathrate structure of silicon $Na_8Si_{46}$ and $Na_xSi_{136}$ ($x$ = 11) // Science – 1965. – V.150, № 3704. – P. 1713–1714.

245. Cros C., Pouchard M., Hagenmuller P. Sur une nouvelle famille de clathrates miucraux isotypes des hydrates de gaz et de liquides. Iuterpretation des rhltats obtenus // J. Solid State Chem. – 1970 – V. 2, – P. 570–581.
246. Yamanaka S., Enishi E., Fukuoka H., Yasukawa M.. High-pressure synthesis of a new silicon clathrate superconductor, $Ba_8Si_{46}$ // Inorg. Chem., – 1999 – V. 39, – P. 56–58.
247. Tse J.S., Uehara K., Rousseau R., Ker A., Ratcliffe C.I., White M.A., MacKay G. Structural principles and amorphouslike thermal conductivity of Na-doped Si clathrates // Phys. Rev. Lett., – 2000 – V. 85, – P. 114–117.
248. H. Kawaji, H. Horie, S. Yamanaka, M. Ishikawa. Superconductivity in the silicon clathrate compound $(Na, Ba)_xSi_{46}$ // Phys. Rev. Lett., – 1995 – V. 74, № 8 – P. 1427–1429.
249. Morito H., Yamada T., Ikeda T., Yamane H., Na-Si binary phase diagram and solution growth of silicon crystals // J. Alloys Compd. – 2009 – V. 480 – P. 723–726.
250. Ramachandran G. K., Dong J., Diefenbacher J., Gryko J., Marzke R. F.,. Sankey O. F, McMillan P. F. Synthesis and X-ray characterization of silicon clathrates // J. Solid State Chem. – 1999 – V. 145 – P. 716–730.
251. Blosser M.C., Nolas G.S. Synthesis of $Na_8Si_{46}$ and $Na_{24}Si_{136}$ by oxidation of $Na_4Si_4$ from ionic liquid decomposition // Materials Letters – 2013 – V. 99 – P. 161–163.
252. Dopilka,1 A. Childs, S. Bobev, C. K. Chan Solid-state electrochemical synthesis of silicon clathrates using a sodium-sulfur battery inspired approach // J. Electrochemical Society, – 2021 – V. 168 – P. 020516-1–020516-7.
253. Courac A., Le Godec Y., Renero-Lecuna C., Moutaabbid H., Kumar R., Coelho-Diogo C., Gervais C., Portehault D. High-pressure melting curve of zintl sodium silicide $Na_4Si_4$ by in situ electrical measurements // Inorg. Chem. – 2019 – V. 58, № 16 – P. 10822–10828.
254. Song Y., Gymez-Recio I., Kumar R., Coelho Diogo C., Casale S., Genois I., Portehault D. A straightforward approach to high purity sodium silicide $Na_4Si_4$ // Dalton Transactions – 2021 – V. 50, № 45 – P. 16703–16710.
255. Morito H., Momma K., Yamane H. Crystal structure analysis of $Na_4Si_{4-x}Ge_x$ by single crystal X-ray diffraction // J. Alloys Compound. – 2015 – V. 623 – P. 473–479.
256. Goebel T., Prots Y., Haarmann F. Refinement of the crystal structure

of tetrasodium tetrasilicide, $Na_4Si_4$ // Z. Kristallogr. NCS – 2008 – V. 223 – P. 187–188.
257. Horie H., Kikudome T., Teramura K., Yamanaka S. Controlled thermal decomposition of NaSi to derive silicon clathrate compounds // J. Solid State Chem. – 2009 – V. 182 – P. 129–135.
258. Reny E., Gravereau P., Cros C., Pouchard M. Structural characterisations of the $Na_xSi_{136}$ and $Na_8Si_{46}$ silicon clathrates using the Rietveld method // J. Mater. Chem. – 1998 – V. 8 – P. 2839–2844.
259. Ammar A., Cros C., Pouchard M., Jaussaud N., Bassat J., Villeneuve G., Duttine M., Ménétrier M., Reny E. On the clathrate form of elemental silicon, $Si_{136}$: preparation and characterisation of $Na_xSi_{136}$ ($x \to 0$) // Solid State Sciences – 2004 – V. 6 – P. 393–400.
260. Vollondat R., Roques S., Chevalier C., Bartringer J., Rehspringerc J., Slaoui A., Fix T. Synthesis and characterization of silicon clathrates of type I $Na_8Si_{46}$ and type II $Na_xSi_{136}$ by thermal decomposition // J. Alloys Comp. – 2022 – V. 903 – P. 163967-1–163967-7.
261. Stefanoski S., Beekman M., Wong-Ng W., Zavalij P., Nolas G. S. Simple approach for selective crystal growth of intermetallic clathrates // Chem. Mater. – 2011 – V. 23 – P. 1491–1495.
262. Kurakevych, O.O., Strobel, T.A., Kim, D.Y., Muramatsu, T., Struzhkin, V.V. Na–Si clathrates are high-pressure phases: a melt-based route to control stoichiometry and properties.// Cryst. Grow. Des. – 2013. – V.13, № 1. – P. 303–307.
263. Jouini Z., Kurakevych O. O., Moutaabbid H., Godec Y. Le, Mezouar M., Guignot N. Phase boundary between Na–Si clathrates of structures I and II at high pressures and high temperatures // J. Superhard Mat., – 2016 – V. 38, №. 1, P. 66–70.
264. Gryko J., McMillan P. F. Marzke R. F., Ramachandran G. K., Patton D., Deb S. K. Sankey O. F. Low-density framework form of crystalline silicon with a wide optical band gap // Phys. Rev. B – 2000 – V. 62, № 12. – P. 7707–7710.
265. Morito H., Shimoda M., Yamane H. Single crystal growth of type I Na–Si clathrate by using Na–Sn flux // J. Cryst. Growth – 2016 – V. 450 – P. 164–167.
266. Morito H., Shimoda M., Yamane H., Fujiwara K. Crystal growth conditions of type I and II Na-Si clathrates by evaporation of Na from a Na-Si-Sn solution // Cryst. Growth Des. – 2017 – V. 18 – P. 351–355.
267. Morito H., Yamane H., Umetsu R. Y. and Fujiwara K. Seeded growth of type-II $Na_{24}Si_{136}$ clathrate single crystals // Crystals –

2021. – V. 11, P. 808-1 – 808-6.
268. Morito H., Futami K., Fujiwara K. Effect of Na–Sn flux on the growth of type-I $Na_8Si_{46}$ clathrate crystals // Crystals – 2022. – V. 12, P.837-1 – 837-8.
269. Cros C., Pouchard M. Sur les phases de type clathrate du silicium et des e.le.ments apparente.s (C, Ge, Sn) : Une approche historique // C. R. Chimie – 2009. – V. 12. – P. 1014–1056.
270. Немошкаленко В. В., Алешин В. Г. Электронная спектроскопия кристаллов // Киев. Наук. думка, 1983, 288с.
271. Macfarlane G. G., McLean T. P., Quarrington J. E., Roberts V. Fine structure in the absorption-edge spectrum of Si // Phys. Rev. – 1958 – V. 111, № 5 – P. 1245–1254.
272. Bletskan D., Vakulchak V., Hapak A. Electronic structure of silicon clathrate $Na_8Si_{46}$ // XI International seminar «Properties of ferroelectric and superionic system» – Ukraine, Uzhgorod, 28 october, 2022. – P. 88–90.
273. Saito S., Oshiyama A. Electronic structure of $Si_{46}$ and $Na_2Ba_6Si_{46}$ // Phys. Rev. B. – 1995 – V. 51, № 4 – P. 2628-2631.
274. Mahammedi N.A., Ferhat M., Tsumuraya T., Chikyow T. Prediction of optically-active transitions in type-VIII guest-free silicon clathrate $Si_{46}$: A comparative study of its physical properties with type-I counterpart through first-principles // J. Appl. Phys. – 2017. – V. 122,№ 20. – P. 205103-1–205103-11.
275. Moriguchi K., Yonemura M., Shintani A. Electronic structures of $Na_8Si_{46}$ and $Ba_8Si_{46}$ // Phys. Rev. B – 2000. – V. 61, № 15 – P. 9859-9862.
276. Koelling D. D., Arbman G. O. Use of energy derivative of the radial solution in an augmented plane wave method: application to copper // J. Phys. F: Met. Phys. – 1975. – V. 5 – P. 2041-2054.
277. Imai Y., Watanabe A. Chemical trends of the band gaps in semiconducting silicon clathrates // Physics Procedia – 2011. – V. 11 – P. 59–62.
278. Yang Y.W. Coppens P. On the experimental electron distribution in silicon // Solid State Comm. – 1974. – V. 15, № 9 – P. 1555-1559.
279. Ley L., Kowalczyk S., Pollak R., Shirley D. A. X-Ray photoemission spectra of crystalline and amorphous Si and Ge valence bands // Phys. Rev. Lett. – 1972. – V. 29, № 16 – P. 1088 – 1092.
280. Майзель А., Леонхардт Г., Сарган Р. Рентгеновские спектры и химическая связь // Киев «Наукова думка» 1981 – 420 с.
281. Simunek A., Polcik M., Wiech G. Si K, Si L, and Cr K X-ray va-

lence-band studies of bonding in chromium silicides: Experiment and theory // Phys. Rev. B. – 1995. – V. 52, № 16 – P. 11865–11871
282. Terauchi M., Sato Y. Chemical state analyses by soft X-ray emission spectroscopy // Jeol News – 2018. – V. 53, №. 1 – P. 1–6.
283. Moewes A., Kurmaev E. Z., Tse J. S., Geshi M., Ferguson M. J., Trofimova V. A., Yarmoshenko Y. M. Electronic structure of alkali-metal-doped $M_8Si_{46}$ (M = Na,K) clathrates // Phys. Rev. B. – 2002. – V. 65, – P. 153106-1–153106-3.
284. Jaussaud N., Pouchard M., Goglio G., Cros C., Ammar A., Weill F., Gravereau P. High pressure synthesis and structure of a novel clathrate-type compound: $Te_{7+x}Si_{20-x}$ ($x \sim 2.5$) // Solid State Sci. – 2003. – V. 5, № 9. – P. 1193–1200.
285. Jaussaud N., Toulemonde P., Pouchard M., San Miguel A., Gravereau P., Pechev S., Goglio G., Cros C. High pressure synthesis and crystal structure of two forms of a new tellurium–silicon clathrate related to the classical type I // Solid State Sci. – 2004. – V. 6,№ 5. – P. 401–411.
286. Jaussaud N., Pouchard M., Gravereau P., Pechev S., Goglio G., Cros C., San Miguel A., Toulemonde P. Structural trends and chemical bonding in Te-doped silicon clathrates // Inorg. Chem. – 2005. – V. 44, № 7. – P. 2210–2214.

Наукове видання

**БЛЕЦКАН Дмитро Іванович**

# КРИСТАЛІЧНІ ТА СКЛОПОДІБНІ ТЕЛУРИДИ КРЕМНІЮ

Синтез, властивості, використання

Монографія

У авторській редакції

Дизайн обкладинки Куклишин Василини